\documentclass[10pt]{wlscirep}
\pdfoutput=1

\usepackage[english]{babel}
\usepackage{authblk}

\usepackage[utf8]{inputenc}
\usepackage[T1]{fontenc}
\usepackage{lineno}
\usepackage{float}
\usepackage{slashed}
\usepackage{rotating}
\usepackage{amsmath,slashed,xcolor,cancel}
\usepackage{array,booktabs,colortbl,multirow,tablefootnote}
\usepackage{makecell}
\usepackage{chngcntr}
\usepackage{relsize}
\usepackage{colortbl,colordvi,color}
\usepackage{subcaption}
\usepackage{todonotes}
\usepackage{textcomp}

\usepackage{eurosym}
\usepackage{amssymb}
\usepackage{graphicx}
\usepackage{wrapfig,mdframed,caption}
\usepackage{comment}
\usepackage{overpic} 
\usepackage{xspace} 

\def\lsim{\mathrel{\rlap{\lower3pt\hbox{\hskip0pt$\sim$}}
   \raise1pt\hbox{$<$}}}         
\def\gsim{\mathrel{\rlap{\lower4pt\hbox{\hskip1pt$\sim$}}
   \raise1pt\hbox{$>$}}}         

\newcommand{\bea}{\begin{eqnarray}}
\newcommand{\eea}{\end{eqnarray}}

\newcommand{\figref}[1]{Fig.~\ref{#1}}
\newcommand{\vphi}{\varphi}
\newcommand{\eps}{\varepsilon}
\newcommand{\vkap}{\varkappa}
\newcommand{\tri}{\pitchfork}
\newcommand{\trm}[1]{\textrm{#1}}
\newcommand{\tsf}[1]{\textsf{#1}}
\newcommand{\mbf}[1]{\mathbf{#1}}
\newcommand{\tbf}[1]{\textbf{#1}}

\usepackage[multidot]{grffile} 

\newcommand{\units}{\,\mathrm} 
\newcommand{\cer}{Cherenkov\xspace } %

\mathchardef\mhy="2D   
\DeclareMathOperator{\Ai}{Ai}

\definecolor{bk1}{RGB}{0,0,153}
\newcommand{\omegaL}{\omega_{L}}
\newcommand{\cL}{\mathcal{L}}
\newcommand{\ess}{u}
\newcommand{\tea}{v}
\newcommand{\coff}{C}
\newcommand{\scal}{\phi^{S}}
\newcommand{\scali}{S}
\newcommand{\alp}{\phi^{P}}
\newcommand{\alpi}{P}
\newcommand{\mcp}{\psi_{\delta}}

\newcommand{\gammaC}{\gamma_{\rm C}}
\definecolor{cccc}{rgb}{0.4, 0.8, 0.3}

\newcommand{\GeV}{\mathrm{GeV}}
\newcommand{\cO}{\mathcal{O}}
\newcommand{\MeV}{\mathrm{MeV}}

\newcommand{\ximax}{\xi_\textrm{nom}}
\newcommand{\xinom}{\xi_\textrm{nom}}

\newcommand{\ximean}{\xi_\textrm{avg}}
\newcommand{\elaser}{$e$-laser\xspace} 
\newcommand{\glaser}{$\gamma$-laser\xspace} 
 
\newcommand{\euxfel}{XFEL.EU\xspace}
\newcommand{\beampipe}{beampipe\xspace}
\newcommand{\beamline}{beamline\xspace}
\newcommand{\geant}{{\sc Geant4}\xspace}
\newcommand{\hiqed}{SFQED\xspace}

\newcommand{\phaseone}{phase-0\xspace}
\newcommand{\phasetwo}{phase-1\xspace}
\newcommand{\nposi}{N_\textrm{pos}}
\newcommand{\ipstrong}{\textsc{IPstrong}}
\newcommand{\zalp}{z_\textrm{ALP}}
\newcommand{\taualp}{\tau_\textrm{ALP}}

\newcommand{\ecrit}{{\mathcal E}_\textrm{cr}}
\newcommand{\enlaser}{{\mathcal E}_\textrm{L}}
\newcommand{\lambdabar}{{\mkern0.75mu\mathchar '26\mkern -9.75mu\lambda}}

\newcommand{\prob}{\tsf{P}}
\newcommand{\rate}{\tsf{R}}

\newcolumntype{L}{>{\arraybackslash}m{6cm}}

\newcommand\tstrut{\rule{0pt}{5ex}}       
\newcommand\bstrut{\rule[-3ex]{0pt}{0pt}} 

\counterwithin{figure}{section}
\counterwithin{table}{section}
\date{}

\title{\bf Conceptual Design Report for the LUXE Experiment}

\author[1]{H. Abramowicz}
\author[2,3]{U.~Acosta}
\author[4]{M. Altarelli}
\author[5]{R.~A{\ss}mann}
\author[6,7]{Z.~Bai}
\author[5]{T.~Behnke}
\author[1]{Y.~Benhammou}
\author[8]{T.~Blackburn}
\author[9]{S.~Boogert}
\author[5]{O.~Borysov}
\author[5,10]{M.~Borysova}
\author[5]{R.~Brinkmann}
\author[11]{M.~Bruschi}
\author[5]{F.~Burkart}
\author[5]{K.~B{\"u}{\ss}er}
\author[12]{N.~Cavanagh}
\author[6]{O.~Davidi}
\author[5]{W.~Decking}
\author[13]{U.~Dosselli}
\author[3]{N.~Elkina}
\author[14]{A.~Fedotov}
\author[15]{M.~Firlej}
\author[15]{T.~Fiutowski}
\author[12]{K.~Fleck}
\author[16]{M.~Gostkin}
\author[5]{C.~Grojean\thanks{also at HU Berlin}}
\author[5,17]{J.~Hallford}
\author[18,19] {H.~Harsh}
\author[17]{A.~Hartin}
\author[5,20]{B.~Heinemann\thanks{corresponding author: beate.heinemann@desy.de}}
\author[21]{T.~Heinzl}
\author[5]{L.~Helary}
\author[5,20]{M.~Hoffmann}
\author[1]{S.~Huang}
\author[5,18,20]{X.~Huang}
\author[15]{M.~Idzik}
\author[21]{A.~Ilderton}
\author[5]{R.~Jacobs}
\author[2,3]{B.~K{\"a}mpfer}
\author[21]{B.~King} 
\author[10]{H.~Lahno}
\author[1]{A.~Levanon}
\author[1]{A.~Levy}
\author[22]{I.~Levy}
\author[5]{J.~List}
\author[5]{W.~Lohmann\thanks{also at BTU Cottbus und RWTH Aachen}}
\author[23]{T.~Ma}
\author[21]{A.J.~Macleod}
\author[6]{V. Malka}
\author[5]{F.~Meloni}
\author[14]{A.~Mironov}
\author[13]{M.~Morandin}
\author[15]{J.~Moron}
\author[5]{E.~Negodin}
\author[6]{G.~Perez}
\author[1]{I.~Pomerantz}
\author[24]{R.P{\"o}schl}
\author[5]{R.~Prasad}
\author[25]{F.~Qu\'er\'e}
\author[5]{A.~Ringwald}
\author[26]{C.~R\"odel}
\author[27]{S.~Rykovanov}
\author[18,19] {F. Salgado}
\author[6]{A.~Santra}
\author[12]{G.~Sarri}
\author[18]{A.~S{\"a}vert}
\author[28]{A.~Sbrizzi\thanks{also at University of Udine, Udine, Italy}}
\author[5]{S.~Schmitt}
\author[2,3]{U.~Schramm}
\author[5]{S.~Schuwalow}
\author[18]{D.~Seipt}
\author[29]{L.~Shaimerdenova}
\author[5]{M.~Shchedrolosiev}
\author[29]{M.~Skakunov}
\author[23]{Y.~Soreq}
\author[12]{M.~Streeter}
\author[15]{K.~Swientek}
\author[6]{N.~Tal~Hod}
\author[21]{S.~Tang}
\author[18,19]{T.~Teter}
\author[5]{D.~Thoden}
\author[16]{A.I.~Titov}
\author[29]{O.~Tolbanov}
\author[3]{G.~Torgrimsson}
\author[29]{A.~Tyazhev}
\author[5,17]{M.~Wing}
\author[13]{M.~Zanetti}
\author[29]{A.~Zarubin}
\author[3]{K.~Zeil}
\author[18,19]{M.~Zepf}
\author[16]{A.~Zhemchukov}

\affil[1]{Tel Aviv University, Tel Aviv, 6997801, Israel}
\affil[2]{TU~Dresden, 01062 Dresden, Germany}
\affil[3]{Helmholtz-Zentrum Dresden-Rossendorf, 01328 Dresden, Germany}
\affil[4]{Max Planck Institute for Structure and Dynamics of Matter, 22761 Hamburg, Germany}
\affil[5]{Deutsches Elektronen-Synchrotron (DESY), 22607 Hamburg, Germany}
\affil[6]{Weizmann Institute of Science, Rehovot, 7610001, Israel}
\affil[7]{Department of Physics, Southern University of Science and Technology, Shenzhen 518055, China}
\affil[8]{University of Gothenburg, SE-41296 Gothenburg, Sweden}
\affil[9]{John Adams Institute at Royal Holloway, Department of Physics, Royal Holloway, Egham, TW20 0EX, Surrey, UK}
\affil[10]{Institute for Nuclear Research NASU (KINR), Kyiv, 03680, Ukraine}
\affil[11]{INFN and University of Bologna, Bologna, Italy}
\affil[12]{School of Mathematics and Physics, The Queen’s University of Belfast, Belfast, BT7 1NN, UK}
\affil[13]{INFN and University of Padova, Padova, Italy}
\affil[14]{National Research Nuclear University MEPhI, Kashirskoe sh. 31, Moscow, 115409, Russia}
\affil[15]{Faculty of Physics and Applied Computer Science, AGH University of Science and Technology,  Krakow, Poland} 
\affil[16]{Joint Institute for Nuclear Research ( JINR), Dubna 141980, Russia}
\affil[17]{University College London, London, WC1E 6BT, UK}
\affil[18]{Helmholtz Institut Jena,  07743 Jena, Germany}
\affil[19]{Friedrich Schiller Universit{\"a}t Jena, 07743 Jena, Germany}
\affil[20]{Albert-Ludwigs-Universit{\"a}t Freiburg, 79085 Freiburg, Germany}
\affil[21]{University of Plymouth, Plymouth, PL4 8AA, UK}
\affil[22]{Department of Physics, Nuclear Research Centre-Negev, P.O. Box 9001, Beer Sheva 84190, Israel}
\affil[23]{Physics Department, Technion—Israel Institute of Technology, Haifa 3200003, Israel}
\affil[24]{Universit{\'e} Paris-Saclay, CNRS/IN2P3, IJCLab, 91405 Orsay, France}
\affil[25]{LIDYL, CEA, CNRS, Universit\'e Paris-Saclay, CEA Saclay, 91 191 Gif-sur-Yvette, France}
\affil[26]{Institute of Nuclear Physics, TU Darmstadt, 64289 Darmstadt, Germany}
\affil[27]{Skolkovo Institute of Science and Technology (Skoltech), Moscow 121205, Russia}
\affil[28]{INFN Trieste, Trieste, Italy}
\affil[29]{National Research Tomsk State University NI TSU, TSU, Russia}

\begin{document}
\maketitle

\begin{abstract}
This Conceptual Design Report describes LUXE (Laser Und XFEL Experiment), an experimental campaign that aims to combine the high-quality and high-energy electron beam of the European XFEL with a powerful laser to explore the uncharted terrain of quantum electrodynamics characterised by both high energy and high intensity. We will reach this hitherto inaccessible regime of quantum physics by analysing high-energy electron-photon and photon-photon interactions in the extreme environment provided by an intense laser focus. The physics background and its relevance are presented in the science case which in turn leads to, and justifies, the ensuing plan for all aspects of the experiment: Our choice of experimental parameters allows (i) field strengths to be probed where the coupling to charges becomes non-perturbative and
(ii) a precision to be achieved that permits a detailed comparison of the measured data with calculations. In addition, the high photon flux predicted will enable a sensitive search for new physics beyond the Standard Model. The initial phase of the experiment will employ an existing 40 TW laser, whereas the second phase will utilise an upgraded laser power of 350 TW. All expectations regarding the performance of the experimental set-up as well as the expected physics results are based on detailed numerical simulations throughout.
\end{abstract}
\thispagestyle{empty}
\newpage
\tableofcontents
\newpage
\section{Introduction}
\label{sec:intro}

The main physics goals of the LUXE (Laser Und XFEL Experiment) experiment are to perform precision measurements to investigate the transition into the non-perturbative regime of quantum electrodynamics (QED), and to search for new particles beyond the Standard Model coupling to photons. In particular, LUXE will 
\begin{itemize}
    \item measure the interactions of real photons with electrons and photons at field strengths where the coupling to charges becomes non-perturbative;
    \item make precision measurements of electron-photon and photon-photon interactions in a transition from the perturbative to the non-perturbative regime of QED;
    \item use strong-field QED processes to design a sensitive search for new particles beyond the Standard Model that couple to photons. 
\end{itemize}
This is achieved by using the high-quality European XFEL electron-beam and a high-power laser, with elaborate diagnostics as well as a powerful detection system.

QED -- the theory, first formulated in 1927, describing the interactions of light and matter in a framework encompassing both quantum mechanics and special relativity -- is the most precisely
tested theory in modern physics. In particular, the precision measurements of the electron's anomalous magnetic moment and the
fine structure constant~\cite{Hanneke:2010au} have been predicted by
QED~\cite{Aoyama:2019ryr} with an accuracy of around 1 part in a billion. 
These past successes have tested the perturbative regime of QED, where phenomena can be accurately described with perturbative methods due to the small value of the fine-structure constant $\alpha$.

The Schwinger limit, $\ecrit=m_e^2c^3/(e\hslash) = 1.32 \cdot 10^{18}$ V/m, is the QED field scale occurring in the rates of a number of novel phenomena.

Of particular relevance are strong-field vacuum polarisation effects leading to pronounced non-linearities in the optical properties of the vacuum \cite{Heisenberg:1935qt}, both dispersive (e.g. vacuum birefringence) and absorptive. The most prominent absorptive phenomenon is the production of electron-positron pairs by field-induced tunnelling out of the vacuum, aka Schwinger pair production. In the weak field regime, i.e. for field strengths well below $\ecrit$, this process is exponentially suppressed and thus basically unmeasurable. Only in the genuine strong-field regime may Schwinger pair production be observed, and this would represent a landmark achievement. 
Experiments, where the Schwinger limit is reached and exceeded, are becoming possible just now, opening up
the regime of strong-field QED to experimental investigation. In recent years, a clean and accessible source of ultra-strong electromagnetic fields has become available in the form of high-intensity lasers. Continuous technological progress over several decades, culminating in the development of ''chirped pulse amplification" (Nobel Prize 2018 \cite{Strickland:1985gxr}) has led to focal intensities beyond $10^{21}$~W/cm$^2$. This corresponds to electric and magnetic field strengths of about $10^{14}$~V/m and $10^5$~T, respectively, which will be typical values for the LUXE experiment in the laboratory frame. 

The field strength value in the laboratory frame, $\enlaser$, accessible today is about more than three orders of magnitude below the critical field given by $\ecrit$. However, one may utilise the laws of relativity (Lorentz invariance) to further boost the field magnitude. A highly energetic electron with relativistic gamma factor $\gamma_e \gg 1$ and collision angle $\theta$ with respect to the laser beam propagation direction will `see' a Lorentz boosted electric field $E_* = \gamma_e \enlaser (1 + \cos \theta)$ in its own rest frame. Thus, with $\gamma_e \simeq 10^4$ corresponding to electron energies $\epsilon_e$ of the order of 10 GeV, the laser field strength in the electron rest frame will be of the order of the critical value, $\ecrit$, for near head-on collisions. This is the physical scenario LUXE is going to realise (with $\epsilon_e \leq 16.5$ GeV and a collision angle of $\theta = 17.2^\circ$). In the following, when we discuss the "field" we always refer to the field in the rest frame of the electron.

This regime, often called strong-field QED (SFQED), is
relevant for several of recently observed astrophysical phenomena
such as the gravitational collapse of Black Holes~\cite{Ruffini:2009hg}, the
propagation of cosmic rays~\cite{Nishikov}, and the surface of magnetars (strongly magnetised neutron stars)~\cite{Kouveliotou:1998ze,Harding_2006,Turolla:2015mwa,Kaspi:2017}. They are
also expected to be present at future linear high energy lepton colliders~
\cite{Yakimenko:2018kih,Bucksbaum:2020} and in atoms with an atomic number
$Z>137$ \cite{pomeranchuk}. 
Moreover, strong fields -- albeit at a scale defined by the ionisation potentials and the Bohr radius -- are of interest in atomic and molecular physics, and the community has been actively investigating this regime, see e.g. Ref.~\cite{ivanov}.

The basic experimental concept of LUXE, as proposed here, is to collide either the electrons directly with the high-power, tightly focused laser beam or to convert the electron beam to a high-energy photon beam ($\gamma$) and then collide these with the laser beam.
In the 1990s, at SLAC in Stanford, the former was conducted in an experiment called E144~\cite{Bula:1996st,Burke:1997ew} using the SPC accelerator. The field achieved in the rest frame of the electron was a factor four below the Schwinger field as the optical lasers available at
that time were about three orders of magnitude less intense than those available today. Nonetheless, this
pioneering experiment explored both non-linear Compton scattering,
\begin{equation}
    e^-+n\gamma_L \to e^- + \gamma ,
\end{equation}
where $n$ is the number of laser photons, $\gamma_L$, participating in the process, and the "two-step trident" process where the photon from the Compton process interacts subsequently with $n'$ laser photons and an $e^+e^-$ pair is produced via the Breit-Wheeler process,
\begin{equation}
\label{eq:trident}
    e^-+n\gamma_L \to e^- + \gamma \ \ \mbox{ and } \ \  \gamma + n'\gamma_L \to e^+e^- .
\end{equation}
In addition to the two-step trident process, 
it can also proceed in one step, $e^-+n\gamma_L \to e^- e^+ e^-$, but 
has a much lower rate and is expected to be negligible for the experiment proposed here, see Sec.~\ref{sec:science}. Experimentally, however, only the total trident production (including both processes and their interference) can be measured. Since then, only two other experimental campaigns in a similar configuration have been performed, using laser-wakefield accelerated electrons with a maximum energy of about 2 GeV with a focused laser with peak electric field $\enlaser\approx 7\times 10^{14}$~V/m~\cite{poder.prx.2018,cole.prx.2018}. Even though quantum effects of radiation reaction in the electron dynamics were observed, the limited statistics available and the significant shot-to-shot fluctuations in the electron and laser parameters prevented a systematic study of the phenomenon and from measuring Breit-Wheeler pair production. 

A key dimensionless parameter characterising such interactions is the intensity of the laser field, $\xi$, defined as $\xi = \frac{e\enlaser}{m_e\omegaL}$
where $\omegaL$ is its frequency of the laser. The region $\xi \ll 1$ corresponds to the perturbative regime; as $\xi$ approaches unity, more and more higher-order terms (eventually all of them), need
to be retained in the perturbation expansion. When $\xi>1$, the truncation causes problems. The other
key parameter is the quantum parameter $\chi_e$. For electron-laser interactions, it corresponds to the ratio of the laser field to the critical field in the rest frame of an electron of energy,
$\chi_e = \frac{E_*}{\ecrit}$. Another useful dimensionless parameter is the energy parameter $\eta=\chi_e/\xi$.

In the E144 experiment, values of $\xi =0.4$ and $\chi_e = 0.25$ were attained for the trident process, still within the perturbative regime, but with observable non-linear effects in the laser electric field. In order to observe the qualitatively different behaviour associated with the strong field it is necessary to achieve values of $\xi \gg 1$ and $\chi_e\sim 1$. With LUXE this regime will be reached for the first time and measurements of the Compton and trident processes will be made.

Colliding high-energy photons as proposed in Ref.~\cite{Hartin:2018sha} enables the Breit-Wheeler process:
\begin{equation}
    \gamma + n\gamma_L \to e^+e^-,
\end{equation}
to be studied for the first time directly using high-energy photons, $\gamma$, produced either via bremsstrahlung or via inverse Compton scattering (ICS) at macroscopic distances away from the strong-field interaction.  In this process one directly converts ``light into matter'' since only real photons are entering the interaction region, as opposed to the two-step process of Eq.~\eqref{eq:trident}.

In the tunneling limit, $\xi \gtrsim 1/\sqrt{\chi_e} \gg 1$, the rate for Breit-Wheeler can be written
\begin{equation}
    \Gamma_\textrm{BW} \propto \left(\frac{\enlaser}{\ecrit}\right) \exp\left[-\frac{8}{3}\frac{1}{\gamma_e(1+\cos\theta)}\frac{\ecrit}{\enlaser}\right].
\end{equation}
In this limit, the rate for the Breit-Wheeler process
resembles that of the rate of the Schwinger process in a static electric field on the field strength ${\mathcal E}_\textrm{S}$ as discussed in Ref.~\cite{Hartin:2018sha}:

\begin{equation}
    \Gamma_\textrm{Schwinger} \propto \left(\frac{{\mathcal E}_S}{\ecrit}\right)^2 \exp\left[-\pi\frac{\ecrit}{{\mathcal E}_S}\right]
\end{equation}

In contrast, at low $\xi$ the rate is expected to follow the power-law ($\xi^{2n}$) expectation based on vertex-counting in perturbation theory. We strive to observe this transition with LUXE, and confront the measurements with the theoretical prediction.

\begin{figure}[htbp]
\centering
\includegraphics[width=0.48\textwidth]{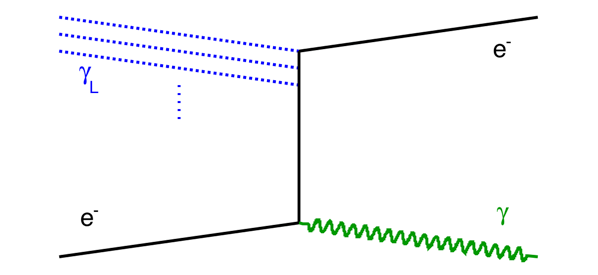} 
\includegraphics[width=0.48\textwidth]{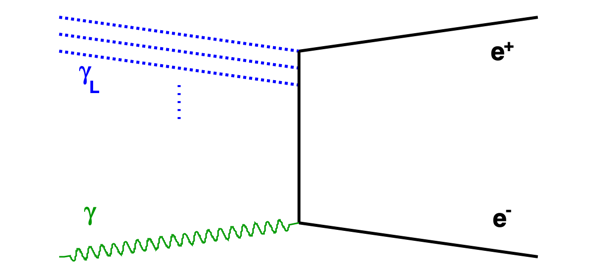} 
\caption{Schematic diagrams for the Compton process ($e+n\gamma_L \to e' + \gamma$) and the Breit--Wheeler process ($\gamma + n\gamma_L \to e^+e^-$).} 
\label{fig:feyn} 
\end{figure}

Figure~\ref{fig:feyn} shows schematically the dominant reactions for the electron-laser and photon-laser modes. 

LUXE also offers new opportunities to directly search for new particles from physics beyond the Standard Model (BSM)~\cite{LUXEBSM}. The Compton process yields a very high flux of high-energy photons which can mix with BSM particles in the mass range between about 10~MeV and 1~GeV. LUXE can also serve as a sensitive beam-dump experiment to search for such BSM particles when placing a detector a few meters behind the photon beam dump. Furthermore, new particles could be produced directly in the beam-laser interactions.

LUXE will use the electron beam of the European XFEL (\euxfel). It is designed to run with energies up to $E_e=17.5$~GeV, and contains trains of 2,700 electron bunches, each of up to $6\cdot 10^9$ electrons but usually operating with $1.5\cdot 10^9$ electrons, that pass at a rate of 10~Hz. One electron bunch per train will be extracted, and guided to the interaction
region. Out of these 10 Hz electron bunch extractions, 1 Hz will collide with the laser beam and 9 Hz will be used for \textit{in-situ} beam background measurements. 

\begin{figure}[htbp]
\centering
\includegraphics[width=0.7\textwidth]{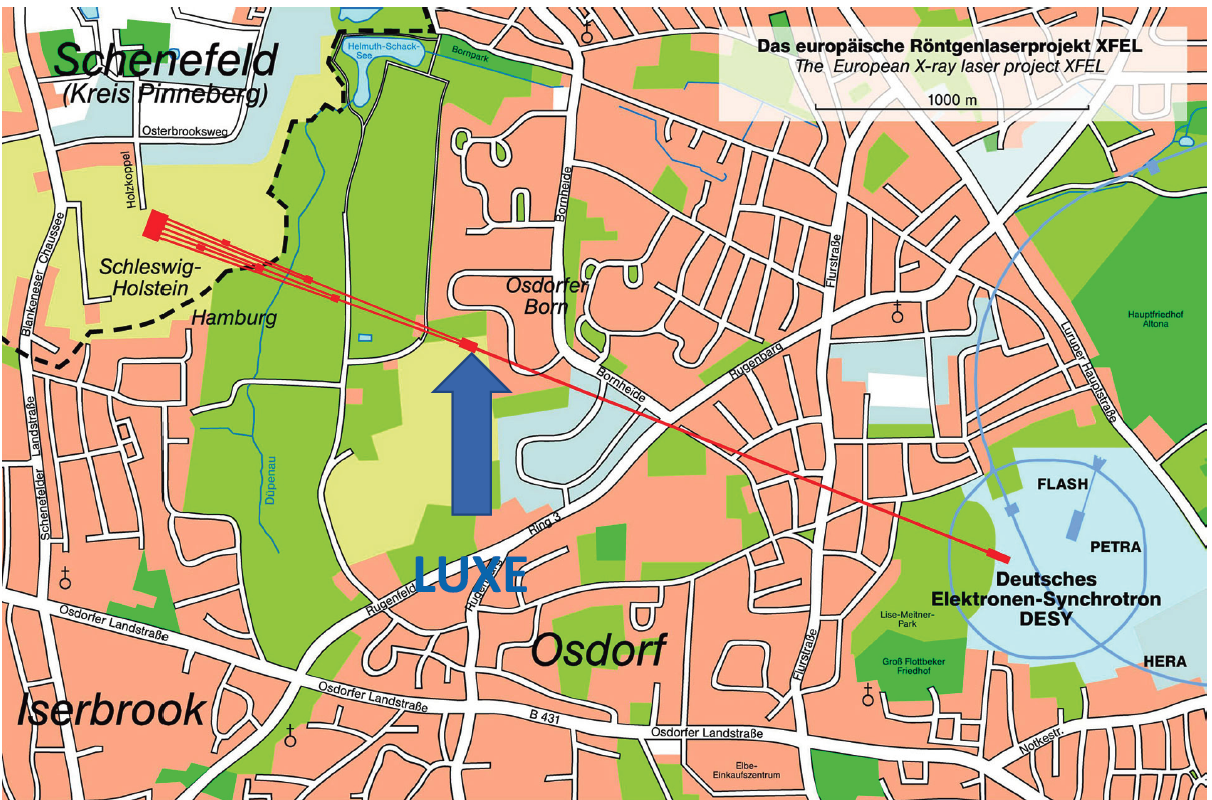}
\caption{Aerial view of the \euxfel location. The linear accelerator starts at the DESY site and ends at the Osdorfer Born location where the \euxfel fan starts with five beamlines going to the \euxfel main site in Schenefeld. The location of the LUXE experiment at Osdorfer Born is indicated.} 
\label{fig:euxfel} 
\end{figure}

An aerial view of the \euxfel is presented in Fig.~\ref{fig:euxfel}. The linear accelerator ends after 1.7~km at which point the "fan" of the \euxfel starts which foresees of a total of five beamlines that serve the photon science programme. At present, three of the five beamlines are operating. 
At this location called "Osdorfer Born", there is a part of a tunnel currently unused which presents an ideal opportunity to install the LUXE experiment. Starting in about 2030 this tunnel will be part of the construction of a second \euxfel fan. 
Above the tunnel there is also a building which provides access to the tunnel, and has significant infrastructure. In this building a laser can be installed, as well as service rooms and other necessary infrastructure for the experiment.

The high electron beam energy, and the chance to access well-defined electron bunches continually at 10 Hz, makes the \euxfel uniquely suitable worldwide for this proposal. While other accelerator-laser combinations can be considered for similar experiments, they have lower beam energies (e.g. SACLA, SLAC or laser accelerators) and/or less access to continuous electron bunches, or a significantly lower beam quality and current (e.g. tertiary electron beam at CERN).

The laser envisaged for the initial phase (\phaseone) of the experiment has a power of 40~TW, and will be focused to about 3~$\mu$m, achieving intensities of up to $1.3 \cdot 10^{20}$~W/cm$^2$. 
The laser photon wavelength is 800~nm, corresponding to an energy of 1.55~eV. It will operate with a steady-state amplifier repetition rate of 1 Hz to ensure high stability. An elaborate state-of-the-art diagnostics system for the laser intensity will be designed with the goal of achieving a precision on the absolute laser intensity below 5\% and an uncertainty on the relative intensity of individual shots below 1\%. In a second phase (\phasetwo) of the experiment a more powerful laser with 350~TW is envisaged to reach intensities up to $1.2 \cdot 10^{21}$ W /cm$^2$.
With \phaseone the critical field will be reached ($\chi_e\lsim 1.2$), but only with the full capacity of \phasetwo
will it be possible to explore the transition to the super-critical regime thoroughly up to $\chi_e\lsim 3$.

\begin{figure}[htbp]
\centering
\includegraphics[width=0.7\textwidth]{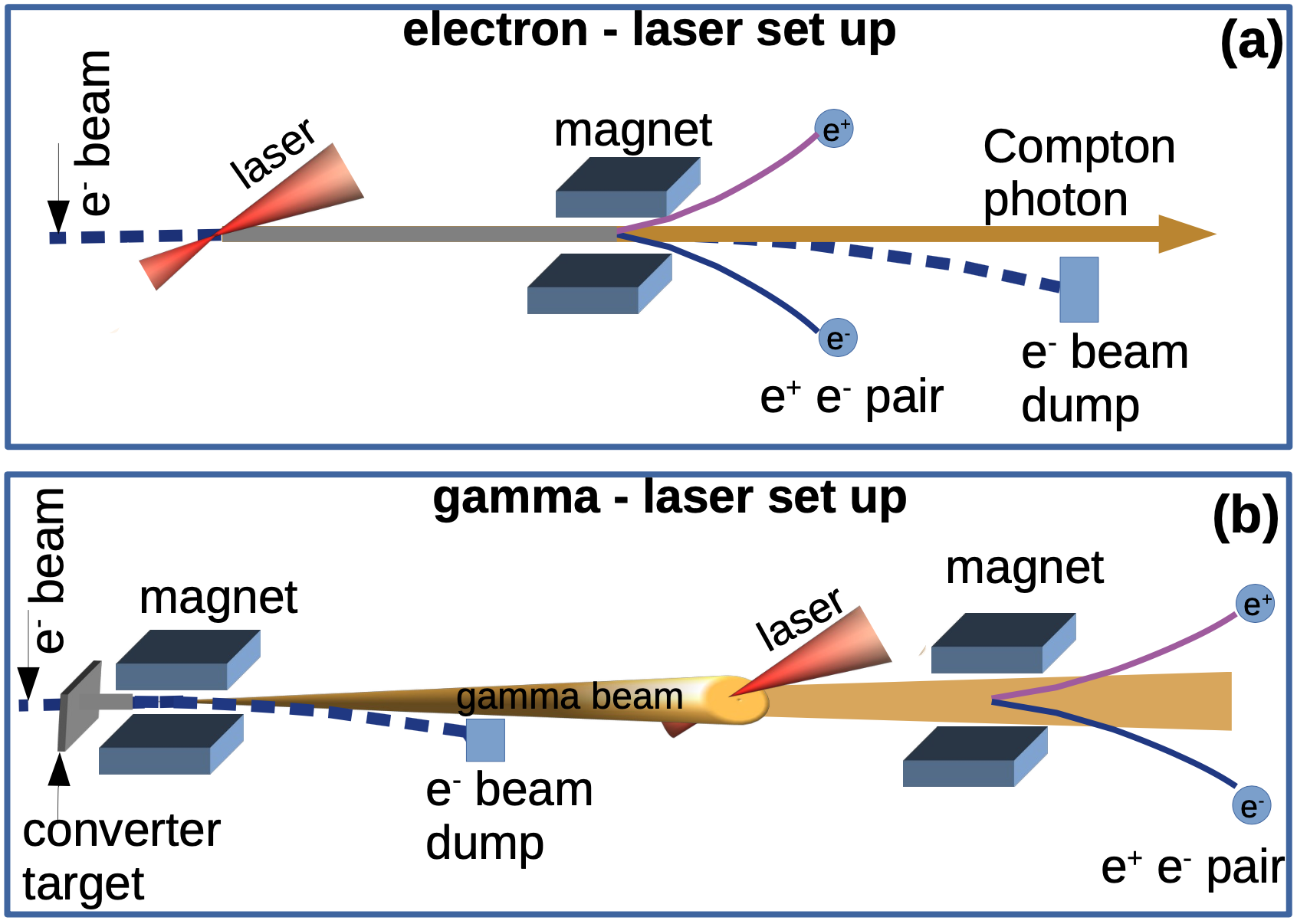}
\caption{Schematic experimental layouts for the a) \elaser and b) \glaser  setup.} 
\label{fig:layout} 
\end{figure}

\begin{figure}[htbp]
\centering
\includegraphics[width=0.48\textwidth]{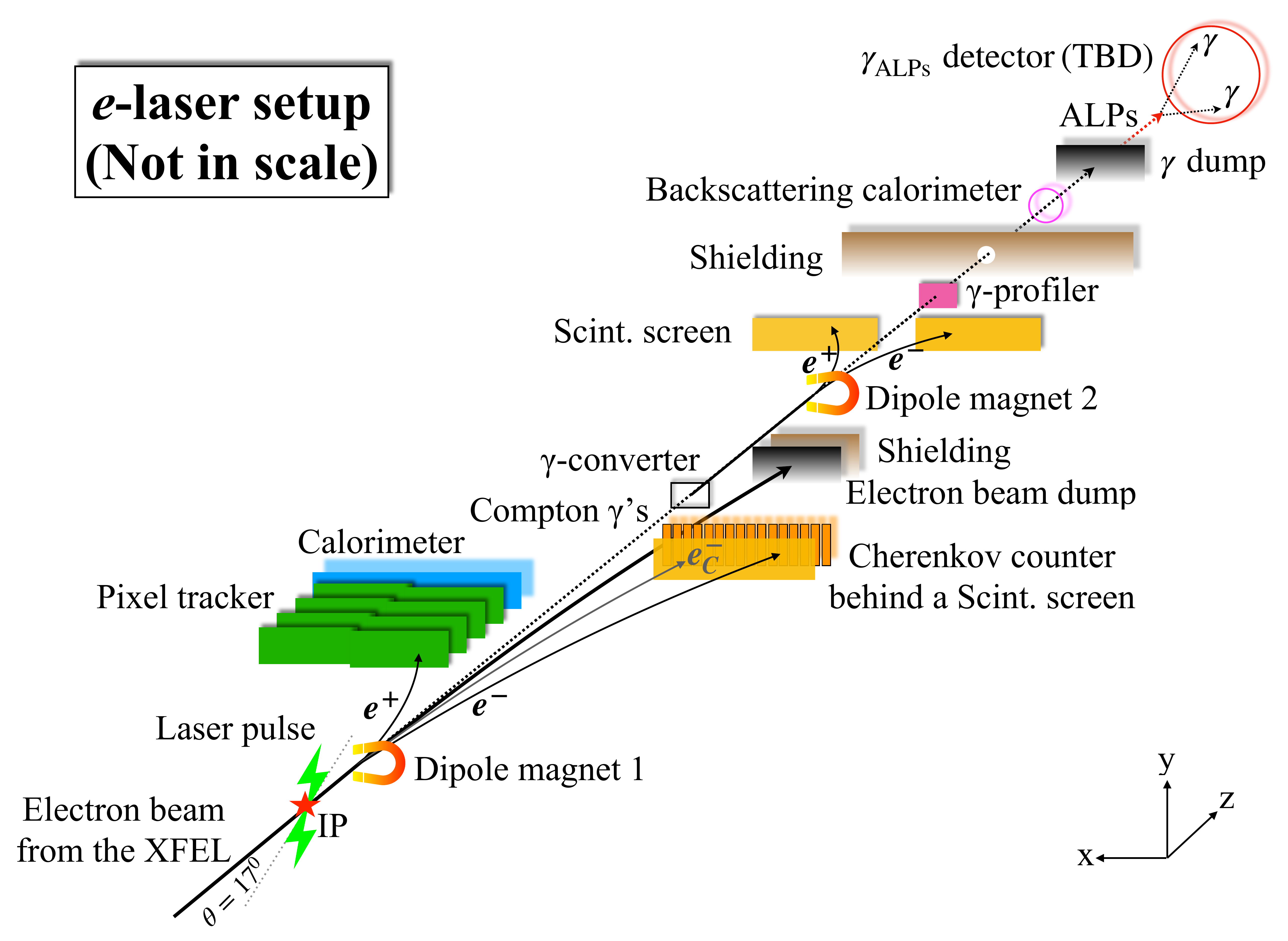} 
\includegraphics[width=0.48\textwidth]{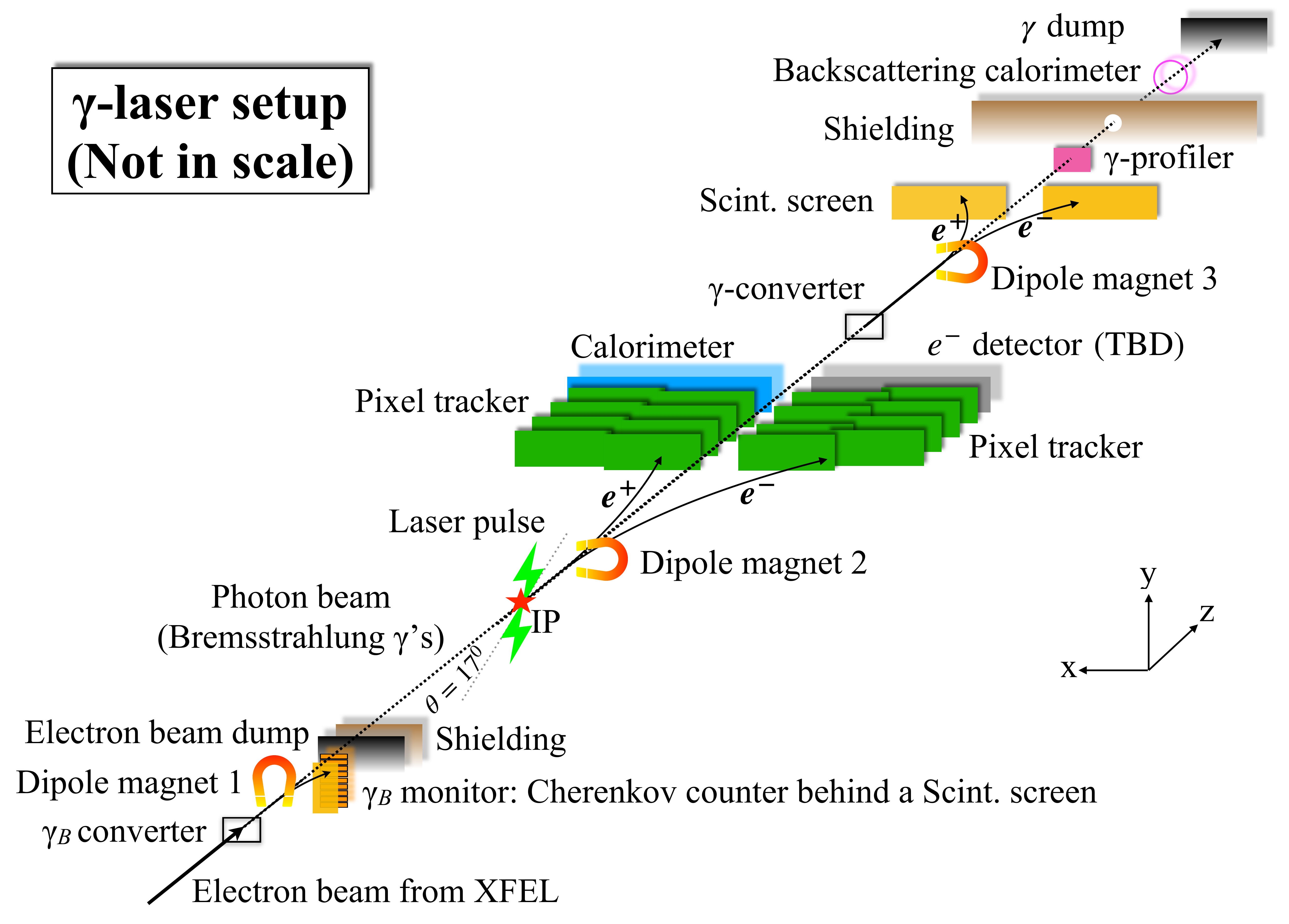} 
\caption{Schematic layouts for the \elaser and \glaser  setup. Shown are the magnets, detectors and main shielding and absorbing elements. The details are explained in Sec.~\ref{sec:detectors}.
} 
\label{fig:layoutdetailed} 
\end{figure}

Schematic layouts of the experiment are shown in Fig.~\ref{fig:layout} for the two configurations envisaged for the \elaser and the \glaser setups. For the \elaser setup the electrons are directly guided to the interaction point (IP), where a laser beam is directed at the same time. The electrons and positrons produced in these collisions are deflected by a magnet and then detected in a variety of detectors optimised for the expected fluxes of particles: positrons are detected by a silicon pixel tracking detector and a high-granularity calorimeter, while electrons are measured by a scintillation screen and gas \cer detectors. Photons that are produced at the IP continue along the beamline towards a photon detection system that is designed to measure their flux, energy and spatial distribution. The \glaser setup is similar, except that either a target is placed into the beam to produce a broadband beam of bremsstrahlung photons or a low-power laser is used to create a monochromatic photon beam via inverse Compton scattering, and the technology for the electron side is adapted due to the lower particle rates expected~\footnote{LUXE uses a right-handed coordinate system: the positive direction of the $z$-axis is defined by the direction of the beam, the positive $y$-direction is upwards towards the sky, and the positive $x$-direction is to the left of the beam direction of travel.}. Fig.~\ref{fig:layoutdetailed} shows a more detailed layout where major shielding structures and the individual detector components are also displayed. 

The expected numbers of particles per bunch crossing (BX) are as high as $10^9$ in the Compton process but as low as $\sim 10^{-3}$ in some cases for the positron production processes. These have been estimated based on a new simulation software developed for LUXE. 
These facts impose strong requirements/high demand on the detection system (in terms of radiation hardness, response-linearity, background discrimination etc.). 

The goal of this conceptual design report is to present the physics case for the experiment and to explain how it can be technically realised so that the physics goals can be achieved. The work presented here builds on the previous Letter of Intent~\cite{Abramowicz:2019gvx}. While the basic principle and physics goal of the experiment remain largely unchanged, a significant number of details have been updated, in particular to ensure that conceptually the experiment can be built on the proposed timescale, and to successfully complete the physics programme suggested. The studies presented are based on state-of-the-art simulations and theoretical understanding. Nonetheless, we made sure that the experiment is designed with a broader scope to be prepared for unforeseeable surprises when exploring this new regime. Redundancy has been built into the detection system to ensure that independent cross-checks with different technologies can be performed. 

\begin{figure}[htbp]
\centering
\begin{subfigure}{0.48\hsize} 
\includegraphics[width=\textwidth]{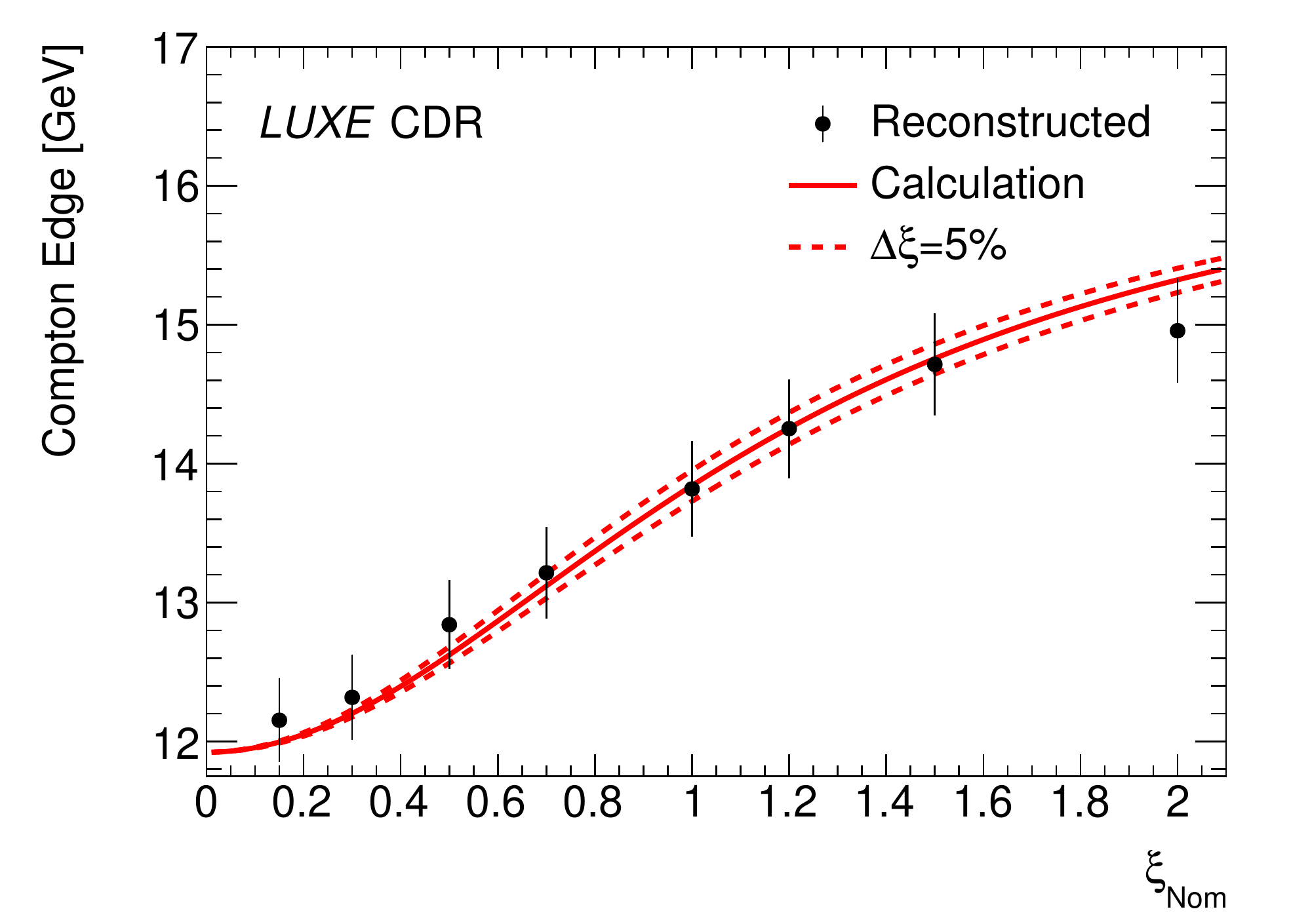} 
\caption{\label{fig:result:compton}}
\end{subfigure}
\hspace{0.5cm}
\begin{subfigure}{0.48\hsize} 
\includegraphics[width=\textwidth]{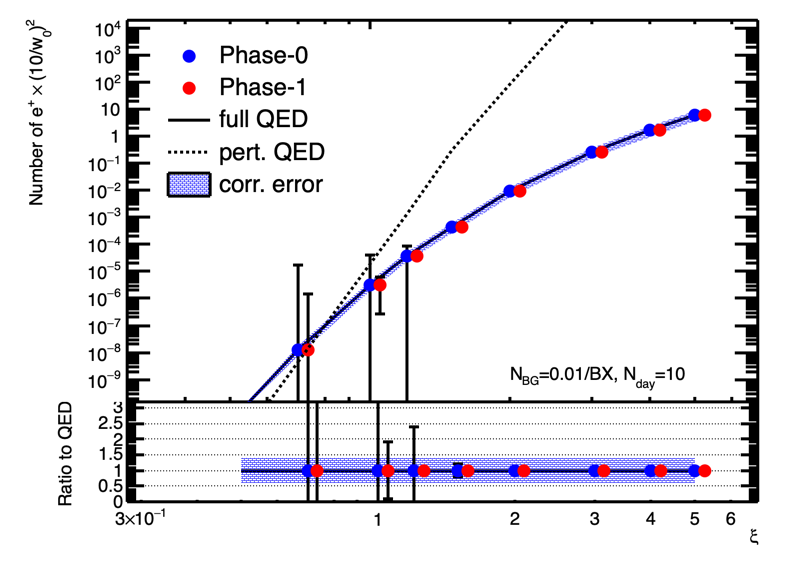}
\caption{\label{fig:result:bw}}
\end{subfigure}

\caption{a) Measured compton edge position as a function of the nominal $\xi$ value compared to the theoretical calculation for \phaseone. The error bars show the expected experimental uncertainties, and the red dashed line shows the uncertainty on the calculation due to a 5\% uncertainty on $\xi$. 
b) The number of positrons per laser shot normalized to a given laser spot size for \phaseone and \phasetwo, compared to the full QED prediction and the purely perturbative QED prediction. The error bars are dominated by statistical uncertainties. A correlated uncertainty of about 40\% is also shown; it arises from a 5\% uncertainty on the laser intensity. The \phasetwo data points are slightly displaced horizontally to make them more easily visible. }
\label{fig:qedresults} 
\end{figure}

\begin{figure}[htbp]
\centering

\includegraphics[width=0.5\textwidth]{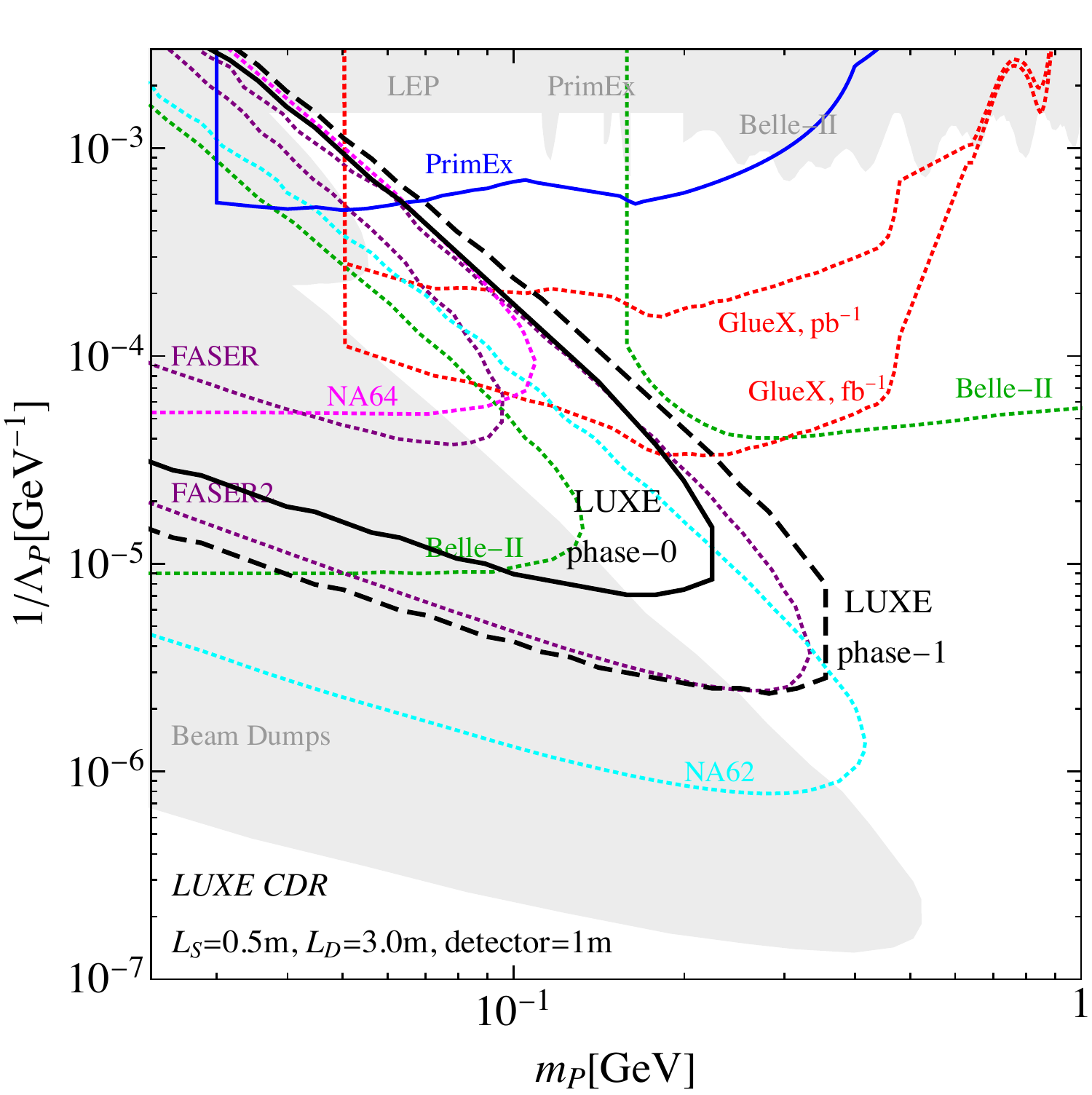}

\caption{Expected 95\% confidence level sensitivity to ALPs for the LUXE experiment in \phaseone and \phasetwo for one year of data taking in the plane of the ALP mass, $m_P$ versus the coupling parameter $1/\Lambda_P$. The grey area shows excluded areas based on LEP~\cite{Jaeckel:2015jla,Knapen:2016moh,Abbiendi:2002je}, {\sc PrimEx}~\cite{Aloni:2019ruo}, NA64~\cite{Dusaev:2020gxi,Banerjee:2020fue}, Belle-II~\cite{BelleII:2020fag}, and beam-dumps experiments~\cite{Bjorken:1988as,Blumlein:1990ay}. In addition, projected sensitivities of future experiment are shown:  NA62,  Belle-II, FASER,  {\sc PrimEx} and {\sc GlueX}~\cite{Dobrich:2015jyk,Dolan:2017osp,Feng:2018pew,Aloni:2019ruo,Dobrich:2019dxc}.}
\label{fig:alpresults} 
\end{figure}

Fig.~\ref{fig:qedresults} shows examples of physics results that can be obtained with LUXE testing strong-field QED. Shown is the shift of the Compton edge with increasing $\xi$ due to an increase of the effective electron mass by a factor $1+\xi^2$. It is seen that already with \phaseone of the experiment the precision will be sufficient to observe this effect clearly. Also shown is the positron rate versus $\xi$ compared to the perturbative QED and the full QED calculation. The two predictions differ by orders of magnitude in the range LUXE will measure, and the precision of the data will probe the calculation to better than 40\% in most of the measured range. Fig.~\ref{fig:alpresults} shows the reach of LUXE in the search for axion-like particles (ALPs). It is seen that it covers regions that have not yet been excluded and is competitive with other ongoing experiments. 
These results are explained in more detail in Sec.~\ref{sec:results}.

Section~\ref{sec:science} discusses the physics opportunities and scientific goals of the experiment. The electron beam transport from the \euxfel to the interaction point is described in Sec.~\ref{sec:machine}. A description of the laser and its diagnostic systems can be found in Sec.~\ref{sec:laser}. Section~\ref{sec:simulation} includes a simulation of both background and typical signal processes for the experiment and provides the basis for Sec.~\ref{sec:detectors} that explains the different detectors required for the experiment. Next, Sec.~\ref{sec:results} gives examples of the physics results that will be obtained, and finally Sec.~\ref{sec:tc} explains the infrastructure and technical requirements to construct and operate the experiment.
The project structure, schedule and cost is provided in Sec.~\ref{sec:orgcost} before conclusions are given in Sec.~\ref{sec:conclude}. 

\clearpage

\section{Physics Opportunities and Scientific Goals }
\label{sec:science}

Quantum electrodynamics is one of the most precise theories of the natural world.  High-precision QED experiments such as the measurement of the electron anomalous magnetic moment (``$g-2$'') have yielded unprecedented agreement between experiment and theory. This agreement is based on the fact that QED is weakly coupled (\emph{perturbative}) at the energy scales of interest. In other words, accurate theoretical results are produced as power series in the small QED coupling, $\alpha = e^2/4\pi\hslash c = 1/137$ (here and elsewhere, the permittivity of vacuum has been set to unity) For the electron anomalous magnetic moment this series has been evaluated up to and including the fifth order in $\alpha$, leading to an agreement between theory and experiment better than 1~ppb (with small electroweak and strong interaction contributions included, see e.g.\ \cite{Aoyama:2019ryr}). 
The experiments in question were carried out at low energy \cite{Hanneke:2008tm} and represent proof that QED works extremely well in this regime. However, it is known that vacuum polarisation effects lead to a running of the charge, hence the coupling $\alpha$, such that both \emph{increase} with energy or spatial resolution. At extremely high energies (of the order of the `Landau pole' \cite{Landau:1965nlt, PhysRev.95.1300, PhysRev.183.1292, PhysRevD.8.1110, PhysRevLett.80.4119} at $10^{286}$\,eV) the running coupling $\alpha$ becomes so large (it actually tends to infinity) that the perturbative expansion breaks down. In other words, QED, viewed as an effective field theory, becomes non-perturbative, which suggests replacing it with a more fundamental theory, its ultraviolet completion \cite{Gies:2020xuh}. The problem remains if QED is viewed as the U(1) component of the electroweak sector of the Standard Model. In this case, the Landau pole experiences a significant reduction, but still remains way beyond the Planck energy of $10^{19}$ GeV.

While energies of the order of the Landau pole, being completely out of reach, are currently only of conceptual interest, it is worth asking whether there are other ways of reaching the non-perturbative regime of QED. One  possibility is to employ an \emph{effective} coupling via introducing an external field that scales up the coupling. If this background field is strong enough, the effective, scaled coupling may become of order unity. But what is a \emph{strong field} in QED? The \emph{typical} field magnitude in QED is obtained by dimensional analysis, i.e.\ by combining the parameters of QED to form an electric (or magnetic) field. This yields the famous 
Schwinger limit, $\ecrit = m_e^2 c^3/e\hslash$, where $m_e$ denotes the electron mass. Expressed in standard SI units it corresponds to $\ecrit = 1.32 \cdot 10^{18}$ V/m or $B_{\trm{cr}} = 4.4 \cdot 10^9$ T. This is the typical field strength for QED processes, which is to be realised over the typical distance of an electron Compton wavelength, $\lambdabar = \hslash/m_ec \simeq 400$ fm. As already discussed by Sauter \cite{Sauter:1931zz}, a classical field configuration of \emph{critical} strength $\ecrit$ will lead to copious pair production from the vacuum (often referred to as the Schwinger effect). The practical challenge is to realise such a field strength over a large enough distance, say of the order of a micron.

Consider, for instance, the Coulomb field surrounding an electron. The 
Schwinger limit will be realised at a distance of about 30 fm from the electron. This distance can be enlarged by employing the Coulomb field of heavy nuclei (atomic number $Z$) which will lead to an effective coupling $Z\alpha$. 
If $Z$ becomes large (of order $10^2$), a series expansion in $Z\alpha$ becomes useless and perturbation theory (i.e.\ the Born approximation) breaks down. Instead, one has to take into account all orders of the interaction with the Coulomb field. This is particularly relevant for the physics of high-$Z$ systems such as heavy atoms \cite{Pomeranchuk:1945} or ions \cite{Ullmann:2017} and their collisions. A disadvantage of high-$Z$ systems is the problem of disentangling QED from nuclear or strong interaction effects which tend to be non-perturbative in their own right. Thus, heavy Coulomb systems do not provide a clean QED environment. One exception is for heavy ions, where one may consider ultra-peripheral collisions (UPCs) at high centre-of-mass energies with impact factor larger than twice the ion radius, $b > 2R$. In this case, fields of the order of the Schwinger limit may be reached in the lab (see e.g.\ the review \cite{Baur:2007zz}). As the nuclear fields are short-ranged, the physics is dominated by QED even for proton collisions \cite{Budnev:1973tz}. In comparison to laser backgrounds, the electromagnetic field in UPCs is extremely short lived, 
and cannot propagate over macroscopic distances. In addition, the fields in question are boosted Coulomb fields rather than plane waves (free photons). Reaching high field strengths requires considerable boost factors. Experiments at RHIC and the LHC (ATLAS, ALICE) employed centre-of-mass energies of 200 GeV and 5 TeV per nucleon pair, respectively. In laser-particle collisions, on the other hand, one is limited by the low energy of the laser photons of about 1 eV. Colliding these with, say, 10 GeV electrons yields a centre-of-mass energy slightly less than 1 MeV, so that a non-linear enhancement is required to produce pairs in this way \cite{Burke:1997ew}. In contrast, at high energies, linear perturbative pair production is relatively straightforward as one is operating above threshold. The ATLAS experiment, for instance, has reported muon pair production via (virtual) photon-photon collisions, $\gamma^*\gamma^* \to \mu^+ \mu^-$ in Pb + Pb scattering \cite{Aaboud:2018eph}. By the same mechanism, ATLAS and CMS have even been able to observe light-by-light scattering, $\gamma^*\gamma^* \to \gamma \gamma$  \cite{Aaboud:2017bwk,2019134826,Aad:2019ock}. As the virtuality of the initial photons is small compared to the large centre-of-mass energy, they can be regarded as `quasi-real' \cite{dEnterria:2013zqi}, which technically amounts to the justified use of the equivalent photon (Weizs\"acker-Williams) approximation \cite{vonWeizsacker:1934nji,Williams:1934ad}. The physics involved has been reviewed in \cite{Baur:2007zz} and \cite{Schoeffel:2020svx}, the latter containing a brief comparison between photon-photon experiments at the LHC and in laser beam collisions. 

Strong electromagnetic fields also occur in astrophysical contexts, the most spectacular example arguably being provided by magnetars with magnetic fields in excess of $10^9$ T \cite{Kouveliotou:1998ze,Harding_2006,Turolla:2015mwa,Kaspi:2017}. Obviously, these fields cannot be utilised for lab based research. Another source of strong electromagnetic fields is in beam-beam interactions as foreseen to occur at future lepton colliders with centre-of-mass energies exceeding 100 GeV \cite{Yakimenko:2018kih,Bucksbaum:2020}. However at present, these are yet to be funded. 

The last few decades have seen the development of ever more powerful laser technology \cite{Danson:2019} that can be employed to realise a relatively clean source of ultra-strong electromagnetic fields. In phase 0 of LUXE, the $40\,\trm{TW}$ JETI (JEna TItanium:sapphire) laser will be employed, and in phase 1, the laser will be upgraded to $350\,\trm{TW}$. The high-energy (order $10\,\trm{GeV}$) electron beam that drives the European XFEL will be used in two modes: i) as a direct probe of the laser pulse in electron-laser collisions; ii) as a source, through bremsstrahlung or inverse Compton scattering, of high-energy photons, which then probe the laser pulse in photon-laser collisions. The primary physics aim is to measure  strong-field QED (SFQED) processes of non-linear Compton scattering \cite{nikishov64,kibble64}, non-linear Breit-Wheeler pair creation \cite{Breit:1934zz,Reiss1962,narozhny69} and the non-linear trident process \cite{ritus72}. (A table of these processes is given in Table~\ref{tab:candproc}.)
\begin{table}[htbp]
\centering
\begin{tabular}{|m{3cm}m{4cm}m{3cm}|}
\hline
\multicolumn{3}{|c|}{LUXE Candidate Processes}\\
\hline\hline
 \centering\includegraphics[width=1.8cm]{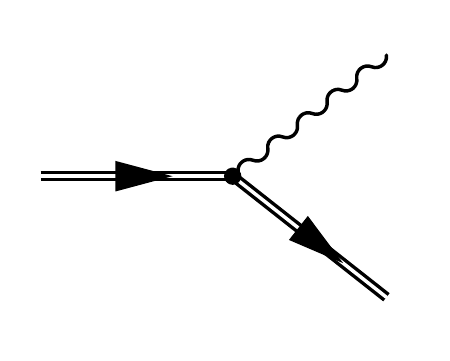} & Non-linear Compton & \[e^{\pm}\to e^{\pm} + \gamma\]\\
 \hline
  \centering\includegraphics[width=1.8cm]{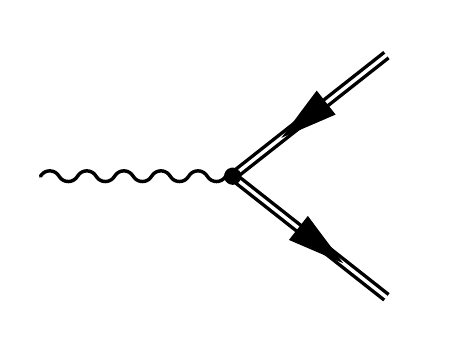} & Non-linear Breit-Wheeler & \[\gamma\to e^{-}e^{+}\]\\
  \hline
   \centering\includegraphics[width=2.5cm]{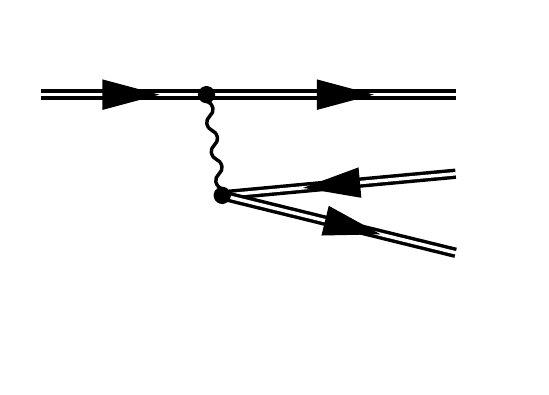} & Non-linear trident & \[e^{\pm}\to e^{\pm} + e^{-}e^{+}\]\\
   \hline
 \end{tabular}   
 \caption{A table of processes to be measured and characterised at LUXE. Double lines indicate fermions in a laser background.} \label{tab:candproc}
\end{table}

In light of this discussion, it is customary to measure the strength of the laser background field of frequency $\omegaL$ in terms of the \emph{classical non-linearity parameter},
$\xi=|e|\enlaser/m_e c\omegaL$. Rewriting $\xi= e\enlaser\lambdabar_e/\hslash \omegaL $, one identifies $\xi$ as the energy transferred to an electron by the laser over a reduced electron Compton wavelength, $\lambdabar_e$, in units of the laser photon energy \cite{ritus85} and thus approximates the number of photons required to achieve this.  In this way, the parameter $\xi$ quantifies the non-linearity of the field-charge interaction. Note that $\xi$ is Lorentz invariant \cite{Heinzl:2008rh} as it may just as well be expressed in terms of the lab field and frequency, $\xi = m_ec^2 \enlaser/\hslash \omegaL \ecrit$. The parameter $\xi$ is often referred to as the \emph{intensity parameter}, which can be seen by writing $\enlaser/\ecrit$ as $\sqrt{I/I_{\trm{cr}}}$, where $I$ is the local laser intensity and $I_{\trm{cr}}=4.6\cdot 10^{29}\,\trm{Wcm}^{-2}$ is the critical intensity (for a circularly polarised laser pulse). The parameter $\xi$ is the effective coupling describing the interaction with the background. Hence when $\xi \sim O(1)$, it is expected that a perturbative approach in $\xi$ breaks down. LUXE will probe the transition of QED processes from the linear to the non-linear, non-perturbative regime with $\xi$ values in the range from $\approx 0.1$ to $\approx 6$ in phase 0 of operation and further explore the non-perturbative quantum regime with $\xi$ values up to $\xi \approx 19$ in phase 1.

The non-linearity parameter $\xi$ is purely classical (it is independent of $\hslash$) as it describes the energy transfer from a classical field to a probe charge. To measure the importance of quantum effects one can use the \emph{quantum non-linearity parameter}, $\chi=e\hslash\sqrt{-(P \cdot F)^2}/m_{e}^{3}c^{4}$, where $F=F^{\mu\nu}$ is the background field tensor and $P$ is the four-momentum of the high-energy probe particle (typically electron or photon). The parameter $\chi$ can be interpreted as the energy transferred from the laser pulse to a probe electron over a reduced electron Compton wavelength, in units of the electron rest energy. In terms of the field strength $E_*$ `seen' by this electron, $\chi = E_*/\ecrit$. Furthermore $\chi$ quantifies the amount of recoil in the interaction e.g. when an electron emits a photon. In a plane-wave background, $\chi$ takes the simple form $\chi = \eta \xi$, where $\eta=\hslash \vkap \cdot P/m_{e}^2c^{2}$ and $\vkap$ is the wave vector of the laser pulse. In terms of probe particle energy $\epsilon$ and collision angle $\theta$, $\eta$ can be written as $\eta=(\hslash\omegaL \epsilon/m_{e}^2c^{4})(1 + \beta\cos\theta)$ and is the energy (or ``lightfront momentum'') parameter (where $\beta$ is the relativistic $\beta$ factor), which becomes maximal for a head-on collision, $\theta = 0$. This ideal case cannot be realised in experiment, so LUXE will have $\theta = 17.2^\circ$ as stated above. A table of these parameters and the range that can be accessed by LUXE is given in Table~\ref{tab:parameters}.

\begin{table}[htbp]
\centering
\begin{tabular}{|m{1cm}cccc|}
\hline
\multicolumn{2}{|c}{\multirow{2}{*}{Theory Parameter}} & \multirow{2}{*}{Definition} & \multicolumn{2}{c|}{Range accessed in LUXE}
\\& & & \phaseone & \phasetwo \\
\hline\hline
\centering $\xi$ & \tstrut\bstrut Classical non-linearity parameter & $\xi=\dfrac{m_e}{\omegaL}\dfrac{\enlaser}{\ecrit}$ & 
$\le 6$ & $\le 19$ \\
  \hline     
\centering $\eta_i$ & \tstrut\bstrut Energy parameter &$\eta_i= \dfrac{\omegaL \epsilon_i}{m_e^{2}} (1+\beta\cos\theta)$ &
\multicolumn{2}{c|}{$\eta_{i} \leq 0.2 $}\\
  \hline
\centering $\chi_i$ & \tstrut\bstrut Quantum non-linearity parameter &
$\chi_i=\dfrac{\epsilon_i}{m_e}\dfrac{\enlaser}{\ecrit}(1+\beta\cos\theta)$ &
$\le 1$ & $\le 3$ \\
  \hline
 \end{tabular}   
\caption{Table of main theory parameters. Here, $\omegaL$ is the laser frequency, $\epsilon$ is the particle (electron, positron, photon) energy, $\theta$ is the collision angle of the particle with the laser pulse such that $\theta=0$ is ``head-on'', $\enlaser$ is the instantaneous laser field strength and $\ecrit$ is the Schwinger limit and $m_e$ is the electron mass ($\hslash=c=1$ has been used, $\beta=1$ for photons and $\beta\approx 1$ for electrons.). Subscripts $i$ are used to denote particle type: ``$e$'' for an electron parameter and ``${\gamma}$'' for a photon parameter, e.g. $\chi_e$ ($\chi_\gamma$) is the non-linearity parameter for the electron (photon). The range of values accessed by LUXE using \phaseone and \phasetwo lasers for the $\xi$ and $\chi$ parameters is also given. 
} \label{tab:parameters}
\end{table}

The classical intensity parameter $\xi$ can be used to denote different physics regimes of strong-field QED in particle-laser interactions. At high field strengths $\xi \gg 1$, and when $\xi\gg \chi^{1/3}$, the total production rates can generally be well-approximated using the parameter $\chi$, which, as remarked, is $\chi=E_{\ast}/\ecrit$. However at LUXE, the \emph{transition} that occurs from perturbative QED at $\xi \ll 1$ through the \emph{intermediate intensity regime} of $\xi\sim O(1)$ into the strong-field regime of $\xi \gg 1$ will be studied. In these other intensity regimes, total production rates no longer depend just on $\chi$, rather on the intensity parameter $\xi$ and energy parameter $\eta$, independently.

In the intermediate intensity regime, two types of non-perturbativity can be identified. The first originates in the fact that $\xi$ quantifies the coupling between the laser background and the probe charge. If $\xi\sim O(1)$, all perturbative orders in $\xi$ must be taken into account. 

The second type of non-perturbativity in LUXE arises in Breit-Wheeler pair-creation (and, as we shall see, also Compton) process. When $\xi \gg1$, for small $\chi$, the probability scales as a typical tunnelling exponent (e.g. in a constant field as $\sim\mbox{e}^{-8/3\chi}$), and therefore does not even permit a perturbative expansion in the field, at any order.

A further type of non-perturbativity in SFQED is associated with the enhancement of radiative (loop) corrections in very intense electromagnetic (EM) backgrounds that can be approximated as a constant crossed  field \cite{ritus1970radiative}. In order for a process in a plane wave background to be approximated with one taking place in a (local) constant crossed field, the intensity parameter must fulfil $\xi\gg1$ and $\xi^{2}\gg\eta$, which can be violated at very high energies \cite{podszus2019high,ilderton2019note}. Assuming its validity, a careful inspection of up to three-loop order corrections \cite{ritus1972vacuum,narozhny1979radiation,narozhny1980expansion} has suggested that such a regime, where radiative corrections dominate, is reached when $\alpha\chi^{2/3}\gtrsim1$. This assertion is referred to as the ``Ritus-Narozhny conjecture'' \cite{fedotov2017conjecture} and has been supported recently by an all-order resummation of bubble-type corrections to the electron self energy \cite{mironov2020re}. For $\alpha\chi^{2/3}\gtrsim1$, the corrections of such type have been conjectured to be dominant, although a formal proof is still missing due to the yet unresolved notorious difficulties in evaluating the actual enhancement of the vertex corrections even at one-loop level \cite{morozov1981vertex,di2020one}, not to mention at higher orders. The future goal of experimentally attaining the range $\chi\gtrsim 10^3$, required to observe this type of non-perturbativity, will be challenging due to the need to mitigate radiative losses, although there are already a few experimental suggestions in the literature for how to overcome these \cite{Yakimenko:2018kih,baumann2019probing,blackburn2019reaching,di2019testing}. In order to realise these longer-term goals of testing high-$\chi$ behaviour, it is crucial to confirm theory with experiment at lower values of $\chi$ leading up to this, and LUXE will provide the first measurements of the transition of SFQED processes from the linear quantum to the non-linear quantum regime. 

For the benefit of easier reading of this document, Table~\ref{tab:symbols} presents the meaning of some of the commonly used symbols.
\begin{table}[ht]
\centering

\begin{tabular}{|cl|}
\hline
\multicolumn{2}{|c|}{Symbol table}\\
\hline\hline
   $e^{-}$ & Electron \\
   $e^{+}$ & Positron \\
   $\gamma_{L}$ & Laser photon \\
   $\gamma_{B}$ & Bremsstrahlung photon \\
   $\gamma_{C}$ & Inverse Compton Scattered photon \\
   $\gamma$ & Radiated photon\\
   \hline
 \end{tabular}   
 \caption{Table of commonly-used symbols.}
 \label{tab:symbols}
\end{table}

New physics~(NP) beyond the Standard Model~(SM) is motivated by questions arising from experimental evidence.
For example, neutrino oscillations, the lack of valid dark matter candidate in the SM and the matter anti-matter asymmetry of the Universe, for recent discussion about NP see e.g.~\cite{McCullough:2018knz,Strategy:2019vxc}. 
Moreover, there are theoretical arguments, such as the ``gauge Hierarchy Problem'' or the flavour puzzle (for a summary, see e.g. Ref.~\cite{Strategy:2019vxc}).
Possible solutions for these problems predict the existence of new light degrees of freedom, which are weakly coupled to the SM and are potentially long-lived.
Among the various possibilities are scalars, pseudo-scalars or milli-charged particles~(mCP)~\cite{Davidson:1981zd,Wilczek:1982rv,Gelmini:1982zz,Kim:1986ax,Feng:1997tn,Graham:2015cka, Gupta:2015uea, Flacke:2016szy, Frugiuele:2018coc,Banerjee:2020kww,Peccei:1977hh,Peccei:1977ur,Weinberg:1977ma,Wilczek:1977pj,Chang:2000ii,Nomura:2008ru,Freytsis:2010ne,Dolan:2014ska,Hochberg:2018rjs,Holdom:1985ag,Gherghetta:2019coi,Abel:2008ai,Abel:2004rp,Aldazabal:2000sa,Batell:2005wa,Abel:2003ue,Dienes:1996zr,Csaki:2020zqz}. We will denote as ``axion-like-particle'' (ALP) any new \emph{pseudo}-scalar particle. Non-perturbative QED, or QED-like, phenomena may enhance event yields for new physics. In such cases, the LUXE experiment provides interesting opportunities to probe ``Beyond the Standard Model'' (BSM) physics. Here, the example of scalars is discussed and a more in-depth analysis will appear soon.

This chapter is organised as follows. In Sec.~\ref{sec:theo1}, some introductory background is given on approximations used to calculate QED processes in a laser background -- more detail is given in App.~\ref{app:theory}. In Sec.~\ref{subsec:cp}, some phenomenology of the LUXE candidate processes given in Table \ref{tab:candproc} are detailed and illustrated. Appendix \ref{app:simtheo} then details  how the numerical framework incorporates classical and quantum effects to simulate the interaction point between probe particles and laser pulse. In the rest of this chapter, we set $\hslash=c=1$ unless $\hslash$ or $c$ explicitly appear. All plots in this Section take the following ``standard parameters'': a 16.5\,GeV electron with a collision angle of $17.2$ degrees to a circularly-polarised, $16$ cycle, sine-squared laser pulse, with frequency
 $1.55\,\trm{eV}$, unless otherwise stated. In reality, it is planned to run mostly with a linearly polarised laser as this has several advantages as discussed in Sec.~\ref{sec:laserpol} but at the time of writing this report the simulation of a linearly polarised laser was not yet available.

\subsection{Theory Background} \label{sec:theo1}
\tbf{The plane wave model:}$\qquad$
LUXE will use a focussed laser pulse to provide the strong EM background. Currently, no exact analytical solutions are known to the Dirac equation in a focussed laser pulse background. Direct evaluation of the Dirac equation, whilst in principle possible, would take a prohibitively long amount of time. Therefore, approximations are required in order to calculate outcomes of experiments.

Two central approximations used in the ``plane wave model'' are,  
\begin{enumerate}
    \item Back reaction on, and depletion of, the intense laser pulse field can be neglected, thereby allowing it to be approximated by a classical background.
    \item Charged particles accelerated by intense EM fields, are highly relativistic: in the particles' rest frame, an arbitrary EM background is well-approximated by a plane wave ``crossed field'' pulse.
\end{enumerate}

Solutions to the Dirac equation are known  in a classical plane-wave background, and the well-known \emph{Volkov} solutions \cite{volkov35} for fermionic wave functions are acquired. Scattering processes are then calculated perturbatively in `dressed' vertices using Volkov wave functions. This is also sometimes referred to as the \emph{Furry picture}. It is then possible to generalise the Feynman rules of standard QED to a plane wave background. Processes will be calculated for individual probe particles of electrons, positrons and photons, and it is assumed that the result for beams of such particles can be acquired by integration at the probability level, of the single-particle result, over distribution functions. In Sec.~\ref{subsec:cp}, we will introduce some of the phenomenology of candidate SFQED processes to be measured at LUXE. 
 
\vspace{0.5cm}

Processes occurring in a plane-wave background have a particular type of kinematics. Whilst the vacuum is invariant under time and space translations and therefore, energy and the three components of momentum are conserved in processes taking place in vacuum, in a plane-wave background, one preferential direction is given by the background wavevector. As a result, only three components of energy-momentum are conserved: those perpendicular to the laser pulse propagation direction, and those projected onto the background wavevector, the so-called ``lightfront momentum''. As a result, it is natural to express the probability for SFQED processes, as integrals over lightfront momentum spectra, rather than energy or momentum separately. Technical details are given in the App.~\ref{app:theory} to this chapter.

A typical approximation employed in numerical simulations of SFQED phenomena in intense laser pulses, is the ``Locally Constant Field Approximation'' (LCFA). This is equivalent to taking the probability of a given first-order dressed process in a purely constant crossed field, forming an instantaneous ``probability rate'', and integrating the rate over an arbitrary pulse shape. For the LCFA to be a good approximation for particle energies considered at LUXE, it is necessary (although not always sufficient) that $\xi\gg1$. Since SFQED phenomena will be probed for a range of intensity parameters $\xi = 0.1 \ldots 6$ in \phaseone of LUXE, the LCFA is insufficient for predicting particle yields in experiment. Furthermore, it will be an experimental goal to measure harmonic structure in particle spectra. Since this structure arises due to interference between emission in different parts of the laser pulse, it cannot be reproduced by a locally-constant approximation. Instead, to model physics at the interaction point, the ``Locally Monochromatic Approximation'' (LMA) will be applied, in which the fast timescale of the laser carrier frequency is handled exactly, but the slow timescale of the laser pulse envelope is approximated. The LMA is valid for many-cycle laser pulses, such as will be employed at LUXE, and when little to no plasma is generated in the interaction. The IP simulation results in Sec.~\ref{sec:simulation}, entirely use the LMA rates given in App.~\ref{app:theory}, where more technical details are given.
\clearpage

\subsection{Compton Scattering} \label{subsec:cp}
\begin{wrapfigure}{l}{5cm}
\centering
\includegraphics[width=3cm]{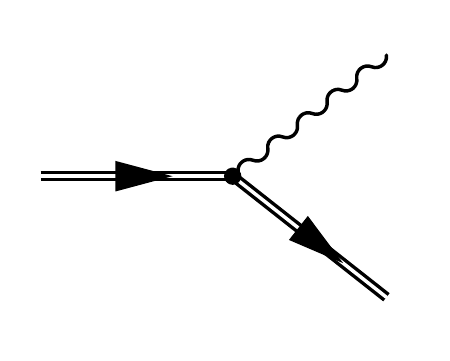}\\
    Compton Process
    \label{fig:NComfeynman}
\end{wrapfigure}
The Compton process in a plane-wave background refers to the process of an electron emitting a high-energy photon. It is often referred to as \emph{non-linear} Compton scattering because of the non-linear dependency on the background intensity parameter. The position of the $n$th harmonic in the photon emission spectrum for an electron in a monochromatic background, can be arrived at using the well-known linear Compton formula and replacing the incoming photon momentum by $n$ lots of the laser momentum and the electron mass by the \emph{effective mass}. Quantum and non-linear effects can be noted by expanding the cross-section in the \emph{quantum} energy parameter $\eta_{e}$ and the classical \emph{non-linearity} parameter $\xi^2$ \cite{heinzl13t}:
\[\sigma_{\gamma}^{\tsf{(Compton)}} \approx \sigma_{\gamma}^{\tsf{(Thomson)}}\left[1 - 2\eta_{e} - \frac{2}{5}\xi^{2}+\frac{14}{5}\eta_{e}\xi^{2}+\ldots\right].
\]
where $\sigma_{\gamma}^{\tsf{(Thomson)}}=6.65\times10^{-29}\,{\rm m}^{2}$ is the classical linear Thomson scattering cross section, and we recall that $\eta_e$ is the $\eta$ parameter for the electron.

\begin{figure}[ht]
\centering
\includegraphics[width=0.625\linewidth]{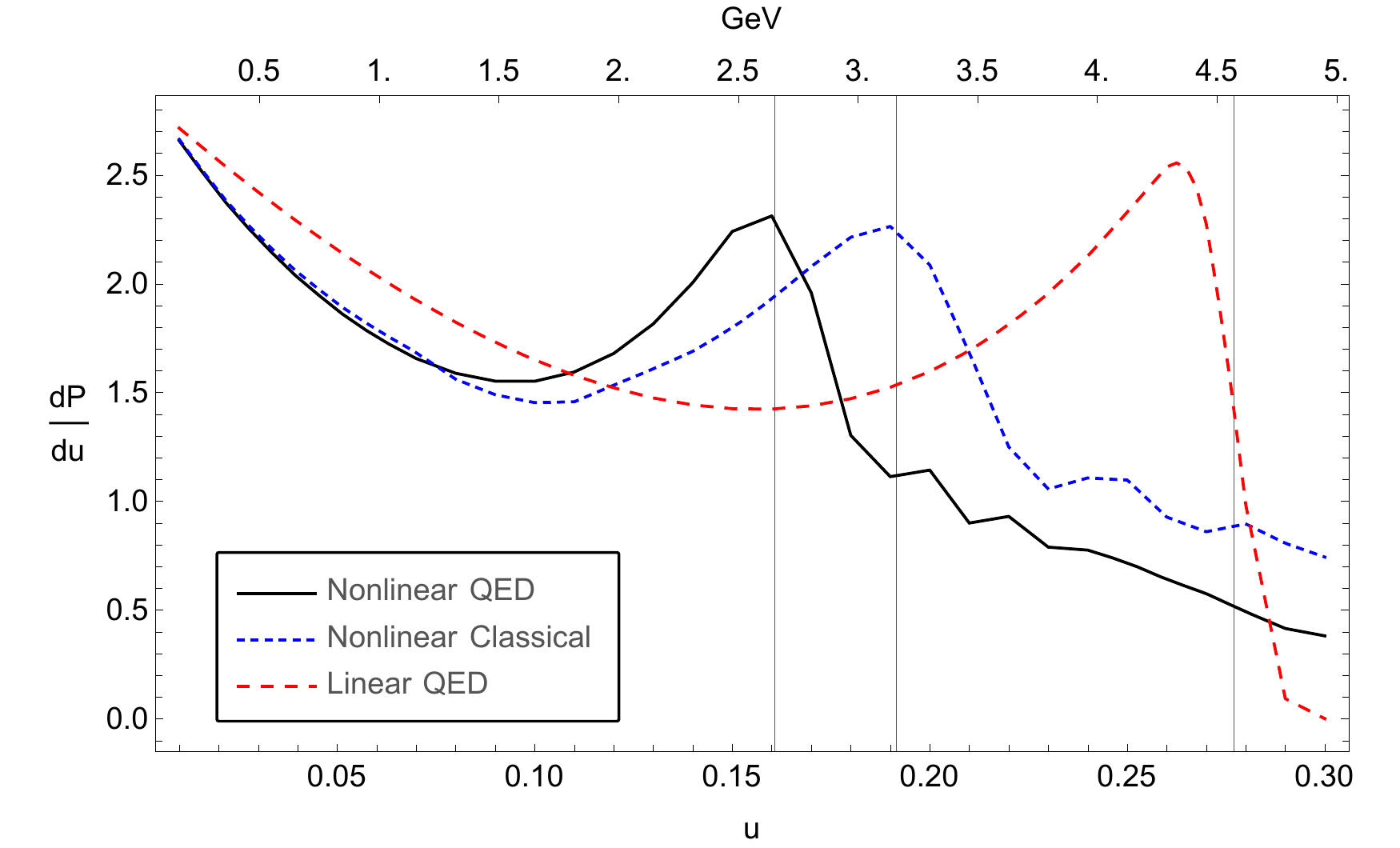}\hfill\includegraphics[width=0.355\linewidth]{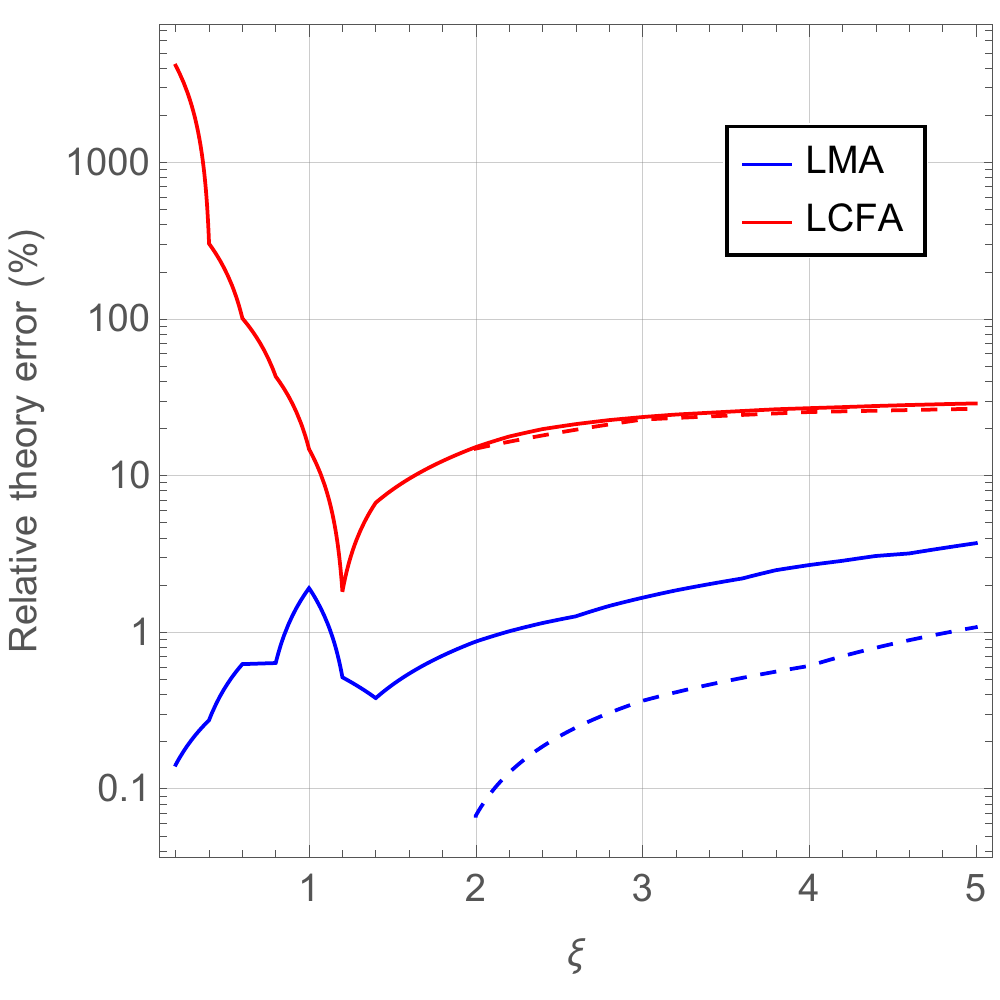}
\caption{
Left: Plot of the first harmonic range for $\xi=1$ (otherwise standard parameters) in the lightfront momentum fraction variable, $u$, (corresponding to energy at fixed detector angle), using various approaches. The grid lines correspond to the position of the first harmonic edge for the three cases. Right: a measure of the error of the LMA and LCFA employed in simulation \cite{lma2} compared to the analytical QED plane wave result for the Compton process.
(Dashed lines correspond to artificially turning off particle recoil in the simulation due to higher number of emissions).}\label{fig:CEdge2}
\end{figure}

One of the manifestations of the effective mass in the Compton process, is in the radiated photon spectrum. Part of the spectrum for the standard parameters of this section, with $\xi=1$, is illustrated in \figref{fig:CEdge2}. The spectrum is plotted in terms of the photon \emph{lightfront momentum fraction}, $u$, where $u=\vkap\cdot K/\vkap \cdot P$, with $K$ ($P$) being the photon (initial electron) four-momentum. The first harmonic is clearly defined by the main peak, which is often referred to as the \emph{Compton edge} \cite{Harvey:2009ry}. Kinematically, the position of the Compton edge in a circularly-polarised background can be derived by assuming one net laser photon is absorbed by the electron and that the electron's four-momentum is replaced by $P=P + (\xi^{2}/2P\cdot\vkap)\vkap$  (sometimes referred to as the \emph{quasimomentum}), with the consequence that its mass is replaced by the \emph{effective mass}, $m_{\ast}$, where $P^{2}=m_{\ast}^{2}=m_e^{2}(1+\xi^{2})$. In a plane-wave pulse, the position of the harmonic is unchanged (although it is slightly smoothed out, see \figref{fig:NLCspec1} and \figref{fig:CEdge2}) and so for a quasi-monoenergetic beam of electrons that propagate through a transversally-homogeneous focussed laser pulse, the position of the Compton edge allows for a determination of the peak intensity of the laser pulse used.

The physics contained in a measurement of the Compton edge is illustrated in \figref{fig:CEdge2}. The position of the harmonic edge gridlines
is given by:
\[
u_{{\sf nonlin. QED}} = \frac{2n\eta}{2n\eta+1+\xi^{2}}; \qquad
u_{{\sf nonlin. class.}} = \frac{2n\eta}{1+\xi^{2}}; \qquad
u_{{\sf lin. QED}} = \frac{2n\eta}{2n\eta+1},
\]
where $n=1$ in the figure. The red-shifting of the Compton edge compared to non-linear Thomson scattering (the classical equivalent of non-linear Compton scattering), is therefore due to the recoil the electron experiences from emitting a photon. The red-shifting of the Compton edge compared to linear QED (where only a single photon is absorbed from the laser background), is therefore due to the effective mass being a non-linear effect involving many orders of perturbation 
theory. 
Also shown is the error obtained by artificially turning off particle recoil in the simulation due to higher number of emissions.

We can obtain an estimate of the error from using the LMA and LCFA, by comparing their use in a numerical simulation with the analytical QED plane wave results. To obtain a measure of the error, we analyse the spectrum around the position of the Compton edge, because this spectral feature is a key experimental observable related to the intensity of the laser pulse in a simple way. We compare the value for the photon spectrum integrated in a region 20\% either side of the Compton edge and form the relative difference with respect to the QED result. Taking the absolute value of this difference, we acquire an estimate of the error, which is shown in the right-hand graph of \figref{fig:CEdge2}, where it is plotted versus laser intensity. It is seen that the result of the LMA agrees with the full calculation to better than 5\% in the entire $\xi$ range while the LCFA result deviates significantly.

\subsubsection*{Non-perturbativity in the Compton Process}
\begin{figure}[ht] 
\includegraphics[width=0.49\columnwidth]{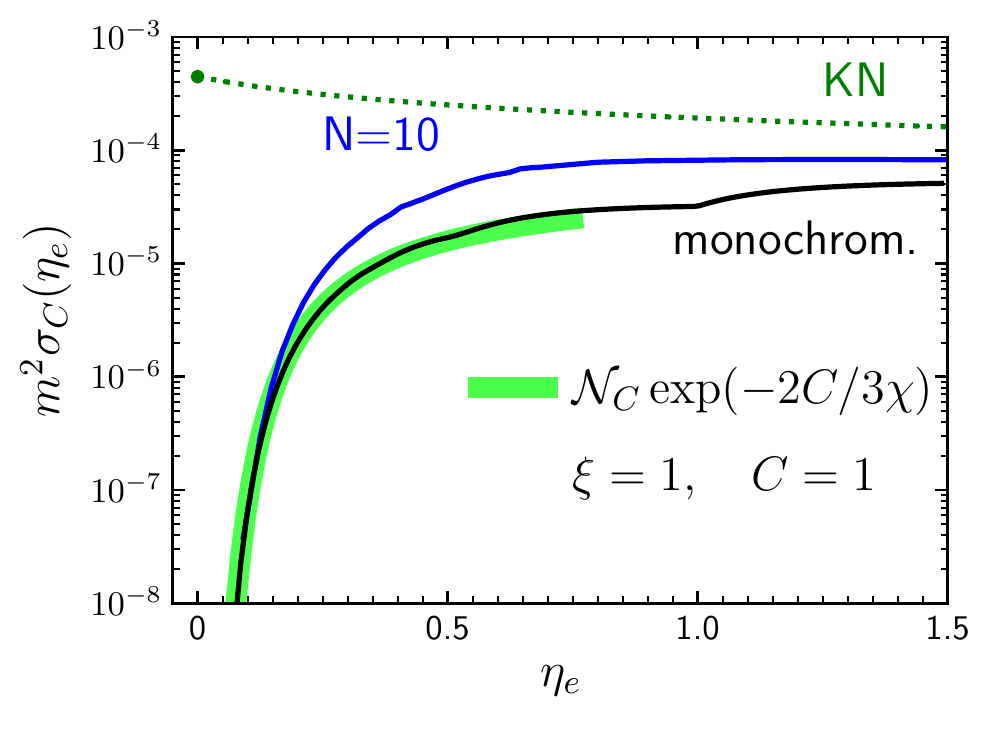}
\includegraphics[width=0.49\columnwidth]{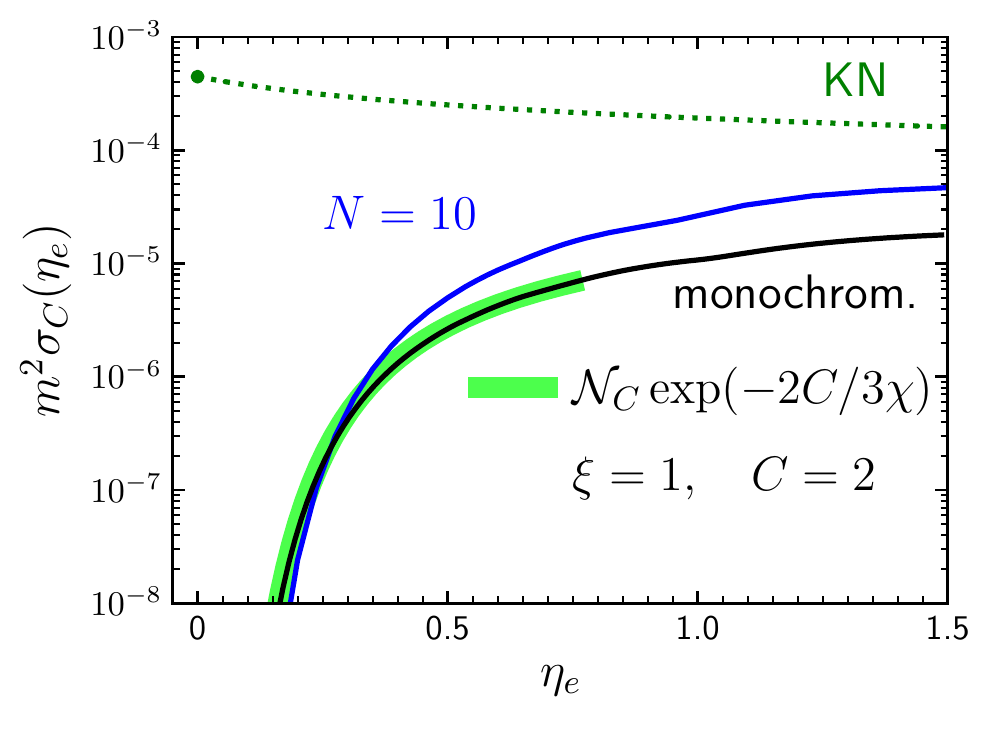}
\vspace*{-1mm}
\caption{Example cross sections $m^2 \sigma_C $ (in this plot $m=m_{e}$ is the electron mass) as a function of $\eta{_e}$ for two models:
model (1) - monochromatic laser beam (solid black curves),
model (2) - laser pulse with envelope $g(\phi) = 1 / \cosh (\phi/\pi N)$
for $N = 10$  (blue curves). 
For the laser intensity parameter $\xi = 1$ 
and cut-off values $C = 1$ (left panel) and $C=2$ (right panel).
The function ${\cal N}_C \exp\{ - 2 C / 3 \chi \}$ 
with normalisations ${\cal N}_1 = 6 \cdot 10^{-5}$ 
and ${\cal N}_2 = 4 \cdot 10^{-5}$
is displayed as wide green bands
to highlight the exponential suppression in the sub-threshold region
relative to the linear Compton (Klein-Nishina, KN) cross section 
(green dotted curves terminating in the Thomson point).
\label{fig:cutoff}}
\end{figure}

In addition to the ``all-order'' dependency on the field non-linearity through $\xi^2$, the Compton process can also display a fully non-perturbative dependence on experimental parameters, similar to the tunnelling dependency of pair-creation. This can be achieved by 
restricting the final electron-photon phase space through detector cuts, which introduces a non-trivial threshold to the Compton process. The cut then leads to an exponential suppression (see Fig.~\ref{fig:cutoff})
typical for tunnelling phenomena.  
  
To illustrate this effect (more details of this procedure are given in \cite{HernandezAcosta:2020agu,Kampfer:2020cbx}), we consider a head-on collision with a circularly-polarised laser pulse with $N$ cycles. Let $\coff$ be a lightfront momentum fraction cut-off
, which then sets a threshold requirement, $\ess \ge \coff$, for detection. Imposing this threshold implies only the final phase space
 of photons in the high-energy tail (see e.g. Fig.~\ref{fig:NLCspec1}) will be detected. Such events are characterised by a large lightfront momentum
transfer from the initial electron to the emitted photon; one could imagine
the process also as tunnelling between remote electron Zel'dovich levels \cite{zeldovich67}
(which suffer some broadening from bandwidth and ponderomotive effects
in laser pulses of finite duration). 

Taking $\eta_{e}=0.15$ as a typical value in the LUXE experiment (corresponding to an electron energy of $12.9\,\trm{GeV}$ at a collision angle of $17.2$ degrees), the cross section $\sigma_\coff$ with imposed cut-off $\coff$ exhibits the paradigmatic transmonomial behaviour
$\sigma_{\coff} \propto \exp\{ - f / \chi \}$, $f = 2 \coff/3$ (see also \cite{Titov:2019kdk}, and e.g. \cite{Dinu:2018efz}). Since $\chi \propto \sqrt{\alpha}$, we see that this is a fully non-perturbative dependency on the field-charge coupling and does not admit any expansion in $\alpha$. 
A similar behaviour is known for the Breit-Wheeler process ($f = 8/3$) 
and two-step trident process ($f = 16 / 3$) in the asymptotic region of $\xi \gg1$, $\chi\ll 1$.
However, remarkably, the results exhibited in Fig.~\ref{fig:cutoff} apply in the non-asymptotic region. 

By employing the high-energy and low-emittance electron beam driving the European XFEL and colliding it with an intense laser pulse, LUXE can measure this non-pertubative signal in the Compton process. A check of the $\coff$-dependence of $\sigma_\coff$ at given $\xi$ and $\eta$ can be performed within the analysis procedure by varying the minimum energy of the emitted photons to be included in the cross section determination. For this, it is necessary that LUXE can measure the photon and electron energy spectra differentially.

\subsection{Breit-Wheeler Pair Creation}
\begin{wrapfigure}{l}{5cm}
\centering
    \includegraphics[width=3cm]{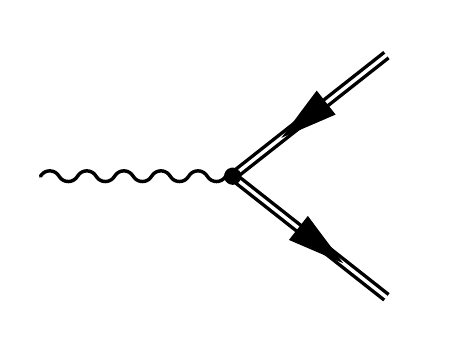}
    Breit-Wheeler Process
    \label{fig:NBWfeynman}
\end{wrapfigure}
Breit-Wheeler pair-creation in a background laser pulse corresponds to the decay of a photon to an electron-positron pair. We first introduce some of the associated phenomenology of the process, and then give details about the two planned photon sources -- a bremsstrahlung source and an inverse Compton scattering source. 

By considering the centre-of-mass energy in the collision of the probe photon with $n$ laser photons, one can derive the threshold harmonic $n_{\ast}$ required, such that the pair can be created. Recalling that the effective mass, $m_{\ast}=m_{e}\sqrt{1+\xi^{2}}$, it is clear that a higher threshold harmonic is required in more intense laser pulses. In the LMA, the threshold harmonic becomes
\bea
n_{*}(\phi) = \frac{2(1+\xi^{2}(\phi))}{\eta_\gamma}, \label{eqn:nastNBW1}
\eea
i.e. the threshold is phase-dependent (we recall quantities with subscript ${\gamma}$ refer to the photon).

\begin{figure}[ht]
\centering
\includegraphics[width=0.475\linewidth]{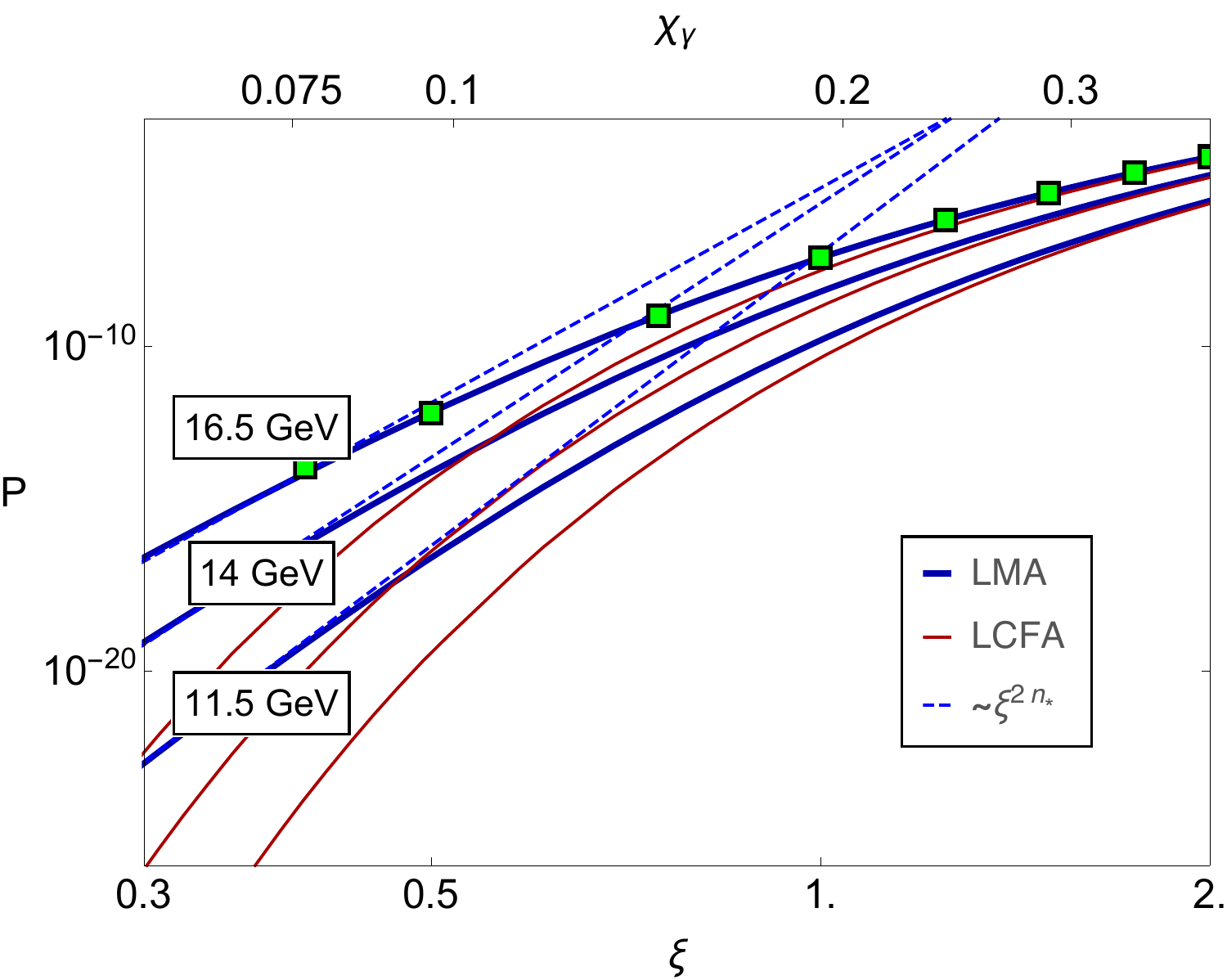}\hfill
\includegraphics[width=0.475\linewidth]{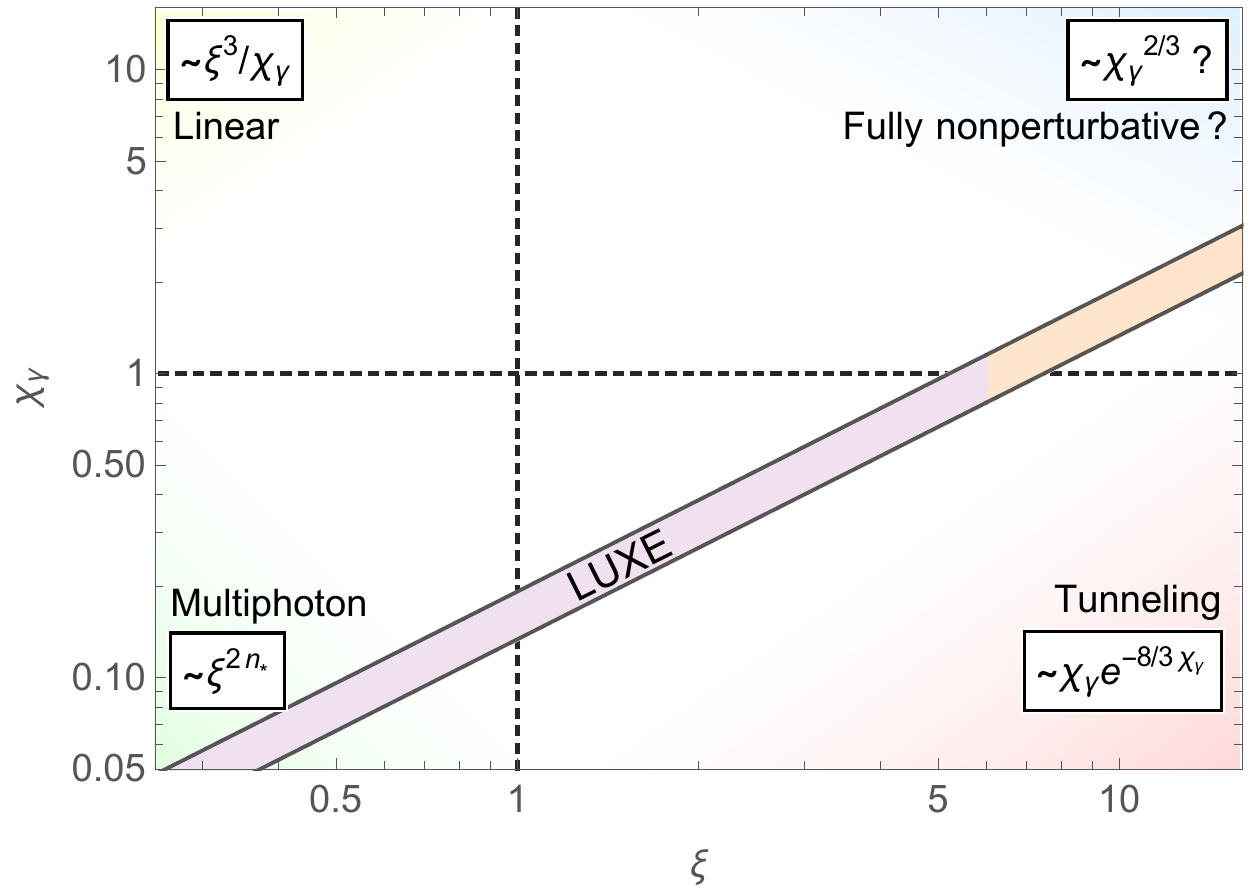}
\caption{Left: The dependency of probability for the Breit-Wheeler process on the intensity parameter $\xi$ for a probe photon colliding at 17.2 degrees with otherwise standard laser pulse parameters. The blue dashed lines indicate multiphoton scaling and the plot markers are the analytical QED plane-wave results for a photon energy of $16.5\,\trm{GeV}$. Right: the parameter region LUXE will probe, compared to the asymptotic scaling of the Breit-Wheeler process at large and small $\xi$ and $\chi$ parameters.
}\label{fig:NBWplot1}
\end{figure}

The effective mass dependency is a signature of non-perturbativity at small coupling. This only becomes apparent when $\xi^{2}\sim1$. When $\xi^{2} \ll 1$, the Breit-Wheeler process proceeds perturbatively, via the ``multiphoton'' process, where the probability scales as $\prob\propto \xi^{2n_{\ast}}$. This is demonstrated by the LMA in Fig.~\ref{fig:NBWplot1}.

As $\xi$ increases past $\xi\approx 1$ in Fig.~\ref{fig:NBWplot1}, the ``turning of the curve'' away from the perturbative multiphoton scaling dependency, is a signature of the non-perturbative dependency on field strength. The LCFA result is plotted as a comparison but only starts to become a good approximation when $\xi$ is large. When $\xi\gg1$ and $\chi_{\gamma} \ll1$ the Breit-Wheeler process demonstrates tunnelling-like behaviour. In a constant crossed field, the scaling of the rate for $\chi_{\gamma}\ll1$ obeys $\sim\chi_{\gamma}\exp(-8/3\chi_{\gamma})$, and since $\chi_{\gamma} \propto \sqrt{\alpha}$, is non-perturbative in the charge-field coupling in an analogous way to the Schwinger effect \cite{Hartin:2018sha}. However, in the Schwinger effect pair-creation is spontaneous whereas in the Breit-Wheeler case the process is stimulated by a high-energy photon. The LUXE experiment will probe an area of parameter space that is somewhere between these different asymptotic scalings, as illustrated in \figref{fig:NBWplot1}.
\begin{figure}[htbp]
\centering
\includegraphics[width=0.625\linewidth]{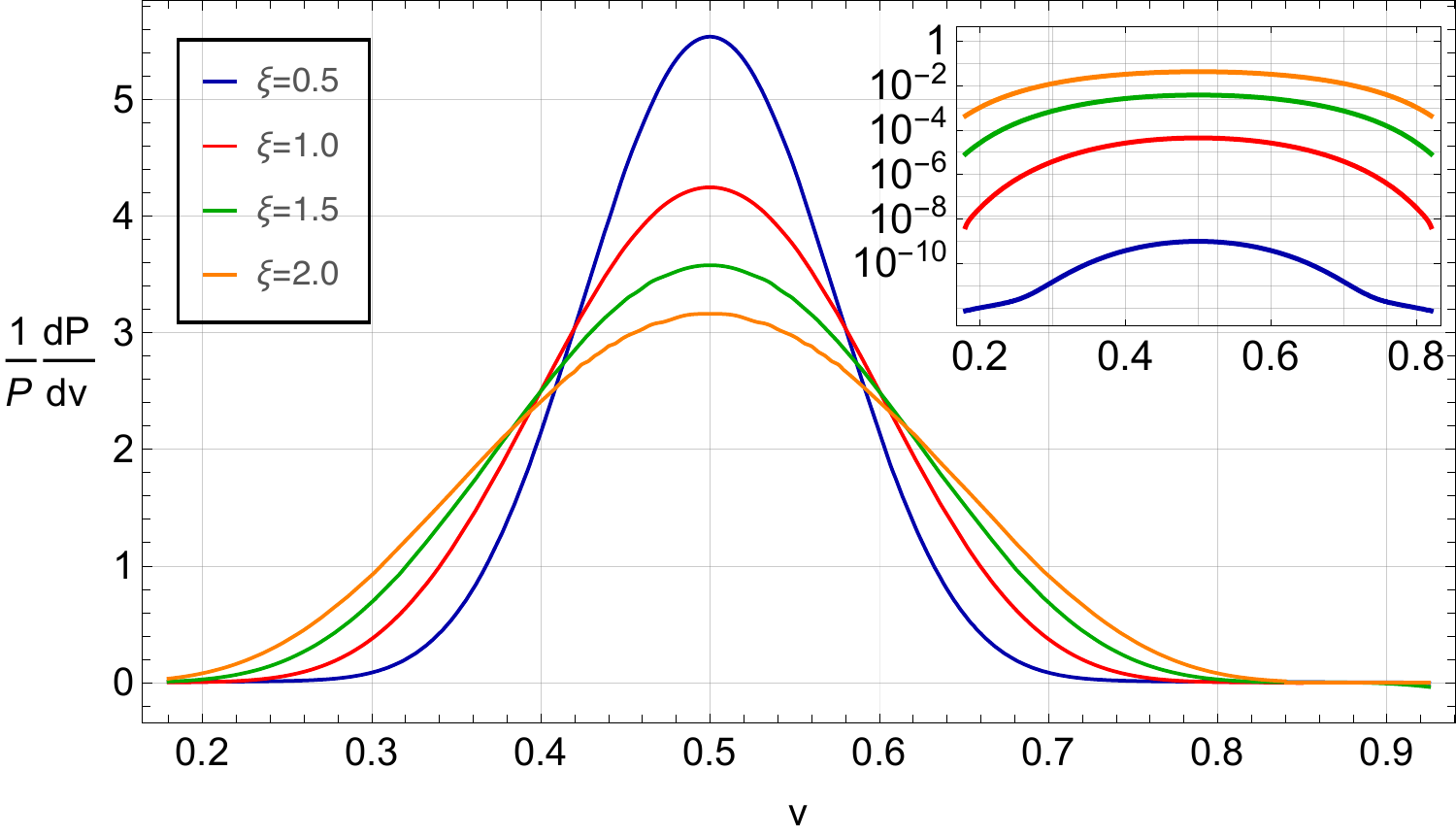}\hfill
\includegraphics[width=0.3\linewidth]{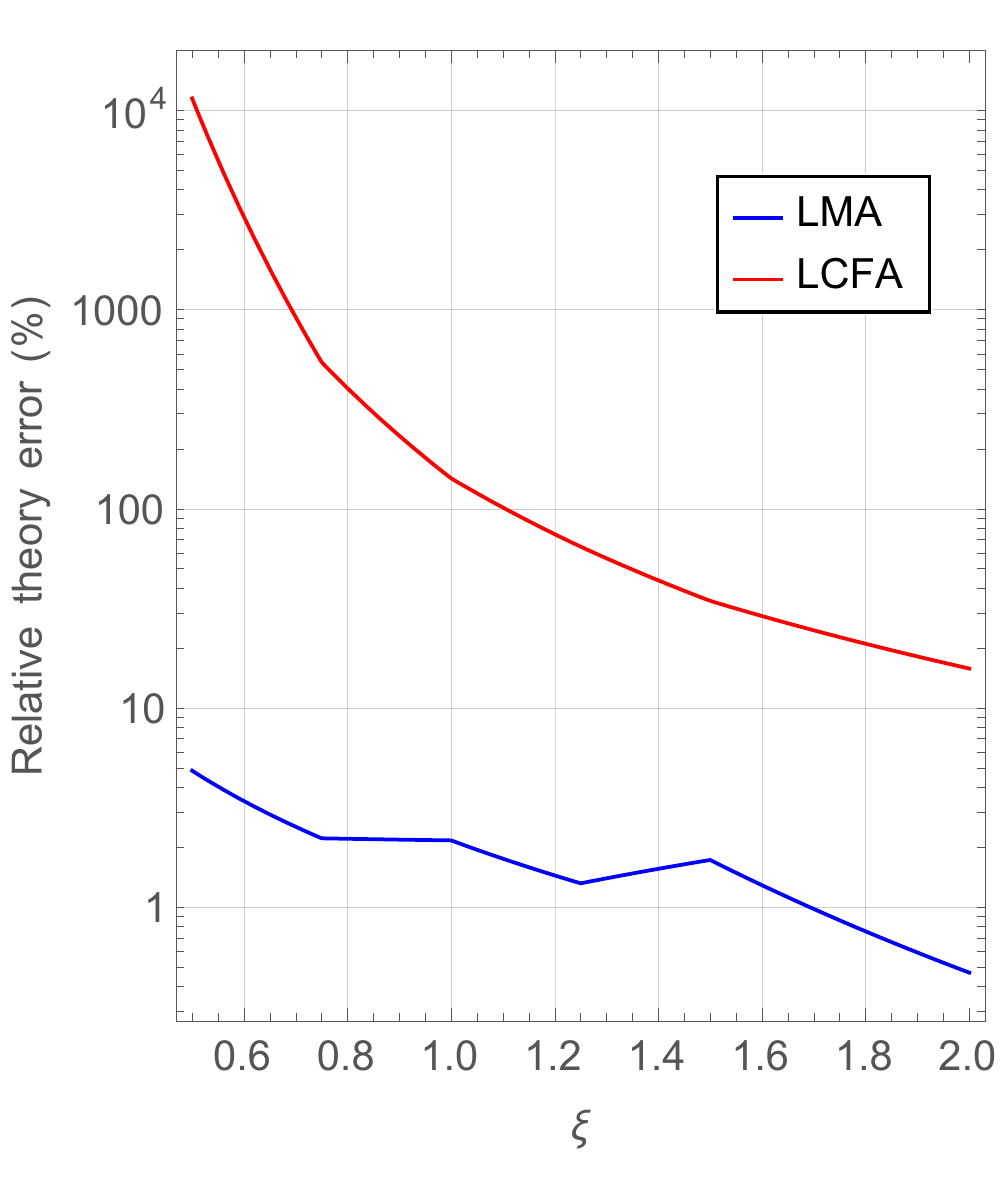}
\caption{
Left: Breit-Wheeler spectra from QED for a $16.5\,\trm{GeV}$ photon, normalised by total probability (we recall $v$ is the lightfront momentum fraction of the emitted electron). Inset: Non-linear Breit-Wheeler spectra without normalisation. Right: Absolute value of the relative error of total yield of photons predicted by the LMA and LCFA from simulation \cite{lma2} compared with the analytical QED plane-wave result.}\label{fig:NBWspec1}
\end{figure}

For the parameters probed by the LUXE experiment, the Breit-Wheeler spectrum is symmetric around $v=0.5$ meaning that the photon's lightfront momentum is shared equally by the electron and positron. However, as the intensity parameter is increased, so too does the width of the spectrum, leading to a broader lightfront momentum distribution of electrons and positrons as shown in \figref{fig:NBWspec1} (where $v=\vkap \cdot P'/\vkap \cdot K'$ is the lightfront momentum fraction of the produced electron where $P'$ ($K'$) is the emitted electron (incident photon) momentum). The LMA is found to be significantly more accurate than the LCFA, particularly when $\xi$ is reduced below $\xi=1$, as expected.

\begin{figure}
\centering
\includegraphics[width=\columnwidth]{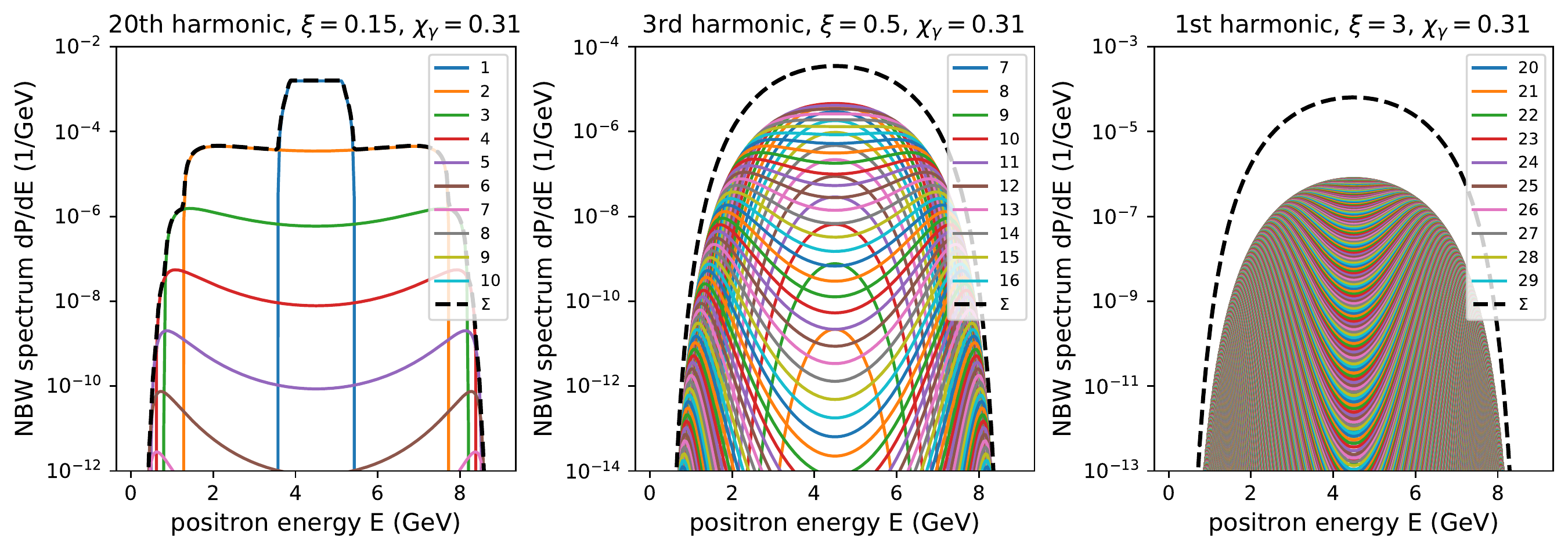}
\caption{
Non-linear Breit-Wheeler pair production spectra 
for different $\xi$ at constant $\chi_\gamma = 0.31$ shows the transition from the perturbative multi-photon regime with distinct ``edges'' (left) to the non-perturbative tunnelling regime (right).
Numbers in the legend refer to the number of laser photons absorbed to produce the pair. It shows that a very large number of laser photons are required in the tunnelling regime.
}
\label{fig:NBWspec2}
\end{figure}

The transition from the multi-photon to the tunnelling regime also shows up in the shape of the spectra, see Fig.~\ref{fig:NBWspec2}. In the multi-photon regime, the spectra show distinct edges related to the $n$th harmonic channels, which is kinematically similar to the first harmonic edge in the Compton process.
In the tunnelling regime, those edges cannot be observed since too many individual channels contribute.
In order to make a direct experimental observation of the Breit-Wheeler edges possible at LUXE it would be necessary to produce pairs at $\xi \lesssim 1$ while keeping  $\chi_\gamma$ reasonably large. This could be achieved, i.e., by using high-order harmonic generation \cite{HHG:Bulanov,HHG:Pukhov,HHG:Dromey}.
Assuming the collision of a 9 GeV gamma-ray photon one would need the 19th laser harmonic to overcome the \emph{linear} pair production threshold. It should be emphasised that the high-harmonic light can be expected to have $\xi\ll1$. However, since in this regime one probes the linear regime, the pair production cross section is independent of $\xi$.
\newline

\begin{table}
    \centering
    \begin{tabular}{cccc} \toprule
    $\tilde{n}$-th harmonic & $\xi$ & $\chi_\gamma$ &  
    estimated pairs per gamma photon
    \\ \midrule
     20  & 0.15 & 0.3 & $3.6 \cdot 10^{-4}$ \\
     3   & 1 & 0.3    & $1.3 \cdot 10^{-5}$ \\
     1  & 3 & 0.3     & $2.3 \cdot 10^{-5}$ \\ \bottomrule
    \end{tabular}
    \caption{
    Pair production probability for the collision of a 9 GeV $\gamma$-ray photon with the $\tilde{n}$-th harmonic of the laser (assuming a 20 cycle pulse) at an angle of $17.2$ degrees at the same value of $\chi_\gamma$, but different $\xi$.}
    \label{tab:NBW}
\end{table}

The Breit-Wheeler process is planned to be measured at LUXE in two different configurations. 
One setup will collide the electron beam with a solid target to generate high-energy bremsstrahlung photons, which will subsequently collide with the main laser pulse and decay into pairs. 
A complementary setup will first collide the electron beam with a weaker laser to generate a narrow bandwidth source of high-energy photons via inverse Compton scattering (ICS) \cite{PhysRevLett.10.75}, and these photons will then collide with the main laser pulse to decay into pairs.
The bremsstrahlung setup has the advantage that it can generate photons with the highest energies, but has a wide angular and energy spread that limits accuracy. In contrast, the ICS setup is better collimated, and can achieve a fuller spacetime overlap with the laser focus at the interaction point, but leads to a lower yield of pairs since it cannot reach as high photon energies.

\begin{minipage}{0.4\linewidth}
\captionsetup{type=figure}
\includegraphics[width=0.99\linewidth]{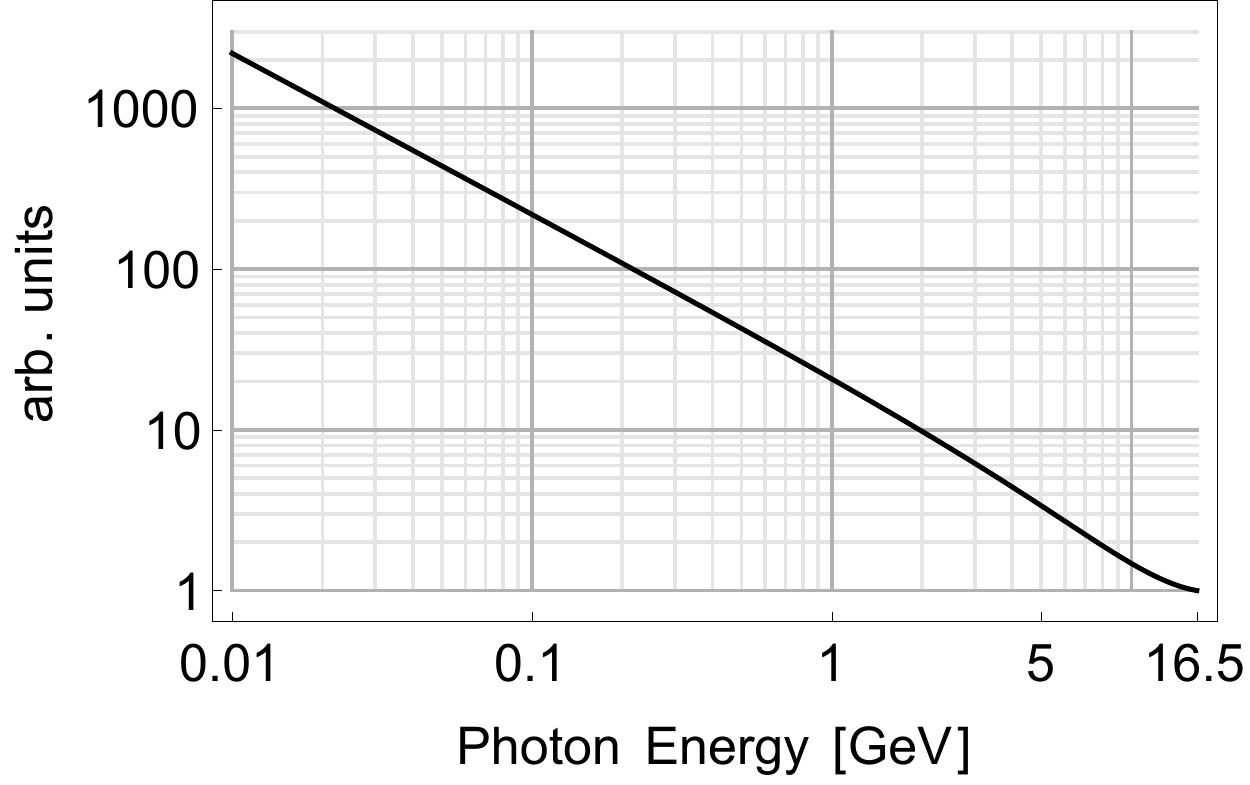}
\captionof{figure}{Example thin-target bremsstrahlung spectrum for $16.5$~GeV electrons.}\label{fig:BremPlot}
\end{minipage}
\hfill
\begin{minipage}{0.5\linewidth}
An example of an approximated thin-target bremsstrahlung spectrum is plotted in the Fig.~\ref{fig:BremPlot}, which takes the form \cite{RevModPhys.46.815}:
\[
\omega_{B} \frac{dN}{d\omega_{B}} = \left[\frac{4}{3}-\frac{4}{3}\frac{\omega_{B}}{\epsilon_{e}}+\left(\frac{\omega_{B}}{\epsilon_{e}}\right)^{2}\right]\frac{X}{X_{0}},
\]
where $\omega_{B}$ is the bremsstrahlung photon energy, $\eps_{e}$ the initial electron energy, $X$ is the target thickness, $X_{0}$ is one radiation length and $X\ll X_{0}$.
\end{minipage}

\subsection{Inverse Compton Source}
\label{sec:science:ics}

Here, we discuss briefly the prospects and benefits of probing the strong focussed laser pulse with multi-GeV $\gamma$-rays generated by an ICS. The benefits of such a Compton source are that it can provide a \emph{mono-energetic} beam of $\gamma$-rays with \emph{tunable energy} at the interaction point
which can be \emph{highly polarised}. 

\begin{figure}[htbp]
    \centering
    \includegraphics[width=0.6\columnwidth]{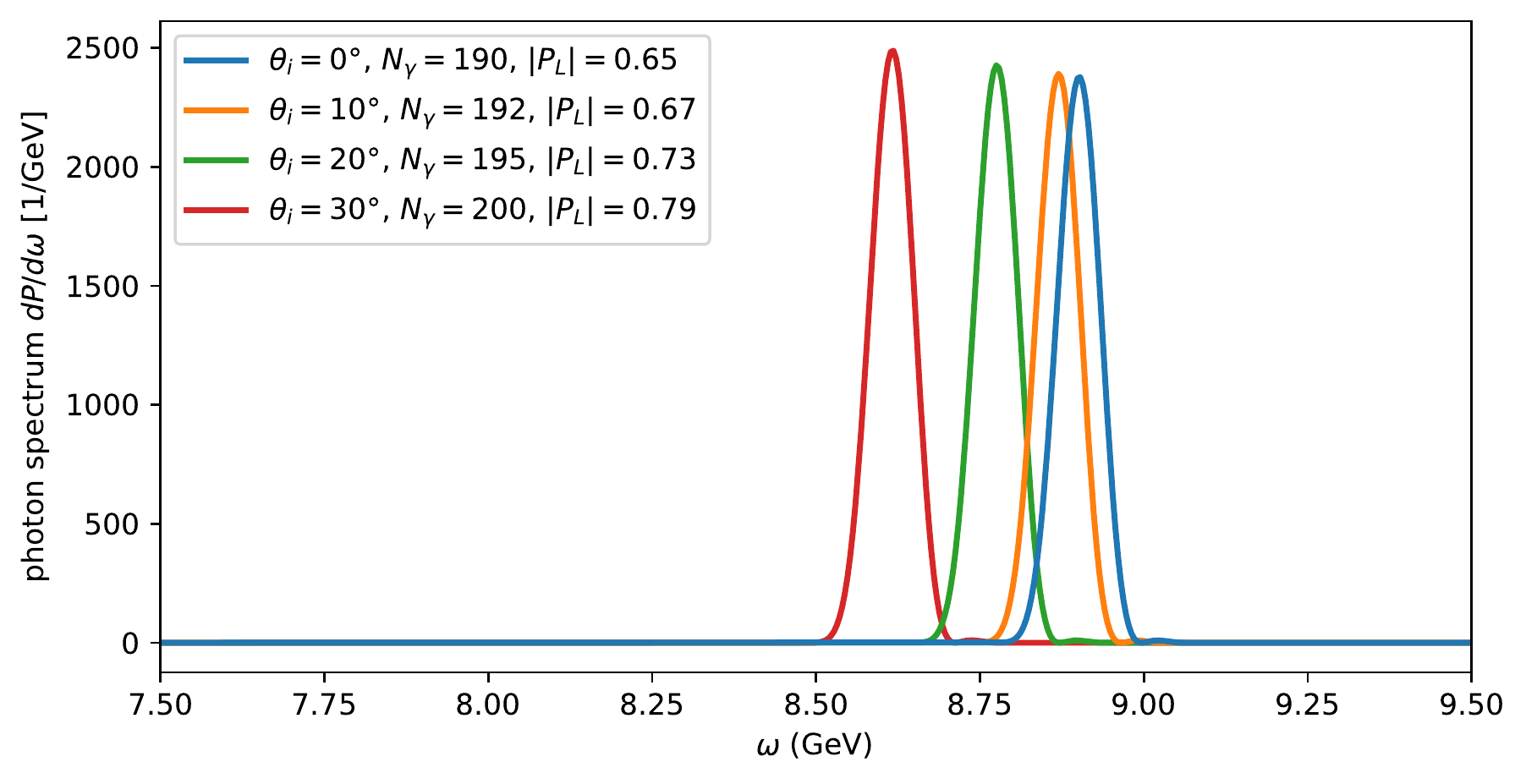}
    \caption{Inverse Compton source photon spectrum at the IP for various electron-laser collision angles $\theta_i$, with $N_\gamma \sim 200$ photons per shot arriving at the strong-field IP (for details see text), and with linear polarisation degree $|P_L|$. 
       } 
    \label{fig:ICS}
\end{figure}

One possibility to realise such a Compton source would be to employ a second laser pulse (possibly split off from the main laser) and utilise the back-scattering of frequency-tripled laser photons off of the XFEL.EU electron beam. We will refer to the main laser pulse that probes the IP as the ``IP laser''. This source would operate in the linear Compton scattering regime, $\xi_\mathrm{ICS}\ll1 $. 
The frequency of the Compton scattered photons is given by

\begin{align}
\omega' = \frac{2 (1+\cos \theta_{i} )\gamma^2 \omega_0 }{1+ 2\eta_\mathrm{ICS} + \gamma^2 \theta^2 + \xi^2_\mathrm{ICS}/2 }\,,
\end{align}
which has a strong dependence on the scattering angle $\theta$, which in turn allows one to select the photon energy by scanning the position of the IP laser (with a focal width of  $\lesssim 10\, \mu$m) inside the gamma-ray beam (which has an angular divergence of $1/\gamma$, thus a typical  size of $\Delta L_\mathrm{ICS-IP}/\gamma \sim 7.5\, \mathrm{m}/\gamma \sim 232\, \mu \mathrm{m}$ at the strong-field IP). For a laser with $\omega_0=1.55$~eV and an electron beam energy of $16.5 \units{GeV}$, one obtains $\eta_\mathrm{ICS}=0.2$, and can reach a photon energy of  $4.6\units{GeV}$ for head-on collisions. However, when using a frequency-tripled Ti:Sa laser ($\omega_0=3\times 1.55 \units{eV}$ and $\eta_\mathrm{ICS}=0.59$), the maximum photon energy achievable is $\simeq 9 \units{GeV}$ for a head-on collision. The dependence of the energy on the electron beam-laser collision angle is relatively weak for $\theta_i \leq 20\deg$,  with the optimum at $\theta_i=0$. In Fig.~\ref{fig:ICS}, example spectra are plotted, to be contrasted with the typical bremsstrahlung spectrum in Fig.~\ref{fig:BremPlot}.
Althouth the ICS spectrum is at a considerably lower than the maximum energy achievable with a bremsstrahlung source, the ICS photons are mono-energetic and highly linearly polarised (above 75 \%), (see also Refs.~\cite{King:2020btz,Tang:2020xlj}).
The quantum energy parameter is $\eta_{\gamma}\simeq 0.1$ (for a head-on collision at the strong-field IP). 

By assuming that a $25\,\trm{fs}$ pulse at the 3rd harmonic ($\lambda = 266$ nm) with $\xi_\mathrm{ICS}=0.1$ is backscattered off 100 pC beam charge, the number of on-axis backscattered photons can be estimated as $dN_\mathrm{ICS}/d\Omega \approx 3.8\times10^{14}\, {\rm photons/sr^{-1}}$. Thus, about 200 ICS photons interact with a $w_0=3\,\mu$m strong-field IP 7.5 m downstream of the ICS (aperture $5\cdot 10^{-13}$ sr).

The on-axis bandwidth of the $\gamma$-rays will most likely be dominated by the laser bandwidth and the electron beam emittance.
The latter contribution is due to the angular spread of the electron beam which in turn depends on the focusing of the electron beam
at the ICS interaction point. Quantifying the emittance contribution thus requires further detailed designs considerations and investigations.
An alternative path for generating mono-energetic ICS photons could be seen in using the higher harmonics from a \emph{non-linear} Compton scattering source with optimised chirped laser pulses \cite{Seipt-Rykovanov}. 

This proposed ICS photon source would have a number of benefits for LUXE to fully harness its physics opportunities and to reach its scientific goals. Operating an ICS would allow for otherwise unattainable levels of precision in strong-field QED experiments by interacting monochromatic gamma-rays with the strong-laser IP, hence having a collision with a well-defined in-state. (As outlined above, the ICS photons have well defined energy-angle correlation which allows the photon energy to be scanned by changing the relative position of the laser focus inside the photon beam.) The precise knowledge of the centre-of-mass energy of the incident photons would allow for a precise investigation, for instance, of the closing of $n$-photon channels in NBW as a function of energy, similar to the observation of non-linear Compton edges, or the intensity dependent effective mass  of the produced pair.
A better-characterised photon source allows for a more precise measurement of strong-field QED.

Since the ICS photons are highly polarised (with potentially tuneable polarisation state) this opens up an avenue to experimentally investigate, for the first time, the polarisation dependence of Breit-Wheeler pair production \cite{Breit:1934zz}. Both in the linear and the multiphoton regime of the Breit Wheeler process, one would expect a strong deviation in particle yield depending whether the ICS photon polarisation is parallel or perpendicular to the strong-laser polarisation. The measurement of the polarisation dependence of the Breit-Wheeler process is directly related, via the Optical Theorem, to the process of photon-photon or light-by-light scattering \cite{Toll:1952rq}. Furthermore, a highly polarised photon source can potentially lead to stronger exclusion bounds of new physics, for spin-dependent processes such as the generation of ALPs.

\subsection{Trident Pair Creation}
\begin{wrapfigure}{l}{5cm}
\centering
    \includegraphics[width=4cm]{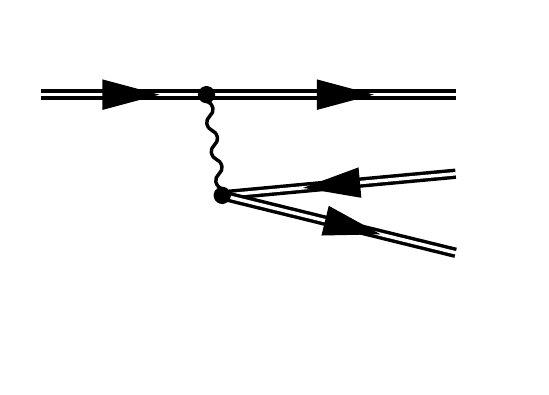}
    Trident Process
    \label{fig:tridentfeynman}
\end{wrapfigure}

Another process that will be studied at LUXE is non-linear trident pair production \cite{ritus72,hu10,ilderton11,king13b,mackenroth18,Dinu:2017uoj,King:2018ibi,Hernandez_Acosta_2019,Dinu:2019pau,Dinu:2019wdw}, which is the production of an electron-positron pair from an electron (or positron), in the laser field.
This includes the inset diagram as well as an exchange diagram, with swapped outgoing electrons and a relative minus sign.  In general there is no unique way of separating this process into different parts (in particular for a short pulse with $\xi \sim \eta \sim 1$). However, for longer pulses, such as at LUXE, there is one separation that depends on the pulse duration $\Phi$ that is useful to consider, and will be adopted in the simulation framework. 

One part of the total probability, $\prob_{\tri}^{(2)}$, scales with $\Phi^2$, and one part, $\prob_{\tri}^{(1)}$, with $\Phi$. When this separation is applied within the LCFA, $\prob_{\tri}^{(2),\tsf{LCFA}}$, an exact factorisation is found, in terms of the rate of first-order processes of the Compton emission of a polarised photon, $\rate_{\gamma,j}^{\tsf{LCFA}}$, followed by the decay of a polarised photon via the Breit-Wheeler process, $\rate_{e,j}^{\tsf{LCFA}}$:

\begin{equation}
    \prob_{\tri}^{(2),\tsf{LCFA}}= \sum_{j=1}^{2}\int_{-\infty}^{\infty}d\phi_{1}\int_{\phi_1}^{\infty} d\phi_{2}\int_{0}^{1} ds ~ \frac{\partial \rate^{\tsf{LCFA}}_{\gamma,j}(\phi_1,s)}{\partial s}\,\rate_{e,j}^{\tsf{LCFA}}(\phi_2,s),
\label{eqn:triLCFA1}
\end{equation}
where the sum is over two transverse polarisation states of the intermediate photons. For this reason, $\prob_{\tri}^{(2)}$ is referred to as the ``two-step''/``avalanche''/``incoherent'' process, whereas 
$\prob_{\tri}^{(1)}$ is the ``one-step''/``coherent'' process. 
The leading-order behaviour for $\xi\ll1$ of each term is $\prob_{\tri}^{(2)}\sim \xi^4$, $\prob_{\tri}^{(1)}\sim \xi^2$,  so the one-step term can in principle dominate the probability. However, if the pulse duration is long enough, the two-step term can compensate for this.

\begin{figure}[ht]
\centering
\makebox[\textwidth][c]{
\hspace{-0.5cm}
\includegraphics[width=0.225\linewidth]{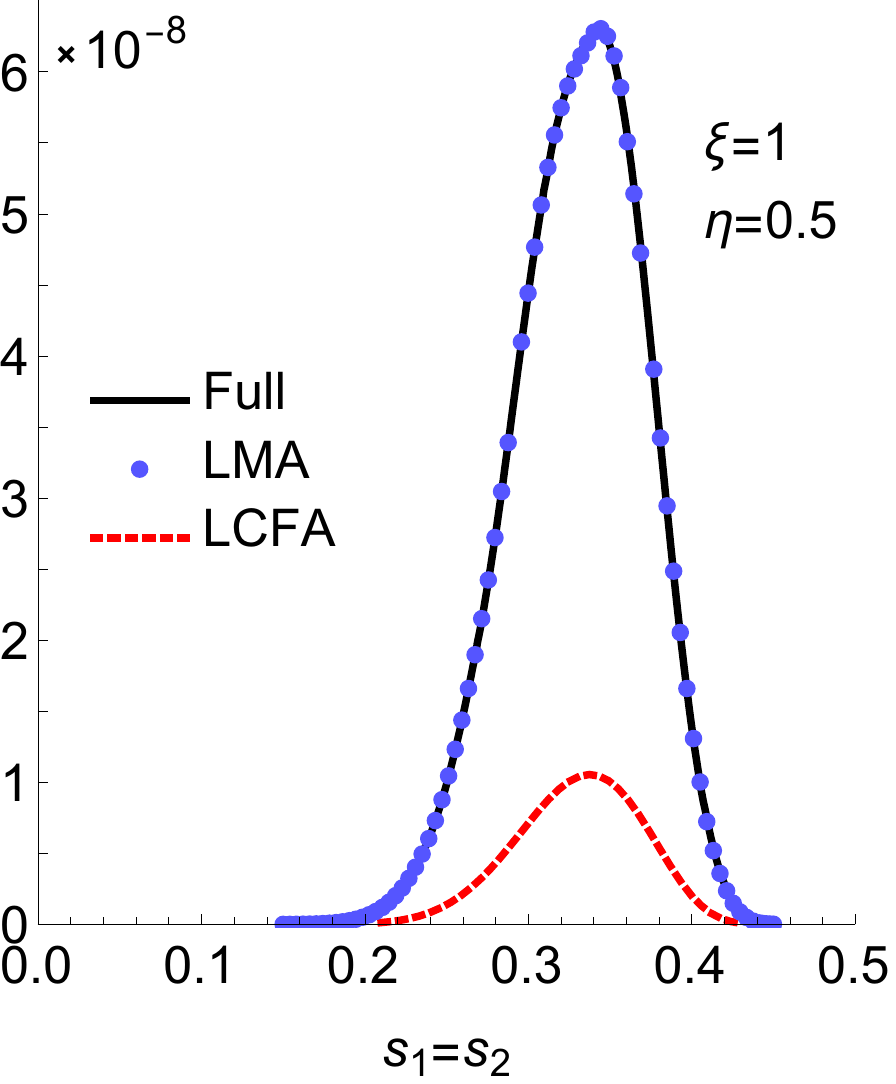}
\includegraphics[width=0.225\linewidth]{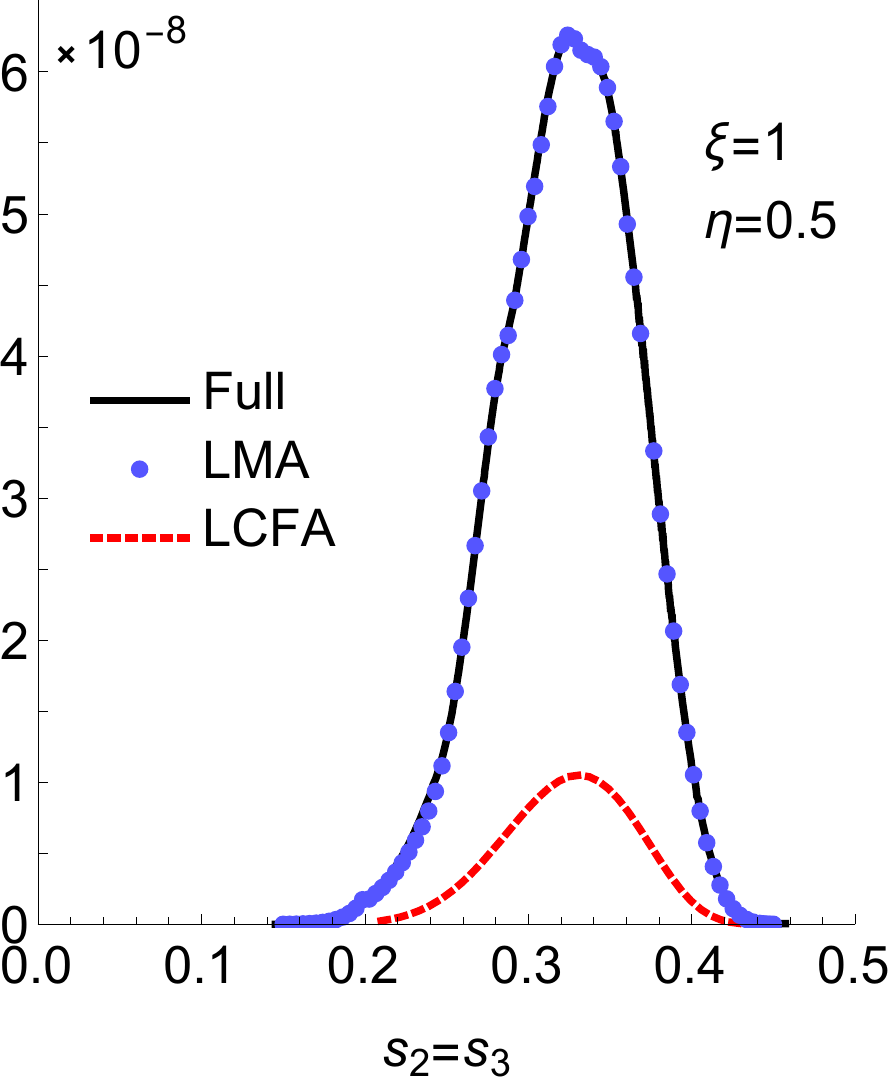}
\includegraphics[width=0.225\linewidth]{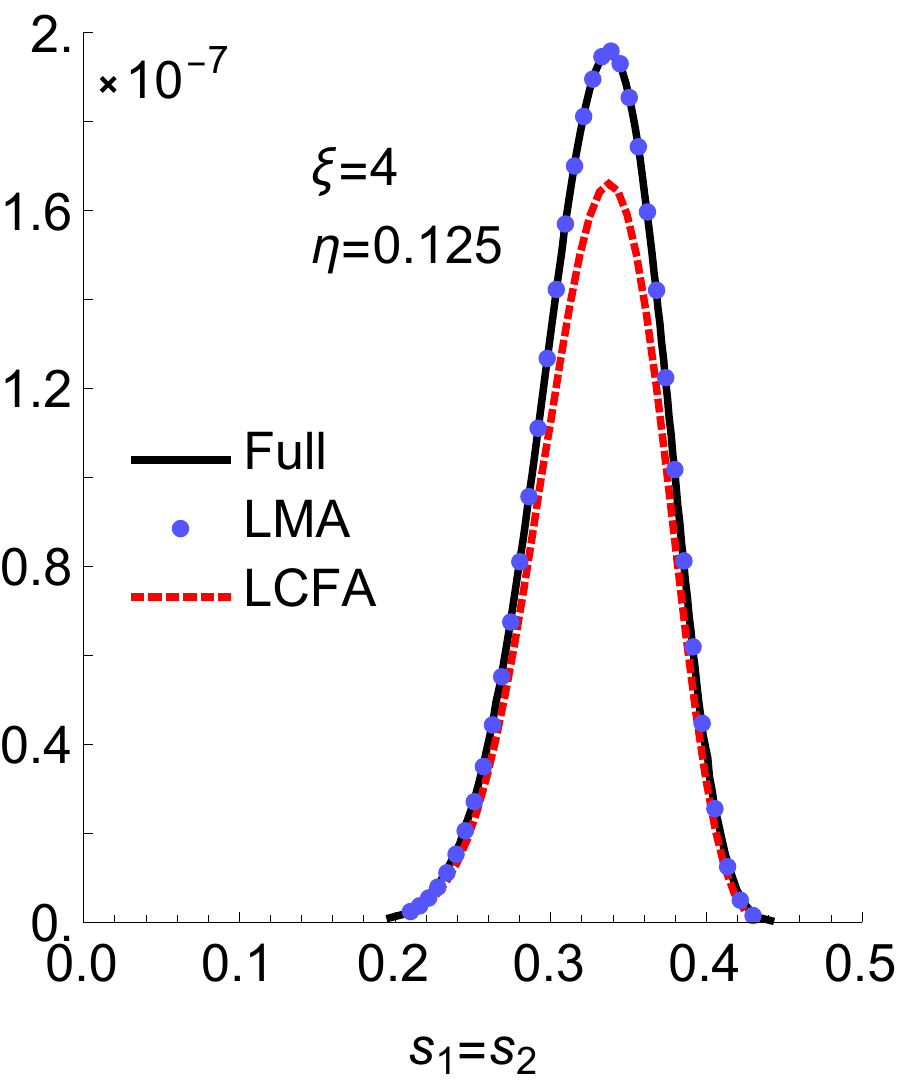}
\includegraphics[width=0.225\linewidth]{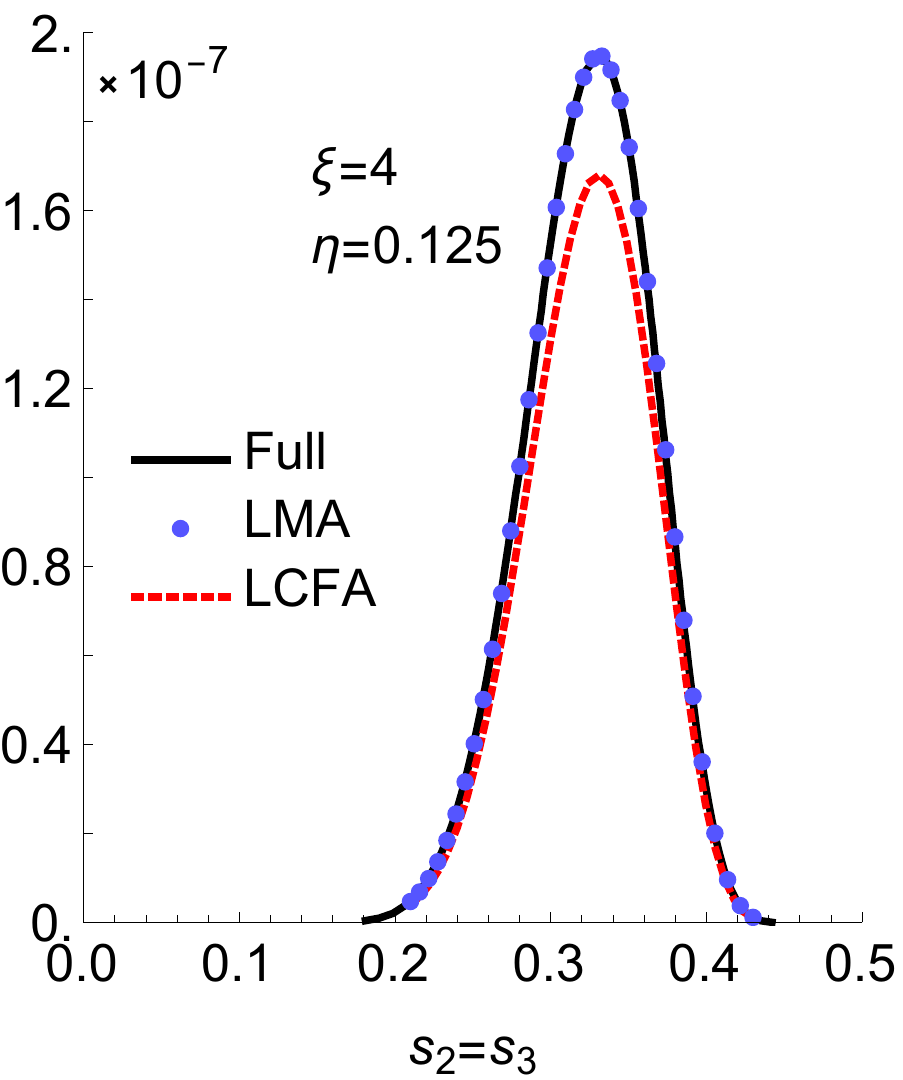}}
\caption{Longitudinal momentum spectrum of trident. Black solid lines show the full QED plane-wave result  (i.e. numerical evaluation of the analytical expression) for two-step $+$ one-step; dashed curves give the LCFA of the two-step; dots give the LMA of the two-step. The full numerical results come from~\cite{Dinu:2019pau}, and the LMA from ~\cite{Torgrimsson:2020gws}.}
\label{tridentSpectrum}
\end{figure}

In Fig.~\ref{tridentSpectrum}, we plot the analytical QED plane-wave result for the two-step term in a circularly-polarised plane wave potential with Gaussian envelope $a(\phi)=(\xi/\sqrt{2})\{\sin\phi,\cos\phi\}\exp\left(-\phi^2/\Phi^2\right)$ with $\Phi=80$ (corresponding to a pulse duration of around $35\,\trm{fs}$) for the two cases $\xi=1$ and $\xi=4$. Since the pulse has many cycles, the two-step part dominates and the one-step terms are negligible.

The ratio of outgoing to incoming particle lightfront momenta is denoted using the energy parameter $\eta$ and $s_i=\eta_{i}/\eta$, with $s_1$ and $s_2$ for the two electrons and $s_3$ for the positron. 
We see that when $\xi=1$, the LCFA makes large errors in estimating the two-step process, whereas the LMA approximates the spectrum very well. At $\xi=4$ the LCFA approximation is better, but even here, the LMA is noticeably more accurate.

On the basis of this analysis, iterated LMA rates for the first-order processes of non-linear Compton scattering and Breit-Wheeler pair creation, will be used to approximate higher-order processes such as trident and multiple Compton emission. (See Appendix~\ref{app:theory} for more technical information about neglecting the one-step process and how a ``gluing'' approach of lower-order processes can be implemented.)

\subsection{Probing Physics Beyond the Standard Model at LUXE}
\label{sec:NPLeff}

Various scenarios for probing BSM physics using the non-perturbative coupling to QED that occurs in strong electromagnetic fields, have been investigated in the literature \cite{gies09,doebrich10,doebrich12,VillalbaChavez:2012bb,King:2018qbq,Villalba-Chavez:2014nya,Dillon:2018ouq, Villalba-Chavez:2015xna,Villalba-Chavez:2016hht,Burton:2017bxi, Dillon:2018ypt, King:2019cpj, Huang:2020lxo}. BSM opportunities specific to the LUXE setup will be presented in Ref.~\cite{LUXEBSM}. Here we focus on scalars and pseudoscalars (ALPs). We denote scalars, $\scal$, with mass $m_{\scali}$ and ALPs, $\alp$, with mass $m_{\alpi}$, which have couplings to photons and electrons. 

Their effective interactions are given by 
    \begin{align}
    	\label{eq:La}
    	\cL_{\alp}
    &=	\frac{1}{4\Lambda_{\alpi}}  \alp F_{\mu\nu} \widetilde{F}^{\mu\nu} 
        + ig_{\alpi e} \alp \overline{\psi} \gamma^5 \psi \, , \\
    	\cL_{\scal}
    &=	\frac{1}{2\Lambda_{\scali}}  \scal F_{\mu\nu} F^{\mu\nu} 
        + g_{\scali e} \scal \overline{\psi}  \psi \, ,
    \end{align}
    with $\widetilde{F}_{\mu\nu}=\frac{1}{2}\epsilon_{\mu\nu\alpha\beta}F^{\alpha\beta}$ and $\Lambda_{\scali}$ and $\Lambda_{\alpi}$ being the scales of new physics. 
        The ALP and scalar decay rates to photons are given by $\Gamma_{\alp\to\gamma\gamma}=m^3_{\alpi}/(64\pi\Lambda^2_{\alpi})$ and $\Gamma_{\scal\to\gamma\gamma}=m^3_{\scali}/(64\pi\Lambda^2_{\scali})$, respectively. 
        
        Two new physics production mechanisms are considered 
    \begin{itemize}
    \item \textbf{Primary production}: 
    	ALPs, scalars and mCPs are produced at the LUXE IP by the processes:
    	\[
    	e^{-} +n\gamma_{L} \to e^{-}+\alp; \qquad e^{-} +n\gamma_{L}  \to e^{-}+\scal; \qquad \gamma \to \mcp^{+}\mcp^{-}; \qquad e^{-}+n\gamma_{L}  \to e^{-}+\mcp^{+}\mcp^{-}.
    	\]
    	The production rate is expected to be enhanced due to non-perturbative interactions with the laser. 
    \item \textbf{Secondary production}: 
    	leverage LUXE as a GeV photon source. 
    	The outgoing Compton photons in the $e$-laser mode are used as a high intensity photon source 
    	and scatter on a nucleus target, $N$,  to produce ALPs or scalars via Primakoff production, 
    	$\gammaC \,N \to a \,N$ or $\gammaC \,N \to \phi \,N$. 
    	For example, assuming the LUXE \phasetwo parameters of long laser pulse, for an initial electron, the number of Compton photons with energy above $1\,$GeV is $\sim 3$, 
    	while the number of bremsstrahlung photons is about $\sim 0.03$, 
    	where a Tungsten target of 1\% radiation length is considered.
    	Therefore, we expect an enhancement of $\cO(10^2)$ in the photon flux, see also Fig.~\ref{fig:photon_hics_brems} for a comparison of the photon energy spectra. 
    \end{itemize}
    The maximum mass accessible in the primary production is limited by the centre-of-mass energy of $\cO(\MeV)$.
    In contrast, in the secondary production, masses of a few hundred MeV can be probed as the centre-of-mass energy is much higher and it is determined by the invariant mass available in the collision of the Compton photon with the particles inside the beam dump.
    
\begin{figure}[t]
    	\begin{center}
    	\includegraphics[width=0.65\textwidth]{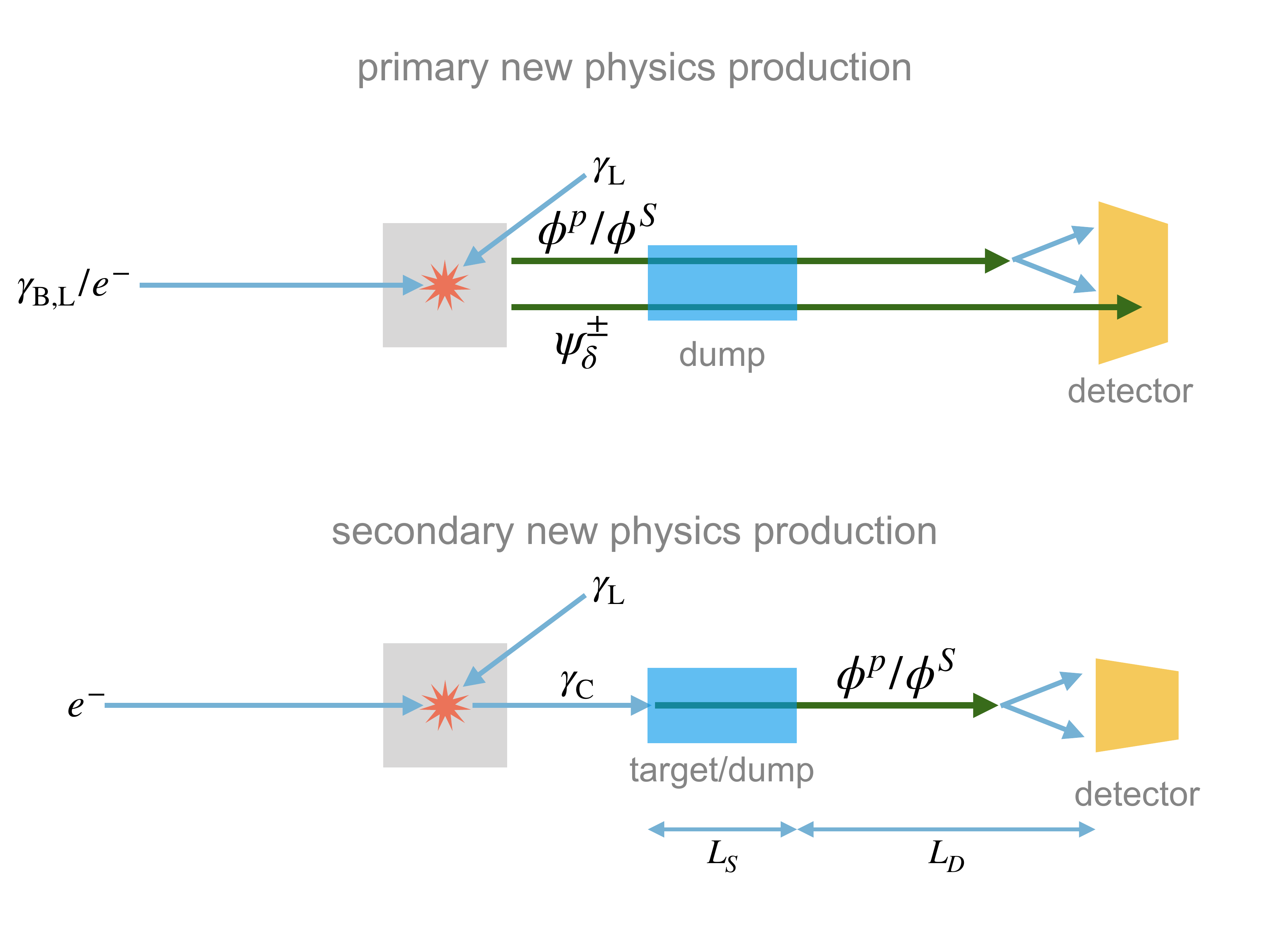}
    	\caption{
    	A schematic design of the LUXE new physics~(NP) search setup. 
    	Top: primary production, where the NP is produced at the IP.
    	Bottom: secondary production, where the high rate Compton photons from LUXE collide with a target/dump of size $L_S$, to produce ALPs/scalars (denoted as $\alp/\scal$ in the figure). 
    	The detector is located at a distance of $L_D$ from the target.
    	}
    	\label{fig:scheme}
    	\end{center}
    \end{figure}

    We are interested in new particles with a long enough lifetime $\gsim 1$~ns, thus, the detector is shielded and located at a sizeable distance from the production point, in a beam-dump like setup.
    Figure~\ref{fig:scheme} shows a schematic design of the proposed setup. 
    Below, we focus on secondary production of ALPs and keep the primary production for future work. All results shown for ALPs also apply to scalars.

    The sensitivity to probe the photon-ALP coupling in the secondary production is estimated using the setup shown in Fig.~\ref{fig:scheme}.
The Compton photons from the primary \elaser collisions collide with a dump of thickness $L_S$,
    which is located at a finite distance from the IP. This dump blocks all the Compton photons. However, ALPs can be generated by a Primakoff process, $\gammaC \, N\to \alp\,N$ and decay to two photons after a finite distance. A detector is placed at a distance of $L_D$ from the target to detect the decay photons. This setup probes ALPs with masses of order MeV-to-GeV, which is a range that has attracted significant attention in recent years~\cite{Marciano:2016yhf,Jaeckel:2015jla,Dobrich:2015jyk,Izaguirre:2016dfi,Knapen:2016moh,Mariotti:2017vtv,Bauer:2017ris,CidVidal:2018blh,Bauer:2018uxu,Harland-Lang:2019zur,Ebadi:2019gij,Mimasu:2014nea,Brivio:2017ije,Aloni:2018vki,Aloni:2019ruo,Bauer:2020jbp,Chala:2020wvs,Florez:2021zoo}.

    The expected number of ALP events can be estimated as, \textit{e.g.}~\cite{Berlin:2018pwi,Chen:2017awl}
    \begin{align}
    	\label{eq:NaSec}
    	N_{\alpi}
    	\approx
    	N_e N_{\rm pulse} \frac{\rho_N X_0}{A_N m_0} \int \!\! d \varepsilon_{\gammaC}
    	\frac{d N_{\gammaC}}{d\varepsilon_{\gammaC}}  \sigma_{\alpi}
    	\left( e^{-\frac{L_S}{L_{\alpi}}} - e^{-\frac{ L_D+L_S}{L_{\alpi}}} \right) A \, ,
    \end{align}
    where $\varepsilon_{\gammaC}$ is the $\gammaC$ energy, the ALP momentum is $p_{\alpi} \approx \sqrt{\varepsilon_{\gammaC}^2-m^2_{\alpi}}$, $L_{\alpi} \equiv  c \tau_{\alpi} p_{\alpi}/m_{\alpi} $ is the ALP propagation length, and $\tau_{\alpi} = 1/\Gamma_{\alp\to\gamma\gamma}$. 
    $\sigma_{\alpi}$ is the ALP Primakoff production cross section as function of $\varepsilon_{\gammaC}$ (see \textit{e.g.}~\cite{Aloni:2019ruo}) and $A$ is the acceptance and efficiency of the detector, which is a function of the final photon momentum.

\subsection{Theory Summary}
\begin{figure}[ht]
    \begin{center}
    \includegraphics[width=0.7\textwidth]{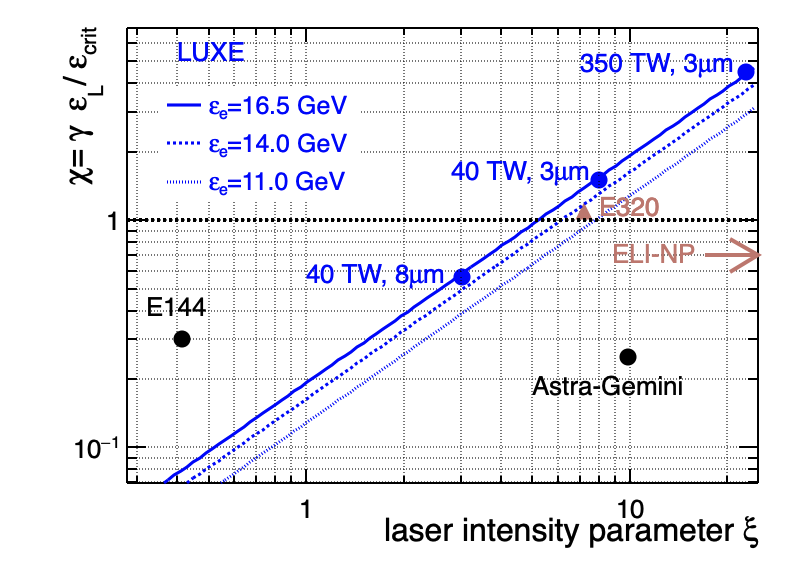}
    \caption{Quantum parameter $\chi_e$ vs the intensity parameter $\xi$ for a selection of experiments and facilities. 
    E144 values were taken from Ref.~\cite{Bamber:1999zt}, E320 parameters from Ref.~\cite{e320}, and Astra-Gemini from Ref.~\cite{Poder:2018ifi}. For LUXE, three beam energies are shown as isolines, and two laser focus spot sizes are highlighted for the \phaseone (40 TW) laser and one for the \phasetwo (350 TW) laser. The regime of ELI-NP is also indicated. ELI-NP and E320 are not yet operating while E144 and Astra-Gemini have already published results. }
     \label{fig:Facilities}
    \end{center}
\end{figure}

The LUXE experiment is planning to investigate strong field QED in a hitherto inaccessible parameter regime and will thus explore an uncharted region of the Standard Model of particle physics. A unique feature of LUXE is the combination of high intensity ($\ximax\lesssim 24$) \emph{and} (probe) energy ($\eta_\mathrm{max} = 0.2$) to reach an unprecedented value for the quantum non-linearity parameter of $\chi\lesssim 4.5$ (see Fig.~\ref{fig:Facilities}). 

The LUXE experiment will focus on the non-linear versions of three basic processes: Compton scattering, Breit-Wheeler pair production, and trident pair production (see Table~\ref{tab:candproc} above). The final states of interest are either the emitted photons, pairs and scattered electrons. The experimental setup will include the possibility to look for `exotic' scalar and pseudo-scalar particles coupling to photons. 

In addition to observing the above processes, a central objective will be to obtain the intensity (i.e.\ $\xi$) dependence of the final state emission and production rates. For non-linear Compton scattering (photon emission), this will allow detection of the photon dependent mass shift through the determination of the Compton edge and higher harmonic peaks. For the non-linear Breit-Wheeler process---to be observed in isolation for the first time---one should be able to see the deviation from perturbative behaviour ($\sim \xi^{2n}$) to its non-perturbative modification 
(see Fig.~\ref{fig:NBWplot1} (left panel) above). A similar modification may be induced in the non-linear Compton cross section by restricting the final-state phase space through detector cuts (thus bounding the lightfront momentum fraction $u$ of the photon from below, $u \ge C$ (see Fig.~\ref{fig:cutoff})). 

For the LUXE parameter regime, the non-linear trident process factorises, to a good approximation, into non-linear Compton and Breit-Wheeler processes. In other words, the one-step trident production can be neglected. However, this means that in a non-linear Compton experiment, there will be a non-zero probability to observe pairs as the emitted Compton photons will undergo final state interactions with the laser, creating pairs via Breit-Wheeler. This mechanism has previously been employed by the SLAC E144 experiment, albeit at lower values of $\xi$. It should be distinguished from the Breit-Wheeler process in isolation (with no electrons in the initial state at all).

\subsection{Theory Outlook}

Calculations in the non-perturbative regime are notoriously difficult. To make progress analytically, one typically relies on idealised settings such as the plane wave and external field approximations. Clearly, these will induce errors, the magnitude of which should be controlled. 

The plane wave model, for instance, neglects transverse focussing.
Adopting a realistic configuration such as a Gaussian beam is possible at the price of losing integrability, hence the Volkov solution and the ensuing Furry picture. This may be partly compensated by employing `adiabatic' approximations such as the LCFA and LMA. The size of the error is somewhat difficult to assess, but one can develop a perturbation theory in the beam `aspect ratio', $\Delta = w_0/z_R \lesssim 1/2\pi$, the ratio of beam waist $w_0$ to Rayleigh length, $z_R$, measuring the beam divergence \cite{Narozhny:2000}. The field invariants then become of order $\Delta^2$ which physically is due to the development of longitudinal field components (along the beam) of order $\Delta$. In the limit $\Delta \to 0$, one recovers the null-field plane wave. 
It is currently unclear how $\Delta \neq 0$ would affect local rates for SFQED processes. Numerical simulations using plane-wave LCFA and LMA rates should mostly account for classical effects of focussing, and it has been shown how a ``high energy'' WKB approximate solution of the Dirac equation in a focussed laser background is related to the plane-wave result \cite{DiPiazza:2015xva,DiPiazza:2016maj,DiPiazza:2016tdf}.
Theoretical work to investigate the effect of focussing is underway.  

Regarding the external field approximation, one expects this to become unfeasible when backreaction effects set in. Depletion of the beam due to radiation effects is known to become relevant only when $\xi \gtrsim 10^3$ \cite{Seipt:2016fyu}. The backreaction of the produced pairs and their Coulomb field on various pair creating background configurations has been studied in some depth and is known to induce plasma oscillations \cite{Kluger:1991ib,Kluger:1998bm,Bloch:1999eu,PhysRevE.71.016404}. (For laser backgrounds, see the recent analysis \cite{Smolyansky:2019yma} and references therein.) In view of the small pair densities to be expected for LUXE, any backreaction effects stemming from $e^+e^-\gamma$ plasma effects should be negligible.

Near-future theoretical efforts will attempt to achieve better control of ponderomotive effects associated with transverse focussing and of higher order effects, in particular multi-particle scattering and emission.     

\subsection{Additional Opportunities}
In this section, we summarise some enhancements and further physics that can be probed at LUXE. The general idea is based on using either high-harmonic generation to extend the accessible parameter space of LUXE, or to exploit and investigate the polarisation dependence of the strong-field QED processes.
The latter is interesting with regard to the polarisation of the laser, as well as the gamma-rays used for pair production, and the spin-polarisation of Compton-scattered electrons. The unique conditions found at LUXE will be optimal for a first experimental study of many of these effects.

\subsubsection{Laser Polarisation}
\label{sec:laserpol}
The main analysis and simulation section is based around using a circularly-polarised (CP) laser pulse. An advantage of using circular polarisation, is that it provides one more symmetry than linear polarisation, namely symmetry in rotations around the propagation axis. Although this symmetry is partially broken for a finite pulse, in the locally-monochromatic approximation \cite{Heinzl:2020ynb} used to model many-cycle pulses, which is suitable for the situation at LUXE, the symmetry around the propagation axis is assumed. A consequence of this symmetry is that the rates for tree-level strong-field QED processes can be written in terms of a single harmonic sum, rather than in the case of a linearly-polarised laser pulse, which contains an extra integral and harmonic sum due to having one fewer continuous symmetries than the circularly-polarised case. This allows quicker calculations and a more exact comparison with theory \cite{lma2}.

\begin{figure}[ht]
\centering
\includegraphics[width=11cm]{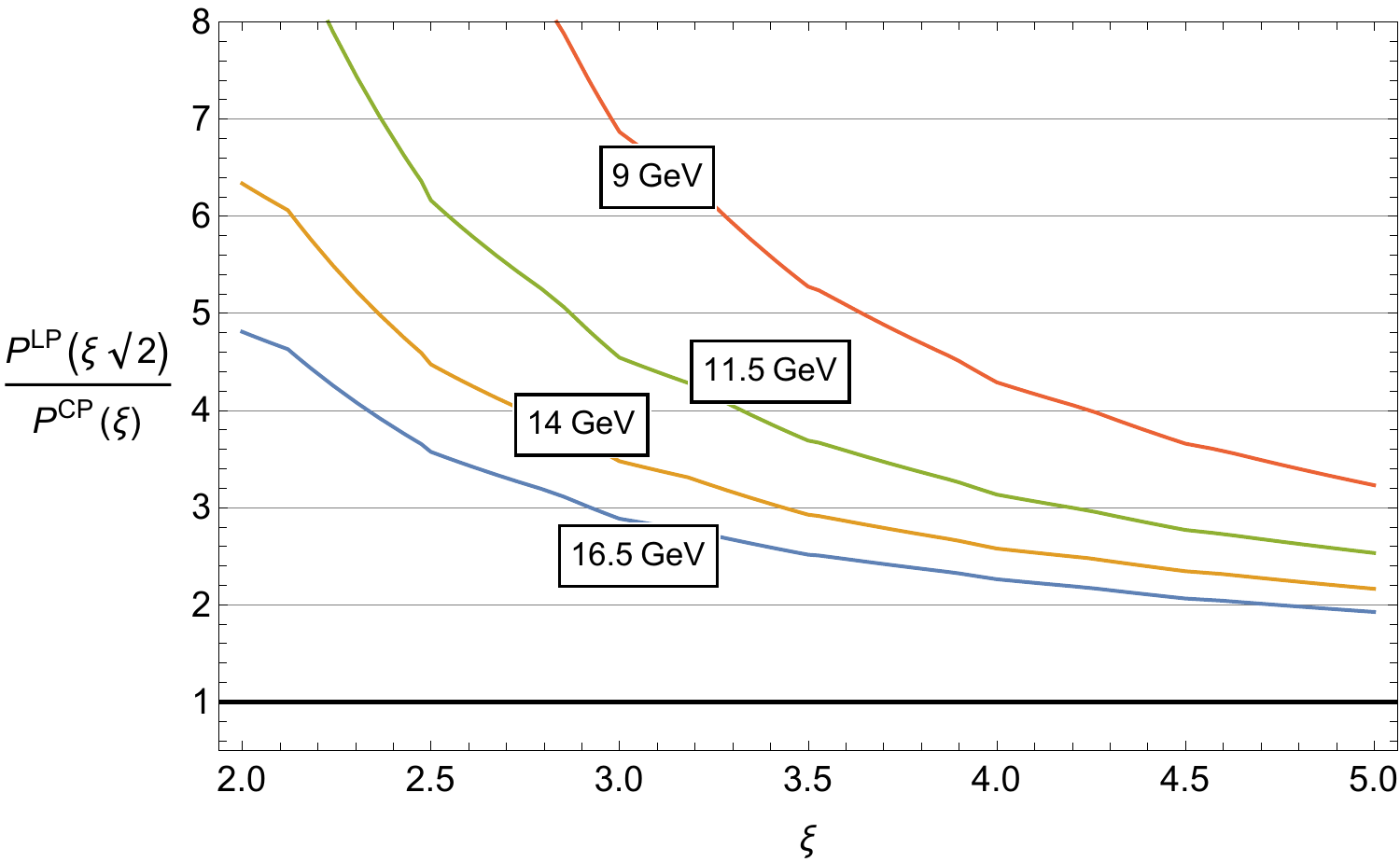}
\caption{The enhancement factors of the pair-creation yield for various photon energies and intensities, due to using an LP rather than a CP laser.
} \label{fig:relratesp1}
\end{figure}

An alternative to using circular polarisation, is to remove the quarter waveplate used to make the pulse circularly polarised, so that the laser pulse becomes linearly polarised (LP). For a constant laser energy, the intensity parameter that can be achieved with LP is a factor $\sqrt{2}$ larger than with CP. This can lead to an enhancement of the pair-creation yield, as shown for different photon energies and intensities. In \figref{fig:relratesp1}, the LCFA is applied to calculate the total probability of pair-creation in a sine-squared plane-wave pulse in a linearly and circularly-polarised background and the ratio formed, for the phase-0 experimental parameters given in the introduction of this chapter.

Integrating the plane-wave result in \figref{fig:relratesp1} over the thin-target bremsstrahlung formula \cite{RevModPhys.46.815}, leads to an enhancement of the pair yield by a factor of around $\approx 2.3$.

\subsubsection{Photon Beam Polarisation}
\begin{figure}[htbp]
\centering
\includegraphics[width=11cm]{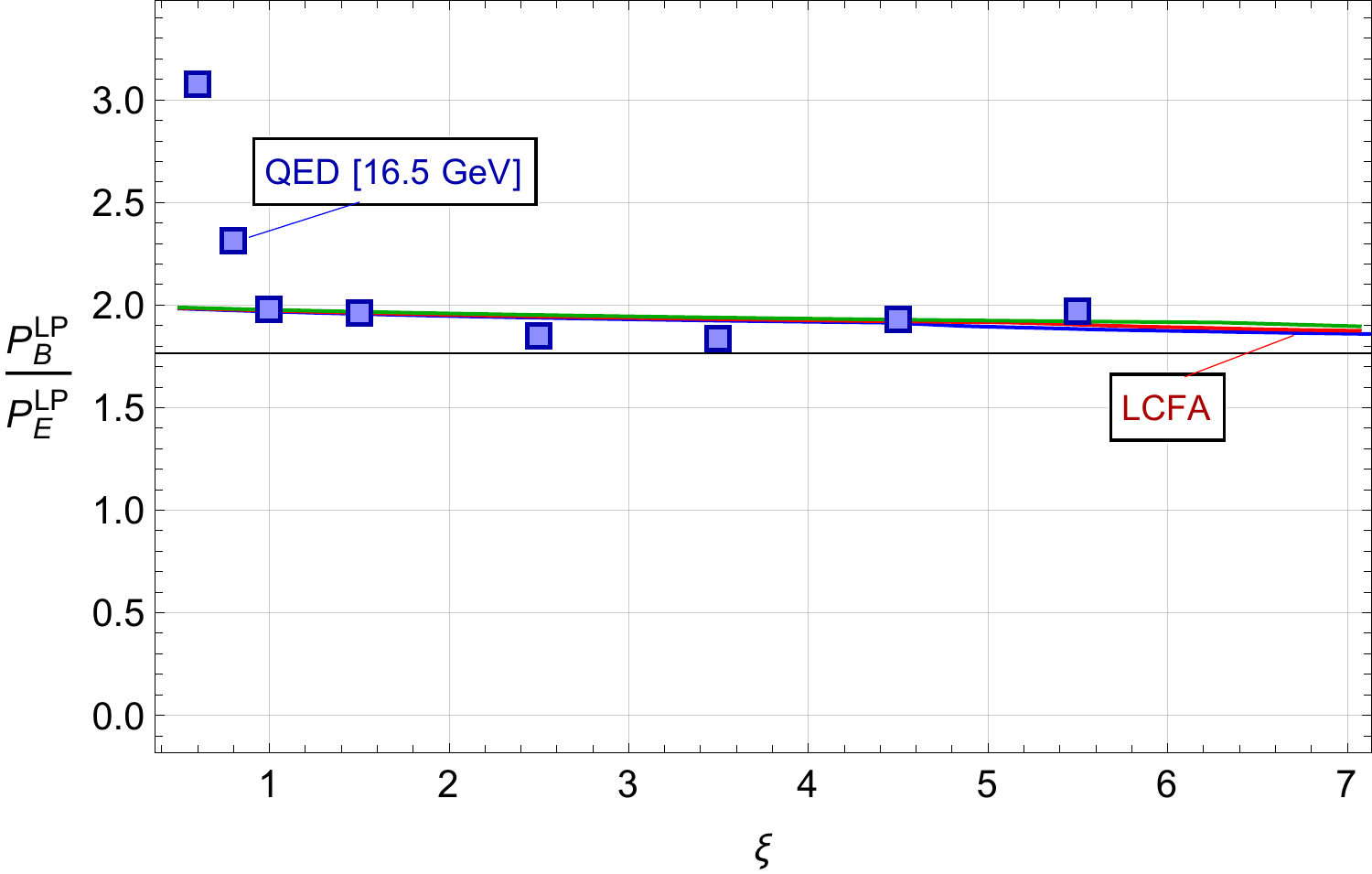}
\caption{Ratio of pair-creation probabilities from photons in different polarisation states decaying in an LP plane-wave pulse (the LCFA curves are for $11.5$, $14$ and $16.5\,\trm{GeV}$ photons).} \label{fig:relratesp2}
\end{figure}

For the ICS source, a photon polarisation of around $85\%$ is expected, which would lead to a comparison factor of $\approx 1.5$. Using an oriented crystal as a target, it would also be possible to partially polarise bremsstrahlung photons \cite{baier94}.  A real propagating photon can take one of two polarisation states. If the photon is propagating in a plane-wave background, one natural choice is to describe its polarisation using one of the two eigenstates of the polarisation operator \cite{baier75a}
\[
\eps_{ \tsf{E}} = \epsilon - \frac{k \cdot \epsilon}{k \cdot \vkap}\,\vkap \qquad
\eps_{ \tsf{B}} = \beta - \frac{k \cdot \beta}{k \cdot \vkap}\,\vkap,
\]
where $\epsilon$ ($\beta$) is the direction of the plane wave's electric (magnetic) field. The LCFA rate for pair-creation from a photon in the electric/magnetic polarisation state is, respectively \cite{King:2013zw,Seipt:2020diz}:
\bea
\rate^{\tsf{LCFA}}_{e} = \frac{\alpha}{\eta_{\gamma}}\int_{0}^{1} d\ess \left\{\Ai_{1}(z) + \left[\frac{2\mp 1 }{z}-\chi_{\gamma}\sqrt{z}\right]\Ai'(z)\right\}\qquad
z = \left(\frac{1}{\chi_{\gamma}}\frac{1}{\ess(1-\ess)}\right)^{2/3}.
\label{eqn:RNBWpollcfa}
\eea
In \figref{fig:relratesp2}, the ratio of the pair-creation yield from the two photon polarisation states is plotted, and a comparison is made between the analytical QED plane-wave calculation in a $16$-cycle sine-squared plane-wave pulse, and the LCFA.

We see from \figref{fig:relratesp2} that the LCFA result agrees rather well with the analytical QED result for $\xi \gtrsim 1$, but, as is expected for the LCFA, disagrees when $\xi \lesssim1$. Furthermore, the result is rather insensitive to photon energy and background intensity in the region that will be probed by \phaseone. One can calculate a \emph{Comparison Factor}, which gives the relative difference in the yield of Breit-Wheeler pairs due to choosing the laser polarisation to be parallel or perpendicular to the photon polarisation (for an equivalent head-on collision where the polarisation state is parallel or perpendicular to the laser field direction).

\subsubsection{Higher Harmonics of the IP Laser}
One of the main aims of the LUXE experiment is to show how the Breit-Wheeler pair creation rate changes as $\xi$ is increased from the region $\xi \ll 1$, where the Breit-Wheeler process  is predicted to proceed via a perturbative (but nonlinear) ``multiphoton'' process, to the region $\xi \sim O(1)$, where the Breit-Wheeler process is predicted to display a non-perturbative dependency on field strength, signified by a ``turning'' of how the yield of pairs depends on $\xi$. Rather than just looking at the dependency of the Breit-Wheeler process on $\xi$, one could also study Breit-Wheeler and the Compton process depend on the energy parameter $\eta$. Although the energy of the electron beam is fixed by the energies provided by the XFEL.EU electron beam, the energy parameter $\eta$, is also linearly proportional to the laser frequency. Therefore, by using the process of harmonic generation, to generate the second or third harmonic of the main laser pulse, one can arrive at energy parameters twice or thrice the fundamental harmonic value. This increase in energy parameter comes at a cost of a reduction in the associated $\xi$ parameter. First, because $\xi\propto 1/\omegaL$, and so the second and third harmonics will have $\xi$ values that are a half or third as large; second, because the generation of a higher harmonic involves losses in the converter crystal. For example, the conversion efficiency of a potassium dihydrogen phosphate (KDP) to $2\omegaL$ is only about $25\%$ \cite{Hornung_2015} (at the level of the intensity).

However, the focus diameter will be reduced by factor 2, for the $2\omegaL$ case.  Moreover, the pulse duration of the converted pulse will be shorter by a factor of $\sqrt 2$. Thus the loss in the intensity  could be partly compensated by reduced focus size and pulse length. Hence, the final intensity in $2\omegaL$ might be only smaller by a factor of 2 or less.

\begin{figure}[htbp]
\centering
\includegraphics[width=10cm]{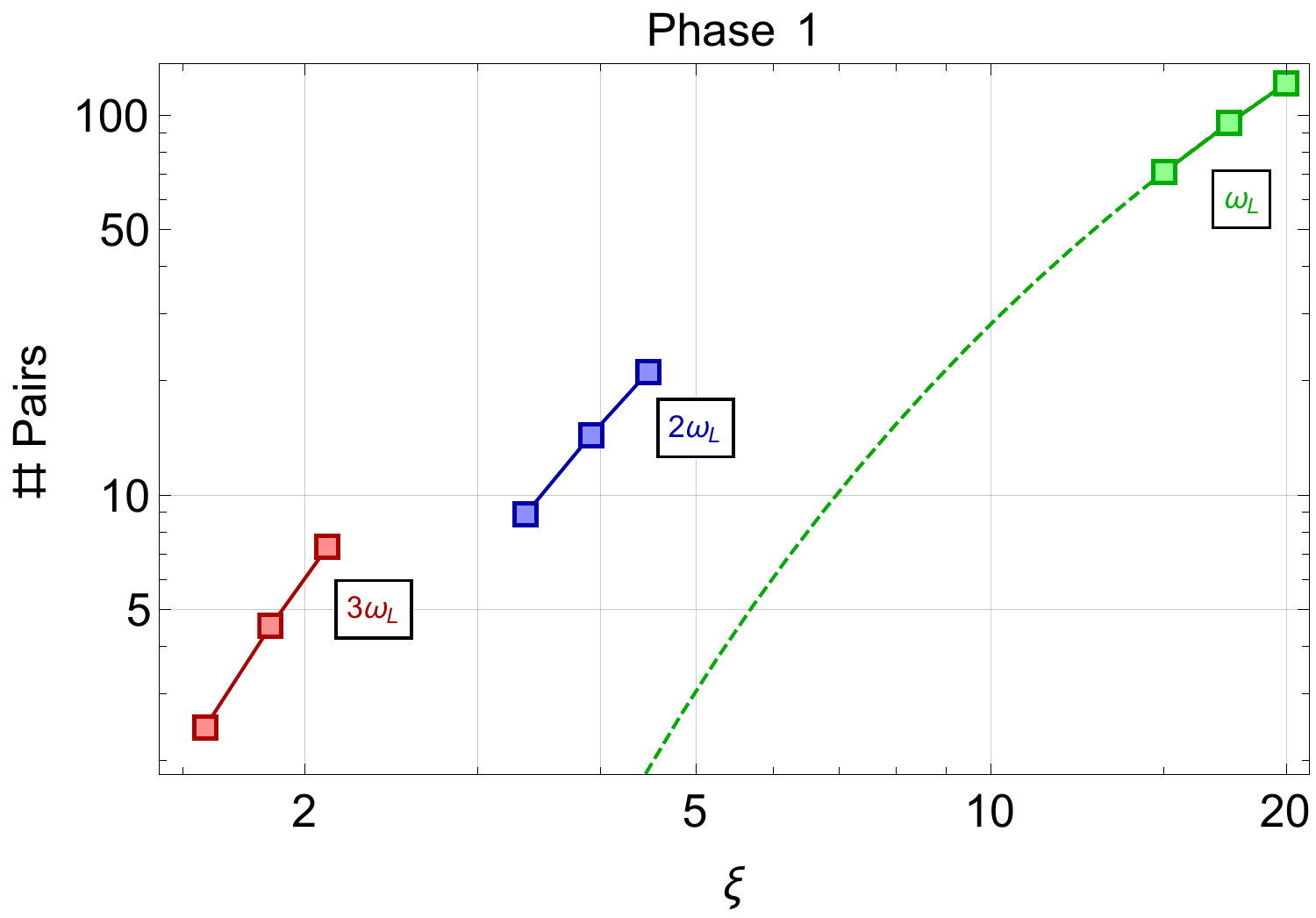}
\caption{Approximate yield of pairs for different harmonics of the IP laser, assuming the simulation result of $\sim 8$ pairs produced by a $\xi=6.5$ laser pulse in phase 1 and a scaling of the yield given by the LCFA for a $16.5\,\trm{GeV}$ photon. The data points on each curve are for equivalent values. (Assumed transmission of laser intensity: $0.2$ into $2\omegaL$ and $0.1$ into $3\omegaL$).}
\label{fig:pairHarms}
\end{figure}

The approximate effect on the yield of Breit-Wheeler pairs for phase-1 parameters in a plane-wave pulse is shown in \figref{fig:pairHarms} for a $16$-cycle sine-squared laser pulse and a photon energy of $16.5\,\trm{GeV}$.
Since the goal detector sensitivity is to measure
$0.001$ positrons per bunch crossing with an accuracy better than 10\% (see Sec.~\ref{sec:posrate}), we conclude that even for the third harmonic with the assumed losses, the number of pairs that are generated can be measured at LUXE.

\subsubsection{Radiative Spin Polarisation of Electrons in Ultra-Intense Laser Pulses}
\label{sec:theory:spin}
By combining the high-quality high-energy primary electron beam from the XFEL.EU linac with a 100 TW class short pulse laser, LUXE provides an opportunity to investigate for the first time electron spin effects in the Compton process. A signal for the influence of electron spin can be found in radiation reaction (RR), which is the back-reaction of radiation emission on the radiating electrons. In a classical picture, RR can be described by a continuous friction force term added to the classical force equation \cite{LL,Blackburn:RevModPlasma2020}. However, in the strong-field quantum regime, $\chi\sim 1$, RR manifests itself in consecutive Compton events, where the electron recoil during each hard photon emission constitutes the stochastic nature of quantum radiation reaction. In addition, because Compton probabilities depend on the spin of the electrons \cite{Seipt:2020diz}, quantum RR is spin-dependent and can thus lead to radiative polarisation of electrons via spin-flip transitions \cite{Seipt:PRA2019,PhysRevResearch.2.032049}.
    
\begin{figure}[ht]
    \centering
    \includegraphics[width=0.47\columnwidth]{{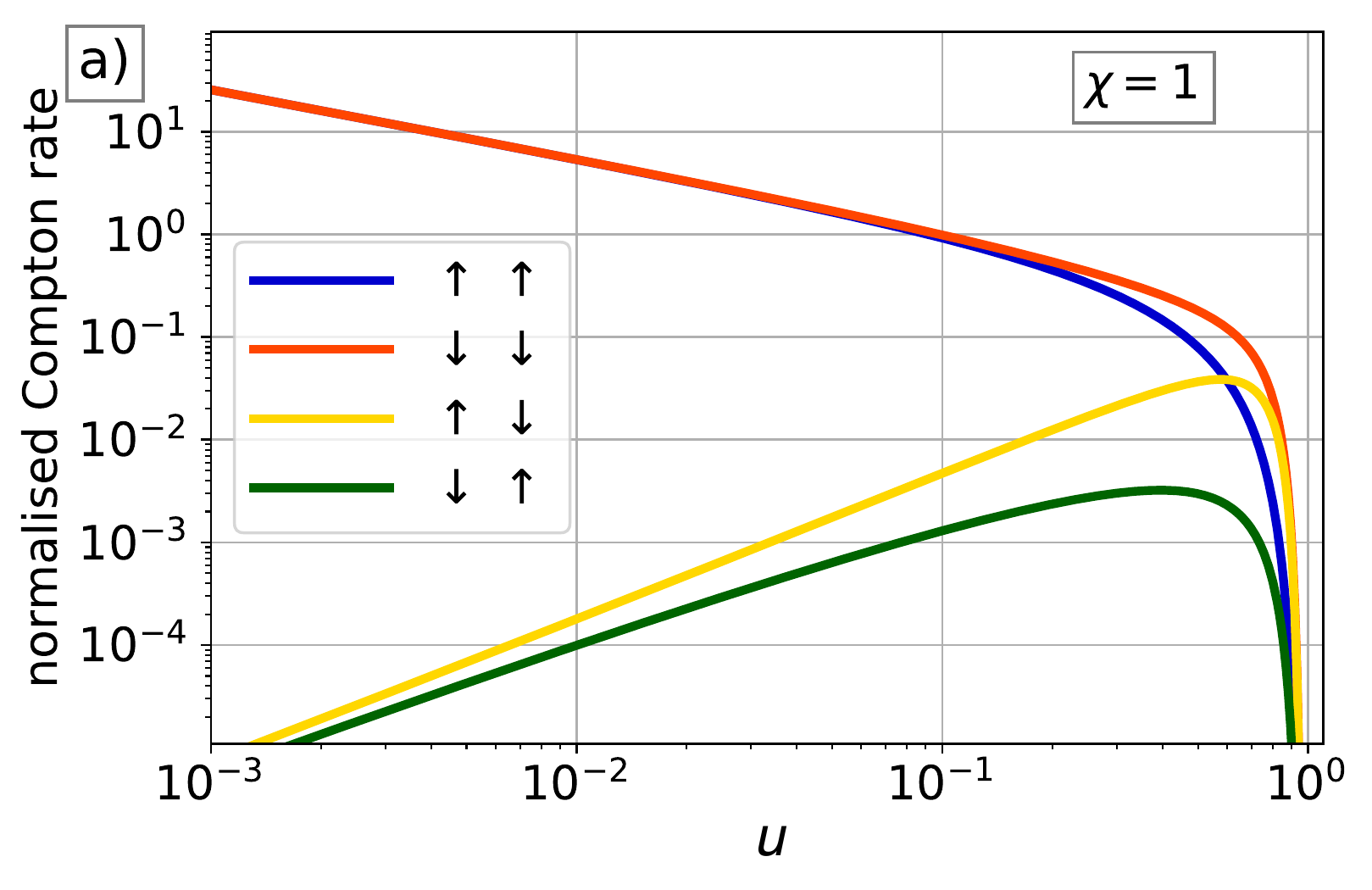}}\hfill \includegraphics[width=0.52\columnwidth]{{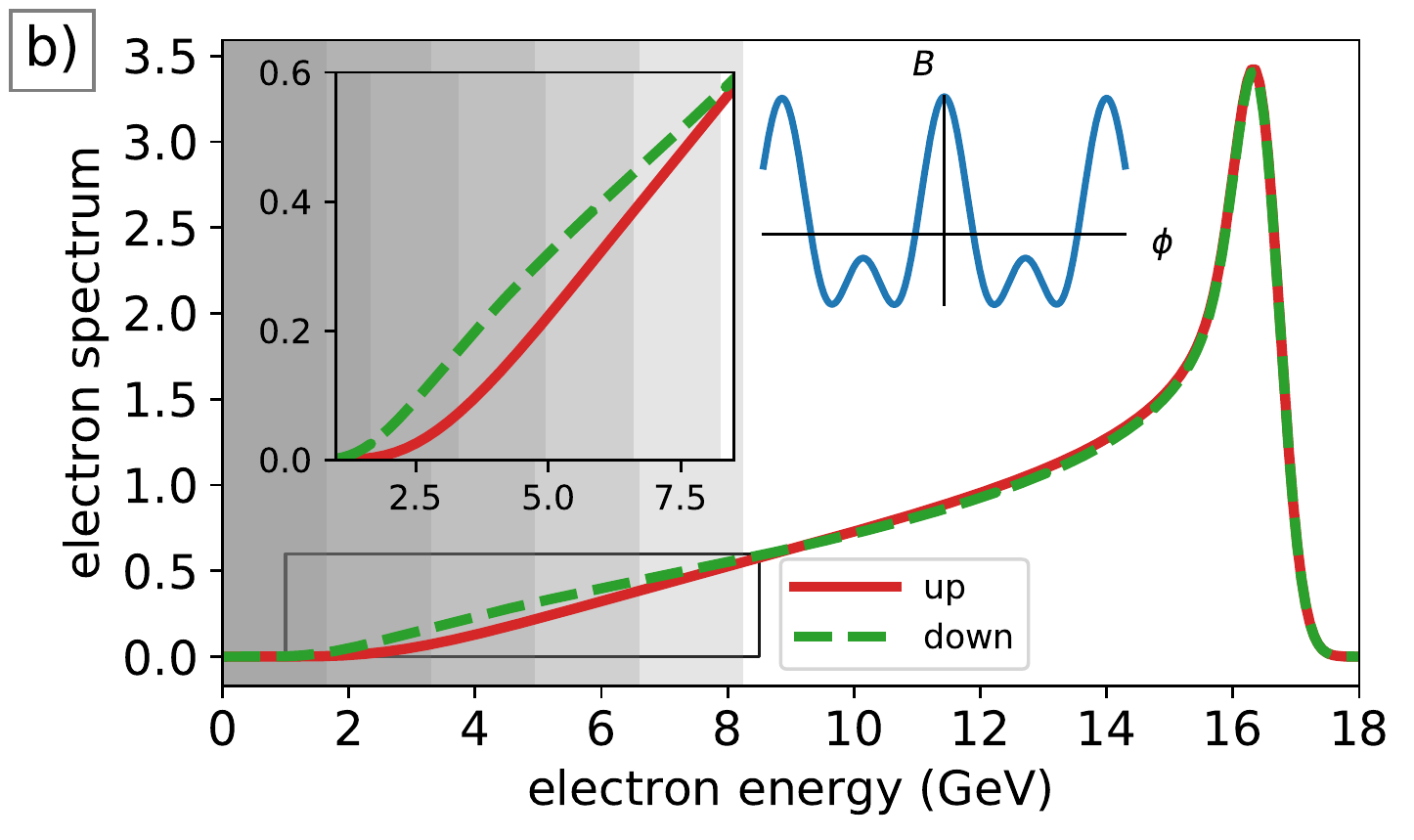}}
    \caption{a) Spin dependent LCFA Compton rates for $\chi=1$ as a function of lightfront momentum transfer $u$. b) Spectra of up and down electrons for $c_2=0.7$ and 16.5 GeV, $\xi = 5$ with an excess of down-electrons in the low-energy part.
    The value of $\xi$ refers to the fundamental harmonic only, thus total laser power has to be approximately doubled. Shaded areas depict various cutoff energies, see also Fig.~\ref{fig:energy}. Insets magnify the low-energy region and show the two cycles of the two-colour laser pulse's magnetic field.}
    \label{fig:PErates}
\end{figure}    
Radiative self-polarisation occurs in synchrotrons through the Sokolov-Ternov effect \cite{book:Sokolov}. Due to the weakness of the synchrotron field, the self-polarisation time is typically very long. In the strong field QED regime, electrons are predicted to become radiatively polarised on fs timescales \cite{DelSorbo:PRA2017}. This ultra-fast spin-dependence of quantum radiation reaction could be observed for the first time at LUXE. 

It is convenient to use  a basis for the spin-states in which the spin vector is parallel to the magnetic field $B$ in the rest frame of the electron. Spin-up, $\uparrow$ (down, $\downarrow$) then refer to the spin four-vector being parallel (antiparallel) to $B$. The spin-dependent rate for an electron scattering from one spin-state to another via the Compton process, in a constant crossed field (the basis of the LCFA approximation Eq.~(3)), is illustrated in Fig.~\ref{fig:PErates}, see also \cite{Seipt:2020diz}.

In a quasi-monochromatic plane wave pulse, the radiative self-polarisation is small due to cancellations every half-cycle of the magnetic field in the spin-polarisation of electrons. However, by frequency-doubling a part of the linearly polarised laser pulse and recombining it with the remainder, so that the magnetic field is given by $B\propto \omegaL \xi (\cos \vphi + c_2 \cos 2 \vphi)$ (where $c_{2}$ is an arbitrary constant) this symmetry is broken. This is necessary to establish a well defined ``spin-down'' direction in the oscillating laser magnetic field to make spin-dependent RR effects visible. Optimally, the strength of the frequency-doubled pulse should be about $70\%$ of the fundamental harmonic \cite{Seipt:PRA2019}.

\begin{figure}[ht]
    \centering
    \includegraphics[width=0.99\columnwidth]{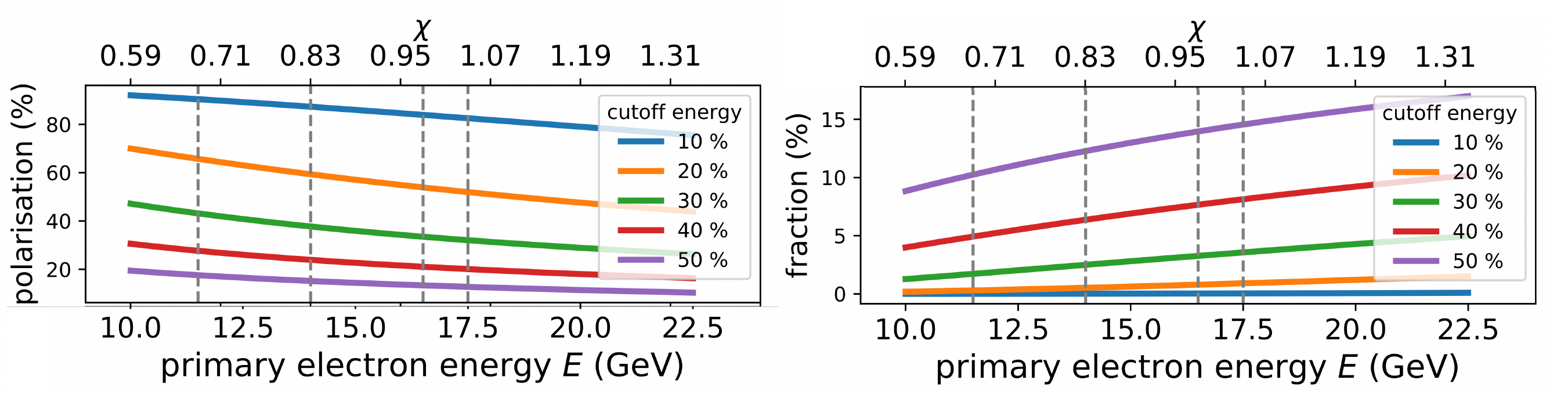}
    \caption{
     Degree of electron polarisation (left) and fraction of electrons below cutoff energy (right) as function of primary beam energy for constant value of $\xi=5$. Vertical lines indicate planned operational energies of the XFEL.EU-linac. The cutoff energy is given as a percentage of the primary electron energy.}
    \label{fig:energy}
\end{figure}

In the energy spectrum of the $16.5\,\trm{GeV}$ LUXE electrons scattered off the laser at $\chi\sim 1$, RR will form a broad low-energy shoulder beneath the primary 16.5 GeV peak.
(In contrast, for classical RR at $\chi\ll1$, there is predicted to be  a complete shift of the primary spectral peak which narrows even further.). The spin-dependence of RR is caused by spin-flip transitions leading to a spin-down state, which are most probable in emissions where the primary electrons lose a large fraction of their energy (corresponding to large $u \lesssim 1$). Consequently, one expects 
    a significant excess of spin-down electrons in the low-energy part of the scattered electron energy spectrum, see Fig.~\ref{fig:PErates}b. This spin-dependent excess can be observed by applying an energy cut to the measured scattered electron spectra. How the expected degree of polarisation in this two-colour set-up depends on the energy-cut is shown in Fig.~\ref{fig:energy}. A polarimetric measurement of this low-energy part of the spectrum would then have to be performed to reveal the effect.

The magnitude of the spin polarisation and optimal interaction parameters can be found using simulations based on a 1D kinetic equation model. The result of these calculations is plotted in Fig,~\ref{fig:energy},  which shows a significant polarisation can be expected for all modes of operation of the XFEL.EU-Linac. (To account for non-ideal effects in the focussing geometry, one needs to perform full 3D Monte Carlo simulations of the evolution of electron spin in the strong-field interaction point \cite{Seipt:PRA2019}). To observe the polarisation effect, it is important to measure the entire electron spectrum down to $\sim 2\,\trm{GeV}$, which can be achieved using  scintillation detectors, see Sec.~\ref{sec:scintDetect}. By a similar method, the production of polarised positrons is possible \cite{Chen:PRL2019}.

\subsubsection{Compton Beam Profiles}
\label{sec:theory:beamprofile}

Besides the energy spectra of the non-linear Compton photons discussed above, also their angular distribution contains valuable information. A measurement of these Compton beam profiles open an additional possibility to infer the value of $\xi$ in the non-linear Compton interaction \cite{Har-Shemesh2011,Yan:NatPhoton2017,Blackburn:PRAB2020}.

Due to relativistic beaming effects, the Compton $\gamma$ ray photons are emitted in a narrow cone around the initial electron beam direction. If the laser is linearly polarised, then the rms emission angle parallel and perpendicular to the laser polarisation scales as $\theta_\parallel\sim \xi/\gamma$ and $\theta_\perp\sim 1/\gamma$, respectively, for $\xi >1$. With 16.5 GeV electrons, the typical gamma beam size at a position 10 m downstream the IP is thus approximately $\sim 300 \mu\trm{m}$ in one direction and $(1\ldots\xi) \times 0.3\,$mm in the other. For the intensity parameters that will be probed at LUXE, the gamma beam width will therefore vary between $300\,\mu\trm{m}$ and $5\,\trm{mm}$. A specialised detector (Gamma Profiler) is envisaged for measuring the transverse beam profiles, see Sec.~\ref{sec:det:gammaprofiler}.

For linearly polarised lasers one expects an elliptic Compton profile with a large eccentricity in the polarisation direction. The widening of the radiation cone can be most easily understood within the LCFA, where it is explained by the wiggling of the electrons in the laser field with an angle $\vartheta(\phi) = ( \xi \sin \phi) /\gamma$ \cite{Blackburn:PRAB2020}. This phenomenon is similar to magnetic wigglers, where the wiggler parameter $K>1$.

For circularly polarised lasers, the widening of the Compton emission cone is symmetric, yielding a widened circular Compton profile. 
But this is not true for ultra-short (near) single-cycle pulses where asymmetries in the Compton profiles have been predicted, related to the absolute (carrier envelope) phase \cite{Seipt:PRA2013,Mackenroth:PRL2010}.  
Blackburn et~al., Ref.~\cite{Blackburn:PRAB2020}, proposed a model independent way of retrieving $\xi$ via $\xi^2=4\sqrt{2} \gamma_i\gamma_f (\sigma_\parallel^2-\sigma_\perp^2)$, by measuring the initial and final electron beam mean energies $\gamma_i,\gamma_f$, and the rms variance of the gamma-ray emission angles $\sigma_\parallel^2$, $\sigma_\perp^2$.

\newpage

\section{Electron Beam Transport and European XFEL Accelerator Aspects}
\label{sec:machine}

In this Section a brief summary of the modifications to the \euxfel necessary for the LUXE experiment is given. A more detailed description is given in Ref.~\cite{beamlinecdr}.

The LUXE experiment aims to be fully transparent to the \euxfel operation, therefore it is foreseen to use the standard beam parameters. The \euxfel runs throughout the year in various configurations. For instance, in 2019 beam was delivered to the experiments for their use for 3648 hours, corresponding to about 40\% of the year. About 80\% of that time the accelerator was running at an energy of 14.0~GeV, and about 10\% each at 16.5~GeV and 11.5~GeV, and in all cases mostly with a bunch charge of 0.25~nC.  

The electron energies and bunch charge at which the \euxfel accelerator is operated are determined by the photon science experimental program.  The frequency of bunch trains is 10~Hz and each train contains up to 2700 bunches. For the LUXE experiment only one bunch is required from each train as the laser frequency is 1~Hz. Assuming the \euxfel operates during a given month 50\% of the time, this corresponds to $10^6$ beam crossings (BX) per month for physics data taking. In addition, 9~Hz of data are taken without laser interaction to measure backgrounds \textrm{in-situ}. For this report, $\epsilon_e=16.5$~GeV and a bunch charge of 0.25~nC ($1.5\cdot 10^9$ electrons) are used as baseline. However, it will be interesting to also operate LUXE at lower energies, e.g. $14$ or $11$~GeV.

An overview of the electron beam parameters for the LUXE experiment is given in Table \ref{tab:xfelepara}. 

\begin{table}[htbp]
\begin{center}

\begin{tabular}{l|r|r}
Parameter          & Value XFEL.EU & Assumed Values for LUXE\\ \hline
Beam Energy [GeV]  & $\leq 17.5$ & 16.5      \\
Bunch Charge [nC]  & $\leq 1.0$ &  0.25      \\ 
Number of bunches/train  & 2700 & 1      \\ 
Repetition Rate [Hz]     & 10   & 10 \\ 
Spotsize at the IP [$\mu$m]        & $-$ & 5  \\ 
Bunch length [$\mu$m]        & 30--50 & 30--50  \\ 
Normalised projected emittance [mm mrad] & 1.4 & 1.4 \\
\end{tabular}
\caption{Electron beam parameters for the \euxfel, as designed and as envisaged for the LUXE experiment. The value for physics data taking for the XFEL.EU is taken from 2019.  The electron beam spot size at the LUXE IP can be increased.
\label{tab:xfelepara}}
\end{center}
\end{table}

\subsection{Location at the \euxfel}
\label{sec:beam:location}
\begin{figure}[ht]
   \centering
   \includegraphics*[height=4.2cm]{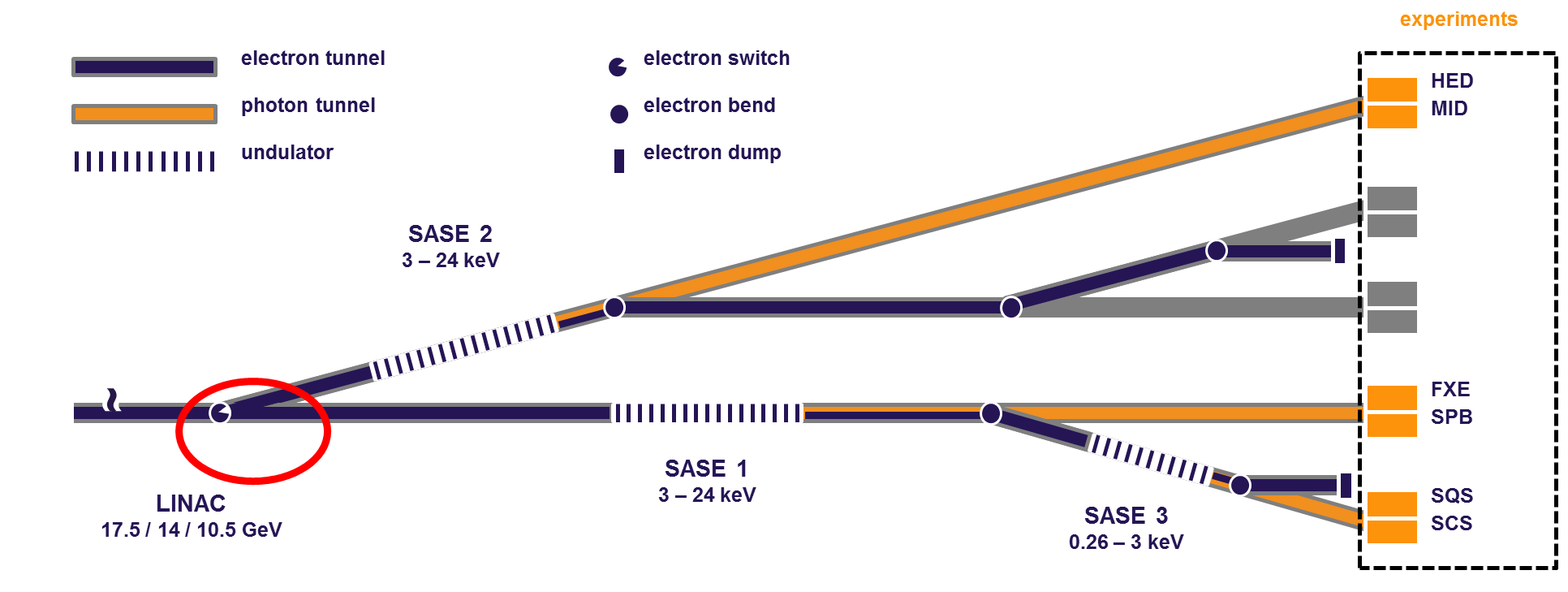}
   \caption{Schematic drawing of the \euxfel fan. The location foreseen of the LUXE experiment is circled in red. The SASE1, SASE2 and SASE3 undulators and the experimental halls are also shown.
   \label{fig:XFEL_sketch}}
\end{figure}

It is planned to install the LUXE experiment in the XS1 annex which is an appendix of the XS1 shaft, located in Osdorfer Born, that has already been foreseen for a possible extension of the distribution fan (termed "2nd fan") of the \euxfel. It is about 60~m long, 5.4~m wide and 5~m high. It is at the end of the main LINAC of the \euxfel (XTL). A drawing of the fan and XTL is shown in Fig.~\ref{fig:XFEL_sketch}. 

\begin{figure}[ht]
   \centering
   \includegraphics*[height=4.2cm]{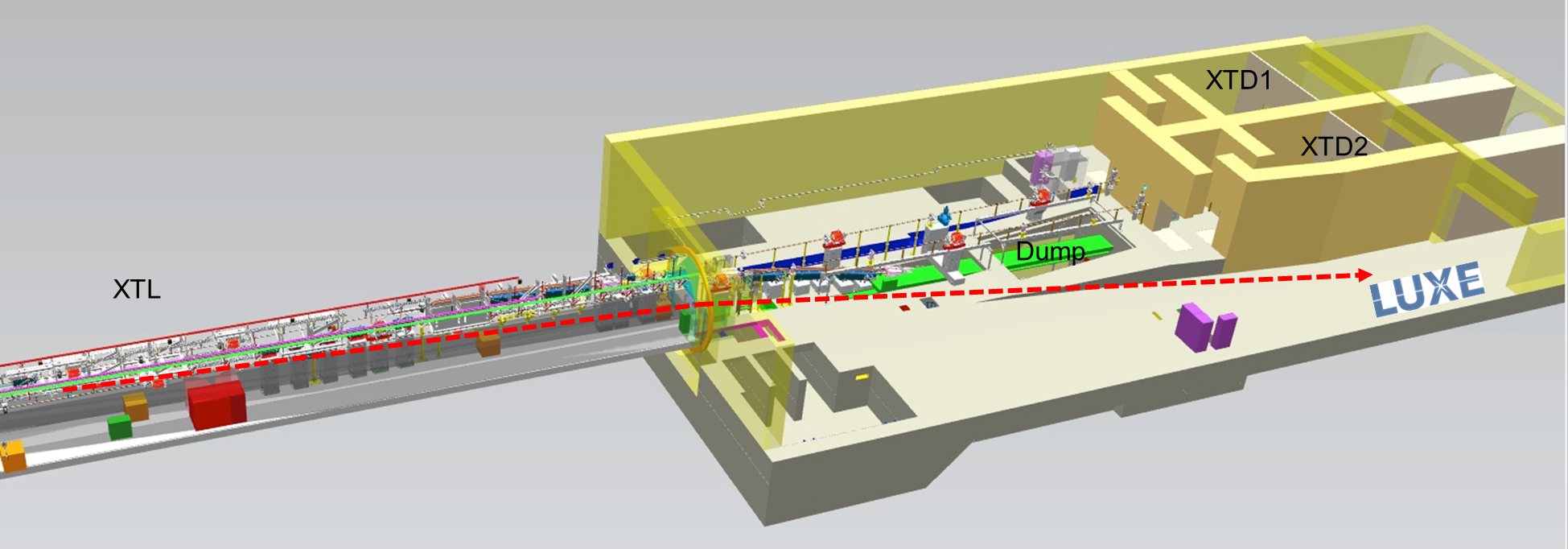}
   \caption{CAD model of the end of the \euxfel accelerator tunnel and the shaft building with the two existing
beamlines XT1 and XT2 to the undulators (SASE1 and SASE2) and the XS1 annex, where the LUXE experiment can be installed. The beam
extraction and the beam line towards the experiment is sketched with the dashed line. }
   \label{fig:LUXE_cad}
\end{figure}

The beam path to the XS1 annex, required for LUXE, and the XTD1 and XTD2 tunnels are also schematically shown in Fig.~\ref{fig:LUXE_cad}. Also shown is the XTL tunnel that hosts the linear accelerator further upstream. The beam transport to the experiment has to be built up with an additional extraction which was already foreseen in the design phase of the \euxfel for the future 2nd fan of the facility. About 40~m of installations in the XTL tunnel are needed to kick out a bunch and guide it towards the XS1 annex. Inside the XS1 shaft building additional components are needed to guide and focus the beam to the experiment. The length of this installation is about 50~m.

\subsection{Lattice Design and Hardware}

\begin{figure}[ht]
   \centering
   \includegraphics*[width=0.8\textwidth]{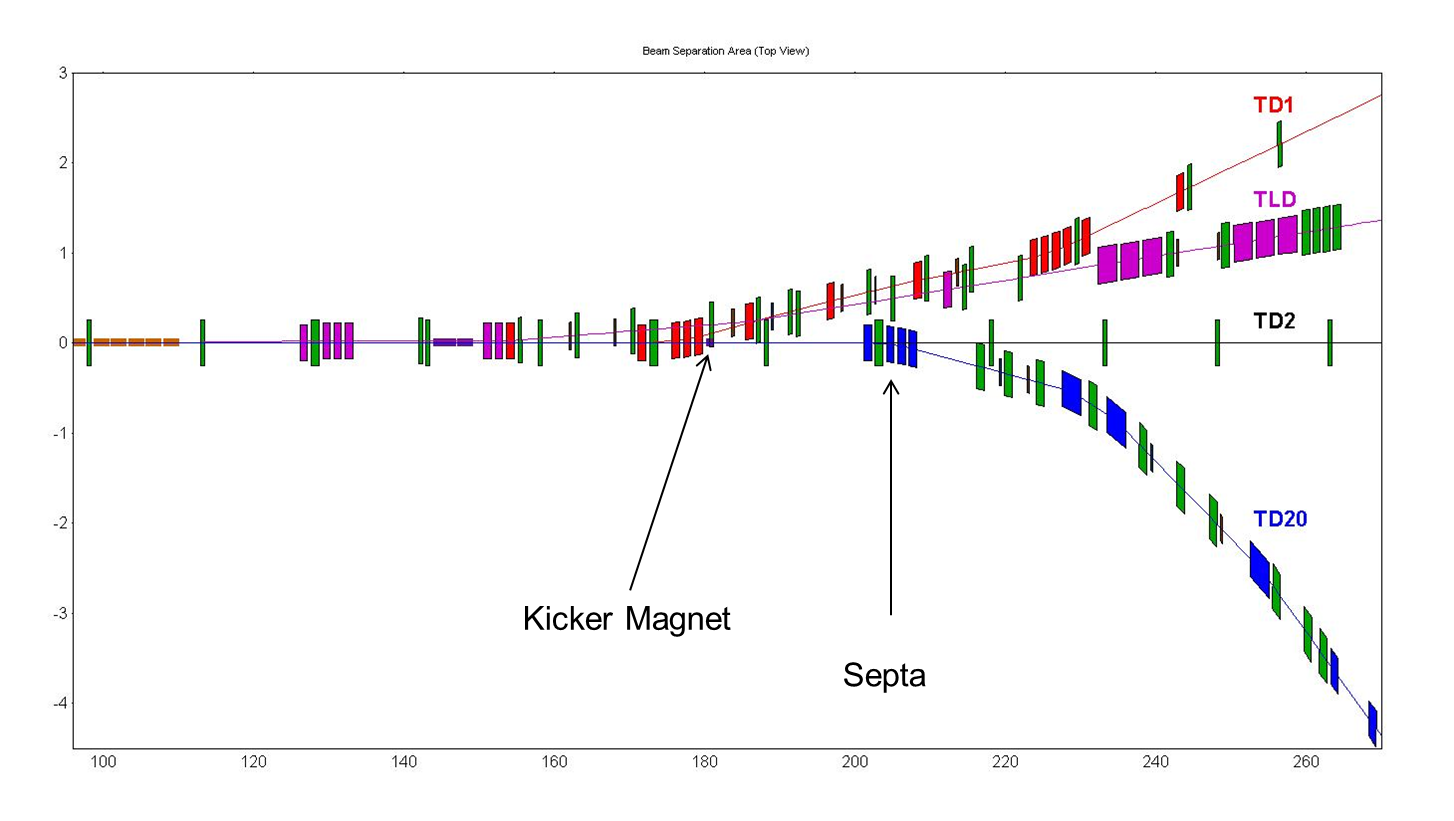}
   \caption{Design of the beamline extraction and transfer line. The different magnets (kicker, dipole, quadrupole) are shown in different colours. The new kicker magnet which kicks out one bunch towards LUXE is indicated. The horizontal and vertical scales are in units of m. The LUXE experiment starts at the end of the area shown ($z=270$~m). The XS1 building starts at $z\approx 235$ m in this drawing, $30$~m upstream from the Septa indicated. 
   \label{fig:lattice}}
\end{figure}

The design for the beamline is shown in Fig.~\ref{fig:lattice}. The part until the last septum is termed \textit{beam extraction} and the part after the septa is termed \textit{beam transfer line}. Part of the beam transfer line ($40 \units{m}$) is located inside the XTL tunnel, whereas the rest has to be installed in the XS1 shaft building and the XS1 annex. A combination of fast kicker magnets, septa, dipole and quadrupole magnets is used for the design of the beam extraction and beam transfer. 
In the beam extraction part two kicker magnets and four septa are used. In the transfer line,  11 dipole magnets, 10 quadrupoles and 10 corrector dipoles are needed. In addition, an array of diagnostic elements are foreseen.  

For the kicker magnets, a new design will be used~\cite{beamlinecdr} while all other components adopt the designs already used in the \euxfel. Standard \euxfel components will be used for power supplies and general infrastructure, e.g. the vacuum system, beam instrumentation, water cooling, cabling, safety systems.

\subsection{Installation Procedure}
The installation of the extraction and transfer line requires major construction work both in the XTL tunnel and in the XS1 shaft. Additional infrastructure has to be installed in the XS1 building. 
Originally it was considered to do the installations over the course of two winter shutdowns~\cite{beamlinecdr}. However, recently it has become clear that in 2024 there will be a 6-month shutdown for mandatory maintenance work on the superconducting accelerator. It would be highly desirable to use this shutdown also for the installation of the beam extraction and transfer lines. 
It should be noted, that during this periods the technical personnel of DESY is fully committed to the \euxfel maintenance and upgrade work. Thus 
this installation work requires the hiring of additional qualified personnel.

The different phases of the installation are listed in the following:
\begin{itemize}
    \item The \textbf{preparatory phase} consists of design, specification, tender, ordering, fabrication, delivery and testing of all necessary components.
    \item For the \textbf{installation in the XTL tunnel} it is foreseen to pre-assemble the vacuum chambers and septum magnets outside the tunnel (as has been done for the installation of the XTD1 extraction). 
    \item The \textbf{installation of the transfer line} concerns all components after the last septum. These components guide the beam towards LUXE. The vacuum system is separate from the already existing beam lines. Nevertheless the installation of the components in the XTL tunnel (about 30~m) is challenging as they are suspended from the ceiling. The installations in the XS1 shaft building are expected to be straightforward. 
    Since the vacuum system of the beam transfer line to LUXE is separate from the rest of the \euxfel, any delays here will only delay LUXE and not pose risks for the \euxfel.
    
\end{itemize}
Since the experiment will need to be installed in the same shutdown, the activities need to be coordinated well to achieve an overall optimal installation schedule. The installation of the experiment is discussed in Sec.~\ref{sec:tc}.

\subsection{Beam Dump}
\label{sec:beamdump}
\begin{figure}[htbp]
   \centering
   \includegraphics*[width=0.8\textwidth]{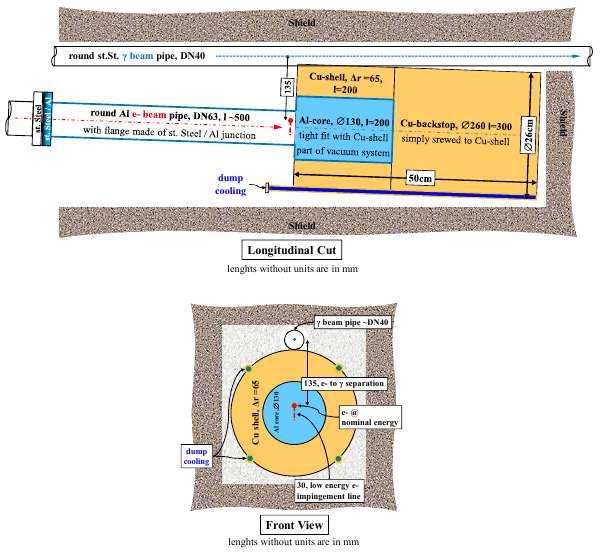}
   \caption{Schematic drawing of the beam dump. The beam pipe in this drawing is made of stainless steel.
   \label{fig:beamdump}}
\end{figure}

Two single-bunch electron beam dumps will be needed to dump the beam safely, one for each of the two running modes. For this purpose a special beam dump has been designed. It can accept a 1~nC beam at 10~Hz, resulting in a maximum power of $P=200 \units{W}$ for a beam energy of $\varepsilon=20 \units{GeV}$ which needs to be dumped. In addition, shielding is required to ensure that there is no contamination of nearby air, water or soil, people can safely work in the adjacent tunnel, nearby electronics equipment is not damaged by radiation, and, last but not least, that the background from stray particles in the LUXE experiment is low enough so that the foreseen physics analyses can be performed (see Sec.~\ref{sec:simulation}). 
A schematic drawing of the beam dump is shown in Fig.~\ref{fig:beamdump}. It consists of an aluminium core with a diameter of 13~cm and a length of 20~cm, which is complemented by copper surrounding it and also extending its length. Water cooling is used to cool it in four cooling pipes. The total length is $50\units{cm}$, and it is integrated into a shielding. 

In the $e$-laser setup, the electron beam is dumped after the interaction point, and the dump is placed $7 \units{m}$ downstream from the IP. In the \glaser setup, the electron beam needs to be dumped after it has traversed the tungsten target; the dump is located $3.24 \units{m}$ downstream from the dipole magnet.

Furthermore a dump is required for the photon beam at the end of the beamline $13.6 \units{m}$ downstream from the IP to reduce backscattering to the sensitive elements of the detector. In principle the same dump as for the electron could be used but for the BSM aspects of the experiment it is important that it absorbs well all the photons and is as short as possible. Thus for the BSM simulation it is assumed to be made of lead and 50~cm long.  

\subsection{Risks and Commissioning}
The risks had been estimated in Ref.~\cite{beamlinecdr} assuming that the installation occurs during the short winter shutdown. If a delay occurs due to equipment being damaged in either the beam transfer or the transfer line, the shutdown period may need to be prolonged. The work plan foresees to mitigate this risk by using trained personnel, proper planning and good workmanship, and estimates the risk to be below 5\% based on prior experience at the \euxfel. Another risk is that the vacuum cannot be closed due to missing or wrong equipment. For the XTL tunnel, this risk is mitigated by keeping the current system fully available so that one could revert to that system if needed. The likelihood is estimated to be below 5\%; it would cause a delay of two weeks. For the transfer line in the XTD20 tunnel, the vacuum is separate and any delays would only affect the LUXE experiment and not the \euxfel at large. The likelihood is estimated to be below 10\%. 
With the possibility to extend the installation period into a 6 month shutdown the impact of these risks on \euxfel operation is further diminished. It should nevertheless be noted that an integrated installation plan does not yet exist.

Operational risks, e.g. failure of any active component (kicker, magnets, etc.) has no impact on the beam to SASE1 and SASE3. A vacuum failure in the transfer line can be quickly recovered by isolating the transfer line from the remaining vacuum system.  The 18~m new vacuum system in the XTL is of comparable complexity to the existing vacuum system and thus poses no additional risk. Only the two additional ceramic vacuum chambers for the kicker magnets add some complexity. \euxfel has already 18 of these chambers installed in different sections of the machine with no recorded issues so far.

Commissioning of the extraction line is considered to be straightforward. Experience with the commissioning of the already existing extraction lines and beam transports lead to a conservative estimate of a maximum of 6 dedicated commissioning shifts needed before the beamline can be operated in a complete parasitic mode.

\clearpage
\section{Laser Specifications and Diagnostics}
\label{sec:laser}

\begin{figure}[ht]
    \centering
    \includegraphics[width=0.5\textwidth]{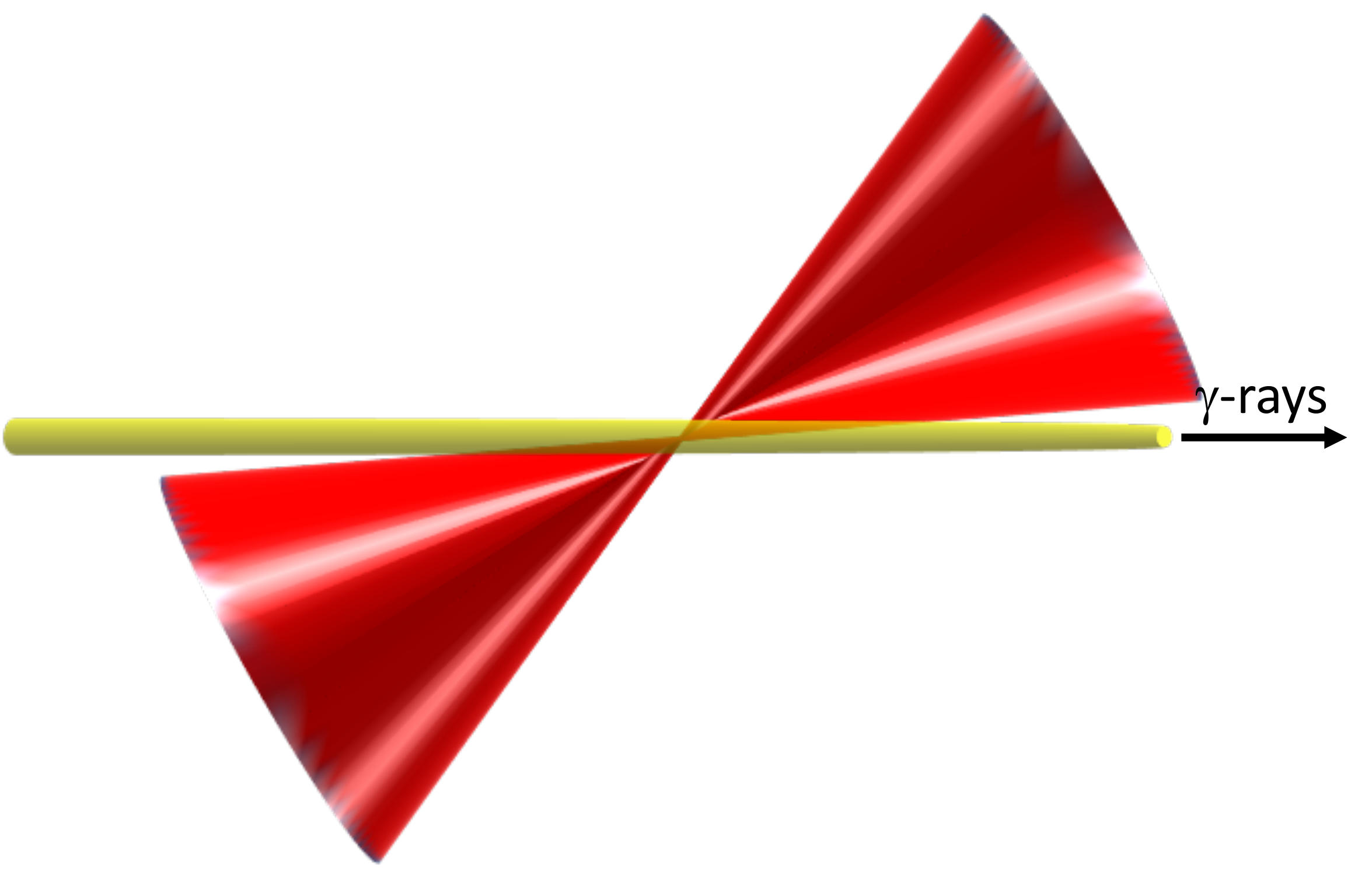}
    \caption{Interaction geometry between a high-power laser and the high-energy electron/photon beam. The laser beam is focussed by off-axis parabolic mirrors to a high intensity point focus and recollimated for on-shot analysis. The typical $\gamma$-ray beam diameter will be around 230 $\mu m$ while the laser spot is as low as 3~$\mu m$ at the highest $\xi$. The longitudinal spot sizes are typically $9$~$\mu$m (30~fs) for the laser and $\sim 40$~$\mu$m for the electron beam. 
    \label{fig:interaction_geometry}}
\end{figure}

Reaching the regime of \hiqed 
requires reaching the Schwinger limit in the centre-of-momentum frame of the colliding particles. As detailed earlier in this report such extreme fields can now be achieved in the laboratory by colliding extremely high energy photons or electrons with energies exceeding 10~GeV with state-of-the-art high-intensity lasers with a peak power of up to 350~TW, and the existing European XFEL infrastructure at the Osdorfer Born facility is very well suited for this. The interaction point (IP) of this ambitious experiment is schematically illustrated in 
Fig.~\ref{fig:interaction_geometry}. 

 A key feature of \hiqed is the non-perturbative, non-linear behaviour of the interaction between the high-energy photons/electrons and the intense laser field as discussed in Sec.~\ref{sec:science}. Additionally, the event rate on each detector will in certain configurations be $\ll 1$ per shot, so that the measurement of a given spectrum or cross section will necessarily need to analyse a large number of bunch crossings. Despite the long-term stable operation  that is typical of current high-power laser systems (with $\sim$percent level stability in the major pulse parameters), the exacting requirements for  high precision measurements following from the non-linearity and low event rate require innovative and precise pulse characterisation. In the following we will describe the laser system and diagnostic equipment envisaged to meet the high requirements.

\begin{table}[ht]
\begin{center}

\begin{tabular}{|l|c|c|c|}
\hline
 & \textbf{40 TW, 8$\mu$m}  &  \textbf{40 TW, 3$\mu$m}  &  \textbf{350 TW, 3$\mu$m}\\
\hline
& & & \\ 
\textbf{Laser energy after compression (J)}  & 1.2 & 1.2 & 10 \\
  \hline
       \textbf{Laser pulse duration FWHM (fs)}  & \multicolumn{3}{c|}{30}\\  
    \hline
    \textbf{Laser focal spot waist $w_0$ ($\mu$m) } & 8  &  3  &  3\\  
    \hline
   \textbf{Fraction of ideal Gaussian intensity in focus (\%)}  & \multicolumn{3}{c|}{0.5}\\   
   \hline
 \textbf{Peak intensity in focus ($\times10^{20}$ ~Wcm$^{-2}$)}  &  $0.19$ & $1.33$ & $12$ \\  
    \hline
    \textbf{Dimensionless peak intensity, $\xi$}  &  3.0 &  7.9 &  23.6\\   
    
    \hline
    \textbf{Laser repetition rate (Hz)}  & \multicolumn{3}{c|}{1}\\  
    \hline
     \textbf{Electron-laser crossing angle (rad)}  &  \multicolumn{3}{c|}{0.35}\\  
     \hline
      \multicolumn{4}{c}{}   \\\hline
    \textbf{Quantum parameter} & \multicolumn{3}{c|}{}\\  
 \hline
    \textbf{$\chi_e$ for $E_e=14.0 \units{GeV}$}  &  0.48 &  1.28    &  3.77\\ 
    \textbf{$\chi_e$ for $E_e=16.5 \units{GeV}$}  &  0.56 &  1.50 &  4.45\\  
    \textbf{$\chi_e$ for $E_e=17.5 \units{GeV}$}  &  0.6 &  1.6 &  4.72\\  
    \hline
    \hline
   
\end{tabular}

\caption{Parameters for different initial laser power and focussing geometries as planned for the different phases of the LUXE experiment. The quantum parameter $\chi_e$ is given for three different electron energies. The peak intensities are conservative estimates which are reliably achievable. For example a pulse duration of $25 \units{fs}$ with a focusing performance equivalent to 70\% of an ideal Gaussian is within normal current operating parameters of JETI40. All values are for circular polarisation with peak field parameters $\xi$ and $\chi$ increasing by a factor of $\sqrt{2}$ for linearly polarised laser irradiation. Note that the $40 \units{TW}$ option with $8 \units{\mu m}$ focal spot and $350 \units{TW}$ $3 \units{\mu m}$ options share the same focal length with the increased beam diameter resulting in the smaller focal spot.}\label{tab:laser}
\end{center}
\end{table}

Table~\ref{tab:laser} shows the key parameters of the laser system envisaged for LUXE. The  laser  in the first and second column with an energy of $1.2 \units{J}$ and a power of $\sim 40 \units{TW}$, is the JETI40 laser and will be used during the initial phase LUXE. This laser allows the regime up to $\chi\sim 1$ to be investigated. During the second phase, the laser system will have a peak power of $350 \units{TW}$ which will enable the $\chi>1$ regime to be reached. Lower values of $\xi$ and $\chi$ can be accessed by defocussing the laser in either the transverse or longitudinal direction. 
The focus in the transverse dimensions is quantified by the waist radius, $w_0$, which is the value at which the intensity falls to $1/e^2$ of its value, and for a Gaussian beam corresponds to $2\sigma$ in intensity.

\subsection{The JETI40 Laser System}
 
\begin{figure}[htbp]
    \centering
    \includegraphics[width=0.7\textwidth]{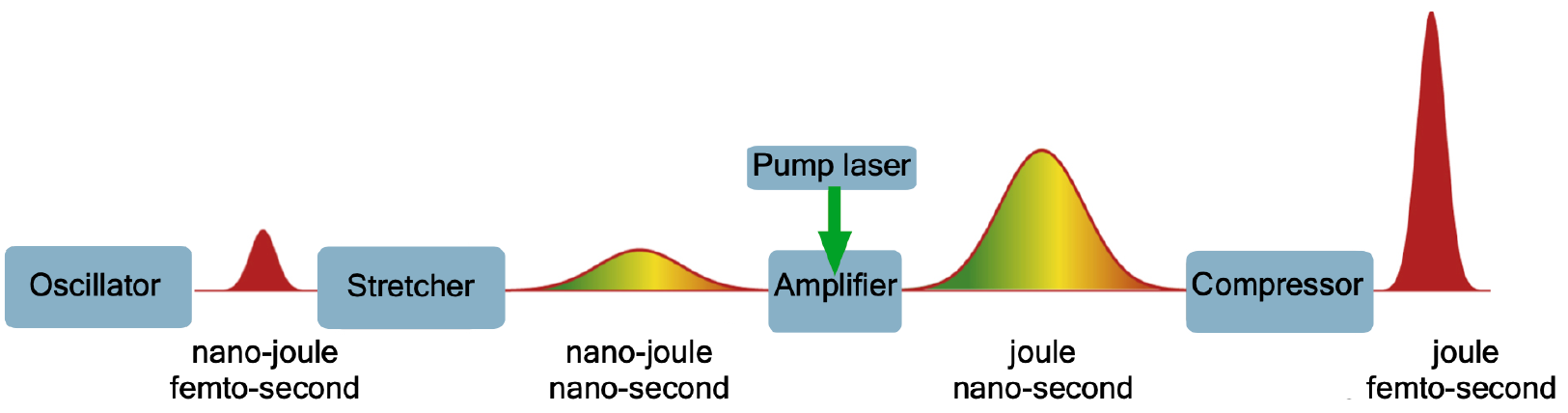}        
    \caption{Cartoon depiction of the chirped pulse amplification technique. }
    \label{fig:CPA-cartoon}
\end{figure}

The initial laser installation will consist of the JETI40 laser system that will be provided by the University of Jena/HI Jena as an in-kind contribution to the LUXE project.  JETI40 is a continuously developed system based on a commercial chirped pulse amplification (CPA) platform (Amplitude Technologies). The laser follows the generic CPA concept, shown in 
Fig.~\ref{fig:CPA-cartoon}, of using pulse stretching  to lower the intensity in the laser chain during the amplification to dramatically increase the maximum energy that can be achieved in a short pulse for a given beam diameter.  

\begin{figure}[ht]
    \centering
    \includegraphics[width=0.3\textwidth]{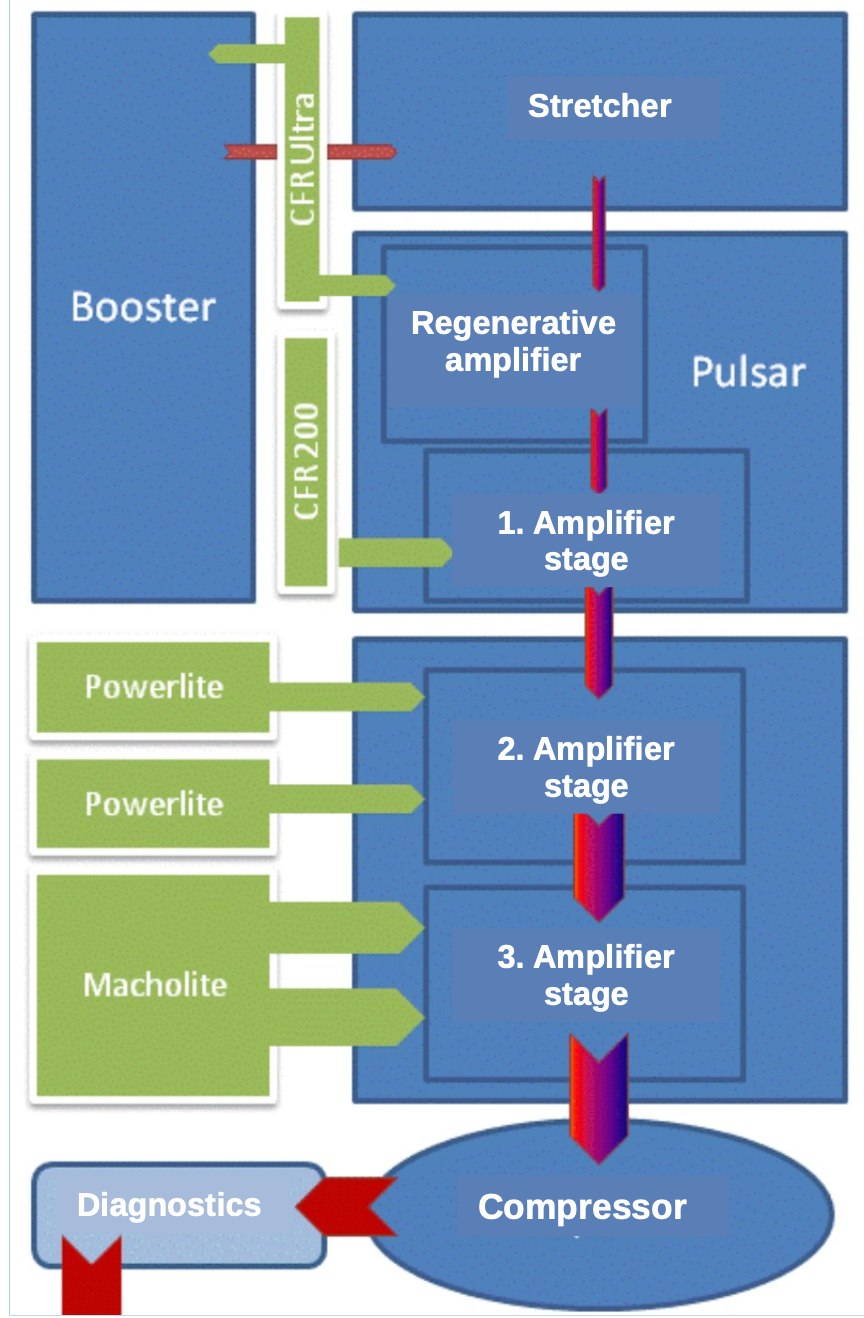}
    \includegraphics[width=0.3\textwidth]{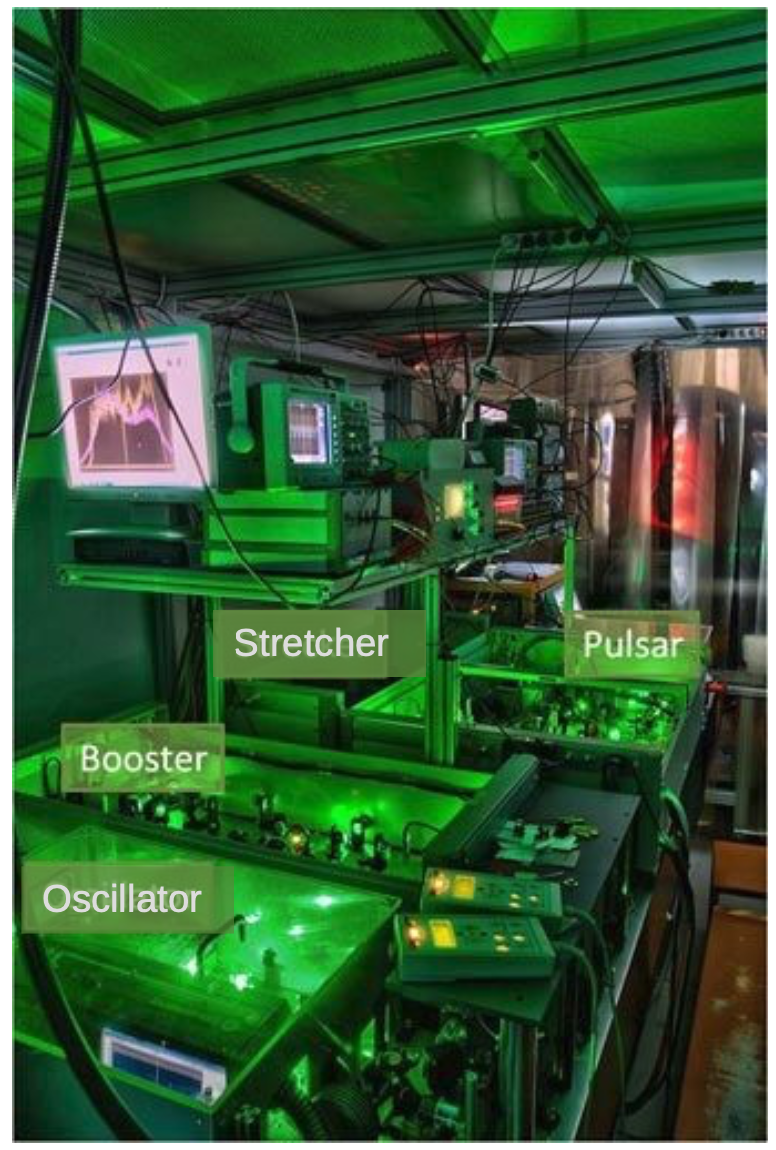}
    \includegraphics[width=0.315\textwidth]{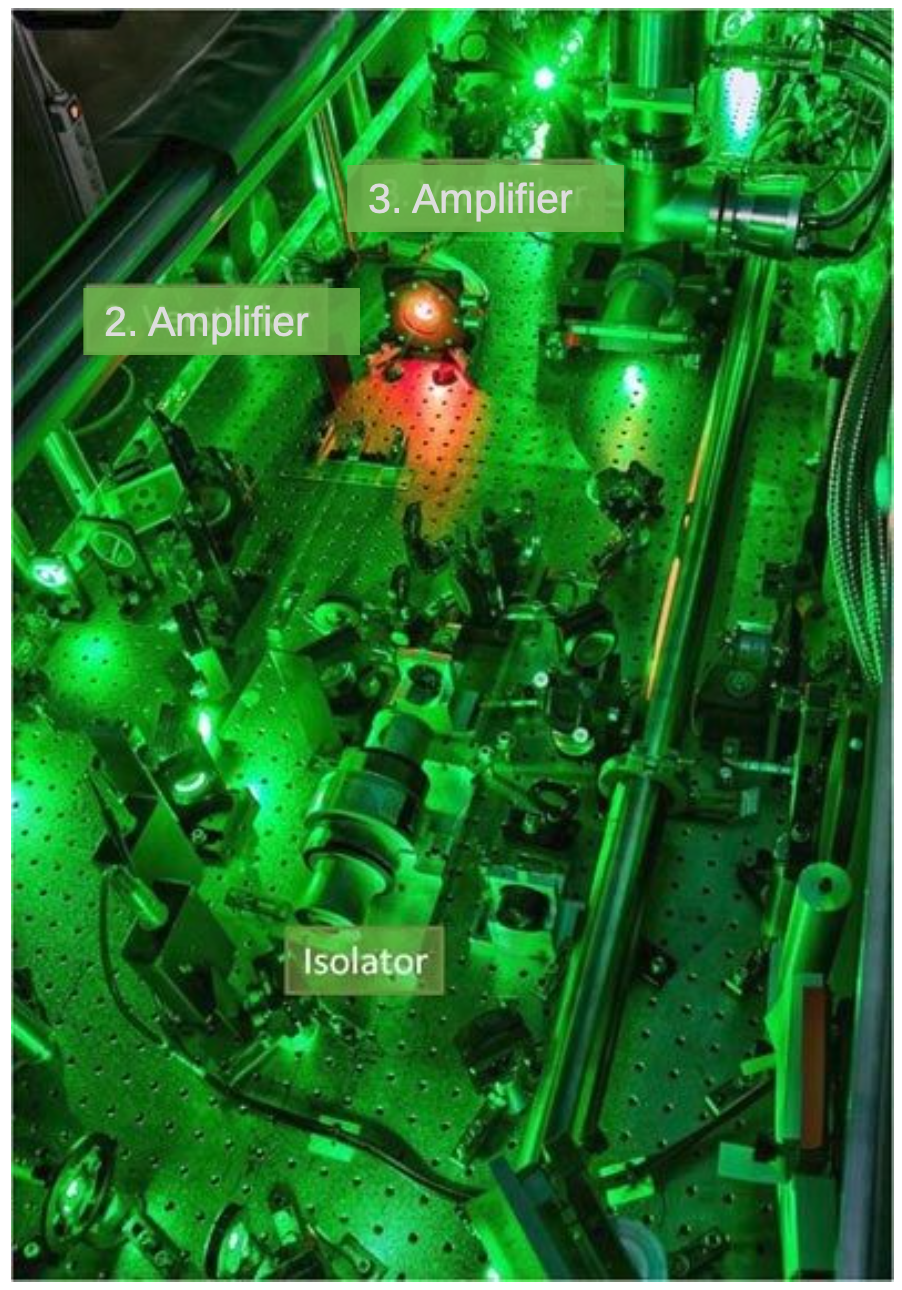}
    \caption{Left: Schematic of the JETI system highlighting major assemblies. Center and right: Current Installation of JETI40 at the University of Jena: the front-end is shown in the centre and and the power amplifiers on the right.}
    \label{fig:JETI}
\end{figure}

The JETI40 laser is a well-characterised system that is in continuous operation for over 10 years to this date and is used for high field science in Jena. 
 It produces pulses down to $25 \units{fs}$ duration of high quality (near the Fourier-transform limit in time and space) with maximum energy after compression exceeding $1 \units{J}$. A schematic layout of the systems is shown in Fig.~\ref{fig:JETI}, and this system will be transferred and installed at the LUXE experiment. This power level is well suited for first stage of LUXE and will allow $\chi$ approaching unity to be achieved with focussing to a 3\,$\mu$m laser spot size. The laser clean room environment that will be installed on level -2  of the building at Osdorfer Born and the beam transport will be specified to include sufficient space for a subsequent installation of a further amplifier and upgrade to $350  \units{TW}$. This will enable LUXE to explore peak intensities of $1.2 \cdot 10^{21}$\,Wcm$^{-2}$, so that data well into the non-linear regime can be obtained at fields significantly exceeding the Schwinger limit  ($\chi_{\rm max}=3.8-4.7$).

The detailed layout of the current JETI40 system and current installation is shown in 
Fig.~\ref{fig:JETI}. The schematic highlights the major components.

\begin{itemize} 
\item \textbf{Front end}\newline
The laser chain starts with a commercial femtosecond oscillator which delivers pulse train at $\approx 72.2$\,MHz (an integer fraction of the \euxfel master clock rate of 1.3\,GHz). A key feature required of this oscillator is that the cavity length can be precisely matched to the \euxfel master clock. The current oscillator will therefore be replaced by a piezo-adjustable oscillator available at the Queen's University of Belfast. Each of these fs pulses contains an energy of a few nJ. A 10\,Hz pulse-train is selected and and amplified to about 0.5\,mJ in the booster. This part of the system is known as the front-end. 

\item \textbf{Optical pulse stretcher}\newline
Following the front end, the pulses are sent into a grating-based pulse stretcher.  There they are steered to hit an all-reflective grating four times.  The stretcher band-pass is set to about 100\,nm to avoid clipping effects that would adversely affect the pulse shape. 
\item \textbf{Multi-pass 10\,Hz Ti:Sapphire power amplifier}\newline  
Following the stretcher the pulses are amplified in a regenerative amplifier with a TEM$_{00}$ spatial mode and amplified in three amplification stages at 10\,Hz repetition rate allowing for stable operation. Each stage consists of a Ti:Sapphire crystal pumped with ns-duration green (532\,nm) laser pulses. Following each stage, the beam is expanded to remain below the damage threshold of the following optics in the optical chain. After a regenerative amplifier and three multi-pass amplifiers, a compressed output energy of 1.2\,J  can be reached. 
\item \textbf{Optical pulse compressor}\newline    
The fully amplified pulses will be expanded and sent into an optical pulse compressor which operates under vacuum.  The compressor  is based on the standard design of four reflections from two large gold-coated diffraction gratings on glass substrates. This compressor design   limits the practical repetition rate for high intensity beam crossings to 1\,Hz due to thermal effects. At higher repetition rates thermal aberrations affect the performance of the laser and higher repetition rates would require costly and complex additional features such as grating cooling or active compensation of grating aberrations. 

\end{itemize}

\subsubsection{Upgrade to $\chi \gg 1$}   

The existing 40\,TW of the JETI-laser is sufficient to achieve the scientific goals of precision measurements in the range of $\chi \leq 1$, but will not allow the regime $\chi>1$ to be accessed in earnest.  The amplifiers after the regenerative amplifier are multi-pass amplifiers in bow-tie geometry. Such amplifiers are conceptually simple and the open set-up of the JETI40 system allows an additional amplifier to be added fairly straightforwardly. Commercially available cooled Ti:Sapphire crystals are available up to the multi-PW level with ns-duration Nd:YAG based pumped lasers with sufficient pulse energy are a standard industry item. Space for an upgrade to the $350  \units{TW}$ level will be left during  installation of JETI40 at Osdorfer Born. The larger energy requires an increased beam diameter from around 50\,mm to 140\,mm to maintain the fluence on the turning mirrors constant.

The vacuum tubes of the beamline and mirror mounts will be prepared for $350  \units{TW}$ from the outset, thus simplifying the upgrade to exchanging the mirrors in the beamline which can be achieved in a standard access period. Similarly, the focussing and recollimation system will be assembled on a pre-aligned and tested breadboard to minimise the required tunnel access time for any change over.

\subsection{Beam-line and Clean-room Installation}

\begin{figure}[htbp]
    \centering
    \includegraphics[width=0.9\textwidth]{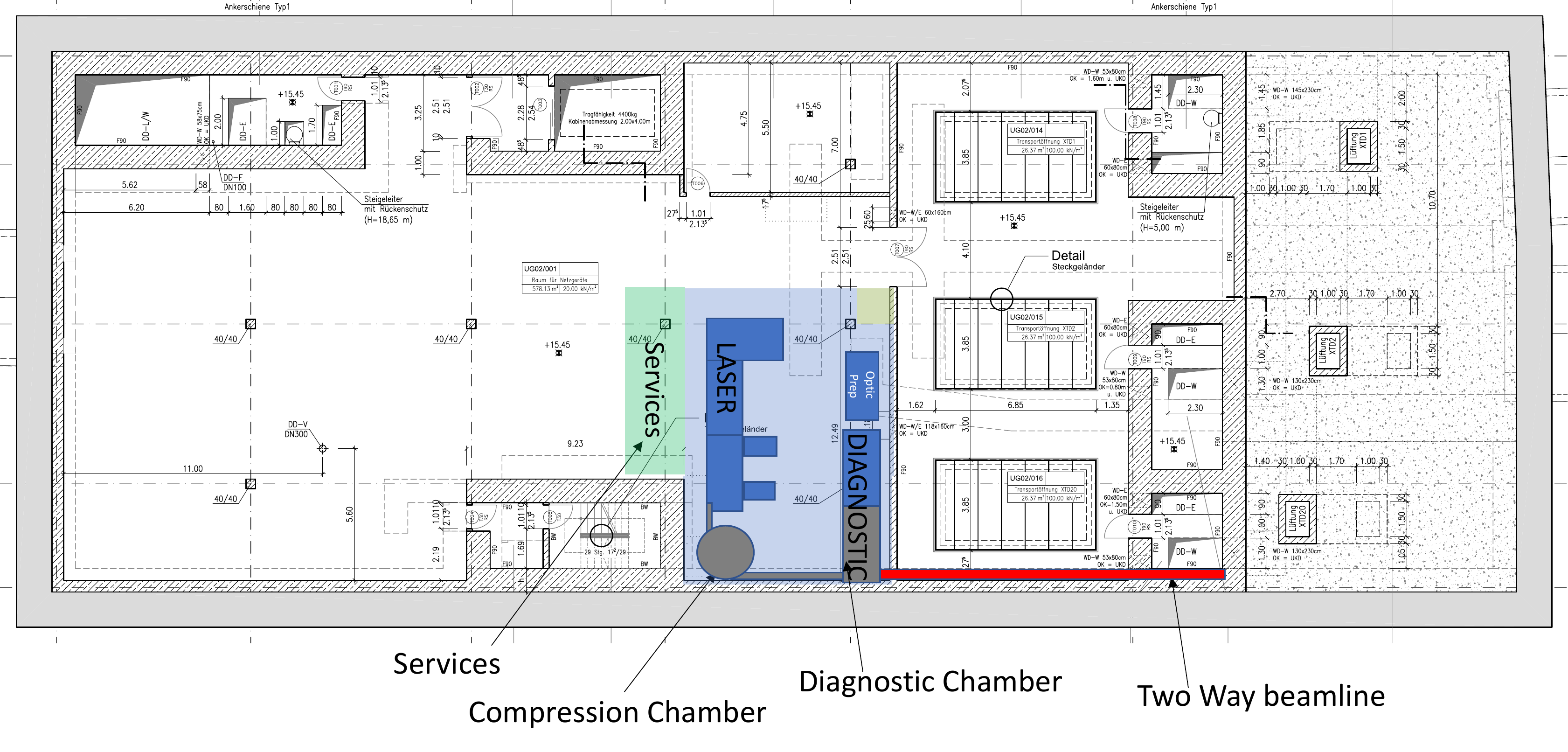}

    \caption{Layout of the proposed laser clean-room in the XS1 building. Services and power supplies with substantial heat load will be outside the clean room to enhance thermal stability and reduce cooling requirements. The two-way beamline transports the high-power beam to the IP and transports the low-power beam to the diagnostic suite for shot-tagging.}
    \label{fig:Cleanroom}
\end{figure}

The existing infrastructure is well suited to installing the LUXE experiment. Sufficient space is available at level -2 with stable floors, access points for deliveries of heavy items (such as laser tables and chambers). A light-weight clean room enclosure (ISO 6) will be set up within the existing floor space to house the laser in a controlled environment. A diagram showing the layout is shown in Fig.~\ref{fig:Cleanroom}.

The laser beam will propagate in a vacuum beamline to the existing area for service feed-throughs to the accelerator level. To future proof the beamline and avoid time-consuming reconstruction, the beamline will be designed to accommodate future upgrades (i.e. large enough for a $350  \units{TW}$ beam) and auxiliary beams. The beamline mirror mounts will be monolithic -- i.e. one single holder will be on the motorised rotation stages capable of holding multiple mirrors. This allows simple transport of multiple beams to the IP and back to the laser clean room for diagnostic purposes. 

After the interaction with the high-energy beam, a small fraction (few percent) of the laser beam energy will be directed back to the laser area for diagnosis to allow shot intensity tagging. The remaining laser energy will be dumped in a beam dump of identical design used on the $350  \units{TW}$ system at the HI Jena~\cite{jena300twlaser}.

An important feature of the beamlines will be relay imaging of laser from the compressor output to the focussing parabola and from the recollimating parabola to the diagnostics to ensure the highest precision of laser beam diagnosis.

More details on the installation are discussed in Sec.~\ref{sec:tc}. 

\subsection{Laser Diagnostics}

In order to perform precision studies, it is crucial to have high quality diagnostics to closely monitor the effects of shot-to-shot fluctuations and long-term drifts on the precision of the data. High-power, femtosecond laser systems are precision tools and respond with significant performance changes to relatively small drifts in alignment. This is because small variations of spatial and spectral phase have noticeable effects on the peak intensity. Such variation can be caused by thermal effects, air currents and mechanical vibrations and drift.
Current commercial systems are engineered to consist of distinct mechanical components set out on a conventional optical table and must therefore be re-aligned routinely and are open to a variety of effects varying performance. This contrasts with sealed, turnkey systems which exist in different parameter ranges in industry.
However, we believe that this will not limit the ultimate precision and reproducibility we can achieve.  Intensity tagging (precisely measuring and assigning an intensity value to each shot during the experiment) will enable high precision measurement of \hiqed phenomena with respect to laser intensity.

In the following we detail the typical level of shot-to-shot fluctuations and approaches with which we aim to tag each shot with a precision of below 1\% (ultimately 0.1\% is believed to be achievable based on currently ongoing developments at the JETI laser system in Jena). 
High energy Ti:Sapphire lasers using flash-lamp technology for the pump-lasers have typical energy fluctuations of around 2--3\% rms. These overall energy fluctuations, while significant, are not the dominant contribution to shot-to-shot intensity fluctuations. The major contributions to such fluctuations are small changes in phase both in real space and in frequency space. Small changes in spatial phase result in the spot radius fluctuating at the few \% level resulting in up to 10\% level intensity fluctuations. Similarly, fluctuations in the spectral phase can lead to 1\% rms fluctuations in the pulse duration. In total, the variation in between the highest and lowest intensity shots $I_0$ on a stable laser can reach $\sim$ 15\%.
To mitigate this, we will set up a state-of-the-art diagnostic system capable of measuring the fluctuations, a schematic of the proposed diagnostic system is shown in Fig.~\ref{fig:diag_layout}. The shots will then be tagged with their precise relative intensity allowing precision relative measurements using the following diagnostics:
\begin{itemize} 
\item \textbf{Energy Tagging}\newline
The full beam energy will be measured by imaging an attenuated beam onto a CCD. For a well exposed image, containing $>10^{10}$ total counts, the energy fluctuations can be measured to $<10^{-5}$ accuracy and the shot-to-shot variation in beam energy eliminated from the data set with high accuracy.

\begin{figure}[htbp]
    \centering
    \includegraphics[width=0.7\textwidth, angle=0]{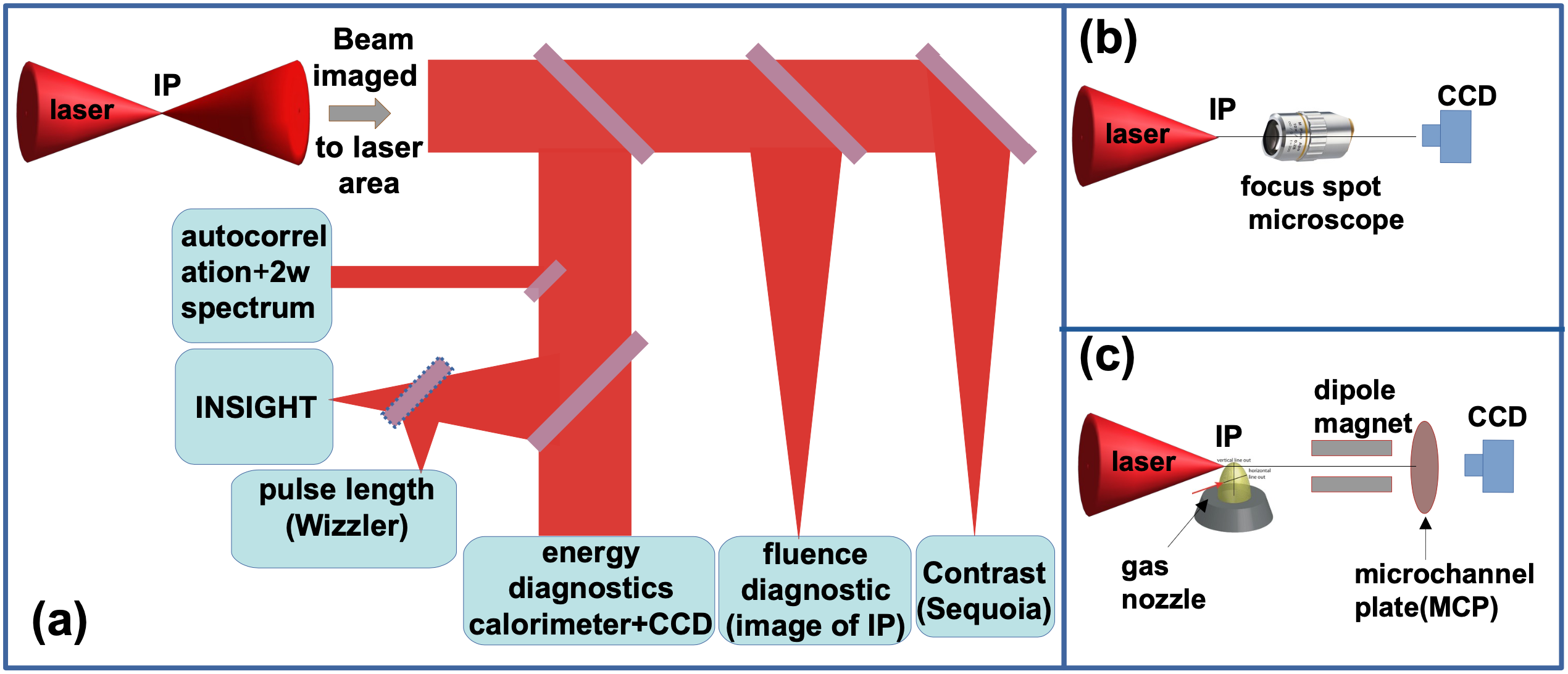}
    \caption{Schematic diagram of the proposed intensity tagging diagnostics. The laser beam will be  attenuated and imaged on the return path to the diagnostics. (a) Layout of the laser shot tagging diagnostics in the laser area. (b) Focus diagnostic for the spot size measurement at the interaction point. (c) Set-up for in situ intensity measurement in the IP. A nozzle valve is used to produce a  tenuous gas jet. The accelerated particle spectra can be recorded employing magnet spectrometer and micro channel plate detectors (MCP) allowing the absolute value of $\xi$ 
    to be constrained by comparison to 3D particle simulations using PIC codes.}
    \label{fig:diag_layout}
\end{figure}

\item \textbf{Fluence Tagging}\newline
A similar approach will be taken to determine the fluence at the laser frequency $F_\omega=\frac{dE}{dA}$
using a high magnification image of the focal spot onto a CCD camera (where $dE$ is the energy incident on a detector pixel of area $dA$). Care will be taken to eliminate any non-linear effects in transmissive optics. The variation of the efficiency of the CCD with wavelength would reduce the effectiveness of the method described above if there are significant shifts or fluctuations of the spectrum. We will employ a spectrometer and colour filtered images to maintain the high precision that is theoretically possible in these measurements.

\item \textbf{Pulse Length Tagging}\newline
Finally, the stability of the pulse duration will be determined by employing two complementary techniques. We will measure the pulse duration on every shot using a state-of-the-art system capable of reconstructing the full pulse shape such as the Wizzler combined with an autocorrelator. The measurement will use the full aperture of the beam to be sensitive to spatio-temporal couplings that can affect the pulse duration. We will employ a simpler, complementary technique to ensure that this measurement is not dominated by fluctuations within the measurement device by producing an image of the frequency converted beam and comparing it to the corresponding image of the fundamental beam. Since the conversion in a thin non-linear crystal far from saturation scales with intensity $I$ as $I^2$,
any fluctuations will appear directly as variations in the ratio of the two images $R_{\omega,2\omega}=F_\omega/F_{2\omega}$. Calibration of this diagnostic should allow precise relative pulse duration tagging. Although we do not anticipate this to achieve the level of precision of the energy and flux diagnostics, a relative pulse duration precision of $<10^{-3}$ is deemed achievable. Finally, we will measure fluctuations in the angular chirp using imaging spectrometers.

\item \textbf{Full Field Reconstruction}\newline
We will implement advanced laser field reconstruction technique known as INSIGHT. This will be led by our collaborators from CEA Saclay who invented this technique \cite{Quere:NPhot}. INSIGHT provides full 3D phase and amplitude reconstruction of a laser beam focus. While this technique requires multiple laser shots, it provides the field distribution with unparalleled precision. This performance check can be run on a daily basis and the obtained field distribution will act as an input for full 3D simulations of the interaction at the IP.

\end{itemize}

Additionally there will be focal spot microscopes at the IP to directly image the focus interacting with the high-energy beam and provide a cross-check for the on-shot diagnostics.
 
The diagnostics will be capable of operating in two distinct modes by switching the beam path from the compressor directly into the diagnostics or alternatively diagnosing the interaction beam after passing through the IP. Imaging of the laser beam ensures that the beam intensity profile at the compressor output, the focussing parabola at IP and at entry of the returning beam have the same phase and amplitude distribution, thus providing the highest possible fidelity of the diagnostic suite. The detailed development and full calibration of the diagnostic suite will take place at the JETI40 laser in Jena prior to its transfer to the new location in Hamburg.

\subsection{Determination of the Peak Electric Field in Focus}

In order to provide precise information on the interaction and therefore perform rigorous comparison with available strong-field QED models, it is necessary to accurately measure the laser peak intensity in focus. 
In principle, the tagging system outlined above will provide an accurate measurement of the peak field in the interaction point. We aim to cross-calibrate our diagnostic suite described above at the Light Intensity Unit of the Physikalisch-Technische-Bundesanstalt (PTB) Braunschweig~\cite{ptbbraunschweig}, which can provide traceable standards for continuous-wave (CW) sources to SI Units with $10^{-4}$ accuracy. This will provide tight limits on the actual value of $\xi$. However, a systematic offset due to small imaging aberrations of the focal intensity distribution at the interaction point and spectral phase errors in the transport will be unavoidable and is not easily accounted for. While it has been recently shown \cite{Quere:NPhot} that spatially resolved Fourier-transform spectroscopy allows one to reconstruct the spatio-temporal profile of the laser electric field in focus, this also is a low power measurement.
It is thus desirable to have a direct measurement of the peak intensity in focus and different methods have been proposed in the literature which constrain the absolute value of $\xi$. Combined with high precision relative tagging and attenuated measurements of the focussed beam this will allow comparison with theory to be well constrained.

\begin{figure}[htbp]
    \centering
    \includegraphics[width=0.45\textwidth, angle=0]{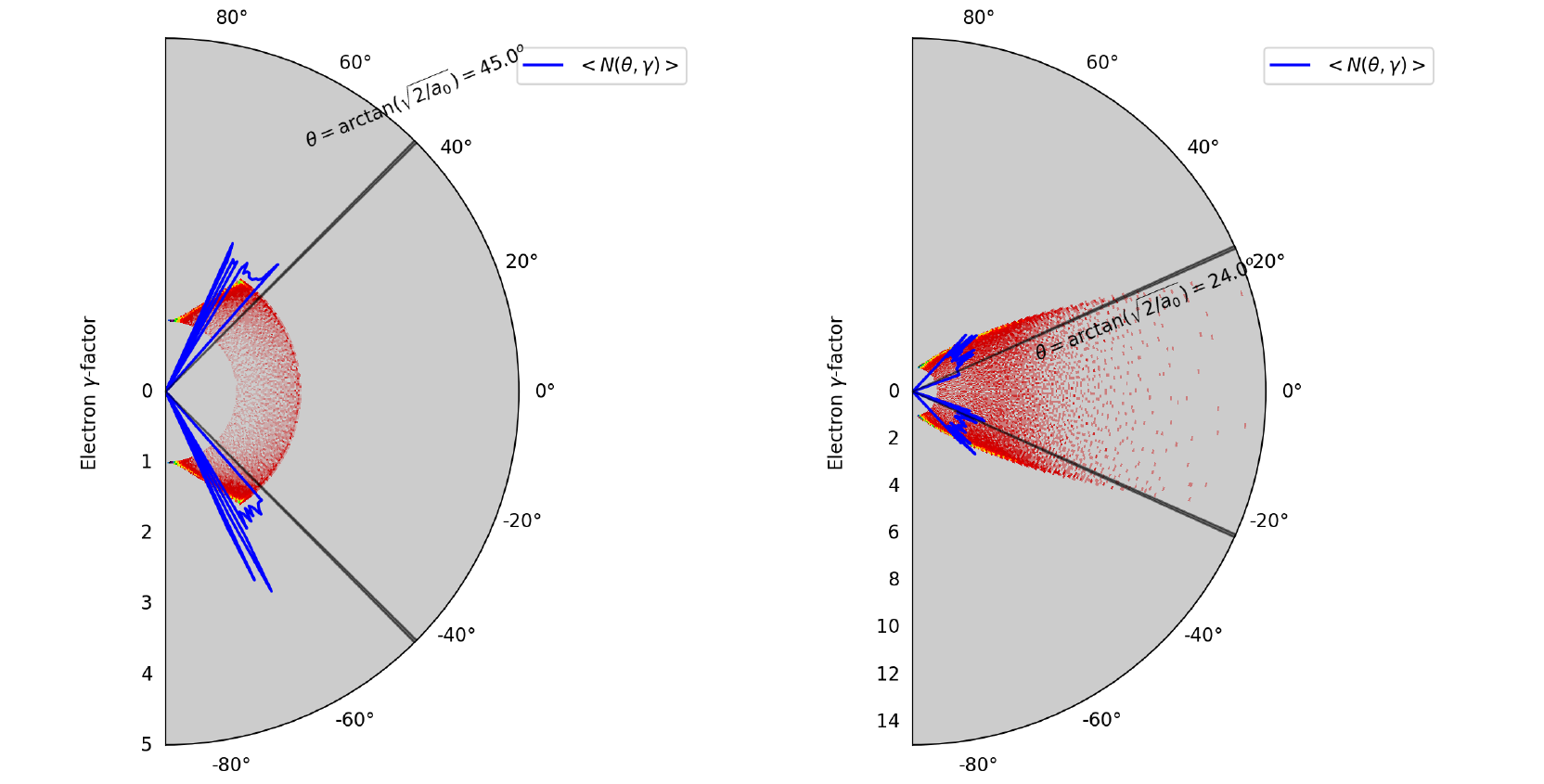}
    \includegraphics[width=0.45\textwidth, angle=0]{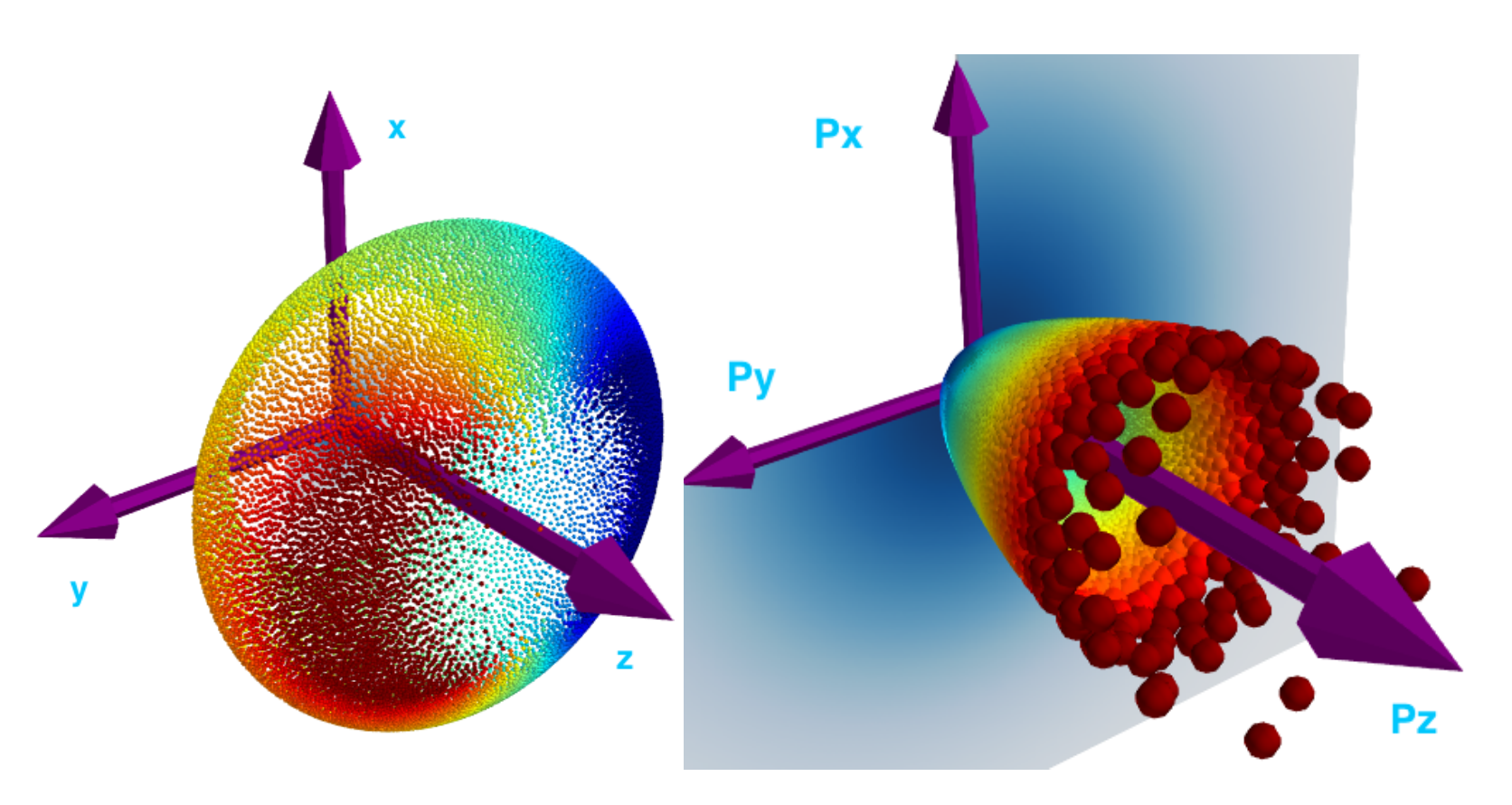}
    \caption{Simulations of in situ approach to determining the laser intensity by electron scattering.   Left panels: Angular distribution of electrons initially at rest for $\xi = 2$ and $\xi = 10$. The blue line shows the energy of the averaged number of electrons scattered under a given angle. A well defined angular edge is visible depending on $\xi$, allowing the effective peak field strength in the interaction point to be cross-calibrated with the remote diagnostics.  Right panels: Angular distribution of electrons scattered by a laser pulse in real and momentum space. In this example, the laser dimensionless parameter is $\xi = 10$ and the focal spot has a width of $4 \,\mu$m.    }
    \label{fig:xi}
\end{figure}

The laser-driven scattering of electrons  from a tenuous Helium target at the IP (for these measurements the IP needs to be vacuum isolated from European XFEL) provide a route to constraining the absolute value of $\xi$ \cite{Vais:2021}. The relative transverse and longitudinal momenta of the  scattered electrons are at well defined angles depending on the laser intensity when the full interaction geometry is taken into account (Fig. \ref{fig:xi}).
These scattering angles are outside the laser beam-cone and can be detected using a compact electron spectrometer optimised for ponderomotive electrons in the MeV energy range without modifying the optical layout. Determination of $\xi$ can be achieved by comparing the measured electron scattering to full 3D simulations. Based on our  simulations  ponderomotive scattering we aim to achieve an absolute calibration of better than 5\% with  relative accuracy of 0.1\% in the laser E-field appearing realistic. 

\subsection {Spatial and Temporal Overlap}
\label{sec:laser:overlap}
Achieving -- and maintaining -- spatio-temporal overlap between the laser focus and with the LINAC electrons (or the bremsstrahlung or ICS photons) at the IP is an essential pre-requisite for precision data acquisition. Overlap in space
requires an alignment, positioning and drift compensation system for the laser focus, to track
the electron bunches. The timing system must
ensure simultaneous arrival at the IP of the laser pulse (with nominal 30~fs duration) and the electron bunch (with typical length of $\sim 40$\,$\mu$m, corresponding to $\sim 130 \units{fs}$
duration). Considering the geometry of the \elaser collisions at an angle about $\theta
\simeq 0.3$\,rad and a spot size of $8\,\mu$m, a timing accuracy of $\sim 100 \units{fs}$ is required.

\subsubsection{Temporal Overlap}
Temporal overlap  between the 30\,~fs laser pulse and the $>100 \units{fs}$ electron or photon pulse is desirable to a precision of ideally half the pulse-width, i.e.\ 50\,fs or better, to cover the most stringent scenario with small spots and short XFEL.EU bunches. The European XFEL has developed world-leading synchronisation capabilities which have demonstrated the stable synchronisation of two RF signals to better than 13~fs and even in the absence of feedback, drifts are limited to 450~fs peak to peak variations over a period of two days~\cite{Schulz}. This technology will be used to synchronise the XFEL.EU master clock oscillator to the oscillator of the JETI40 laser system. The repetition rate of the laser oscillators will be set to be precisely matched to a sub-harmonic of the master clock by adjusting the cavity length. Fine adjustment to compensate for any drift will be achieved using piezo-actuators at the cavity end mirror to fine-tune the cavity of the JETI-laser. This method is successfully used across the XFEL.EU to synchronise lasers to the accelerator.

\begin{figure}[htbp]
    \centering
    \includegraphics[width=0.7\textwidth]{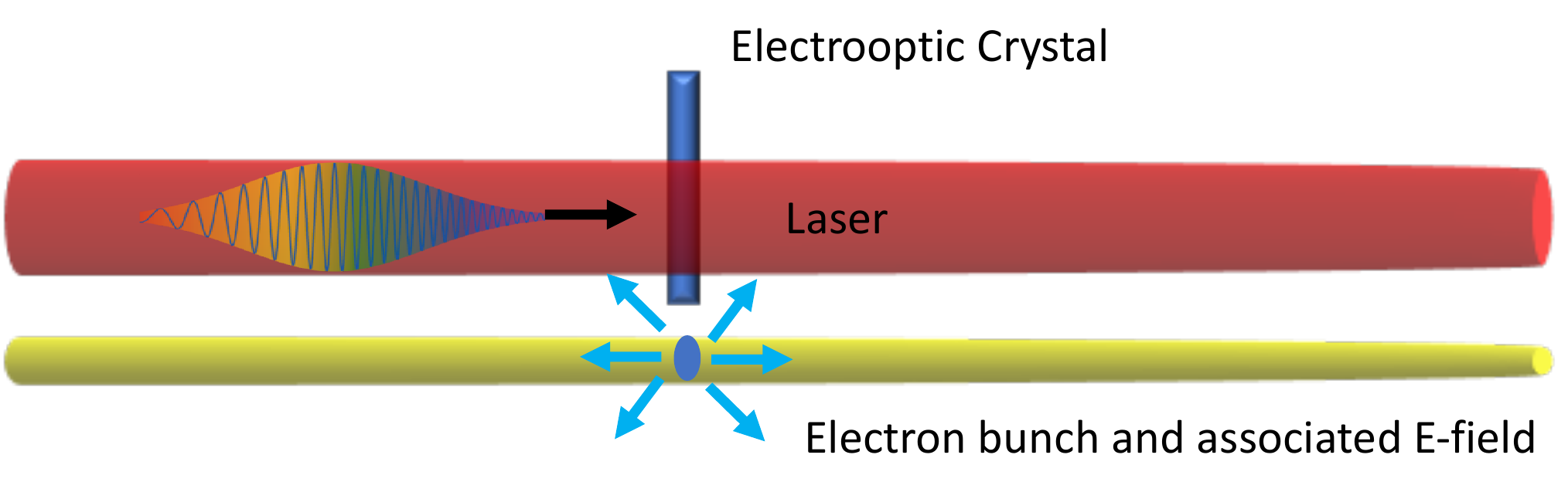}
    \caption{The laser oscillator and the XFEL.EU oscillator will be synchronised at 10~fs level using the XFEL.EU timing system. To compensate for drift and jitter in the beam transport at low repetition rate, an electro-optical timing tool will be implemented which provides a signal proportional to the electron-beam induced polarisation change.}
    \label{fig:timingtool}
\end{figure}
However, this high precision synchronisation only synchronises the two sources of RF signal to each other. The exact overlap in the interaction point is still subject to variations due to temperature drift that may change the length of the laser beam path. The aim is to achieve sufficiently stable operation passively by isolating optical components from sources of vibration using commercially available  vibration isolation and good thermal stability. At the same time we will plan for active feedback to maintain highest timing accuracy at the interaction point.
Drift and jitter can be compensated by directly measuring the relative timing of the electron beam and the laser in the chamber housing the target, described in Sec.~\ref{sec:tchamber}. As shown in Fig.~\ref{fig:timingtool}, a portion of the laser beam will be sent through an electro-optic crystal adjacent to the electron beam. The electro-optic effect will lead to a depolarisation of the probe light that can be detected on every shot. By placing a chirp on the probe laser spectrum to achieve a pulse-length of a few 100\,fs, the instantaneous frequency of the laser can be used to temporally encode \cite{Dromey:NComm} the relative timing on a single shot basis. High-speed piezo-translation stages can respond to timing signals and adjust the length of the dog-leg to keep the timing signal constant. This level of feedback stabilisation will be possible at the 10\,Hz pulse repetition rate of the laser system, which is sufficiently rapid to control low frequency vibrations and any temperature induced drift. 

Previous experience from synchronising lasers suggests that the shot-to-shot residual jitter due to mechanical vibrations will be at most $\sim 10-20 \units{fs}$ and hence tolerable. 

In the unlikely event that 10\,Hz stabilisation will not be sufficient we will use the high repetition rate front end to characterise the beam-line vibrations. Using high CMOS cameras, kHz vibrations can be analysed by interfering the pulse that has propagated through the beam-line beam path with a leakage pulse from the oscillator. As the oscillator is synchronised to $<10 \units{fs}$  to the bunch frequency of the accelerator, this high frequency signal can be used to stabilise  mechanical vibration using fast delay stages if required.

\subsubsection{Spatial Overlap}
The vibration isolation of lasers on their optical tables is generally sufficient to achieve a vibration level such that the spot-movement $\Delta_{xy}<w_0$,  where $w_0$ is the laser waist size. However, the beamline vibrations discussed above can be the source of spot vibrations such that  $\Delta_{xy}= m w_0$, with $m>1$. This level of vibration is not critical for the overall performance of the system (spot quality) but needs to be considered in terms of experimental stability. We will characterise the pointing stability of the electron and laser beams using the in-situ alignment tools: The electron beam stability will be measured by recording the x-ray production by electrons impinging on the needle alignment tool or a thin Ce:YAG scintillation screen. The laser pointing stability will be quantified by recording its far field images at low power, using the in-situ microscope, over a few-hundred shots.

As shown in Table \ref{tab:laser} the waist of the laser spot will be $w_0=8 \units{\mu m}$ or less in LUXE. We identify the following regimes of sensitivity relative to our measurement goals:

\begin{itemize}
\item{Laser-electron Interactions\\
Nominally the electron beam diameter can be focussed as small as 10~$\mu$m and thus of comparable size to the laser beam. Small displacements between the two can have a significant effect and it might be desirable to focus the electron beam less strongly. For relaxed electron beam focussing such that $\sigma_{xy} \gg w_0$ at the interaction point it is required that the electron beam focal size be sufficiently larger than any spot movement, $\sigma_{xy} > \Delta_{xy} = 2m w_0$, thus guaranteeing good overlap. Care will be taken to minimise any spot movement and  beam pointing monitoring in the IP chamber will be implemented to confirm overlap on each shot.}

\item{Laser-photon Interactions}\\
The high-energy photons have an angular spread of $\theta = 1/\gamma$ implying that the central intense peak of the $\gamma$-ray distribution is $\sim 230\,\mu$m (see Sec.~\ref{sim:geant4}).
This is substantially larger than the laser spot-size and consequently small vibrations in the spot position can be tolerated without affecting measurement precision. 
\end{itemize}

\subsection{Interaction Chamber}
\label{sec:ic}

\begin{figure}[htbp]
    \centering
    \includegraphics[width=0.85\textwidth]{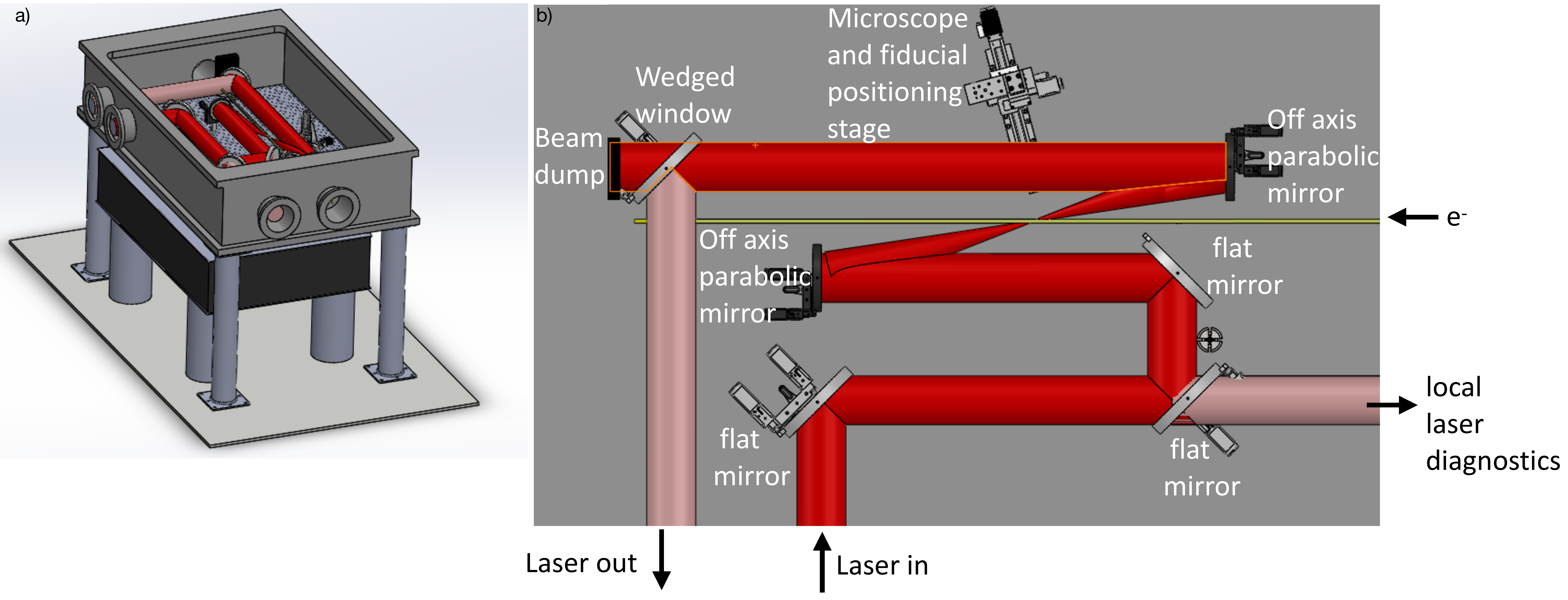}
    \caption{CAD drawings of the interaction point chamber in (a) perspective and (b) top views for the 350\,TW beam configuration with the parabola position shown also corresponding to the $8\,\mu m$ spot configuration for JETI40. The linear dimensions of the chamber are 2452$\times$1500$\times$504\,mm.}
    \label{fig:ip-chamber-schem}
\end{figure}
CAD drawings for the interaction chamber are presented in Fig.~\ref{fig:ip-chamber-schem}. 
The incoming beam will propagate over three bend mirrors and an $f/\#=f/D_\mathrm{Laser}=f/3$  off-axis parabolic mirror to focus at the interaction point (where $D_\mathrm{Laser}$ is the laser beam diameter). 
A second parabolic mirror will re-collimate the beam and direct it to a beam-dump outside the chamber. Before reaching the beam-dump, about $1\%$ of the laser's energy will be sampled and sent back to the pulse diagnostics suite. 
All optics will be motorised for fine manipulation under vacuum. A motorised in-vacuum microscope will be used to optimise the mirrors alignment, and to evaluate the laser focal spot size. We will monitor the incoming laser pointing stability by imaging the near- and far-fields of its leaking fraction behind the second bend mirror.

We will mechanically decouple the optics from the chamber's body by fixing the breadboard to an optical table beneath the chamber. The chamber itself, with its set of legs, will be set directly on the floor. Flexible bellows will form the contact surface between the vacuum vessel and the optical table.

\subsection{Target Chamber}
\label{sec:tchamber}

\begin{figure}[htbp]
    \centering
    \includegraphics[width=0.7\textwidth]{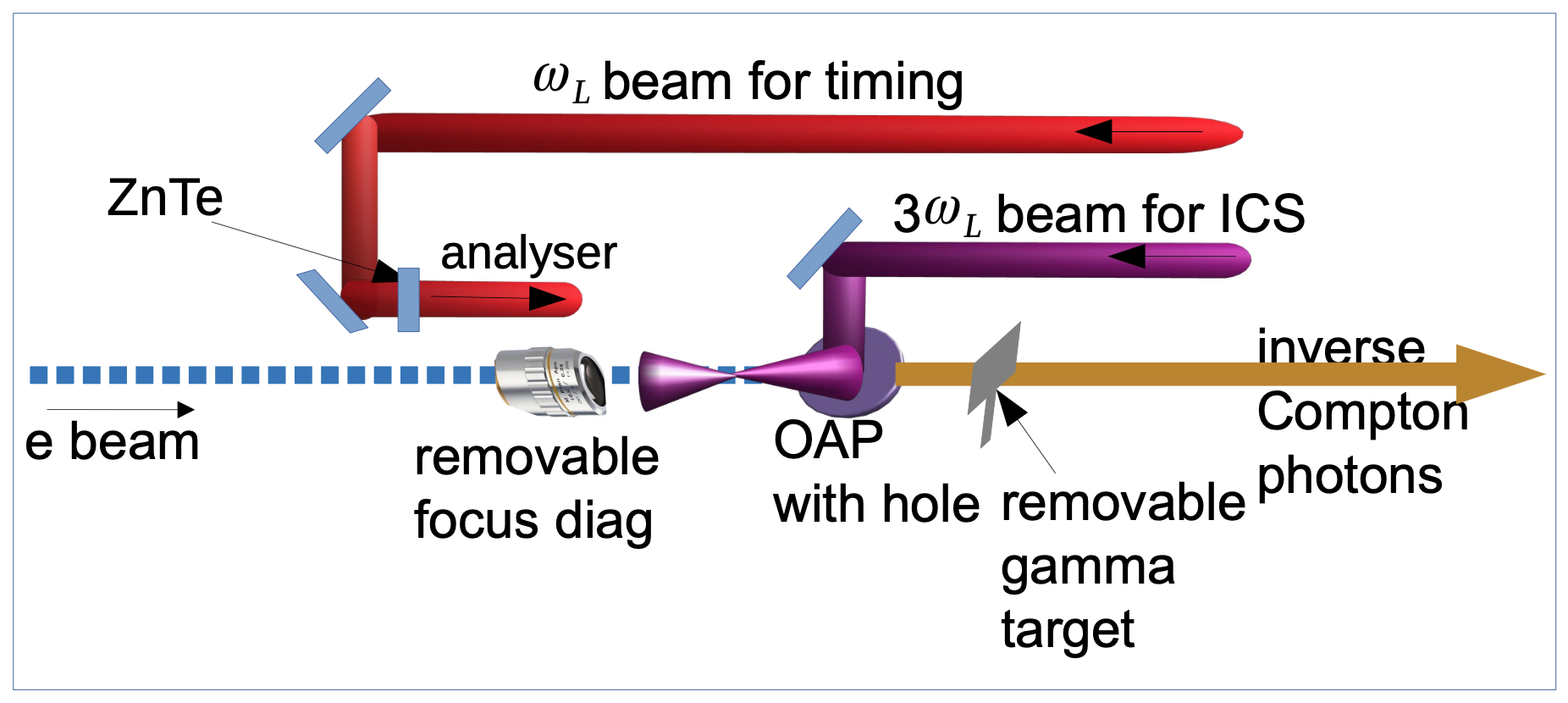}
    \caption{Schematic of target chamber for bremsstrahlung photons and inverse Compton photons. In addition, the timing unit will also be housed in this chamber.}
    \label{fig:ics_setup}
\end{figure}

A smaller "target chamber" will be located at the position of the bremsstrahlung target. This chamber will house the electro-optical timing tool described in Sec.~\ref{sec:laser:overlap}. The probing of the timing crystal will be done using a small probe line in $\omegaL$. The schematic layout is shown in Fig.~\ref{fig:ics_setup}. 

In addition, this chamber will also host the set-up for the production of high energy, narrow band ICS photons (as discussed in Sec.~\ref{sec:science:ics}). A head-on collision geometry will be used (see Fig. \ref{fig:ics_setup}). Here the electron beam will interact with a low intensity $3\omegaL$ laser pulse derived from the same laser system. The conversion to $3\omegaL$ and delay compensation will take place in the laser hall.  Both the $\omegaL$ and $3\omegaL$ probe will travel through the main beamline, however spatially separated.   An off axis parabola (OAP) with a hole will be used. This will allow the passing of the ICS photons towards the main interaction chamber. 
With a few 10s of mJ energy at $3\omegaL$, a focus spot size of $\sim 5~\mu$m and an interaction length of $2z_R=2\pi w_0^2/\lambda\sim 300-400\,\mu$m, one can easily reach $\xi$ values of $0.1$. The electron beam will be focussed to a matching spot size as detailed in the spatial overlap criteria above.

\clearpage
\section{Simulation}
\label{sec:simulation}

The Monte Carlo (MC) simulation of the physics at realistic strong-field interaction point geometries is first described. It is based on the theoretical concepts discussed in Sec.~\ref{sec:science} and serves as an important tool to optimise the design of the LUXE experiment. In particular, based on the MC, the event rates for the signal processes are determined and presented in Sec.~\ref{sec:mc}.

To study and optimise the performance of the detectors, the experimental layout is implemented in detail in {\sc Geant4} as discussed in Sec.~\ref{sim:geant4}. Simulation of the signal and beam induced background is then performed for various configurations of the planned measurements programme to optimise the overall layout. In particular, based on these simulations, hot-spots of secondary particle production were identified and minimised or mitigated through relocation of passive and active components or through the addition of shielding~\footnote{In some cases, due to time constraints the background particles were removed by software cuts when they originated from a region which will be shielded effectively in the experiment but was not yet in the simulation.}. 

Finally, in Sec.~\ref{sim:sig_bkg}, the relative rates of signal and background at the face of the planned detectors are compared and discussed.  Beyond the geometry, these rates are reasonably independent of the detector technology and 
highlight the challenges faced by the detectors and motivate their choices. The purpose of this section is to motivate why a given technology is chosen to cope with the expected signal and background in the various locations. However, details on the detector technologies are only given in Sec.~\ref{sec:detectors}.

\subsection{Monte Carlo Event Generation}
\label{sec:mc}

A custom-built strong-field QED MC computer code, named \textsc{PTARMIGAN}~\cite{ptarmigan},
is used to simulate the strong field interactions for LUXE for the relevant physics processes using the LMA for cross-section calculations as described in Sec.~\ref{app:theory}.
The task of MC generator is to simulate the physics processes in a realistic scenario with the laser focusing and electron beam sizes taken into account properly. Several cross-checks have been made to ensure the validity of the programme~\footnote{These cross-checks also uncovered that the \textsc{IPSTRONG}~\cite{ipstrong}  simulation, used for the Letter of Intent~\cite{Abramowicz:2019gvx}, had several features which caused an overestimate of the pair production rate in \glaser collisions. An earlier version of this CDR~\cite{luxecdrv1} also suffered from small inconsistencies and problems in the \textsc{IPSTRONG} simulation which however had no impact on the conceptual design of the experiment.}. Details on the simulation are given in App.~\ref{app:sim}.

\begin{figure}[htbp]
    \centering
    \includegraphics[width=0.5\textwidth]{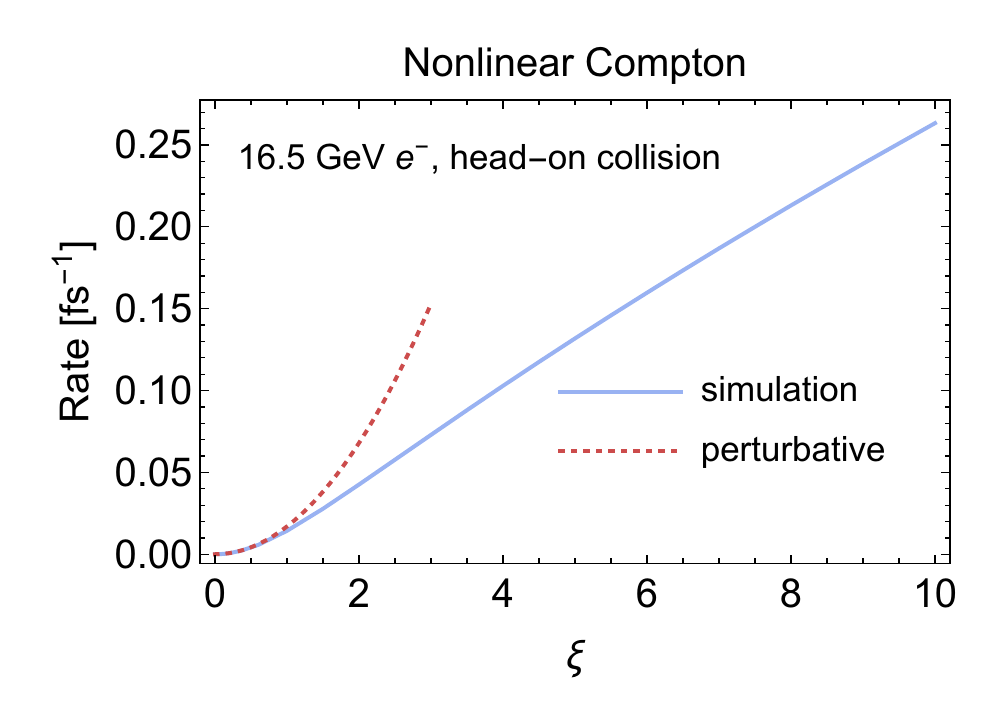}
    \caption{Comparison of the nonlinear Compton rate, as implemented in PTARMIGAN~\cite{ptarmigan} (blue), with the theoretical, perturbative result $\propto \xi^2$ (red, dotted), for an electron-beam energy of 16.5\,GeV.}
    \label{fig:hics-analytic-sim}
\end{figure}

\begin{figure}[htbp]
    \centering
    \includegraphics[width=0.5\textwidth]{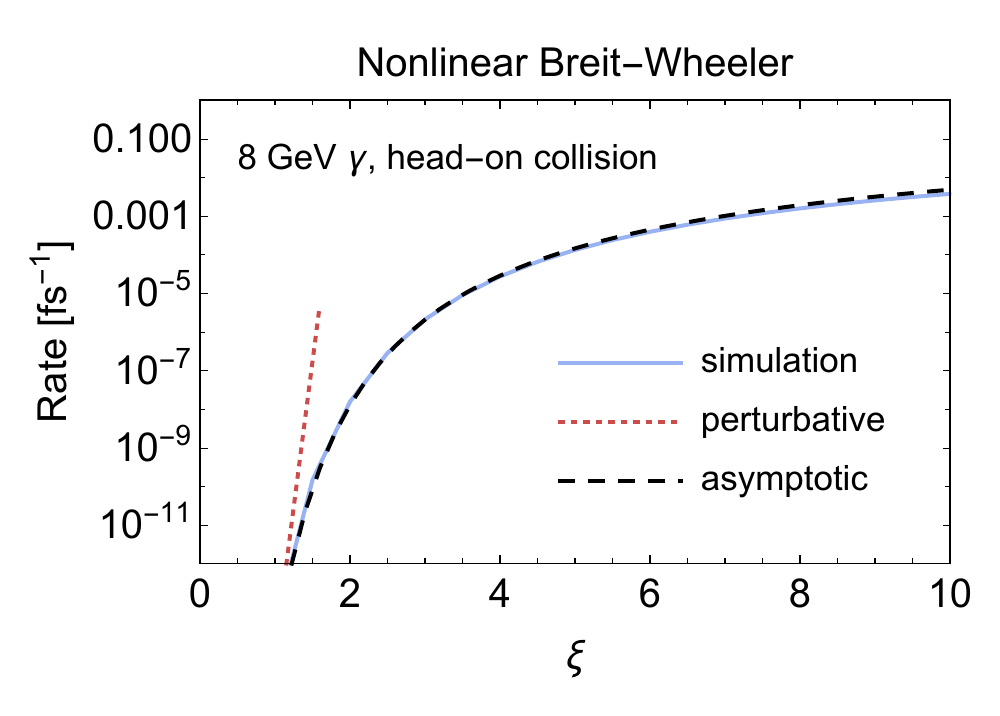}
    \caption{Comparison of the nonlinear Breit-Wheeler rate, as implemented in PTARMIGAN~\cite{ptarmigan} (blue), with the theoretical, perturbative result $\propto \xi^{2 n_\star}$ (red, dotted), where $n_\star$ is given by Eq. \ref{eqn:nastNBW1}, and the asymptotic result $\propto \xi \exp[-8/(3\chi_\gamma)]$ (black, dashed), for an photon-beam energy of 8\,GeV.}
    \label{fig:oppp-analytic-sim}
\end{figure}

The MC has been used to generate datasets for all foreseen LUXE experimental configurations. These include the non-linear Compton process, the two-step trident process, and the non-linear Breit-Wheeler pair production process. The results of the MC are validated by comparing with analytic distributions (see also App.~\ref{sim:montecarlo}). 
Figures~\ref{fig:hics-analytic-sim} and ~\ref{fig:oppp-analytic-sim} show a comparison of the nonlinear Compton and Breit-Wheeler rates, as implemented in the simulations using the locally monochromatic approximation, with theory.

For the \elaser set-up the electron beam parameters are as stated in Table~\ref{tab:xfelepara}, particularly $\varepsilon_e=16.5$\,GeV, the beam spot $\sigma_x=\sigma_y=5$\,$\mu$m, $\sigma_z=24$\,$\mu$m, and the normalised emittance $1.4$\,mm$\cdot$mrad. 
For the \glaser set-up the photon energy and angular spectra are taken from the \geant simulation as described in Sec.~\ref{sim:geant4}. The laser is assumed to be circularly polarised. 

\begin{figure}[htbp]
    \centering
\includegraphics[width=0.5\textwidth]{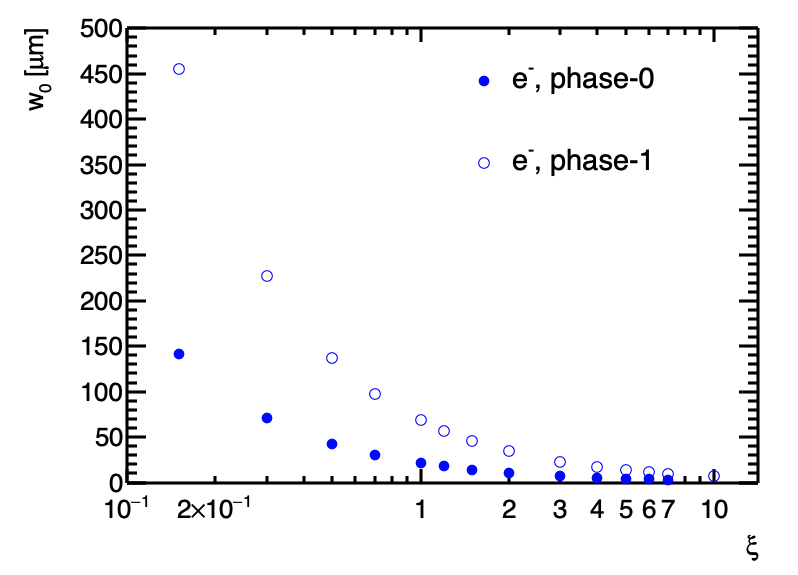} 
    \caption{Values of $w_0$ and $\xi$ simulated for the \elaser set-up for \phaseone and \phasetwo.}
    \label{fig:w0vsxi}
\end{figure}

The simulation of the laser uses the parameters of Table~\ref{tab:laser} with two exceptions for the JETI40 laser: an energy after compression of 0.8~J was used and a pulse length of 25~fs. This has no practical impact on the design of the experiment. And, in order to study the dependence on $\xi$ the laser spot size is varied. Specifically, the value varied is the waist, $w_0$, which for a Gaussian pulse corresponds to $2\sigma$ in intensity. Figure~\ref{fig:w0vsxi} shows the $w_0$ and $\xi$ values simulated for the \phaseone and \phasetwo lasers. 

\begin{figure}[htbp]
\centering
\begin{subfigure}[t]{0.48\textwidth}
\centerline{\includegraphics[width=\textwidth]{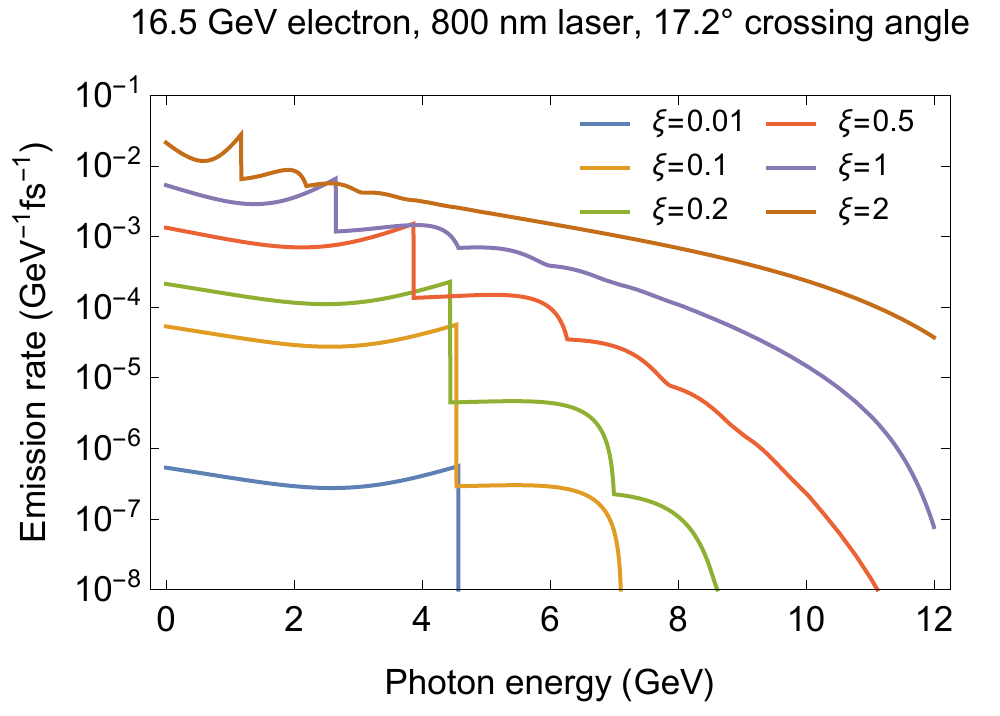}}
\caption{}
\vspace{0.1cm}
\end{subfigure}
\begin{subfigure}[t]{0.48\textwidth}
\centerline{\includegraphics[width=\textwidth]{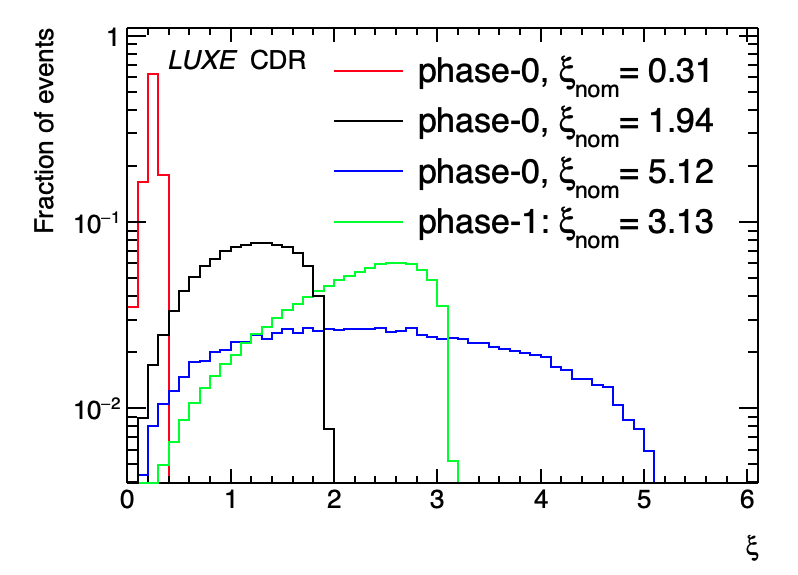}}
\caption{}
\end{subfigure}
\caption{a) The photon emission rate for the Compton process as function of the photon energy, based on calculations for a selection of $\xinom$ values. 
b) The distribution of the actual $\xi$ values in the Compton process in the MC simulation for a selection of $\xinom$ values.
}\label{fig:sim:hics}\end{figure}

The laser pulse is modelled as having a Gaussian profile both in the longitudinal (or time) direction and in the transverse direction. Thus the highest intensity is only present at the peak of the pulse while in the tails of the pulse the intensity, and correspondingly the actual $\xi$ value is lower. Figure~\ref{fig:sim:hics} shows the photon energy spectrum for the calculation at fixed $\xi$ values, where sharp Compton edges are observed at low $\xi$. Also shown is the distribution of the actual $\xi$ values of the interactions for four cases of nominal values,  $\xinom$, which is defined as the nominal value calculated based on the laser parameters (see Sec.~\ref{sec:science}). It is seen that the maximum value of $\xi$ distribution corresponds to $\xinom$ but there is a large tail towards lower values. Since the location of the edge in the energy spectrum depends on the actual $\xi$ value of the interaction, the edges in the observed energy distribution are smeared out. However, the point where the edge is expected to start is the maximum value of $\xi$. 
In the following, the term $\xinom$ is used when discussing the nominal (and maximum) $\xi$ value corresponding to the laser configuration, $\xi$ for the actual $\xi$ value of the interaction and $\ximean$ for the average $\xi$ value. 
The impact on the Breit-Wheeler rate is seen in  Fig.~\ref{fig:oppp-analytic-sim}. 

\begin{figure}[htbp]
    \centering
    \includegraphics[width=0.48\textwidth]{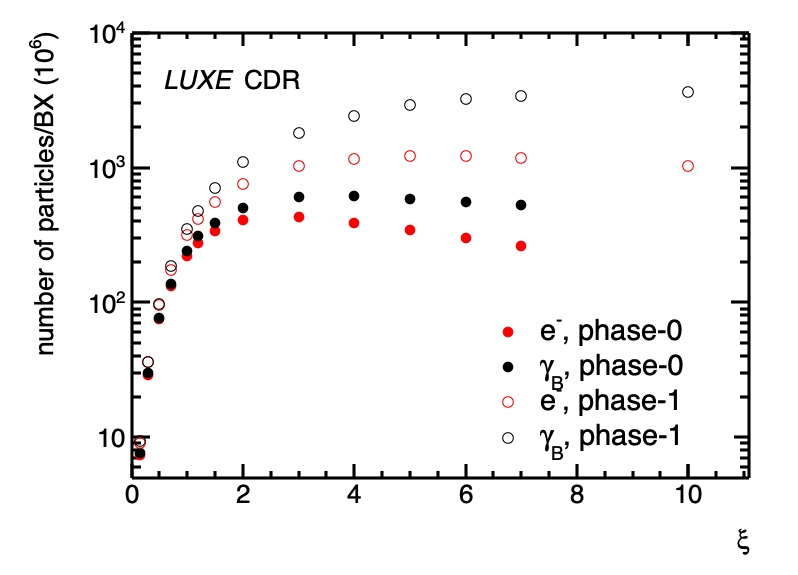}
    \includegraphics[width=0.48\textwidth]{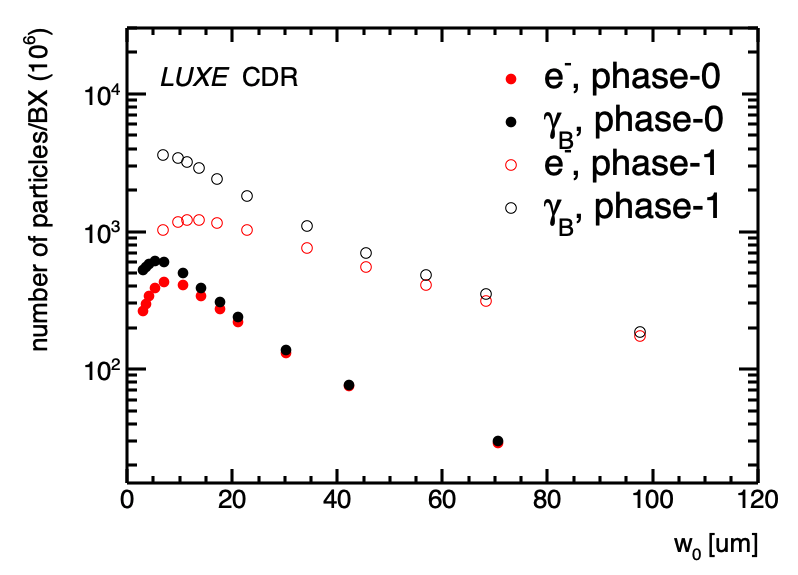}
    \\
    \includegraphics[width=0.48\textwidth]{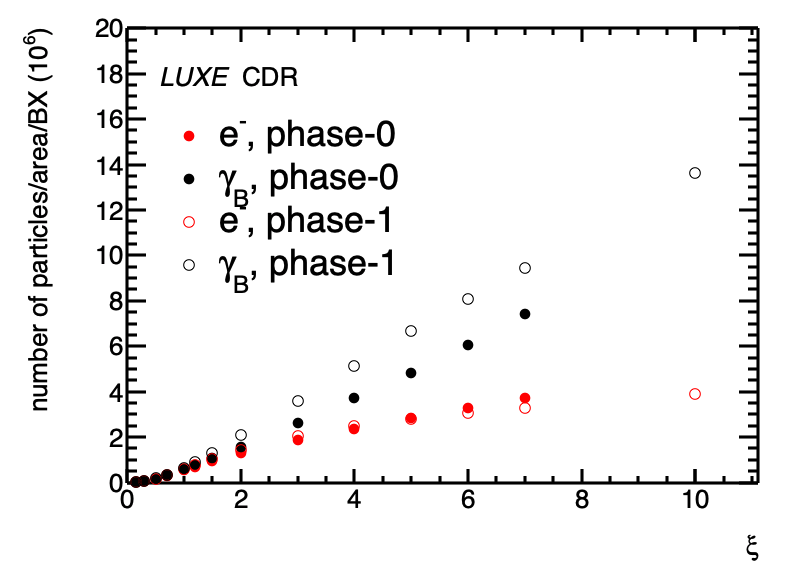}
    \includegraphics[width=0.48\textwidth]{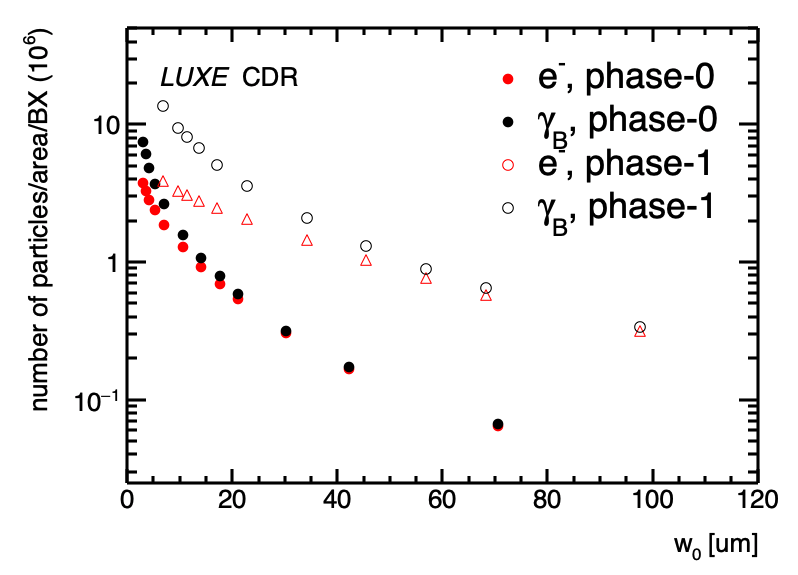}
    \caption{Number of outgoing electrons and photons in the Compton process for the \elaser set-up versus $\ximax$ and $w_0$. The top plots show the rates as expected for LUXE. The bottom plots show rescaled values of these rates that account for the geometric effect due to the changing laser spot size (see Eq.~\ref{eq:normspot}). In all figures \phaseone and \phasetwo are shown. 
    Beam electrons that have not interacted are not shown. All estimates are based on \textsc{PTARMIGAN}.
    }
    \label{fig:rates_elaser}
\end{figure}

The predicted numbers of electrons and photons per bunch crossing are shown in Fig.~\ref{fig:rates_elaser} as function of $\ximax$ and $w_0$. In the simulation, the different $\ximax$ values are reached by changing the $w_0$, i.e. by focusing or de-focusing the laser.

Since the electron beam itself has a finite size of $\sigma_\textrm{el}=5$\,$\mu$m and the laser pulse size changes, the overlap area between the electron beam and the laser pulse changes, causing the rates of interactions to change purely based on geometry. In particular, at high $w_0>10$\,$\mu$m the overlap area is determined by the electron beam and the geometric overlap is constant, but as the laser beam size gets smaller the geometric overlap decreases. 
It can be shown that the dependence is expected to be proportional to $a$ with 
\begin{equation}
\label{eq:normspot}
    a= \frac{w^2_0}{\sqrt{w_0^2+2\sigma_\textrm{el}^2}} .
\end{equation}

When rescaling the electron and photon rates using this geometric factor, the values shown in the bottom panels of Fig.~\ref{fig:rates_elaser} are obtained. It is seen that the turnover behaviour observed in the rates before rescaling at $w_0\approx 7$\,$\mu$m and $\xi \approx 3$ has disappeared, instead they continue to increase. It is also seen that the rescaled electron rates versus $\xi$ agree well between the two phases. The rescaled number of photons is higher for \phasetwo than \phaseone at high $\xi$ where a single electrons can participate in multiple Compton interactions. 

\begin{figure}[htbp]
    \centering
    \includegraphics[width=0.48\textwidth]{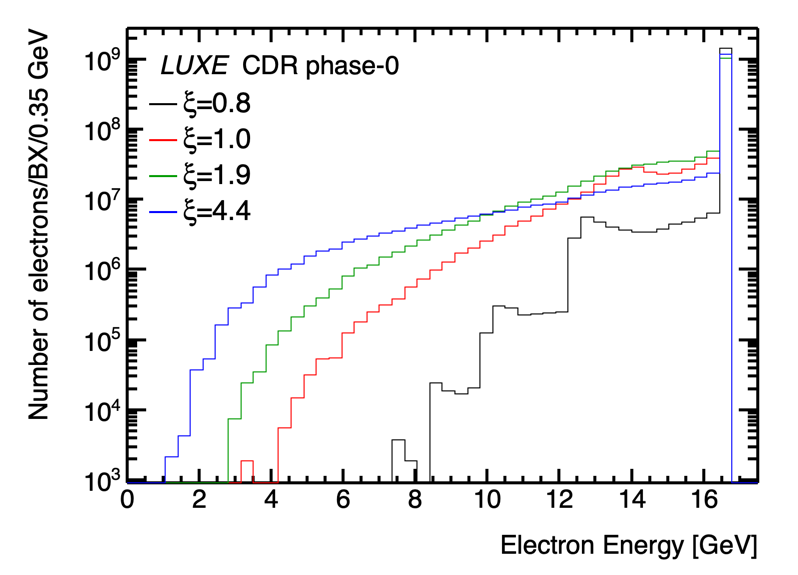}
    \includegraphics[width=0.48\textwidth]{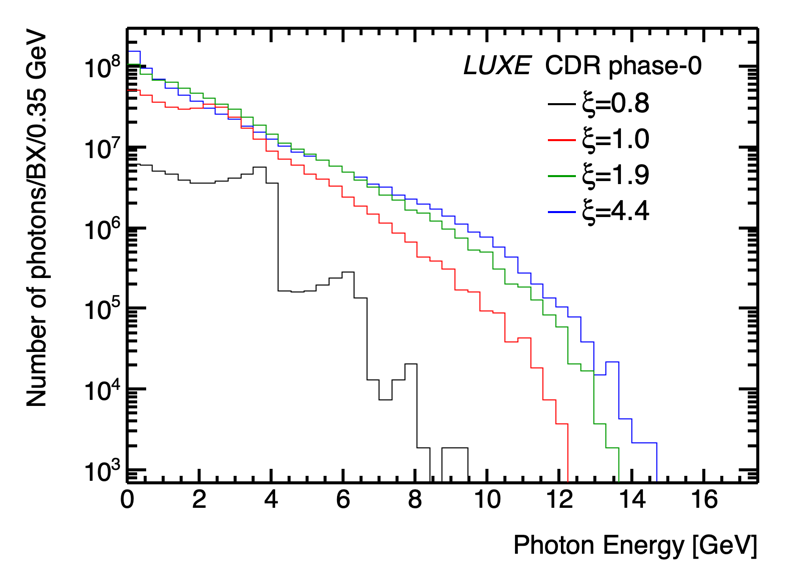}
    \caption{Electron and photon energy spectra for the Compton process predicted by MC simulation. A selection of four configurations for $\xinom$ values is shown. The peak in the electron energy spectrum at 16.5~GeV is due to beam electrons that have not interacted.
    }
    \label{fig:spectra_elaser}
\end{figure}

The energy spectra of the electrons and photons are shown in Fig.~\ref{fig:spectra_elaser} for four examples as predicted by the MC simulation. At the nominal $\xi$ value of 0.31, clear edges are observed in the electron and photon energy spectra. At $\xi=1.94$ there is still a hint of edges while at higher $\xinom$ values the spectra are very smooth and show no structures. 

\begin{figure}[htbp]
    \centering
    \includegraphics[width=0.5\textwidth]{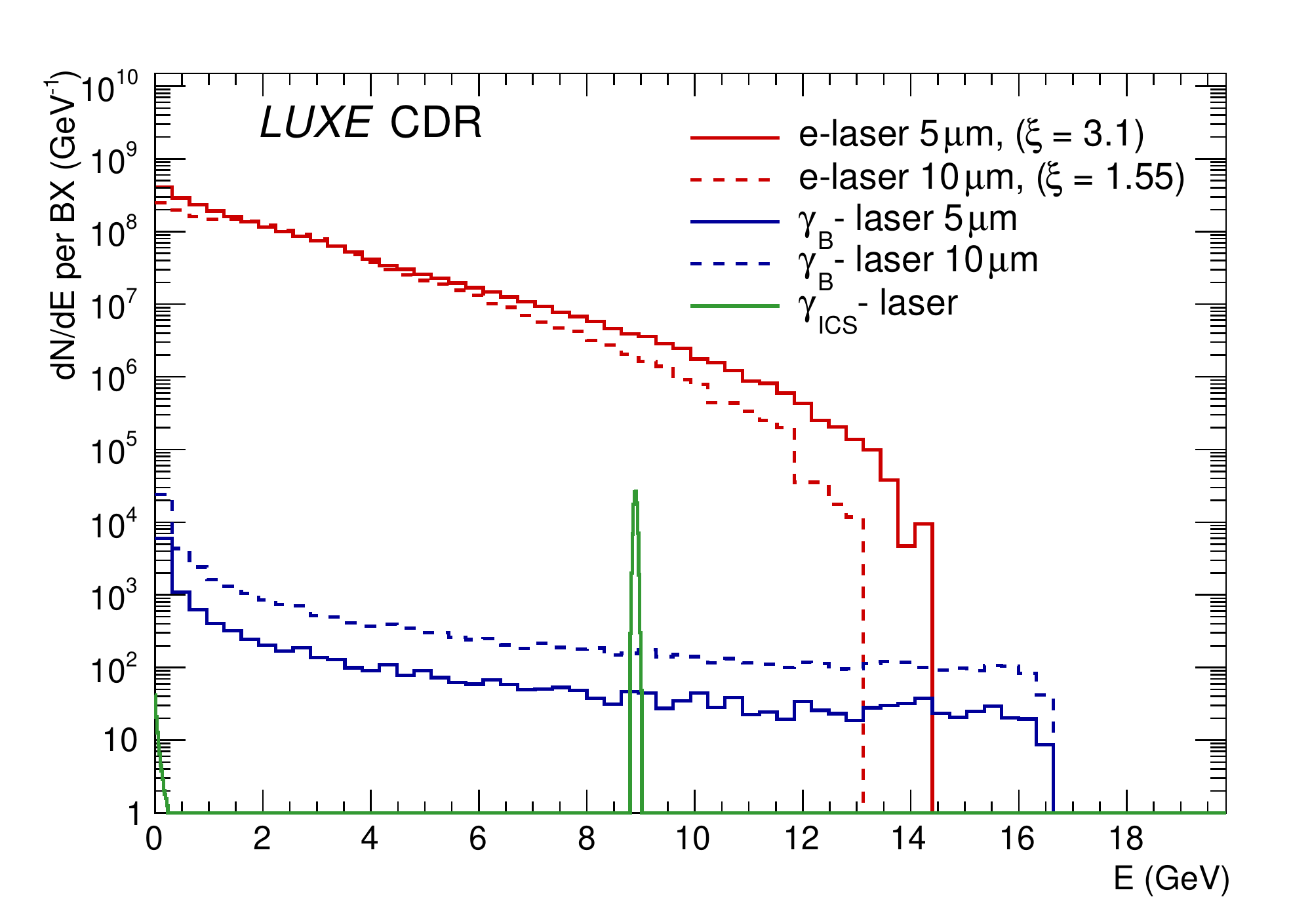}
    \caption{Photon energy spectra for the Compton, bremsstrahlung and ICS processes for two different laser spot sizes $w_0$. For the ICS process the $w_0$ used here is 8\,$\mu$m. Shown are the numbers of photons that arrive at the IP within the given laser spot size.}
    \label{fig:photon_hics_brems}
\end{figure}

The expected photon energy spectra for the \elaser and \glaser modes, from the Compton, bremsstrahlung and ICS processes, are compared in Fig.~\ref{fig:photon_hics_brems}
for two laser spot sizes. 
The expectations clearly show that although the 
spectrum from bremsstrahlung is harder than the Compton photon spectrum, there are many more photons available in \elaser than in the \glaser set-up, even at high energies.
At low energies, there are up to a factor of $10^6$ more photons; at higher energies this factor is reduced but it is still about $10^4$ at 12\,GeV. 
When these photons interact with the laser field, $e^+e^-$ pairs are created via the Breit-Wheeler process. Given the much higher number of photons in the \elaser set-up, the number of $e^+e^-$ pairs is also expected to be higher for that set-up. The photon spectrum for the ICS process peaks sharply at 9\,GeV.

\begin{figure}[htbp]
    \centering
    \includegraphics[width=0.48\textwidth]{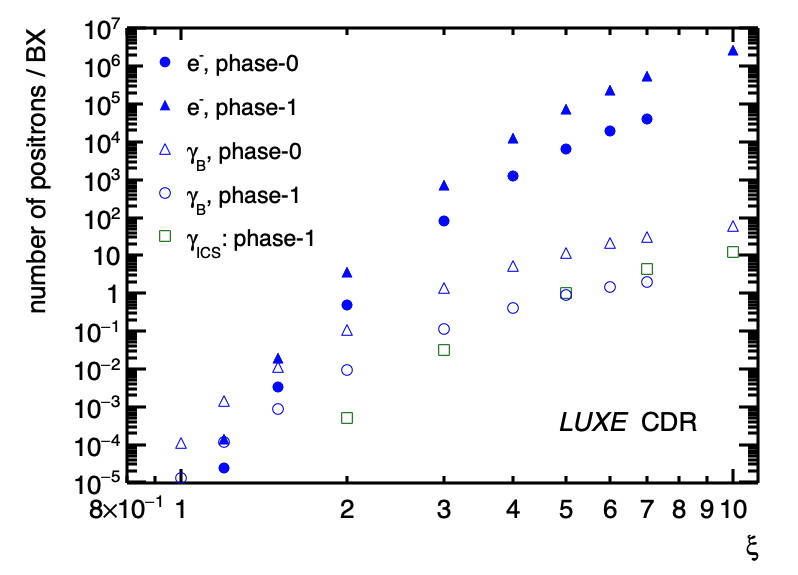}
    \includegraphics[width=0.48\textwidth]{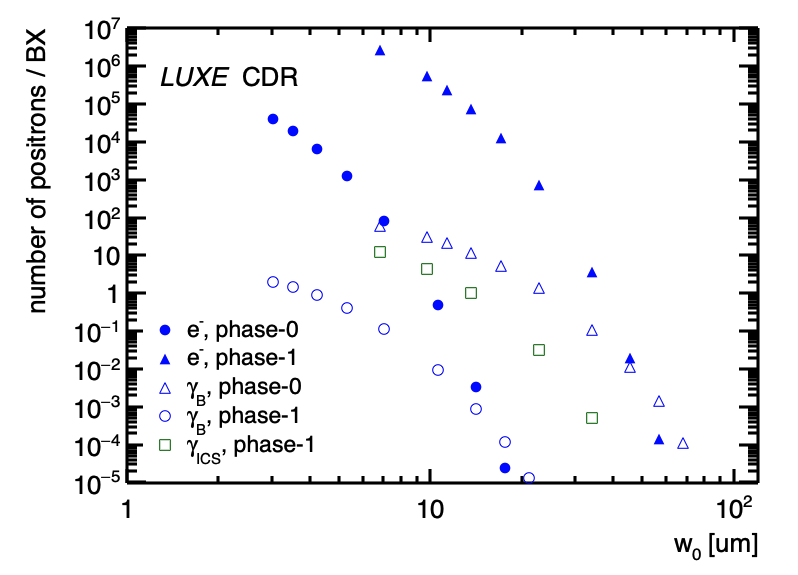}   
    \includegraphics[width=0.48\textwidth]{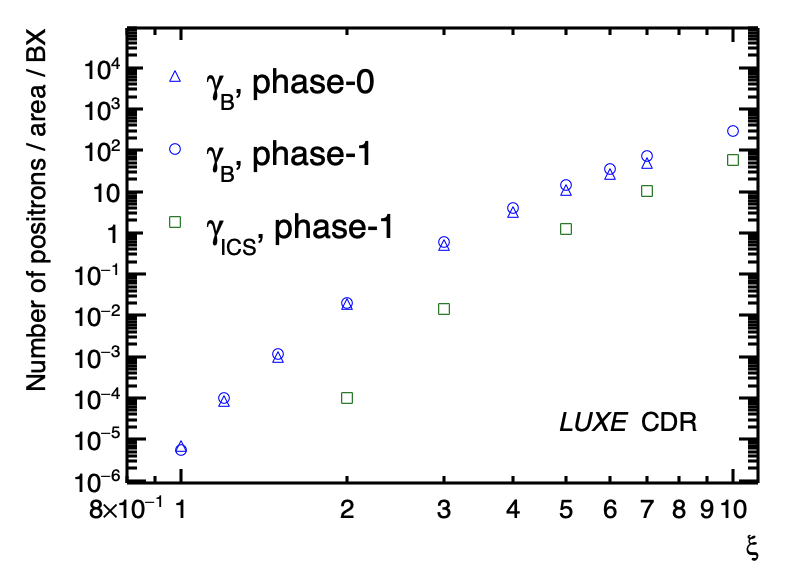}
    \caption{Top left: Number of positrons per beam crossing produced in the \elaser and \glaser set-ups for \phaseone and \phasetwo, as a function of $\xi$. Top Right: Number of positrons per beam crossing produced in the \elaser set-ups for \phaseone, as a function of $w_0$. The legend in the left plot also applies to the right plot. Bottom: Number of positrons per beam crossing normalized to area ($w_0^2$) produced in the \elaser and \glaser set-ups for \phaseone and \phasetwo, as a function of $\xi$. All estimates are based on \textsc{PTARMIGAN}.}
    \label{fig:rates_elaser_posi}
\end{figure}

The positron rates for \elaser and \glaser interactions are shown in Fig.~\ref{fig:rates_elaser_posi} as function of $\ximax$ and $w_0$. It is seen that for the \elaser running the positron rate is larger than for the \glaser running by a factor of $\sim 10 - 10^{4}$, depending on $\xi$. For \phaseone the \elaser run, up to $10^4$ positrons are expected per bunch crossing, while for the \glaser running only $\sim 1$ are expected even at the highest $\ximax$. For \phasetwo the rates are typically 10 times larger. For the ICS \glaser process the rates are about a factor 10 lower than the $\gamma_\textrm{B}$-running. The desire to study all these scenarios places high demands on the design of the detectors as they need to function at large occupancy and achieve a very strong background suppression (see discussion in Sec.~\ref{sim:sig_bkg} and Sec.~\ref{sec:detectors}).

Also shown in Fig.~\ref{fig:rates_elaser_posi} are rescaled rates, normalized by $1/w_0^2$, where \phaseone and \phasetwo are seen to be consistent.  

\begin{figure}[htbp]
    \centering
    \includegraphics[width=0.68\textwidth]{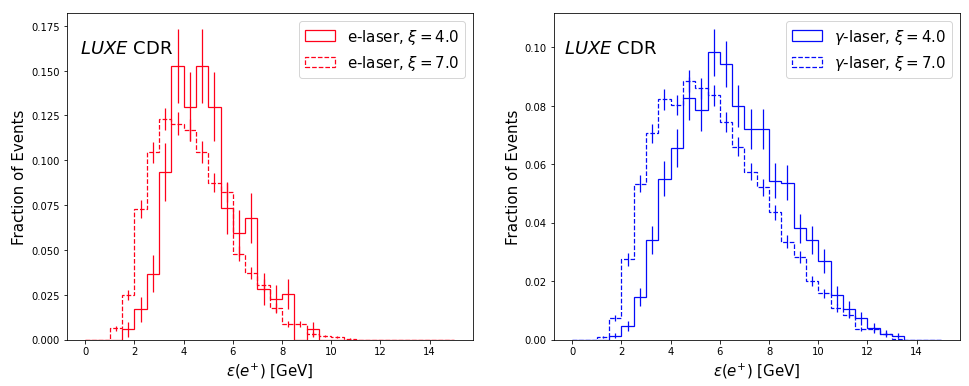}
    \caption{The positron energy spectrum for the \elaser and the \glaser set-up for \phaseone for two values of $\xi$ for \elaser and \glaser running.}
    \label{fig:elaser_posi_espectrum}
\end{figure}

The positron energy spectrum is shown in Fig.~\ref{fig:elaser_posi_espectrum} for both processes for selected $\ximax$ values for \phaseone. The energies are typically between 2 and 14~GeV and are somewhat lower for the \elaser mode. 

\clearpage
\subsection{Geant4 Simulation}
\label{sim:geant4}
A simulation of particle fluxes considering all components of the experiment is performed to estimate the background due to secondary particles from the beam. Through an iterative process the layout, schematically shown in Fig.~\ref{fig:layoutdetailed}, was optimised to reduce these backgrounds. The simulation is also used to estimate the ionisation dose and to aid the development of parametric fast emulation of the response of the detectors to the various particles.

\begin{figure}[htbp]
  \centering
  \includegraphics[width=0.95\textwidth]{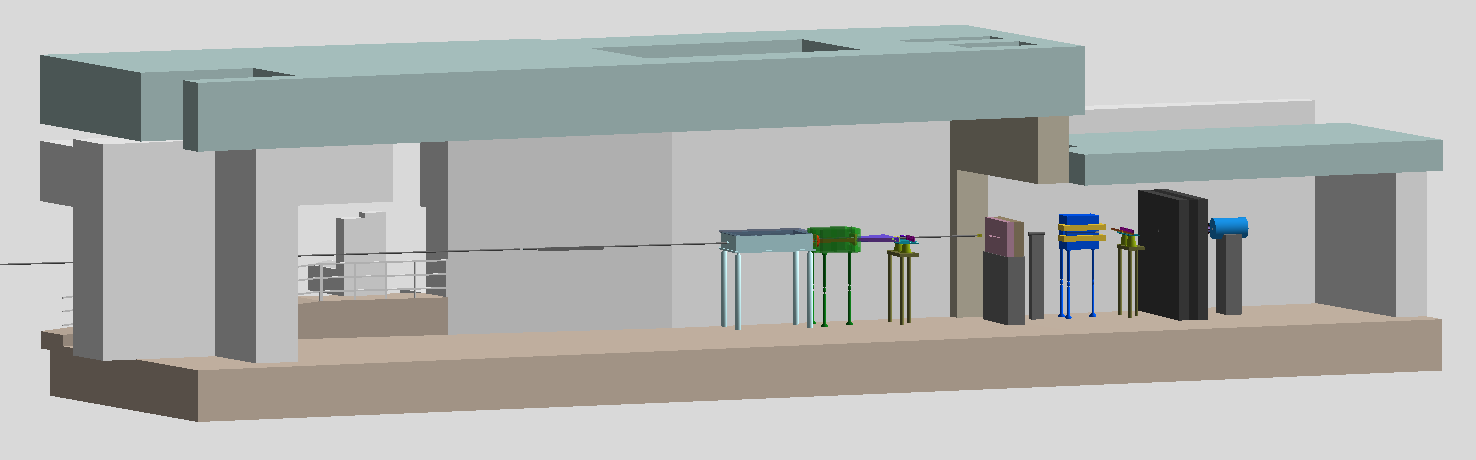}
  \includegraphics[width=0.95\textwidth]{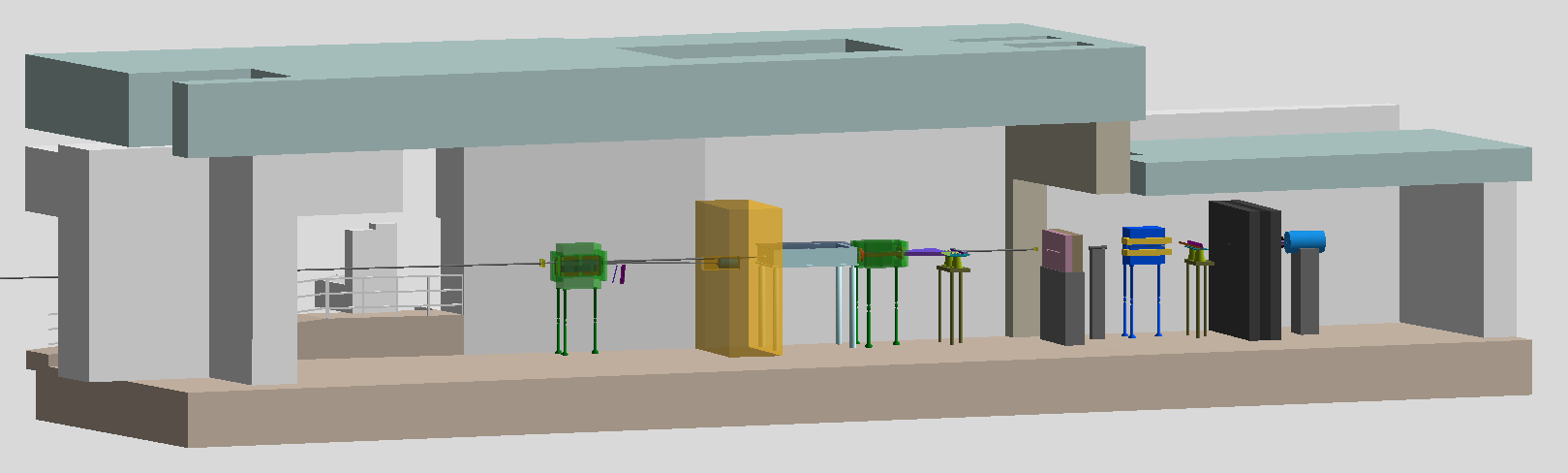}
  \caption{The layout of LUXE as modelled in the \geant simulation for the \elaser (top) and \glaser (bottom) set-up. Dipole magnets after the tungsten target and the interaction chamber are shown in green, the yellow and black blocks are shielding elements. Behind the shielding in light blue is the interaction chamber. A C-shaped dipole magnet is shown in blue/yellow upstream from the gamma ray spectrometer. }
  \label{fig:luxe_geant4_geom}
\end{figure}

The geometry model of the LUXE experiment is implemented in \geant~\cite{Allison:2016lfl, Allison:2006ve} version 10.06.p01 using the EM opt0 physics list.  Figure~\ref{fig:luxe_geant4_geom} shows a general view of the LUXE 
simulation model for the \elaser and \glaser modes which implements the layout shown schematically in Fig.~\ref{fig:layoutdet}. It includes beam instrumentation components, detector systems and infrastructure of the XS1 cavern. The infrastructure is imported 
using an existing 3D CAD model of the building. The magnet models are based on the existing devices used at the DORIS accelerator (see Sec.~\ref{sec:tc}). Magnetic fields are considered as uniform and the volume they occupy is based on the documentation for the magnets as well as their material. 

The implementation of the interaction and target chambers (see Sec.~\ref{sec:ic} and Sec.~\ref{sec:tchamber}), beam dumps (see Sec.~\ref{sec:beamdump}) and detector support structures are implemented according to the design envisaged for LUXE and described in the relevant sections of this document. The locations of the key components are given in Table~\ref{tab:sim_keypar}. Some simplification has been made for supporting structures in the areas not directly exposed to the signal or main beam particles. 

\begin{table}[htbp]
    \centering
    \begin{tabular}{|l|c|c|L|}
    \hline
     \textbf{Component}    & distance from IP  &  transverse size & purpose  \\
        & (m) & (m) &  \\\hline
     \multicolumn{4}{|l|}{\textbf{Bremsstrahlung system area}}\\\hline
     $W$ target & $-7.5$ & small & produce $\gamma$\\
     Dipole magnet ($B=1.8$~T)& $-[7.2,5.8]$ & $[-0.3,0.3]$ & deflect $e^\pm$ \\
     Scintillator & $-5.4$ & $[0.05,0.55]$ &  measure $e^-$\\
     \cer & $-5.3$ & $[0.05,0.55]$ & measure $e^-$ \\\hline
     \multicolumn{4}{|l|}{\textbf{IP area}}\\\hline
     Shielding & $-[3.0,1.5]$ & $[-0.6,2.3]$ & protect IP \\ 
     IP & $0.0$ & $0.0$ & interactions\\
     Interaction chamber & $[-1.25,1.25]$ & $[-0.75,0.75]$ & final steering and focus of laser\\
     Dipole magnet ($B=1-2$~T) & $[1.4,2.8]$ & $[-0.6,0.6]$ & deflect $e^\pm$ \\
     Pixel tracker & $[3.9,4.2]$ & $[0.05,0.55]$ & measure $e^+$ (and $e^-$ for \glaser case) \\
     Calorimeter & $[4.3,4.4]$ & $[0.05,0.60]$ & measure $e^+$ \\
     Scintillator & $3.9$ & $-[0.05,0.55]$ &  measure $e^-$ for \elaser \\
     \cer (Ar gas) & $4.3$ & $-[0.05,0.20]$  & measure $e^-$ for \elaser (*)\\ 
     \cer (Quartz) & $4.3$ & $-[0.05,0.55]$ & measure $e^-$ for \glaser\\
     Beam dump & $[7,7.8]$ & $[-0.05,0.23]$ & dump primary $e$ beam\\
     Shielding & $[6.8,7.3]$ & $[-0.4,1.5]$ & reduce background in GDS \\
     \hline
     \multicolumn{4}{|l|}{\textbf{Photon Detection System} (GDS)}\\\hline
     Gamma profiler & $6.5$  and $12.5$ & small & measure $x-y$ profile of $\gamma$'s \\
     Kapton foil target & $6.5$ & small & convert $\gamma$\\
     Dipole magnet ($B=1.4$~T) & $[9.0,10.3]$ & $[-0.6,0.3]$ & deflect $e^\pm$ \\
     Scintillators & $10.9$ & $[-0.5,0.5]$ & measure $e^\pm$ \\
     Backscattering calorimeter & $[13.0,13,5]$ & $[-0.15,0.15]$ & backscatter of $\gamma$'s \\
     Beam dump & $[13.6,14.1]$ & $[-0.3,0.3]$& dump photons \\\hline
     Final wall & $17.4$ & large &- \\\hline
    \end{tabular}
    \caption{Table of the key components simulated. Given are the locations and purposes. The transverse direction here is the plane in which the charged particles are bent by a dipole magnet. In cases where the objects are sizeable, intervals are given. When a sign is given before the interval it applies to both values.
    (*) For the simulation with $B=1$~T the \cer detector spans the range of 0 to -0.15~cm.}
    \label{tab:sim_keypar}
\end{table}

For both the \elaser and the \glaser studies, the primary electrons are generated in accordance with the XFEL.EU beam parameters, see  Table~\ref{tab:xfelepara}.

For the \elaser simulation the primary electron beam directly enters the IP, and after the IP it is deflected by a magnet towards a dump. Additionally, when the laser is fired, a large rate of \elaser interaction occurs resulting in a large number of lower energy electrons and photons. While the photons continue in the \beampipe towards the end of the cavern, the electrons are deflected by the magnet towards the detectors designed to detect these. 

For the \glaser set-up, the simulation of the bremsstrahlung process at the target is important. The energy spectrum of the bremsstrahlung photons was approximated using the following formula~\cite{Tanabashi:2018oca}: 
\begin{equation}{
   \frac{dN_{\gamma}}{dE_{\gamma}} = \frac{X}{E_{\gamma} X_{0}} 
                     \left( \frac{4}{3} - \frac{4}{3} \frac{E_{\gamma}}{E_{e}} + \left(\frac{E_{\gamma}}{E_{e}}\right)^{2}\right) \, ,
  }\label{eq_bremsstrahlung_pdg}
\end{equation}
where $E_{e}$ is the energy of the incident electron, $X_{0}$~the radiation length of the target material and $X$~is the target thickness. It is valid for thin targets. 

\begin{figure}[htbp]
  \begin{subfigure}{0.48\textwidth}
    \includegraphics[width=\textwidth]{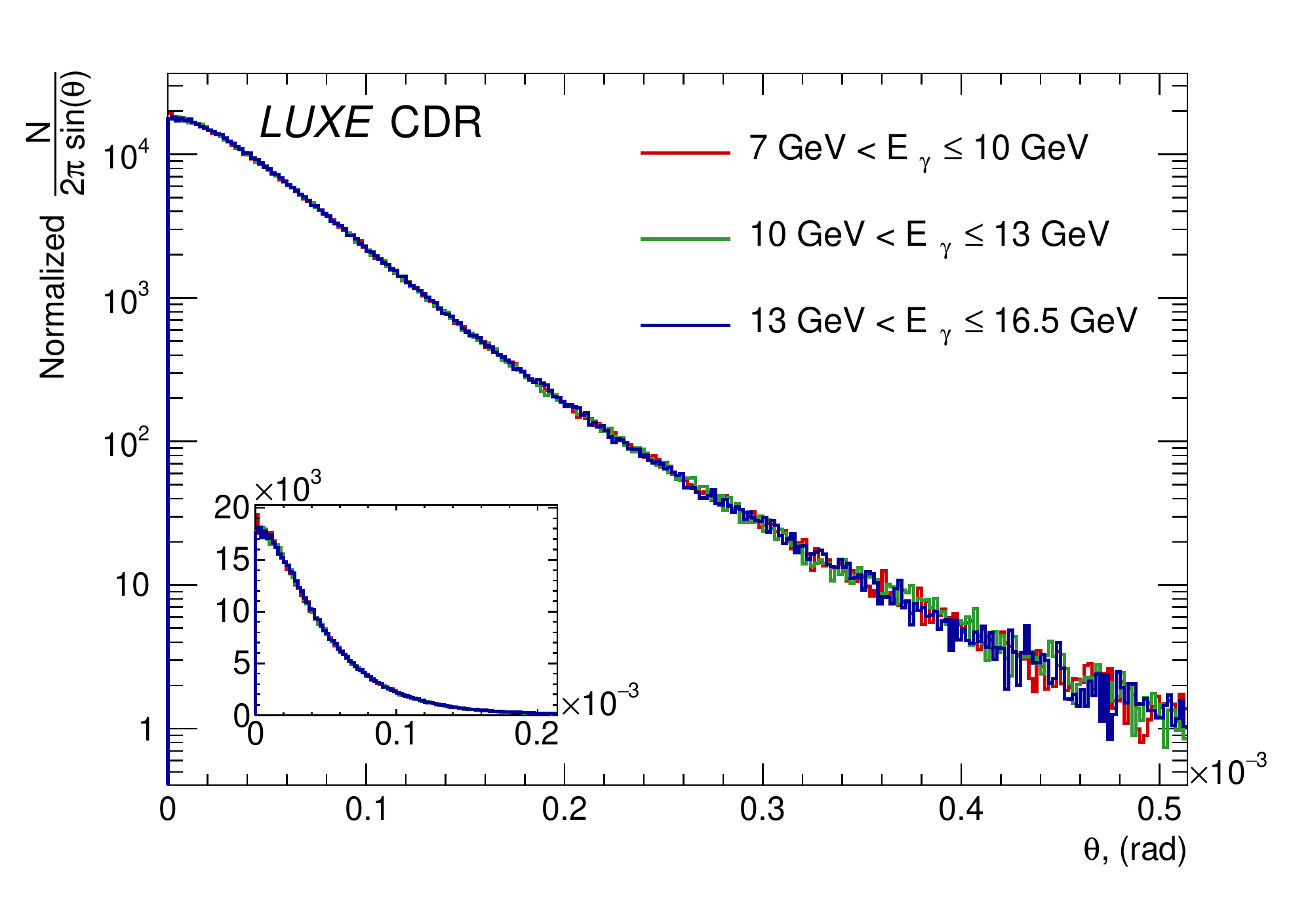}
        \caption{}
    \end{subfigure}%
  \begin{subfigure}{0.48\textwidth}
    \includegraphics[width=\textwidth]{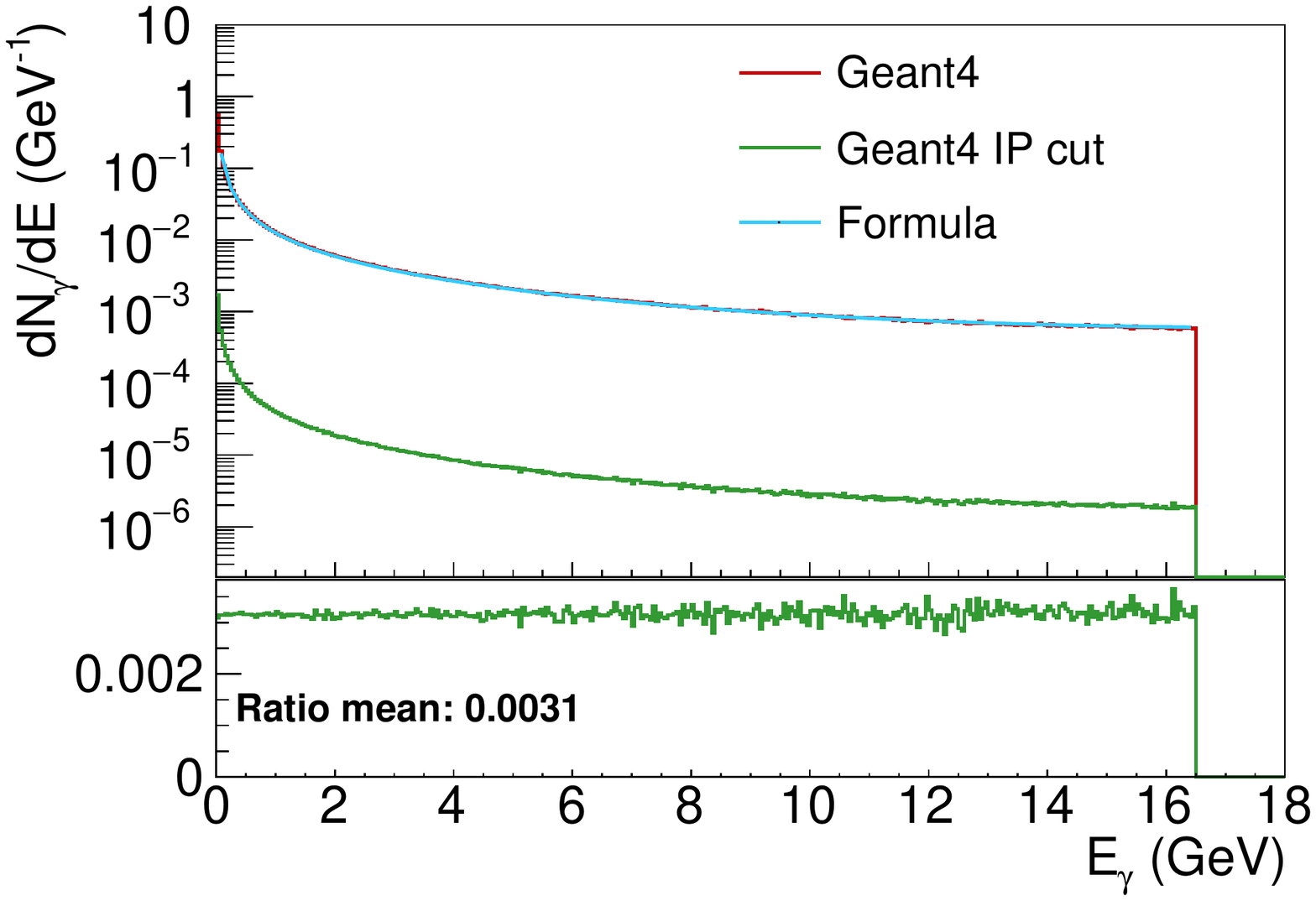}
        \caption{}
    \end{subfigure}%
\\
\begin{subfigure}{0.48\textwidth}
    \includegraphics[width=\textwidth]{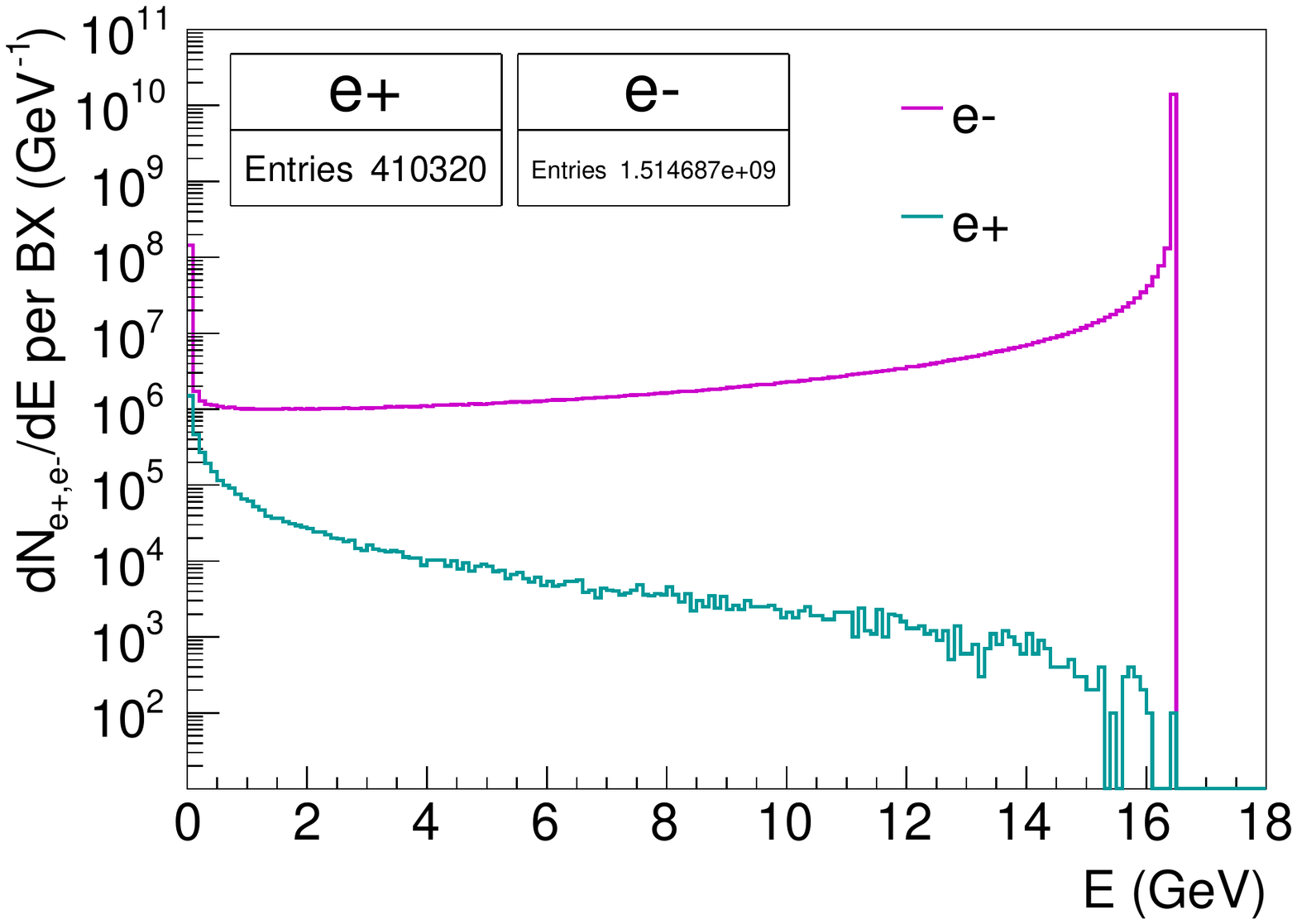}
    \caption{}
    \end{subfigure}%

\caption{
a) Polar angle distribution of bremsstrahlung photons in different energy ranges for one BX normalised by area.
b) Energy spectrum of bremsstrahlung photons obtained by analytical calculation~(Eq. (\ref{eq_bremsstrahlung_pdg})) and by \geant simulation. The green line shows the $\gamma$ spectrum after imposing limits on position in the transverse plane to~$\pm$25\,$\mu$m. The bottom plot shows the fraction limited by the interaction area. 
c) Energy spectra of electrons and positrons produced in the tungsten target. In all cases, a tungsten target with a thickness of 35\,$\mu$m (1\%$X_{0}$) is simulated. 
}
    \label{fig:sim:brem}
\end{figure}

The angular spectrum of the photons is expected to follow $1/\gamma$, i.e. about $\sim 30$\,$\mu$rad for $\gamma=16.5~\textrm{GeV}/m_e$. It is independent of the photon energy as seen in Fig.~\ref{fig:sim:brem}a). As the IP is at a distance of 7.5\,m from the target, the beam size at the IP is about 230\,$\mu$m, much larger than the envisaged laser spot size. 

Figure~\ref{fig:sim:brem}b) shows the spectrum of bremsstrahlung photons produced in the \geant simulation for a 35\,$\mu$m (1\%$X_{0}$) thick tungsten target. The simulation agrees well with the calculation based on Eq.~(\ref{eq_bremsstrahlung_pdg}). Also shown is the fraction of photons that are within $\pm 25$\,$\mu$m in both the $x$ and $y$ direction at the IP. 

The fraction of bremsstrahlung photons that are within $\pm 25$~$\mu$m at the IP is 0.3\%. This fraction depends quadratically on the laser spot size and the distance between the target and the IP, e.g. for a $5 \units{\mu m}$ laser beam spot it is 0.01\% (and this is taken into account in the simulations described in Sec.~\ref{sec:mc}). If the distance could be decreased from 7.5\,m to 5\,m the rate in the IP would be enhanced by a factor $2.25$. This distance is mostly determined by the requirement that the primary electron beam is steered by the magnet to a beam dump.

For an electron bunch containing $1.5\cdot 10^9$ electrons, the average number of bremsstrahlung photons produced per bunch crossing (BX) is about~$1.45 \cdot 10^8$.
Even though the target is thin there is a finite chance that these photons interact again in the target which gives rise to pair production of $e^+e^-$ pairs at a rate of $4.1 \cdot 10^5$ per BX. 
The spectra of positrons and electrons and their average numbers for one BX are presented in Fig.~\ref{fig:sim:brem}c).
It is seen that the secondary positron rates are small, particularly in the high-energy region. Based on the measurement of the electron rate the photon energy distribution can be determined, as discussed in Sec.~\ref{sec:detectors:glaser}. While it would be useful to measure also the positron rate, it seems not critical and at present no detectors are foreseen for this purpose. 

\begin{figure}[htbp]
\centering
    \includegraphics[width=0.49\columnwidth]{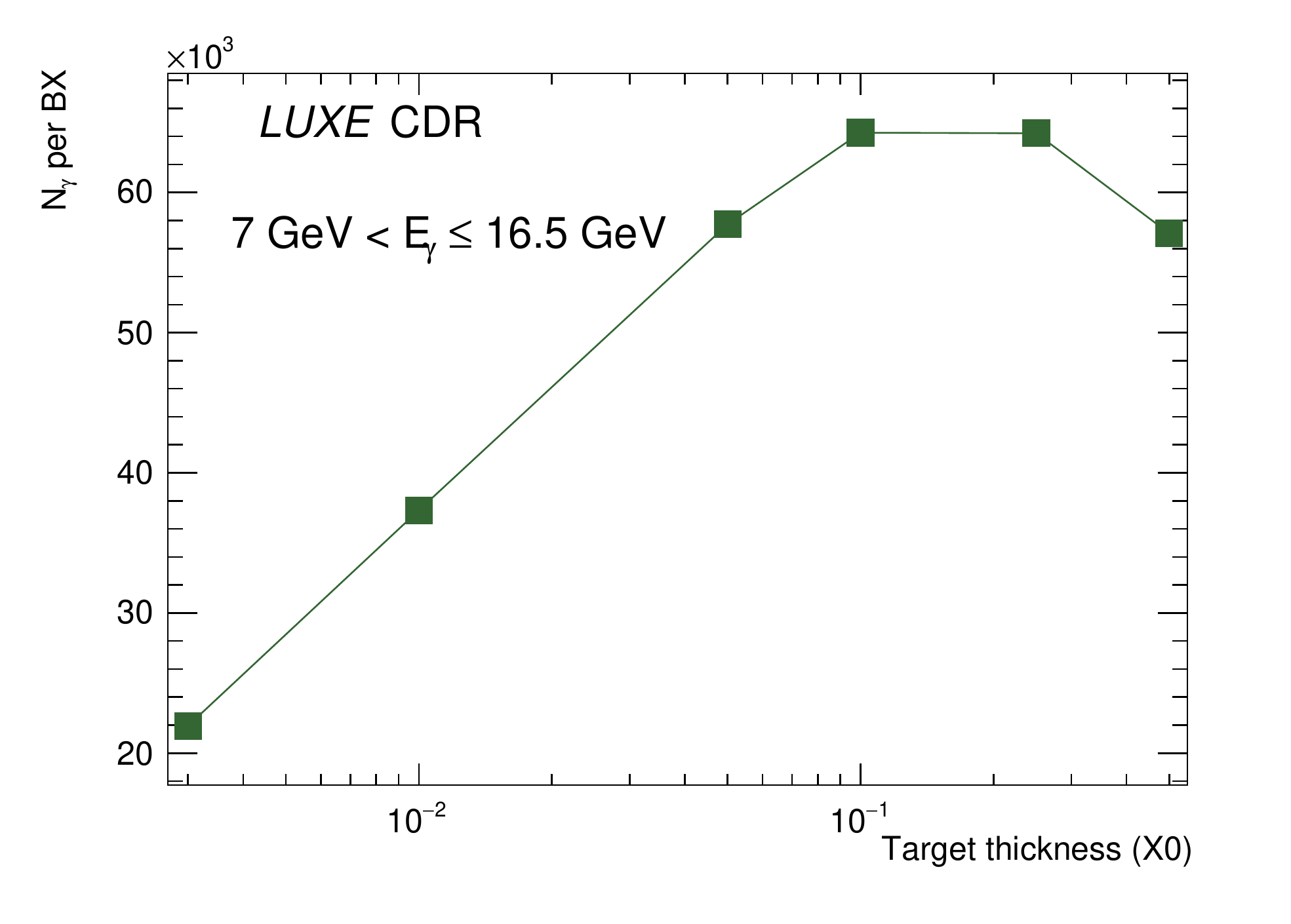}
    \caption{Average number of bremsstrahlung photons produced in the energy range~7\,GeV~$< E_{\gamma} \leqslant $~16.5\,GeV expected at the LUXE IP as a function of thickness of the tungsten target from \geant simulation of 50 bunches of \euxfel electron beam of~16.5\,GeV.}
    \label{fig_brems_production_target_thickness}
\end{figure}

The dependence of bremsstrahlung production on the material properties of the target, such as atomic number and mass density, is well approximated by a single parameter~-- the radiation length $X_{0}$ and described by Eq.~(\ref{eq_bremsstrahlung_pdg}). 
Tungsten was chosen due to its high melting point temperature, high thermal conductivity and high sputtering resistance. The average number of photons as a function of the target thickness is shown in Fig.~\ref{fig_brems_production_target_thickness}. Here, only photons with $E>7$\,GeV and within $\pm25$\,$\mu$m of the IP are considered. The simulation demonstrates that the number of high energy bremsstrahlung photons increases as the target becomes thicker up to about~25\%$X_{0}$ where it nearly doubles compared to~1\%$X_{0}$. For thicker targets, the rate of pair generation probably becomes significant and in combination with multiple scattering the number of high energy photons observed at the IP is reduced. As stated above, the default simulation used a target with 1\%$X_{0}$, and this simulation shows that in principle a factor two could be gained in the photon flux by optimising the target thickness. 
An optimisation of the thickness taking all factors into account will be done for the final design. 

In summary, with the present configuration, about $5\cdot 10^5$ photons per BX arrive at the IP within $\pm 25$\,$\mu$m, and $1.5\cdot 10^4$ arrive within $\pm 5$\,$\mu$m, and this is the basis for the projected number of positrons and electrons discussed in Sec.~\ref{sec:mc} for the \glaser set-up. By optimising the distance between the target and the IP and the target thickness these rates can potentially be increased by a factor of 4 to 5. 
\subsection{Expected Signal and Background Rates}
\label{sim:sig_bkg}

The challenges of measuring the energy spectra and fluxes of the $e^-$, $e^+$ and photons produced at the IP as a function of the QED quantum parameters will be highlighted in this chapter. As shown in Sec.~\ref{sec:mc}, the particle rates vary significantly. In some regions they are as low as $0.1$ particles per bunch crossing,
placing a high demand on the background rejection capabilities of the detector, and in other regions there are $>10^7$ particles, placing high demand on the linearity and the radiation tolerance of the device. In the low-rate regions the detection relies on a silicon pixel tracker and a high-granularity compact calorimeter (low Moli\'ere radius), while in the high-flux regions scintillation screens and \cer detectors are the technologies selected. These detector technologies are discussed in detail in Sec.~\ref{sec:detectors}. Here, the background in all the detector areas is presented and compared to the expected signal yield.  

In the following, the number of electrons, positrons and photons are shown for background and signal. These results are largely independent of the detector technology and highlight the challenge and requirements of potential detector solutions. 

It is important to note, that the background due to the beam alone will be measured continuously \textrm{in-situ} during data taking as 9~Hz of the 10~Hz of electron bunches will pass through the experiment without the presence of a laser shot.  

In the following the term background is used for all secondary particles which are created through interactions of the electron or photon beam with material. Not included in the background are secondary particles in primary physics events as their estimate is normally part of a given analysis.  

\subsubsection{Expected Signal and Background Rates in Electron-laser Collisions}
\label{sec:elasersim}
First the background purely due to the electron beam is discussed, and then the signal and background rates in \elaser collisions are compared.

\begin{figure}[htbp]
  \centering
  \includegraphics[width=0.49\textwidth]{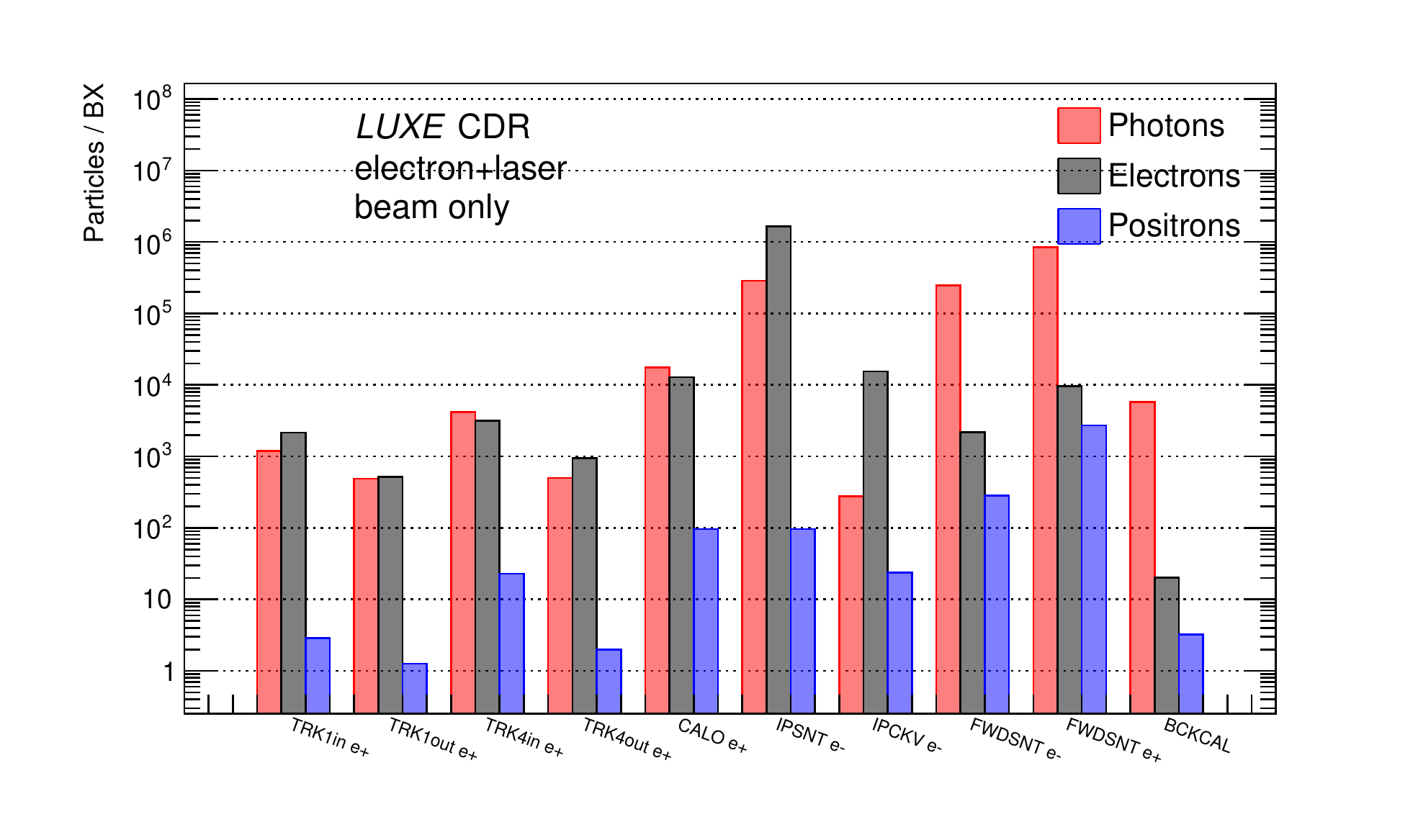}
  \put(-120,10){\makebox(0,0)[tl]{\bf (a)}}
  \includegraphics[width=0.49\textwidth]{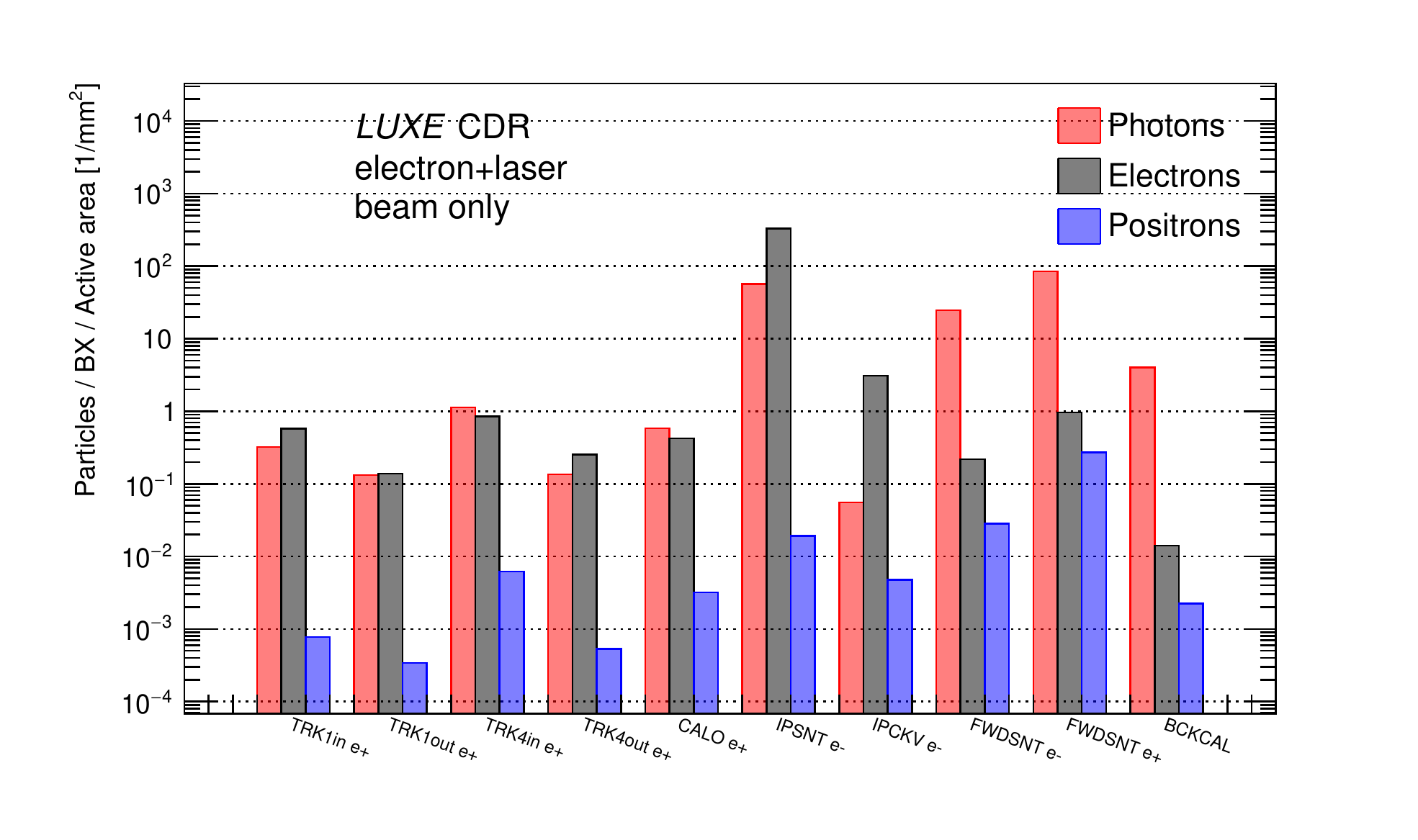}
  \put(-120,10){\makebox(0,0)[tl]{\bf (b)}}\\
  \includegraphics[width=0.49\textwidth]{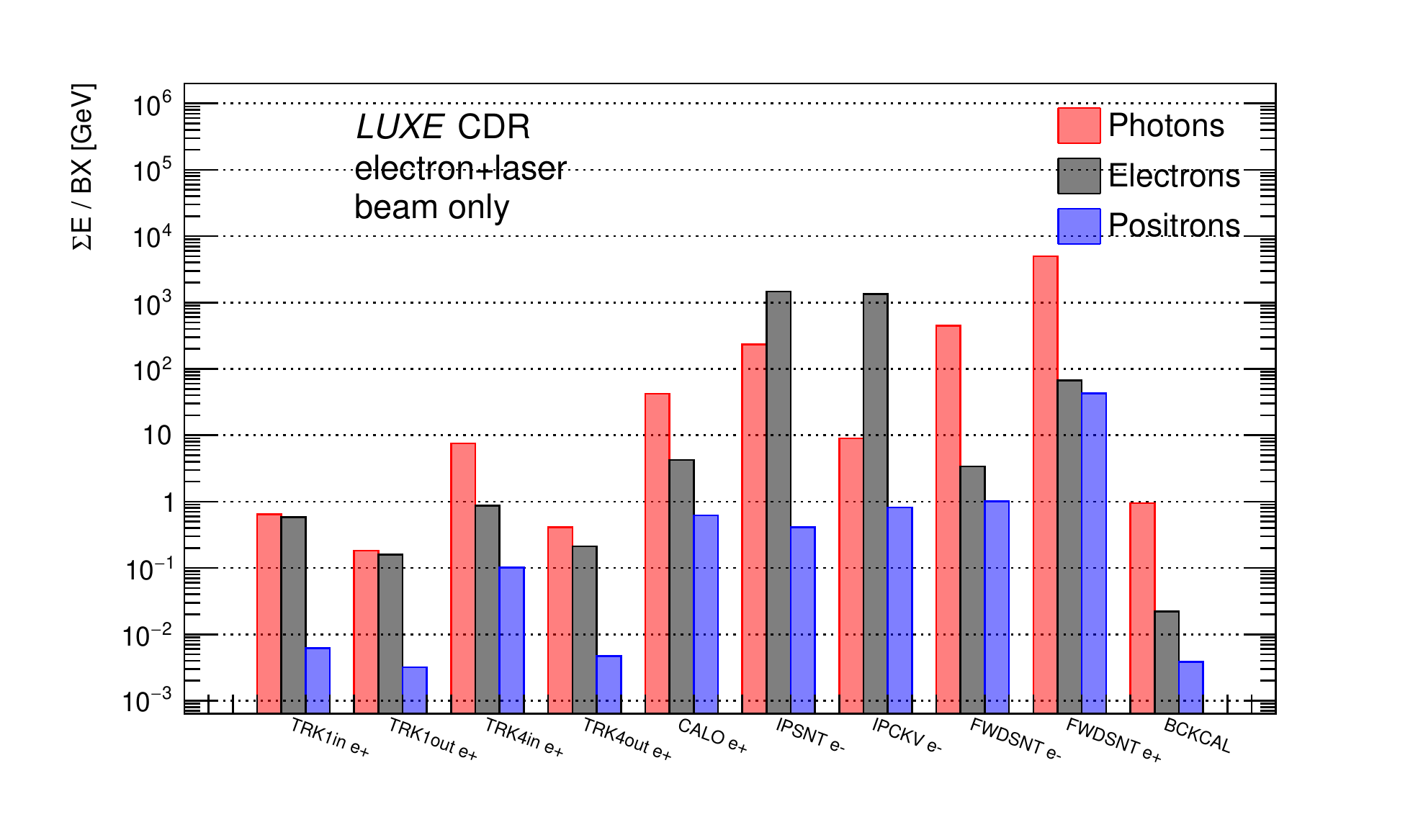}
  \put(-120,10){\makebox(0,0)[tl]{\bf (c)}}
  \includegraphics[width=0.49\textwidth]{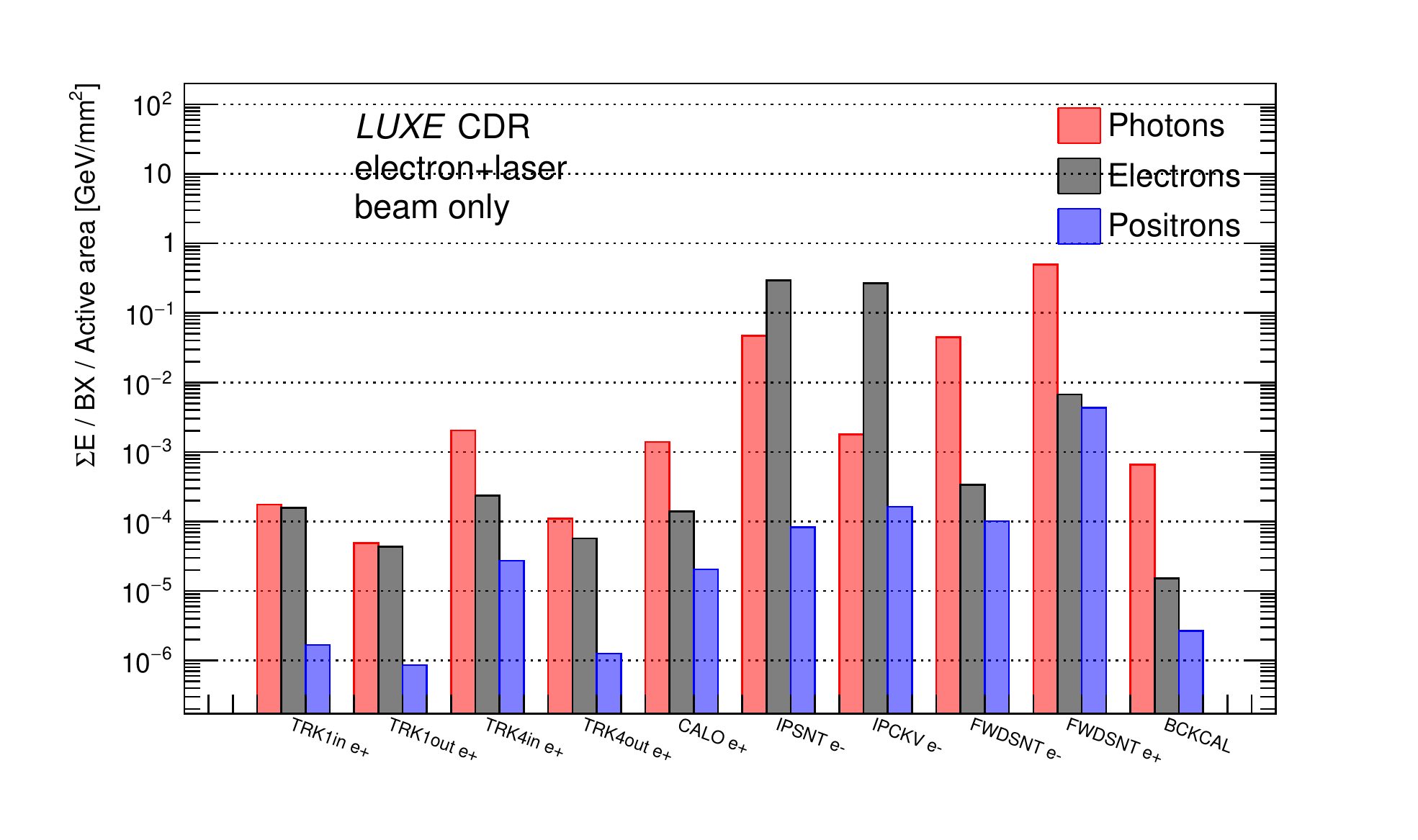}
  \put(-120,10){\makebox(0,0)[tl]{\bf (d)}}\\
  \caption{Distributions of (a) number of particles, (b) number of particles per unit area, (c) the summed energy and (d) the summed energy per unit area 
  in front of each detector system for \elaser collisions for photons, electrons and positrons per BX. The first 4 entries show expectations for the first and last tracker layers, followed by the calorimeter on the 
  positron side and the scintillator screen and \cer detector on the electron side.  The next entries show the scintillator screens in the forward photon spectrometer for the electron and positron sides and finally the backscattering calorimeter.}
  \label{fig:elaser-barcharts}
\end{figure}

In Fig.~\ref{fig:elaser-barcharts}
the number of particles and summed energies impacting the detectors
are shown at the front face of the detectors 
for the background in the \elaser set-up. 
Here, the laser is off and all the particles originate from secondary interactions due to the electron beam. As the detectors are positioned after a dipole magnet to separate electrons and 
positrons, the results are shown for both the electron side and positron side. 
On the positron side, the expected rates for signal are $\sim 10^{-2}-10^4$ depending on $\xi$ (see Fig.~\ref{fig:rates_elaser_posi}). In particular for the positron rate measurement at low $\xi$ it is important to reduce the background ideally to levels below 0.01 as discussed in Sec.~\ref{sec:results}. 

On the $e^+$ side, in the silicon tracking detector the background electrons and photons is similar, with about 1000 particles/BX. Photons are naturally suppressed as they rarely produce ionisation signals in a semiconductor tracker. The mean energy of the particles is low, typically $\sim 1$\,MeV. These low-energy electrons can cause hits in the silicon pixel tracker but they are mostly confined to just one of the four planes and efficiently rejected by the tracking algorithm, as discussed in Sec.~\ref{sec:detectors_tracker}. 
The rate of electrons and positrons with energies $>1$\,GeV is below 0.1 per bunch crossing. Behind the tracker, the calorimeter is exposed to about $10\times$ higher rates of electrons and photons. The dominant background originates from low-energy photon flux from the area of the beam-pipe. The energy is dominated by photons but rather low, about 1~MeV/mm$^2$. 

On the $e^-$ side, $\sim 10^5$  background electrons per BX are present in the scintillation screen and $\sim 10^4$ in the \cer detector. 
This is lower than the expected signal of
$10^6-10^8$ electrons (see Fig.~\ref{fig:rates_elaser}) but not negligible, and it is important that the measurement can be performed in-situ and then subtracted. For the scintillation screen the electron background is suppressed by selecting only the part of the screen in the plane of the beam ($\pm 5$~mm of the vertical beam position, see below for discussion of the fiducial volume used), while for the \cer detector it is suppressed by designing it such that only particles with energies above 20~MeV produce \cer radiation. 
The photon rate is subleading and photons induce a lesser response in the scintillator and none at all in the Cherenkov detector.

In the photon detection system, the background in the scintillator screen is about $10^3$ for electrons and $10^6$ for photons, compared to an expected signal of typically $10^5$. It is thus important to choose a technology that rejects photons effectively and that is adequate for the high flux of particles. The energy due to background in the backscattering calorimeter is low.    

\begin{figure}[htbp]
\includegraphics[width=0.49\textwidth]{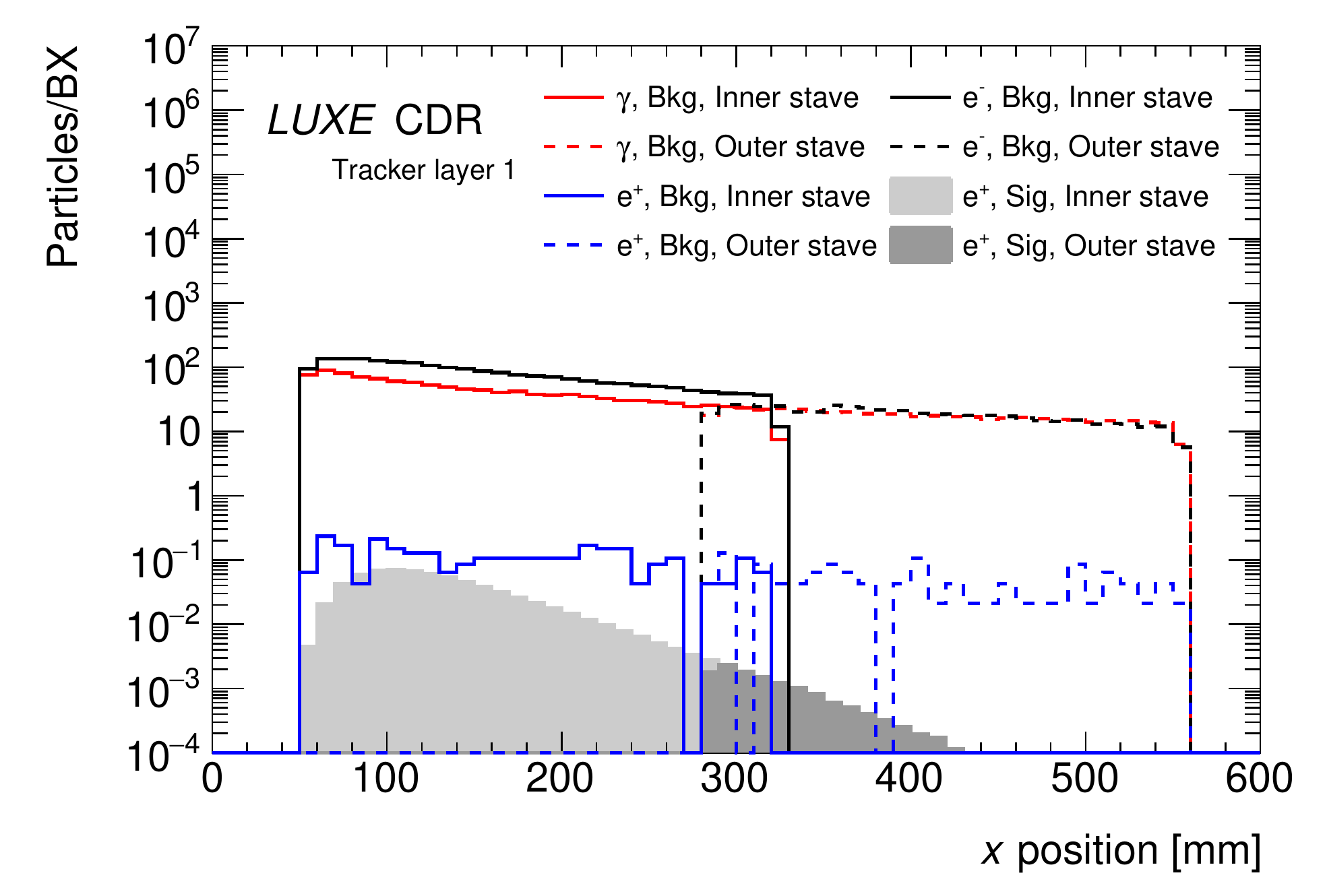}
\put(-120,10){\makebox(0,0)[tl]{\bf (a)}}
\includegraphics[width=0.49\textwidth]{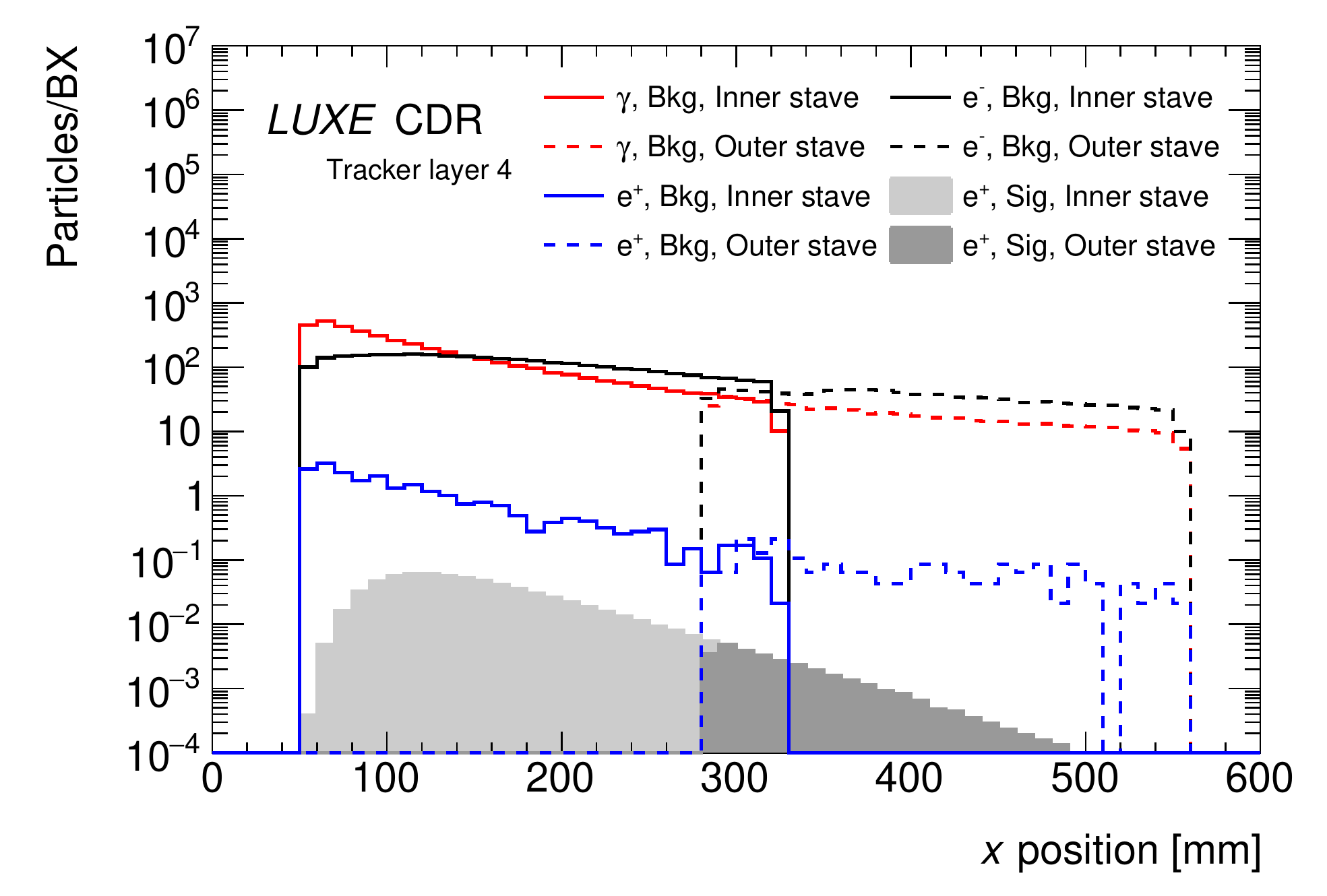}
\put(-120,10){\makebox(0,0)[tl]{\bf (b)}}\\
\includegraphics[width=0.49\textwidth]{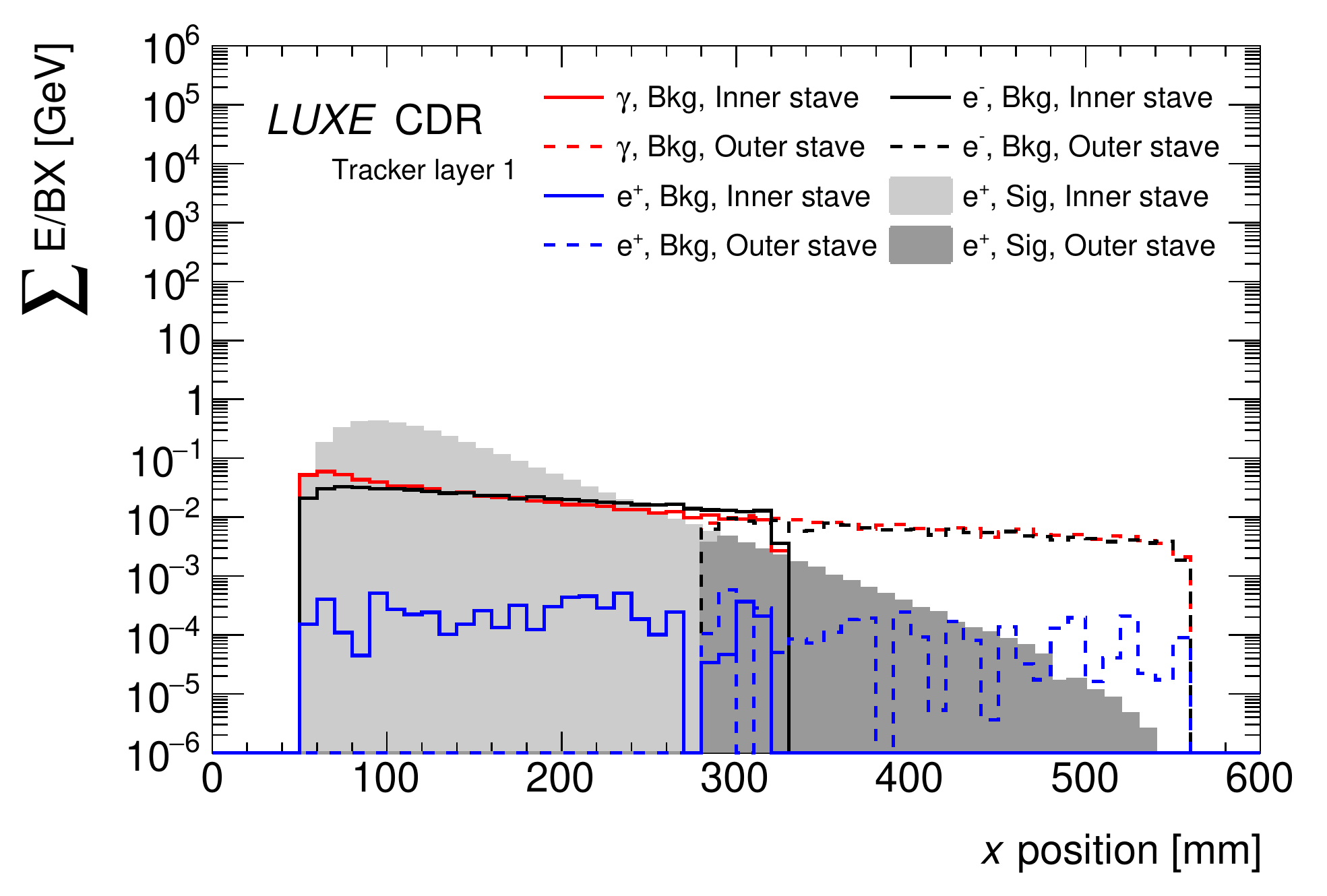}
  \put(-120,10){\makebox(0,0)[tl]{\bf (c)}}
\includegraphics[width=0.49\textwidth]{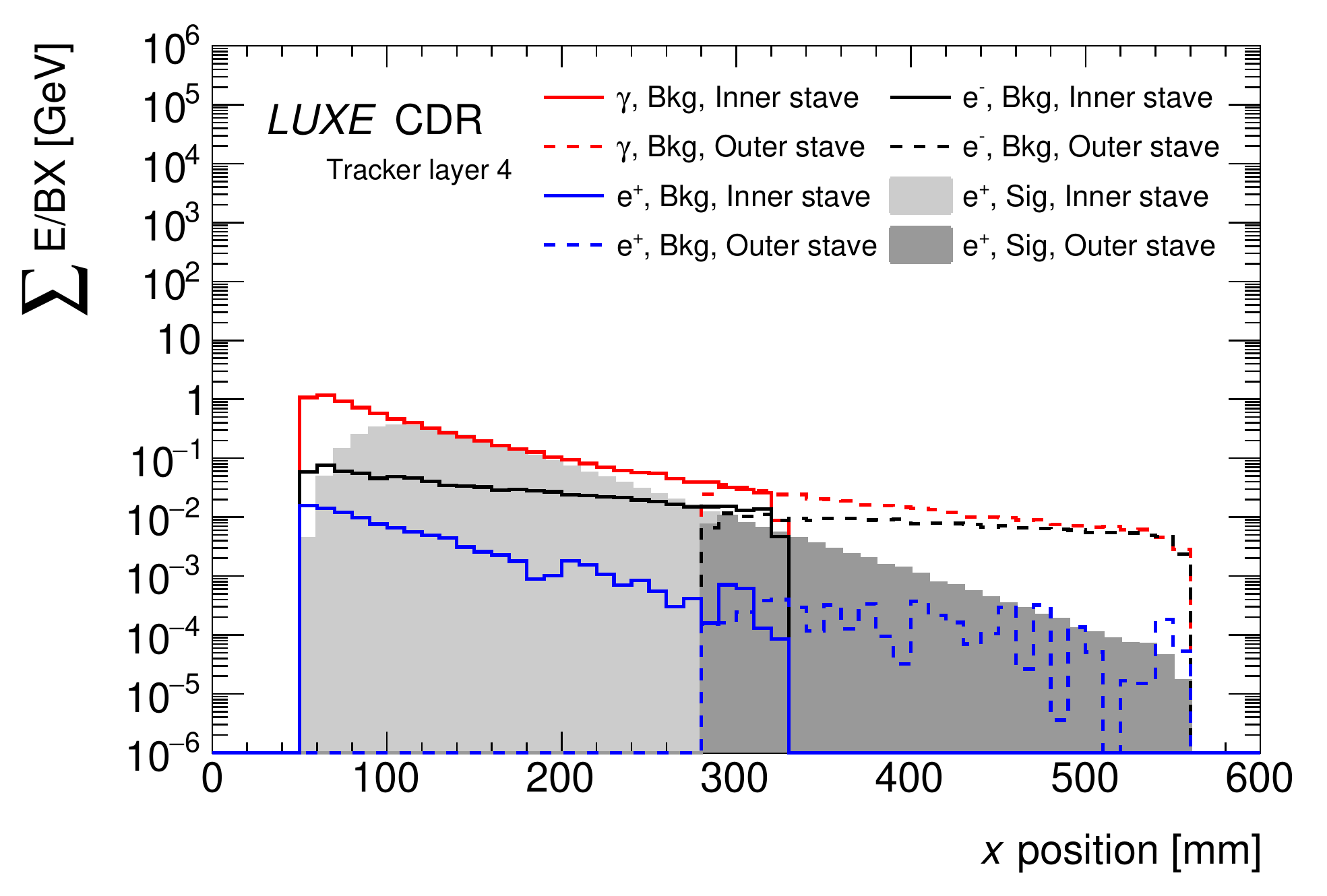}
\put(-120,10){\makebox(0,0)[tl]{\bf (d)}}\\
\caption{The (a, b) number of particles and (c, d) sum of energy of particles per bunch crossing versus $x$ position of hits for the (a, c) first 
layer and (b, d) last layer (layer 4) for background and signal. All particles intersecting the tracker layers are considered regardless of the actual energy deposition in the sensitive volume. The solid lines (light grey fill for the signal) correspond to the tracks in the inner stave  and the dashed lines (dark grey fill for the signal) correspond to the tracks from the outer stave. The positron signal, shown in grey, corresponds to the JETI40 laser with $w_0=6.5$~$\mu$m, corresponding to $\xi=2.4$.
}
\label{fig1-tracker}
\end{figure}

Next, the background is compared to the signal to aid the discussion of the various detector requirements. Figure~\ref{fig1-tracker} shows the  signal and background per BX in the first and fourth layer of the tracking detector for the positron side in \elaser collisions. Each layer consists of two staves that are placed next to each other, an inner and an outer one, which overlap by 4~cm (see Sec.~\ref{sec:detectors_tracker}).
The number of particles and summed energy are shown as a function of $x$ along the detector face. The signal is illustrated for $\xi=2.4$ where $0.7$ positrons per BX are expected. The signal is small compared to the background of electrons and photons, although the $e^+$  generally have higher energy. The overall difference in rates illustrates the challenge that the tracking detector faces and why a single plane of e.g.\ scintillator will not suffice.  The performance of the tracker, and how the background can be suppressed while retaining the signal efficiently, is discussed in Sections~\ref{sec:detectors} and~\ref{sec:results}.

\begin{figure}[htbp]
\includegraphics[width=0.49\textwidth]{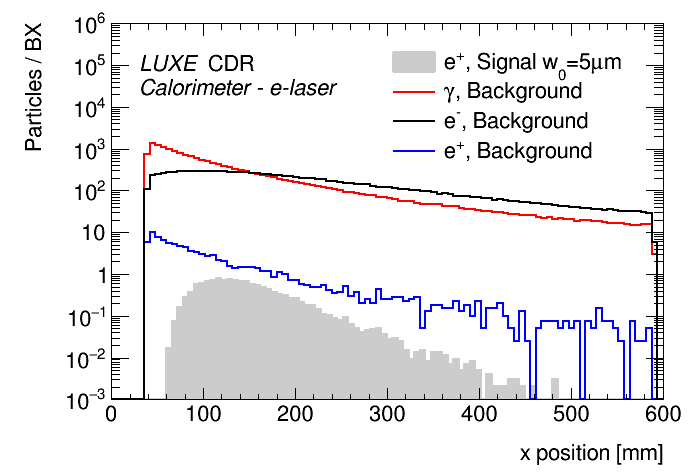}
\put(-120,10){\makebox(0,0)[tl]{\bf (a)}}
\includegraphics[width=0.49\textwidth]{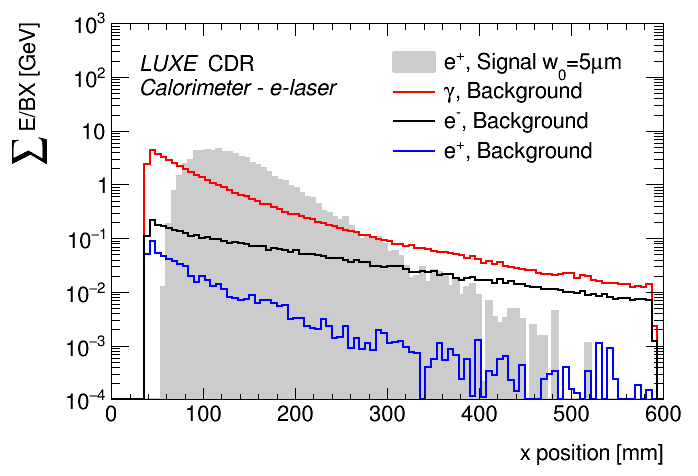}
\put(-120,10){\makebox(0,0)[tl]{\bf (b)}}\\
\caption{(a) Number of particles and (b) their energy sum per BX, split into different background particle types and signal positrons as described in the legend, as a function of the $x$ position in the calorimeter for the \elaser set-up for JETI40 with $w_0=5\units{\mu m}$, corresponding to $\xi=3.1$. 
}
\label{fig:ecal-elaser-bkg-rates}
\end{figure}

A picture similar to that observed in the tracker emerges for the flux of particles in the calorimeter. As an example, the expected particle rates and the energy sum of particles impacting the calorimeter per BX as a function of the $x$ position in the calorimeter are shown in Fig.~\ref{fig:ecal-elaser-bkg-rates} in the \elaser set-up for JETI40 and $\xi=3.1$. The background contribution is split into different particle types and compared to the expected rates of signal positrons. 

 Much of the background originates from a flux of very low energy particles created by the beam, which are partly suppressed by adding shielding on the beam side of the tracker and the calorimeter. This background constitutes a pedestal which can be measured in BX without laser and subsequently subtracted, pad by pad. 

\begin{figure}[htbp]
\includegraphics[width=0.49\textwidth]{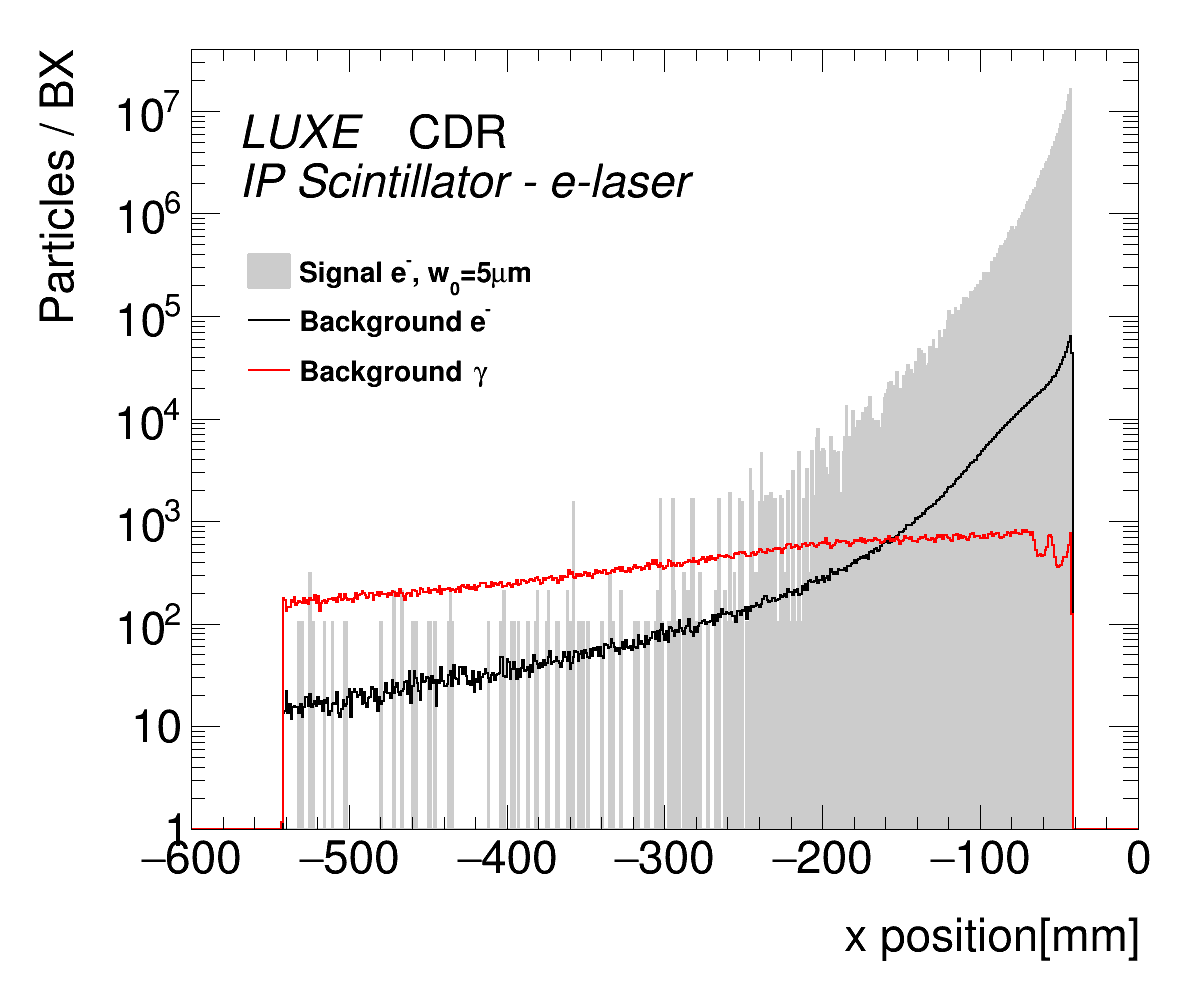}
\put(-120,10){\makebox(0,0)[tl]{\bf (a)}}
\includegraphics[width=0.49\textwidth]{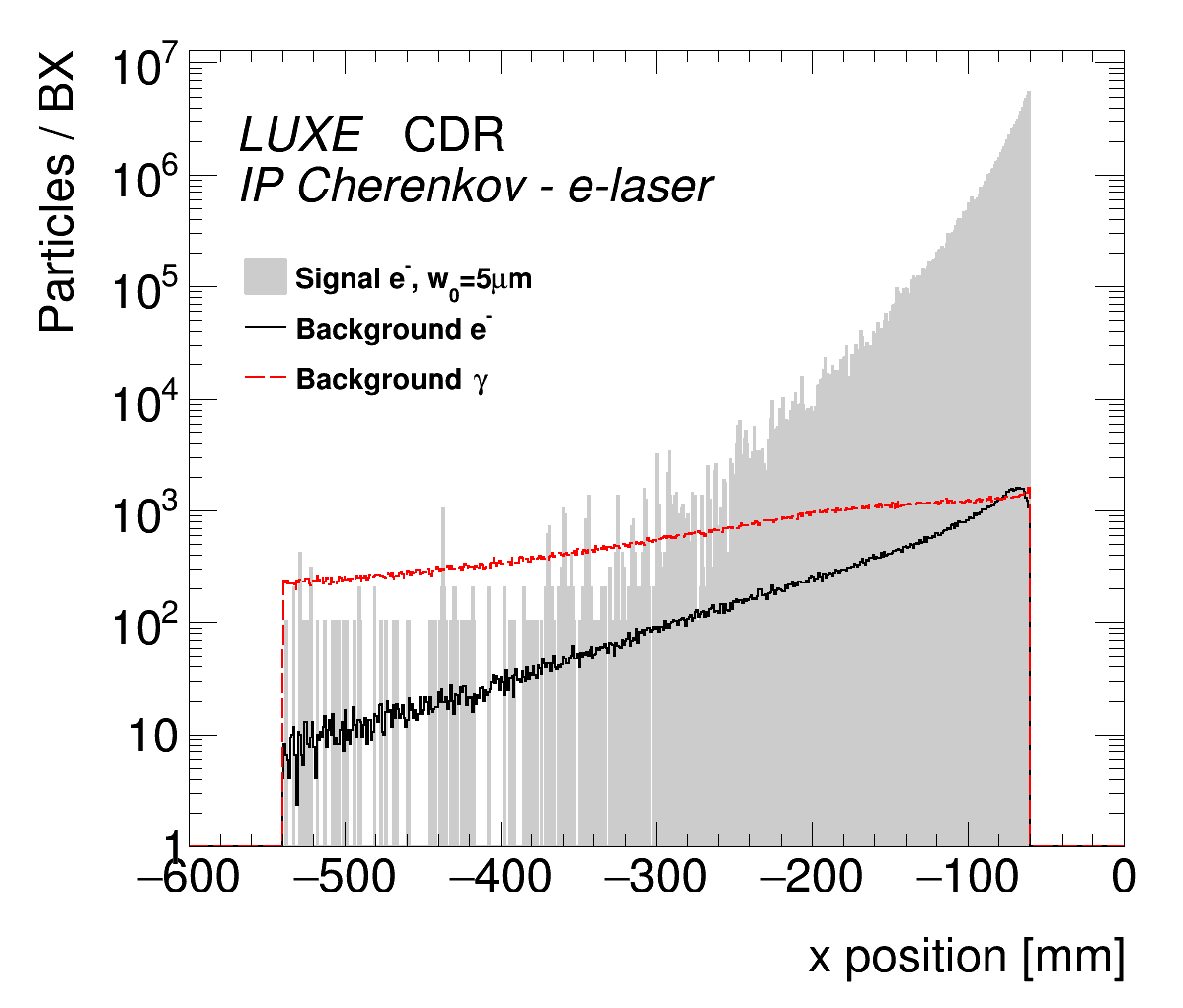}
\put(-120,10){\makebox(0,0)[tl]{\bf (b)}}\\
\includegraphics[width=0.49\textwidth]{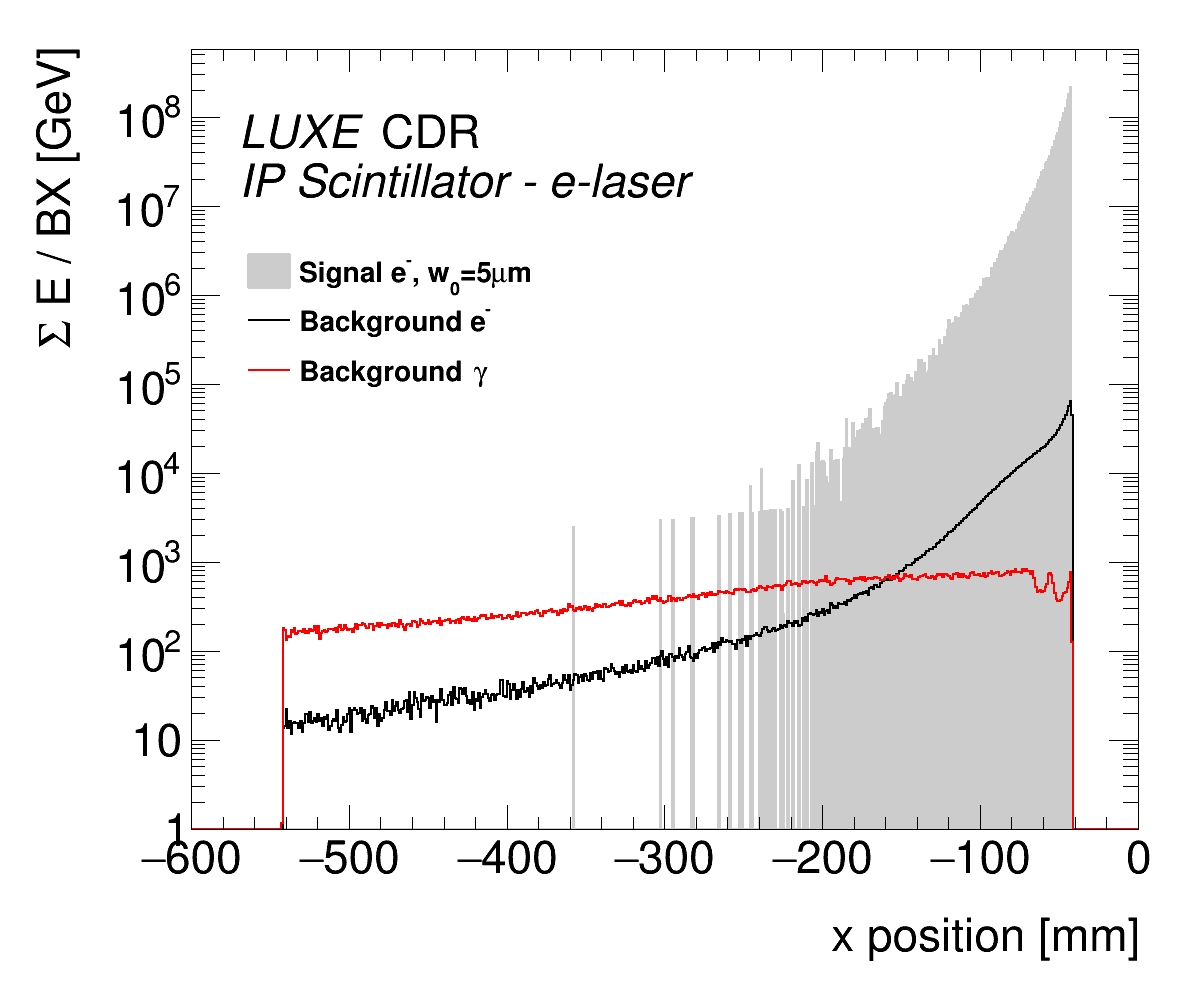}
\put(-120,10){\makebox(0,0)[tl]{\bf (c)}}
\includegraphics[width=0.49\textwidth]{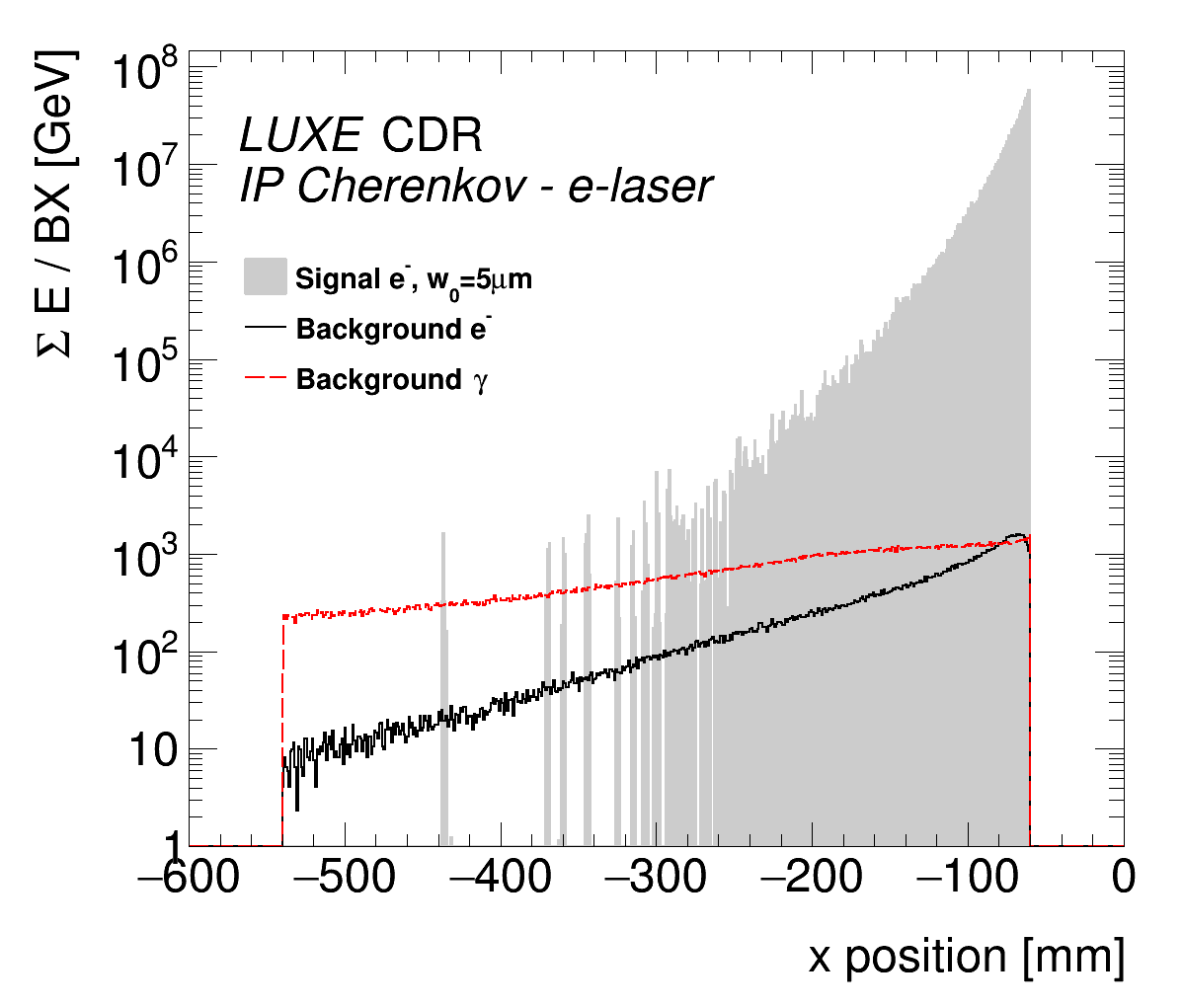}
\put(-120,10){\makebox(0,0)[tl]{\bf (d)}}\\
\caption{
The number of electrons versus the $x$-position for (a) the  scintillator screen and (b) the \cer detector at the post-IP electron measurement for  
\elaser scattering with the JETI40 laser (using a spot-size of 5\,$\mu$m). The sum of the energy of the particles is shown in (c) and (d) for the scintillator and \cer detector, respectively. For both detectors only the fiducial volume is considered. 
For the \cer detector, signal and background are selected with the requirement of a particle energy greater than 20\,MeV as for the IP \cer detector.
}
\label{fig:ip_lanex_ckv}
\end{figure}

On the electron side, the signal particle rate is too high for the use of tracking detectors and instead a detector system consisting of a scintillator screen and a \cer detector is planned, see also Section~\ref{sec:detectors}.
Both of which can cope with the high particle rates. Figure~\ref{fig:ip_lanex_ckv} shows the occupancy for signal and background in the scintillator screen. It is seen (in Sec.~\ref{sec:det:scintillator}) that the signal is concentrated at $|y|<5$~mm while the background is rather spread out. This motivates the definition of a fiducial volume for the screen in $y$ of $\pm 5$~mm around the central value. The larger $y$ values are useful for background estimation. After having selected this fiducial area in $y$, the $x$ positions 
are shown in Fig.~\ref{fig:ip_lanex_ckv} for one example of laser parameters. Close to the beam, the ratio of  signal to background electrons is about $10^2$ and $10^4$ for the scintillator screen and \cer detector respectively, but reduces with decreasing $x$. 
The rate of photons is rather high (see Fig.~\ref{fig:elaser-barcharts}) and the \cer technology has been chosen in part as it is insensitive to photons. The scintillator also is expected to be less sensitive to photons than electrons and since electrons dominate in this region photon background are not a major concern here. The background from positrons is not shown as it is negligible. Qualitatively similar results are obtained for different laser parameters.

\begin{figure}[htbp]
  \begin{center}
  \begin{subfigure}{0.48\textwidth}
      \includegraphics[width=\textwidth]{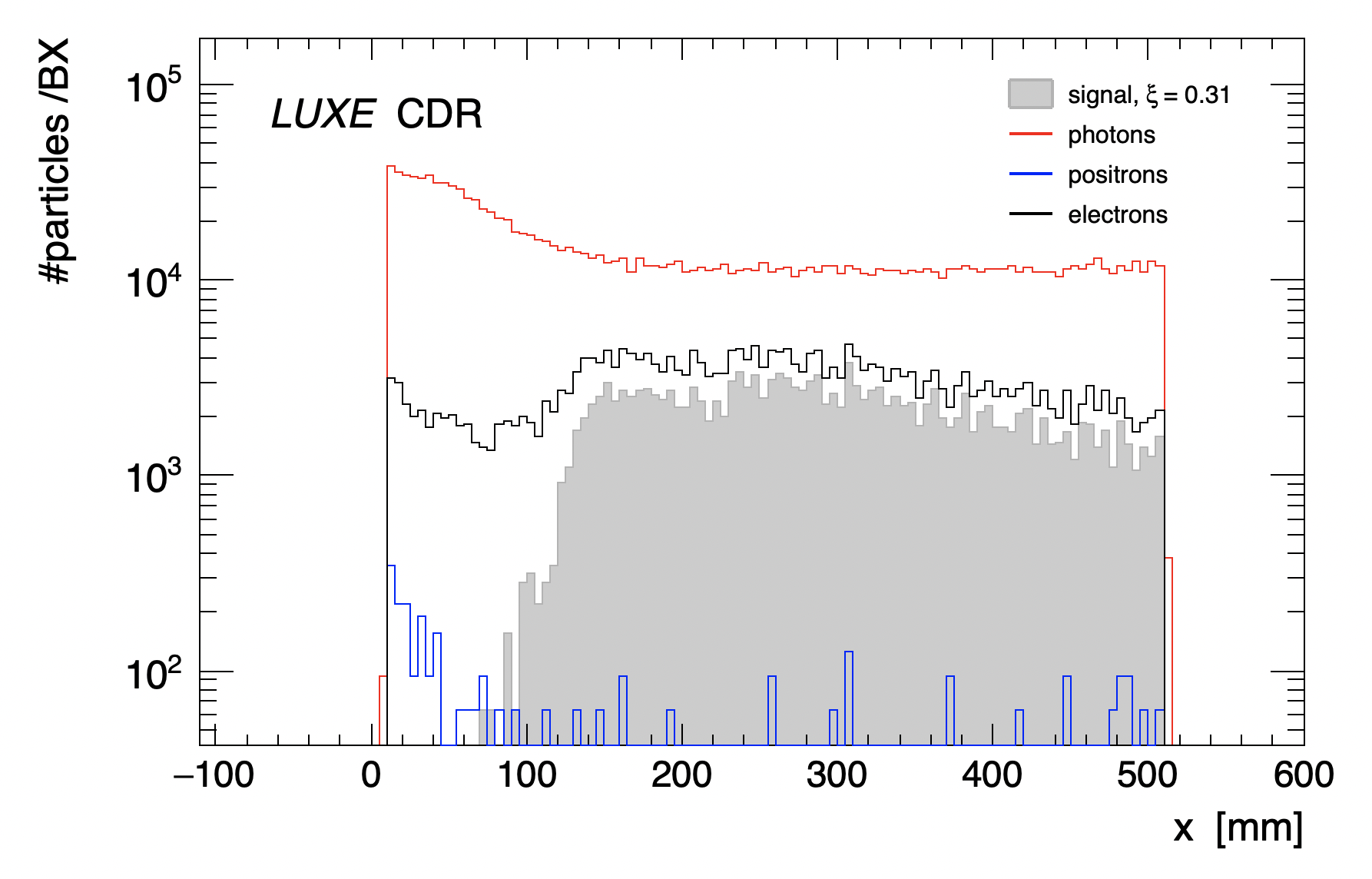}
      \caption{Scintillator screen in Photon Detection System.}
  \end{subfigure}
  \begin{subfigure}{0.48\textwidth}
      \includegraphics[width=\textwidth]{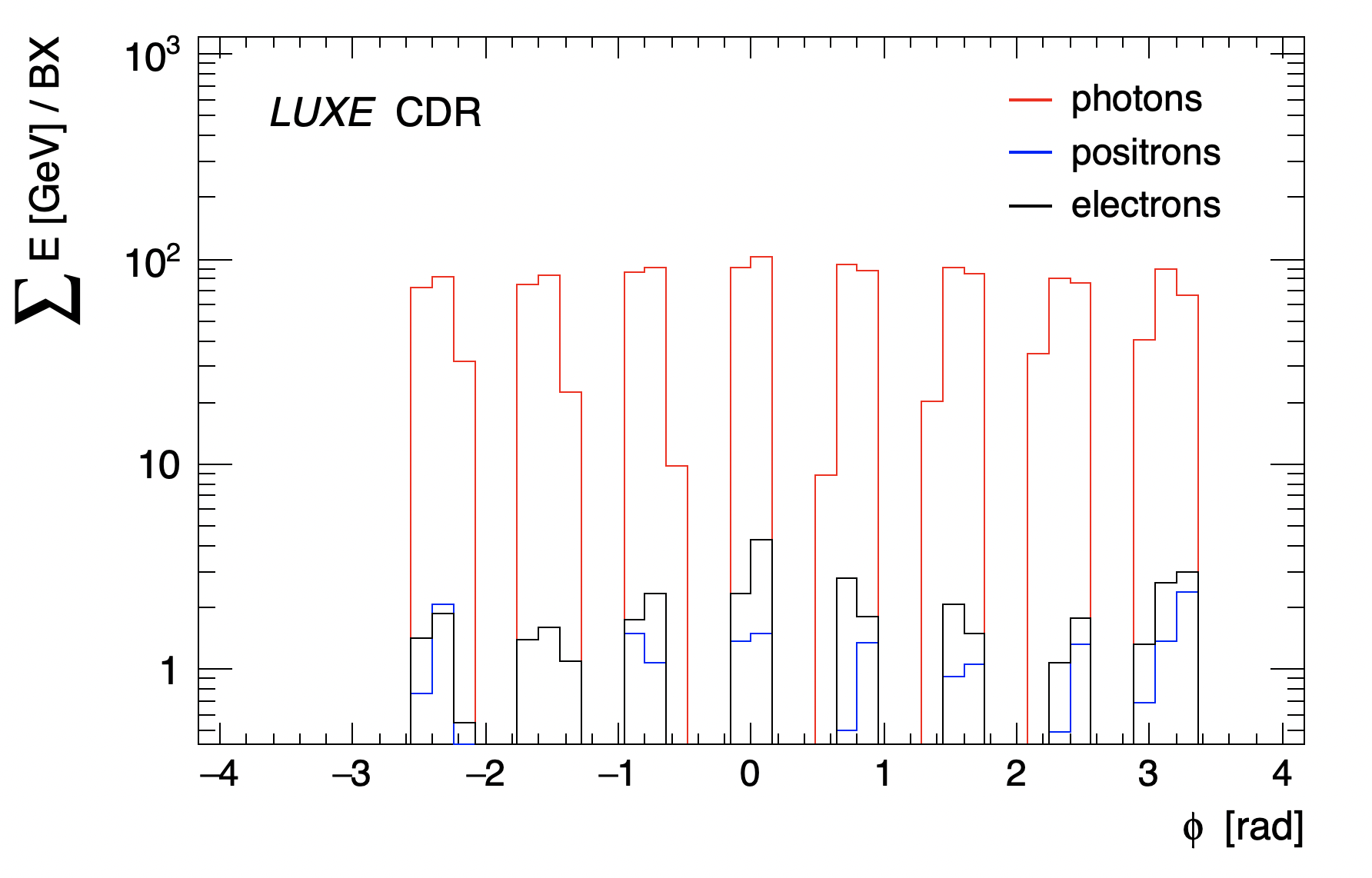}
      \caption{Gamma Flux Monitor in Photon Detection System.}
  \end{subfigure}
     \end{center}
      
    \caption{(a) The number of particles per bunch crossing versus the $x$ position of particles for the electron side of the scintillating screen of the forward photon spectrometer. The signal and background particles are estimated from the simulation of the \elaser set-up. All particles intersecting the scintillating screen are considered regardless of the actual energy deposition in the sensitive volume. The signal is shown for the JETI40 laser using a spot-size of $w_0=50\,\mu$m., corresponding to $\xi=0.31$. (b) Sum of energies of particles per bunch crossing versus azimuthal angle of tracks of forward gamma flux monitor. No signal is shown as all particles are secondary in a sense that they result from backscattering.}
    \label{fig:FWDLanex_B}
\end{figure}

A spectrometer will also be employed in the very forward region in order to monitor the number of photons; this again uses a scintillator screen. 
In Fig.~\ref{fig:FWDLanex_B}(a), 
the particle rates 
are shown at the face of the scintillator screen for the electron side. The number of positrons is negligible compared to the number of electrons. In contrast, the number of photons is about a factor of 100 larger than the number of electrons.
While the response of the screen to photon is generally lower than that to electrons, given the larger number of photons, both background sources are important here. 

Distributions for the backscattering calorimeter are shown in Fig.~\ref{fig:FWDLanex_B}(b). It is seen that the energy is dominated by photons. The energy is about 100~GeV per module, compared to the background of typically $1$~GeV.

\subsubsection{Expected Signal and Background Rates in Photon-laser Collisions}

\begin{figure}[htbp]
  \centering
  \includegraphics[width=0.49\textwidth]{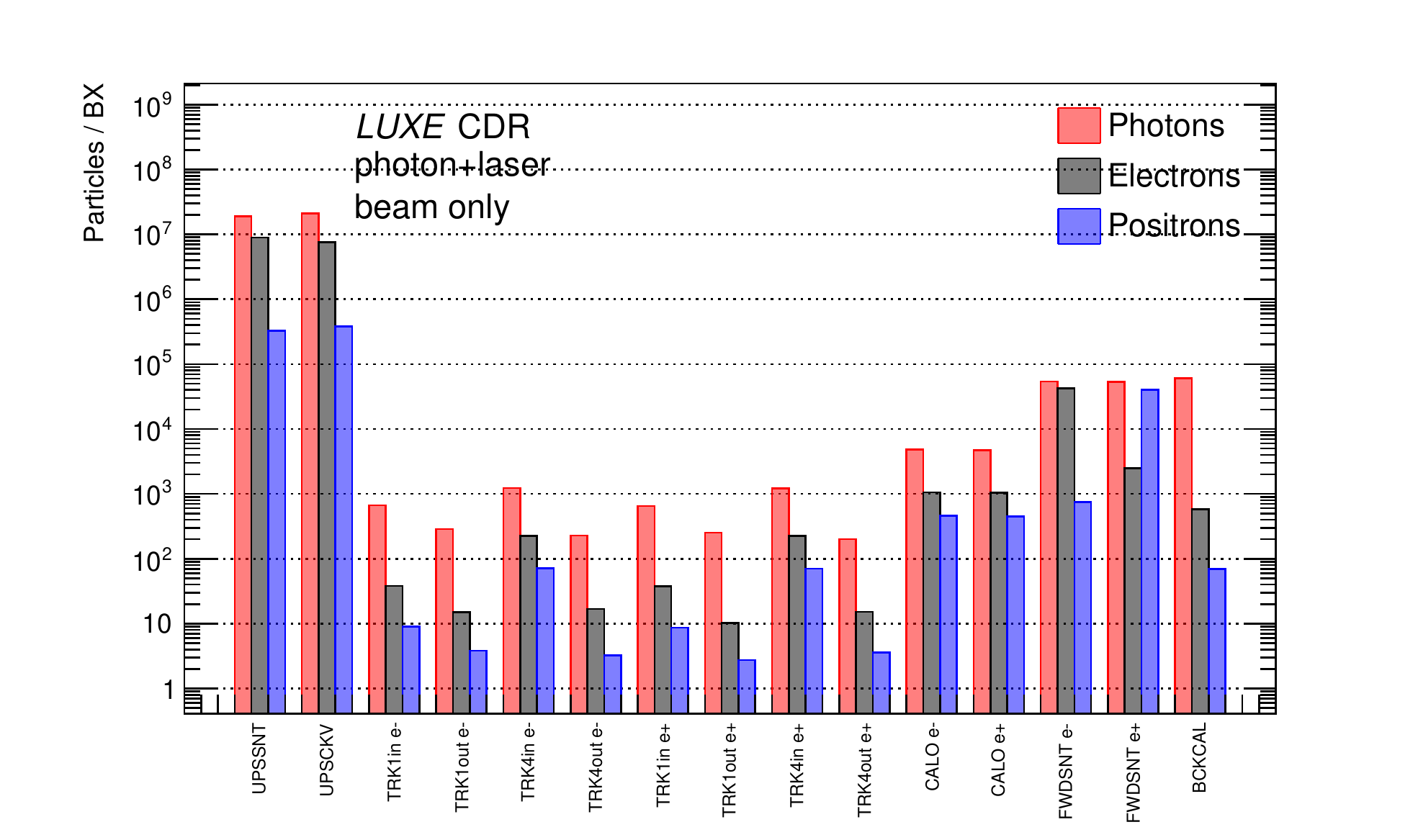}
  \put(-120,0){\makebox(0,0)[tl]{\bf (a)}}
  \includegraphics[width=0.49\textwidth]{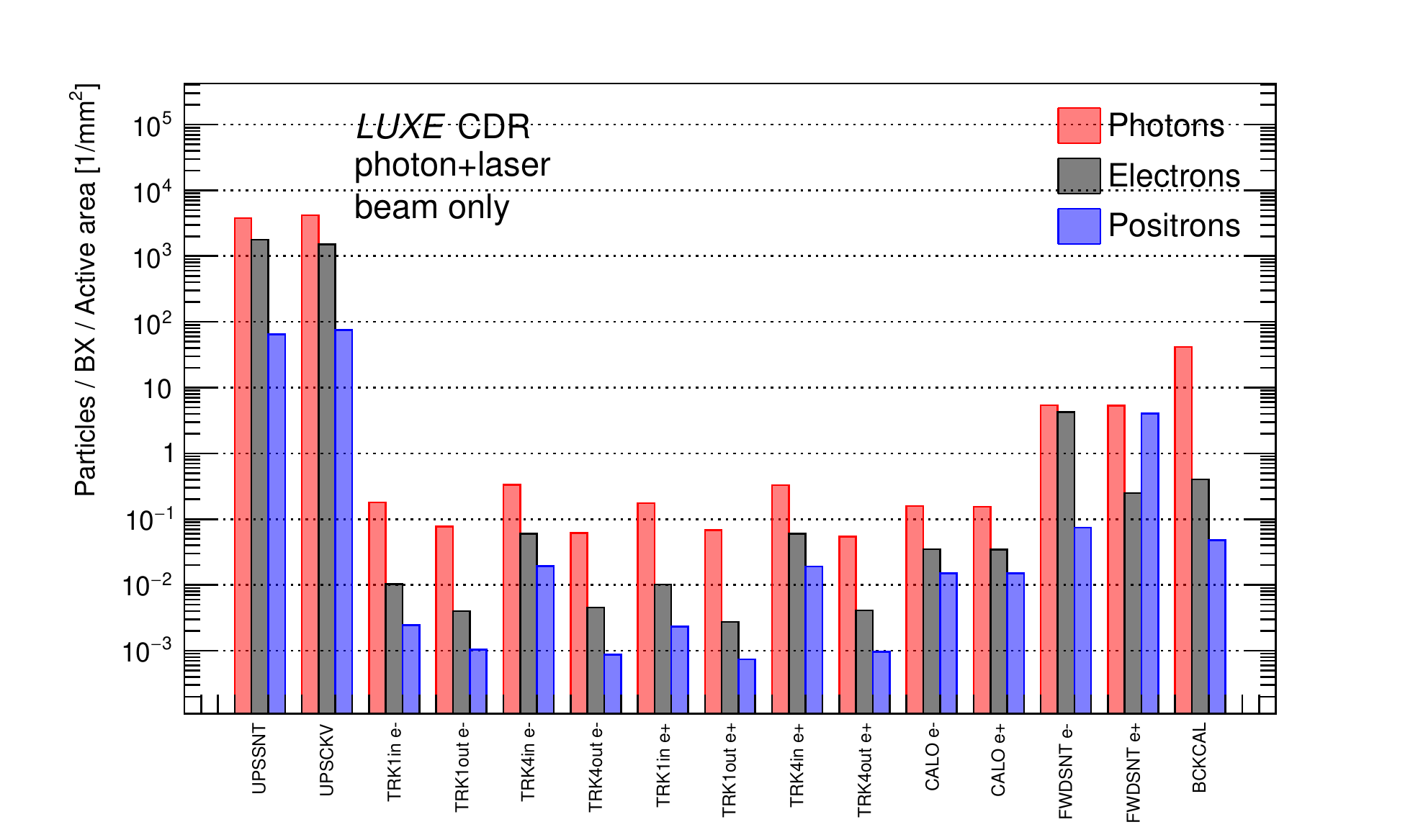}
  \put(-120,0){\makebox(0,0)[tl]{\bf (b)}}\\
  \vspace{0.25cm}
  \includegraphics[width=0.49\textwidth]{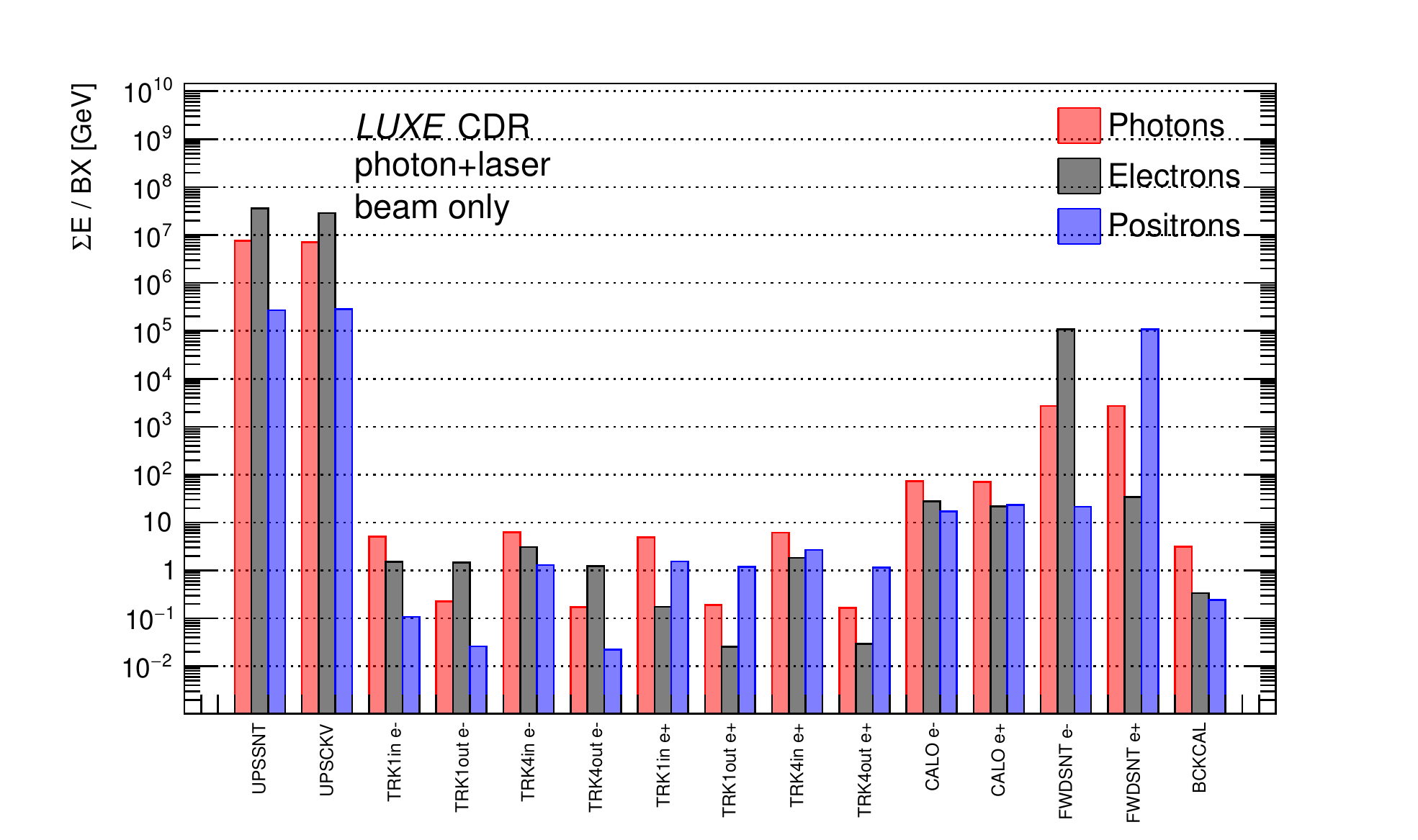}
  \put(-120,0){\makebox(0,0)[tl]{\bf (c)}}
  \includegraphics[width=0.49\textwidth]{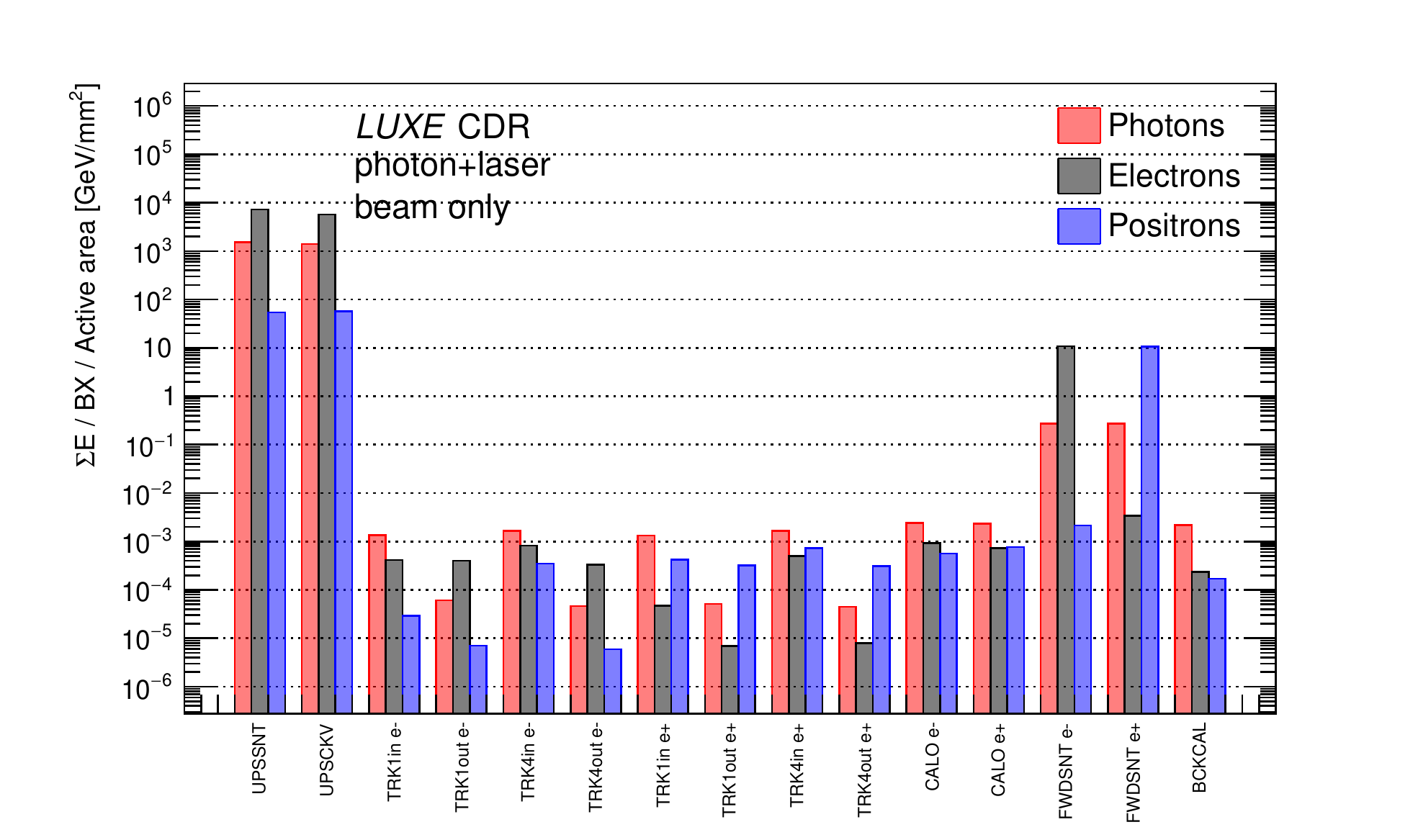}
  \put(-120,0){\makebox(0,0)[tl]{\bf (d)}}\\
  \vspace{0.25cm}
  \caption{Distributions of (a) number of particles, (b) number of particles per unit area, (c) the summed energy and (d) the summed energy per unit area at the front face of each detector system for \glaser collisions. The numbers and energies are shown separately for photons, electrons and positrons.  All 
  values are normalised to one bunch crossing.  The first two entries show the scintillator screen and \cer detector after the bremsstrahlung target. The next eight entries show the first and last tracker layers, followed by the calorimeter on the $e^-$ and $e^+$ sides. The next two entries show the forward scintillator screens. The final entry shows the backscatttering calorimeter.}
  \label{fig:glaser-barcharts}
\end{figure}

A major difference in the \glaser set-up compared to the \elaser set-up is the addition of a target into the electron beam to produce high-energy 
photons and the spectrometer after the target. As shown in Fig.~\ref{fig:glaser-barcharts}, the rates for detectors positioned after the target are high, 
with $> 10^7$\, particles and so a system of a scintillator screen in combination with a \cer detector is proposed. 

\begin{figure}[ht]
\includegraphics[width=0.45\textwidth]{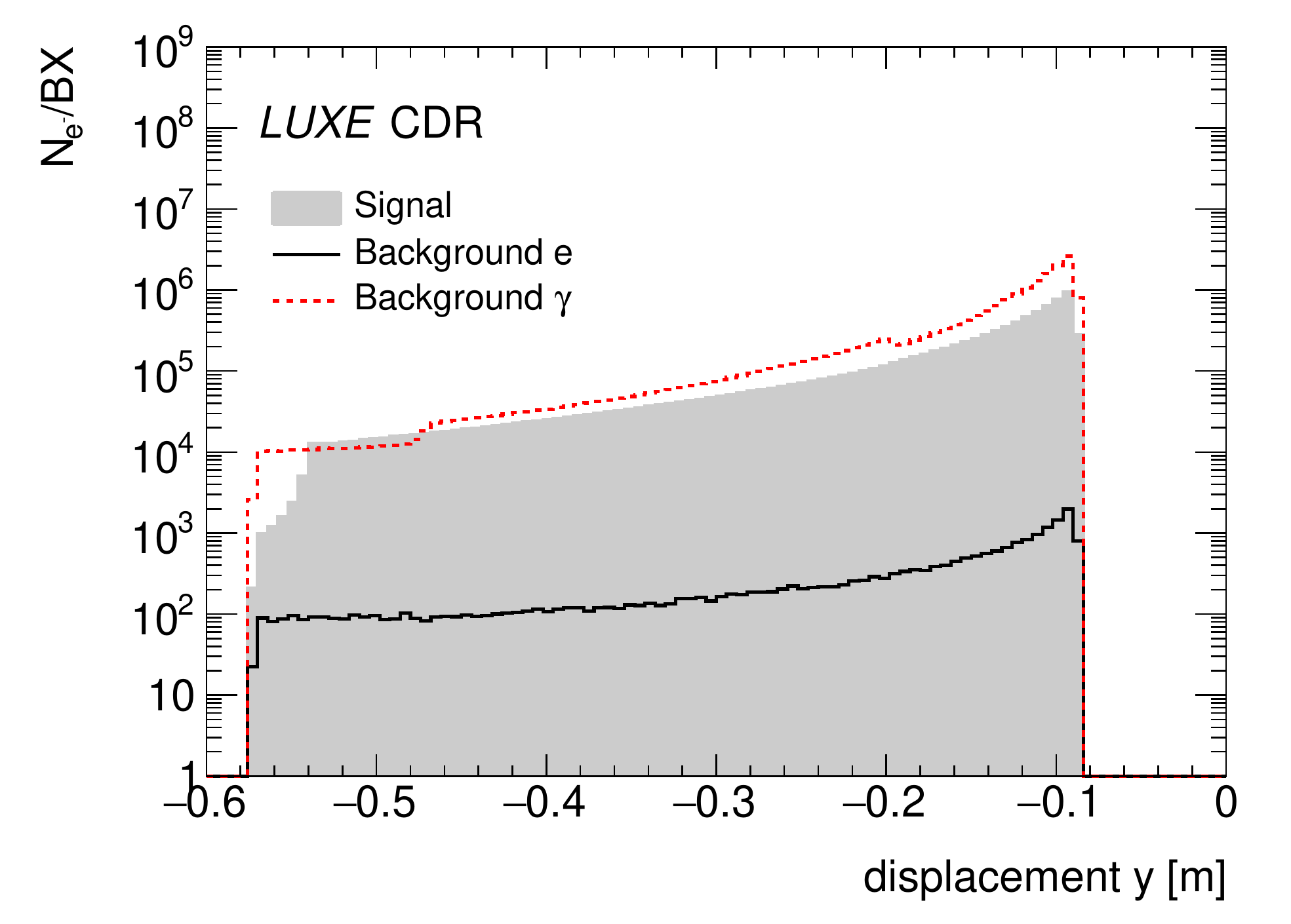}
\includegraphics[width=0.45\textwidth]{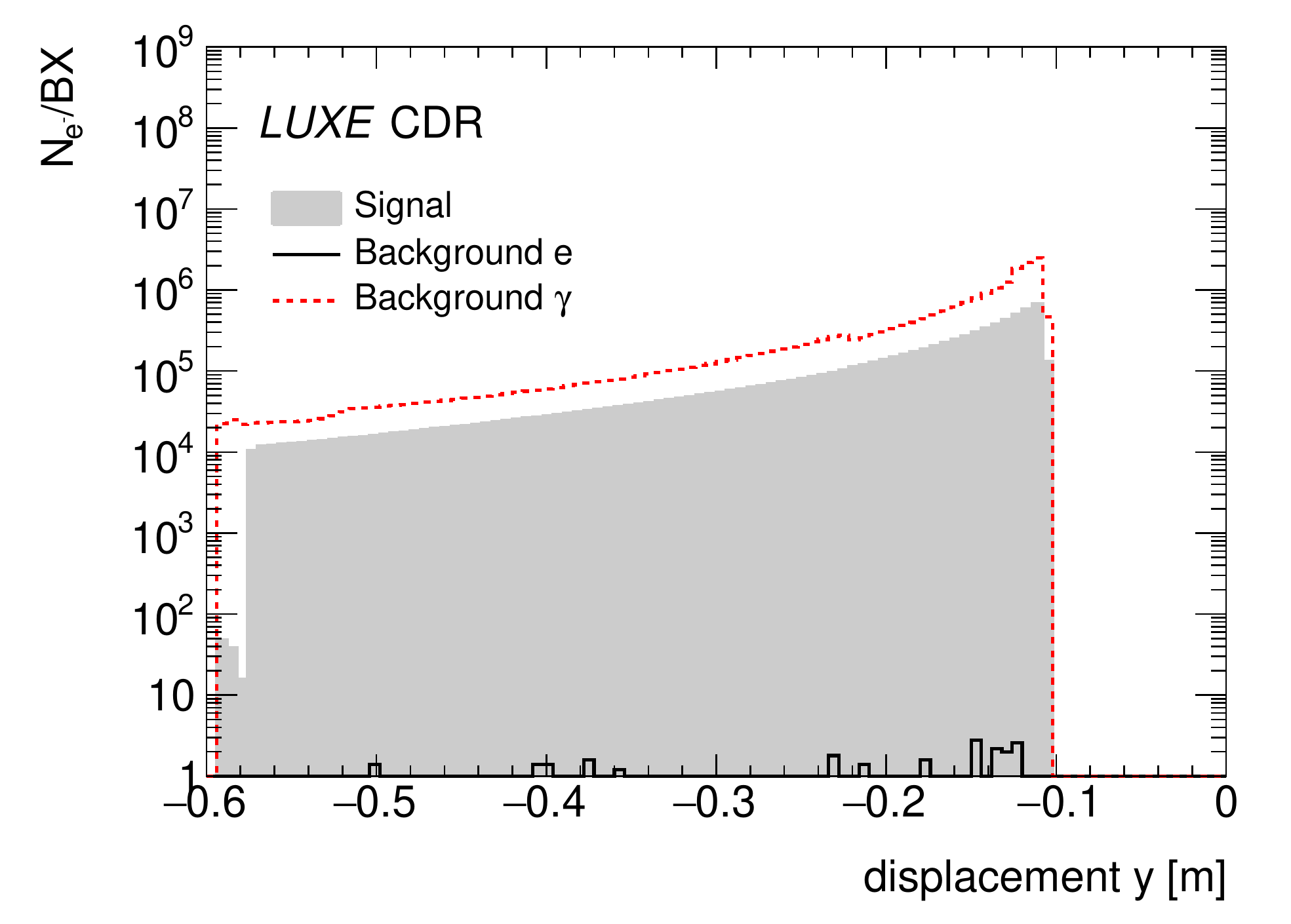}\\
\includegraphics[width=0.45\textwidth]{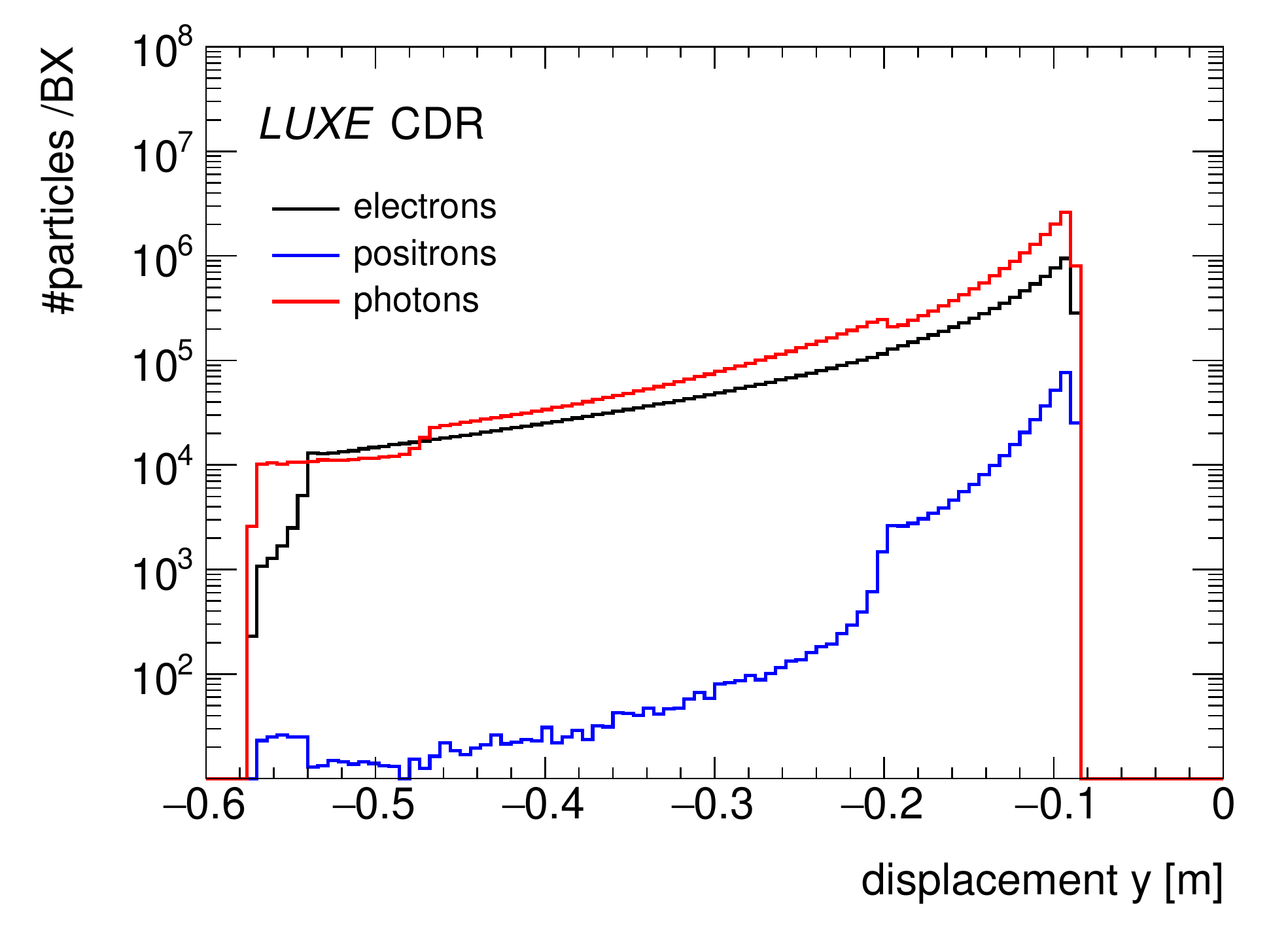}
\includegraphics[width=0.45\textwidth]{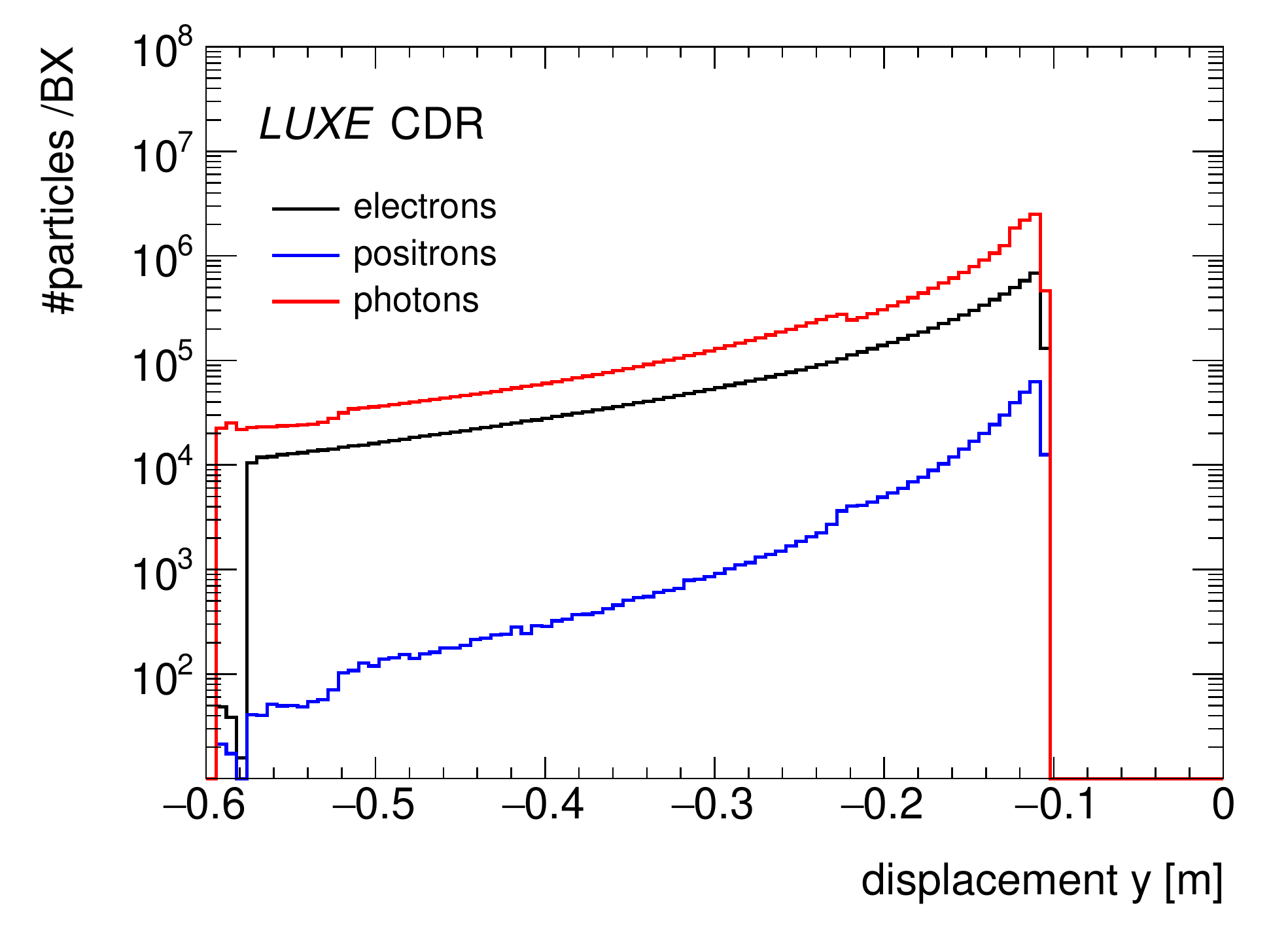}\\
\includegraphics[width=0.45\textwidth]{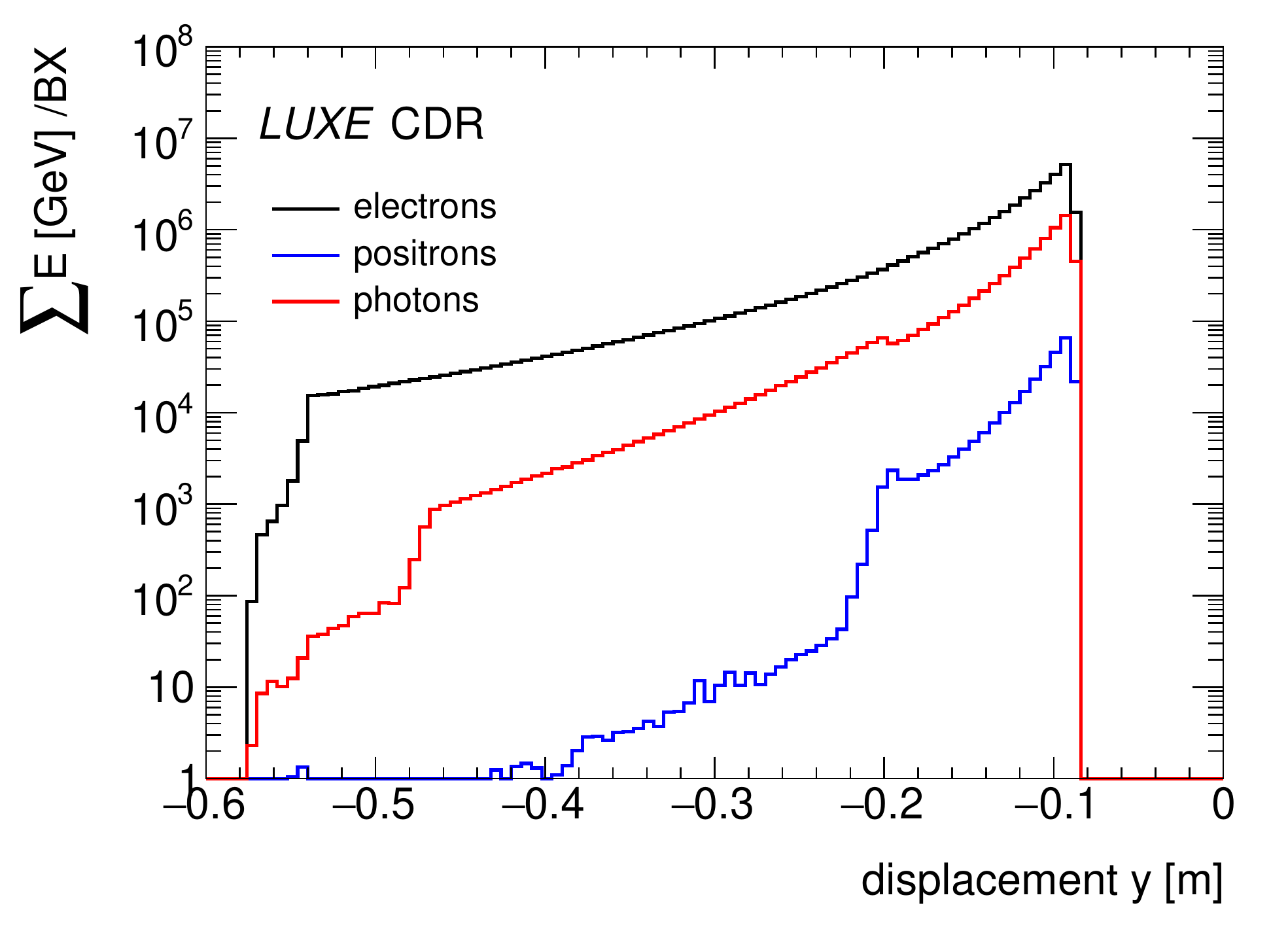}
\includegraphics[width=0.45\textwidth]{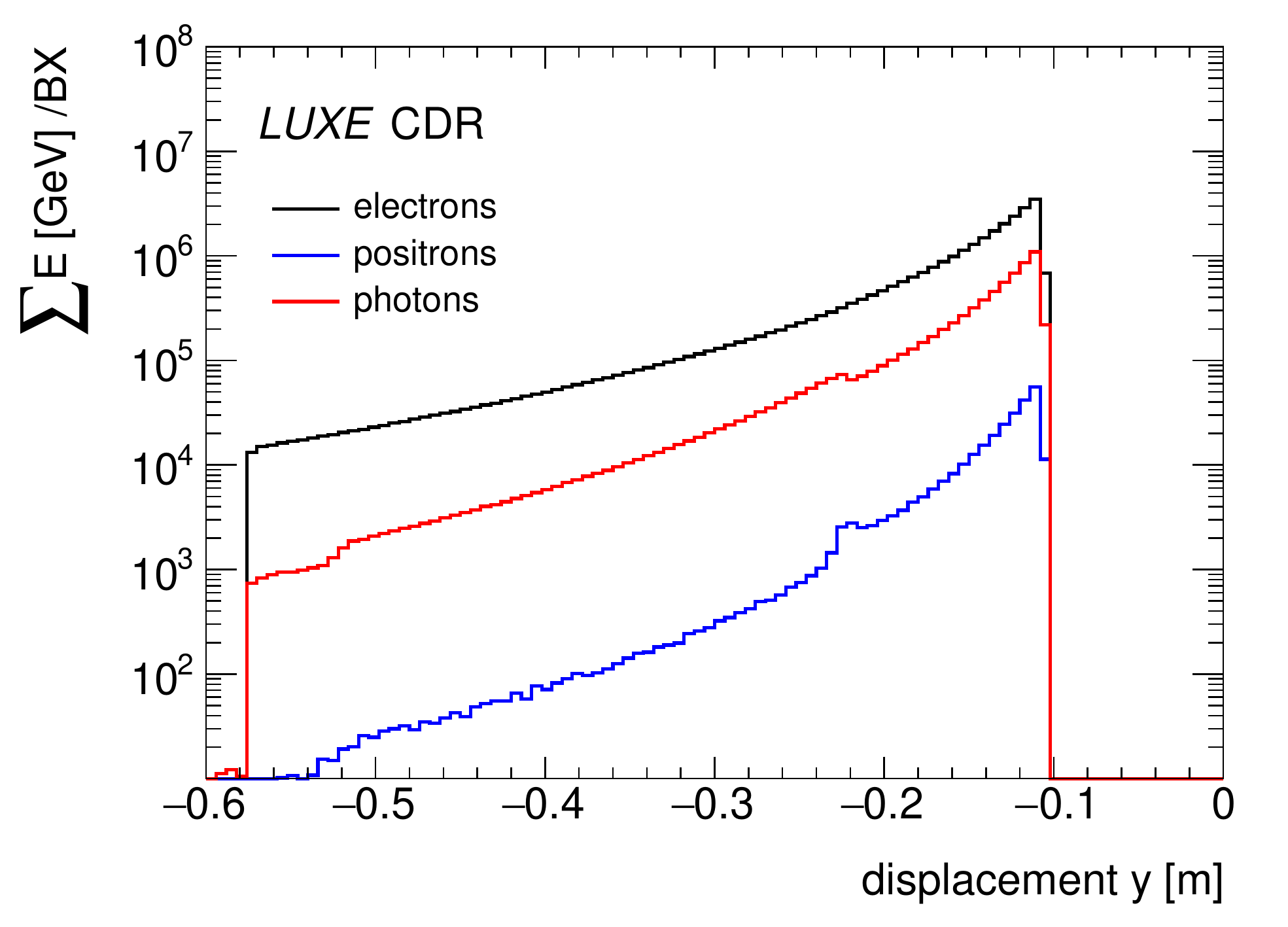}
\caption{Occupancy and summed energy per mm for the scintillator screen (left) and \cer detector (right) after the bremsstrahlung target for  
\glaser scattering. The occupancies shown are for signal electrons and background electrons (top), and for electrons, positrons and photons (middle). Bottom) The sum of incident energy per mm of electrons, positrons and photons is shown.}
\label{fig:g_target_lanex_ckv}
\end{figure}

The $y$ positions of the signal and background particles hitting the scintillator and \cer detector are shown in Fig.~\ref{fig:g_target_lanex_ckv} in the fiducial volume. 
The fiducial volume for both detectors is the same as for the detectors behind the IP case: the full detector in the $y$ direction and $\pm 5$\,mm around the beamline in the $x$ direction~\footnote{Here, the dipole magnet is oriented such that it bends the particles in the vertical plane.}. For the \cer detector, both signal and background 
are selected with the requirement of a particle energy greater than 20\,MeV, which corresponds to the \cer radiation threshold energy for electrons/positrons in argon at atmospheric pressure.  
The signal to background ratio for electrons is high, $>10^2$ for the scintillator screen and $>10^4$ for the \cer detector. The contamination due to positrons is small. Photons also contribute a large rate but are generally lower energy and the detector technologies chosen are less sensitive to photons than electrons. 

The other difference is that the signal rates on the electron side behind the IP now is the same as that on the positrons side, thereby opening the possiblity to use the same technology. It is seen that background is also similar on both sides for the \glaser mode. Typically 10-100 charged background particles are present in the tracker, and about 1000 in the calorimeter. The signal rates we would like to measure are as low as $10^{-3}$, thus placing a high demand on the detector to suppress background. 

\subsection{Ionisation Dose}
The simulation was also used to estimate the ionisation dose for the detector locations. For this estimate the \textsc{QGSP\_BERT\_HP} physics list was used as it also simulates thermal neutrons. Table~\ref{tab:dose} summarises the average annual dose assuming $10^7$~s of operation. Also shown is the dose tolerable for each detector type.

\begin{table}[htbp]
    \centering
    \begin{tabular}{|l||c||c|}
    \hline
         Detector &  avg. annual dose [Gy] & tolerable dose [Gy]\\\hline
         IP Pixel tracker&  $10$ & $27\cdot 10^3$ \\
         IP Calorimeter& $1$ & $10^6$\\
         IP Scintillator& $2\times 10^4$ & $10^8$ \\\hline
         GDS GadOx Scintillator & $10$ & $10^8$ \\
         GDS LYSO Scintillator & $10^4$ & $10^4-10^7$ \\
         GDS Gamma Profiler & $<10^7$ & $10^7$ \\
         GDS Backscatter Calorimeter& $0.5$ & $300$ \\
         \hline
    \end{tabular}
    \caption{Annual average dose at the various detector locations for the \elaser set-up based on \geant simulations with the \textsc{QGSP\_BERT\_HP} physics list. The calculation assumes $10^7$\,s of operation. Also shown is the dose that can be tolerated by each detector type without performance degradation.}
    \label{tab:dose}
\end{table}

\clearpage
\section{Detectors, Monitors and Data Acquisition Aspects}
\label{sec:detectors}

In this section, the detector technologies chosen for the various regions of the experimental setup are discussed. To achieve precision measurements of \hiqed in the transition from the perturbative to the non-perturbative regime, an excellent performance is required in the range $\ximax\sim 1-10$. There are two qualitatively very different environments in LUXE, and in both regimes it is important that there is sufficient redundancy in the measurements to ensure that systematic uncertainties can be estimated and minimised.

First, there are regions of very high rates of electrons where typically $10^6-10^{9}$ particles are present, placing high demand on the radiation hardness and on the response-linearity of the exposed detectors.
In these regions the signal is typically 10--100 times larger than the background (see Sec.~\ref{sim:sig_bkg}), so that background suppression is not of primary concern, in particular as it can be subtracted based on in-situ measurements. Similarly, at the very end of the experiment the photon flux is up to $10^{9}$ photons per laser shot.

Second, in the region behind the IP, the number of positrons per BX needs to be measured between $10^{-3}$ and $10^3$ during the initial phase, and up to values of about $10^4$ for \phasetwo. 
In particular at low $\ximax$, the signal rates are much lower than the background (see Sec.~\ref{sim:sig_bkg}) and thus powerful detectors are required with a strong capability to reject the latter. The better the background rejection the lower $\ximax$ values can be measured and thus the fully perturbative regime of QED can be examined. Furthermore, the detectors need to function at fluxes of $\sim 1,000$ particles per laser shot and allow the reliable reconstruction of the number and the energy spectrum of particles in this regime where QED is expected to become non-perturbative. Having the lever-arm to observe the particle rates in the entire $\xi$ range is critical to explore how exactly the transition into the non-perturbative regime occurs.  

\begin{figure}[htbp]
\centering
\includegraphics[width=0.48\textwidth]{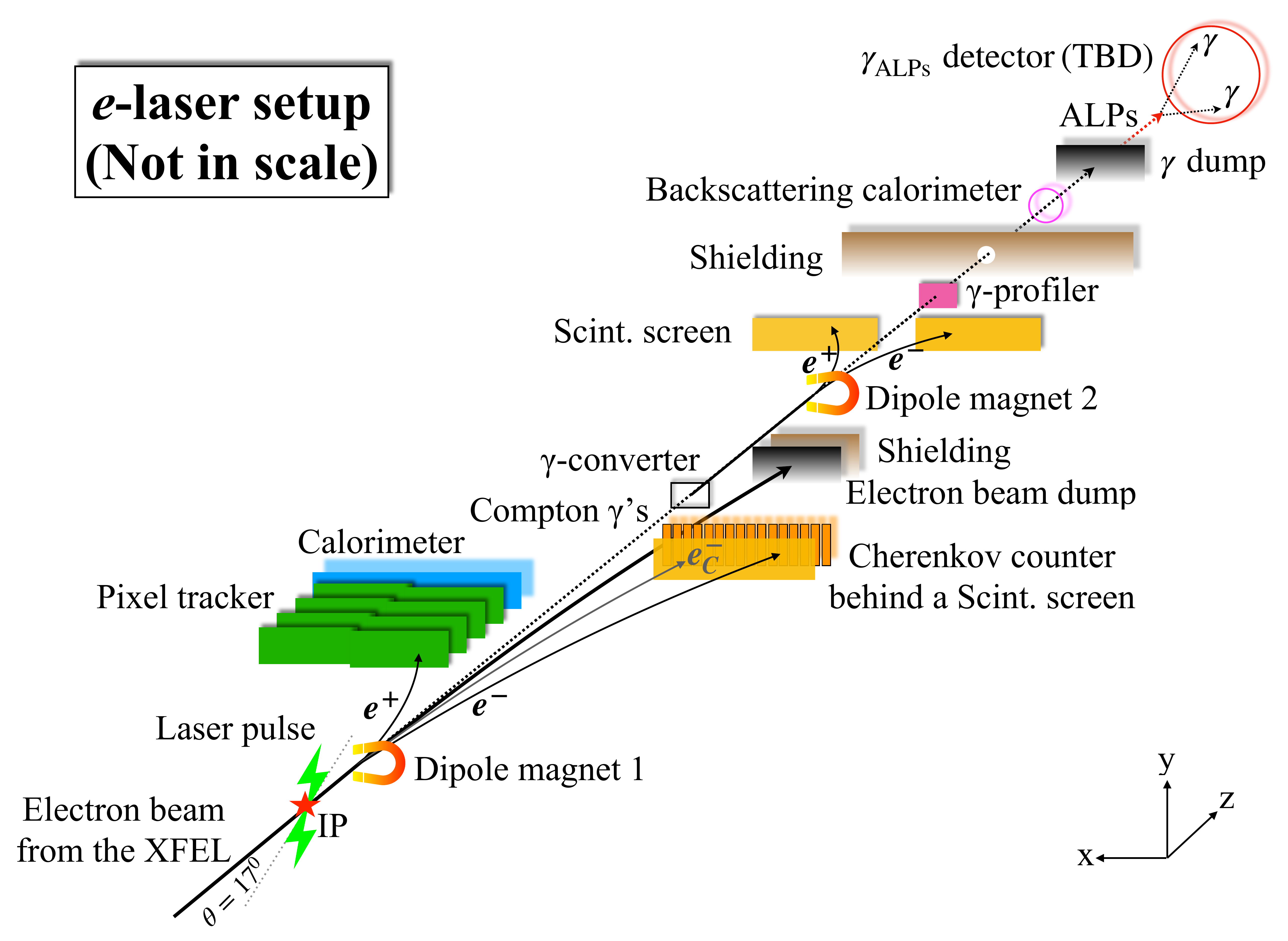} 
\includegraphics[width=0.48\textwidth]{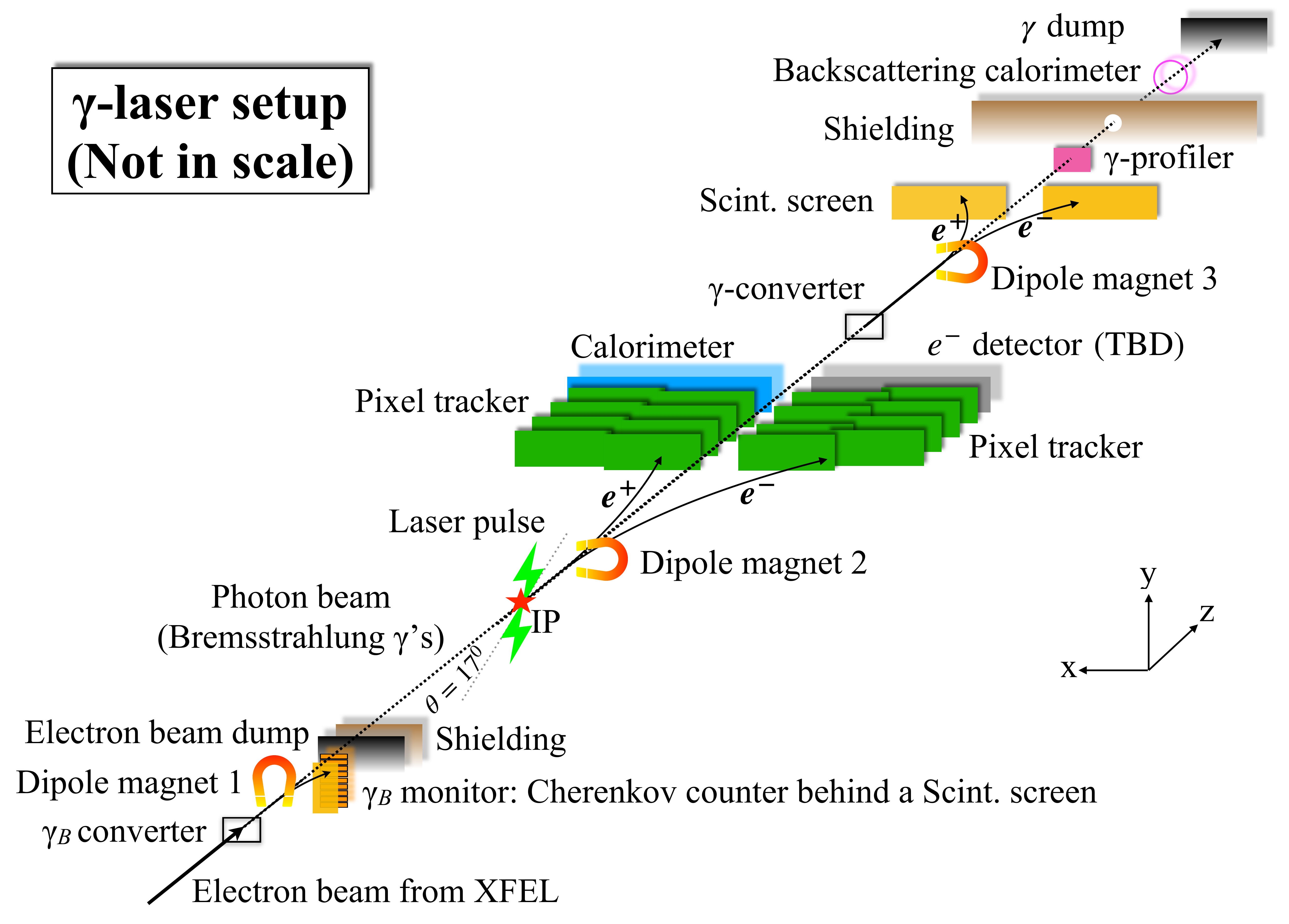} 
\caption{Schematic layouts for the \elaser and \glaser  setup. Shown are the magnets, detectors and main shielding and absorbing elements. 
} 
\label{fig:layoutdet} 
\end{figure}

Conceptual layouts of both the \elaser and \glaser setup are shown in Fig.~\ref{fig:layoutdet}.

For the electron and positron measurements a magnet is used to divert the particles from the beamline and their distance to the beamline is then a measure of the energy. The energy $E$ and horizontal position $x$ of the particle in the detector are roughly related to each other through
\begin{equation}
\label{eq:spectrometer}
    x=B c z_m\times\frac{z_m/2+z_d}{E} \, ,
\end{equation}
where $B$ is the magnetic field strength, $z_m$ and $z_d$ are the length of the magnet and the distance of the detector from the exit of the magnet, respectively. In both cases two complementary technologies are used to enable the possibility of cross-calibration and alignment as well as assessment of systematic uncertainties. 

In the high electron-rate regions, a scintillation screen and a \cer detector are chosen, making up the electron detection system (EDS). The scintillator technology has an excellent position resolution and can reliably measure fluxes at intermediate values but it is susceptible to saturation behaviour at very high fluxes. The segmentation of the \cer detector is limited by the size of individual channels but it has an excellent linearity up to the highest fluxes, and also superior rejection against low energy particles.  

The positron detection system (PDS) consists of a silicon pixel tracker and a silicon-tungsten sandwich calorimeter. The tracker has an excellent position resolution and with its ability to reconstruct charged particle trajectories can reconstruct their origin, providing powerful rejection against spurious signals from background sources. The calorimeter suppresses low-energy electrons via the shower shape information and can correlate the calorimetric energy measurement with the position to achieve background rejection. The tracker and calorimeter can also be used in combination by correlating tracks and clusters to further reject backgrounds. 

In the high-flux photon region, a photon detection system (GDS) is designed based on three complementary technologies: a Gamma Ray Spectrometer (GRS) uses converted photons to determine the photon flux as function of their energy based on measuring electrons and positrons, a Gamma Profiler (GP) measures the spatial profile of the photons, and a back-scattering calorimeter determines the overall photon flux (Gamma Flux monitor, GFM). 
The PDS is followed by a photon beam dump, and behind that dump a BSM detector is envisaged to search for scalar and pseudo-scalar particles predicted by BSM models. 

In the following sections, the design and performance of each detector is discussed. 
We discuss first the electron detection system, then the positron detection system and finally the photon detection system. Next a description of the planned DAQ architecture is provided. Then the changes necessary to adapt the layout for the \glaser data taking are explained before the ideas for detectors for the purpose of probing BSM physics are discussed. 

\subsection{Electron Detection System: \cer and Scintillation Detectors}
\label{sec:scintDetect}

Due to the high charged particle fluxes created in the beam-target interactions, of the order of $10^4-10^9~e^-$ (see Sec.~\ref{fig:rates_elaser}), the detector technologies chosen for the EDS are gas \cer detectors in combination with a scintillation screen. 
The use of both \cer and scintillation screen detectors is motivated by the complementarity of the two technologies. The scintillation screen offers a simple (and inexpensive) setup with a high resolution in position. The \cer device, for an appropriate choice of refractive medium, will offer a greater resistance to low-energy non-signal particles, and also will be more robust to the challenges of high particle flux. As a result, the scintillation screen is used mainly to measure the lower-flux (and lower electron energy) parts of the spectrum.

\begin{figure}[htbp]
\centering
\includegraphics[width=0.6\textwidth]{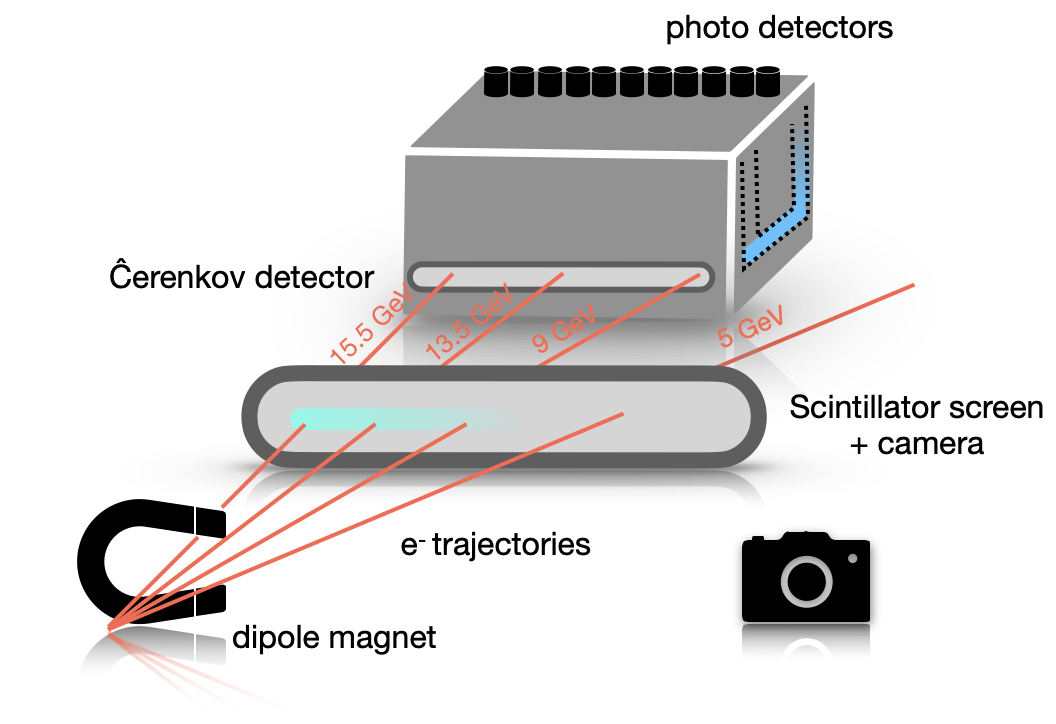}
\caption{ A visualisation of the set-up of the electron detection system. Electrons from the IP, deflected by a dipole magnet, interact first with a thin screen of scintillating material and then the \cer detector. The light profile from the screen is imaged remotely with an optical camera, and light produced in the \cer medium is reflected towards an array of photodetectors.}
\label{fig:ceds-diagram}
\end{figure}

A schematic drawing of the EDS is presented in Fig.~\ref{fig:ceds-diagram}. The \cer and scintillator detectors are placed behind each other, displaced from the nominal \beamline by about $5 \units{cm}$. The \cer detector is $15 \units{cm}$ wide while the scintillator screen is $50 \units{cm}$ wide. 

The presence of a thin screen with thickness $0.5 \units{mm}$ in front of the \cer detector will not affect the spectrum of electrons seen by the latter as the loss of the electron  velocity $\beta$ is negligible. 

\begin{figure}[htbp]
    \centering
      \includegraphics[width=0.4\textwidth]{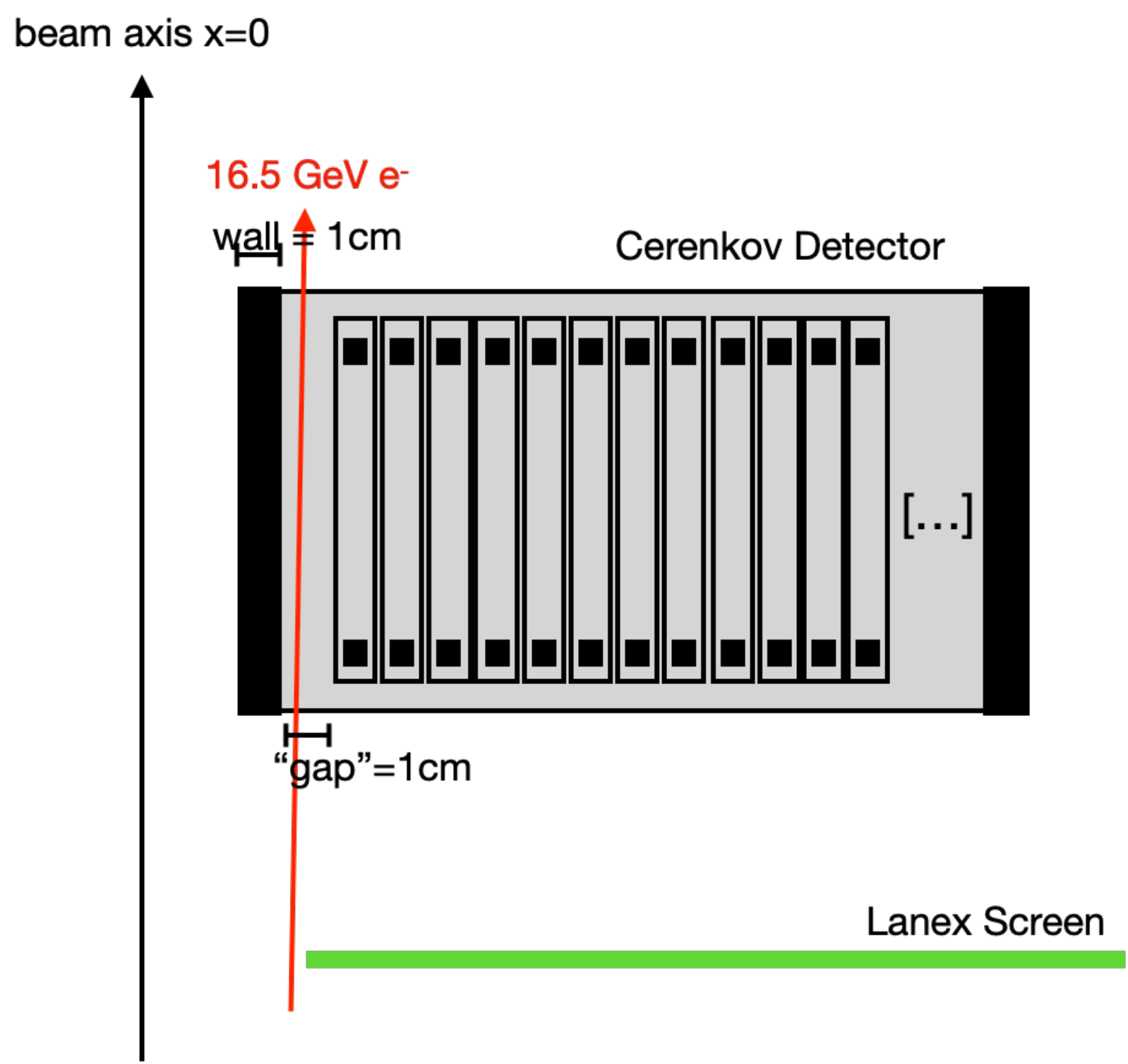}
      \caption{Sketch of the electron detection system: shown are the locations of the \cer detector and the scintillation screen with respect to the nominal beam axis. The path of beam electrons is also shown. The scintillation screen extends further to the right.}
      \label{fig:cere:positionbeam}
 \end{figure}

The aim for the EDS is to measure precisely the Compton edge position at low $\xi$, but also more generally to measure the electron energy spectrum at all $\xi$ values. 
At high $\xi$ values, the most important system is the tracker on the positron side, which measures the rate of positrons from the trident process. Since the energies of the trident positrons are of the order of a few GeV, the optimal magnetic dipole field to bend the positrons into the tracker acceptance is $B=1 \units{T}$. However, to resolve the Compton edges in the electron spectrum at high energy, a magnetic field of $B=2 \units{T}$ is optimal. 
To enable running with different magnetic field settings, it is necessary to make the electron arm of the post-IP spectrometer movable by about $5\units{cm}$ in $x$. This is to ensure that when  the magnetic field is changed, the $16.5\units{GeV}$ beam electrons do not impinge directly on the scintillation screen and the aluminium wall of the \cer gas-tight box, which would result in a significant beam background. The displacement of the detector is chosen such that the beam misses the screen but still passes through the beam window of the \cer detector, as shown in Fig.~\ref{fig:cere:positionbeam}. A displacement by $5\units{cm}$ fits within the dynamic range of the stages foreseen for the post-IP detector system. 
The energy ranges the scintillation screen and the \cer detector are sensitive to, for $B$-field values of 1\,T and 2\,T, are given in Table~\ref{tab:ip-el-energy-acceptance}. 

The low-energy part of the spectrum is important for measuring spin-dependent effects as discussed in Sec.~\ref{sec:theory:spin}. The scintillator covers the entire energy spectrum but the \cer detector only covers energies above 5-8~GeV. At lower energies, it is not critical to have such a high resolution, and it is under consideration to use 5 \cer detectors with 1~cm diameter to determine the electron flux at low energies. These would be the same as those proposed for the bremsstrahlung measurement for the \glaser setup, see Sec.~\ref{sec:detectors:glaser}.

\begin{table}[htbp]
\centering
\begin{tabular}{|c|c c|c c|}
\hline
\hline
   & \multicolumn{2}{c|}{Scintillator} & \multicolumn{2}{c|}{\cer detector}\\
Energy & $B=1$\,T & $B=2$\,T & $B=1$\,T & $B=2$\,T \\
\hline
$E_{max}$ (GeV) & $15.6$ & $16.0$ & $15.5$ & $15.5$\\ \hline
$E_{min}$ (GeV) & $1.1$ & $2.1$ & $5.7$ & $8.4$\\ \hline
\hline
\end{tabular}
\caption{Energy ranges within the acceptance of the scintillation screen and \cer detectors in the electron IP measurement.}
\label{tab:ip-el-energy-acceptance}
\end{table}

\subsection{The \cer Detector}
\label{sec:detectors:cerdet}

The \cer detector can follow the design developed for polarimetry measurements at future lepton colliders, of which a two-channel prototype has been built and successfully operated in test-beam~\cite{Bartels:2010eb}, and a calibration system for this detector has been described in Ref.~\cite{Vormwald:2015hla}.
With the prototype and its calibration system, a linearity up to a few per mille has been demonstrated over a dynamic range spanning a factor of $10^3$. As active medium a gas with a refractive index $n$ very close to unity (e.g.\ Argon with $n=1.00028$) has the advantage of being very robust against backgrounds from low-energy charged particles due to the \cer threshold of the order of $20 \units{MeV}$. If required by radiation conditions, the gas can be exchanged frequently or even cycled continuously.

\begin{figure}[htbp] 
\begin{subfigure}{0.45\hsize} 
\includegraphics[width=\textwidth]{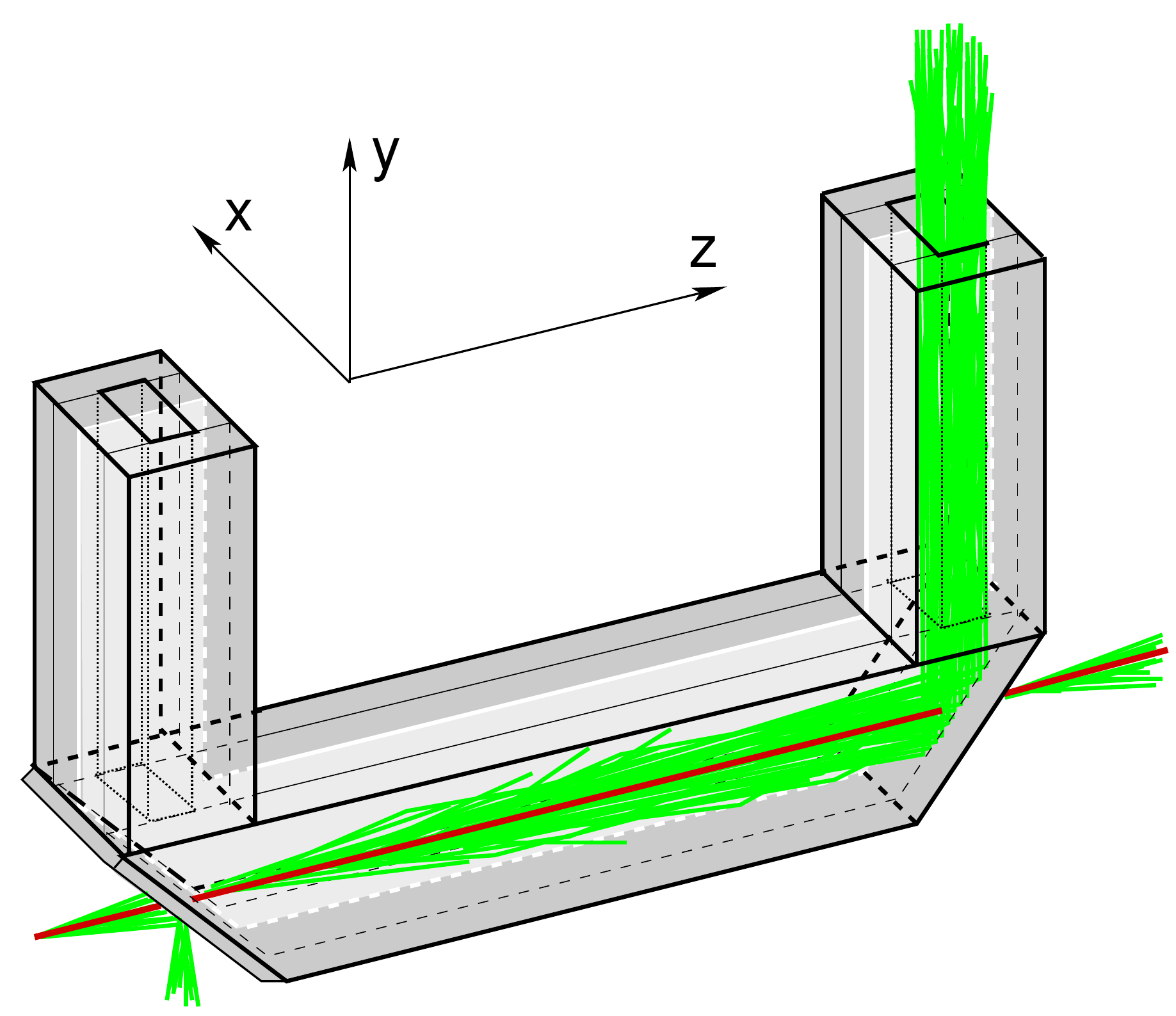}
\caption{ \label{fig:cerdet:simulation} \geant simulation of \cer light created by an electron passing through the detector.}
\end{subfigure}
\hspace{0.6cm}
\begin{subfigure}{0.48\hsize} 
\includegraphics[width=\textwidth]{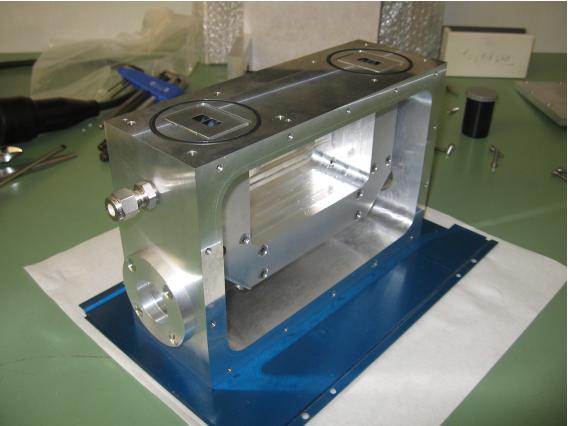}
\caption{  \label{fig:cerdet:assembly} Picture taken during the assembly of the prototype.}
\end{subfigure}
\caption{Two-channel prototype of a gas \cer detector developed for Compton polarimetry~\cite{Bartels:2010eb}, which serves as basis for the \cer detectors for LUXE.}
\label{fig:cerdet}
\end{figure}

Figure~\ref{fig:cerdet:simulation} shows the two-channel prototype in a \geant simulation with an electron passing through the horizontal part of the gas-filled ``U''-shaped structure. A mirror in the corner reflects the \cer light upwards to the photodetectors (PDs), which are placed outside of the plane of the main beam and of the dipole-magnet in order to minimise their radiation load. A picture taken during the assembly of the prototype is shown in Fig.~\ref{fig:cerdet:assembly}. The channels have a square cross section of about $1\times1 \units{cm^2}$. This cross section can be increased by a factor of about two to match the granularity required for LUXE, and the surface of the selected PDs. While in principle a single-anode PD would be sufficient, it has been shown in the test-beam campaign that PDs with a segmented anode can be very useful in order to align the detector~\cite{Bartels:2010eb}.

The particle rates expected in the LUXE experiment are up to five orders of magnitude higher than those in polarimetry at an $e^+e^-$ collider. To prevent a saturation of the readout electronics due to high signal charge of the PD, 
the following adjustments are made to reduce the signal per primary electron:
\begin{itemize}
    \item{\textbf{Active medium:} The number of \cer photons per primary electron, $N_\gamma$, scales as a function of the refractive index of the active medium as $1-1/n^2$. Choosing as active medium a gas with a very low refractive index (e.g.\ argon with $n=1.00028$) reduces $N_\gamma$ by a factor of 5 compared to $C_4F_{10}$ with $n=1.0014$. An additional advantage of argon is its slightly higher \cer threshold at $20 \units{MeV}$ which further reduces the low-energy background.} 
    \item{\textbf{Active medium length:} $N_\gamma$ scales linearly with the length $l_z$ of the active medium traversed by the primary electron. Reducing the length of the horizontal part of the gas-filled U-shaped structure to $1 \units{cm}$, compared to $15 \units{cm}$ in the polarimetry prototype reduces $N_\gamma$ by a factor of 15. } 
    \item{\textbf{Neutral density optical filter:} A neutral density optical filter inserted in front of the PD can further reduce the amount of \cer light. Neutral density filters have a flat transmission spectrum and are available for different transmission coefficients. For the \cer detectors in LUXE, filters with a transmission of $1\%$ and $0.1\%$, respectively, are foreseen.}
 \end{itemize} 
A combination of these adjustments leads to a total reduction of the signal charge by a factor of $7.5\times10^4$, which is sufficient to prevent the readout system saturation. 
The total \cer detector efficiency is commonly described by the parameter $\kappa$, which includes the photon-wavelength-dependent photodetector quantum efficiency, $QE(\lambda)$, the photodetector wavelength sensitivity range $[\lambda_{\text{min}},\lambda_{\text{max}}]$, the reflectivity, $\epsilon_{\text{refl}}$ of the inner channel walls, the channel length, $l_{z}$, and the refractive index of the \cer medium, $n$,
\begin{equation}
\label{eq:kappa}
\kappa=2\pi\alpha l_{z}\frac{1}{n^2} \epsilon_{\text{refl}} \int_{\lambda_{\text{min}}}^{\lambda_{\text{max}}} \frac{QE(\lambda)}{\lambda}\, d\lambda
\end{equation}
where $\alpha$ is the fine structure constant. For the chosen design it amounts to 0.076.


The design parameters of the \cer detector for the measurement of Compton electrons after the IP are summarised in Table~\ref{tab:cere:comptgeom}. The channel size is driven by the requirement to measure the Compton electron energy distribution with a precision sufficient to resolve its edges.
\begin{table}[htbp]
\begin{center}
\begin{tabular}{|l|c|l|}
\hline
\hline
  \textbf{Parameter}   & \textbf{Value} & \textbf{Comment}  \\
  \hline
    $N_{\text{channels}}$ $x$  & 50 & number of channels in $x$\\ 
    $N_{\text{channels}}$ $y$ & 1 & number of channels in $y$ \\  \hline
    $l_x$ & 1.5\,mm & channel length in $x$\\
    $l_y$ & 1.5\,mm & channel length in $y$\\
    $l_z$ & 10\,mm & channel length of active medium in z\\\hline
    Total detector volume & $150\times 150\times 100$~mm$^3$ &$x \times y\times z$, total size of the gas-tight volume\\\hline
    $\Delta x$ & 88~mm & displacement relative to beam axis \\\hline
    Active medium & Ar gas& \\\hline
    PD configuration & $(1.5\times1.5)$~mm$^2$ & 50 tiled SiPM \\ \hline
    PD gain & $7\cdot 10^5$ & \\\hline
    $T_{\text{ND}}$ & 0.001 & fractional transmission of ND-filter \\ \hline
    $\kappa$ & 0.076 & measure of total efficiency, see Eq.~(\ref{eq:kappa})\\
    \hline\hline
    \end{tabular}
\caption{ \label{tab:cere:comptgeom} Geometry and parameters of the \cer detector for the Compton electron measurement. The $\Delta x$ value is the displacement of the first sensitive channel relative to the beam axis. The gas tight volume is larger than the volume of the channels.}
\end{center}
\end{table}

\begin{figure}[htbp] 
\begin{subfigure}{0.48\hsize} 
\includegraphics[width=\textwidth]{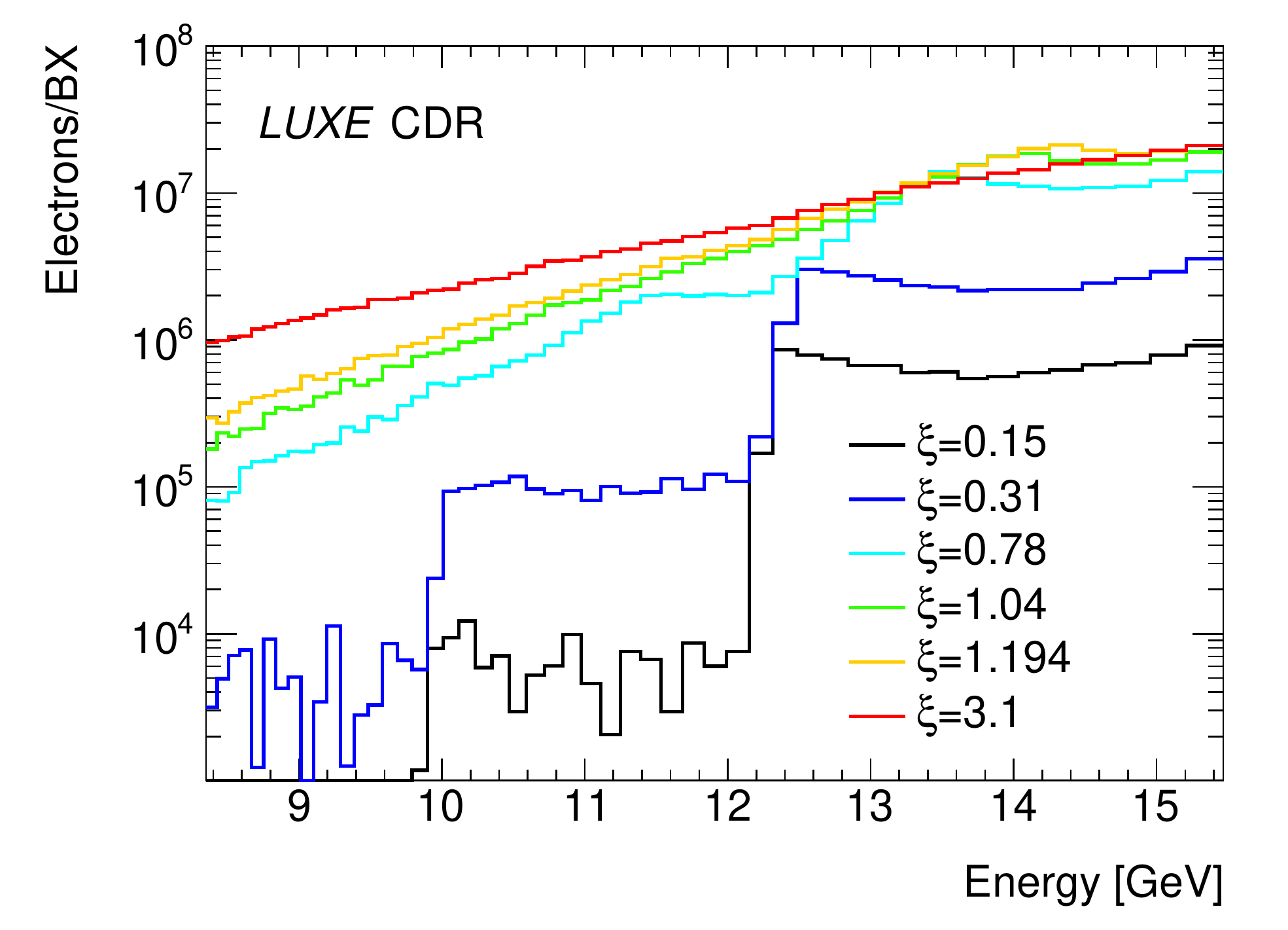}
\caption{ }
\end{subfigure}
\begin{subfigure}{0.48\hsize} 
\includegraphics[width=\textwidth]{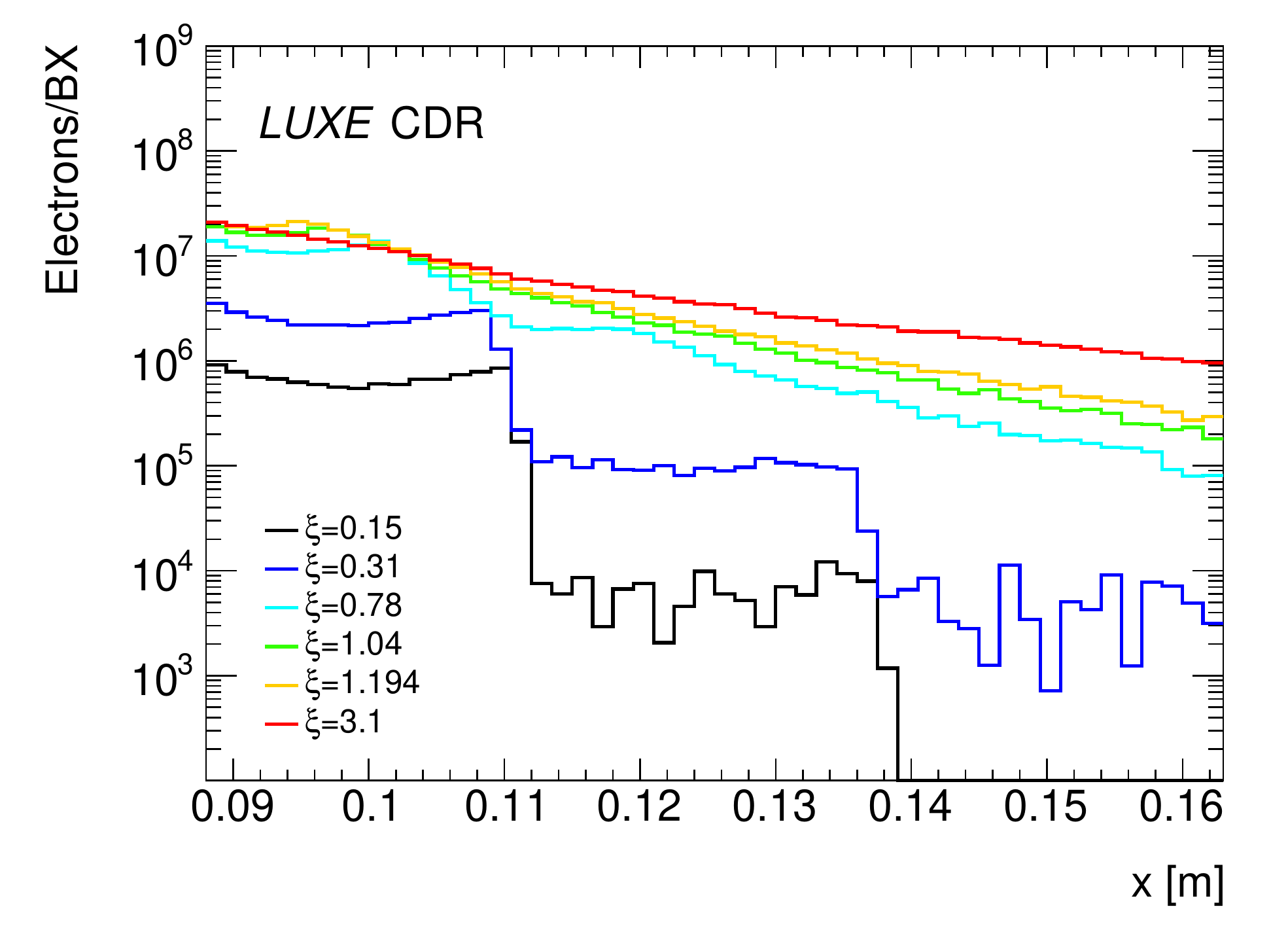}
\caption{ }
\end{subfigure}
\caption{
a) Simulated energy distribution of electrons after the $e^-+\gamma_L$ interaction point. b) Zoom of the simulated distribution of the electron position $x$ in the detector plane $1.69\,$m after the magnet. The distribution is cut off at $x=0.165\,$m to show the Compton edge region. The propagation through the magnetic field is done using fast simulation.}
\label{fig:ecompt}
\end{figure}

Figure \ref{fig:ecompt}a) shows the energy distribution of electrons after the IP for different laser intensities. The most important features of interest are the positions of the Compton edges which are expected to change with $\xi$ (see Sec.~\ref{sec:science}).

The other important aspect is to ensure that the number of particles can be counted reliably, and this requirement drives the choice of gas, filter etc. 

A zoom of the electron spatial distribution in the \cer detector plane is shown in Fig.~\ref{fig:ecompt}b). The \cer detector position after the IP is assumed to be $1.69\,$m after the magnet. It is visible 
that resolving the first-order Compton edge requires a compact detector with a sensitive area spanning $7.5\,$cm in $x$ and positioned $8.8\,$cm away from the beam axis. To resolve the edge position, a fine channel segmentation of $1.5$mm per channel in $x$ is necessary. The channel segmentation of the \cer detector is limited mainly by the photodetector granularity. However, silicon PMs with a surface of $1.5\times 1.5\units{mm^2}$ are commercially available, which makes a segmentation of the order of $1.5\units{mm}$ possible. Mechanically, the channel size is only limited by the thickness of the inter-channel foil, which in the available prototype is $0.3 \units{mm}$, which could be further reduced.

One important aspect for the positioning of the \cer detector is its placement with respect to the $16.5\units{GeV}$ electron beam trajectory. To reduce backgrounds from scattering of beam electrons in the \cer detector material, the amount of material crossed should be minimal. The main scattering was shown to originate from the aluminium side-wall of the gas-tight volume. To mitigate this, the \cer sensitive channels within the box should be placed according to Fig.~\ref{fig:cere:positionbeam}, leaving a gap between the wall of the gas-tight box for the beam to pass through. The beam electrons then only pass the thin entrance and exit volumes, avoiding the thicker outer wall. 

\subsubsection{Integration and Readout}
\label{sec:cer_integration_readout}

 \begin{figure}[htbp]
    \centering
      \includegraphics[width=0.7\textwidth]{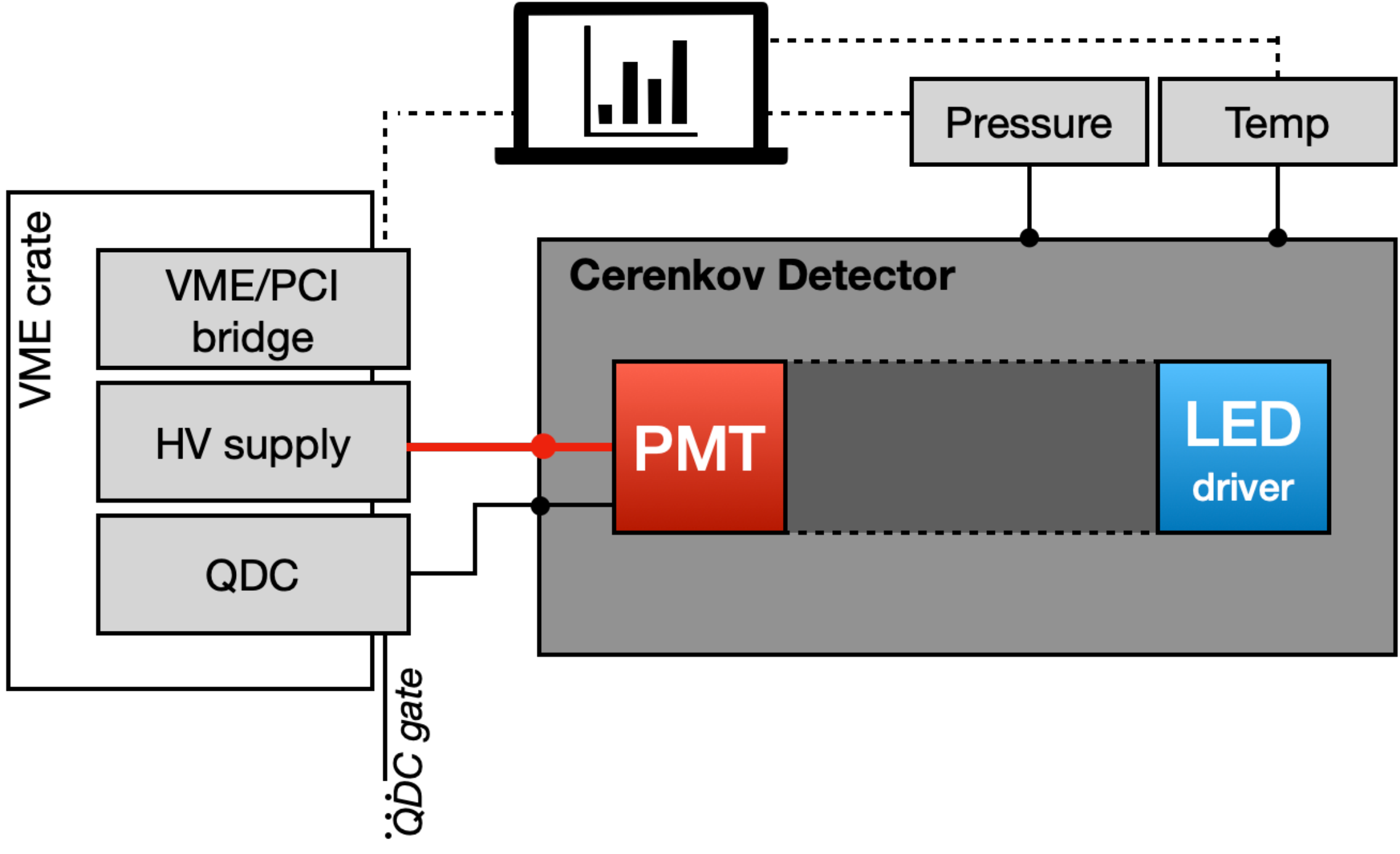}
      \caption{Readout schematics of a single \cer detector channel (channel view from the top). Solid (dashed) lines indicate analogue (digital) data flow.}
      \label{fig:cere:readout}
 \end{figure}

Figure~\ref{fig:cere:readout} shows a schematic diagram of the readout electronics developed for the \cer prototype described in \cite{Bartels:2010eb}. The same readout scheme can be used for \cer detectors in the LUXE experiment.

The core element of the \cer detector readout system is a CAEN V965 16-channel dual-range charge-to-digital converter (QDC), which integrates and digitises the PMT voltage signal from the individual detector channels. With a 12-bit resolution the QDC dynamic range is limited by $819.2 \units{pC}$ in the high range ($200\,$fC/channel) and $102.4 \units{pC}$ in the low range ($25\,$fC/channel). In the high electron rate environment of LUXE the QDC dynamic range strongly affects the detector design to avoid saturation. Using measures described at the beginning of Sec.~\ref{sec:detectors:cerdet} the amount of charge from the PMT is reduced sufficiently to comply with the QDC dynamic range.

The QDC readout is triggered by a NIM logic gate signal which can be synchronised with the beam clock for the bremsstrahlung photon flux \cer system, or the LASER clock for the Compton \cer system. The digitised QDC signal is transferred via a VME bridge with an optical link to a computer where it is processed by the central data acquisition (DAQ) system. A CAEN multi-channel high voltage power supply for the photodetectors is steered from the detector control software via the VME bridge. Furthermore, one pressure sensor per gas volume and several temperature sensors provide data to the detector monitoring system. 

One important property that has to be quantified for the \cer counting detector is the linearity of both the photodetector and the QDC. For the available \cer prototype, extensive studies measuring the differential and integrated non-linearity of the QDC and different photodetectors were performed~\cite{Vormwald:2015hla}, where in a larger light-intensity regime, non-linearities of $(1.0 \pm 0.1)\%$ were found for the combined system of PMT and QDC. An analogous LED-based calibration system to the one described in~\cite{Vormwald:2015hla} can be used for in-situ determination of the detector linearity in LUXE in breaks between data taking.

\subsubsection{Performance}
The objective for the \cer detectors is to measure the Compton edge position as a function of the laser intensity parameter $\xi$. It is expected that the Compton edge is shifted to larger values with a functional behaviour of
\begin{equation}
\label{eq:edgexi}
 \frac{E_\text{edge}(\xi)}{E_\text{beam}}=1-\frac{2\eta}{2\eta+1+\xi^2} \,
\end{equation}
where $\eta(16.5 \units{GeV}) \approx 0.192$.

The performance of the LUXE \cer detector is estimated using a fast detector simulation based on a package developed for polarimetry \cite{Eyser:2007ks}. It uses as input the particles from the {\sc IPstrong} MC (see Sec.~\ref{sec:simulation}) or from beam-target interaction simulation in \geant (see Sec.~\ref{sim:geant4}). The package emulates the propagation of charged particles through a uniform magnetic dipole field, as well as the production of \cer photons in the medium. The package furthermore emulates the photodetection and the QDC digitisation step.

\begin{figure}[htbp] 
\begin{subfigure}{0.48\hsize} 
\includegraphics[width=\textwidth]{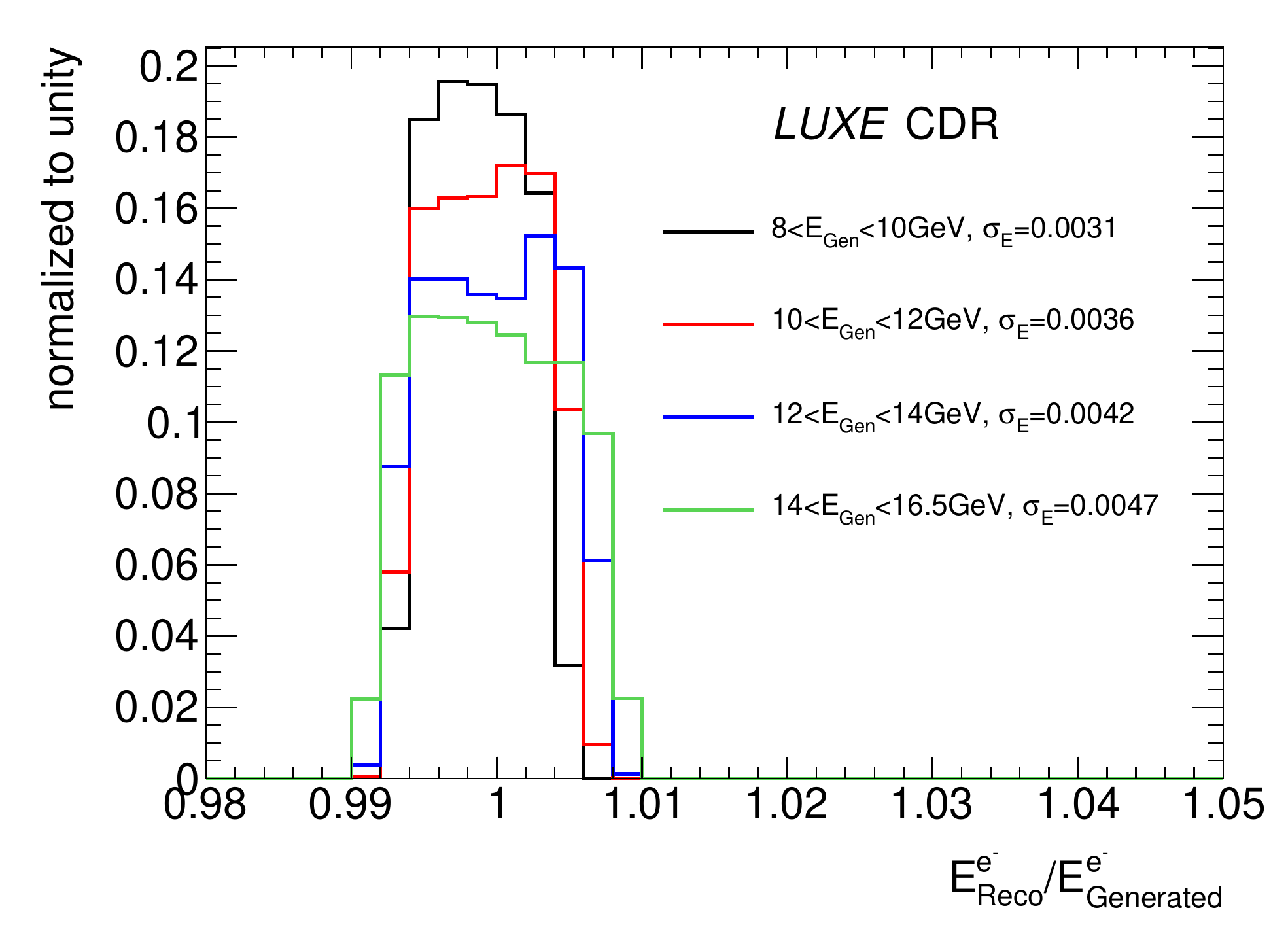}
\caption{ }
\end{subfigure}
\begin{subfigure}{0.48\hsize} 
\includegraphics[width=\textwidth]{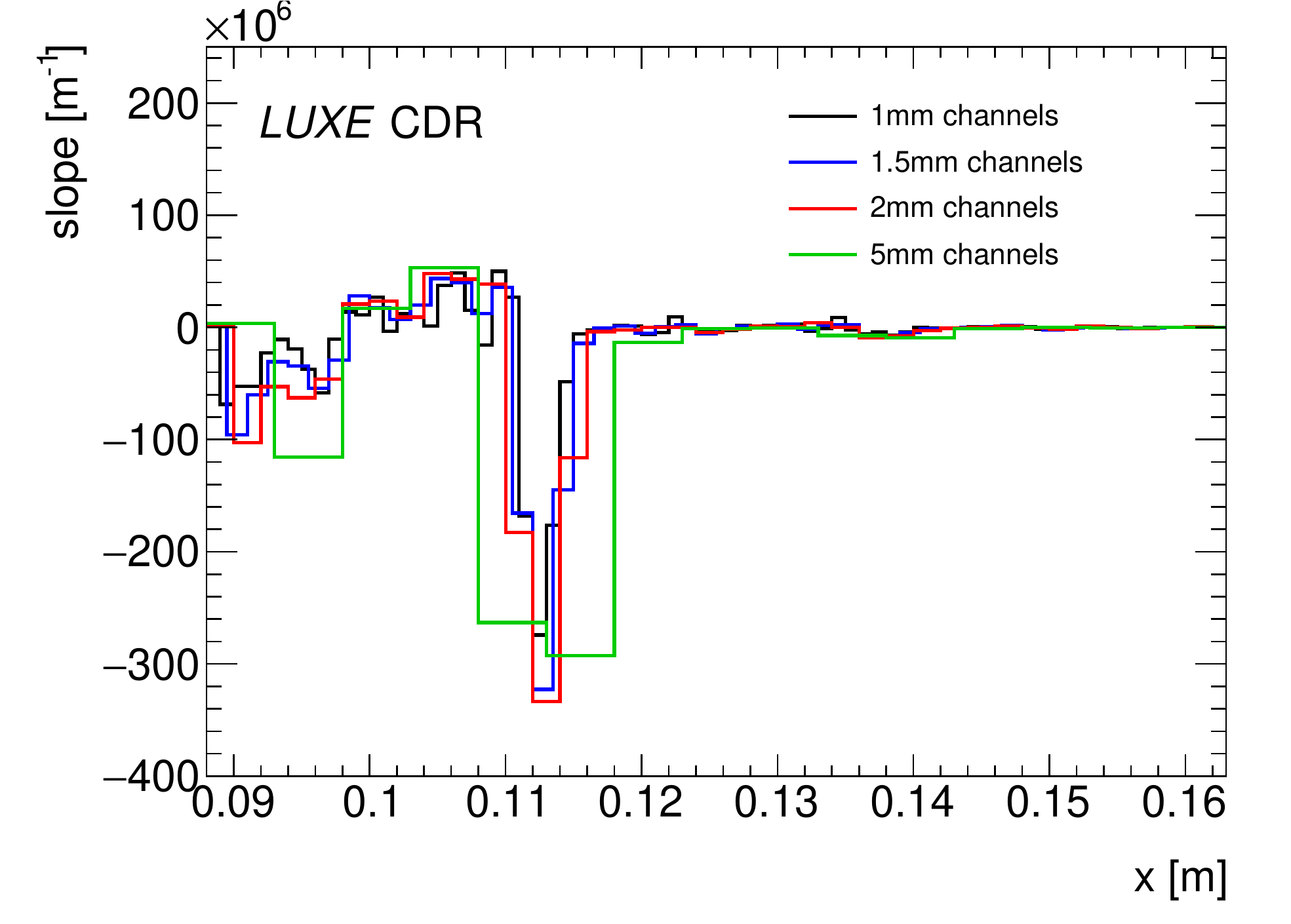}
\caption{ }
\end{subfigure}
\caption{Performance estimation of the \cer detector. a) Ratio of the electron energy reconstructed in the \cer detector and the generated energy, for different generated electron energy ranges. b) Bin-by-bin slope of the simulated true electron distribution in the detector plane with different binning corresponding to different \cer channel segmentation. The minima in the slope parameter correspond to the Compton edge position. }
\label{fig:cerdet:comptperf}
\end{figure}

Figure \ref{fig:cerdet:comptperf}a) shows 
the ratio of the reconstructed electron energy divided by the generated energy, for different ranges of the generated electron energy for the \cer detector with $1.5\units{mm}$ channel segmentation. The resolution varies from $0.3\%$ for the low-energy regime to $0.5\%$ in the high-energy regime. 

Figure \ref{fig:cerdet:comptperf}b) shows the slope between bin contents of the $x$ distribution for the Compton electrons, calculated bin-by-bin. Local minima of the slope (i.e.\ the steepest local drop in the Compton spectrum) are used as an estimate for the Compton edge positions. It is clearly visible that the first Compton edge can be resolved well with a \cer detector channel segmentation of  $1\units{mm}$, and the second-order Compton edge is also visible. Finally, Fig.~\ref{fig:cerdet:comptperf}b) shows that for coarser channel segmentation the local minimum in the slope gets broader, and therefore the uncertainty on the edge position would be larger. This leads to less separation between the edge positions for different laser intensity configurations. A segmentation of $1.5\units{mm}$ is small enough so that the Compton edge can be measured well, and is technically possible. More details on the results of the Compton edge analyses are presented in Sec.~\ref{sec:results}. 

\subsection{Scintillation Screen and Camera}
\label{sec:det:scintillator}
Scintillation screens, in comparison to the segmented approach of the \cer device, offer a measurement of the same electrons with a superior position resolution. The large flux of ionising radiation expected in the Compton process will induce enough light in a scintillating material to be imaged by a remote camera, while the electrons will pass through a thin screen mostly unperturbed on their path to the \cer detector. It is expected that the screen will be sensitive in the entire flux range relevant here ($\sim 10^4-10^8$ particles). At high fluxes it was demonstrated by the AWAKE experiment that it has a linear response up to charges of 350\,pC~\cite{BAUCHE2019103}, i.e.\ also adequate for the primary electron beam.

There is no practical difficulty or high cost associated with extending the range in the $x$ direction, so in contrast to the initial design of the \cer detector in this region, with width $7.5 \units{cm}$, the scintillation screen is engineered to cover the full possible energy spectrum of electrons exiting the magnetic field (which is expected to be constrained by the size of the aperture of the preceding magnet). In Table~\ref{tab:scintcamera} these dimensions and key parameters are listed.

\begin{table}[htbp]
\centering
\begin{tabular}{|l|c|l|}
\hline
\hline
Parameter & Default Value & Comments \\
\hline
\textbf{Scintillation screen} & & \\
Length ($x$) & $500$~mm & Length in direction of deflection of electrons.\\ 
Height ($y$) & $100$~mm & Height orthogonal to the deflection plane. \\ 
Thickness & $0.5$~mm & Thickness of screen. \\ \hline
\textbf{Camera} & & \\
Nominal position sensitivity & $125 \units{\mu m}$  & Nominal distance covered by one pixel. \\ 
Estimated final position sensitivity & $500 \units{\mu m}$ & Estimate for final effective position resolution. \\ 
Dynamic range & $39.7 \units{dB}$ & Above dark current available from a suitable camera. \\ \hline
\hline
\end{tabular}
\caption{Key parameters of the scintillation screen and the camera used in the electron detection system.}
\label{tab:scintcamera}
\end{table}

The mechanism of scintillation is complex and specific to the material used. Empirically it is consistent with the picture in which sufficiently energetic radiation which deposits energy within the material induces fluorescent photon emission. These photons are released in a relatively isotropic manner allowing flexibility in the direction of the optical path to a camera. The light emission is also at a fairly monochromatic wavelength and is released over some short time after the deposition of energy.

A scintillation material such as Terbium-doped Gadolinium Oxysulfide, $\mathrm{Gd}_2\mathrm{O}_2\mathrm{S:Tb}$, (GadOx) is envisioned for the screen. Analysis of the material's properties ensures it is suitable for use in this detector: 
\begin{itemize}
  \item \textbf{Emitted light wavelength profile:} Best chosen to be within the optical wavelength band of non-specialised optical equipment (mirrors, cameras). The characteristic light emission wavelength of GadOx doped with Terbium is around $545\units{nm}$, which easily fulfils this requirement.  
  \item \textbf{Light emission time profile:} The maximum bunch frequency expected at LUXE is $10 \units{Hz}$. The relatively long emission evolution of GadOx compared to most common materials is still comfortably within $0.1\units{s}$.
  \item \textbf{Scintillation photon yield:} Greater light output gives more statistical significance in general and above the ambient light background, and GadOx offers some of the highest outputs in terms of the number of photons for energy deposited \cite{MORLOTTI1997772}.
  \item \textbf{Shape:} Given the density and corresponding radiation length of GadOx, and the requirement that the screen perturb the electron energy spectrum as little as possible before measurement by the \cer device, the screen must be adequately thin. Use of this material in thin screens in industry applications \cite{Bakerhughesds} show that this is possible.
  \item \textbf{Radiation hardness:} GadOx screens are typically referred to as radiation hard, which is important due to the high-flux region(s) at LUXE. 
  A material's response after deposition of energy from ionising radiation is always specific to the particle type and kinematics of the radiation, as well as to the flux and frequency of exposures. This makes it difficult to quantify the radiation hardness with absolute confidence. The material's response to electrons of order $10 \units{keV}$ in the context of electron microscopy has been studied \cite{ScintiMaxdatasheet} and a threshold of noticeable change in the scintillation mechanism as a result of radiation damage is calculated in terms of dose of around $10^8 \units{Gy}$.
\end{itemize}

\begin{figure}[htbp] 
\centering
\includegraphics[width=0.4\textwidth]{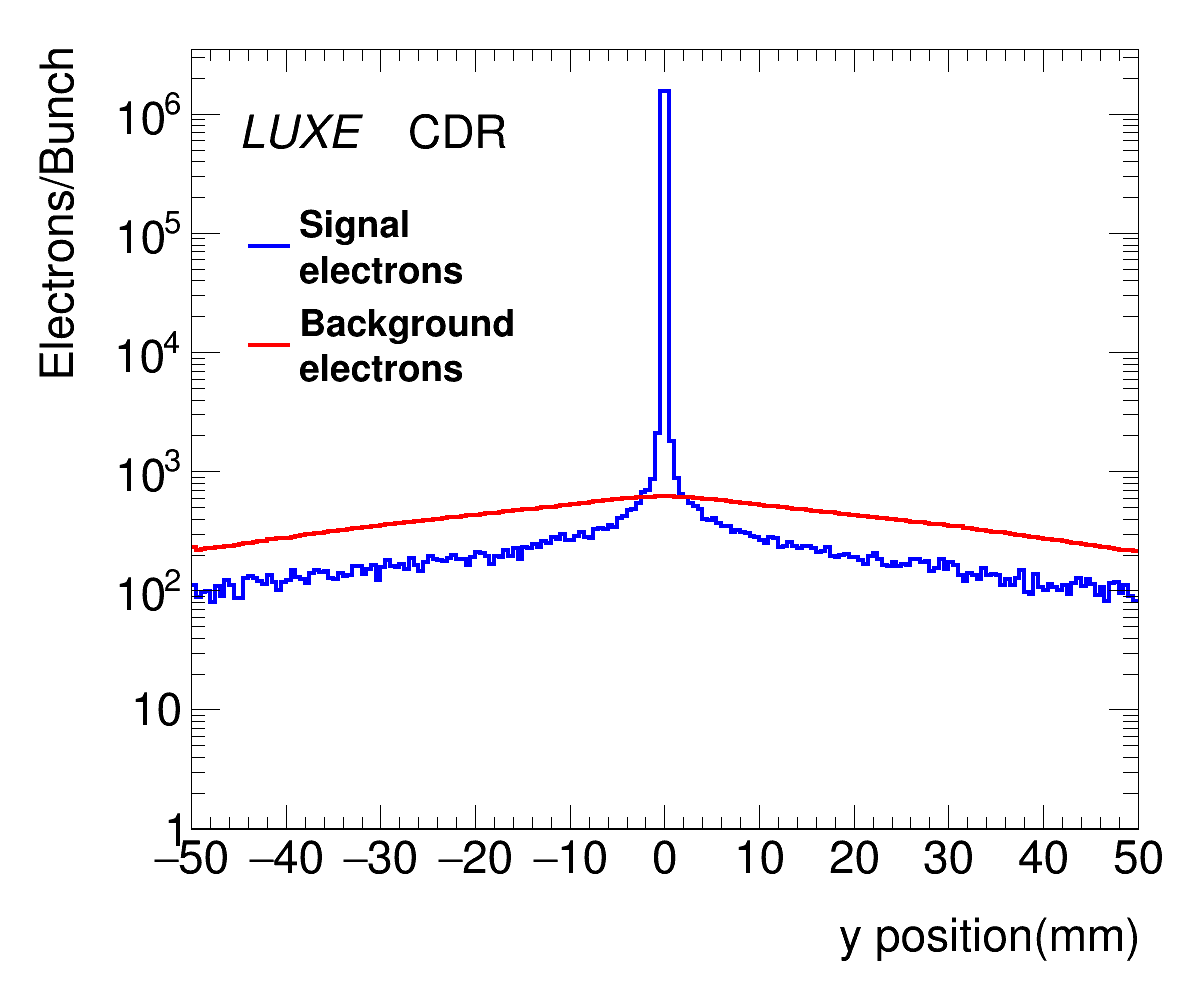}
\caption{Distribution of electrons in $y$ for the signal and background in the scintillation screen for a sample interaction with $\ximax = 3.1$. The background is that determined from beam-only simulation. The signal measurement includes secondaries from Compton-scattered electrons within the screen's acceptance, which can be corrected for as part of signal measurement. 
}
\label{fig:lanex_sigbkg}
\end{figure}

From the Bethe--Bloch formula for energy deposition over distance travelled in matter, the magnitude of signal response in this detector is related to the velocity of the particle, $\beta$, 
\begin{equation}
\frac{dE}{dx} \approx \frac{n e^6}{4 \pi m_e c^2 \beta^2 \epsilon_0^2}\left[ \ln\left( \frac{2 m_e c^2 \beta^2}{I(1-\beta^2)} \right) - \beta^2 \right] ,
\end{equation}
for electron number density $n$ and mean excitation potential $I$, both constants specific to a material. 
Like in the case of the \cer detector, the response of the screen is independent of the energy of the electrons, and thus the detection efficiency is also independent of the incident energy. The energy is solely determined by where the particle hits the screen, i.e. by the spatial distribution of the scintillation light observed by the camera.
Unlike the \cer device, the scintillator is rather sensitive to low-energy particles as in this regime $\frac{dE}{dx} \propto 1/\beta^2$. The background can, however, be determined in-situ based via two methods. The first method uses the $9\units{Hz}$ of electron-beam data where there is no laser shot. The second method uses the data from the side-bands of the scintillation screen: the scintillation screen covers $10\units{cm}$ in the vertical plane and the signal is confined to just the most central $1\units{cm}$ while the background is rather flat in that plane, see Fig.~\ref{fig:lanex_sigbkg}.

In order to protect the camera(s) from damage from high-energy background it is placed within lead shielding and if necessary optical mirrors will be used to reflect the light from the screen to camera, ensuring that the open aperture is not directly facing the \beamline. 

\subsubsection{Scintillator Data Acquisition}
An optical camera and computer DAQ system is used here. The use of a chromatic filter, that allows only the appropriate wavelength of light to pass, is used to ensure high signal over ambient optical-light background. A standard pixel resolution of 2000 pixels covering the screen of $50$~cm in one direction can result in the nominal positional sensitivity of $250$~$\mu$m. By choosing more than one camera, the resolution can be further enhanced. As default, the screen covered by two cameras ($\Delta x = 125 \units{\mu m}$) is used. With reference to appropriate previous detection efforts \cite{Keeble:2019} with a similar detector an estimate for final position resolution after accounting for systematic and optical effects is $500 \units{\mu m}$ -- this value is used for following analysis and reconstruction.  

With a reasonable choice of imaging camera \cite{Baslercamera}, more default parameters for the scintillation screen and camera system are given in Table \ref{tab:scintcamera}.

\subsubsection{Performance}

Reconstruction of the photon energy spectrum involves correlating the electron energy to the position in the scintillation screen in order to relate the light output to a given energy interval. 

\begin{figure}[htbp] 
\begin{subfigure}{0.49\hsize} 
\includegraphics[width=\textwidth]{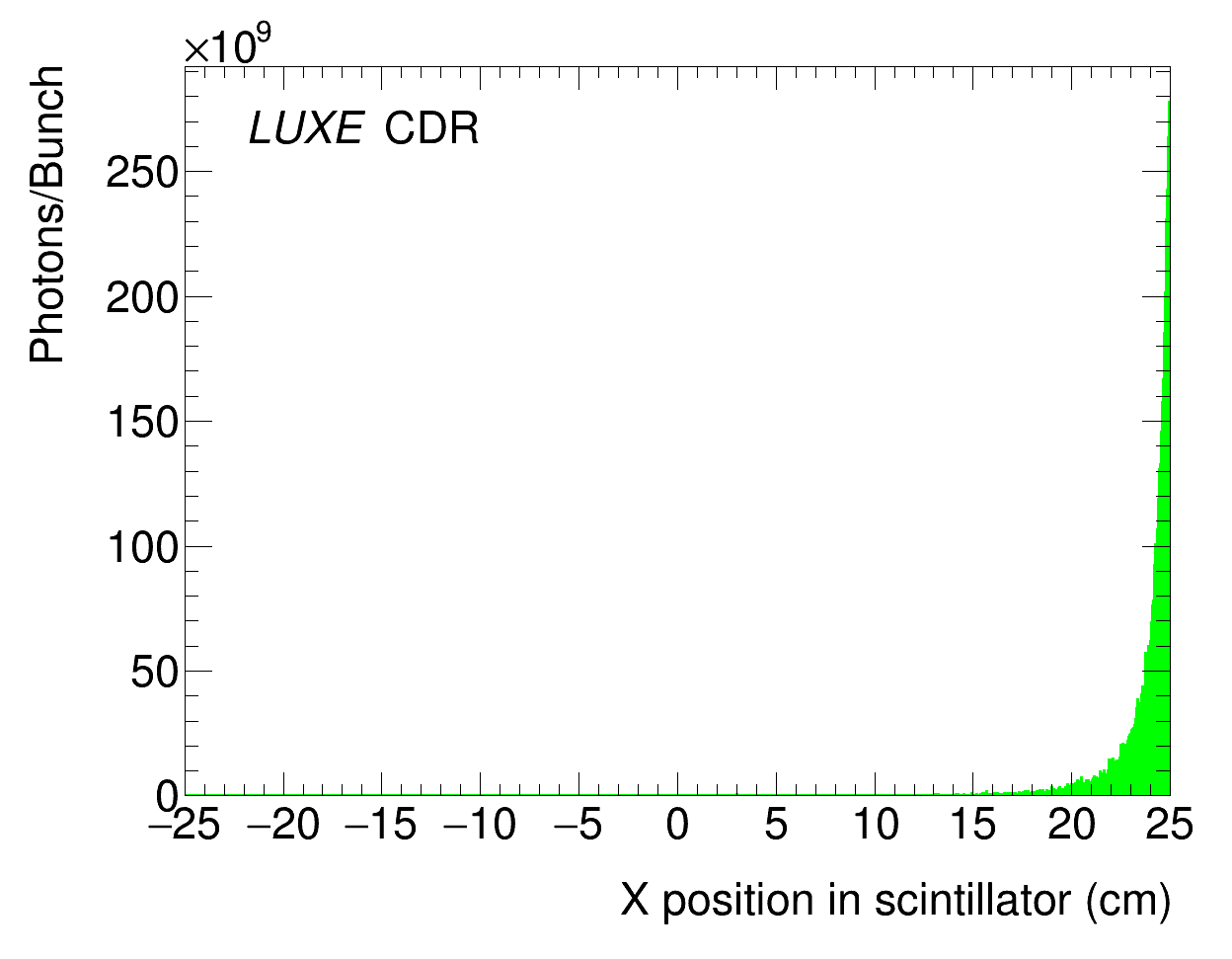}
\caption{\label{fig:scint-light}}
\label{fig:photon-before-IP-scint-light}
\end{subfigure}
\begin{subfigure}{0.49\hsize}
\includegraphics[width=\textwidth]{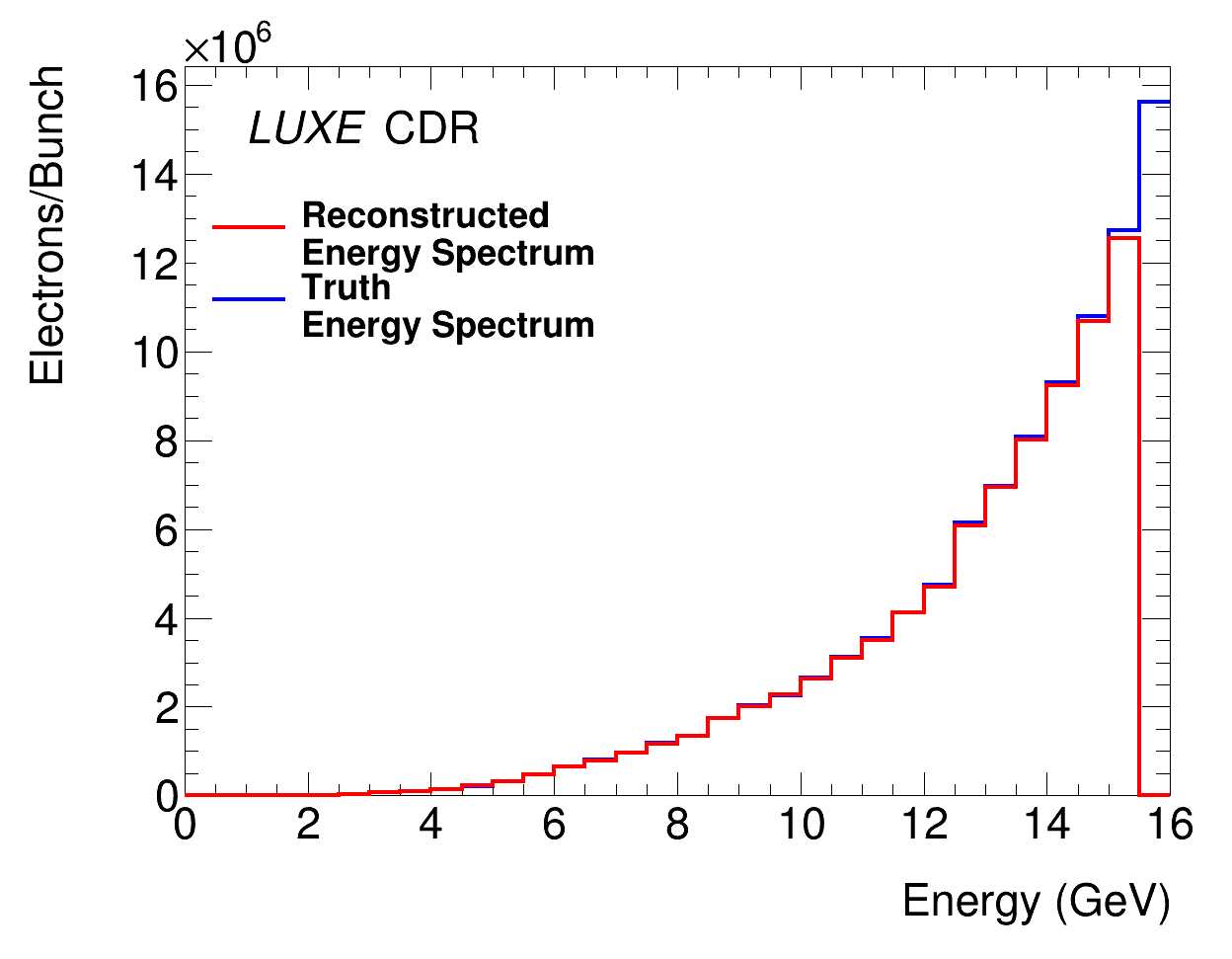}
\caption{\label{fig:scint-recon-electron} }
\end{subfigure}
\caption{\label{fig:scint-recon} (a) Expected number of scintillating photons emitted per bunch crossing as a function of the $x$ position in the scintillating screen, in local coordinates, as obtained from \geant. This uses MC simulation for a laser intensity parameter of $\ximax = 5.1$, propagated through \geant including deflection in a dipole magnetic field of $1 \units{T}$. (b) Comparison of the true input electron energy distribution to the reconstruction of this using Fig.~\ref{fig:scint-light}.
}
\end{figure}

It also requires an accurate calibration of the scintillation light response w.r.t. a known flux of electrons. This will be done via a dedicated test-beam measurement. The impact of the electrons travelling some distance through beam window material and surrounding air can also be implemented in a test-beam environment. 

The scintillation emission and detection are simulated with \geant to emulate the response of the screen to the Compton electrons. The distribution of photons created in the scintillation process are shown in Fig.~\ref{fig:scint-light} 
versus the position in the scintillation screen, in coordinates local to the centre of the screen. The reconstructed electron energy distribution and the comparison to the incident spectrum is presented in Fig.~\ref{fig:scint-recon-electron}.  The reconstructed spectrum matches the incident spectrum well at all energies; the small difference at highest energies is to be expected as this is at the edge of the detector (15.5\,GeV corresponds to a distance of 0.15\,cm from the screen's edge).

\subsection{Positron Detection System: Tracking Detector and Calorimeter}

The number of positrons per bunch crossing depends strongly on $\xi$ and the goal is to be sensitive to values between $10^{-3}$ and $10^4$.  Therefore rather powerful high-granularity detectors are deployed here to ensure that a small signal can be extracted and that at high multiplicity the number of particles can be determined reliably. 
Behind the vacuum chamber, four layers of silicon pixel detectors are staggered followed by a calorimeter. 

\subsection{Tracking Detector}
\label{sec:detectors_tracker}
The tracking detector has to reconstruct positrons with high efficiency while efficiently rejecting background from stray particles (see discussion in Sec.~\ref{sec:simulation}). 

The technology selected for the LUXE tracking detector is based on the ALPIDE~\footnote{ALPIDE stands for ``ALice PIxel DEtector''} silicon pixel sensor~\cite{Senyukov:2013se,Mager:2016yvj} developed for the upgrade of the inner tracking system (ITS) of the ALICE experiment at the LHC~\cite{Abelevetal:2014dna,ALICE-PUBLIC-2018-013}.
It is a monolithic active pixel sensor (MAPS) integrating the sensing volume and the readout circuitry in one cell.

The ALPIDE chip is a $30\times 15~{\rm mm}^2$ large MAPS containing approximately $5\cdot 10^5$ pixels, each measuring about  $27\times 29 \units{\mu m^2}$ arranged in 512 rows and 1024 columns that are read out in a binary hit/no-hit fashion~\cite{YANG201561}.
The chip's time resolution is $\sim 2 \units{\mu s}$.

The ALPIDE prototypes have been extensively tested in various laboratories as well as at a number of test-beam facilities in the past few years.
The chips have demonstrated very good performance both before and after irradiation. 

In~\cite{Mager:2016yvj}, the sensor has shown a detection efficiency above 99\%, a noise hit rate much below $10^{-5}$, a spatial resolution of around 
$\sim 5 \units{\mu m}$, while being able to tolerate an ionisation dose of up to $2.7 \units{Mrad}$~\cite{Senyukov:2013se}.
Though this level of irradiation is not relevant in the LUXE experiment, the high radiation tolerance goes hand in hand with good time response which is desirable for LUXE.

The basic unit of the detector is a $\sim 27$~cm long stave which is based on the following elements:
\begin{itemize}
\item \textbf{Space Frame}: a carbon fibre structure providing the mechanical support and the necessary stiffness;
\item \textbf{Cold Plate}: a sheet of high thermal-conductivity carbon fibre with embedded polyimide cooling pipes, which is integrated into the space frame and is in thermal contact with the pixel chips to remove the generated heat;
\item \textbf{Hybrid Integrated Circuit (HIC)}: an assembly consisting of a polyimide flexible printed circuit (FPC) onto which the pixel chips and some passive components are bonded.
\end{itemize}

\begin{figure}[!ht]
\centering
\begin{overpic}[width=0.49\textwidth]{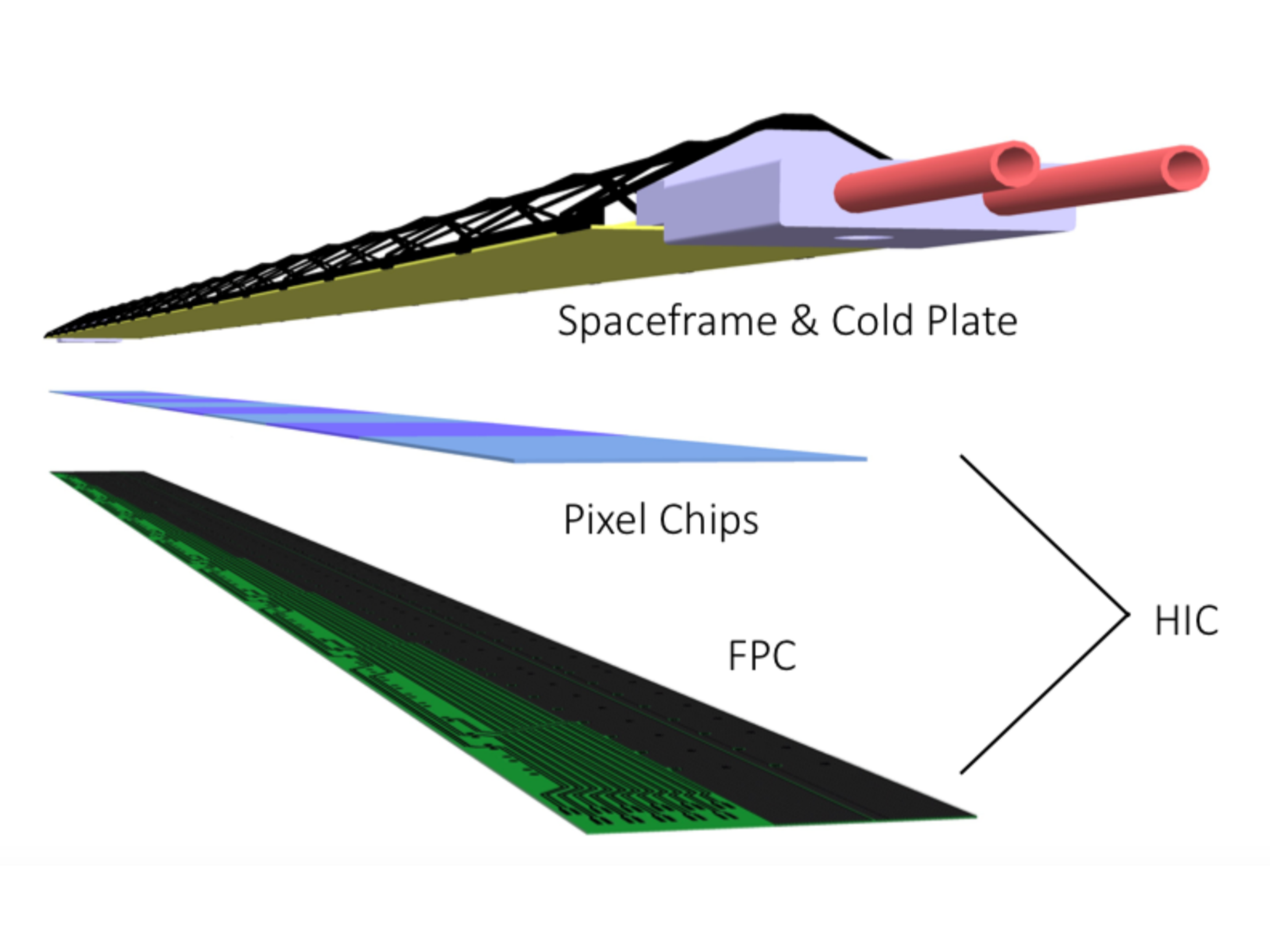}\end{overpic}
\begin{overpic}[width=0.49\textwidth]{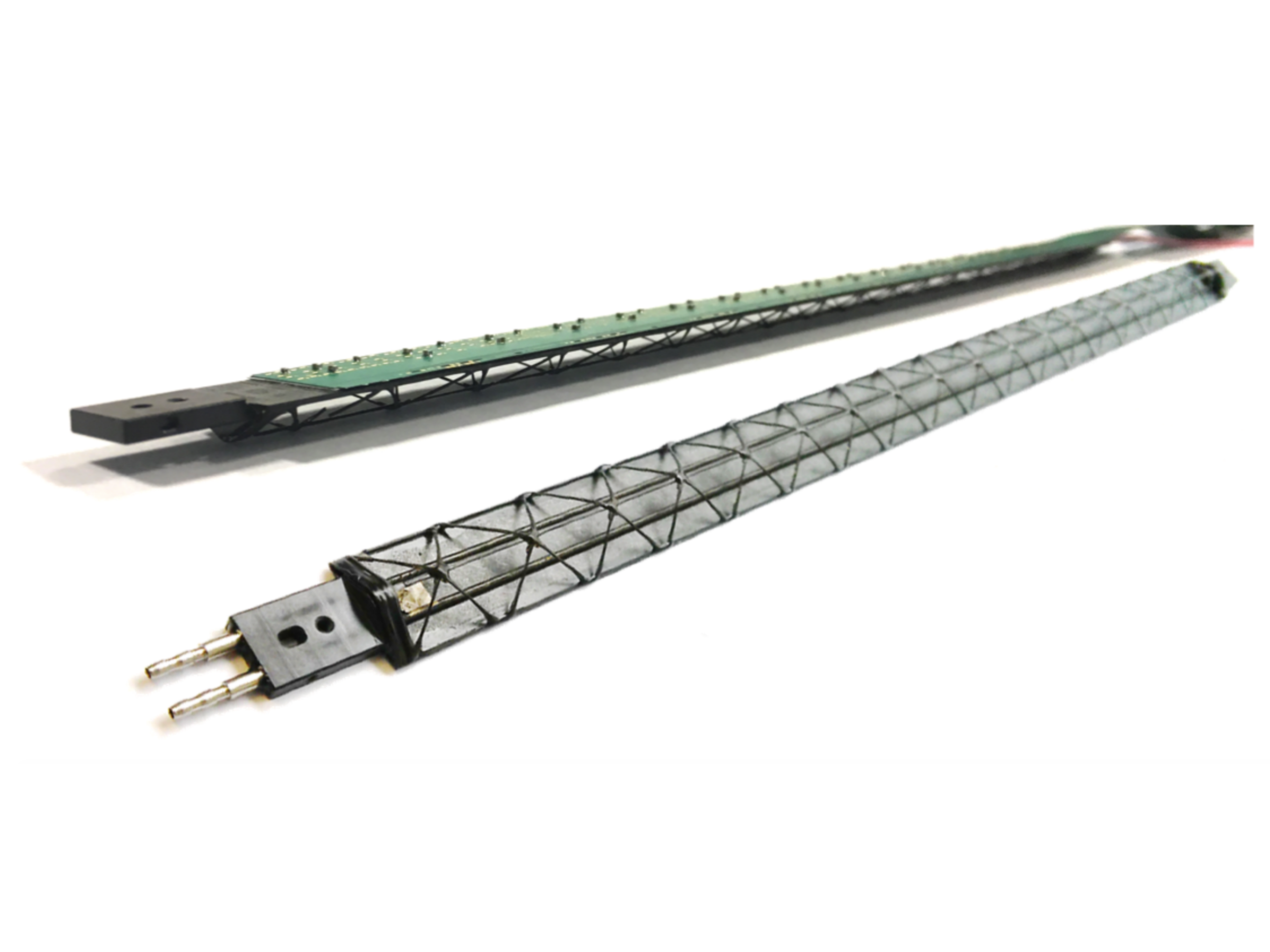}\end{overpic}
\caption{Left: schematic layout of a single stave. Nine pixel sensors are flip-chip mounted on a flexible printed circuit (FPC) to form a hybrid integrated circuit (HIC). 
Right: picture of a production sample of a single stave.
\label{fig:stave}}
\end{figure}

Each stave is instrumented with one HIC, which consists of nine pixel chips in a row connected to the FPC, covering an active area of $\sim 270.8\times 15 \units{mm^2}$ including $100 \units{\mu m}$ gaps between adjacent chips along the stave.
The interconnection between pixel chips and FPC is implemented via conventional aluminium wedge wire bonding.
The electrical substrate is the flexible printed circuit that distributes the supply and bias voltages as well as the data and control signals to the pixel sensors.
The HIC is glued to the ``cold plate'' with the pixel chips facing it in order to maximise the cooling efficiency.
The polyimide cooling pipes embedded in the cold plate have an inner diameter of 1.024~mm and a wall thickness of 25~$\mu$m.
These are filled with water during operation. A schematic drawing and a picture of the stave are shown in Fig.~\ref{fig:stave}.
An extension of the FPC, not shown in the figure, connects the stave to a patch panel that is served by the electrical services entering the detector only from one side. The average thickness of the stave is $X/X_0 = 0.357 \%$.

The tracker in LUXE will have four layers on the positron side of the \beamline. Each stave serves as a ``half-layer'' such that the total transverse length of one layer would be roughly $50~\units{cm}$. The two staves are placed such that they are slightly staggered (by a distance of $\sim 1 \units{cm}$)) and overlap in horizontal direction in the middle by $\sim 4\units{cm}$. 

\begin{figure}[!ht]
\centering
\begin{overpic}[width=0.49\textwidth]{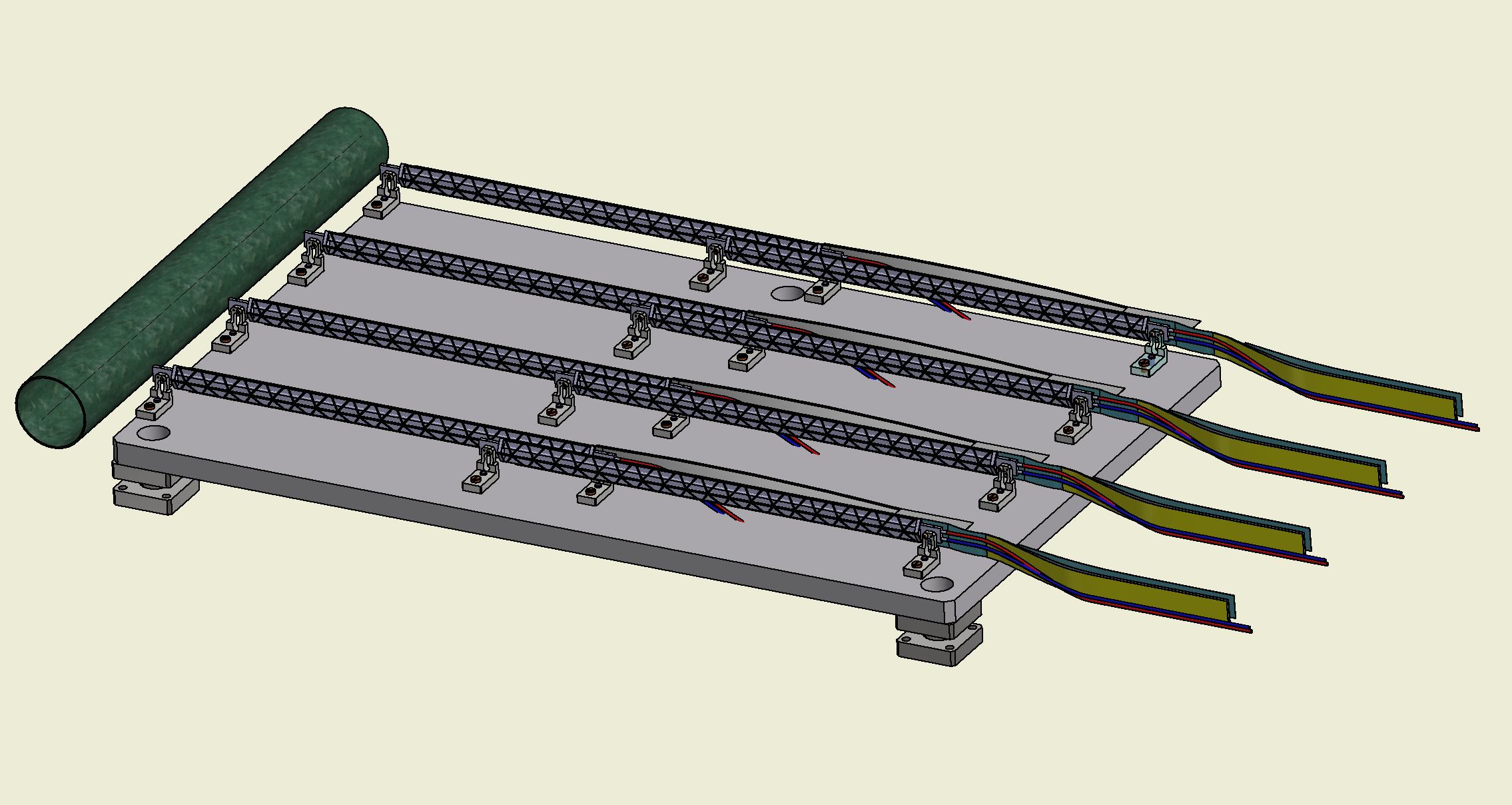}\end{overpic}
\begin{overpic}[width=0.49\textwidth]{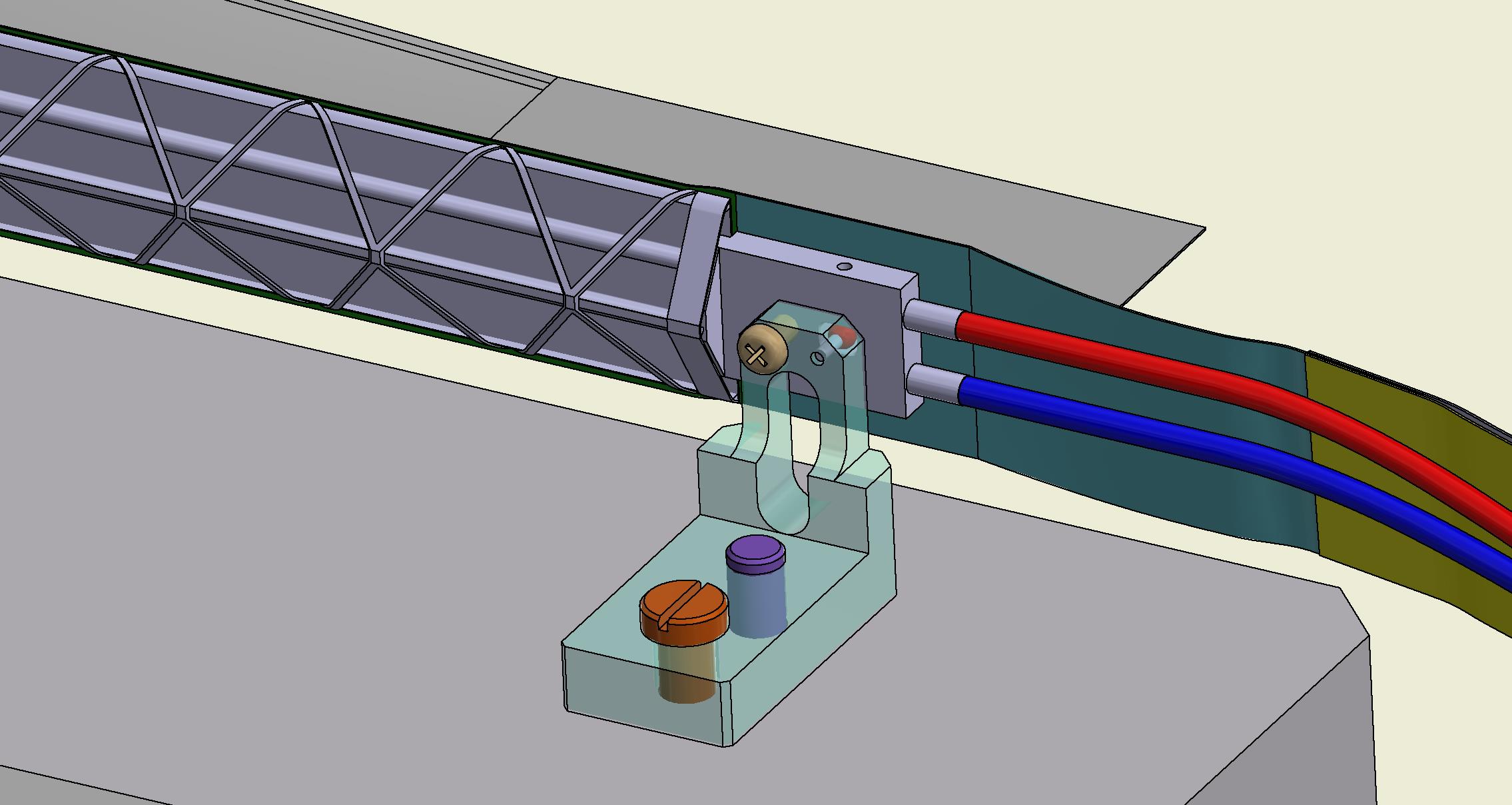}\end{overpic}
\caption{Left: one full tracker arm showing the flat tray and the L-shape bars used to fix the staves in their positions
Also shown in the picture is the \beampipe. Right: A zoomed-in drawing of the L-shape bars.}
\label{fig:tracker_tray1}
\end{figure}

A graphical view of the tracker is shown in Fig.~\ref{fig:tracker_tray1}.

The eight staves will be fixed in space to form the four layers by using precision holes at the plastic edges of the space frame as depicted in Fig.~\ref{fig:stave}. This is done using precision pins mounted on vertical L-shaped bars which are connected to $\sim 60\times 50 \units{cm}^2$ tray flat that supports the entire structure from beneath.
The staves are surveyed by the manufacturer with respect to the precision holes so the position of all chips within a stave is known with a good precision.

The \beampipe outer radius and the fixation of a stave on its two edges dictates the minimum distance in the horizontal axis between the stave's active edge and the \beamline axis.
This distance is about $5.7\units{cm}$.
The distance along the \beamline between the exit of the magnet and the front face of the first layer is $\sim 100 \units{cm}$, while the distances between subsequent layers is $\sim 10$~cm providing a good lever-arm for precise tracking.

The staves on the tray should be aligned relative to each other to better than $20 \units{\mu m}$ in all three dimensions.
This precision is normally achievable via survey but even if this cannot be achieved, an optical alignment system should be available in order to determine the in-situ position of the staves.
Since the staves spaceframe is already a rigid precision object, where the exact position is determined by only two precision holes on the spaceframe fixation ears, it is sufficient that the optical alignment system determines the position of the L-shapes holding the stave.
This optical alignment system should also allow the entire tray to be precisely positioned with respect to the beam axis and the IP. Furthermore, it is also possible to align the detector in-situ based on the hit-residuals of the observed tracks.


\subsubsection{Performance}
\label{sec:detectors_tracker_performance}

The performance is studied using a parameterised sensor response of the ALPIDE sensors which results in effective pixel-clusters, to which a high-level tracking algorithm is applied. In addition, the geometry is simplified in two aspects:
\begin{itemize}
	\item the sensitive length of one detector arm is assumed to be $50 \units{cm}$, i.e.\ without the overlap due to the staggering of two staves per layer.
	\item it is assumed that there is no extra material between the staves, i.e.\ no fixations, L-shapes, etc.
\end{itemize}
The magnetic field is assumed to be uniform with a value of $1.9 \units{T}$. The reconstruction proceeds through a Kalman filter (KF) track fitting algorithm~\cite{BILLOIR1990219} which considers all clusters from all tracks from the four layers, including clusters from toy background tracks resembling the distribution of background from full simulation.
The KF algorithm reduces most of the combinatorial backgrounds, while a loose selection is applied to reject the remaining reconstructed tracks built out of one or more non-signal clusters. This selection requires clusters in all four layers and includes e.g.\ a cut on the $\chi^2/N_\textrm{DoF}$ of the track fit or the fitted vertex-to-IP significance.

\begin{figure}[!ht]
\centering
\includegraphics[width=0.5\textwidth]{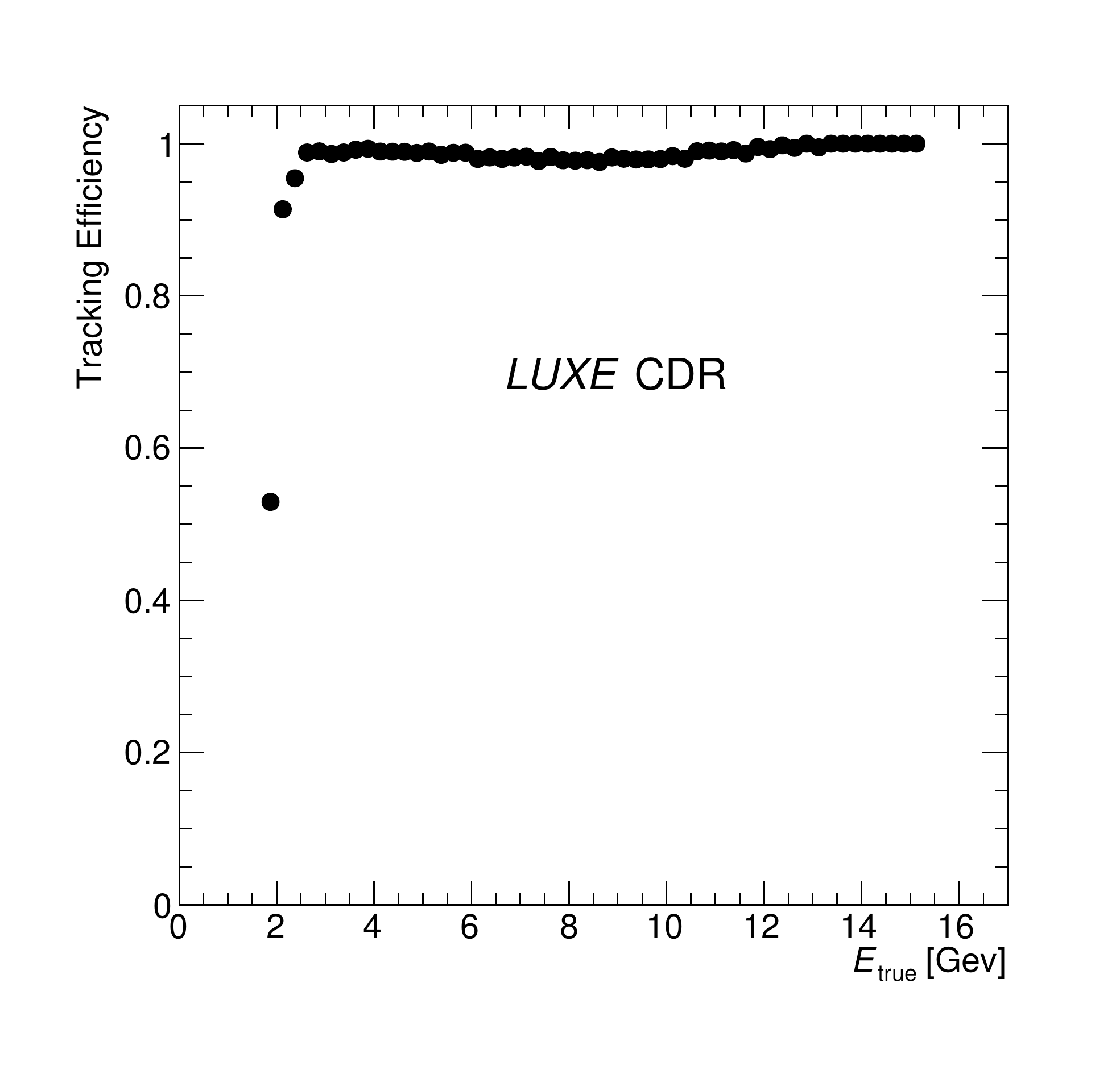}
\caption{Tracking efficiency as  a function of  true track energy (at these high energies the $e$ mass is negligible, so that $E=p$). The denominator are all true particles produced in the interaction. The numerator are all tracks that enter the detector, are reconstructed and pass the selection cuts.
}
\label{fig:eff_bppp}
\end{figure}

The tracking efficiency is shown in Fig.~\ref{fig:eff_bppp} and includes the acceptance, reconstruction and selection efficiency. It is $>95\%$ for momenta above 2.5~GeV. The turn-on observed at low energies is mostly due to the geometrical acceptance of the detector, as at low energies tracks do not cross the last layer which at present is required for the seeding algorithm. However, this causes no major problem for the analysis since less than 5\% of the positrons are expected to have an energy below 2~GeV (see Fig.~\ref{fig:elaser_posi_espectrum}). And, with further studies it might be possible to regain some of the inefficiency by adapting the seeding algorithm.

\begin{figure}[ht]
\centering
\begin{overpic}[width=0.49\textwidth]{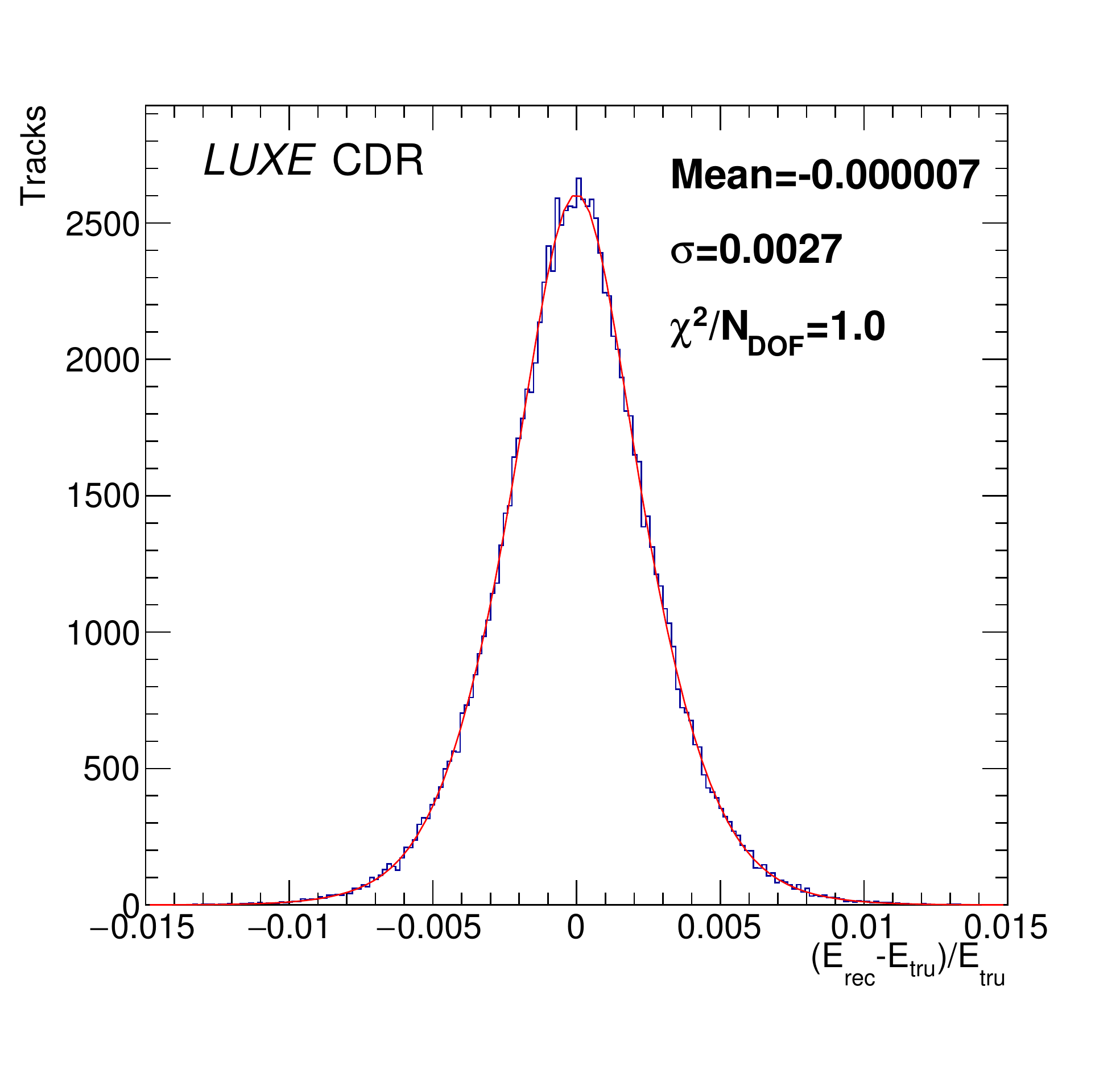}\end{overpic}
\begin{overpic}[width=0.49\textwidth]{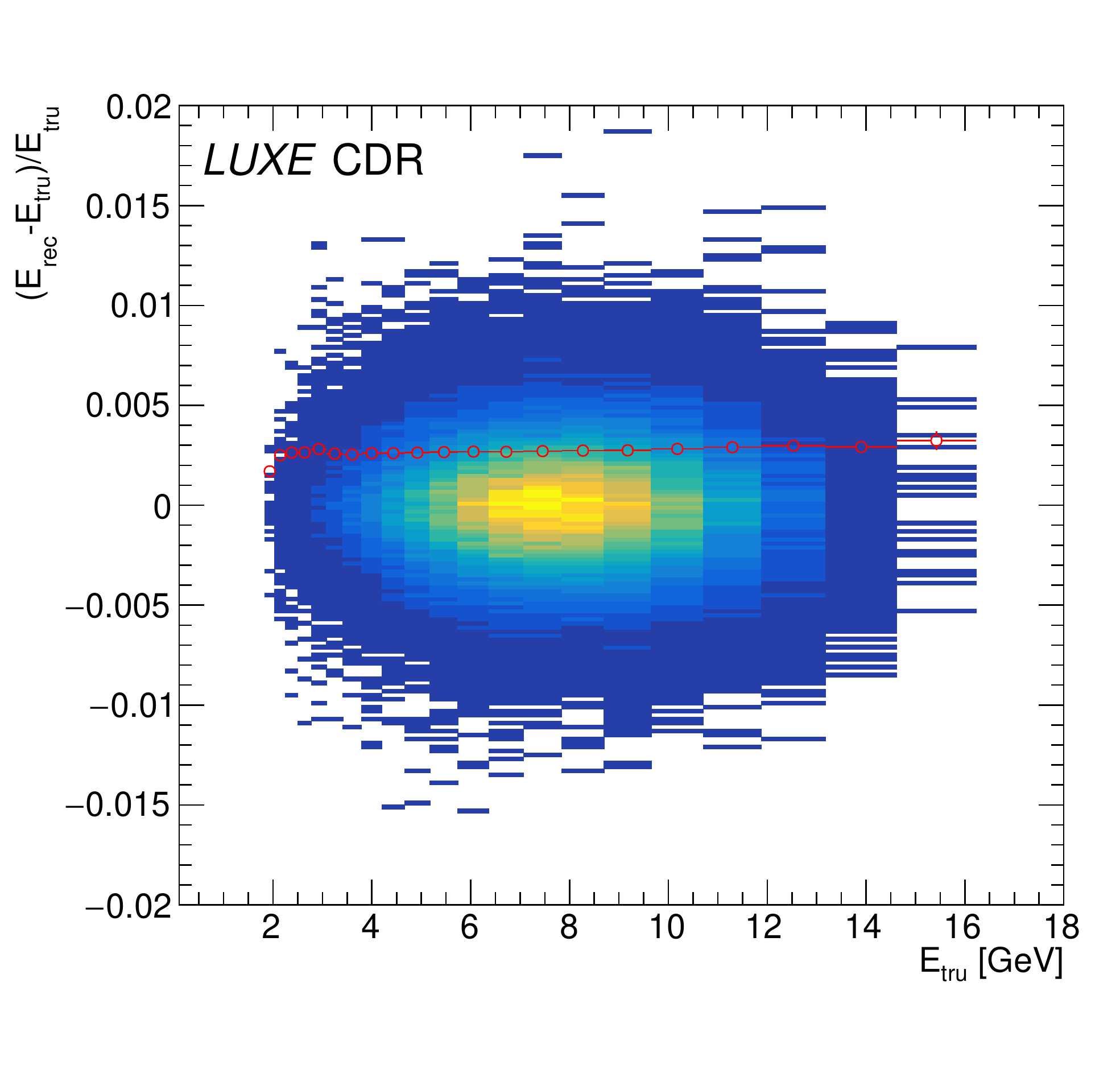}\end{overpic}
\caption{The inclusive energy resolution for the \elaser setup on the left. This resolution is approximately independent of the true track energy as can be seen on the right. The red points are the RMS profile of the energy response (in the y-axis), which is a measure of the resolution.
\label{fig:energy_resolution}}
\end{figure}

The energy resolution is about 0.27\% as determined 
from a triple Gaussian fit of the energy response distribution, $(E_{\rm rec}-E_{\rm true})/E_{\rm true}$, as shown in Fig.~\ref{fig:energy_resolution}.
The energy resolution is found to be independent of the positron energy. 
This will be slightly degraded in the experiment due to the small multiple scattering, imperfect knowledge of the magnetic field non-uniformity, uncertainties on the materials map and residual misalignments. For the physics studies it is assumed that the energy resolution is 1\%.

The background was estimated using the \geant simulation for the \elaser and \glaser setups with $1.5\cdot 10^9$ electrons per BX. As was shown in Section~\ref{sim:sig_bkg} background particles produce a significant number of hits, and it is important to ensure that the reconstruction algorithm is able to reject those. To that end, a basic tracking algorithm was developed and used to analyse the reconstruction of electrons and positrons in the simulated signal and background events. The tracking algorithm uses the first and last layer to build seed tracks, and then tries to attach hits in the two inner layers which are then fit to reconstruct a particle trajectory. Quality cuts are made to loosely constrain the track to originate from the \beamline, to have no significant slope in the $y$ direction, and on the fit quality. 

\begin{table}[htbp]
    \centering
    \begin{tabular}{|c|c|c c|}
    \hline
         selection & \elaser  & \multicolumn{2}{c|}{\glaser} \\
          & $\nposi$ & $\nposi$ & $N_\textrm{ele}$\\\hline
         Pre-selection & $9.3\pm 0.2$  & $1.7\pm 0.2$ & $1.8\pm 0.2$ \\
         3 hits &  $1.06 \pm 0.08$ & $0.17 \pm 0.07$ & $<0.1$ \\
         4 hits &  $<0.02$ & $<0.1$ & $<0.1$ \\\hline
    \end{tabular}
    \caption{Backgrounds for positrons and electrons in the tracking system for the \elaser and \glaser setup for three selections: a pre-selection and for tracks with 3 and 4 hits. Shown is the background per bunch crossing. The upper limits are given when no event is observed in the sample of simulated BXs and correspond to 95\% confidence level. The number of simulated BX is 160 for the \elaser and 30 for the \glaser case. The present study is based on hits, rather than clusters of hits.}
    \label{tab:posirecbg}
\end{table}

The background is shown in Table~\ref{tab:posirecbg} for positrons for the \elaser and \glaser setups; for electrons it is shown for the \glaser setup. It is seen that it is generally higher for the \elaser than the \glaser setup~\footnote{This is due to the electron beam exiting and hitting several components causing sprays of background. We hope to optimise this to reduce the background further.}
For the \glaser setup it is similar for electrons and positrons. 
When requiring four hits the background is $0.01$.
This is low enough so that also the very low positron rates in the \glaser process can be measured as discussed in Sec.~\ref{sec:results}. 
This also illustrated the power of the fourth layer, reducing the background by about a factor of 10--100 compared to having just three layers. 
Additionally, given that it is possible that individual components malfunction or become inactive at the \% level, it seems prudent to have one additional layer to ensure that at least three layers are active.
With a more sophisticated analysis it might well be possible to further reduce the background also for the 3-hit tracks when allowing missing hits just in modules which are known to be inactive.

\begin{figure}[!ht]
\centering
\includegraphics[width=0.49\textwidth]{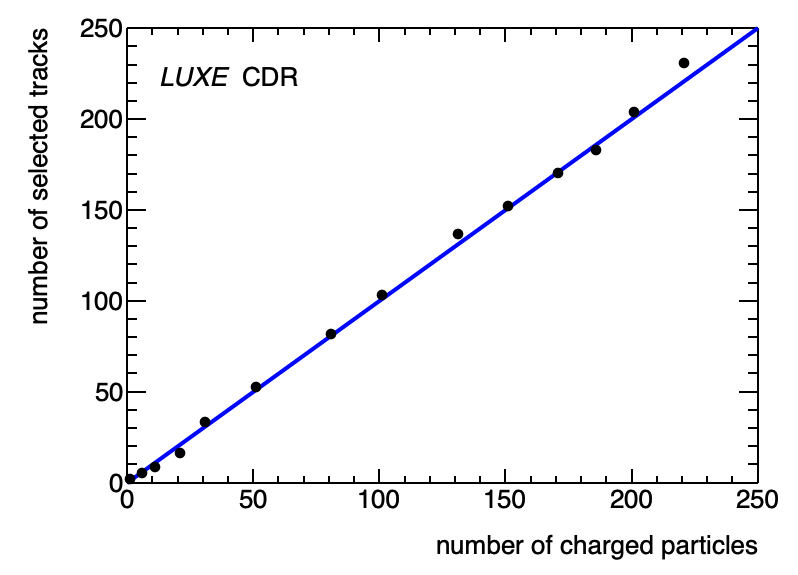} 
\includegraphics[width=0.35\textwidth]{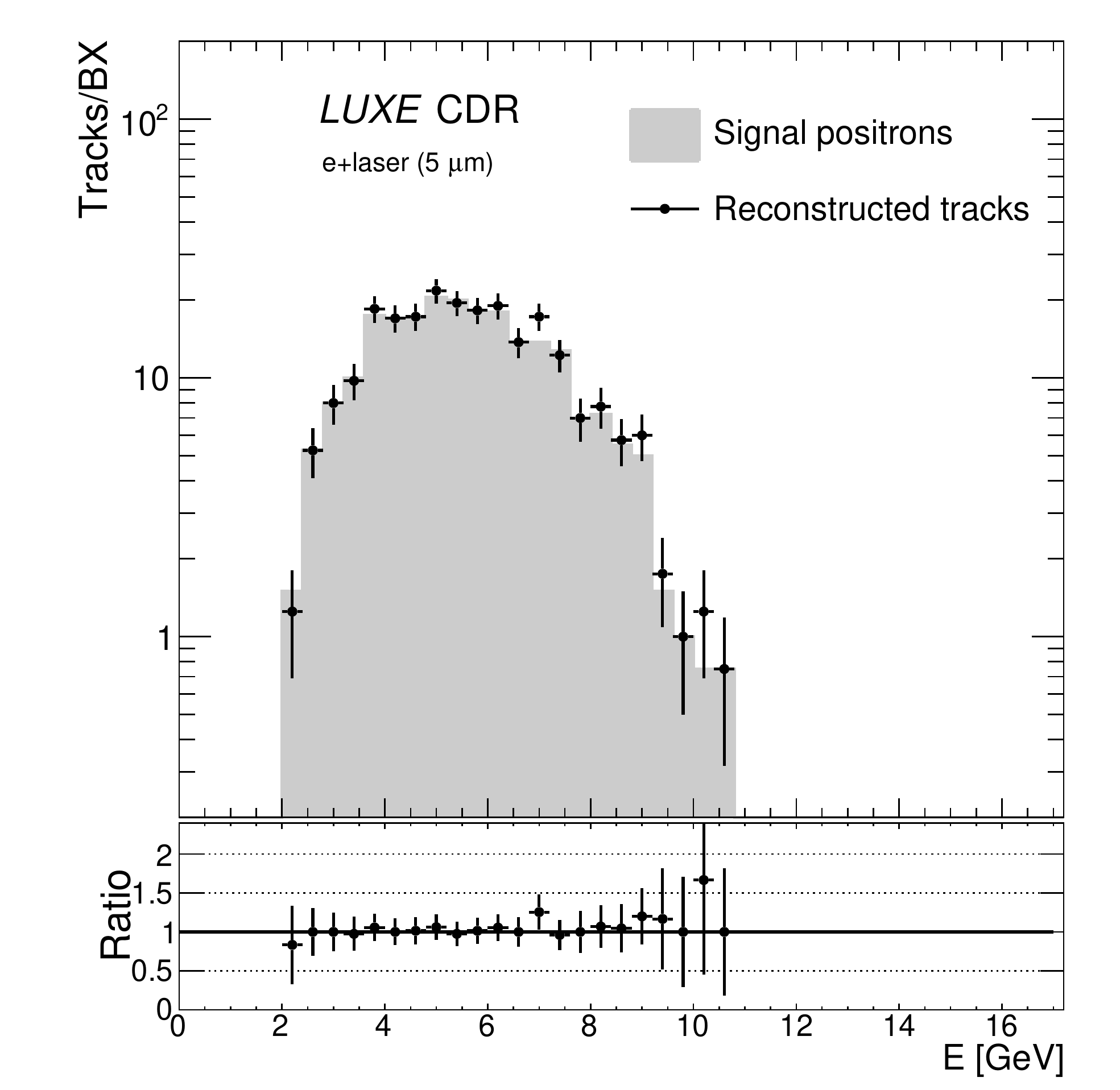}
\caption{Left: Number of reconstructed and selected tracks versus the number of charged particles for the \elaser setup. Right: Number of true and reconstructed charged particles in the tracking detector for MC simulation.
\label{fig:positrackrec}}
\end{figure}

Another concern is to which extent the detector is adequate for the high multiplicity region. Here, the concern is not beam background but rather combinatorial background due to hits being wrongly combined. This was also studied and Fig.~\ref{fig:positrackrec} shows the number of selected tracks versus the number of charged particles. It is seen that a good linearity is observed up to 200 positrons. Also shown is he reconstructed positron energy spectrum in the simulation including background, compared to the true spectrum for \phaseone with $w_0=5$~$\mu$m, corresponding to $\xi=3.1$. Good agreement is observed. Studies at higher multiplicity are not yet available but in progress. However, even with 10,000 signal particles the occupancy per pixel is $\sim 10^{-3}$ and based on experience e.g.\ at the LHC experiments it seems likely that it will be possible to still resolve the particles. Furthermore, the combination with the calorimeter will also improve the ability to reconstruct the number of positrons correctly. 

\subsubsection{Integration and Readout}
\label{sec:detectors_tracker_integration}

\begin{figure}[!ht]
\centering
\begin{overpic}[width=0.60\textwidth]{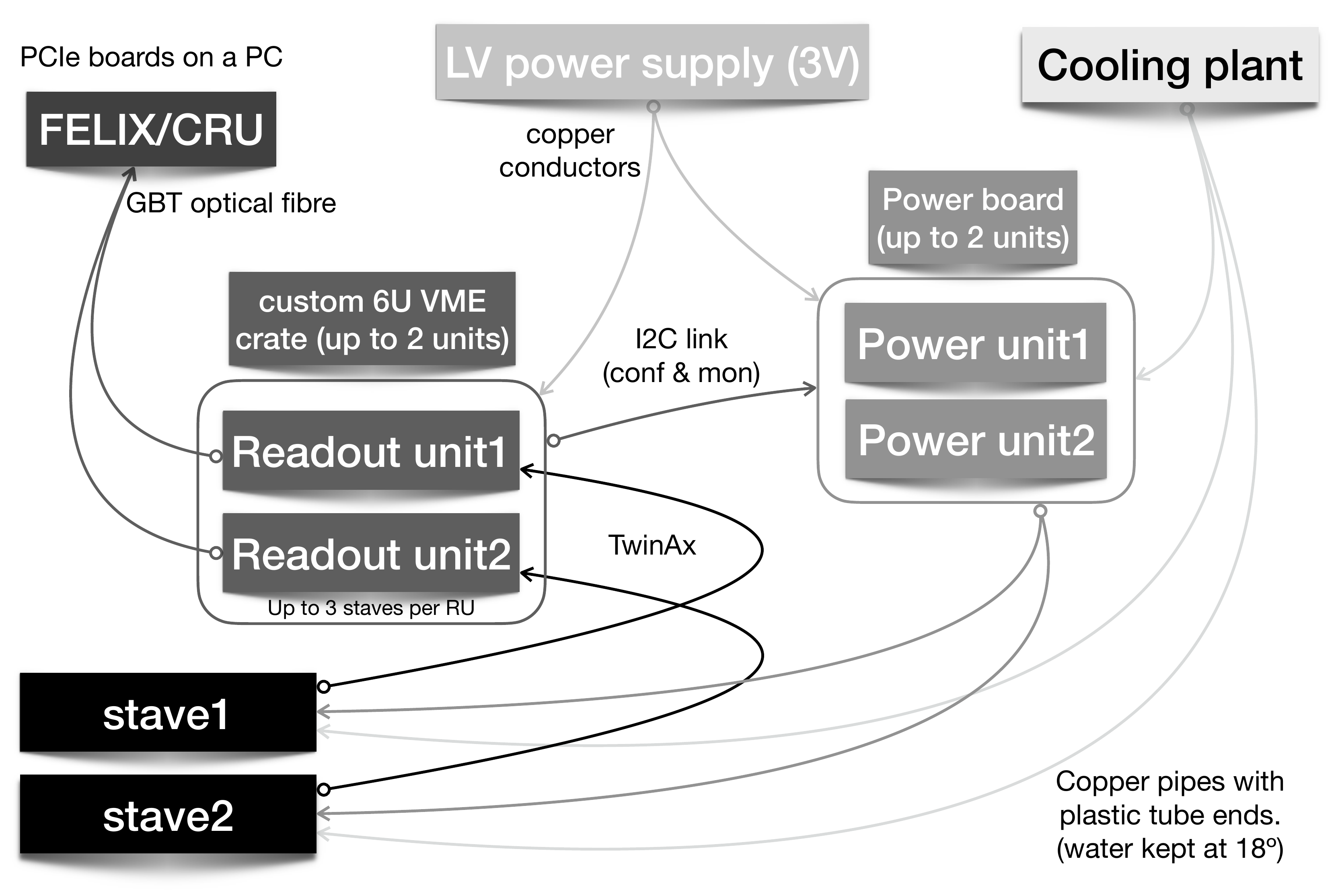}\end{overpic}
\caption{System layout, showing the staves as the end point client of all services and with the power, readout and cooling interfaces.}
\label{fig:system_layout}
\end{figure}

A schematic and conceptual layout of the entire tracker system beyond the staves is shown in Fig.~\ref{fig:system_layout}.
The cooling (at $18^\circ$ ) is provided to the staves at $0.3 \units{l/h}$ via small pipes with outer diameter of $\sim 2 \units{mm}$ as shown in Fig.~\ref{fig:tracker_tray1} (right) to keep the staves in a temperature of roughly up to $21^\circ$.

Each readout unit (RU) serves one stave and has up to three uplinks, and one downlink. The staves are connected to the RUs with relatively rigid, halogen free TwinAx cables (of about $5 \units{m}$ length).
The recommended setup is to use two FELIX boards~\cite{Anderson_2016} (Front-End LInk eXchange, essentially a PCIe card) installed on a single PC, with each managing ten staves.

Finally, one DAQ PC can in principle handle 24 staves although some redundancy should be considered.
In LUXE, 2 DAQ PCs are foreseen.

The ALPIDE chip needs the following power supply: the chip itself needs 1.8~V, the power board channel 3.3~V and reverse bias --5~V. The power to the RUs and power-boards is supplied by a CAEN power system. These components can be easily shielded in LUXE while still placed close enough to the detector.

\subsection{The Electromagnetic Calorimeter}
\label{sec:detectors_calorimeter}

The tracker will be followed by a highly granular and compact electromagnetic calorimeter, hereafter denoted as ECAL. The ECAL will allow to measure independently of the tracker the number of positrons, and their energy spectrum. These measurements will be robust, since they use the total deposited energy by positrons, and are hence not affected by background particles of low energy. 

Essential for the performance of ECAL are high granularity for very good position resolution, compactness, i.e.\ a small Moli\`{e}re radius, to ensure a high spatial resolution of local energy deposits, and good energy resolution to measure and infer the spectrum of positrons. To match these requirements, the technology developed by the FCAL collaboration is proposed~\cite{Abramowicz:2018vwb}.

\subsubsection{The Structure of ECAL }
The LUXE ECAL is designed as a sampling calorimeter composed of 20 layers of $3.5\units{mm}$ ($1X_0$) thick tungsten absorber plates, and assembled silicon-sensor planes placed in a 1\,mm gap between absorber plates. 

\begin{figure}[htbp]
\begin{center}
  \includegraphics[width=0.5\textwidth]{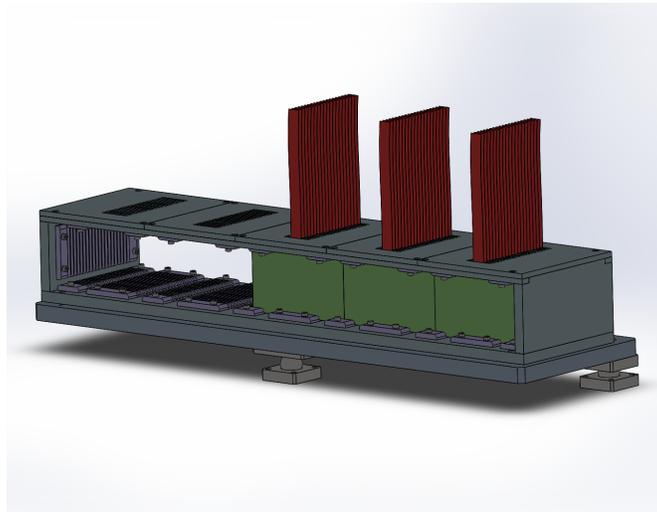}
    \caption{A sketch of LUXE ECAL. The frame holds the tungsten absorber plates, interspersed with assembled silicon detector planes. The front-end electronics will be positioned on top of ECAL.} 
    \label{ECAL_layout}
    \end{center}
\end{figure}

A sketch is shown in  Fig.~\ref{ECAL_layout}.
The thickness of the silicon sensor is $320 \units{\mu m}$ and the pad size is $5\times 5 \units{mm^2}$.   The sensor is embedded in a carbon fibre structure and covered by flexible Kapton PCBs on both sides, to supply bias voltage and to read out the signals via copper traces, resulting in the assembled sensor plane thickness of less than $700 \units{\mu m}$. More details are given below. The fiducial volume of the calorimeter is $55 \times 5.5 \times 9 \units{cm^3}$. Each complete sensor plane will consists of five adjacent silicon sensors. 

\subsubsection{The Fully Instrumented Detector Plane }
\begin{figure}[htbp]
\begin{minipage}{0.5\textwidth}
\includegraphics[width=\textwidth]{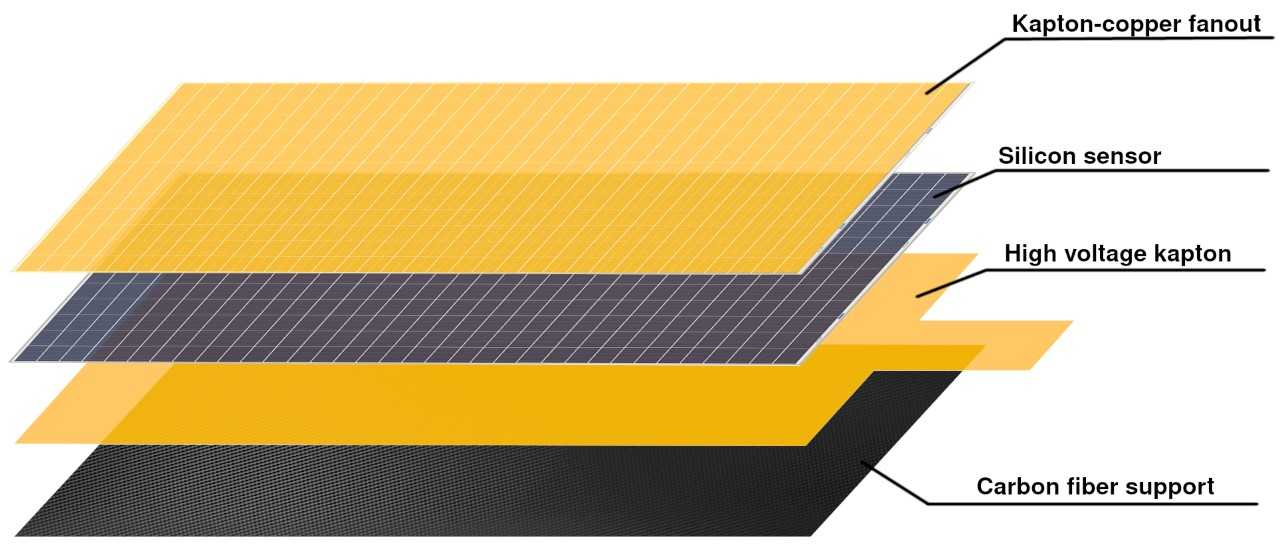}
\end{minipage}
\begin{minipage}{0.5\textwidth}
      \includegraphics[width=\textwidth]{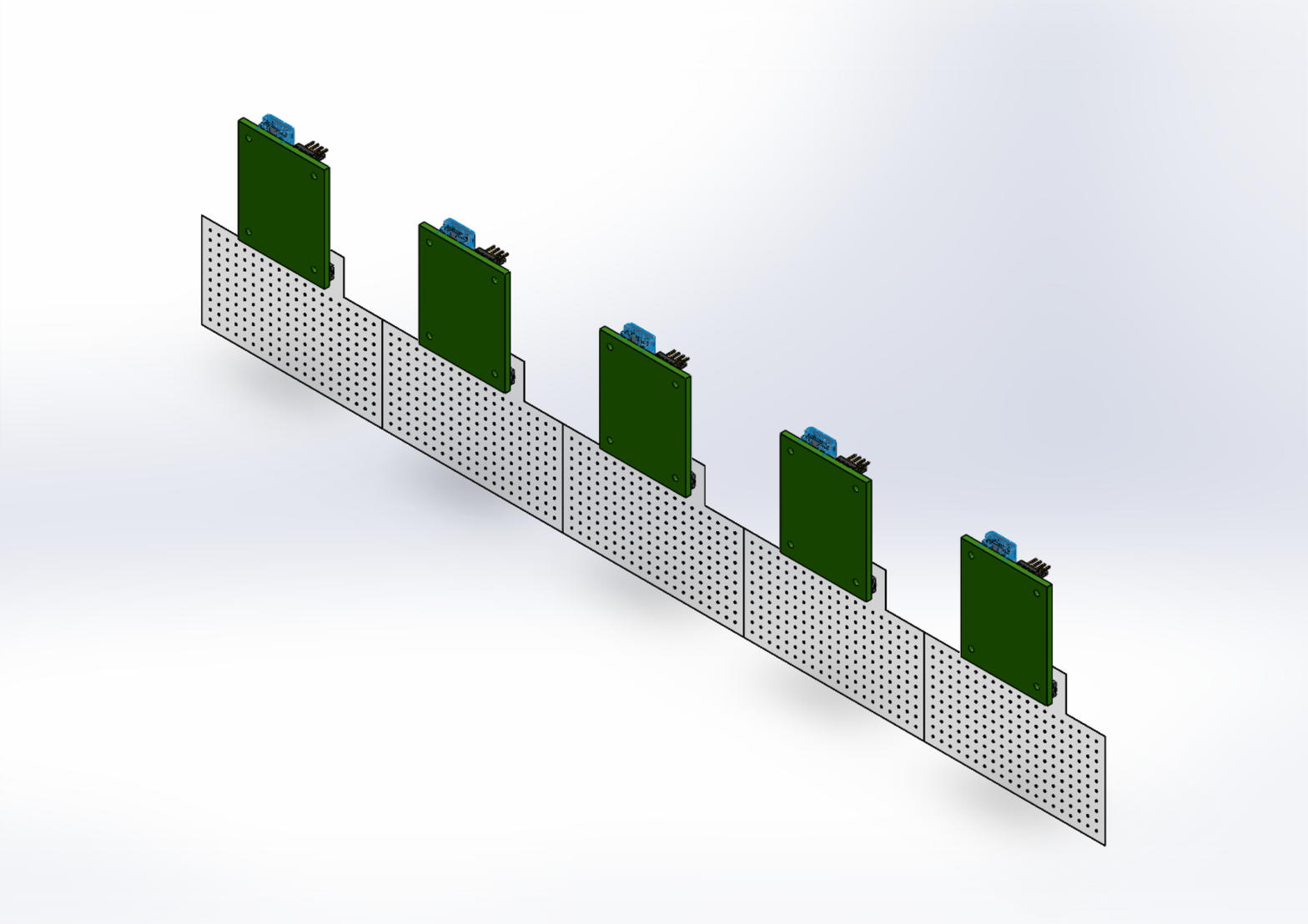}
\end{minipage} 
     \caption{Left: A sketch of a detector plane structure. The total thickness is below $700 \units{\mu m}$. Right: Sketch of the readout scheme of one calorimeter plane.}
    \label{fig:plane_and_read_out}
 \end{figure}

The silicon pad sensor is embedded in a sub-millimetre detector plane. The bias voltage is supplied to the n-side of the sensor 
by a $70 \units{\mu m}$ flexible Kapton-copper foil, glued to the sensor with a conductive glue.
The 231~pads of the sensor are connected to the front-end electronics using a fan-out made of $120 \units{\mu m}$ thick 
flexible Kapton foil with copper traces. Ultrasonic wire bonding will be used to connect conductive traces on the 
fan-out to the sensor pads. A support structure, made of a carbon fibre composite with a thickness of $150 \units{\mu m}$, provides mechanical stability for the detector plane. Special jigs will be  designed and produced to ensure the necessary thickness and uniformity of three glue layers between different components of the detector plane all over the area of the sensor. A sketch of the structure of the detector plane is shown in Fig.~\ref{fig:plane_and_read_out} (left).

An new idea for the signal fanout is also under consideration at present. In that scenario, the sensors will be covered with another layer of metallic traces that will be used for signal fanout. This will remove the need for an external Kapton layer as well as the need for fragile wire-bonding between the sensor pads and fanout, reducing the module thickness by about $150 \units{\mu m}$ and making it more durable and easier to handle in the detector assembly. This idea will be explored on GaAs sensors to which the LUXE team has an easy access through the facilities existing in the Tomsk State University. If successful, GaAs sensors will be implemented instead of the silicon-sensors. Prototypes of GaAs pad-sensors were investigated by the FCAL collaboration in a test beam~\cite{Abramowicz:2014gdq} and the response was found to be very similar to the one of silicon pad-sensors.

\subsubsection{Front-end Electronics and DAQ }
Each detector plane is read out by front-end ASICs mounted on a PCB and positioned on top of the calorimeter frame, as sketched in Fig.~\ref{fig:plane_and_read_out} (right) for one calorimeter layer. Connectors on top of the PCBs are foreseen for cables to transfer digital signals from the ASICs to FPGAs installed at a larger distance.

\paragraph{FLAME front-end electronics}
Front-end electronics is based on dedicated readout ASIC called FLAME (\textbf{F}ca\textbf{L} \textbf{A}sic for \textbf{M}ultiplane r\textbf{E}adout), designed for silicon pad detectors in the LumiCal calorimeter of a future linear collider.
The main specifications of the FLAME ASIC are shown in Table~\ref{tab:flame}.
\begin{table}[htbp]
\begin{center}
\begin{tabular}{|l|l|}
\hline
Variable & Specification \\
\hline
Technology & TSMC CMOS 130\,nm \\
Channels per ASIC & 32 \\
Power dissipation/channel & $\sim 2\units{mW}$ \\
\hline
Noise & $\sim$1000 e$^-$@10\,pF + 50e$^-$/pF \\
Dynamic range & Input charge up to $\sim 6 \units{pC}$ \\
Linearity & Within 5\% over dynamic range\\
Pulse shape and & T$_{peak} \sim$50\,ns \\
\hline
ADC bits & 10 bits  \\
ADC sampling rate &  $\sim 20$\,MSps \\
Calibration modes & Analogue test pulses, digital data loading\\
Output serialiser & serial Gb-link, $\sim 5$\,GBit/s\\
Slow controls interface & I\textsuperscript{2}C, interface single-ended\\
\hline
 \end{tabular}
 \end{center}
 \caption{Summary of the specifications of the FLAME ASIC.}
 \label{tab:flame}
\end{table}

\begin{figure}[htbp]
 \begin{center}
  \includegraphics[width=0.7\textwidth]{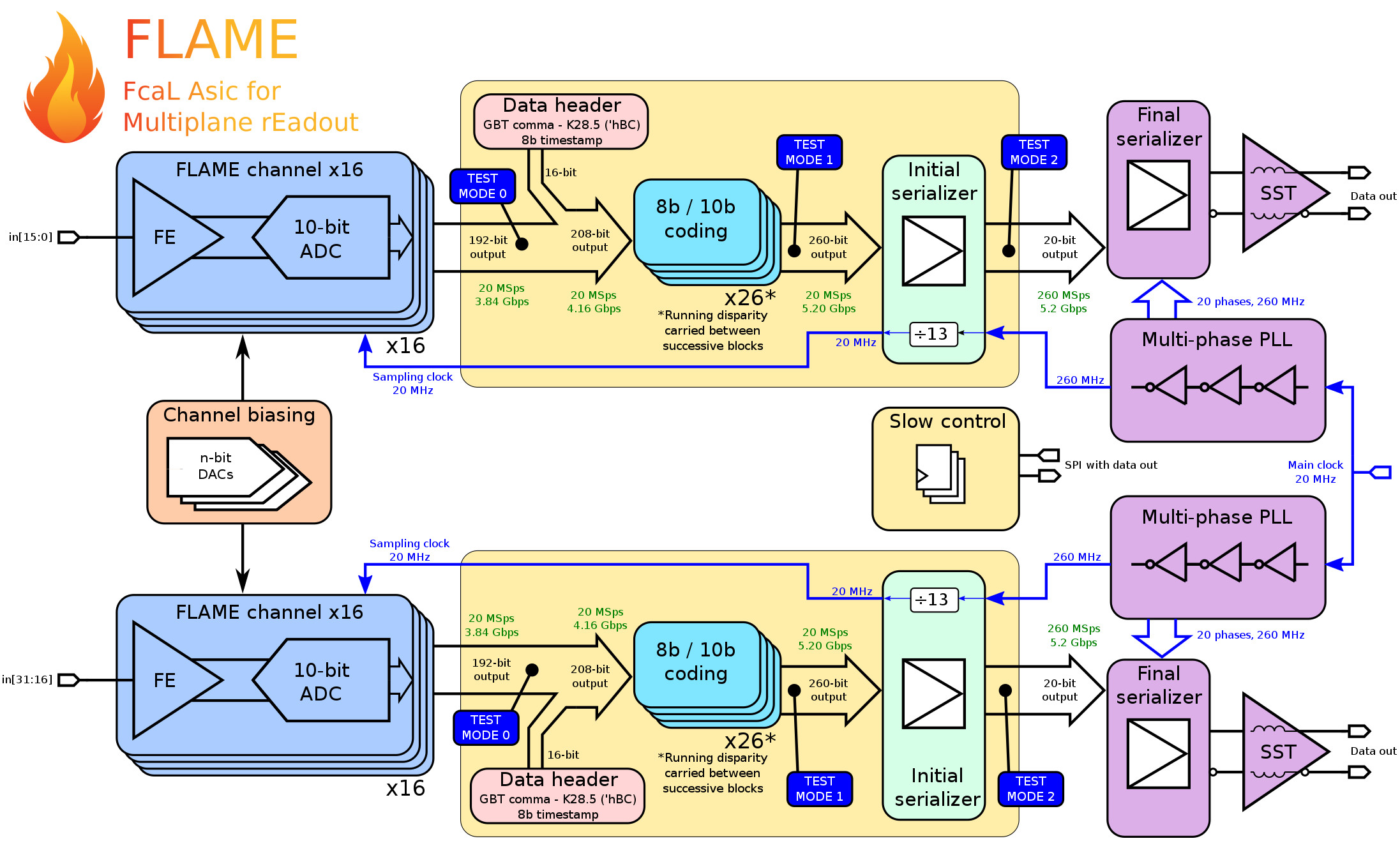}
    \caption{Block diagram of 32-channel FLAME ASIC.} 
    \label{fig:ECAL_flame}
    \end{center}
\end{figure}

A block diagram of FLAME, a 32-channel ASIC designed in TSMC CMOS 130nm technology, is shown in Fig.~\ref{fig:ECAL_flame}. FLAME comprises an analogue front-end and 10-bit ADC in each channel, followed by a fast data serialiser. It extracts, filters and digitises analogue signals from the sensor, performs fast serialisation and transmits serial output data.

As seen in Fig.~\ref{fig:ECAL_flame}, the 32-channel chip is designed as a pair of two identical 16-channel blocks. Each block has its own steriliser and data transmitter so that during operation two fast data streams are continuously sent to an external DAQ. The biasing circuitry is common to both 16-channel blocks and is placed in between. Also the slow control block is common and only one in the chip.
During standard operation, the 10-bit ADC samples the front-end output at a rate of $\sim$ 20\,MSps.

The analogue front-end consists of a variable gain preamplifier with pole-zero cancellation (PZC) and a fully differential CR--RC shaper with peaking time $\sim 55 \units{ns}$. 
The shaper includes also an 8-bit ADC for precise baseline setting. The analogue front-end consumes in total 1\,--\,1.5\,mW/channel.
The ADC digitises with 10-bit resolution and $\sim$ 20\,MHz sampling rate. 
The power consumption is below 0.5\,mW per channel at 20\,MS/s. 
In order to ensure the linearity of the DAC, the input switches are bootstrapped, reducing significantly their dynamic resistance.

To adopt the current FLAME design for LUXE, a few modifications are needed.
The fast data transmission components will be replaced by a simpler and slower data transmitter to reduce the costs. In addition, zero suppression will be implemented in the chip to decrease the data volume. 

\paragraph{Back-end electronics and DAQ }

The five detector planes in each ECAL layer will be equipped with five front-end boards (FEB), as shown in Fig.~\ref{fig:plane_and_read_out} (right).
The signals from the FEBs will be sent to an FPGA board. One FPGA board will manage ten FEBs, comprising 2 calorimeter layers. The signals from the FEBs to the FPGA are fully digital, hence the FPGA boards can be positioned at a larger distance from the calorimeter, e.g.\ near or in the calorimeter rack. 
The communication between the FPGA boards and the DAQ computer is done through the user datagram  protocol (UDP). All FPGAs and the computer will be connected with a switch (1\,Gb/s) located in the calorimeter rack. 

\subsubsection{Performance}

\begin{figure}[!ht]
 \begin{minipage}{0.45\textwidth}
   \includegraphics[width=\textwidth]{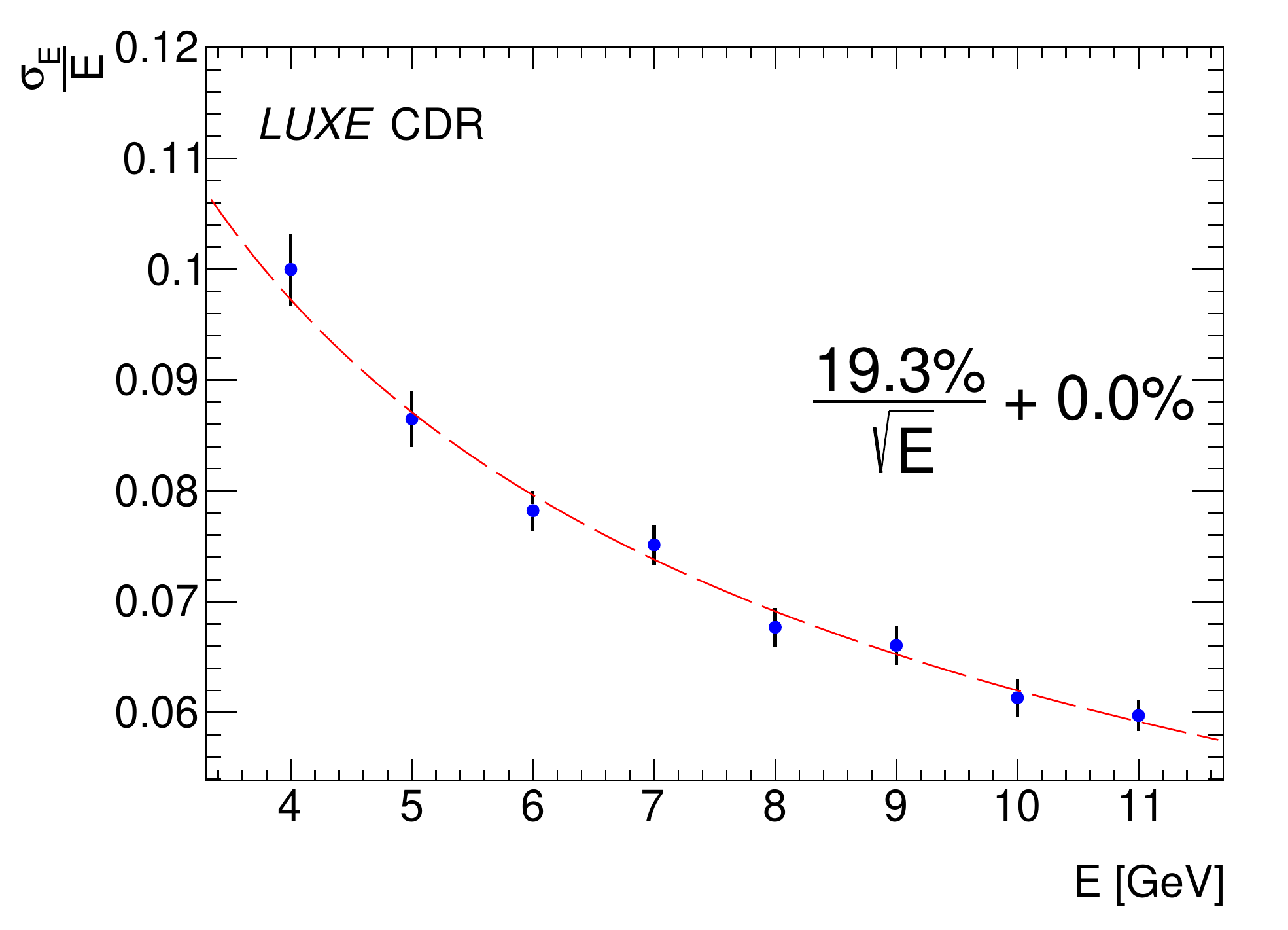}
  \end{minipage}
 \hspace{0.05\textwidth}
  \begin{minipage}{0.45\textwidth}
   \includegraphics[width=\textwidth]{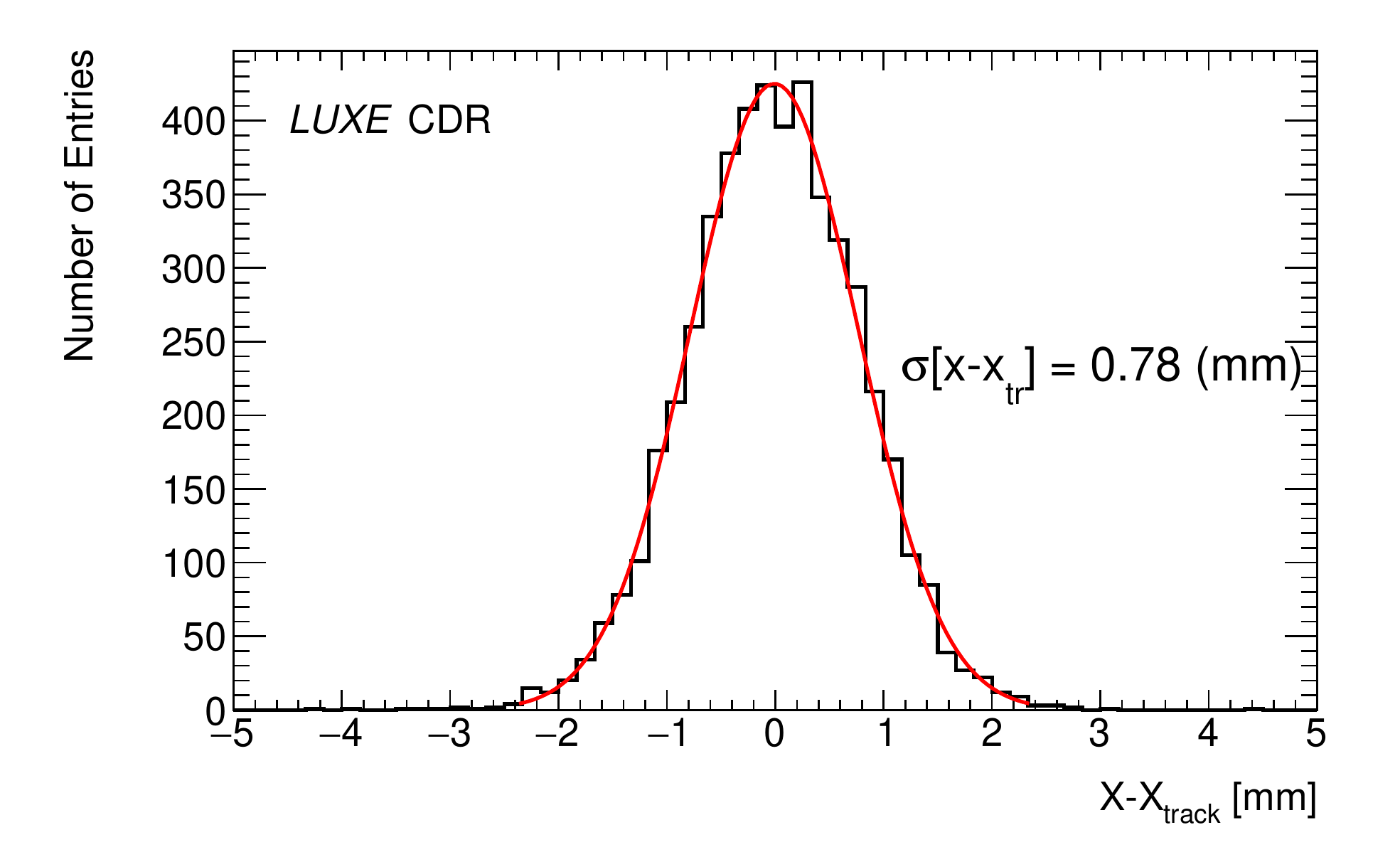}
  \end{minipage}
  \caption{The energy resolution as a function of the positron energy (left), and the shower position resolution in the $x$-coordinate for positrons of 10 GeV (right).} 
  \label{ECAL_performance}
\end{figure}

The performance of the ECAL was originally studied with \geant simulations for electrons. The same performance applies to positrons.
For single electrons hitting the calorimeter perpendicular to the front face, and after reconstructing the shower, the energy and position resolution are estimated to be 
\begin{equation}
    \frac{\sigma(E)}{E} = \frac{19.3\%}{\sqrt{E}},
\end{equation}
and 
\begin{equation}
    \sigma(x) = 780\units{\mu m}.
\end{equation}
These results are illustrated in Fig.~\ref{ECAL_performance}.

\begin{figure}[htbp]
\centering
    \includegraphics[width=0.7\textwidth]{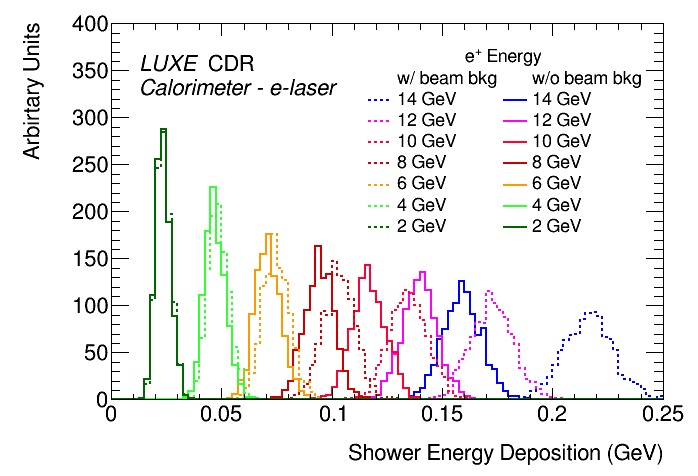}
    \caption{Distribution of total energy deposited in the sensor-pads of the ECAL for selected energies of positrons (full lines) and for the same positron showers with added background (dashed-lines) in the \elaser setup. The various energies of the positrons listed in the legend are identified by different colours.}
    \label{fig:ecal-singlep-edep-elaser}
\end{figure}

\begin{figure}[!ht]
\includegraphics[width=0.45\textwidth]{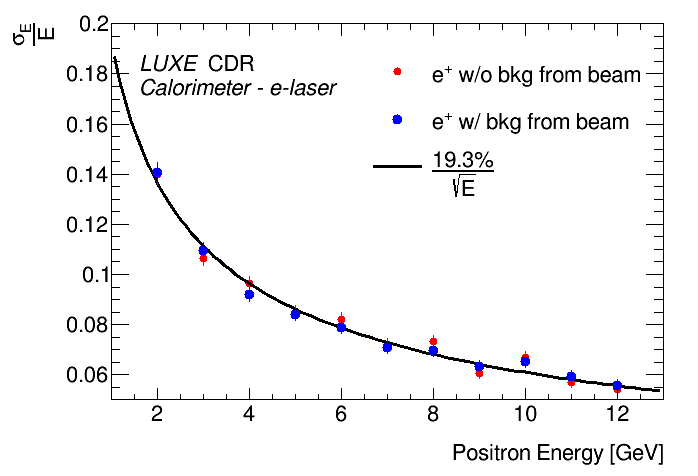}
\includegraphics[width=0.45\textwidth]{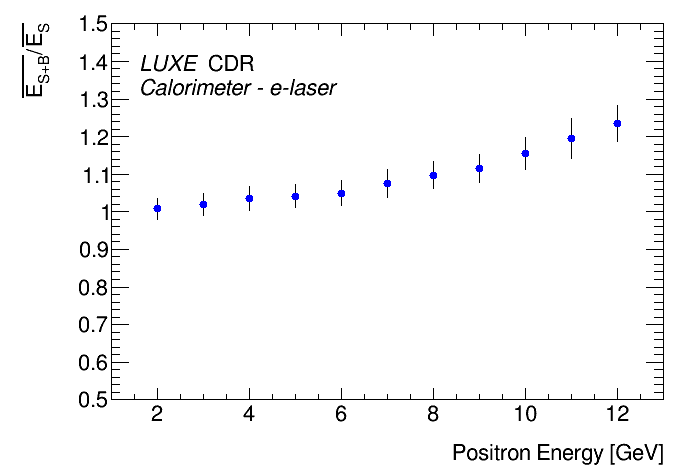}
\caption{Left - ECAL resolution with and without background as a function of positron energy for \elaser setup. Right - ratio of the expected energy deposition in the ECAL with background to the one without background as a function of positron energy for \elaser setup.}
\label{fig:ECAL-elaser-sig+bgr}
\end{figure}

\begin{figure}[!ht]
\includegraphics[width=0.45\textwidth]{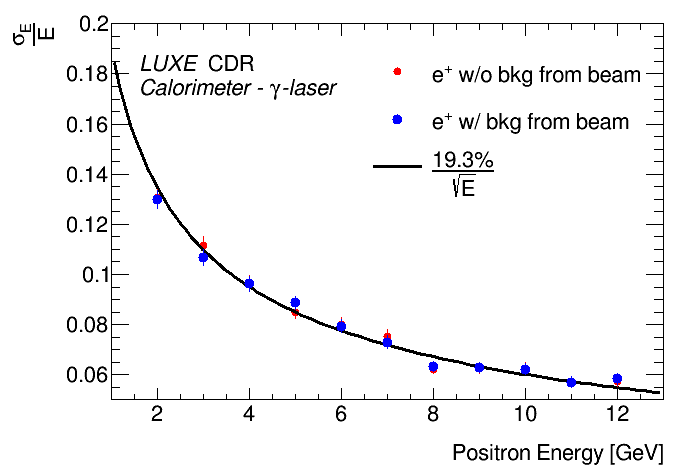}
\includegraphics[width=0.45\textwidth]{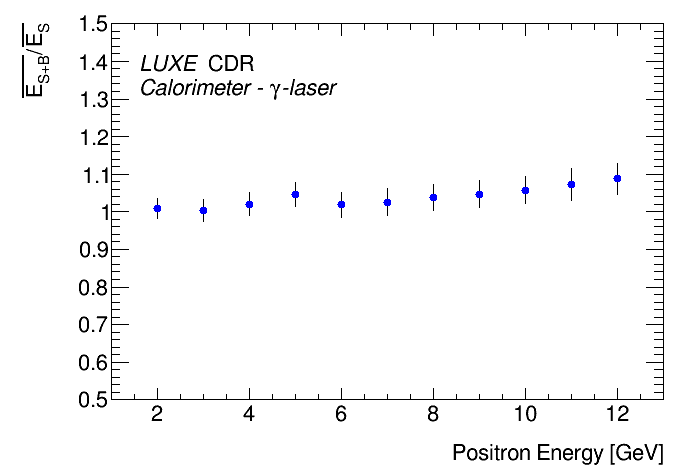}
\caption{Left - ECAL resolution with and without background as a function of positron energy for \glaser setup. Right - ratio of the expected energy deposition in the ECAL with background to the one without background as a function of positron energy for \glaser setup.}
\label{fig:ECAL-glaser-sig+bgr}
\end{figure}

It is important to ensure that this performance can also be achieved in the presence of background which is particularly sizeable near the beamline.
The effect of the beam-related background on the energy deposited in the ECAL by a single positron from the IP in the \elaser setup is shown in Fig.~\ref{fig:ecal-singlep-edep-elaser} for a sample of positron energies. The distribution of the energy deposited by positrons of a given energy is compared to the distribution obtained after adding the background to all pads containing the electromagnetic shower. 
One clearly observes a shift in the total deposited energy but no significant deterioration in the resolution. As expected, the shift in energy is more pronounced the higher the energy of the positron, the closer it gets to the beam-line. This is illustrated in Fig.~\ref{fig:ECAL-elaser-sig+bgr}, where the resolution with and without background is shown as a function of positron energy, as well as the ratio between the expected energy deposition with and without background. 
A similar effect is observed for the \glaser setup as shown in Fig.~\ref{fig:ECAL-glaser-sig+bgr}. The expected shift in the deposited energy is smaller than in the case of \elaser setup due to smaller background level.
The observed shift in the energy deposited in the ECAL due to background can be corrected for by direct background measurements with the laser off, with minimal effect on the resolution for most of the ECAL fiducial volume.

\subsubsection{Phase-0 Options}

In order to perform energy measurements in the \phaseone early runs, an already available high granularity electromagnetic calorimeter from the CALICE collaboration~\cite{Kawagoe:2019dzh} together with a partly instrumented LUXE ECAL may be used. This way both measurements for physics at low laser power as well as background will be performed. 

The CALICE calorimeter has a fiducial area of $18 \times 18 \units{cm^2}$ and sensor pads of the same size as foreseen for the LUXE ECAL. The silicon-sensor structure of the ECAL, five silicon-sensors per sensor-plane, allows to assemble the calorimeter in a modular way~\cite{CALICE-ECAL}.  At least one module of the LUXE calorimeter will be fully instrumented, extending the range of the CALICE calorimeter in the $x$-coordinate, and allowing to test the performance under realistic conditions. 

\newpage
\subsection{Photon Detection System: Gamma Ray Spectrometer, Gamma Profiler and Gamma Flux Monitor} 
\label{sec:detectors_gammaray_sepctrometer}

\begin{figure}[htbp]
    \includegraphics[width=\textwidth]{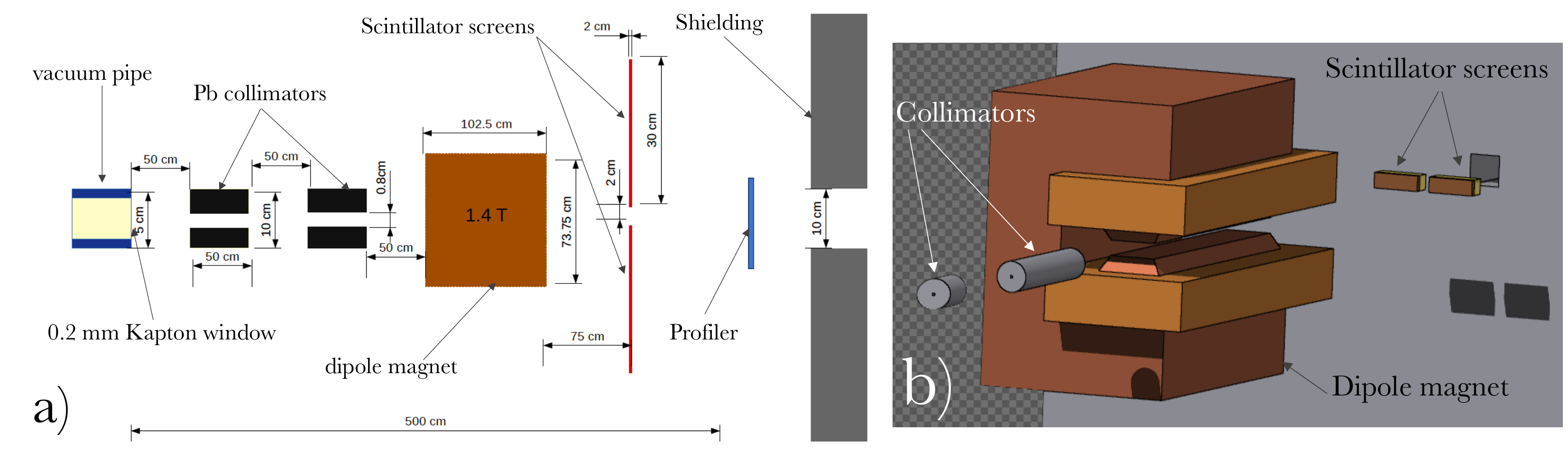}
    \caption{(a) Bird's eye view of the spectrometer. (b) Rendering of the setup as an input for the \textsc{Fluka} simulations. 
    \label{fig:fig1_gammaray_spectrometer}}
  \hfill
  \hspace{0.05\textwidth}
\end{figure}

For \elaser interactions, important information is contained in the energy spectrum and the spatial distribution of photons downstream of the interaction area. Several distinctive features can be identified and studied in the spectrum of the photon beam generated by the Compton process, $e^-+n\gamma_L\to e^-+\gamma$, in the laser field, in particular the Compton edges in the energy spectrum. 
A precise determination of the angular distribution of the photons produced in the interaction will also allow an independent determination of the 
laser dimensionless intensity $\xi$. For a linearly polarised laser pulse, which is the preferred operation mode (see sec.~\ref{sec:laserpol}), and assuming a negligible divergence of the primary electron beam, the photon beam divergence along and transversely to the laser polarisation axis are of the order of $\theta_{||} = \textrm{max}(\xi,1)/\gamma$ and $\theta_{\perp} = 1/\gamma$, respectively \cite{Har-Shemesh2011,Yan:NatPhoton2017} (see also Sec.~\ref{sec:theory:spin}). The ellipticity of the photon beam is then a direct measure of the $\xi$ value 
experienced by the electron beam. 

These diagnostics will not only provide precious information for the study of non-linear Compton scattering and the radiation reaction but will also provide a useful shot-by-shot tagging mechanism to monitor the stability of the laser shots and to potentially discard shots with e.g. non-optimal electron-laser overlap or other undesired features. The total photon flux as function of energy is another quantity that can be measured and compared with the predictions in the \elaser collisions, and is important input for the BSM physics programme. In \glaser collisions it serves as an independent measure of the flux of bremsstrahlung photons.

Three detector systems are foreseen for this purpose: 
\begin{itemize}
    \item a gamma ray spectrometer (GRS) that determines the photon energy spectrum based on the measurement of the spectrum of secondary electrons and positrons generated during the interaction of the photon beam with a solid converter;
    \item a gamma profiler (GP) that measures the energy-integrated spatial distribution of the photons
    \item a gamma flux monitor (GFM) that measures the relative flux of photons via a backscattering calorimeter.
\end{itemize}

All detector systems will be placed between the end of the vacuum pipe, after the positron and electron detection systems, and the end of the tunnel. A top view of the sketch of the spectrometer, together with its rendering for the \textsc{Fluka}~\cite{fluka} simulations, is given in Fig.~\ref{fig:fig1_gammaray_spectrometer}. For the simulations presented here it is assumed that the vaccum pipe ends before the photon detection system components, causing significant scattering of particles in air. While it seems that such a setup is workable, it is not clear if it is optimal. Studies are ongoing to evaluate the advantages and disadvantages of extending the vacuum pipe further, at least beyond the gamma ray spectrometer.

\subsubsection{Gamma Ray Spectrometer}
The main purpose of the spectrometer is to measure the spectrum of the electron-positron pairs generated during the interaction of the photon beam with the window at the end of the vacuum pipe, which is hereafter assumed to be $200 \units{\mu m}$ of Kapton. Deconvolution of the electron and positron spectra, as described in Ref.~\cite{Fleck:2020} for similar experimental conditions, allows retrieval of the spectrum of the photon beam, with an energy-dependent resolution at the few-percent level. 

A summary of the specifications of each element in the spectrometer is given in Table~\ref{tab:table1_grs}.
\begin{table}[t!]
    \centering
    \begin{tabular}{ |c|c| }
        \hline
        \multicolumn{2}{|c|}{Technical Specifications}\\
        \hline 
        \multicolumn{2}{|c|}{Converter target}\\
        \hline 
        Material & Kapton \\
        Thickness ($z$) & 200 $\mu$m\\
        Width ($y$) & 5 cm\\
        Height ($x$) & 5 cm \\
        \hline
        \multicolumn{2}{|c|}{Collimators}\\
        \hline
        Material & Pb\\
        Length & 50\,cm\\
        Inner Radius & 0.4\,cm \\
        Outer Radius & 5.0\,cm\\
        Separation & 50\,cm \\
        \hline
        \multicolumn{2}{|c|}{Magnet}\\
        \hline
        Field Strength & Up to 1.5\,T \\
        \multicolumn{2}{|c|}{Scintillator screens}\\
        \hline
        Material & GadOx Scintillator\\
        Effective Position Resolution & 0.5\,mm $\times$ 0.5\,mm\\
        Screen Size & 5\,cm $\times$ 50\,cm $\times$ 0.05\,cm\\
        Off-axis Displacement & 2\,cm (symmetric)\\
        \hline
        
    \end{tabular}
    \caption{Summary of the main technical specifications of the elements in Fig. \ref{fig:fig1_gammaray_spectrometer}. More details on the magnet are given in Sec.~\ref{sec:tc}.
    }
    \label{tab:table1_grs}
\end{table}

\begin{figure}[htbp]
    \includegraphics[width=\textwidth]{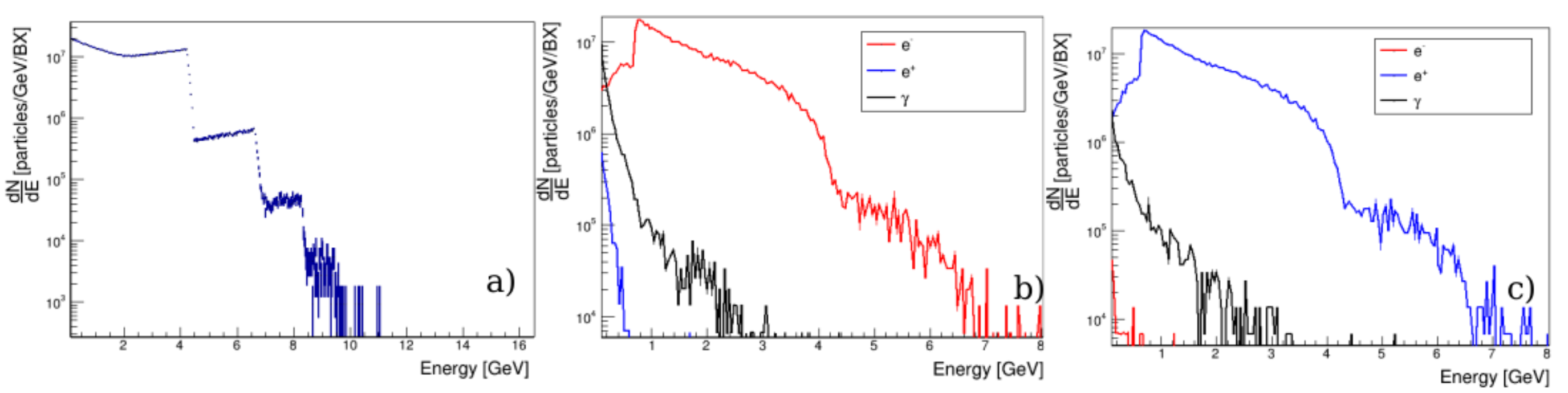}
    \caption{a) Spectrum of the photon beam used as an input for the \textsc{Fluka} simulation of the spectrometer in the \elaser configuration as predicted by \textsc{IPstrong} MC. b) Spectra of electrons (red), positrons (blue), and photons (black) onto the left scintillation screen. c) Spectra of electrons (red), positrons (blue), and photons (black) onto the right scintillation screen. Spectra are in units of particles per GeV per bunch crossing (BX). N.B. \textsc{Fluka} and {\sc Geant} are equivalent MC simulation programs, whose consistency has been extensively benchmarked, and further validated in LUXE.
    \label{fig:fig2_gammaray_spectrometer}}
  \hfill
  \hspace{0.05\textwidth}
  \end{figure}

A series of \textsc{Fluka} simulations were performed to test the performance of the spectrometer, assuming the expected photon output from the \elaser configuration. To test the response of the spectrometer, a representative photon spectrum from ICS is assumed, as plotted in Fig. \ref{fig:fig2_gammaray_spectrometer}a. In Figs. \ref{fig:fig2_gammaray_spectrometer}b and \ref{fig:fig2_gammaray_spectrometer}c the corresponding spectra of the electrons, positrons and photons reaching the two scintillator screens are shown. 

\begin{figure}[htbp]
    \includegraphics[width=\textwidth]{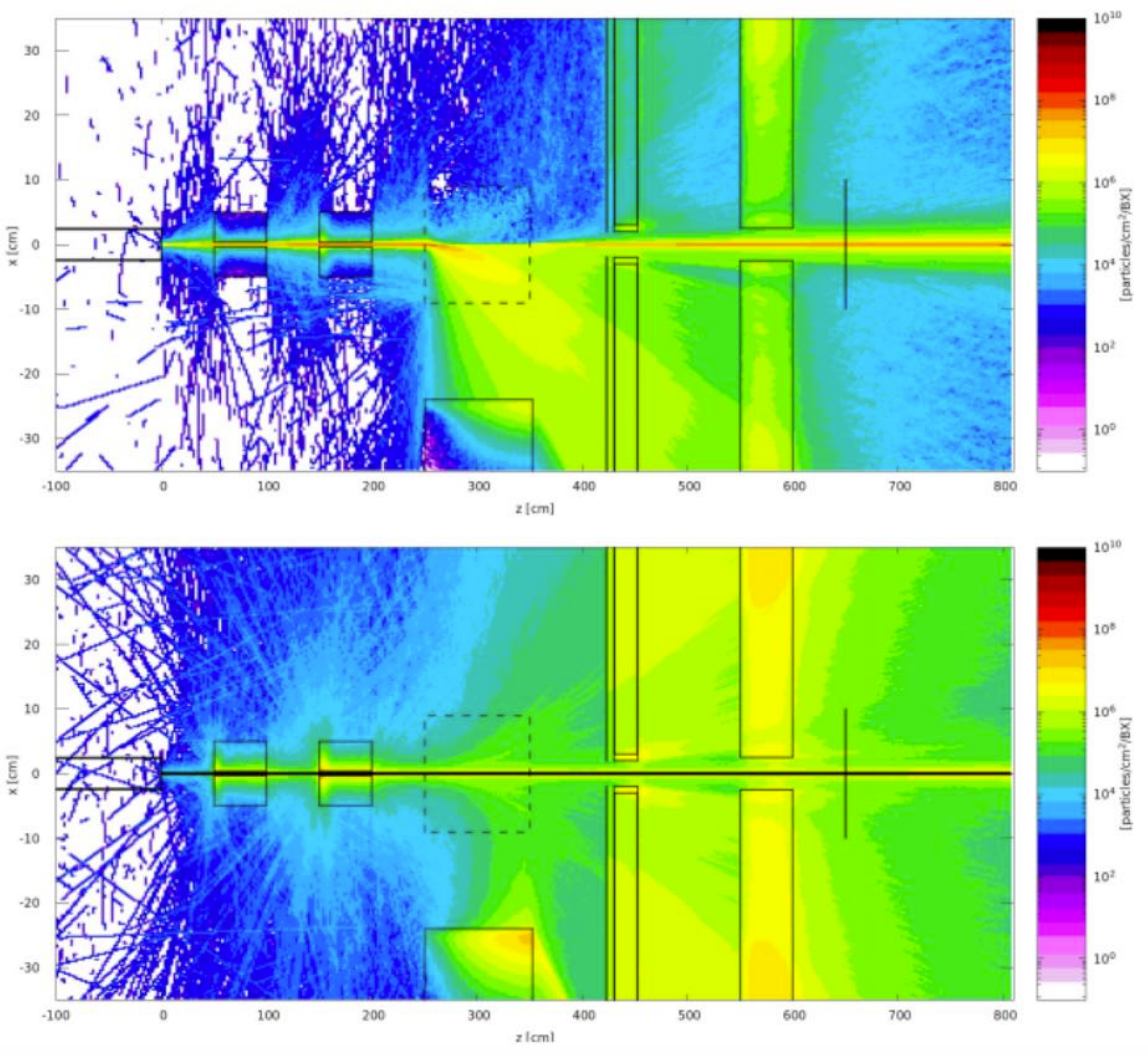}
    \caption{(Transverse spatial distribution (in units of particles per cm$^2$ per bunch crossing) of the (top) electrons and (bottom) photons for the \elaser configuration. The $z$ axis represents the propagation axis of the primary electron beam and the $y$ axis is the transverse coordinate. 
    \label{fig:fig3_gammaray_spectrometer}}
  \end{figure}

A side view of the spatial distribution of the electrons and photons through the spectrometer is given in Fig. \ref{fig:fig3_gammaray_spectrometer}. The positron distribution (not shown) is identical to the electron distribution, with the only difference of being flipped along the main electron beam propagation axis ($z$-axis).
  
The main sources of background in the system are secondary particles from events involving an interaction with air or any component of the spectrometer other than the converter, such as dipole magnets and collimators, off-axis photons and low-energy (tertiary) electrons exiting the converter. At the detector plane ($z=420$ cm), the signal electron flux is of the order of a few times $10^5$ particles/cm$^2$/BX, while the flux of background photons is $10^3 - 10^4$ particles/cm$^2$/BX and that of background electrons $\sim 10-10^2$ (see also Fig.~\ref{fig:elaser-barcharts}). Thus the signal/background is at least 10-100, and probably better as the light yield of the scintillator for incoming photons is lower than for incoming electrons.

Different scintillation materials can be used for the detectors, with the main candidates currently investigated being GadOx (Terbium doped gadolinium oxysulfide, $\mathrm{Gd}_2\mathrm{O}_2\mathrm{S:Tb}$), LYSO, and Caesium Iodide. The main properties of these scintillators are summarised in Table \ref{tab:table2_grs}. The GadOx screen is the preferred option due to the combination of a relatively high yield of scintillation photons ($\approx 
6\cdot 10^4$ photons per MeV of deposited energy), high radiation hardness ($>10^4$ Gy), reasonably fast response (600 $\mu$s), 
and emission wavelength in the optical range ($545 \units{nm}$), but far from any harmonic of the laser frequency. Furthermore it is advantageous that the same screens are used at the IP and in the bremsstrahlung detection system (see Sec.~\ref{sec:det:scintillator} and~\ref{sec:detectors:glaser}). However, the LYSO does offer the advantage that it has a much lower decay time which would offer an additional handle against background. However, it is less radiation hard and is at the edge of being able to withstand the dose expected in LUXE. At present, both options are still being considered. Caesium Iodide has a too low radiation hardness for the LUXE environment.

\begin{figure}[htbp]
  \centering
    \includegraphics[width=0.7\textwidth]{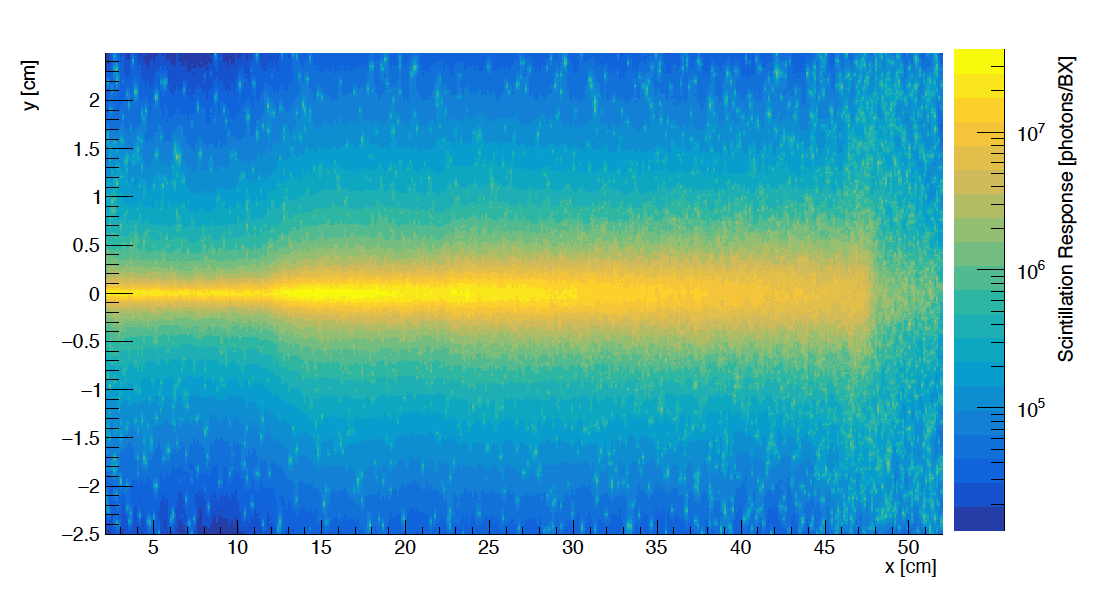}
    \caption{Simulated scintillation output at the electron detector plane for the representative example of the \elaser case, including both signal and background particles. The nominal beam axis is at $x=y=0$. 
    \label{fig:fig4_gammaray_spectrometer}}
  \end{figure}

The simulated scintillation output of the scintillation screens at the electron detection plane is shown in Fig. \ref{fig:fig4_gammaray_spectrometer}. It is seen that the signal is concentrated within $\pm 0.05$~cm around zero vertically and extends over the entire screen length. A similar output is obtained for the positron side (not shown).

\begin{table}[htbp]
    \centering
    \begin{tabular}{|c|c|c|c|}
    \hline
      & GadOx ($\mathrm{Gd}_2\mathrm{O}_2\mathrm{S:Tb}$) & LYSO & Caesium Iodide  \\
      \hline
      Crystal Density [g/cm$^3$]   & 7.32 & 7.1 & 4.51\\
      Decay Time [ns] & $\sim 6\cdot 10^5$ & 40 & 1000\\
      Light Yield [$\gamma$/MeV] & $6\cdot 10^4$& $3\cdot 10^4$ & $5.9\cdot 10^4$\\
      Peak Emission Wavelength [nm] & 545 & 420 & 560\\
      Refractive Index at Peak Wavelength & 2.2 & 1.81 & 1.84\\
      Afterglow [\%]& $<0.1$ after 3~ms & & $0.5-5$ after 6~ms \\
      Quantum Efficiency (at Peak Wavelength) &$\sim 0.15$ & 0.13 & 0.07\\
      Radiation Hardness [kGy] & $\sim 10^5$ & $> 10 - 10^4$  & $1$\\
      \hline
    \end{tabular}
    \caption{Comparison of three different scintillator crystals. 
    Data taken from Refs.~\cite{Krus:1999}, \cite{Gobain:2018}, \cite{Rihua:2007} and \cite{Eijk:2002}.  }
    \label{tab:table2_grs}
\end{table}

Typically 1 GeV of energy are deposited per mm, see Fig.~\ref{fig:elaser-barcharts}, and given the light yield of $6\times 10^4$ photons/MeV this results in approximately $10^{7} - 10^{8}$ scintillation photons for the signal along the main axis. This can be compared with a background of approximately $10^5$ scintillation photons. The signal and background can be compared in Fig.~\ref{fig:fig4_gammaray_spectrometer} where the signal is concentrated at the centre (at $y=0$) while the background is rather independent of $y$.  
The dimension of the screen is chosen as $5 \times 50 \times 0.05 \units{cm^3}$: in the $x$-direction this ensures that electrons at all energies are detected and in the $y$-direction the region $|y|>1$~cm can be used to constrain the background. 

For the LYSO scintillator, the dose deposited on the screen per bunch crossing is estimated from the simulations to be $3\times 10^{-4} \units{Gy}$, which corresponds to an annual dose of $1-10 \units{kGy}$, compared to a radiation hardness of $10-10^4$ kGy. 
This implies a risk that the LYSO would need to be replaced each year. For the GadOx, the dose per BX is only $10^{-6} \units{Gy}$ and the material is more radiation hard (up to 10 MGy), so that there are no concerns in this case. 

The scintillation photons can then be collected by a fast-focus lens and directed onto a i-CCD camera for imaging. As an example, assuming an F/2 large aperture lens and an isotropic scintillation emission, up to 5\% of the scintillation photons can be collected and imaged down onto an i-CCD camera. As an example, a gated and amplified i-STAR Andor camera will have a quantum efficiency of the order of 10-20 \% in the range of $400 \units{nm} - 500 \units{nm}$, implying a readout of the order of 10$^6$ photons per pixel. An advantage of LYSO over the GadOx scintillation screen is the fast response time of the LYSO ($40 \units{ns}$) which would allow for optical gating with a manageable loss in signal. 
Assuming e.g. a gating time of $5 \units{ns}$ one eighth of the signal is observed, i.e. $10^5$ scintillation photons, still large enough for a robust detection. Due to the larger decay time of the GadOx screen such a tight gating is not possible without large loss in signal. Once the scintillation photons are collected, they can also be optically transported to a quiet area, avoiding issues of electromagnetic noise onto the camera. 

\begin{figure}
    \centering
    \includegraphics[width=\textwidth]{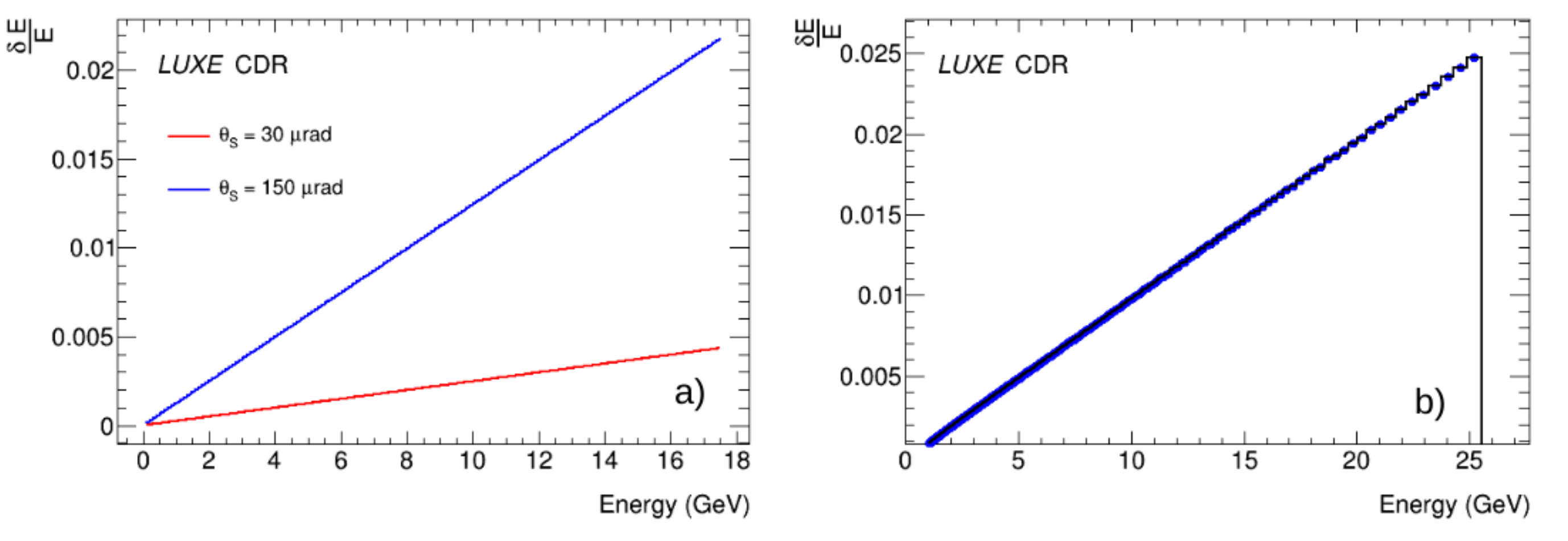}
    \caption{a) The relative energy resolution given by the divergence of the incident photon beam before conversion using the spectrometer parameters mentioned above. The values of 150 and 30~mrad correspond to the expected angular dispersion for photons with energies of 3 and 16 GeV, respectively. b) Relative energy resolution given by the non-linear relationship between spatial location on detector and the energy of the incident particle due to dispersion by the magnetic field (see also Eq.~(\ref{eq:spectrometer})) assuming a spatial resolution of 0.5~mm.}
    \label{fig:fig5_gamma_ray_spectrometer}
\end{figure}

The energy resolution at the detector is mainly determined by the pixel size of the scintillator and the combination of the magnetic field dispersion and the divergence of the incident photon beam on the spectrometer. The relative energy resolution \cite{Glinec:2006}, computed using the initial photon beam divergence, the magnetic field and the distance of propagation of the beam and is shown in Fig. \ref{fig:fig5_gamma_ray_spectrometer}(a) for two examples corresponding to photon energies of 3 and 16~GeV. 

Additionally, the magnetic dispersion of the particle beams causes a non-linear relationship between the position on the detector and the energy of the impacting particle. In particular, more energetic particles are dispersed less and so fixed spatial bins in the detector correspond to smaller energy bins; as the particle energy decreases, the width of the energy bin increases with decreasing particle energy. The relative energy resolution in this case is shown in Fig. \ref{fig:fig5_gamma_ray_spectrometer}(b). In both cases, there is a clear linear relationship between the electron/positron energy and the energy resolution, which, using the parameters mentioned above, gives an expected energy resolution on the few percent level. The deconvolution of the spectrum to obtain the photon energy spectrum is discussed in Sec.~\ref{sec:results}. 

\begin{figure}[htbp!]
    \centering
    \includegraphics[width=0.6\textwidth]{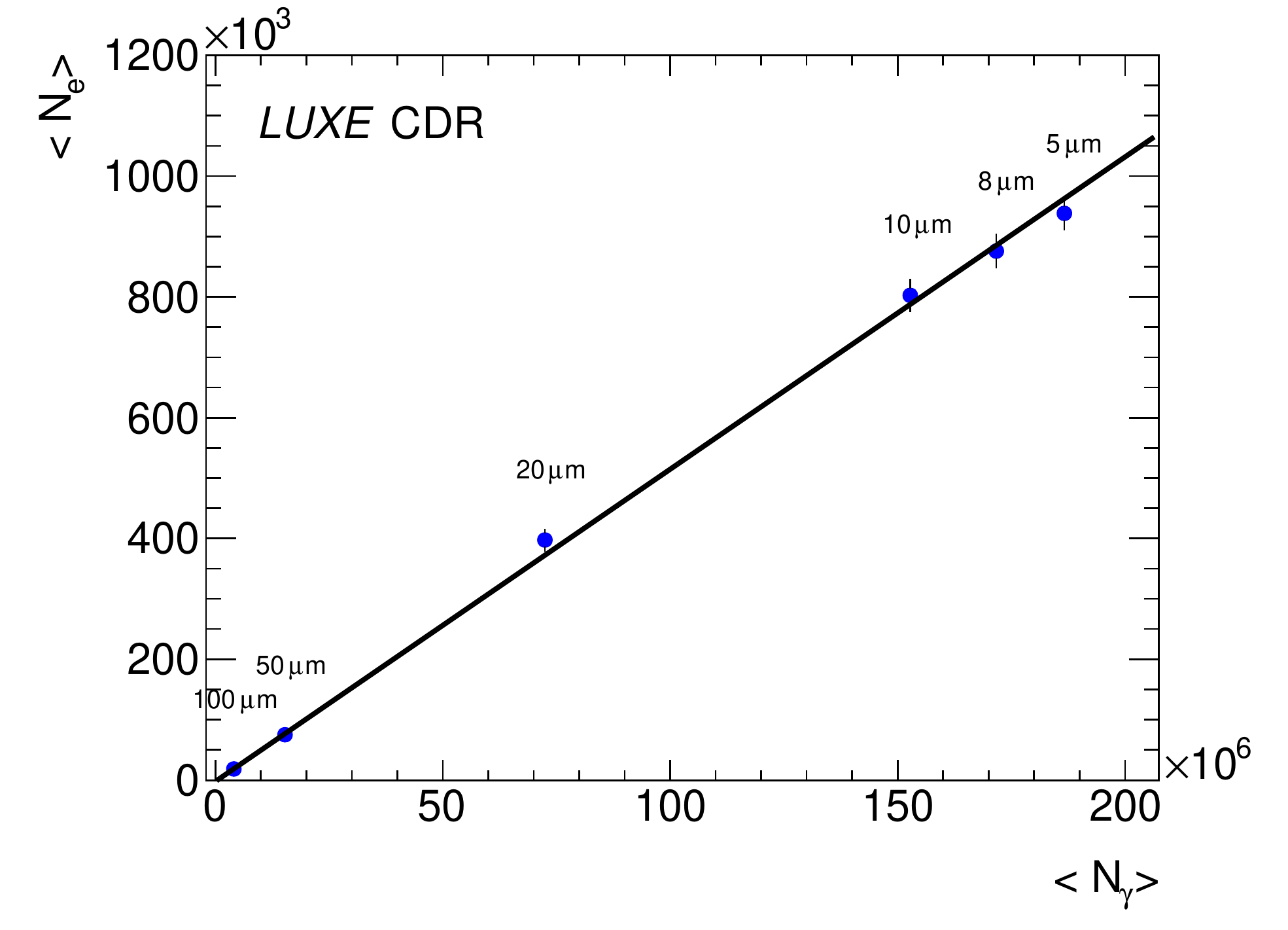}
    \caption{Mean number of electrons in the scintillator screen of the GRS versus the number of photons produced in the interaction. Each data point is labelled by the laser spot size. The uncertainties are statistical as expected for $\sim 500-1000$~BX.
    } 
    \label{fig:fig7_gamma_ray_spectrometer}
\end{figure}

The GRS can also be used to determine the total number of photons in a given BX. Figure~\ref{fig:fig7_gamma_ray_spectrometer} shows the number of electrons observed versus the true number of photons for various laser spot sizes. It is seen that there is a very good correlation over the entire dynamic range.

\subsubsection{Gamma Ray Profiler}
The purpose of the Gamma Ray Profiler (GP) is to measure the angular spectrum of the photons, with a precision corresponding to about $5$~$\mu$m in $x$ and $y$ at the location of the profiler. It is foreseen to be a set of two sapphire strip detector with a pitch of $100$~$\mu$m staggered behind each other with a relative rotation of $90^\circ$, so that they effectively represent a lattice-like structure. Two positions for the profiler are being explored: either at $z\sim 6$~m (before the GRS, near end of the vacuum beam pipe) or at $z\sim 11$~m behind the GRS. The former is at present considered the better option as it suffers less from scattering of particles in the air before they arrive at the GP.
The possibility to have two such stations, one placed just after the vacuum pipe and one just before the beam dump wall is also under consideration, but detailed Monte Carlo simulations, not available at this stage, will be needed to decide the usefulness of such a setup.

The profiler must operate in a stable and reliable way in presence of a very intense high energy gamma ray flux. The typical photon energy is in the GeV range, where electron-positron pair production is the dominant process in the detector. 

The expected beam spatial distributions in the directions perpendicular to the beam direction ($z$): $x$ (parallel to the laser beam polarisation axis) and $y$ (perpendicular to the laser beam polarisation axis), have been simulated using \textsc{Fluka}.
The simulation shows that most of the profile is contained in a $2\times 2\;\text{cm}^2$ square, although there are long tails.
In the case of the beam profiler placed just out of the vacuum pipe, at about 6\,m from the interaction point, 
the central high-energy component of the gamma beam can be approximated with a Gaussian having standard deviations $\sigma_{x}=\textrm{max}(1,\xi)\times 180\;\mu\text{m}$ for $\xi>1$,  $\sigma_y=180\;\mu\text{m}$ (see also Sec.~\ref{sec:theory:beamprofile}). For instance, for $\xi=5$, $\sigma_{x}=900\;\mu\text{m}$ while for $\xi<1$ the spot size remains $\sigma_x=180\;\mu\text{m}$. 
By measuring $\sigma_x$ with a precision of 5~$\mu$m the $\xi$-value can be determined with a precision better than 1\% for $\xi\gsim 2$, exceeding the precision expected for the laser diagnostic of 5\%. 
The total intensity per bunch crossing (BX) of the gamma beam is typically $\phi=10^7-10^9$ photons/BX. Given this very high flux that the detectors will need to withstand, a sapphire strips detector is proposed since this material very radiation hard (up to 10 MGy~\cite{Karacheban:2015jga}).
The choice of a strip readout will allow to keep the readout electronics in a low irradiated area. Two successive strip detectors (with strips placed perpendicular between themselves) are used to measure both coordinates with good precision. 
Scattering between the two detectors is negligible. 
The second detector will be submitted to twice the radiation dose and will have higher signal amplitude, but the benefit to save the readout electronics by high irradiation level will compensate these drawbacks which are anyway expected to be tolerable.
The detectors should be as thin as possible to minimise the absorbed dose while still preserving an acceptable signal level.

These requirements are met by a sapphire detector having $2\times 2\;\text{cm}^2$ area, $dz=100\;\mu\text{m}$ thickness and $\Delta x(\Delta y)=100\;\mu\text{m}$ strip pitch, resulting in 400 readout channels for both detector planes together. 
The digital resolution of such a detector would be $100/\sqrt{12}$~$\mu$m but it can be improved reading out the charge for each strip (analogue readout) and using that to determine the centre-of-gravity (COG) via a clustering algorithm.  

These detector specifications were optimised assuming a Gaussian beam with a width of $(1 .. \xi)\times 180$~$\mu$m as described above. However, an expected broader distribution of low-energy photons might require a somewhat larger detector area, in order to fully capture the photon beam. A full numerical study, based on QED MC, \geant or \textsc{Fluka} will be performed in order to finalise the design parameters.

\paragraph{Sapphire detector characteristics} 

Artificial sapphire (aluminium oxide) is a wide bandgap (9.9\,eV) isolator material that since the last decade started to be used as a particle detector in high energy physics~\cite{Karacheban:2015jga}. 
 It has excellent mechanical and electrical properties and it is produced industrially in large amount. 
 It was shown recently that optical grade sapphire can be used directly as a detector material even without extra purification with the advantage of low cost and superior radiation hardness.
 
The signals collected from sapphire detectors are a relatively small 22 electron-hole (eh) pairs per micron of a MIP track and low ($\sim 10\%$) charge collection efficiency, but they are suitable for large fluxes of particles simultaneously hitting the detector. The extremely low leakage current ($\sim$pA) at room temperature even after high dose irradiation makes sapphire detectors practically noiseless.
 
The detector design includes a thin sapphire plate with continuous metallization at one side and a pattern of electrodes (pads or strips) at the opposite side. For the metallization aluminium is used with a typical thickness of a few microns. The operational voltage of the detector is few hundreds volt, the signal duration is several nanoseconds, and the charge collection is completely dominated by electrons.

\paragraph{Detector performance} 

The performance is estimated for a GP station with two such sapphire strip detectors, placed at 6m distance along the beam direction, with the $y$-strip detector upstream and the $x$-strip detector downstream.

Assuming that the photon beam mainly produces electron-positron pairs in the detector material~\footnote{For GeV photons pair-production dominates compared to the photoelectric and Compton effect.}, the relevant quantities to characterise the detector performances can be calculated analytically, at least to first approximation. 
Table~\ref{tab:output} summarises the results for a $y$--strip detector.

\begin{table}[htbp]
  \centering
    \begin{tabular}{cccl}
    \toprule
    Quantity & Value & Units & Description \\
    \midrule
    $D_{MAX}$ & $1$ & Gy/BX & Maximum dose absorbed \\
    \midrule
    $N_{MAX}$ & $15\cdot 10^7$ & eh pairs/BX & Maximum number of eh pairs \\
    \midrule
    $Q_{MAX}$ & $25$ & pC/BX & Maximum collected charged \\
    \midrule
    C &  $2$ & pF & Strip capacitance \\
    \bottomrule
    \end{tabular}%
      \caption{Relevant characteristics of the proposed sapphire strip detector assuming a photon flux of $10^9$ photons per BX. } 
  \label{tab:output}%
\end{table}%

Using a COG algorithm, a spatial resolution of about $11 \;\mu\text{m}$ in a single BX can be achieved with a detector energy resolution $R \simeq 1\%$. 
Since the spatial resolution scales as $1/\sqrt{\text{BX}}$, the desired spatial resolution of $5 \;\mu\text{m}$ can be achieved with $\sim 4 \; \text{BX}$. Strips of $50 \;\mu$m would allow to get the target spatial resolution for each individual BX.

The annual radiation dose is calculated as $10$~MGy assuming a flux of $10^9$ photons per BX, $10^7$~s of operations and and the maximum dose reported in Table~\ref{tab:output} ($D_{MAX}$), and corresponds approximately to the dose the detector can withstand. Thus the detector would need to be replaced each year. However, in \phaseone the average photon flux will be below about $2\times 10^8$/BX (see Fig.~\ref{fig:rates_elaser}) so that this detector should suffice for \phaseone.
If needed, the lifetime can be increased of a factor one hundred by providing the detectors support with two independent sub-millimetre movements in the strip direction ($100\;\mu\text{m}$ steps).

This is possible with sapphire detectors (not with silicon) since the leakage current of the irradiated sensor remains at negligible levels (few pA) after intense irradiation.

\paragraph{Readout electronics} 
Given the low BX rate (1\,Hz), the speed of the readout electronics is not a concern.
However, the charge produced in the strips with higher occupancy is huge and must be accurately measured to describe the tails of the beam profile (12 bits seems necessary).
The energy resolution of the readout electronics should be at the level of few percent.
This will allow to reconstruction the beam profile with a spatial resolution better than $5\;\mu\text{m}$ after collecting some tens BX and will not impose limitations to the detector capacitance, which in any case is quite low as shown in Table~\ref{tab:output} ($C \sim 2$\,pF).
The strip signals will be transmitted to the LUxe Beam Profile ReadOut board (LUBPRO) via Cat-5 cables (four signals per cable, a total of 100 cables).
The LUBPRO accepts the signals from the individual lines and feeds them into a sample and hold capacitor.
A multiplexer will present the signals to the single output of this card. 
The single output of this card will be fed into a general purpose FADC card which will process the amplitudes and store them in a circular buffer for readout. 
The LUCROD~\cite{Bruschi:2015kev} card developed for the readout of the ATLAS LUCID detector is a valid option for the readout.
Each LUBPRO card will have 10 RJ45 inputs and 1 analog output on LEMO connector.
The entire beam profiler will be read with 10 LUBPRO cards and one LUCROD card with 16 input channels.

\subsubsection{Gamma Flux Monitor}
\label{sec:detectors_photflux_GammaMonitor}

The overall flux of photons created in the \elaser interactions is high, e.g. for values $\xi\gsim 1$ the number of photons exceeds $10^{8}$. This large number makes counting of photons and measuring their spectrum challenging. 
A solution complementary to the GRS and GP is proposed here based on measuring the energy flow of particles back-scattered from the dump that absorbs the photon beam at the end of the beam-line. 
This detector provides a simple and robust way to monitor the variation in the photon flux with time. It may also be used at the beginning of the run to optimise the collisions. It plays a role very similar to a luminosity monitor in collider experiments.

\begin{figure}[ht!]
  \begin{center}
    \includegraphics[width=0.45\textwidth]{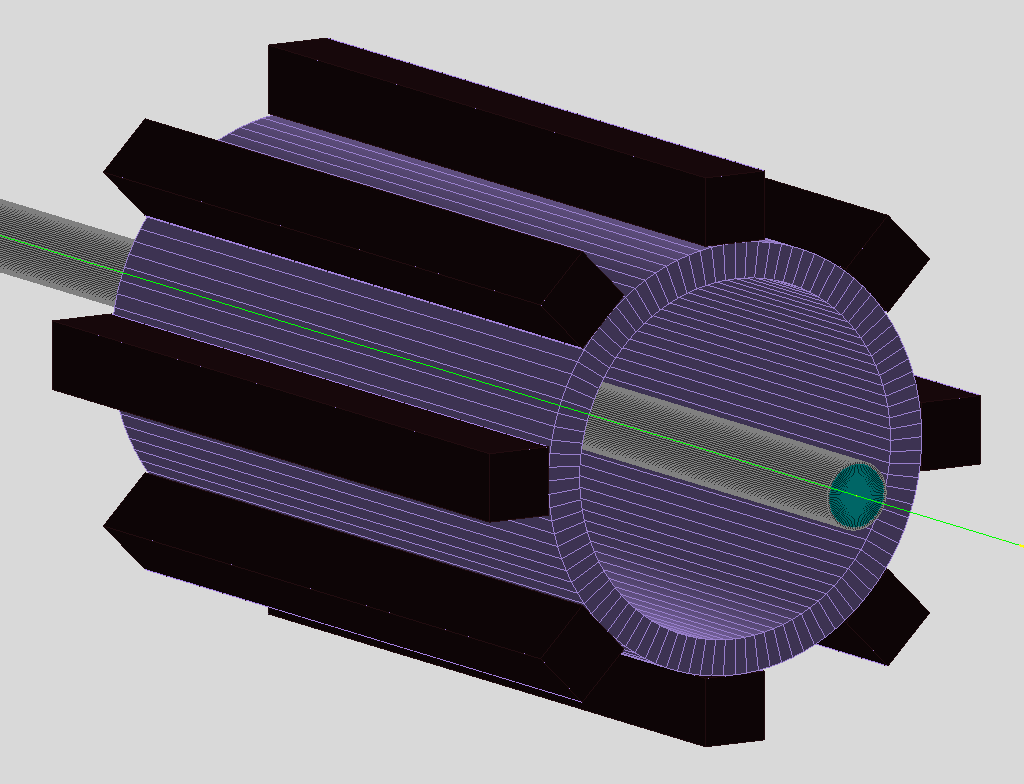}
      \end{center}
    \caption{Design of the back-scattering calorimeter built of eight TF-101 type lead-glass blocks, each of size $3.8 \times 3.8 \times 45 \units{cm^3}$. }
    \label{fig:GM_conf}
\end{figure}

The Gamma Flux Monitor (GFM) consists of a simple calorimeter assembled from eight blocks of lead-glass, oriented longitudinally with respect to the beam line with PMTs looking upstream, as shown in Fig.~\ref{fig:GM_conf}. This configuration was selected to ensure the radiation load is tolerable, and the eight blocks were arranged in this way to provide some sensitivity to the transverse location and dimension of the photon beam. The blocks consist of $3.8 \times 3.8 \times 45 \units{cm^3}$ TF-101 type lead-glass modules. Each module is $18 X_0$ deep. The chemical composition and relevant physical properties of the lead-glass are listed in Table~\ref{tab:LGproperties}.

\begin{table}[htbp]
  \begin{center}
    \begin{tabular}{|l|l|l|}
      \hline
      \textbf{Chemical composition  } & weight  &Fractions atomic units  \\
      \hline
       $\mathrm{Pb}_{3}\mathrm{O}_{4}$   &  51.23 &	Pb - 0.0795  \\
       $\mathrm{SiO}_{2}$  & 41.53	 &  O - 0.6223  \\
       $\mathrm{K}_{2}\mathrm{O}$   & 7.0   &  Si - 0.2450 \\ 
       $\mathrm{Ce}$  &  0.2 &	K - 0.0527  \\
         &   &	Ce - 0.0005  \\
            \hline
            \hline
       Radiation length (cm) &2.78 & \\
       Density ($\mathrm{g/cm^3}$) & 3.86 & \\
       Critical energy (MeV) & 17.97 & \\
       Refraction index     & 1.65 & \\
       Moliere radius (cm)     & 3.28 & \\
       Thermal expansion coefficient ($C^{-1}$)     & $8.5\cdot 10^{6}$ & \\
      \hline
    \end{tabular}
    \caption{Chemical composition and physical properties of the lead-glass TF-101.}
     \label{tab:LGproperties}
  \end{center}
\end{table}

\begin{figure}[ht!]
  \begin{center}
    \includegraphics[width=0.45\textwidth]{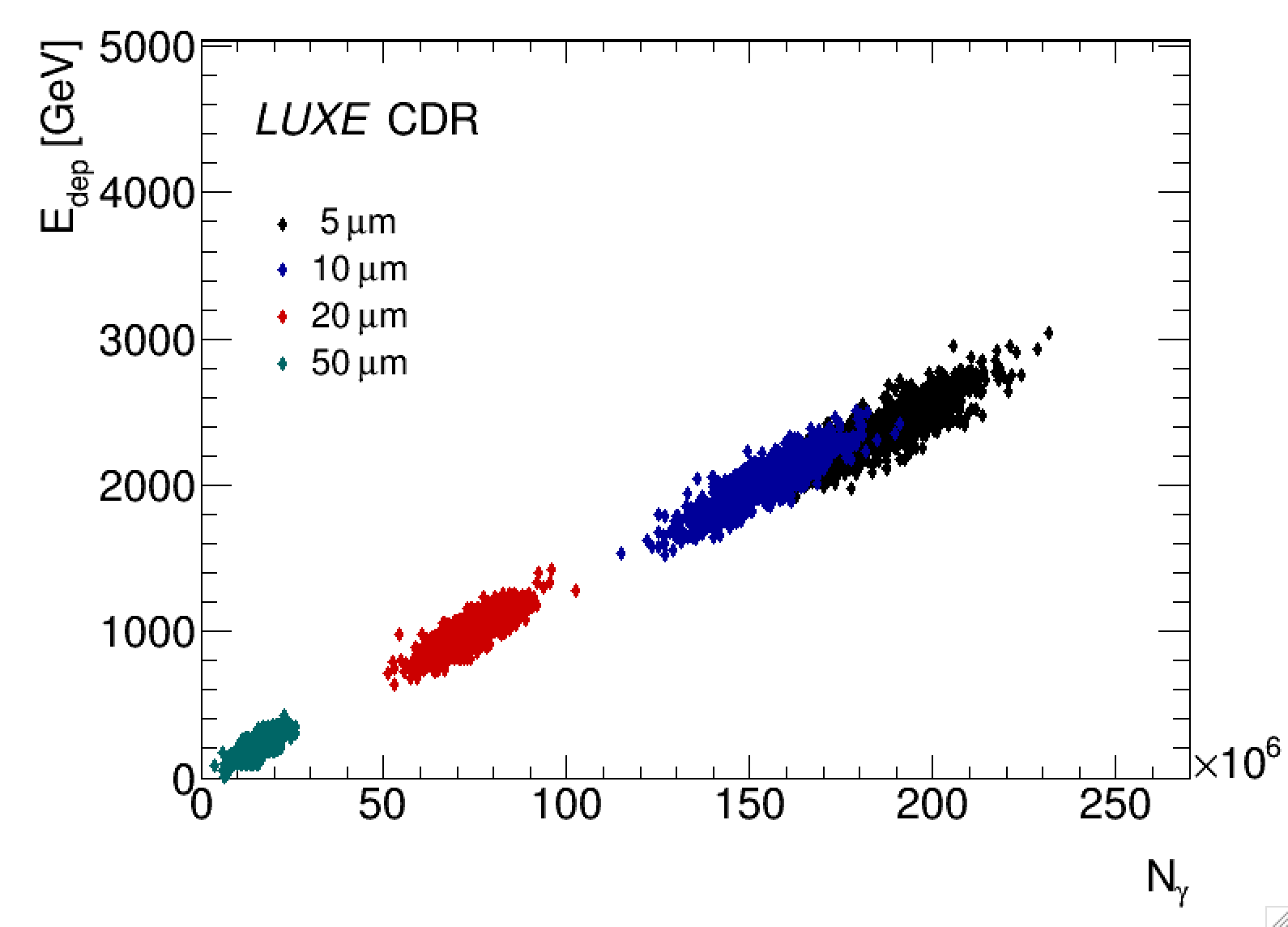}
    \includegraphics[width=0.45\textwidth]{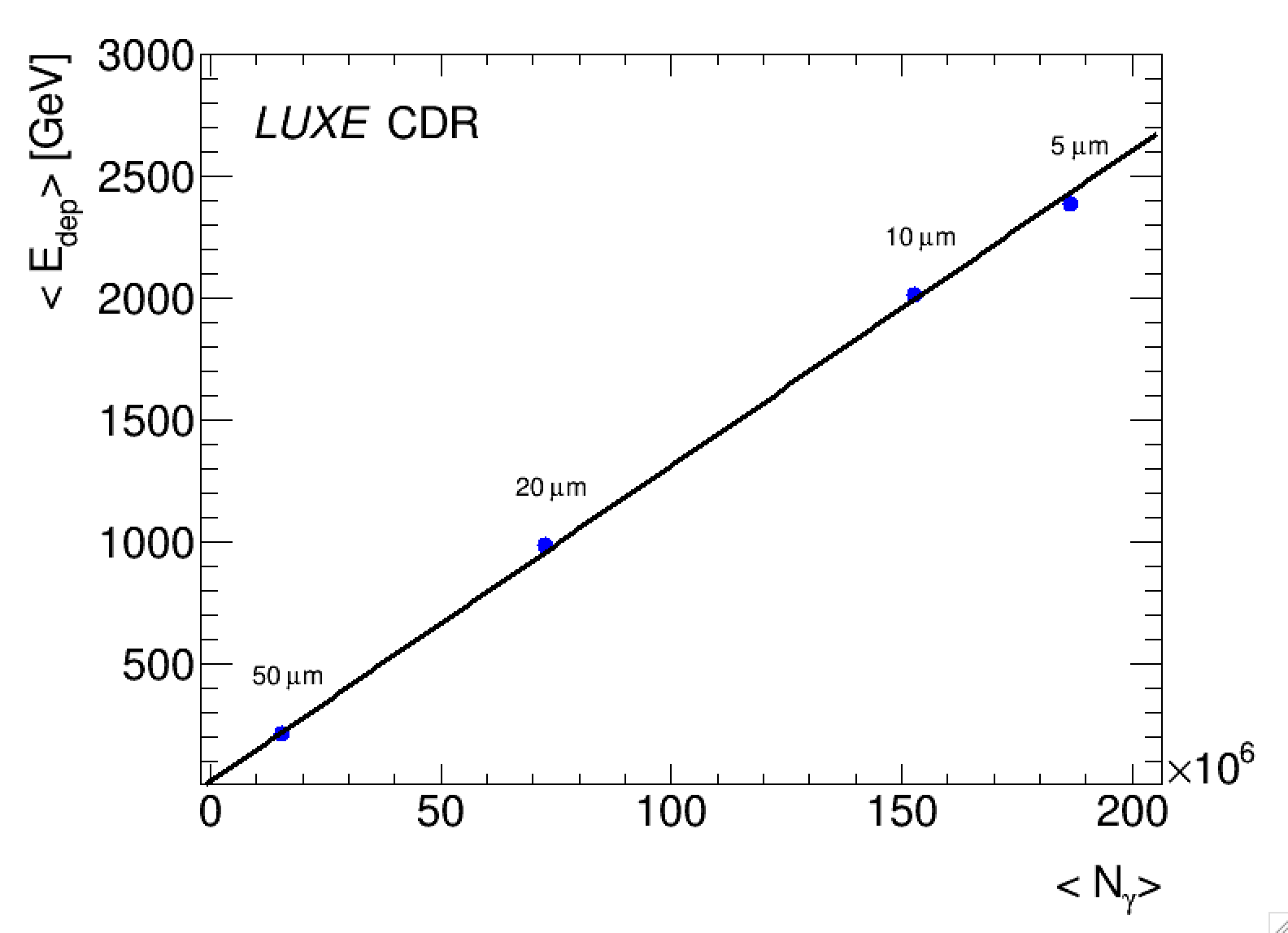}
      \end{center}
    \caption{Left: Deposited back-scattered energy, $E_\mathrm{deposited}$, as a function of the number of photons, $N_\gamma$. Each point corresponds to one bunch-crossing. Right: Average energy deposition versus the mean number of photons.}
    \label{fig:GM_perform}
\end{figure}

\geant simulation is used to study the performance of the GFM. 
The expected back-scattered energy deposit, $E_{dep}$, in the GFM  as a function of the number of dumped photons, $N_{\gamma}$, is shown in Fig.~\ref{fig:GM_perform}, for different laser intensities (expressed in terms of the laser spot size).
A clear correlation is seen between the deposited energy and the number of incident photons. 

The precision with which the number of photons can be determined is estimated as
\begin{equation}
\label{eq:deltaN}
\Delta N_\gamma = \frac{\partial N_\gamma}{\partial E_{dep}} \Delta E_{dep} \, ,
\end{equation}
 where $\frac{\partial N_\gamma}{\partial E_{dep}}$  is the value of the slope in Fig.~\ref{fig:GM_perform} (right) and $\Delta E_{dep}$ is the measure of the inherent fluctuations of the back-scattered particle flow.
The uncertainty on the number of measured photons 
is in the range of 5 to 10\%, decreasing with increasing photon flux.
To achieve this level of precision, a system of standard candles will be implemented to monitor the stability-response of the crystals, in particular because some deterioration with time may be expected due to exposure to high radiation dose

The radiation dose deposited in the calorimeter per bunch-crossing is estimated from the simulations to be $\sim 7\cdot 10^{-8}\units{Gy}$ per lead-glass block. For a year with $10^7$~s this corresponds to an annual dose of $0.7$~Gy. 
According to Ref.~\cite{Kobayashi:1994kf} a dose of $20$~Gy can be tolerated with $<1\%$ degradation for a device of 2.8~cm length. For the present device with 45~cm length the tolerance is 320 Gy. Thus the detector can withstand the radiation expected in LUXE for many years.

\subsection{Data Acquisition and Control System}
\label{sec:daq}

As the maximum data-taking frequency will be 10\,Hz and all detectors are relatively
small, current data acquisition (DAQ) solutions are appropriate for LUXE.
Calibration data will also be needed, e.g.\ when there is an electron bunch but no
electron–laser events or when there are no electron bunches, in order to measure
pedestals, noise and backgrounds. When running with electron bunches only, the
well-known energy of the electrons will allow their use for detector alignment
and cross checks of the magnetic field.  These calibration and alignment data
again should not yield large data rates or high data volume. The DAQ system will
need to be bi-directional as control data will need to be sent to the detectors,
e.g.\ to distribute timing information, to control motorised detector stages, etc..

Typically, the data rate from one tracking detector is expected to be $<$ 0.5\,MB/s.  This assumes that all chips of all staves see the same number of hits from 100 background particles at 9\,Hz (with no laser) and 500 signal particles at 1\,Hz.  Under the most extreme conditions, at the highest laser intensities (and under the same assumption of all chips seeing the same number of particles) the data rate is expected to reach at most 5\,MB/s mainly due to a factor of $\sim$\,10 higher signal-particle rate. In reality, the distribution of hits will not be uniform. Hence, this number should be seen only as an upper bound which would be relevant for a short period in the overall data-taking.  Assuming a camera used to image a scintillation screen has 1\,Mpixels with 8\,bits, then at 10\,Hz, reading out all pixels gives a rate of 10\,MB/s.  These rates, $< O(10\,{\rm MB/s})$, are typical of all sub-detectors.

\begin{figure}[hbt]
\centering
    \includegraphics[width=0.8\textwidth]{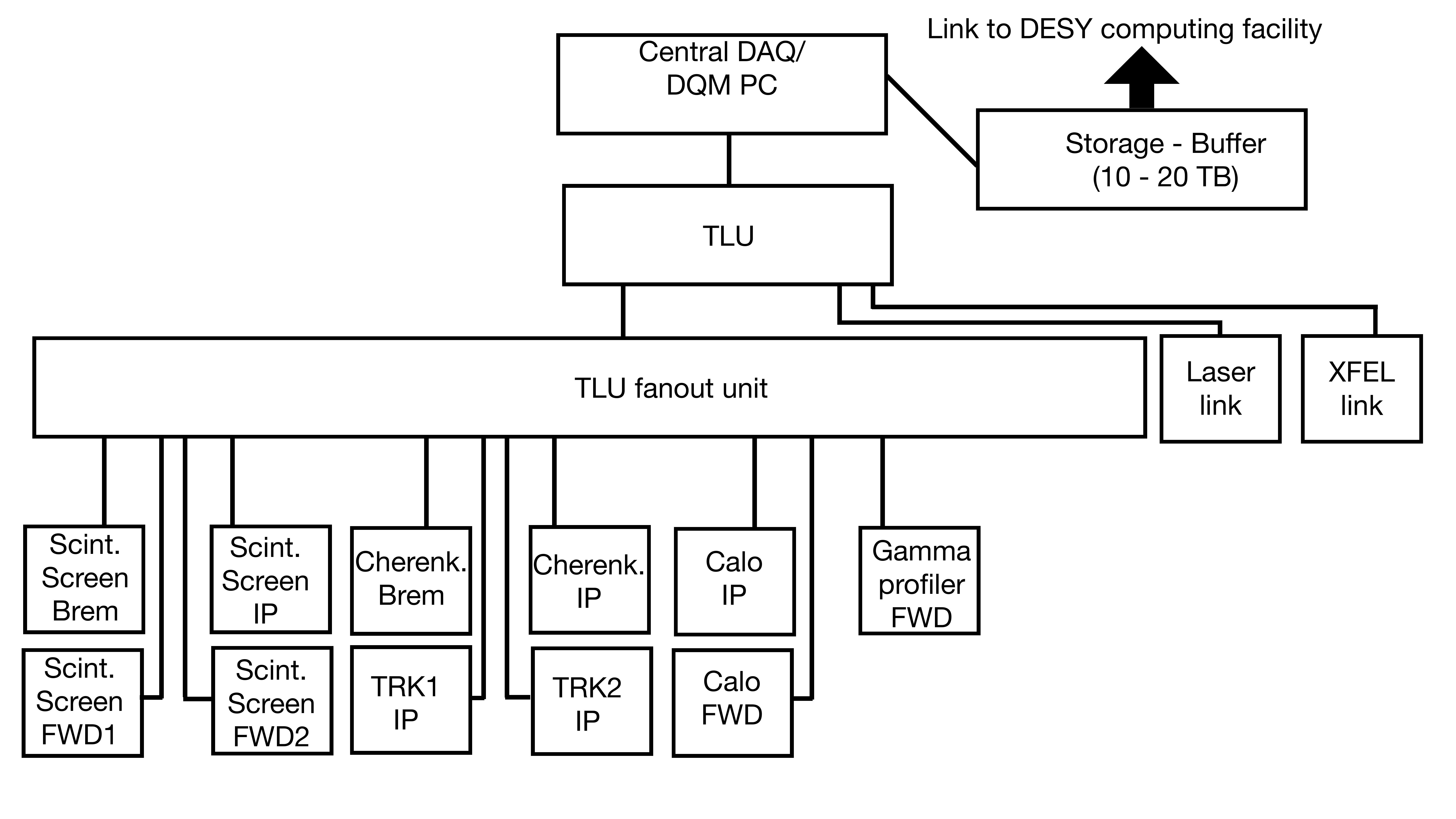}
    \caption{Proposed architecture of the data acquisition.}
    \label{fig:DAQ-diagram}
\end{figure}

Therefore each detector component, e.g.\ silicon tracker after the IP or \cer
detector can be read out and controlled by one front-end computer (PC) which is
interfaced to the front-end electronics.  Note more than one PC may be used for
contingency, although the actual data rates would not require this. Each of the
detector PCs will then be connected to a central DAQ PC, based in the LUXE control
room, as well as a trigger and control system. Further PCs will be required in the
control room to display detector information for e.g. data quality and slow control
monitoring purposes. The proposed architecture of DAQ is shown in Fig.~\ref{fig:DAQ-diagram}.

The Trigger Logic Unit (TLU)~\cite{Baesso:2019smg,TLU:www} developed most recently within the EU AIDA-2020
project has been designed to be a flexible and easily configurable unit to provide
trigger and control signals to devices employed during test beams and integrating them
with pixel detectors.  It has been used extensively at DESY as well as CERN beam lines
by a number of different detectors.  The unit can accept signals from some detector and
can generate a global trigger for all LUXE detectors to indicate the start and end of
data taking.  The unit can also act as a master clock unit, receiving synchronisation
and trigger commands from the software DAQ, as well as a precise clock reference.
Therefore the timing from the electron bunch and/or the electron--laser timing will be
fed to the TLU which will then synchronise the LUXE detectors.  The detectors can also
send signals back to the TLU to indicate that they are busy and, for example, data
taking to be paused until the busy is removed.  Communication between the DAQ system
and the TLU uses the IPBus protocol, a well-established and reliable protocol widely
used in the CMS and ATLAS experiments at CERN.

A DAQ software will be required which could be used for individual components or,
if they come with their own software, a central DAQ software will be needed to
interface to the detector software.  Many different DAQ softwares exist with
different levels of complexity and scale, with some designed for specific
experiments and others as generic developments. At LUXE, we propose to use the
EUDAQ2~\cite{Liu:2019wim,EUDAQ2:www} software which has been developed for tests of high energy physics
prototype detectors (in beam tests), a similar setup to LUXE.  It can cope with
different triggering and readout schemes and has been used by several different
detectors and projects.  Specifically, EUDAQ2 has been used in beam tests to read
out several detectors like those to be used in LUXE, such as calorimeters and 
pixel detectors~\cite{Liu:2019wim}, including the ALPIDE sensors.  EUDAQ2 is also
fully compatible with the TLU and the two have been used as a basis for DAQ
systems in several situations.  The software has been used extensively in DESY
test beams, as well as other facilities.  DESY is also the lead developer of the
EUDAQ2 software and so we will have access to in-house expertise.

The hardware, firmware and software designs of the TLU~\cite{TLU:www} and the EUDAQ2 software~\cite{EUDAQ2:www}
are freely available and so work can start quickly on their implementation in LUXE.

\subsection{Modifications Required for Photon-laser Interactions}
\label{sec:detectors:glaser}
In addition to the study of \elaser interactions, an important part of the physics programme is the study of \glaser interactions using either bremsstrahlung photons generated in a tungsten target or photons generated via inverse Compton scattering as discussed in Sec.~\ref{sec:science:ics}. 
This has never been studied directly and provides a unique possibility to directly measure the Schwinger process. 

\begin{figure}[hbt]
\centering
    \includegraphics[width=0.8\textwidth]{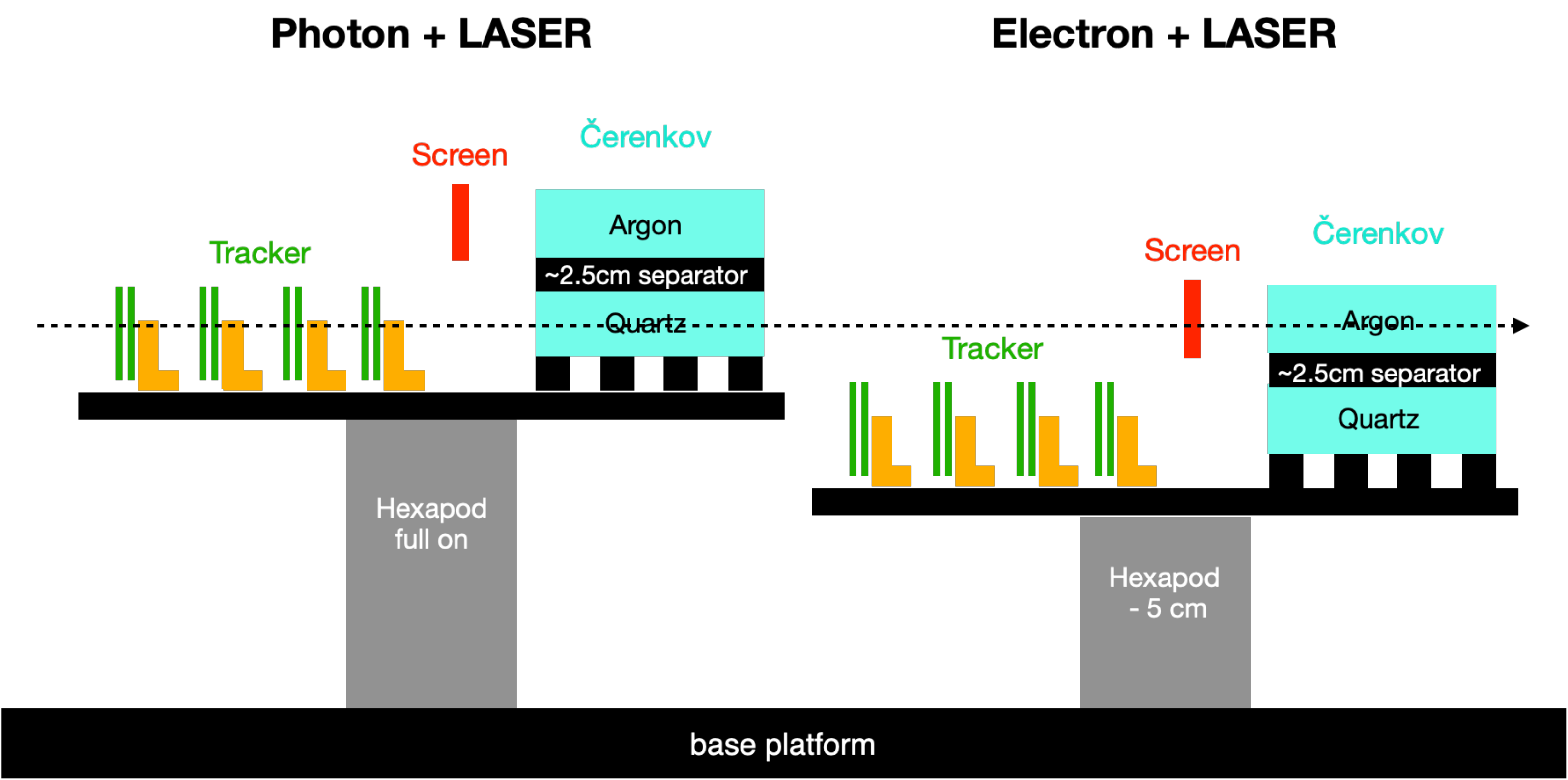}
    \caption{Sketch of the hybrid IP detector setup for option 2 (see text). Shown are the tracker and the \cer detector on a stage that moves vertically depending on whether \glaser or \elaser data taking is desired. }
    \label{fig:hybridsetup}
\end{figure}

However, a technical challenge is that in this case the rate of physics events is rather low, so that it is difficult to control the experiment (e.g. data quality monitoring etc.) at the level required for physics analyses \textrm{in-situ} based on the \glaser collisions alone. Thus the solution envisaged is a hybrid setup that allows to switch between \elaser and \glaser collisions via remote operations as described below and sketched in Fig.~\ref{fig:hybridsetup}. 
    
The following modifications are needed:
\begin{itemize}
    \item For monitoring and measuring the flux of photons at the tungsten target, a system to measure the electrons from the bremsstrahlung process is needed. The fluxes here are high and the \cer and scintillator screens are also adequate for this purpose. The scintillator setup is identical to the one behind the IP in the \elaser setup. The \cer channel size is increased to $1\units{cm}$ as the energy resolution requirement is more relaxed here. The magnet is oriented such that the electrons are deflected towards the floor, and the detectors are also oriented in the vertical plane. They are about 0.5~m long. The setup looks very similar to Fig.~\ref{fig:ceds-diagram} except that both detectors are in the vertical plane and have the same length. 
    \item Behind the IP, the detectors on the positron side are identical to those in the \elaser setup. The electron side, however, now has the same low flux as the positron side and needs to be modified. And, in order to be able to switch between the \elaser and \glaser mode it is placed on a stage that can move vertically to switch between the two setups (see discussion below).
    The rates of signal electrons and positrons are below $0.1$ for the \phaseone setup, compared to typical rates for backgrounds of 10 (see Fig.~\ref{fig:glaser-barcharts}) thus requiring a background rejection better than 1/100. 
    Three solutions are being considered
    \begin{enumerate}
        \item duplicate the setup of the positron side, i.e. use a silicon pixel tracker and a calorimeter;
        \item use the silicon pixel tracker as on the positron side but use a Quartz \cer detector (sensitive to individual particles) behind it;
        \item use only the silicon pixel tracker.
    \end{enumerate}

The three solutions have different implications on the background rejection power and the financial resources required. 
    At present, the statistics for background simulations is not yet sufficient to evaluate backgrounds below 0.1 
    particles/BX, an thus no decision could be made. Furthermore, this hybrid setup is somewhat complicated mechanically, and will need to be 
    engineered professionally, which due to limited resources, has not yet been possible. 
    \item The three detectors of the photon detection system are unchanged. They are now mostly used for monitoring the stability of the bremsstrahlung photon flux. 
\end{itemize}

Operationally, a \glaser run would commence as follows. At the start of each run first \elaser collisions are established, and data are taken to ensure that the laser and beam have good overlap and the signals observed in the detectors are as expected. Then, the target in the target chamber will be moved via remote control into the $e$-beam. The bremsstrahlung detectors will then see the signals and once operation is stable, the stage that holds the IP detectors on the electron side will be lowered, so that the electrons are measured on that side by the system described under item 2 above. 

\subsection{Detection of BSM Particles}
\label{sec:det:bsm}
As discussed in Sec.˜\ref{sec:NPLeff}, LUXE also has potential for probing physics beyond the Standard Model directly, in particular by using the very high photon flux to search for anomalous production of new scalar or pseudo-scalar particles that couple to photons, have lifetimes of ${\cal O}(1)$~ns  and masses between about 10~MeV and 1~GeV .

\begin{figure}[hbt]
\centering
    \includegraphics[width=0.5\textwidth]{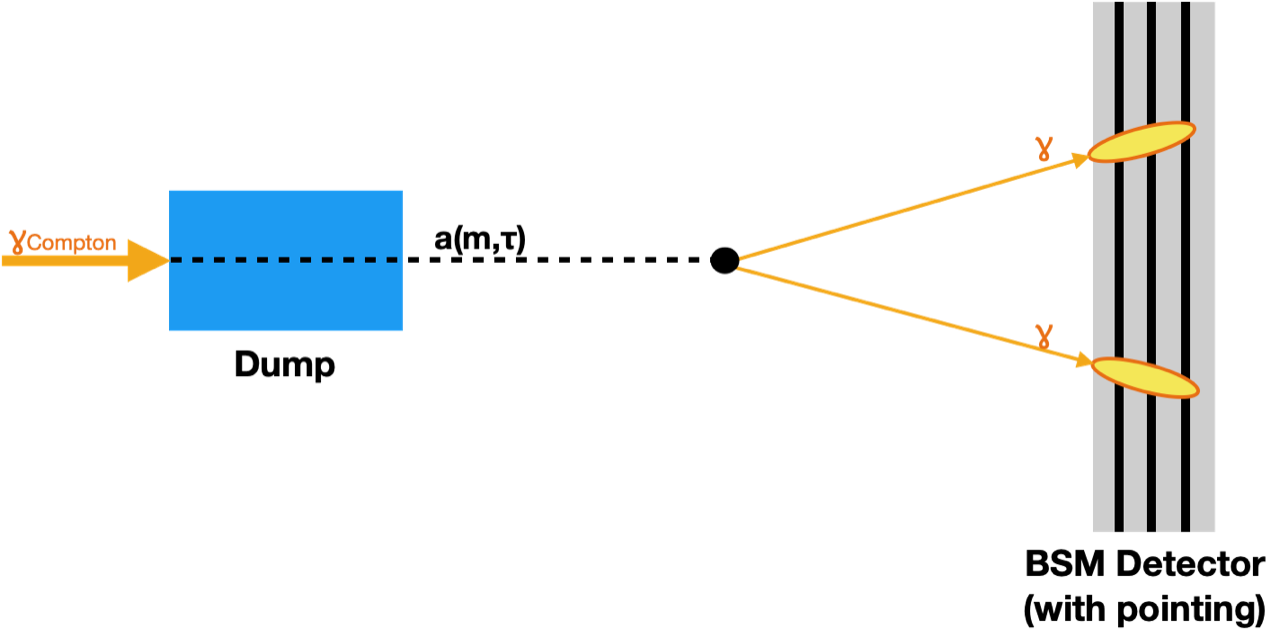}
    \caption{Illustration of the ALP decay. The ALP with mass $m$ is produced in the dump and then decays to two photons after it exits the dump with an opening angle $\alpha$ at a location $\zalp$}
    \label{fig:det:bsm}
\end{figure}

For this purpose, behind the photon dump situated about 13.5~m behind the IP and the final wall at $z$=17.4~m, a decay tunnel (with length 3.3~m) is considered at the end of which detectors can be placed to search for photons from the ALP decays. This is illustrated in Fig.~\ref{fig:det:bsm}. 
The typical opening angle, $\alpha$, between the two photons for the mass range considered is below $15^\circ$, thus if all decayed right after exiting the beam dump at a distance of 3.3~m they would be spread by up to about 1~m. This is roughly the size of the detector needed. 

The technology for the detectors has not yet been worked out in detail. A plausible idea would be an electromagnetic (EM) calorimeter followed by a hadronic calorimeter to reject neutrons. 

The invariant mass is given by $M=\sqrt{2\varepsilon_1\varepsilon_2(1-\cos\alpha)}$ where $\varepsilon_{i}$ are the energies of the two photons and $\alpha$ is the angle between them. A mass resolution of 10\% could be achieved for instance with an energy resolution of 10\% and an angular resolution of $\lsim 20$~mrad. 

Backgrounds from charged particles can be rejected via an additional scintillator veto system, or they can be actively deflected by a permanent magnet placed between the photon dump and the BSM detector. If an EM calorimeter is used as a BSM detector, the electromagnetic shower shape can also be used to reject non-photon backgrounds. 

Both for estimating the mass and 
to obtain information about the ALP lifetime, $\taualp$, it is beneficial to measure the photon entrance angle in the detector in order to point back to the ALP candidate decay vertex, $\zalp$ (see Fig.~\ref{fig:det:bsm}). A sampling calorimeter with sufficiently fine segmentation in the transverse and longitudinal planes could provide information about the electromagnetic shower direction. Such pointing information is also useful to reject backgrounds. 

The angular resolution that is deemed technically possible using a sampling calorimeter with shower direction reconstruction capability is $\sigma_\theta=16\units{mrad}/\sqrt{E/\units{GeV}}$, see Ref.~\cite{Bonivento:2018eqn}. For a typical energy of $4$~GeV, this corresponds to $8\units{mrad}$, and thus a resolution on the ALP lifetime of about $\sigma(\taualp)/\taualp\lsim 6-20\%$ for $\alpha$-values of $15-5^\circ$. 

An actual design of this detector based on these requirements is not yet available.

\clearpage
\section{Results}
\label{sec:results}
In this section examples are provided for the measurements LUXE is planning to make. The primary goals are to measure the Compton and Breit-Wheeler processes as a function of the laser intensity and to compare them with theoretical predictions, as well as to use the experiment for a search for new particles. The results are illustrative of what LUXE can achieve. Also given in this section is a list of systematic uncertainties which are of relevance to these measurements. 

\subsection{Measurement of the Compton Process in Electron-laser Interactions}

The interesting observables of the Compton process are the energy spectra and rates of electrons and photons. The electrons are measured by the scintillation screen and the Cherenkov detector, while the photons are measured in the gamma ray spectrometer which again uses a scintillation screen. The possibility to make three independent measurements of the spectra will be of great value to reduce systematic uncertainties.

\subsubsection{Electron Measurements}
\begin{figure}[htbp]
\centering
\includegraphics[width=0.48\textwidth]{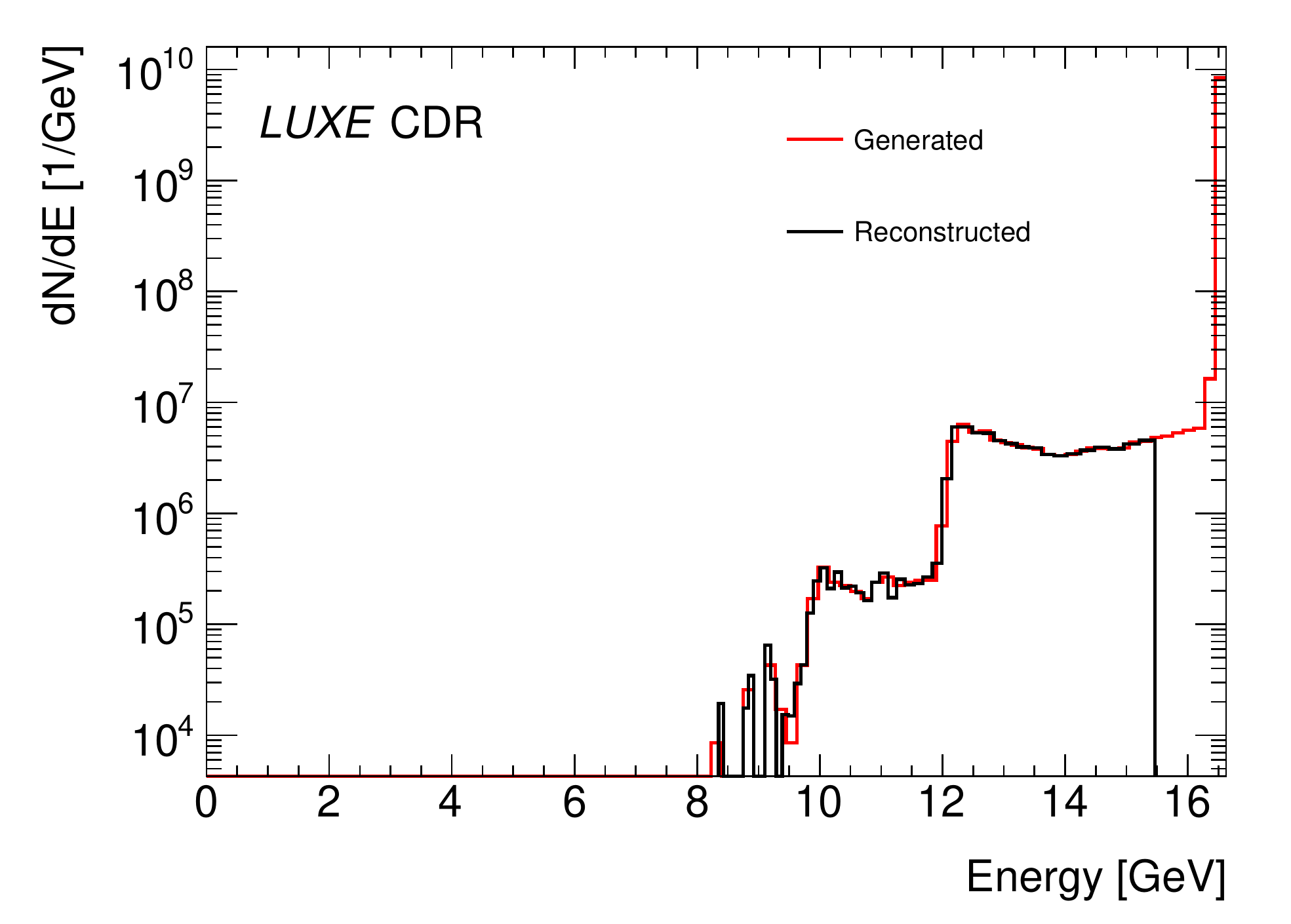}
\includegraphics[width=0.48\textwidth]{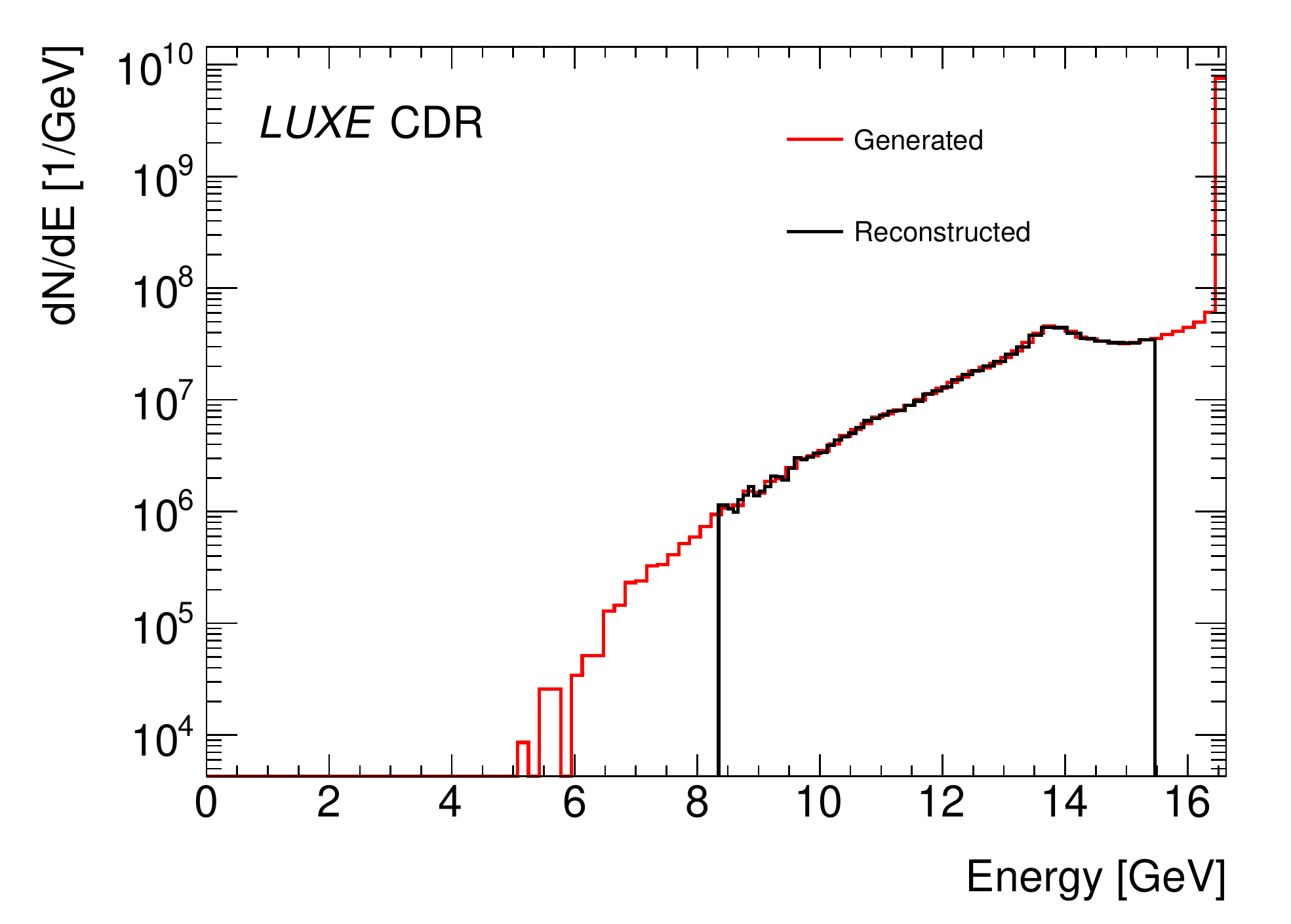} 
\caption{Simulated and reconstructed energy spectra for $\xinom=0.3$ (left) and $\xinom=1.0$ (right). The red histogram shows the simulated spectrum and the black histogram shows the spectrum as it is expected to be reconstructed with the \cer detector. The acceptance of the detector is restricted to $8.4<E_e<15.5$~GeV.} 
\label{fig:erecch} 
\end{figure}

A simulation of the reconstructed electron energy spectrum is shown in Fig.~\ref{fig:erecch} for the \cer detector for two selected $\xi$-values for signal. The background is expected to be negligible, see Sec.~\ref{sec:simulation}.

For the Compton edge analysis, a robust method for finding the Compton edge is the so-called \textit{finite-impulse-response-filter (FIR)} method, which which was proposed in the context of kinematic edge reconstruction at the International Linear Collider (ILC)~\cite{Berggren:2020gna}. The position of the edge in the electron energy distribution $\mathbf{g_d(i)}$ is given by the maximum in the so-called \textit{Discrete Response Function}, $\mathbf{R_d(i)}$, defined as the discretised convolution of $\mathbf{g_d(i)}$ and a filter $\mathbf{h_d(i)}$ : 
 
 \begin{equation}
     R_d(i)=\sum_{k=-N}^{k=N}h_d(k) \cdot g_d(i-k)
 \end{equation}

The choice of the filter function is in principle arbitrary. A common choice, also employed in this analysis is the so-called \textit{first derivative of a Gaussian (FDOG)}:

 \begin{equation}
     h_d(k)=-k\exp{-\frac{k^2}{2\sigma^2}} \;\; \text{for}\; -N\le k\le N 
 \end{equation}

\begin{figure}[htbp]
\centering
\includegraphics[width=0.48\textwidth]{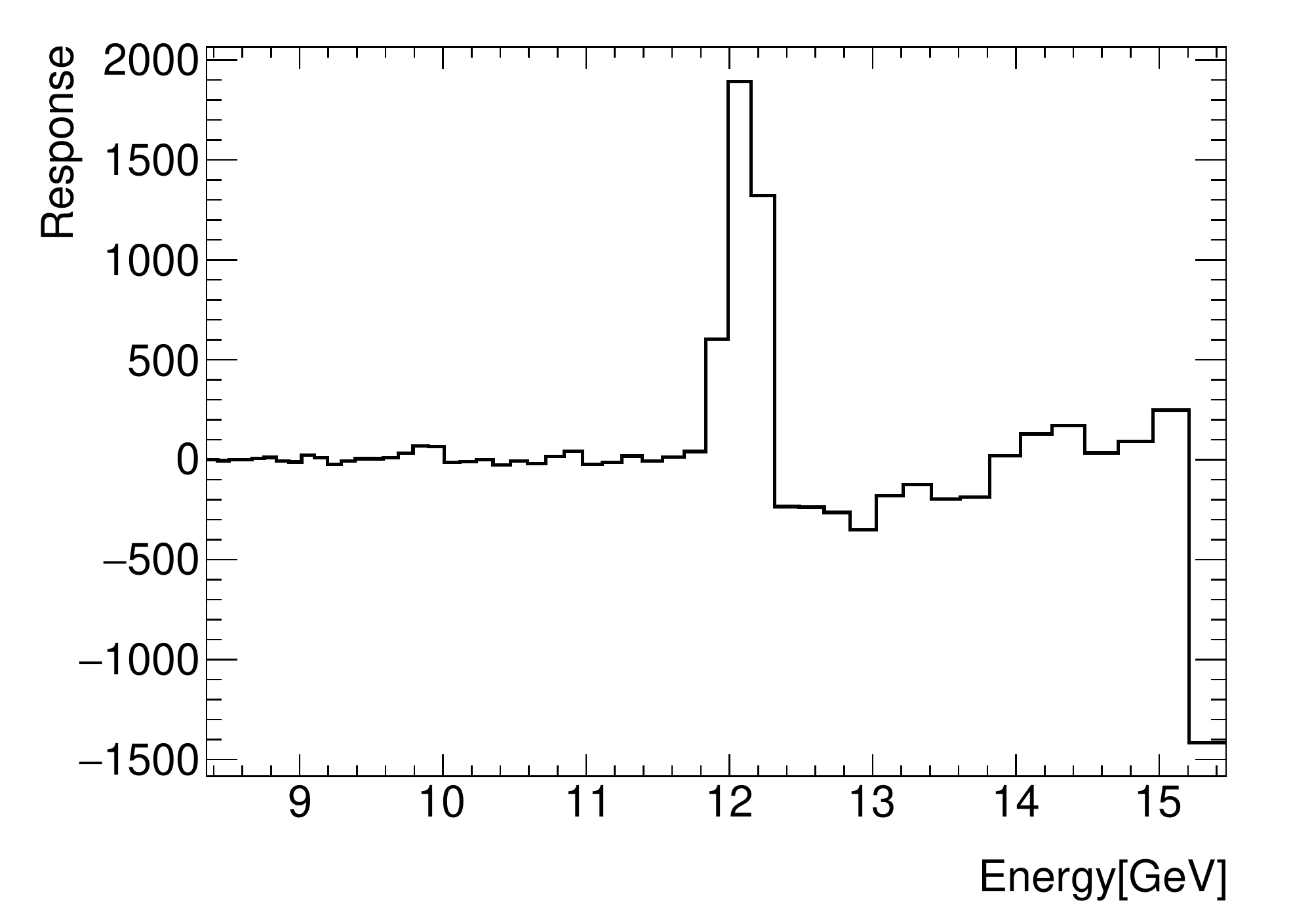}
\includegraphics[width=0.48\textwidth]{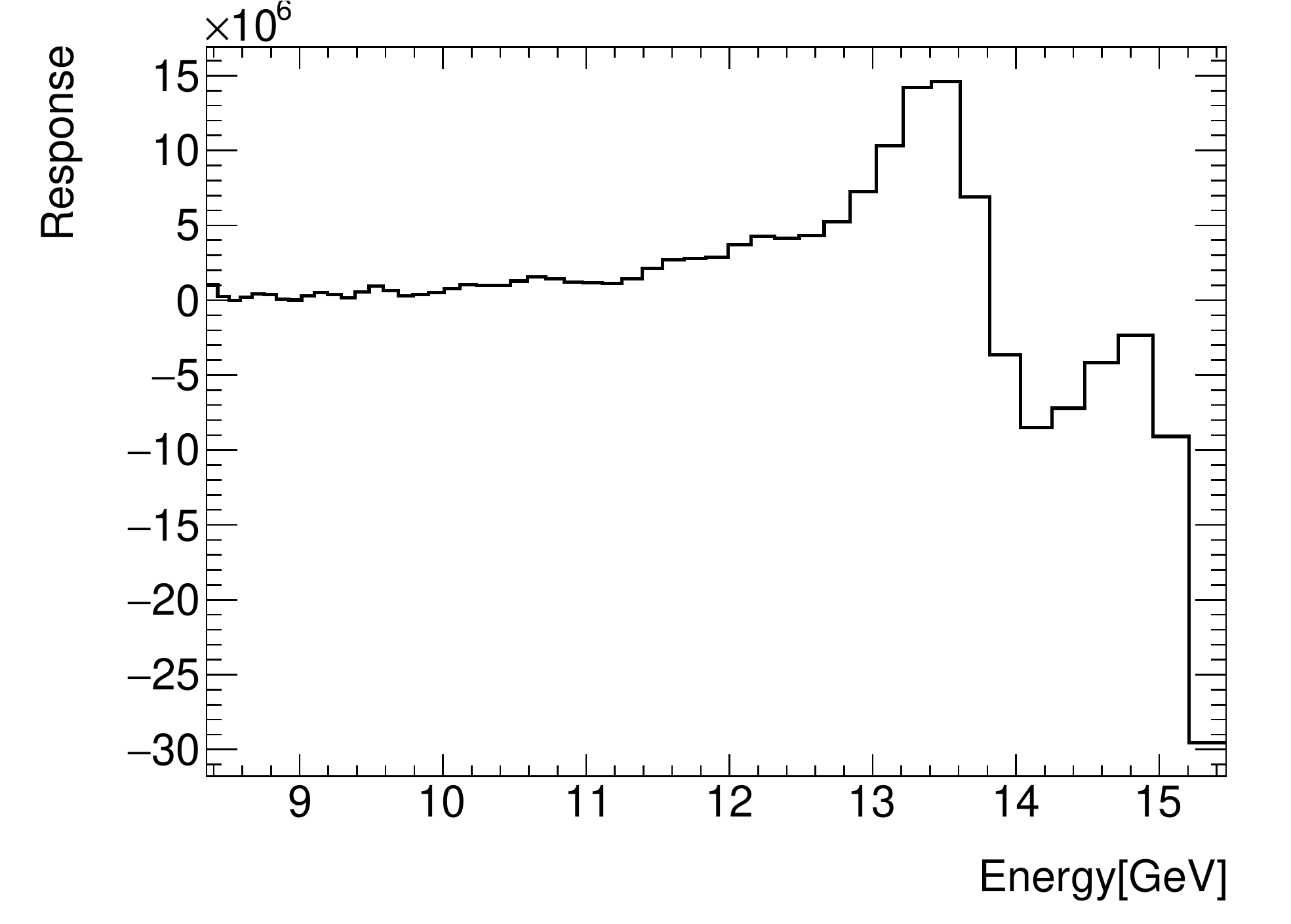} 
\caption{
Discrete Response function for Compton edge finding for $\xinom=0.3$ (left) and $\xinom=1.0$ (right) based on the reconstructed energy spectrum shown in Fig.~\ref{fig:erecch}.} 
\label{fig:xiresponse} 
\end{figure}

Figure \ref{fig:xiresponse} shows the response distributions corresponding to  Compton electron energy distributions shown in Fig.~\ref{fig:erecch} for two different values of $\ximax$. It is calculated for an FDOG filter with $\sigma=1$ and $N=10$. The correspondence between the edge location and the maximum of the response function is clearly visible. Higher order edges are also observed as maxima at lower energies.

Due to the Gaussian shape of the laser pulse in both transverse and longitudinal coordinates (see Sec.~\ref{sec:laser}) the spectra shown in Fig.~\ref{fig:erecch} correspond to a distribution of different true $\xi$-values with a maximum value of $\ximax$. Here, the position of the edge of the $\ximax$-value in the laser pulse is taken as a reference point. This position corresponds to the position at the upper-energy limit of the edge, where the slope becomes negative due to the shape of the Compton edge. This position is referred to as the \textit{upper kink position}. It corresponds to the first zero-crossing of the discrete response function after the response maximum.
 
\begin{figure}[htbp]
\centering
\begin{subfigure}{0.48\hsize} 
\includegraphics[width=\textwidth]{KinkFunctionBen_TomMC_TomMC_FIR_statunc2.5sys_varfdog.pdf} 

\caption{\label{fig:cerdet:comptresultedge}}
\end{subfigure}
\hspace{0.5cm}
\begin{subfigure}{0.48\hsize} 
\includegraphics[width=\textwidth]{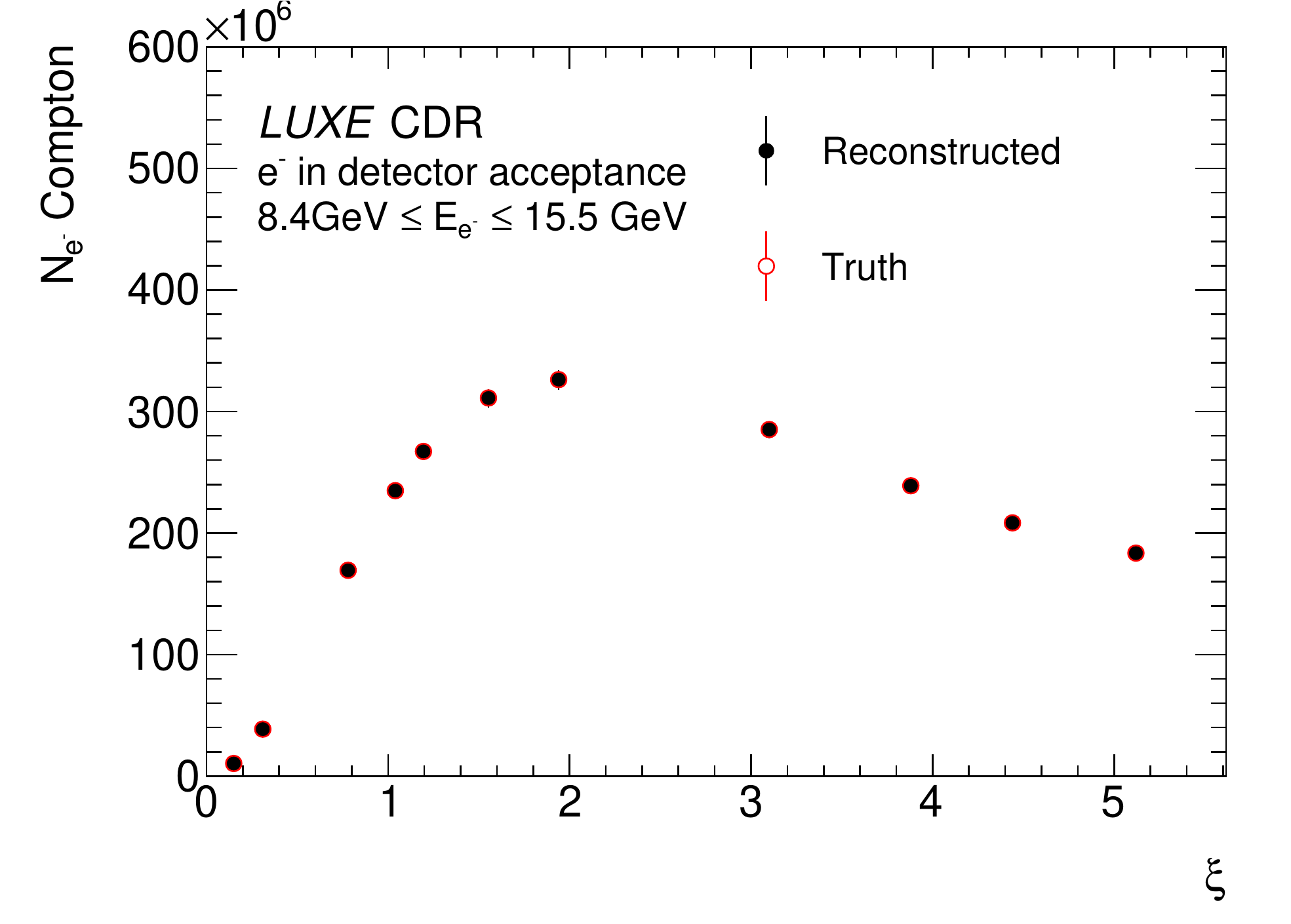}
\caption{\label{fig:cerdet:comptresultnorm}}
\end{subfigure}
\caption{a) Compton electron edge upper kink position from simulated analysis as a function of the nominal $\xi$ value. The red solid line shows the theoretical prediction. The dashed lines show the impact of a 5\% uncertainty on the laser intensity. The black circles show the anticipated data result based on the analysis discussed in the text. The uncertainty shown is dominated by an energy scale uncertainty of 2.5\%, a statistical uncertainty corresponding to 1\,h of data taking is also included but negligible. b) Number of Compton electrons as a function of $\xinom$. The uncertainty shown corresponds to the statistical uncertainty for 1\,h of data taking and a systematic uncertainty on the response calibration of 2.5\%. Due to the finite acceptance of the \cer detector only the number of electrons within a fiducial volume of $8.4<\varepsilon_e<15.5$~GeV is shown.}
\label{fig:ce} 
\end{figure}

In Fig.~\ref{fig:cerdet:comptresultedge}, the Compton edge upper kink position is shown as a function $\ximax$, compared to the theoretical prediction based on Eq.~(\ref{eq:edgexi}). Good agreement is observed between the reconstructed position and the theoretical prediction.

Figure~\ref{fig:cerdet:comptresultnorm} shows the total number of Compton electrons within the \cer detector acceptance as a function of $\ximax$. Since the energy spectrum gets skewed towards lower electron energies at high $\xi$, which are outside of the \cer acceptance, the total number of electrons reaches a maximum and then decreases.

This measurement is shown here based on the \cer detector but a very similar performance should be achieved independently by the scintillation screen. 

\subsubsection{Photon Measurements}
The photon energy spectrum is best measured by the gamma ray spectrometer (GRS) discussed in Sec.~\ref{sec:detectors_gammaray_sepctrometer}. Based on the measurement of electron and positron energy spectra from photon conversions, the photon energy spectra can be reconstructed as follows.

\begin{figure}[htbp]
    \centering
    \includegraphics[width=\textwidth]{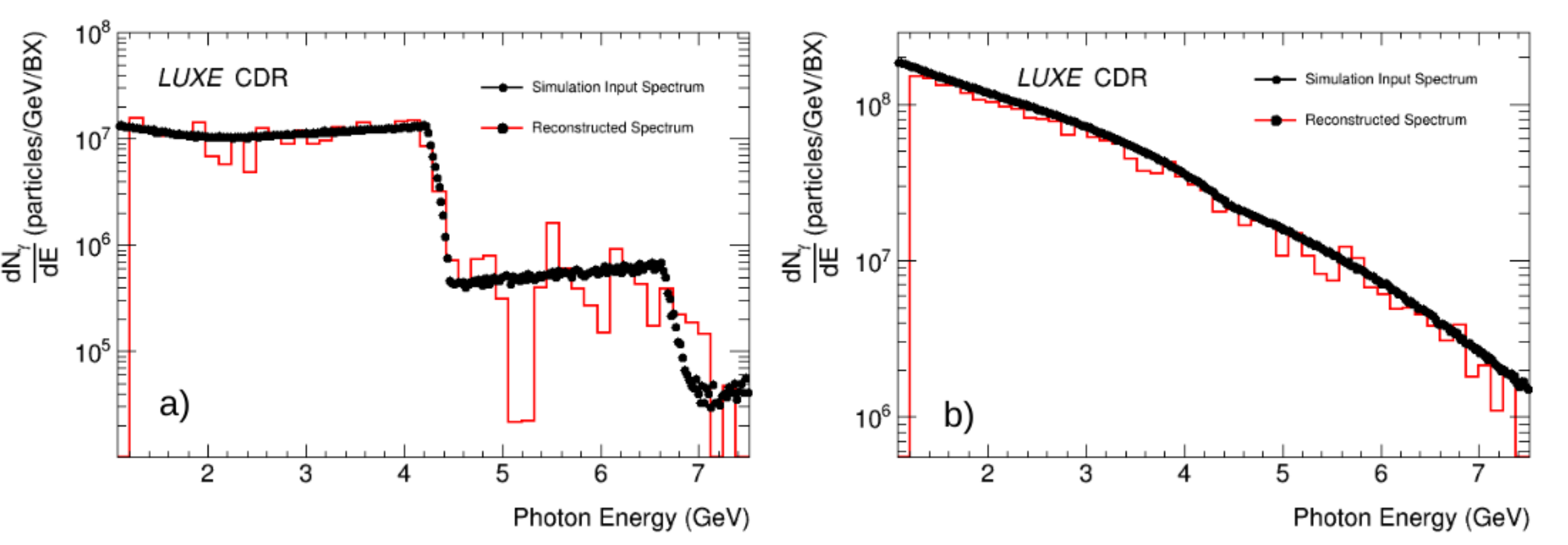}
    \caption{Input spectrum (black) and spectrum reconstructed from the deconvolution of the spectrometer output (red) for a) $\xi=0.31$  and b) $\xi=3.1$. The statistics used in this simulation is  low and causes the fluctuations in a) for energies above 4.5~GeV.} 
    \label{fig:fig6_gamma_ray_spectrometer}
\end{figure}

The electron/positron energy spectra, as shown in 
Fig.~\ref{fig:fig2_gammaray_spectrometer}, can be extracted from the scintillation response like in Fig. \ref{fig:fig4_gammaray_spectrometer}. From either of these spectra, the energy spectrum of the photon beam incident on the spectrometer set-up can be reconstructed using a deconvolution algorithm (see Ref.~\cite{Fleck:2020} for details). Taking the electron spectrum as the input, it is assumed that the highest energy electrons, say in the interval $[E_{max} - \Delta E , E_{max}]$, in the distribution are produced by photons of equal energy passing through the converter. Using the corrected Bethe-Heitler formula in \cite{Tsai:1974}, the expected number of produced electrons for a beam of mono-energetic photons given the conversion parameters is calculated. The number of photons within the energy range $[E_{max}-\Delta E, E_{max}]$ in the incident spectrum is then the ratio of the measured electron number and the theoretical single photon result. However, a beam of mono-energetic photons will contribute to the electron spectrum at all energies and so this response must be subtracted from the measured electron spectrum. This process is repeated, decreasing the maximum energy in steps of $\Delta E$ until the lower energy limit of interest is reached. Dividing each calculated photon number by $\Delta E$ gives the energy spectrum of the incident photon beam. Fig. \ref{fig:fig6_gamma_ray_spectrometer} shows two examples of the result of the deconvolution, for a typical linear ($\xi<1$) and non-linear ($\xi>1$) ICS interaction, respectively, and its comparison to the original photon spectrum. As one can see in Fig. \ref{fig:fig6_gamma_ray_spectrometer}.a, the primary Compton edge can be clearly identified, together with an indication of perturbative contributions of higher non-linearities (secondary Compton edge at approximately 6.8 GeV). For high $\xi$ the monotonic shape of the photon spectrum is closely reproduced by the deconvolution routine. 

And, then the same algorithm as used for the reconstruction of the edges in the electron spectrum described above can be used to determine the kink positions in the photon energy spectrum. 

Furthermore, the flux of photons can be measured for photons with energies above 1~GeV with the GRS, and for all photons with the backscattering calorimeter. The goal is to also achieve here a precision of 5\%, dominated by systematic uncertainties. The ratio of the number of Compton photons to electrons is a particularly interesting quantity as it directly measures the number of Compton interactions per laser shot. 

\subsection{Measurement of the Positron Rate in Electron- and Photon-Laser Interactions} \label{sec:posrate}

The measurement of the positron rate is the primary way to characterise the trident process. It is measured using the pixel tracker and the electromagnetic calorimeter. As was shown in Sec.~\ref{sim:sig_bkg}, the backgrounds in this region are typically 100 charged particles for the \elaser mode and 10 for the \glaser mode. The signal rate varies strongly with $\xi$, starting at $0$ and increasing to $\sim 10^4$. The goal of LUXE is to be able to measure any rates larger than 0.001 per BX with an accuracy better than 10\%.

\begin{figure}[htbp]
    \centering
    \includegraphics[width=0.48\textwidth]{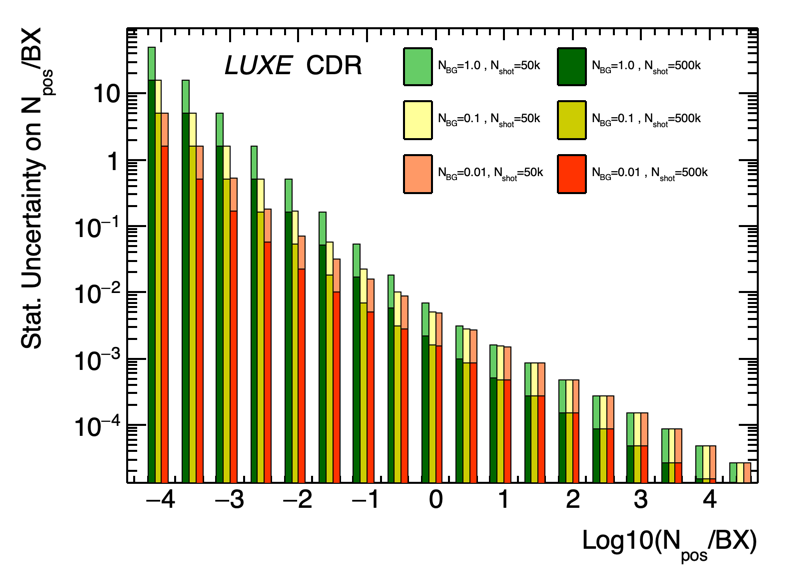}
    \caption{Statistical uncertainty on the number of positrons, $\nposi$, per bunch crossing as function of  $\nposi$/BX. Shown are the uncertainties for three assumptions on the background and for two assumptions on the number of shots, corresponding roughly to one day (50k shots) and 10 days (500k shots).
    }
    \label{fig:results:nposerr}
\end{figure}

The measurement of the number of positrons, $N_\textrm{pos}$, at a given $\xi$ value is given by 
\begin{equation}
    N_\textrm{pos}(\xi)=N_\textrm{Data}(\xi)-N_\textrm{BG}
\end{equation}
where $N_\textrm{Data}$ is the number of reconstructed positrons and $N_\textrm{BG}$ is the number of background events which is measured using bunches where the laser is not fired. Assuming a rate of laser shots of 1~Hz and 9~Hz of background-only data the statistical uncertainties on the measurement can be estimated as a function of the expected number of positrons for various assumptions on the background.   
Figure~\ref{fig:results:nposerr} shows the statistical uncertainty on the rate of positrons per bunch crossing, $\nposi$/BX, as function of the rate assuming backgrounds of $1$, $0.1$ and $0.01$ charged particles. 
For $\nposi$/BX$>1$ the uncertainty is below 1\% with just one day of data and for all background assumptions. However, for lower values of $\nposi$ the uncertainties increase rapidly. For $\nposi=10^{-3}$, an uncertainty of 10\% can only be achieved with 10 days of data if the background is below $0.01$/BX. With the tracker, for the \glaser setup, it was estimated that the background is $<0.1$/BX based on the currently available simulation statistics, see Sec.~\ref{sec:detectors_tracker_performance}. With further refinement of the tracking requirements and the addition of the calorimeter information, it seems plausible that it will be possible to measure rates as low as $\nposi=0.001$/BX with an accuracy of $\sim 10\%$. 

\begin{figure}[htbp]
    \centering
    \includegraphics[width=0.48\textwidth]{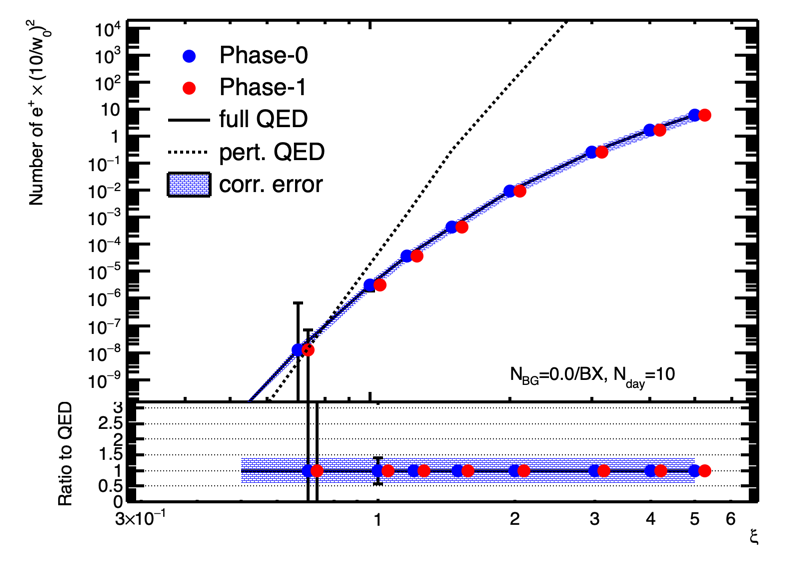}
    \includegraphics[width=0.48\textwidth]{glaser_err_ratio_bg001.png}
    \caption{Number of positrons per laser shot, normalised by $10/w_0^2$, versus $\xi$ for \phaseone and \phasetwo. The uncertainties show the expected statistical precision of the data assuming 10 days of data taking. The \phasetwo points are slightly displaced horizontally for better visibility, they actually have the same $\xi$-values as the \phaseone data. The left (right) plot shows the uncertainties assuming no background (a background of 0.01 particles per BX). Also shown are the theoretical expectation for the full QED prediction and assuming perturbative QED only. The bottom panels show the ratio to the QED prediction, and include a display of the correlated uncertainty dominated by the uncertainty on the laser intensity.
    }
    \label{fig:results:nposresult}
\end{figure}

Fig.~\ref{fig:results:nposresult} shows the result one could obtain based on 10 days of data taking per $\xi$ value and assuming either no background or $0.01$ particles per bunch crossing. Also shown are the theoretical predictions based on full QED and on perturbative QED only. This illustrates that in order to make a meaningful measurement in the perturbative regime it is important to suppress the background to below $0.01$ particles. At high $\xi>2$ the statistical precision is well below 5\% and the difference to the perturbative prediction is very significant.

The ECAL with its fine granularity and a small Moli\`{e}re radius of $2 \units{cm}$ allows reconstruction of the number of positrons impacting the calorimeter as well as their energy distribution. The number of produced positrons can be either measured as the number of reconstructed showers, in case of low expected positron rates, or by analysing the energy flow as a function of the $x$-coordinate, by exploiting the relation between the positron momentum and the $x$-coordinate of the impact point. In general, because of the linear response of the calorimeter, the energy flow  method can be applied either to single reconstructed showers or to fragments of showers. In particular, when many particles originating from the IP appear at the same location in the ECAL, which can only happen if they all have the same energy, their number can be estimated by dividing the measured energy by the energy expected at this location.

\begin{figure}[!ht]
 \begin{minipage}{0.65\textwidth}
  \includegraphics[width=0.9\textwidth]{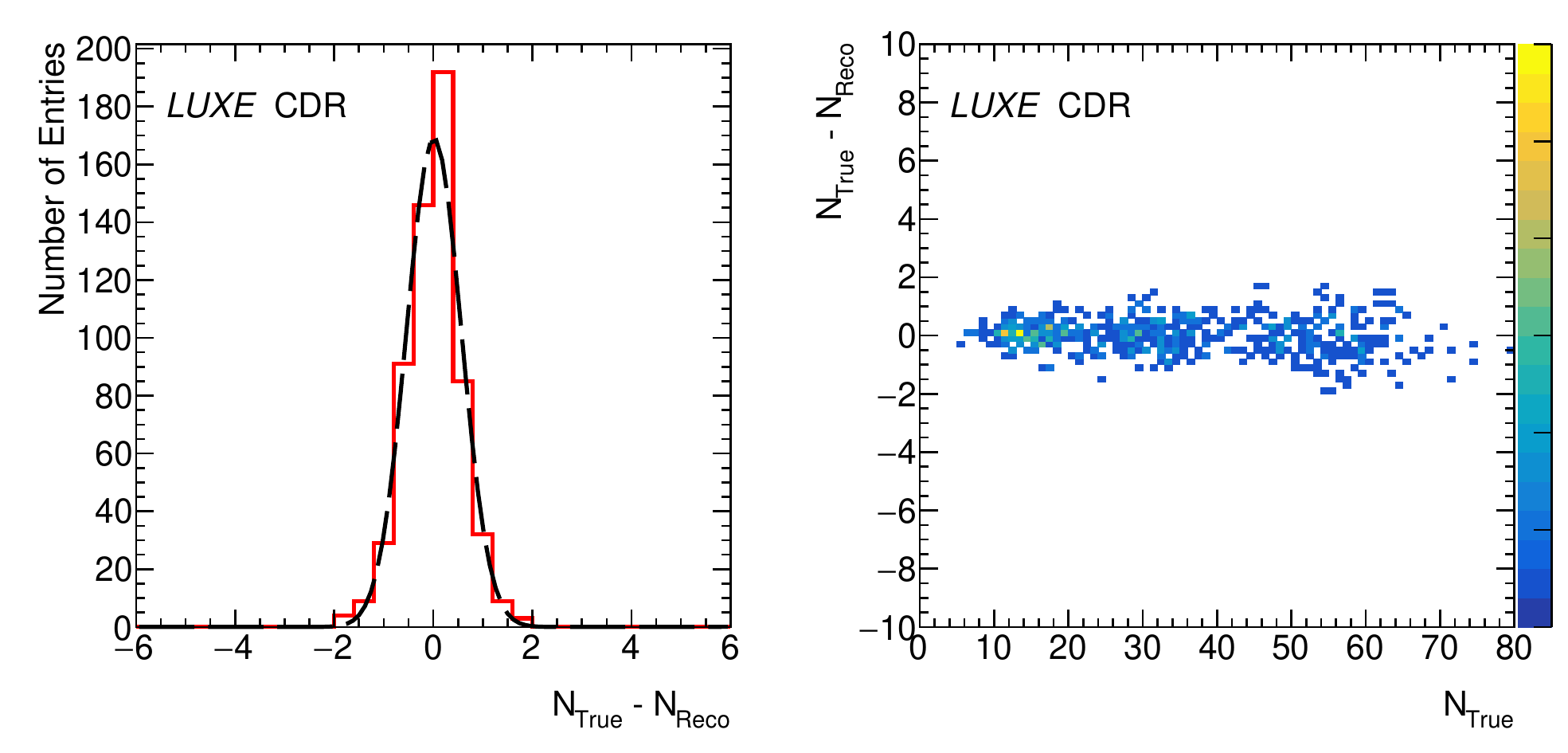}
 \end{minipage}
 \begin{minipage}{0.3\textwidth}
\includegraphics[width=1.2\textwidth]{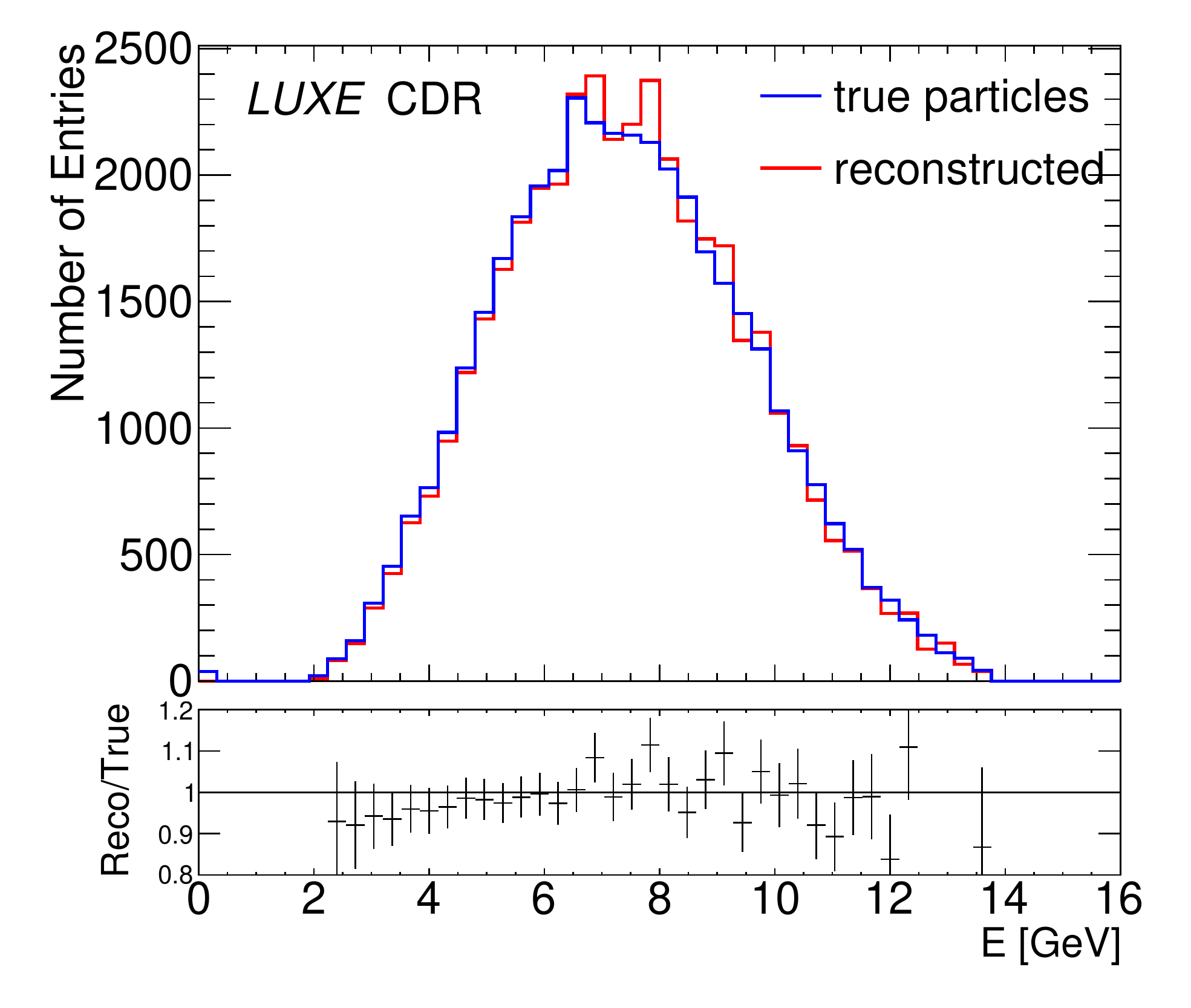}
 \end{minipage}
    \caption{The resolution of the number of positrons measurement (left), the spread of the number of reconstructed positrons as a function of the number of incident positrons (middle), and the comparison between the reconstructed and generated energy spectrum of the positrons (right). In the latter case, the ratio of the two spectra as a function of energy is shown in the lower panel. 
    }
  \label{ECAL_performance_more}
\end{figure}

The results, for a simulation without background, is illustrated in Fig.~\ref{ECAL_performance_more}. In this particular case, where up to 80 positrons impact the ECAL per event, the number of positrons is determined with an uncertainty of 0.7, only slightly dependent on the number of impacting particles.
As shown in Fig.~\ref{ECAL_performance_more} (right), also an excellent agreement with the generated spectrum is found. 
Since the energy resolution is not significantly affected by background (see Sec.~\ref{sec:detectors_calorimeter}), the performance should be largely retained also in the presence of background, after correcting for it.

\subsection{Measurement of Electrons in Photon-Laser Collisions}
\begin{figure}[htbp]
    \centering
    \includegraphics[width=0.45\textwidth]{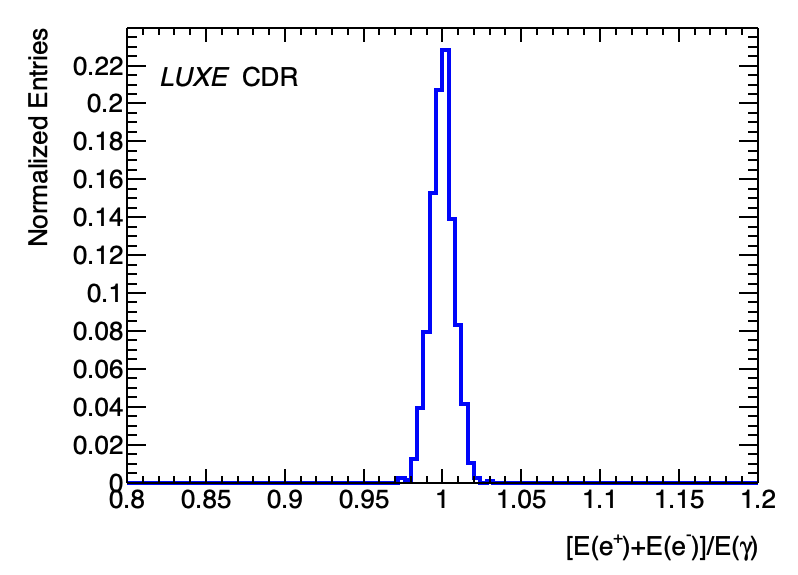}
    \includegraphics[width=0.45\textwidth]{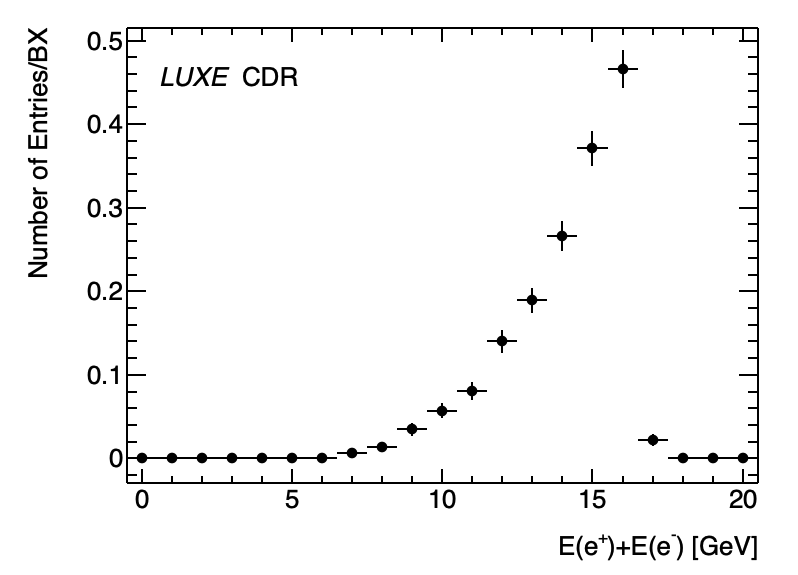}
    \caption{Left) Ratio of the measured energy sum of the electron and positron and the energy of the incoming photon. Right) Example for a reconstructed photon energy spectrum for \phasetwo for $\xi=5.6$. The uncertainties represent the statistical errors expected for 1000 BXs with a background of 0.01 tracks/BX.}
    \label{fig:results:massvse}
\end{figure}

In the \glaser mode, the detector is designed such that both the positrons and electrons from the pair production process can be measured. For low rates $\nposi < 1$ they can be matched uniquely, and the sum of the two energies can be determined. Figure~\ref{fig:results:massvse} shows the ratio of the energy sum to that of the incoming photons. It is seen that they are very well correlated, enabling a measurement of the cross section versus the photon energy. Since the momenta of the electrons and positrons can be determined with a precision of 1\% the photon energy is also determined with that precision. An example spectrum is shown in Fig.~\ref{fig:results:massvse}.

\subsection{Search for Scalar and Pseudo-scalar Particles Decaying to Photons}

In Sec.~\ref{sec:NPLeff} it is discussed that scalar or pseudo-scalar particles (called ALPs thereafter) could be produced via the Primakoff effect when the large flux of Compton photons interacts in the beam dump. A full simulation of this process and detector setup is underway but here a basic estimate of the sensitivity is presented. 

For this estimate a beam dump of $L_S=0.5\,$m with a lead core is used, and a detector at a distance of~3~m behind the end of that dump for detecting photons. The number of electrons per bunch is  $N_e=1.5\cdot 10^9$, and for the study it is assumed that one year of data taking corresponds to $10^7$ live seconds and thus $10^7$ laser pulses.  

The differential Compton flux, $dN_{\gammaC}/dE_{\gammaC}$, includes both Compton photons from the LUXE IP, resulting from electron-laser interactions, as well as photons from EM showering processes inside the dump, which are simulated 
with \geant version~10.06.p01~\cite{Agostinelli:2002hh,Allison:2006ve,Allison:2016lfl}. The Compton photon flux is determined using the MC simulation with \textsc{IPstrong} and \textsc{PTARMIGAN} (see Sec.~\ref{sec:mc}). The flux will ultimately be measured in LUXE in-situ using the GRS discussed in Sec.~\ref{sec:detectors_gammaray_sepctrometer}. 

The angle in which 95\,\% of the Compton photons fall determines the minimum target cross section. It is at present envisaged that the dump is at a distance of $\sim 13$~m from the IP (see Table~\ref{tab:sim_keypar}); at that distance the Compton photons will fall into a circle of radius $\lesssim 5$~cm. 

A detector of size $1$~m$^2$ is estimated to have a good acceptance at a distance from the dump of 3~m, see Sec.~\ref{sec:det:bsm}. For this basic first sensitivity estimate the acceptance of the detector is estimated assuming tat the ALPs decay right after the beam dump. In reality they will decay along the decay tunnel, and the latter they decay they more likely will fall into the acceptance of the detector. In addition, we required that the energy of each photon needs to be larger than $0.5\,$GeV to be detected. 
    
The dump length, $L_S$, is designed such that it completely blocks all beam photons, so that no background is present. Other background sources are expected to be negligible and a detailed study is beyond the current scope. One possible SM background is muons which are produced during the EM shower inside the dump.  
Most of these muons are expected to be very soft and can be vetoed or bend away by using a magnetic field. 
Therefore, we consider a background-free ALPs search and in order to estimated the sensitivity to probe ALPs we set the discovery threshold to $N_{\alpi}\approx3\,$. 
    
\begin{figure}[htbp]
    \centering
    \includegraphics[width=0.5\textwidth]{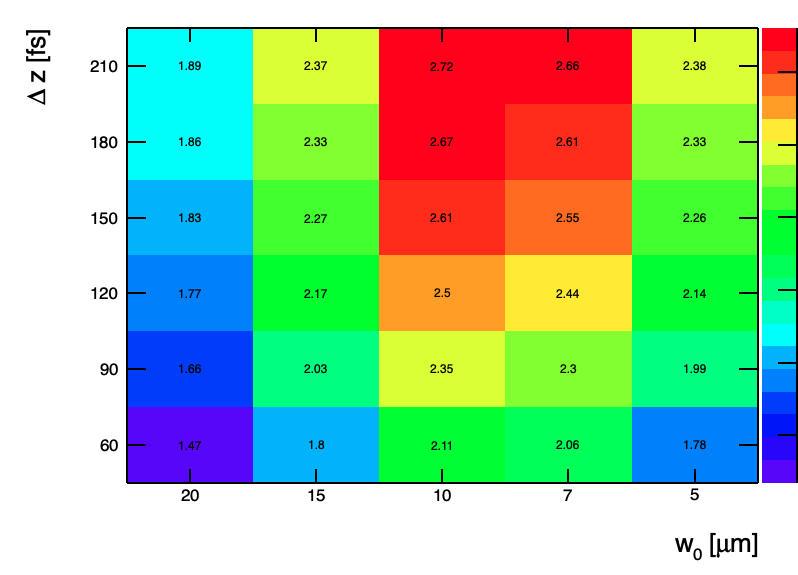}
    \caption{Number of photons in units of $10^9$ with energy $\varepsilon_\gamma>1$~GeV per bunch crossing as function of $\xinom$ and $\Delta z$ for \phasetwo based on MC simulation. }
    \label{fig:gamflux_e2}
\end{figure}

The sensitivity of LUXE is estimated using two benchmarks for the \phaseone and \phasetwo runs.  
In \phaseone the electron energy beam is $E_e=16.5\,$GeV, the laser pulse length is 25\,fs with a spot size of 6.5\,$\mu$m, corresponding to $\xinom=2.4$. For phase-1, we consider a long laser pulse of 120\,fs with $w_0=10$~$\mu$m, corresponding to $\xinom=3.4$. The longer pulse length of 120~fs has been chosen as to maximise the interaction rate. By stretching the pulse longer the $\xi$-value decreases but with the longer pulse the interaction cross section increases. The sensitivity can be maximised by optimising these variables. While these use a realistic simulation of the interactions, it is important to note that these are just benchmarks that serve as examples and do not show the ultimate reach. In principle any $\xi$ value can be used as long as it is high enough so that the flux of photons with energies above $\sim 1$~GeV is large. Figure~\ref{fig:gamflux_e2} shows the flux of photons with energies above 1~GeV versus $\xinom$ and $\Delta z$ for \phasetwo, where $\Delta z$ is the longitudinal size of the beam. For the \phasetwo benchmark the flux is $2.5\times 10^9$ photons with $\varepsilon_\gamma>1$~GeV. For the \phaseone benchmark it is about $10^8$.

\begin{figure*}[htbp]
    	\includegraphics[width=0.45\textwidth]{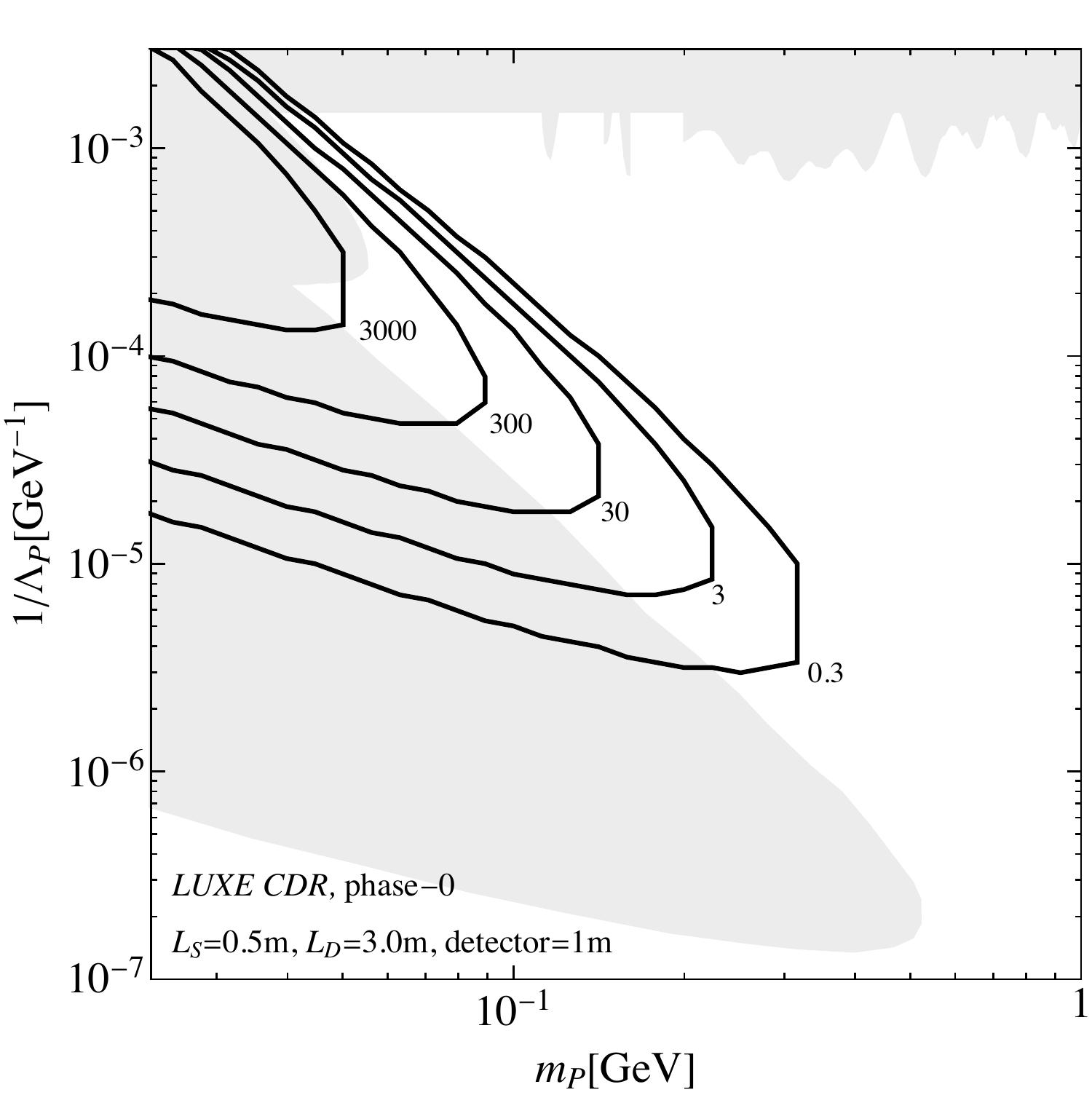}~~~
    	\includegraphics[width=0.45\textwidth]{Reach_LUXE_0530_phase1_phase2.pdf}
    	\caption{
    	Left: contours (black) of the expected number of ALP events, $N_{\alpi}$ for LUXE phase-0;
    	Right: the projected LUXE reach for phase-0 and phase-1 in solid and dashed black compared to current bound (grey) from LEP~\cite{Jaeckel:2015jla,Knapen:2016moh,Abbiendi:2002je}, {\sc PrimEx}~\cite{Aloni:2019ruo}, NA64~\cite{Dusaev:2020gxi,Banerjee:2020fue}, Belle-II~\cite{BelleII:2020fag}, and beam-dumps experiments~\cite{Bjorken:1988as,Blumlein:1990ay}. In addition we compare to the projection of other future runs of experiments as NA62,  Belle-II, FASER,  {\sc PrimEx} and {\sc GlueX}~\cite{Dobrich:2015jyk,Dolan:2017osp,Feng:2018pew,Aloni:2019ruo,Dobrich:2019dxc}.
    	}
    	\label{fig:ReachSec}
    \end{figure*}

In the left panel of Fig.~\ref{fig:ReachSec}, contours of $N_{\alpi}$ in the $m_{\alpi}$--$1/\Lambda$ plane are shown for \phaseone. The discovery contour at $3$ corresponds to the discovery for a background-free search. The typical lifetimes of the particles LUXE is sensitive to is $1.5$~ns at the exclusion limit. 

The right panel of Fig.~\ref{fig:ReachSec} contains the sensitivity projections for both phases compared to other experiments. 
Shown are the current bounds from LEP~\cite{Jaeckel:2015jla,Knapen:2016moh,Abbiendi:2002je}, {\sc PrimEx}~\cite{Aloni:2019ruo}, NA64~\cite{Dusaev:2020gxi,Banerjee:2020fue}, Belle-II~\cite{BelleII:2020fag}, and beam-dumps experiments~\cite{Bjorken:1988as,Blumlein:1990ay}.
    
In addition, the future projections of NA62, Belle-II, FASER (planned for run3 at CERN from 2022-2024), {\sc PrimEx} and {\sc GlueX}~\cite{Dobrich:2015jyk,Dolan:2017osp,Feng:2018pew,Aloni:2019ruo,Dobrich:2019dxc} are presented. 
As we can see the LUXE experiment can probe unexplored parts of the parameter space with $20\,\MeV\lesssim m_{\alpi} \lesssim 500\,\MeV$ and $1/\Lambda_{\alpi} > 2\cdot 10^{-6}\,\GeV^{-1}$. 
With \phaseone the reach is similar to that of FASER-2, considered at CERN to take data starting in about 2027. With \phasetwo it is comparable to the projection by NA62 with $10^{18}$ protons on target at CERN, expected to be collected during LHC Run\,3. In both cases LUXE is more sensitive at higher coupling values. 
    
It is worth noting that the LUXE estimate is based on a single year of data taking, and the configuration has not yet been fully optimised. Thus it is particularly encouraging that it provides a sensitivity that is similar to other ongoing or planned experiments, and in some cases unique.

\subsection{Systematic Uncertainties}

For the above measurements the following sources of systematic uncertainty have to be considered, and in many cases motivate the requirements on the experimental setup.

\begin{itemize}
    \item An uncertainty of 5\% on the absolute intensity of the laser, and thus the value of $\xi$ causes an uncertainty on the positron rate prediction of about 40\% for both $e$-laser and $\gamma$-laser collisions, and a shift of the Compton edge by about 2\%.  
    \item An uncertainty on the relative intensity from shot to shot of 1\% basically results in a finite resolution on the $\xi$ value. This will smear out the Compton edge and the widen the $\xi$-distribution. It will need to be taken into account in simulation to be able compare with data. If the goal of 0.1\% is achieved, it will have a negligible impact. 
    \item A jitter of 10--20~fs in the timing of the laser from shot to shot w.r.t. the electron beam is expected. This also contributes to the uncertainty on the relative intensity shot to shot. Since the electron bunches have a length of $\sim 100$~fs this will change the $\xinom$ value by only about 1--2\%. 
    \item Energy scale: for the Compton edge measurement it is important to have an energy scale uncertainty below 2.5\% to see the dependence of the edge position on $\xi$, see Fig.~\ref{fig:CEdge2}. The main effects that impact the energy scale are possible misalignments of the detector w.r.t. the magnet and miscalibrations of the the magnetic field. Misalignments of the detector elements in the transverse plane translate into miscalibrations of the energy. For a magnet with a length of $1$~m the relation the relative energy shift is given by 
    \begin{equation}
    \label{eq:xshift}
        \Delta E/E=0.0163\times \frac{\delta x}{1~\rm{mm}}\times \frac{B}{2~\rm{T}}\times \frac{E}{1~\rm{GeV}}
    \end{equation}
    For instance, for a magnetic field of 2~T, a misalignment of 100~$\mu$m results in an energy shift of up to 2.7\%. Normally, it should be possible to achieve an alignment better than 50~$\mu$m, i.e.\ an uncertainty on the energy scale of 1.4\%. Based on Eq.~(\ref{eq:xshift}) can also be used to determine the impact of a miscalibration in the magnetic field. A 1\% miscalibration will lead to a 1\% uncertainty in the energy scale. 
    \item Energy resolution for Compton measurements: the energy resolution will determine the width of the Compton edge. It is driven by the channel size of the detectors but would also be increased if there is a misalignment between the individual channels. However, for the \cer detector the channel size (1.5~mm) is large compared to the possible misalignments (0.1~mmm) and thus this is not expected to be a large effect. 
    \item Energy resolution for Breit-Wheeler measurements: in the tracker the positron energy resolution is expected to be 1\% and would be degraded by misalignments. However, with four layers it is possible to perform an in-situ alignment. In particular with the additional external constraint from the calorimeter it seems plausible that this can be done, but studies are ongoing to demonstrate this. 
    \item The measurement of the total particle flux for each detector technology will have an uncertainty that depends on the technology. For instance, for the \cer detector the observed charge needs to be calibrated in a test beam. Similarly, for the scintillator screen the brightness of the light needs to be calibrated in a test beam. For the tracking detector, the flux depends on the tracking algorithm and how well the combinatorial background can be understood/suppressed, and is non-linear. The independent measurement of the calorimeter will measure the flux based on the calorimetric energy measurement and the position. Given the built-in redundancy and the ability to test the technologies in test beam prior to installation, it seems reasonable that a 2.5\% uncertainty on the particle counting can also be achieved. 
\end{itemize}

\newpage
\section{Technical Coordination}
\label{sec:tc}

In this section, the plan for the construction, installation and operation of the LUXE experiment will be presented. In all cases the procedure is designed to minimise as much as possible any impact on the \euxfel operations.

As discussed in Sec.~\ref{sec:beam:location}, it is currently planned to install the experiment in the annex of the XS1 access shaft that is scheduled to host the future TD20 line for the 2nd fan of the \euxfel. 

\begin{figure}[htbp]
   \centering
   \includegraphics*[height=8.4cm]{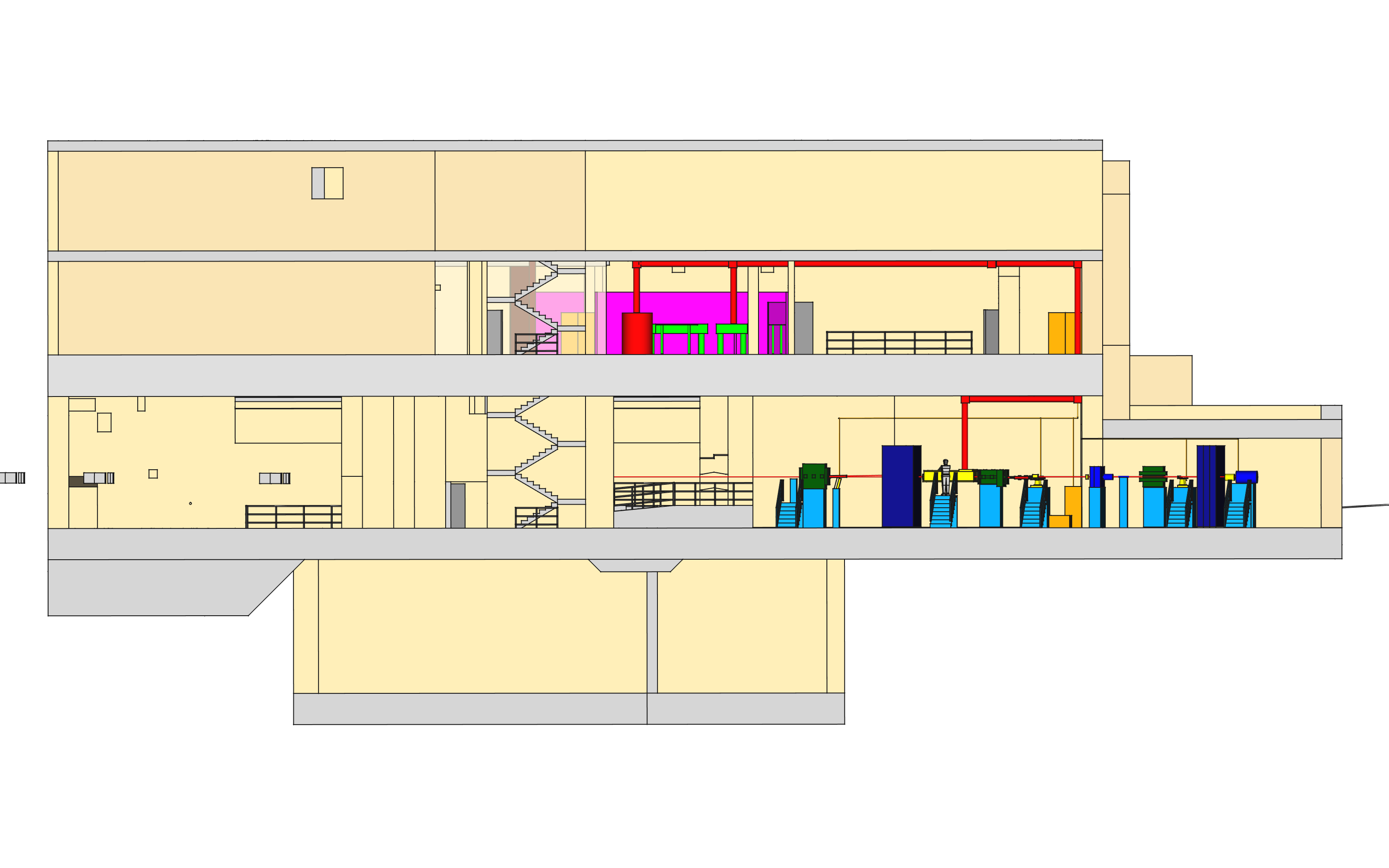}
   \caption{Side-view CAD drawing of the XS1 access shaft building, including the LUXE experiment. The building is shown in beige and grey, while the different experimental elements are shown coloured. The laser clean room is shown in magenta, the laser and electron beam pipes are shown in red, the supporting structure are shown in light blue, the magnets are shown in green, the shielding and dumps are shown in dark blue, the different active components (interaction chambers (IC), and detectors) are shown in yellow, and the services are shown in orange.
   \label{fig:LUXE_CAD_Building}}
\end{figure}

Figure~\ref{fig:LUXE_CAD_Building} shows a 3D CAD view of the XS1 shaft building and various components. The conversion of the experimental components from the {\sc Geant4} simulation described in Sec.~\ref{sec:simulation} to the XS1 CAD model was done using the software described in Ref.~\cite{keith_sloan_2020_4008390}. The experiment is built on two different floors. The experimental area where the electron-laser or photon-laser collisions will happen is located in underground level --3 (UG03), with all the other XFEL.EU \beamline passing through XS1. This area is not accessible during XFEL.EU operations for radio-protection control purposes. 
The laser and most services will be located in underground level --2 (UG02) to allow access at any time, in particular also during \euxfel operations.

The installation of the elements that are located in UG03 needs to be done during a long shutdown of the \euxfel accelerator, as planned for the first half of 2024. Other elements that need to be installed, especially if they are located in UG02 or on the surface, can happen at any time as long as they do not impact the accelerator operations.

The experimental area has been designed to allow data-taking for the different phases of the experiment. In the \elaser mode, the \euxfel electron beam will be directly collided with the laser beam. In the \glaser mode, in addition to the \elaser collisions, the electron beam will also be converted into a photon beam allowing $\gamma$-laser collisions. The $\gamma$-beam will be obtained by using a tungsten target (see Sec.~\ref{sim:geant4}), or by taking advantage of inverse Compton scattering, as is described in Sec.~\ref{sec:science:ics} and Sec.~\ref{sec:tchamber}.


\subsection{Laser}
\label{sec:tc:laser}
\subsubsection{Construction}
The laser system will be installed in a clean-room of ISO-6 standard that will need to be constructed. The room will be isolated from the rest of the building using walls made of special material for fire insulation as per requirements of the \euxfel
safety standards. The clean-room will cover an area of 100~m$^2$ and have a height of 3~m. 
In order to control precisely the laser environment to a temperature of $21\pm 0.5 ^{\circ} $C and a humidity of $40 \pm 5 \%$, an air conditioning control system will be installed.

\begin{figure}[htbp]
   \centering
   \includegraphics*[height=12.cm]{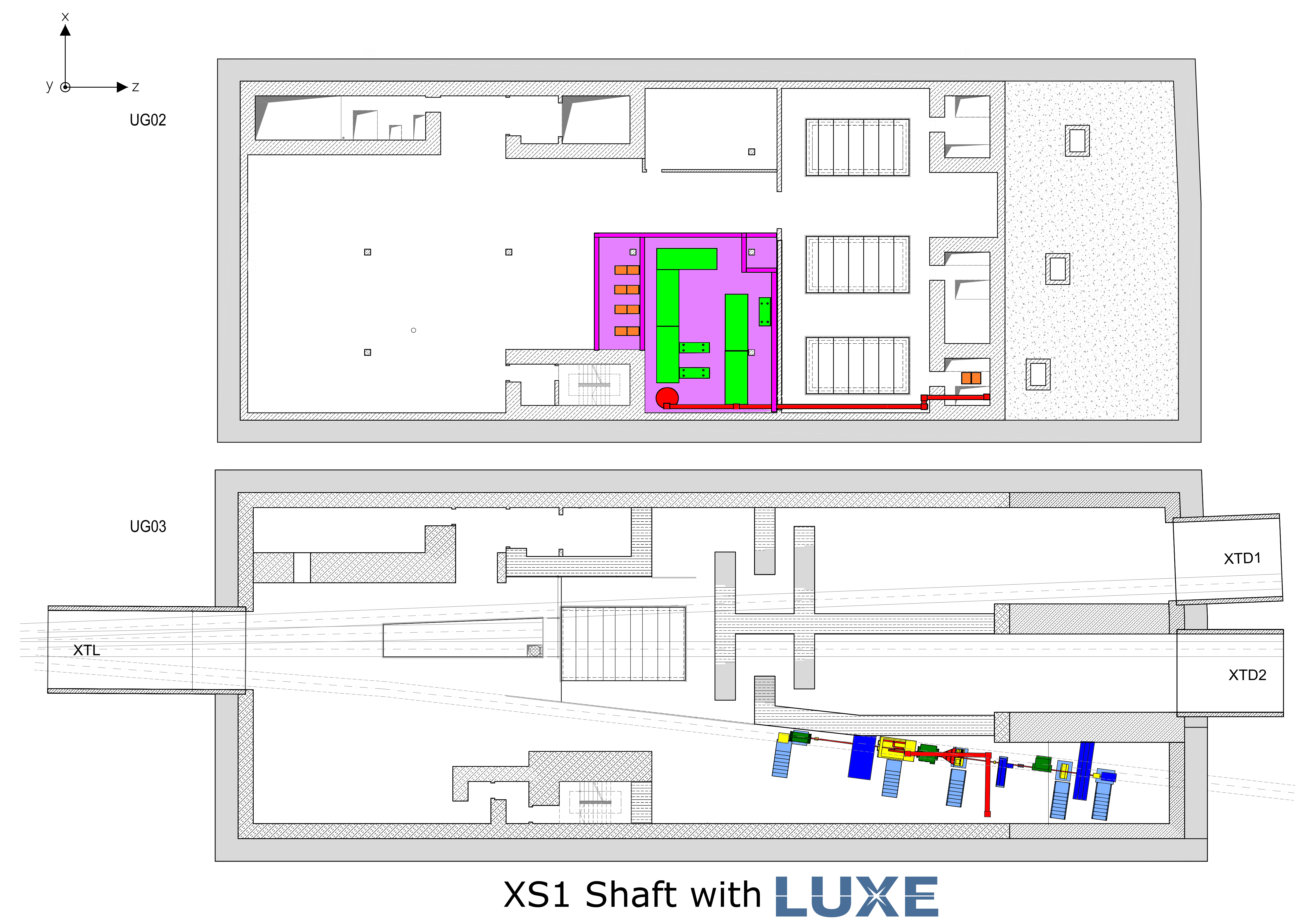}
   \caption{2D drawing of the XS1 access shaft building, in UG02 and UG03 including the LUXE experiment. The building is shown in grey, while the different experimental elements are shown coloured.
   \label{fig:LUXE_2D_Building}}
\end{figure}

For the initial phase of the project, the footprint of the JETI40 laser fits well as shown in Fig.~\ref{fig:LUXE_2D_Building} on floor UG02. The space constraints required by the elements of a 350 TW laser for \phasetwo have also been taken into account in the design of the clean room, such that enough space is available for both. Moreover, a gowning room attached to the main laser room is also planned. For the laser installation and upgrade, a portable clean tent will be deployed in front of the clean-room to allow cleaning and preparation of the different hardware elements before their final installation.
 
\begin{figure}[htbp]
   \centering
   \includegraphics*[height=4.2cm]{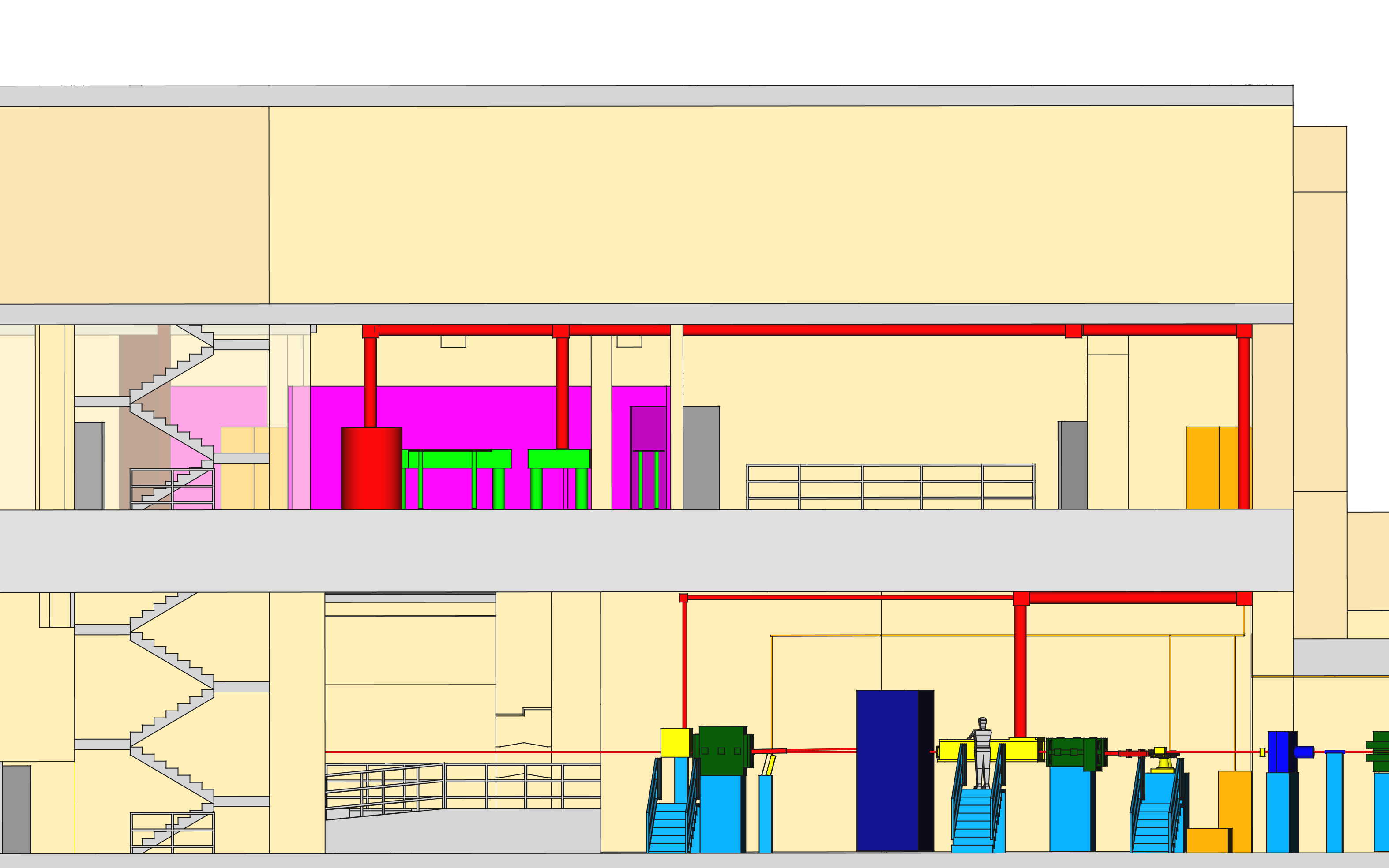} \hspace{2 cm}  
   \includegraphics*[height=4.2cm]{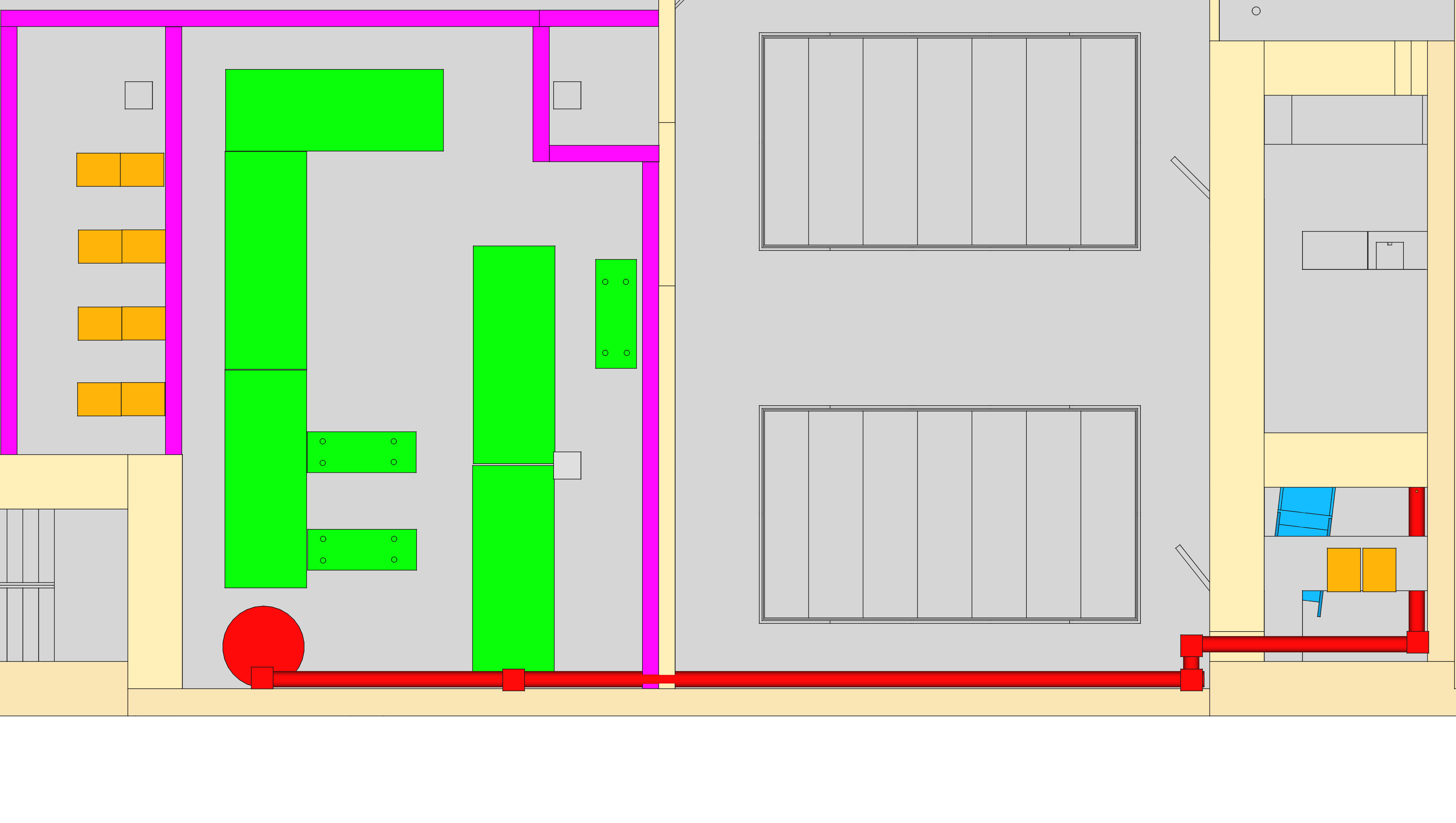}   
   
   \caption{CAD drawing of the XS1 access shaft building, including the LUXE experiment, zoomed on UG02 with a focus on the laser (left) sideview, (right) topview. The building is shown in beige and grey, while the different experimental elements are shown coloured.
   \label{fig:LUXE_CAD_Building_laser}}
\end{figure}

\subsubsection{Installation}
For \phaseone, the JETI40 laser will be moved from Jena and installed in UG02. The installation includes the assembly of components of the laser system such as front end, pre-amplifiers, main amplifier and compressor (see also Fig.~\ref{fig:JETI}). In addition, as mentioned in Sec.~\ref{sec:laser}, a dedicated diagnostics station will also be installed in the laser room. 

After the compressor, the high-power laser beam needs to be guided to the interaction chamber (IC). After the interaction, the attenuated part of the beam will be brought back to the laser clean-room for post diagnostic purposes. Furthermore, one laser beam is required for the inverse Compton scattering and another laser beam is required for measuring the spatio-temporal overlap of the laser and the electron beams (see Sec.~\ref{sec:laser:overlap}). For this purpose a vacuum pipe made of stainless steel with diameter 45~cm and a length of about 40~m, will be constructed. This pipe will pass through the 2~m high concrete floor via a feed-through hole, already present, to the floor UG03, as illustrated in Fig.~\ref{fig:LUXE_CAD_Building_laser}. All laser beams will propagate through the same pipe, however, via different guiding optics. The diameter is sufficient for the laser transport for both the JETI40 and the 350 TW laser.

The IC and associated diagnostics will be installed in the interaction area in UG03. 
An optical breadboard where all optical components are pre-aligned will be placed inside the IC using the access shaft available in the cavern.   

For \phasetwo, a few upgrades have to be carried out to ensure a power of 350~TW can be achieved. Firstly, a new amplification stage will be added. Secondly, as a consequence of the increased energy in the pulse, the size of the laser beam will be increased to 14~cm, compared to 5~cm for the JETI40 laser, to keep the fluence below the damage threshold for the downstream optics. This implies that the compressor also will have to be upgraded to be larger.

\begin{figure}[htbp]
   \centering
   \includegraphics*[height=4.2cm]{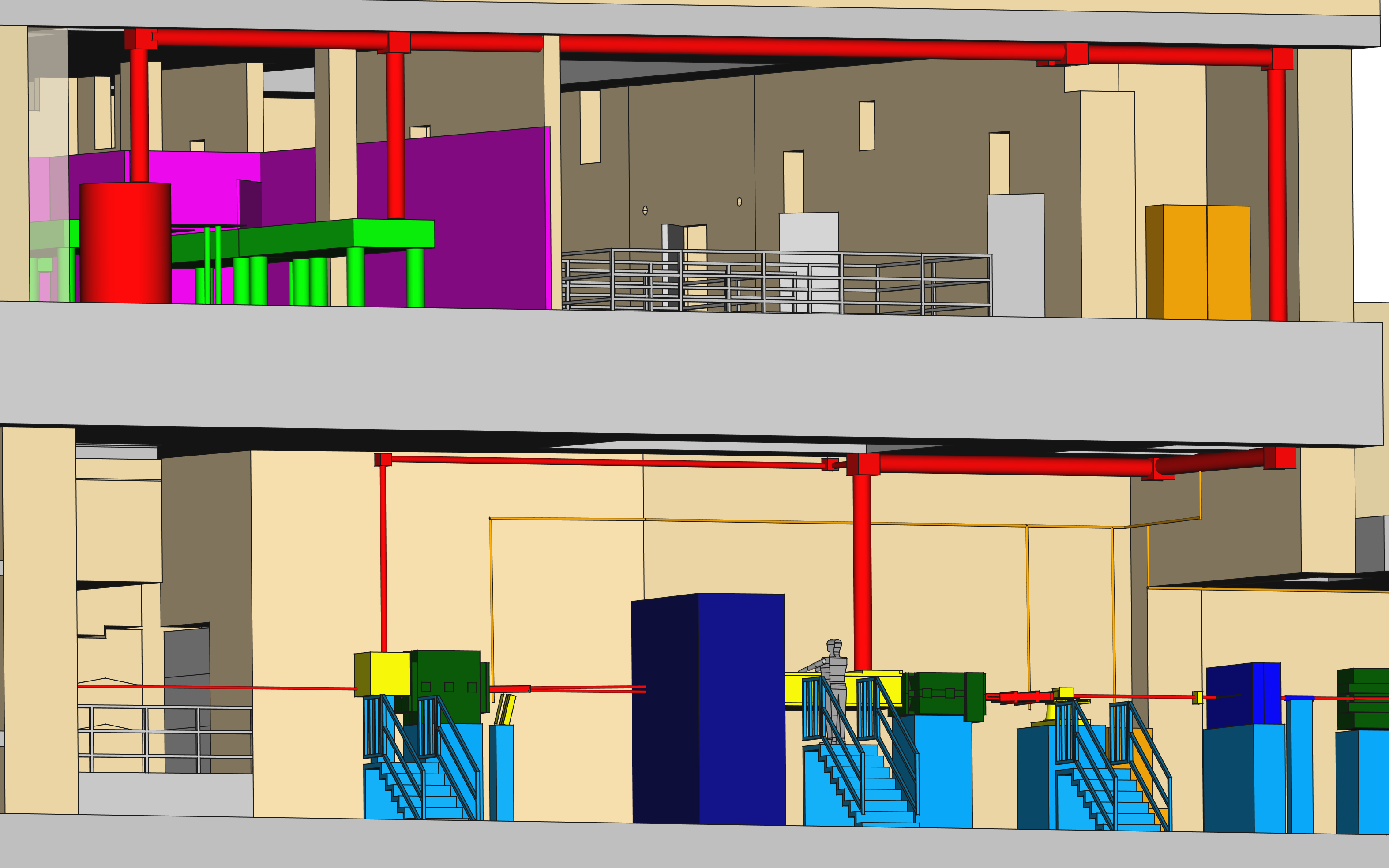}   
   \caption{CAD drawing of the XS1 access shaft building, including the LUXE experiment, zoomed on the laser transport pipe from the clean-room to the interaction point. The ICS setup is also shown. The building is shown in beige and grey, while the different experimental elements are shown coloured.
   \label{fig:LUXE_CAD_Building_HICS}}
\end{figure}

In addition, in the experimental area a small target chamber will be installed upstream of the IP. As discussed in Sec.~\ref{sec:laser}, this smaller vacuum chamber will provide \glaser collisions through the creation of photons via either a gamma converter target, or the ICS setup and timing tool. Utilities to move the target in and out of the electron beam path will be installed. For the ICS and timing tool, two additional laser beams will be required. 
The schematic of the extra beam-line carrying these beams is shown in Fig.~\ref{fig:LUXE_CAD_Building_HICS}. 

\subsubsection{Services} 
A service room is foreseen to be placed adjacent to the laser clean-room and will require an area of about 20 m$^2$. It will host all the power supplies and cooling units for the pump lasers in both phases. Moreover, this separate service room helps reduce the unnecessary heat generated through the power units in the laser area.  

\subsection{Construction in the Experimental Area}

\subsubsection{Construction of Supporting Structure} 
\label{sec:TC:supportStructure}
\begin{figure}[ht]
   \centering
   \includegraphics*[height=4.2cm]{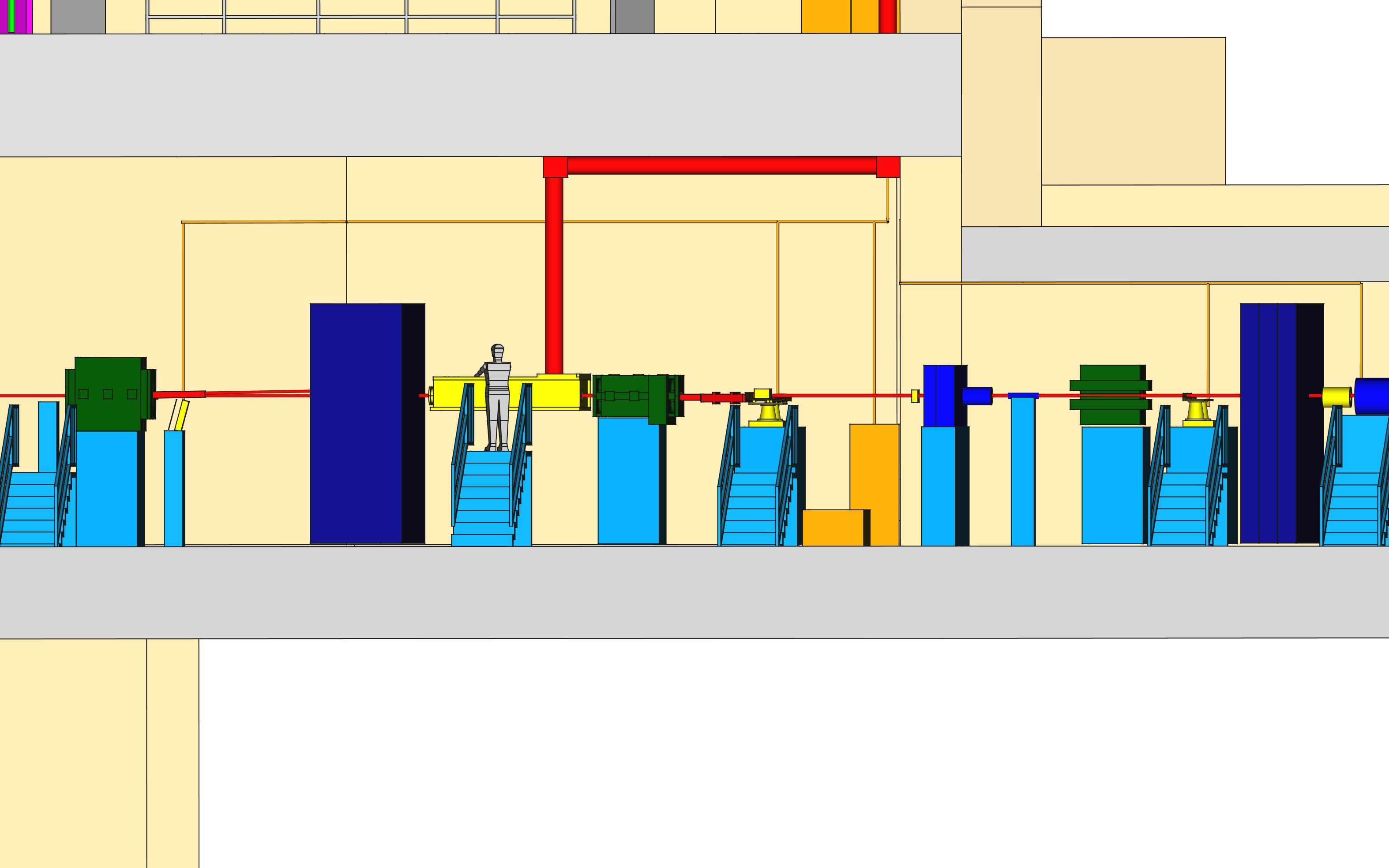} \hspace{2 cm} 
   \includegraphics*[height=4.2cm]{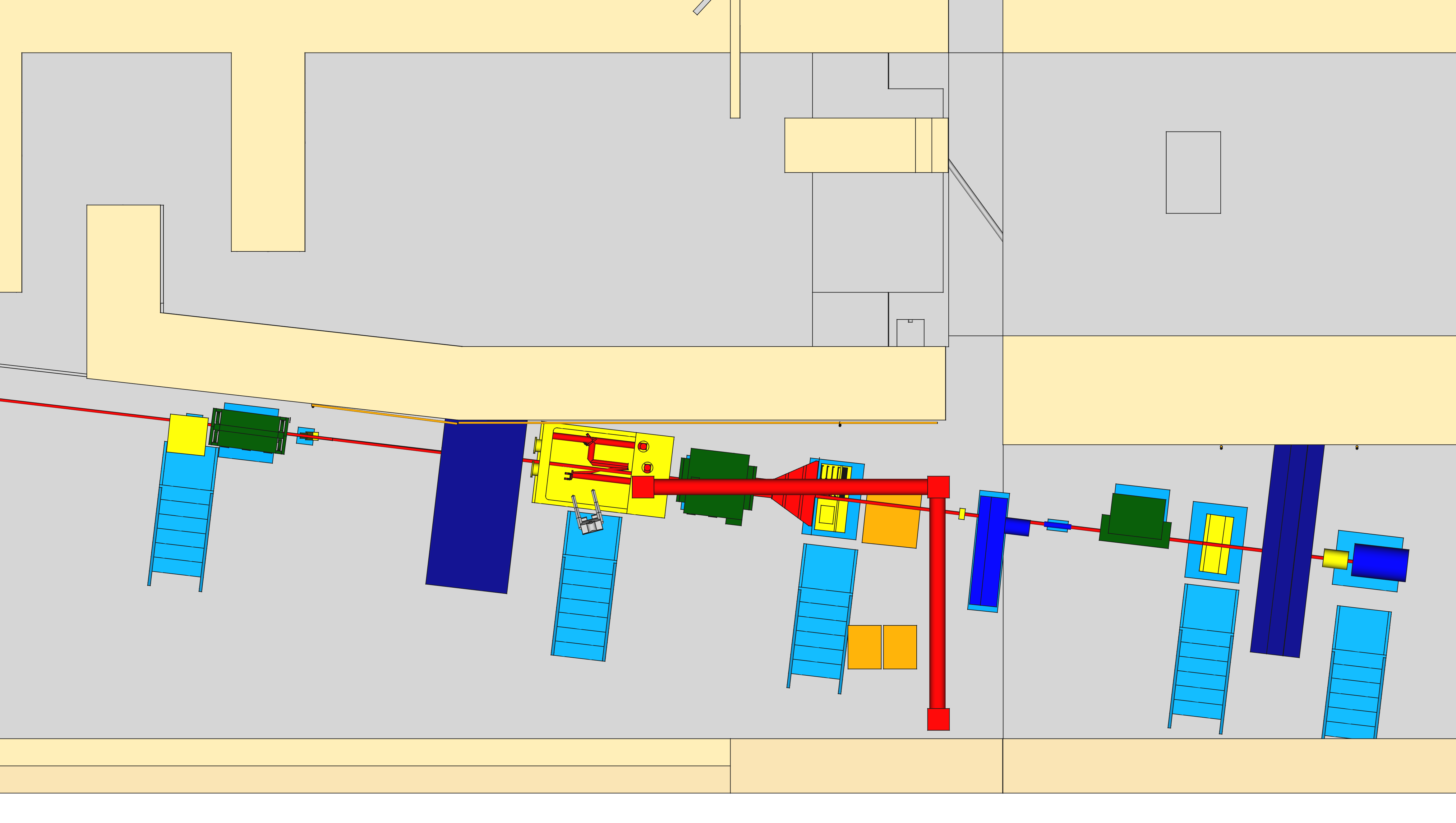}
   
   \caption{CAD drawing of the XS1 access shaft building, including the LUXE experiment, zoomed on UG03 with a focus on the experiment(left) sideview, (right) topview. The building is shown in beige and grey, while the different experimental elements are shown coloured.
   \label{fig:LUXE_CAD_Building_detectors}}
\end{figure}

As can be seen in Fig.~\ref{fig:LUXE_CAD_Building_detectors}, the \euxfel beam-pipes are located about 2.5~m above the floor of XS1. In order to support and access the different experimental elements used in LUXE, support structure with access gateways, shown in light blue, will need to be built.

The other massive structures that will require attention are the shielding elements, shown in dark blue, that are required to avoid secondary particles to reach the IP or the detectors. These shielding and supporting structures will be made of concrete elements. Table~\ref{tab:TC:supportAndShieldings} summarises the number of blocks and their characteristics in the configuration of the LUXE experiment as presented in Fig.~\ref{fig:LUXE_CAD_Building_detectors}. The shielding elements are currently rather massive and based on a very preliminary study aiming to drastically reduce the particle flux around the sensitive detectors; further studies will be conducted to see what is required and technically feasible.

\begin{table}[htbp]
\begin{center}
\begin{tabular}{|l|c|c|}
	\hline
	Type & Volume $(m^{3})$ & Number needed\\
	\hline
Small support & 0.2 & 3 \\
Medium support & 2 & 4 \\
Large support & 2.6 & 4\\
Large Shielding & 17 & 2  \\

	\hline	

\end{tabular}
\caption{Characteristics of the supporting structure and shielding that will be required by the experiments.
\label{tab:TC:supportAndShieldings}
}
\end{center}
\end{table}

This construction might interfere with construction work required for the \euxfel maintenance in 2024, and/or with the construction of the beam extraction and transfer line for LUXE. In particular the cranes in XS1 are important for much of the installation work. It thus is critical that all the work is well planned to ensure that it can be accomplished in time during the \euxfel shutdown.

\subsubsection{Electron Beam-line and Magnets}

The TD20 beam-line construction and installation are explained in detail in Ref.~\cite{beamlinecdr} and summarised in Sec.~\ref{sec:machine}, and not subject of this section. Here, only the aspects of the \beamline which are part of the LUXE experiment are discussed.

The installation of the electron \beamline elements that are specific to the experiment are described in the following. The \beampipe will be made of aluminium with 5~cm diameter. Ion getter pumps will allow to keep an average pressure of $5\cdot{}10^{-9}$~mbar, comparable to what is currently achieved in the rest of the beam distribution system of the \euxfel~\cite{XFELTDR}.

Up to three dipole spectrometer magnets will be used in the two data-taking phases of the experiment. In the $e$-laser collision mode, the electron beam arrives at the IP where it interacts with the laser beam. Just after the IC,  a first dipole magnet is placed that allows to modify the trajectory of the electrons that have not interacted with the laser beam, to guide them to the dump. This magnet will also modify the momenta of the electrons and positrons that have been created in the $e$-laser collisions, to allow their measurements by the detectors (Cherenkov, scintillation screen, tracker and calorimeter) placed after the interaction point. A second magnet is placed downstream in the forward $\gamma$-spectrometer to allow the detection of  electrons and positrons that will be created after the beam interacts with a target.

In the $\gamma$-laser collisions, photons are created upstream of the IP, either via bremsstrahlung or via ICS. The electrons from the beam that have not interacted are dumped in the shortest possible distance,  using another magnet which is placed before the IP. As in the $e$-laser setup, the other two magnets are then used as spectrometer to measure the energy of electrons and positrons that are created in the setup.

The main characteristics of the 3 magnets that have been considered are summarised in Tab.~\ref{tab:TC:Magnets}. The aperture and length are based on a magnet~\cite{MBMagnet} which had been used at the DORIS accelerator but these magnets have a very high power consumption (450kW per magnet) and water cooling is needed (150l/min). Thus
new magnets will be built following similar specifications. Ideally the three magnets are C-shaped, increasing the aperture effectively on one side. The stability and magnetic field of each magnet will need to be measured precisely by the accelerators and experiments assembly (MEA) group before their installation in the cavern. 

All these elements will be installed by the MEA group at DESY, and their precise position will be surveyed and monitored with a precision up to $100~\mu{}$m. 

\begin{table}[htbp]
\begin{center}
\begin{tabular}{|l|c|c|c|c|}
	\hline
Magnet position     & Characteristic     											& \elaser mode, low-$\xi$& \elaser mode high-$\xi$& \glaser mode \\
	\hline
Bremsstrahlung target & \makecell{ B-field (T)\\ Magnetic length (m) \\Aperture (m)} & \makecell{ n.a.}    &\makecell{ n.a.}     & \makecell{ 2\\ 1.08 \\0.33} \\
	\hline
Interaction Point 			 & \makecell{ B-field (T)\\ Magnetic length (m) \\Aperture (m)} & \makecell{ 2\\ 1.08 \\0.33} & \makecell{ 1\\ 1.08 \\0.33} & \makecell{ 2\\ 1.08 \\0.33} \\
	\hline
Forward photon detection& \makecell{ B-field (T)\\ Magnetic length (m) \\Aperture (m)} & \makecell{ 1.5\\  0.98\\c-shaped}& \makecell{ 1.5\\ 0.98\\c-shaped} & \makecell{ 1.5\\ 0.98\\c-shaped}\\
	\hline

\end{tabular}
\caption{Characteristics of the spectrometer dipole magnets used in LUXE. The aperture is in the bending plane of the magnet.
\label{tab:TC:Magnets}
}
\end{center}
\end{table}

\subsection{General Infrastructure and Extra Services}

\subsubsection{General Infrastructure}

During the installation, commissioning, and data-taking phase, various activities will take place on the experiment. Since the XS1 access shaft is located about 2~km away from the main DESY campus, it is mandatory to build a control room and up to three offices on site. These rooms will be  created either inside the XS1 building on the surface or trailer space will be installed near the building. The control room will be equipped with PCs allowing up to four persons, shift leader,  detector shifter, laser shifter, and expert to monitor all components of the experiment. Ideally the control-room dimension should be about 40~m$^{2}$.

The different elements used by the detectors and the laser will benefit from the presence of an access shaft equipped with a crane sitting directly above the experiment. However, an additional small portable crane will be made available in UG03 in order to allow easy access and maintenance of the experimental items.\

\subsubsection{Data Acquisition Infrastructure}

In Fig.~\ref{fig:LUXE_CAD_Building}, the services used by the detectors and the data acquisition system are shown in orange. Since UG03 will not be accessible during data-taking, it is essential to deport as much as possible equipment outside the experimental cavern to the service floor in UG02.

On this floor, the room used to route the laser pipe to the interaction point will also be used to host most electronic services and the DAQ. Two standard 50U racks are currently scheduled to house all the electronic back-end, power supply and DAQ PCs used in the experiment. An extra 50U rack has been added in UG03 to allow the essential electronics that must be close to the detector to be installed there.

The routing length from the service room to the different detectors area are summarised in Tab.~\ref{tab:TC:lengthService}.

\begin{table}[htbp]
\begin{center}
\begin{tabular}{|l|c|}
	\hline
	
Area & Length \\	\hline

Bremsstrahlung target & 22m \\
Interaction Point  & 10m\\
Forward photon detection & 18m\\
	\hline

\end{tabular}
\caption{Distance separating the service cavern to the different experimental area.\label{tab:TC:lengthService}
}
\end{center}
\end{table}

\subsection{Detectors}

\subsubsection{Scintillation Screen}

In the \elaser setup, scintillation screens will be placed behind the IP on the electron arm of the spectrometer, and on each arm of the spectrometer in the forward region. In the \glaser setup, an extra screen will be added near the bremsstrahlung target. Each scintillation screen will be read out by high-resolution cameras that are directly connected to a computer used to process the images. Each camera will be installed around 1~m away from the scintillation screens on a mounting bracket with a placement optimised to maximise the field of view on the screen and minimise the radiation received by the cameras. 

One important necessity for the scintillation screen setup is to operate in a dark region to observe the light emission off the screen. Since the interaction between particles and the screen emits a specific wave length, an optical filter will be used to mitigate the potentially problematic ambient light present in the experimental area.

It is expected that the total electrical consumption of this system is minimal and driven by the PCs (2~kW) used to readout and process the images.

\subsubsection{\cer Detector}

\cer systems are expected to be placed behind the scintillation screens at the IP in the electron arm of the spectrometer and at the bremsstrahlung target (in the \glaser setup).
Since the system uses silicon photon multiplier to read out the signal, as for the scintillation screen setup, it is expected that the total energy consumption will be  driven by the two DAQ PCs (1~kW) used to read out the system and by the VME crates used to house the photo multiplier links (1.6~kW). About 50 DAQ and low voltage links will be used by every \cer system. Each system will be read out by a single DAQ PC. 

This \cer detector requires about 25~l of argon gas to run. The detectors will be built gas-tight, and it is not expected that argon needs to be refilled during operation. Based on tests of the current prototype detector it has been demonstrated that it can run with an over-pressure of $140~$mbar for more than 3 days, with a leak measured to be less than 0.1~mbar/h. Longer tests have been done in the past~\cite{Bartels:2010eb} and are ongoing. The detector will be equipped with a pressure sensor for monitoring of the pressure allowing to calibrate properly the \cer signal.

\subsubsection{Silicon Pixel Tracker}

The tracker will be installed behind the IP, on the positron-side of the spectrometer in the \elaser setup, and on both sides of the spectrometer in the \glaser setup.

The detector requires a water cooling system to keep the temperature of the staves around $21^\circ$ Celsius. This can be achieved for the whole system using a commercial water chiller that will be placed in the experimental area close to detector. As explained in Sec.~\ref{sec:detectors_tracker}, one of the main limitations of the system concerns the readout that has to be done by units that needs to be placed close to the system, as they are connected to the detector by expensive cables, and to minimise cost these cables need to be as short as possible. The signal is then converted to an optical signal and transmitted to the DAQ PCs in the service room. The readout units will be placed in the unique electrical cabinet scheduled to be in UG03, the rest of the equipment will be placed in UG02.

The electrical power consumption of the tracker will be driven by the full power supply system (5.5~kW), the chillers(2~kW) used to cool down the staves, the two DAQ (1~kW) PCs, and the two readout-units placed in the VME crate in UG03 (3~kW).

In order to keep the tracker electronic dry, it is expected to run a continuous flux of nitrogen in the direct vicinity of the tracker.

\subsubsection{Electromagnetic Calorimeter}

The calorimeter will be installed behind the pixel tracker on the positron-side of the spectrometer.

Part of the front-end electronics must be close to the detector, before the signal is sent optically to the single DAQ PC used for the readout in the service room. This front-end electronics will be placed in the electrical cabinet foreseen in UG03, the rest of the equipment will be placed in UG02.

The electrical power consumption of the calorimeter will be driven by the full power supply system (1.25~kW), the front end (1.25~kW), and the DAQ PC (0.5~kW).

As for the tracker, the system needs to run in a dry environment, and thus a constant flux of about 2~l/h of nitrogen will be run in the direct vicinity of the tracker and calorimeter. 

\subsubsection{IP Stage and Vacuum Chamber}
\begin{figure}[htbp]
   \centering
   \includegraphics*[height=4.2cm]{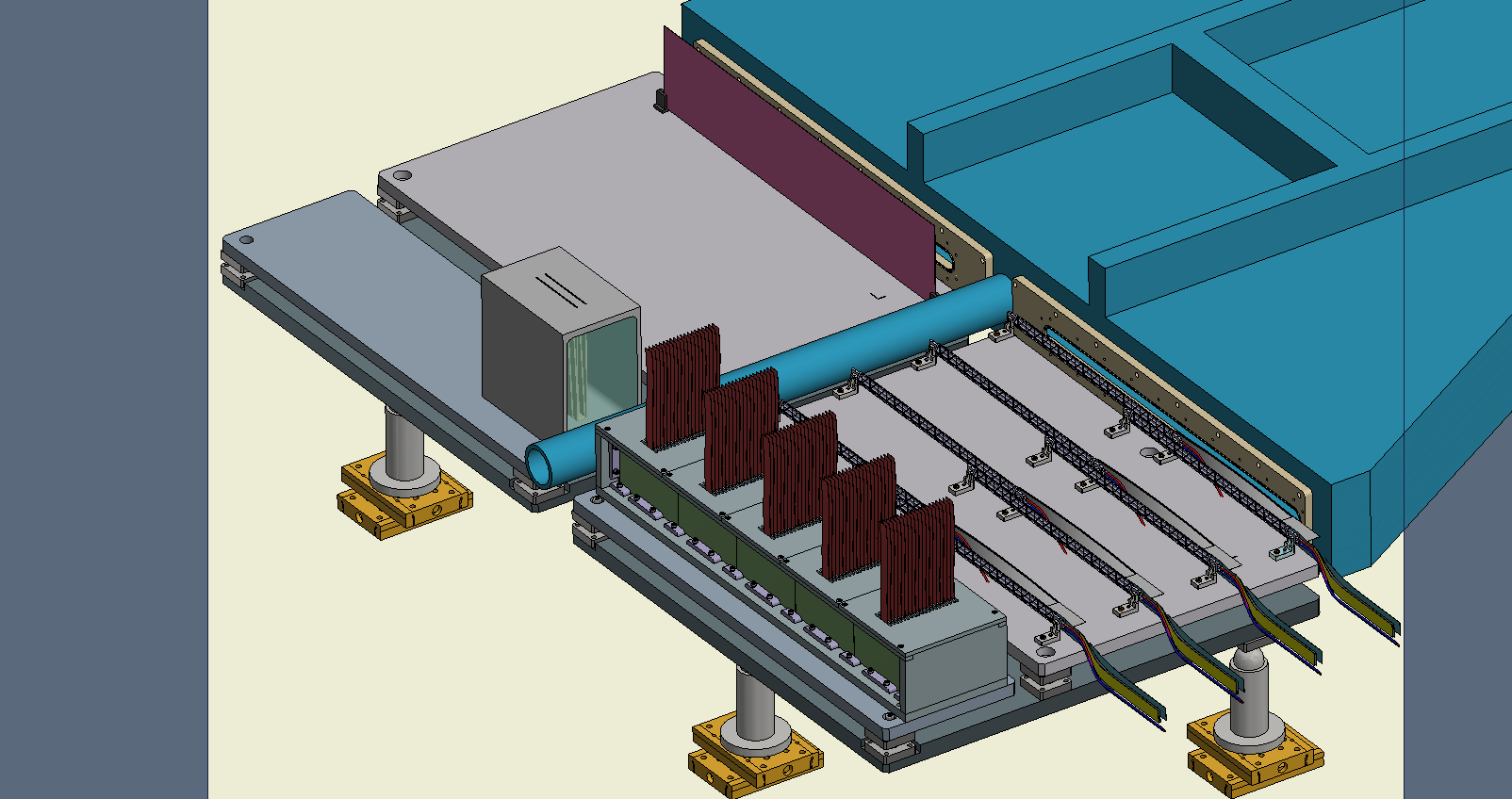}
    \includegraphics*[height=4.2cm]{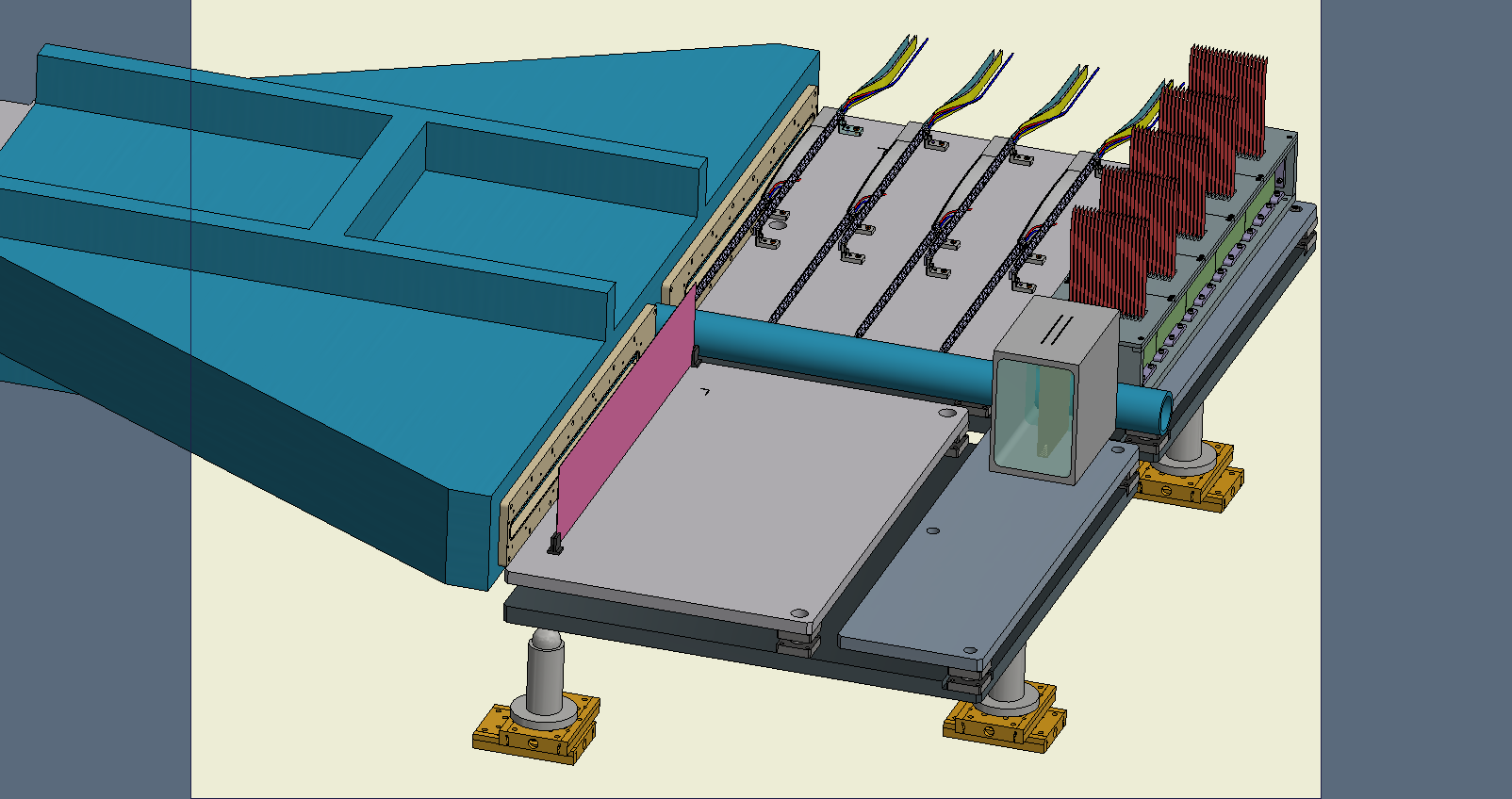}
   \caption{3D CAD drawing of the anti-vibration stage (in grey) where the detectors will be installed at the IP. The detectors are also shown (trackers in grey, calorimeter in green, \cer in grey and scintillation screen in burgundy). A vacuum chamber (in blue) to minimise the interaction of the electrons and positrons with the beam-pipe and air is also shown.
   \label{fig:CAD_hexapodStages}}
\end{figure}

All the systems directly behind the IP will be installed on two high-precision manual mechanical stages that will be placed on each side of the beam-pipe and located on an anti-vibration stage placed on one of the concrete pillars presented in Sec.~\ref{sec:TC:supportStructure}. The installation on the stages will take place before the overall apparatus is lowered down in the experimental cavern. In order to minimise the material between the IP and the detectors, a vacuum chamber with aluminium windows is also installed as shown in Fig.~\ref{fig:CAD_hexapodStages}. 

The precise position of the stages will be surveyed with the same technique as the other elements on the beam-line with a precision up to $100~\mu$m. The mechanical position of the elements on the stage will be known with a precision of $<20~\mu$m for the trackers and $<50~\mu$m for the calorimeter and the \cer detectors. 
 
\subsubsection{Gamma Flux Monitor}

A back-scattering calorimeter made of lead-glass blocks read-out by photo-multipliers is envisaged to be installed between the shielding and the beam-dump to estimate the total photon flux from that (see Sec.~\ref{sec:detectors_photflux_GammaMonitor}).

The electrical power consumption of the back-scattering calorimeter will be driven by the DAQ PC (0.5~kW) used to readout the signal and the full power supply system (1~kW). 

\subsubsection{Gamma Profiler}
\label{sec:det:gammaprofiler}
A Gamma Profiler made of sapphire strips will be installed at the forward spectrometer target to measure the total photon flux. 

The electrical power consumption of the Gamma Profiler will be driven by the VME crate hosting the front-end electronic (3~kW) and the DAQ PC (0.5~kW) used to readout the signal.

\subsection{Commissioning and Data Taking}
During their developments, the detectors will undergo testing at the DESY (or other) test-beam facilities to demonstrate that they achieve the performance needed and to be fully integrated in the LUXE DAQ. Once all the detectors are constructed, and if time allows it, they (or a subset of them) will be tested in a combined test-beam including magnets at the DESY test-beam facility, where electrons of up to 5~GeV are accessible~\cite{DIENER2019265}. This would provide additional calibration of the detectors, and reduce the commissioning time required after their final installation. The detectors will then be lowered down in UG03, where they will be further commissioned. Detectors that can sustain a high-flux of particles such as \cer or scintillation screens will be further commissioned using electron bunches from \euxfel. Other detectors such as the tracker and the calorimeter will also be commissioned using cosmic ray muons. 

After these tests, the DAQ will be integrated with the LUXE DAQ for data taking. The following triggers are anticipated to be used for cosmic muons (mostly for commissioning), electron beam (for noise measurement and physics data-taking), laser beam (for physics),random (for noise measurements). During data-taking, the following information needs to be monitored by a shifter, sensor and Front-End Board currents (critical with alarm to the shifter if above certain threshold and will stop the LV and HV power supplied), data quality and Temperature/humidity inside the box (which will send an alert to the shifter).

\subsubsection{Summary of Detector Service Requirements}

The most important services needed by the different detectors are summarised in Tab.~\ref{tab:TC:detectorService}.

\begin{table}[htbp]
\begin{center}
\begin{tabular}{|l|c|c|c|c|}
	\hline
	
Detector             & Electrical consumption (kW)& Cooling & Gas       & N DAQ output PCs\\   
	\hline
 
Scintillation screen & 2 & --    & --      & 4\\               
Cherenkov            & 2.6 & --   & Argon     & 4\\               
Tracker              & 11.5 & Water   & Nitrogen  & 2\\              
Calorimeter		     & 3 & Air     & Nitrogen  & 1\\               
Back-scattering	Calorimeter	 & 1.5 & --    & --      & 1\\  
Gamma Profiler       & 3.5 & --    & --      & 1\\
	\hline
             
\end{tabular}
\caption{Summary of the main service requirements needed by each detectors.
\label{tab:TC:detectorService}
}
\end{center}
\end{table}

\subsection{Power Consumption}
The total electric power consumption expected by the experiment is found to be of the order of 140~kW in \phaseone and will grow to about 170~kW in \phasetwo when the laser will be upgraded. The different contributions to these numbers are summarised in Tab.~\ref{tab:TC:detectorServiceConsumption}.

\begin{table}[htbp]
\begin{center}
\begin{tabular}{|l|c|c|c|c|}
	\hline
	Type                            &\phaseone &\phasetwo & Place\\
		\hline
Laser Clean room AC	                & 10000  & 10000    &  UG02 \\
Laser Front-End	                    & 5000	 & 5000     &  UG02 \\
Laser power amplifier+pump	        & 25000  & 50000    &  UG02 \\
Laser pulse compressor	            & 3000	 & 5000     &  UG02 \\
Laser additional flow box	        & 5000	 & 5000     &  UG02 \\
Laser Computers, Controller, etc..	& 5000	 & 5000     &  UG02 \\
	\hline

Laser beam-line (pumps)	            & 2000	& 2000       &  UG02 \\ 
IP Box (pumps)	                    & 1000	& 1000       &  UG03 \\
Target chamber (pumps)	            & 0	    & 1000       &  UG03 \\
Electron beam-line (pumps)      	& 2000	& 2000       &  UG03 \\
Magnet vacuum chamber pump	        & 1000	& 1000       &  UG03 \\
	\hline

Magnet power supplies	            & 60000	& 60000      &  UG02 \\
	\hline

Gamma Profiler FE	                & 3000 & 3000  &  UG03 \\
Gamma Profiler DAQ	                & 500  & 500  &  UG02 \\
	\hline

Back-scattering Calo HV	            & 1000 & 1000  &  UG02 \\
Back-scattering Calo DAQ	        & 500  & 500         &  UG02 \\
	\hline

Cherenkov VME	                    & 1600 & 1600           &  UG02 \\
Cherenkov DAQ	                    & 2000 & 2000         &  UG02 \\
	\hline

Scintillation screen DAQ	        & 2000 & 2000        &  UG02 \\
	\hline

Calo DAQ	                     &   500   &  500         &  UG02 \\
Calo HVPS	                     &   1250  &  1250        &  UG02 \\
Calo FEB+VME	                 &   1250  &  1250         &  UG03 \\
	\hline

Tracker HV PS	                &    5500&	5500        &  UG02 \\
Tracker DAQ PC	                &    1000&	1000        &  UG02 \\
Tracker RU+VME	                &    3000&	3000        &  UG03 \\
Tracker chiller 	            &    2000&	2000        &  UG03 \\
			\hline
Total in W                      &    144100&	172100 & \\
			\hline

\end{tabular}
\caption{LUXE scheduled power consumption (in Watts) for the two running phases of the experiment.
\label{tab:TC:detectorServiceConsumption}
}
\end{center}
\end{table}

\subsection{Safety}

Since the experiment will be built underground in the \euxfel complex, its safety concept must comply with all the safety requirements that were developed for the \euxfel and are described in details in Ref.~\cite{XFELTDR}. 

In particular, new supporting structures will be built with fire-resistant construction materials that can resist fire for 90 min. All cable and insulation material that will be used in the underground area will be halogenated-free, contain Flame-retardant materials and be in line with the DESY cable specification. The usage of gas when necessary will be monitored for gas-leak and the environment where it is used will be controlled with oxygen deficiency monitoring devices, even if the amount of gas used in the experiment is expected to be low. The level of radiation created in the experiment will be monitored and should not exceed the limits set within the \euxfel complex. 
All sensitive equipment that could be damaged from failures of the main power supply will be equipped with battery allowing to slowly turn them off in order to save them.

Finally, the work in the experimental and laser area in UG03 and UG02 will be restricted and limited to essential access. In UG03, the access to the experimental equipment will only be possible during stops of the \euxfel, the area being interlocked to the electron beam is not accessible outside these breaks. 
In UG02, a daily access to the laser will be authorised for laser calibration and tuning purposes. 

Recent measurement of the neutron activity correlated to the beam dump present in XS1 showed that a non-negligible radiation level is present in UG02: if the entire beam is dumped at the XS1 beam dump up to 15 $\mu$Sv per hour for a beam power of 80 kW are observed~\footnote{The laser is in a radiation controlled area, defined as an area with an exposure of 6-20 mSv/year. The limits for people working there is up to 6 mSv/year, corresponding to about 400 working hours assuming a dose of 15 $\mu$Sv/h.}. However, the highest level of radiation measured are limited to small part of the area in UG02, and the extent of this is currently being investigated. The design of the clean-room described in Section~\ref{sec:tc:laser} might therefore evolve to reduce the potential exposure of laser operators to radiation as much as possible. Radiation measurement devices will also be installed to monitor the flux of neutrons in the area. While the area will always be accessible with a dosimeter, the access routine will take into account known breaks in the accelerator operation.

40 TW to 350 TW lasers are classified in class 4. Such apparatus crosses the maximum permissible exposure limit by more than $10^6$. Therefore, required safety regulations will be followed. Protective personal equipment (PPE) and safety interlocks will be implemented as per legal regulations.
Safety officers will be appointed to restrict and control the access to the high power laser and interaction areas. 


\newpage
\section{Organisation}
\subsection*{Project organisation, schedule and resources}
\label{sec:orgcost}

The project is organised according to principles that proved their validity in many examples of successful international scientific collaborations. The tasks to be performed, under the leadership of an experienced scientist as Project Leader, are divided into Work Packages, with a clear line of responsibility. Regular consultations, updates and operative decision-making take place at meetings with weekly or bi-weekly periodicity. Quarterly status reports and annual monitoring by an independent expert advisory board are also foreseen.

The schedule of the project is centred around the planned 6-month shutdown of the \euxfel in the first half of 2024. 

The costing of the project, including design, assembly and operation phases was established in a bottom-up process.

\subsection{Project Structure}

We foresee a scientist as LUXE Project Leader (PL), which exercises scientific leadership, steers and coordinates the overall project towards implementation, allocates resources and acts as interface to funding agencies, participating institutions, and the science community at large. The PL can appoint a Deputy Project Leader, to whom he/she can delegate, occasionally or regularly, a part of the tasks. 

The project is broken down in Work Packages (WP), trying to clearly identify homogeneous groups of tasks that are to be performed by people with specific areas of expertise, to create well-defined interfaces and to avoid overlaps of responsibility. Following a common practice for EU-supported projects, management and administration of the project are considered as one of the WPs. The (initial) set of LUXE WPs could be
\begin{itemize}
    \item WP1: Theory and Science Case
    \item WP2: Electron beam transport
    \item WP3: Lasers and diagnostics
    \item WP4: Detectors
    \item WP5: Simulation and results
    \item WP6: Technical Coordination
    \item WP7: Project Management
\end{itemize}
Each WP has a Work Package Leader (WPL), who reports to the PL. Each WPL should 
have a Deputy PL, to represent him/her occasionally in some tasks and in the project meetings (see below).

The task of WP7 is to monitor the progress of the project and the use of available resources, by creating appropriate tools to compare the current status to the project schedule and deadlines, and to the planned flow of resources, as quantitatively and systematically as is practical. The PL should decide on the number and competence area of members of WP1 (organisational, administrative, technical, etc.). It will include for instance a Technical Coordinator and a Resource Coordinator.

The tasks and area of competence of the WP2 to WP6 should be self-explanatory.

\subsection{Schedule}
The following considerations impact the schedule
\begin{itemize}
    \item The only longer shutdown planned is in the first half of 2024: 6 months. 
    \item Normally there is a 5-6 week shutdown each winter.
    \item The \euxfel will be upgraded with a 2nd fan in the part of the tunnel where the LUXE experiment is situated. While the schedule for this \euxfel upgrade is uncertain, it will not occur before 2030 and thus at present 2029 is considered to be the last year of the experiment\footnote{Depending on the scientific results obtained with LUXE and the schedule of the \euxfel upgrade at some point in the future it could be considered to extend the run and/or to explore if LUXE could run after the upgrade in a modified form.}. 
    \item The resources of technical personnel available during the long shutdown periods are limited, and thus it is advantageous to do whatever is possible outside that period.
\end{itemize} 

\begin{figure}[htbp]
    \centering
    \includegraphics[width=0.98\textwidth]{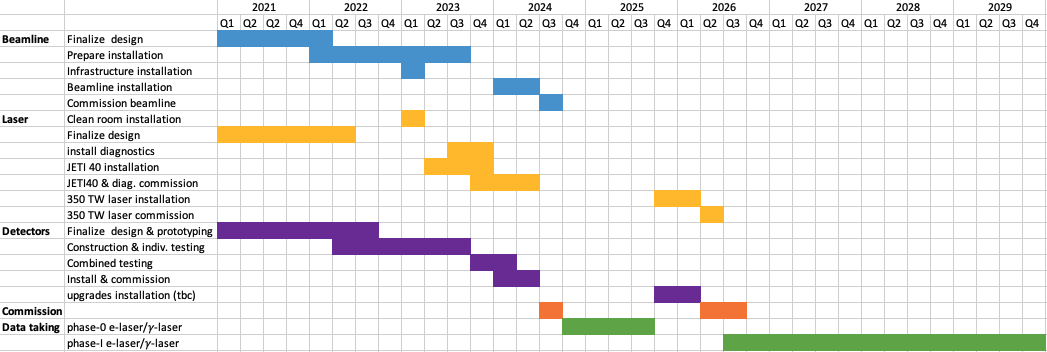}
    \caption{Schematic view of the schedule envisaged for the beamline and infrastructure, laser aspects, detectors, combined commissioning and data taking.}
    \label{fig:schedule}
\end{figure}

The schedule is illustrated in Figure~\ref{fig:schedule}. 

\paragraph{Main installation:}
It is foreseen that the installation of the new beam extraction, including the kicker magnet that directs a bunch to LUXE, occurs in 2024. During this time the installation of the entire beamline, as well as the active elements, such as the laser and the detectors will also occur. This is challenging since many activities need to be coordinated in detail, and the manpower allocation needs to be managed carefully.

\paragraph{Preparatory work:}
Until early 2022 the design of the beamline will need to be finalised so that the installation can be prepared and the installation can commence in 2024. In order to be ready for installation in 2024, it is necessary to order long lead-time items (e.g. the magnets for the beam line) in early 2022. 

Prior to the main installation in 2024, in the winter shutdown 22/23 the laser clean room needs to be constructed, so that the installation of the laser can occur during 2023. Furthermore, support structures will be installed as much as possible prior to the main installation in 2024. The installation of other passive components such as shielding structures is also envisaged. This is termed ''infrastructure" installation in Fig.~\ref{fig:schedule}

The laser diagnostics designs will be finalised during 2021, and the JETI40 laser installation (incl. diagnostics) will start as soon as the clean room is ready. The laser is then commissioned during 2023 before the main installation in 2024. 

The design of the various detectors will be finalised between end of 2021 and mid 2022, including also prototyping in some cases. The construction starts in 2022 and will in all cases finish by the third quarter of 2023. It is then planned to do a combined system test of the IP detector system, the Bremsstrahlung detector system and the Gamma detection systems between mid 2023 and early 2024 at the DESY main site. During 2021-2023 a variety of test beam campaigns are also planned to test detector components individually. 

\paragraph{Phase-0:} 
After installation an extensive commissioning will need to occur, initially without beam for some components but after summer 2024 with beam, and very limited access to UG03. It is envisaged to have a 3-month commissioning phase before stable physics data taking commences in the \elaser setup in autumn of 2024. The WS 24/25 will be used to fix any issues that will come up and then during 2025 \elaser and \glaser data taking in phase-0 will take place. 

\paragraph{Phase-1:} 
During the last quarter of 2025 the high-power laser will be installed and commissioned with the goal of being commissioned by April 2026. Then \phasetwo of LUXE commences and will run for four years until the end of 2029. An upgrade of some of the detectors might also be necessary prior to \phasetwo, or at a later stage, depending on the experience in \phaseone.

\subsection{Cost}

In principle the estimate of the resources needed to achieve the goals of the LUXE project should include three items:

\begin{enumerate}

 \item The capital investment needed to build up (design, procure, assemble, commission) the experimental apparatus.
 \item The cost of personnel for the build up phase
 \item Operation costs: recurrent and personnel costs to actually perform the experiments (including maintenance, repairs, etc.)
\end{enumerate}

At the present point in time, a preliminary estimate of the capital investment (Item 1. above) is available. The estimates vary in precision, in some cases quotes were obtained or it was obtained by comparing to recent similar purchases. We estimate an uncertainty of about 20\% on the values.

A summary of the capital investment contributions is give in Table~\ref{tab:cost}. For the initial phase, including both \elaser and \glaser running, the investment needed is \EUR{8.2~M}. For the \phasetwo experiment an additional investment of \EUR{3.05~M} for the higher power laser is needed. If only performing the \elaser setup the cost would be reduced by \EUR{0.6~M}.
Not included are the costs for the BSM detector.

It is expected that many components, particularly of WP3 and WP4, will be contributed as in-kind contributions by the collaborating institutions. 

\begin{table}[htbp]
    \centering
    \begin{tabular}{l|r}
         Component & cost K\EUR{} \\\hline
         \textbf{WP2: Beam extraction and transfer} & \textbf{2,715} \\
         Magnets & 1,000 \\
         Vacuum system & 640 \\
         Power and Cooling & 500 \\
         Magnet supports & 225 \\
         Kicker magnet & 60 \\
         Beam instrumentation & 190\\
         infrastructure & 100 \\\hline
         \textbf{WP3: Laser and diagnostics} & \textbf{4,636}\\
         \phaseone components & 775\\
         \phasetwo components & 3,050\\
         Beamline optics & 370 \\
         Diagnostics & 386 \\
         Interaction and target chambers & 55\\\hline
         \textbf{WP4: Detectors} & \textbf{1,257}\\
         Calorimeter & 381 \\
         Trackers & 480 \\
         Scintillation Screens and cameras & 67 \\
         Cherenkov detectors & 156 \\
         Gamma flux monitor & 23 \\
         Gamma profiler & 150 \\\hline
         \textbf{WP6: Technical Coord., Infrastructure} & \textbf{2,628}\\
         Infrastructure Exp. & 128\\
         Installation Exp. & 1,200\\
         Service Exp & 372 \\
         Infrastructure Laser & 805 \\
         Installation Laser & 123 \\\hline\hline
         \textbf{Total \phaseone \elaser} & \textbf{7,600} \\
         \textbf{Total \phaseone \elaser and \glaser} & \textbf{8,200} \\
         \textbf{Total \phaseone and \phasetwo} & \textbf{11,200 }\\\hline
    \end{tabular}
    \caption{Capital cost for each of the work packages. In bold are the total costs of the relevant WPs, and detailed below are the values for the individual aspects of a given WP. Finally the total costs are given for three configurations. For the total costs the numbers are rounded.}
    \label{tab:cost}
\end{table}

\section{Conclusions}
\label{sec:conclude}
In conclusion, LUXE is a novel experiment that could be undertaken at the \euxfel and that would pioneer a regime of physics that has so far not been explored. The excellent quality of the \euxfel beam, together with a high-power precision laser and well-designed state-of-the-art detectors provide a unique opportunity to perform this experiment in the 2020s. 

\section*{Acknowledgements}
We thank the members of various DESY groups without whom this work would not have been possible: MVS (vacuum modification), MIN (kicker magnet, beam dump), D3 (radio protection), MEA (installation and magnets), ZM1 (construction), MKK (power and water), IPP (CAD integration).

We also thank the DESY directorate for funding this work through the DESY Strategy Fund. The work by B.~Heinemann and C.~Grojean was in part funded by the Deutsche Forschungsgemeinschaft under Germany's Excellence Strategy -- EXC 2121 ``Quantum Universe" -- 390833306. H.~Abramowicz and B.~Heinemann would like to thank the German-Israel-Foundation (GIF) (grant number 1492). H.~Abramowicz would also like to thank the Israel Academy of Sciences.
The work by A.~Hartin and M.~Wing was supported by the Leverhulme Trust Research Project Grant RPG-2017-143 in the UK. The work of G. Perez is supported by grants from the BSF, ERC-COG, ISF, Minerva, and the Segre Research Award. A.~Ilderton, B.~King, S.~Tang and A.J.~Macleod acknowledge support from the Engineering and Physical Sciences Research Council (EPSRC) (grant no. EP/S010319/1). G.~Sarri wishes to acknowledge support from EPSRC (grant Nos: EP/N027175/1 and EP/P010059/1). T.~Blackburn acknowledges the use of resources provided by the Swedish National Infrastructure for Computing (SNIC) at the High Performance Computing Center North (HPC2N). The work of N.~Tal Hod is supported by a research grant from the Estate of Moshe Gl\"{u}ck, the Minerva foundation with funding from the German Ministry for Education and Research, the Israel Science Foundation (ISF) (grant No. 708/20), the Anna and Maurice Boukstein Career Development Chair and the Estate of Emile Mimran. Y.~Soreq and T.~Ma are supported by grants from the ISF, BSF (NSF-BSF program), Azrieli Foundation and the Taub Family Foundation. A.~Fedotov and 
A.~Mironov were supported by the MEPhI Academic Excellence Project (Contract No. 02.a03.21.0005) and by the Russian Foundation for Basic Research (under grants No. 19-02-00643 and No. 20-52-12046). B.~Heinemann, H.~Harsh, X.~Huang, R.~Prasad, F.~Salgado, U.~Schramm, K~Zeil and M.~Zepf thank the Helmholtz Association and the Bundesministerium f{\"u}r Bildung und Forschung for the support via the 2018 Helmholtz Innovation Pool. 

This work has benefited from computing services provided by the German National Analysis Facility (NAF).


\providecommand{\href}[2]{#2}\begingroup\raggedright\endgroup

\newpage
\appendix
\section{Appendix on Theory}
\subsection{Theory Background}
\label{app:theory}
\noindent\tbf{Light-front kinematics:}$\qquad$ 
Coordinates and momenta in a plane wave background are more conveniently expressed using light-front variables. We choose the plane wave to propagate along the $z$-axis with increasing $z$ co-ordinate, so the plane wave only depends on the light-front co-ordinate $x^{-}$, where $x^{\pm}=t\pm z$. It will be useful to write the $x^{-}$ co-ordinate dependence in terms of the laser pulse \emph{phase}, $\vphi=\vkap\cdot x$, where $\vkap^{\mu} = \delta^{\mu +}\vkap^{+}= \omega(1,0,0,1)$ is the laser pulse wave vector. It follows that only three components of energy and momentum are conserved: $P^{-}$ and $\mbf{P^{\perp}}$, where $\mbf{P}^{\perp}$ is the particle momentum transverse to the laser wave vector. In particular it is useful to analyse particle spectra in the  ``\emph{light-front momentum fraction}'', $\ess=K^{-}/P^{-}$, where $K$ is the momentum of the particle to be detected and $P$ is the momentum of the incoming particle. For a detector at fixed angle, a spectrum in $\ess$ is equivalent to an energy spectrum; for a detector at fixed energy, a spectrum in $\ess$ is equivalent to a spectrum in emission angle.

Since $P^{+}$ is \emph{not} conserved in a plane wave, this allows for processes to occur that are kinematically forbidden in vacuum. Example processes that LUXE will measure include the (non-linear) Compton process of emission of a high-energy photon by an electron, and the (non-linear) Breit-Wheeler process of the production of an electron-positron pair from a photon. Both of these are often referred to as ``$1\to2$'' processes and as such, kinematically they are \emph{emission} or \emph{decay} \emph{processes}. However, in this report, we will introduce a variable, $n$, such that ``$n\vkap$'' represents the momentum contributed by the field to these processes. In a monochromatic laser background, $n$ is an integer, and is interpreted as the number of laser photons.

\noindent\tbf{Scattering in a plane wave:}$\qquad$  A key aspect of particle scattering mediated by a laser pulse background is the contribution of interference effects over space-time scales much larger than the probe particle's Compton wavelength. 
\begin{figure}[htbp]
\centering
\includegraphics[width=8cm]{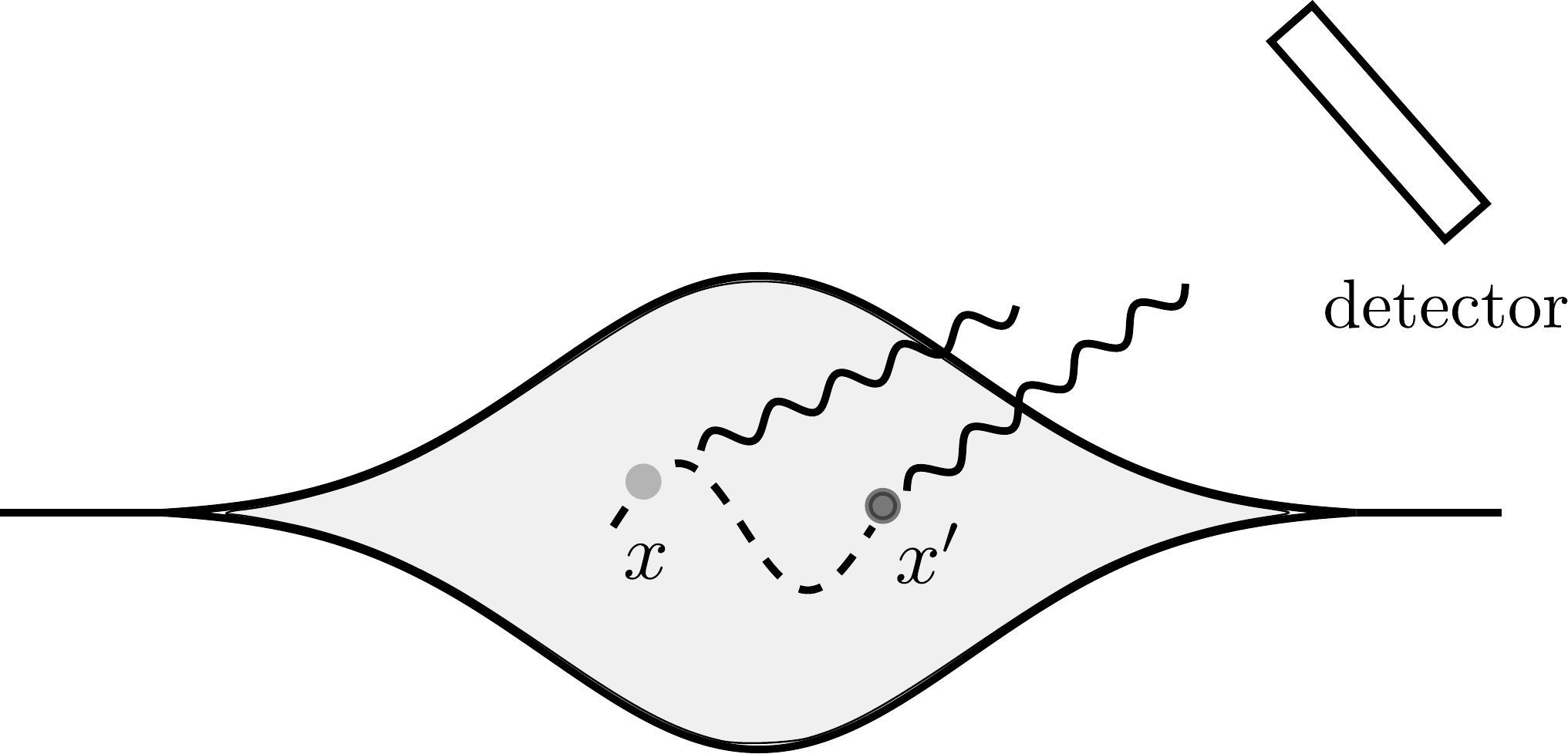}
\caption{Sketch illustrating interference in emission from the same electron at different points of its trajectory in a laser pulse.} \label{fig:pulseSketch}
\end{figure}    

Taking a first-order process such as non-linear Compton scattering or non-linear Breit-Wheeler pair-creation (illustrated in Table~\ref{tab:candproc}), probabilities $\prob$
are proportional to a mod-squared integral over all space-time of the interaction Lagrangian, $\mathcal{L}_{I}$. This leads to interference between processes occurring at different spacetime points, as illustrated in Fig.~\ref{fig:pulseSketch}, becoming important in determining spectral features. Due to the plane-wave nature of the background, this integral can be written in terms of the laser pulse phase i.e.  $\prob\sim \int d\vphi\,d\vphi' |\mathcal{L}_{I}|^{2}$. Writing in terms of the \emph{average phase}, $\phi=(\vphi+\vphi')/2$ and the \emph{interference phase}, $\vartheta=\vphi-\vphi'$, one can define a  \emph{formation phase length}, $L_\vartheta$, as being the interference phase over which one must integrate to attain a good approximation to the total probability, i.e. by only considering $|\vartheta|<L_{\vartheta}$.

By way of example, we focus on the Compton process. We use Volkov wave functions, $\psi$, that satisfy the Dirac equation:
\bea
\psi=u_{\pi_p,\sigma} \mbox{e}^{iS_{p}};\qquad u_{\pi,\sigma}=\left(1+\frac{\slashed{\vkap}\slashed{a}}{2 \vkap\cdot p}\right)u_{p,\sigma}; \quad S_{p}=-p\cdot x - \int^{\vphi} \frac{2 p\cdot a(x) - a^{2}(x)}{2\vkap\cdot p}\,dx, \label{eqn:Volkov1}
\eea
where
$u_{\pi,\sigma}$ is the spinor of an electron in spin state $\sigma$ with kinetic momentum:
\[
\pi_p = p - a + \vkap~ \frac{2\,p\cdot a - a^{2}}{2 \vkap\cdot p},
\]
equal to the instantaneous \emph{classical} momentum of an electron in a plane wave, and $S_{p}$ is the corresponding classical action ($a=eA$ is the (normalised) vector potential and $e<0$ is the charge of an electron). The resulting probability for the Compton process in a plane-wave pulse can be written as \cite{Dinu:2013hsd}:
\bea
\prob = -\frac{\alpha}{\pi\eta}\int_{0}^{1}d\ess \int_{-\infty}^{\infty} d\phi \int_{0}^{\infty} d\vartheta \, \left[\frac{1}{\mu}\frac{d\mu}{d\vartheta}+g\langle a'\rangle^{2}\vartheta\right] \sin (x_{0}\vartheta\mu) \label{eqn:NLCQED}
\eea
\[g = \frac{1}{2}+\frac{\ess^{2}}{4(1-\ess)}; \qquad x_{0}=\frac{1}{2\eta}\frac{\ess}{1-\ess}; \qquad \mu = \mu(\phi,\vartheta)= 1+\langle\mbf{a}_{\perp}^{2}\rangle-\langle\mbf{a}_{\perp}\rangle^{2},
\]
where $\langle f \rangle$ is the window average of the function $f$: $\langle f \rangle = \vartheta^{-1}\int_{\phi-\vartheta/2}^{\phi+\vartheta/2} f(x)dx$. The exact QED plane-wave result Eq.~(\ref{eqn:NLCQED}) can be evaluated numerically and benchmarked against simulations.

In Eq.~(\ref{eqn:NLCQED}), $\mu$ is the (Kibble) \emph{effective mass} squared over the squared electron mass. (In the absence of the laser background,  $\mu=1$, so the effective mass reduces to the usual rest mass, $m$.) The effective mass depends in general on the electron's average phase position, $\phi$ and its interference phase $\vartheta$. It can have an important effect on dynamical quantities, depending on the form of the background field~\cite{Harvey:2012ie}. For example, in a circularly-polarised monochromatic background, its effect is particularly straightforward to understand. In the frame in which the initial electron is at average at rest, the photon spectrum produced in the Compton process in a circularly-polarised monochromatic background demonstrates a first harmonic at exactly the position given by the Klein-Nishina formula, but for an electron with a heavier mass, $m_{\ast} = m\sqrt{1+\xi^{2}}$, where $m_{\ast}$ is the cycle-averaged (i.e. $\vartheta$ is equal to the phase difference in one laser cycle) effective mass. This concept will reoccur in the Compton and Breit-Wheeler processes when studied in long (approximately monochromatic) pulses.

\begin{figure}[htbp]
\centering
\includegraphics[width=0.6\linewidth]{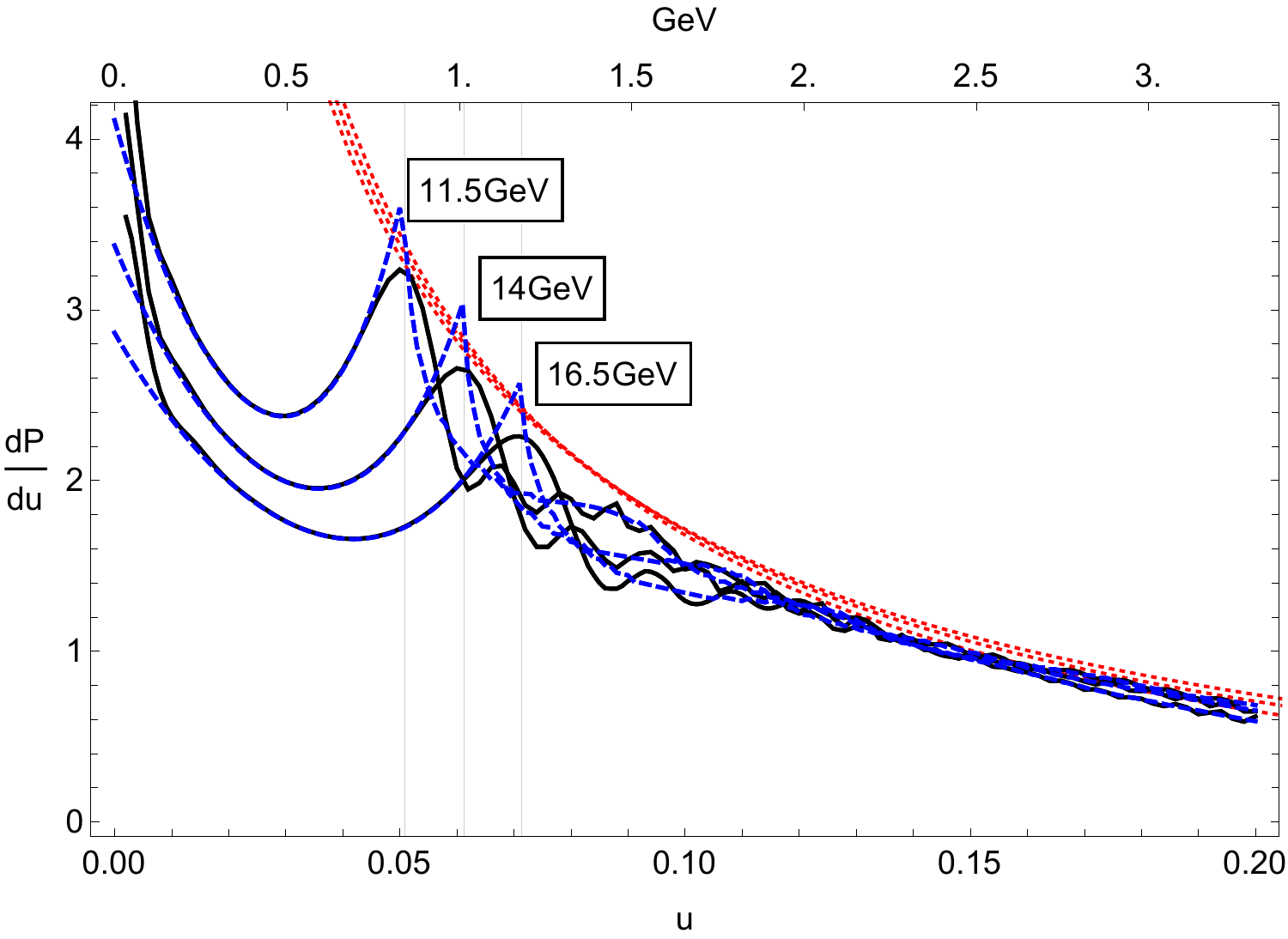}
\caption{Compton photon spectrum for three different electron energies and a $\xi=2$, laser pulse. The solid lines represent the full QED result [Eq.~(\ref{eqn:NLCQED})], the blue dashed (red dotted) lines are the integration of the LMA [Eq.~(\ref{eqn:Rnlma})] (LCFA [Eq.~(\ref{eqn:Rlcfa})]) rates. The vertical grid lines signify the position of the edge of the first harmonic (Compton edge) at intensity parameter $\xi$. (The photon energy scale in GeV on the upper horizontal axis is for $16.5\,\trm{GeV}$ electrons.)
}\label{fig:NLCspec1}
\end{figure}

\noindent \tbf{Local approximation to SFQED processes:}$\quad$ To apply the QED plane-wave results to the situation in LUXE of having a focussed laser pulse background, the electron motion due to the laser pulse will be solved using the Lorentz equation. A standard technique to solve for the electron motion due to the Compton process, as well as calculate the radiation spectrum in the laser pulse background, is a Monte-Carlo sampling of a probability \emph{rate}, $\rate$. This rate is defined e.g. per unit phase $\phi$, such that the total probability $\prob$ for the process is $\prob=\int d\phi\, \rate(\phi)$. There are two local definitions of the rate which will be employed in the numerical code to simulate LUXE: the locally constant field approximation (LCFA) \cite{ritus85,harvey.pra.2015,dipiazza.pra.2018,ilderton.pra.2019,dipiazza.pra.2019} and the locally monochromatic approximation (LMA) \cite{Heinzl:2020ynb}. Both approximations assume that the formation phase length of the QED process satisfies $L_{\vartheta} \ll 1$ (although the LMA assumes this only in the pulse envelope and not the carrier frequency, which is treated exactly), allowing a Taylor expansion of Eq.~(\ref{eqn:NLCQED}) in $\vartheta$, and a subsequent integration out of the $\vartheta$ variable. This procedure leaves just the average phase remaining, which is then used to define the rate. 
By defining a rate that only depends upon ``local'' quantities, SFQED processes can be simulated in more complicated and experimentally-relevant laser pulse shapes. Furthermore, the probability of multiple SFQED events per probe particle can then be approximated, which is more challenging to handle from the exact QED calculation.
In the following,  we introduce the approximations for the case of the Compton process, and in Fig.~\ref{fig:NLCspec1} give an example benchmark against the QED result in a plane wave for typical LUXE parameters.  

\noindent\tbf{LCFA: }$\quad$ Taylor-expanding the integrand in Eq.~(\ref{eqn:NLCQED}) directly in $\vartheta$, retaining up to cubic terms in the $\sin$ function, and leading-order terms in the coefficient, allows the integral over $\vartheta$ to be performed. The LCFA for the Compton process is then:
\bea
\rate^{\tsf{LCFA}}_{\gamma} = \frac{\alpha}{\eta}\int_{0}^{1} d\ess \left\{\Ai_{1} (z_{\gamma}) + \left[\frac{2}{z_{\gamma}}+\chi_{\gamma}\sqrt{z_{\gamma}}\right]\Ai'(z_{\gamma})\right\} \label{eqn:Rlcfa}
\eea
\[
z_{\gamma} = \left(\frac{1}{\chi_{e}}\frac{\ess}{1-\ess}\right)^{2/3}; \qquad \Ai_{1}(x)=\int_{x}^{\infty}\Ai(y) dy.
\]
This rate is so-called ``locally constant'' because the integrand is the same as for a constant crossed EM field. The LCFA for the Compton process is quantitatively a good estimate when $\xi\gg1$ and when~\cite{Khokonov:2002cf},
\bea
\left(\frac{\xi^{2}}{8\eta}\frac{\ess}{1-\ess}\right)^{1/3}\gg 1. \label{eqn:lcfacond}
\eea
The accuracy of the LCFA for calculating the photon yield around the first harmonic, is illustrated in the right-hand plot of \figref{fig:CEdge2}. Therefore, the LCFA is more accurate at lower particle energies and higher field strengths. It is therefore applicable and widely employed in intense laser-plasma physics where field strengths are higher ($\xi \gg1$) and particle energies lower ($\eta \ll 1$) than in LUXE, and is the standard method for including QED effects to model experiments at the very high field intensities to be produced by lasers at e.g.  ELI \cite{Turcu:2016dxm}. However, among the shortcomings of the LCFA is that it cannot produce harmonic structure in the spectrum of emitted photons and it is not accurate when $\xi \approx 1$ (see \figref{fig:NLCspec1}). Since it is planned at LUXE to make measurements in both of these cases, it is necessary to develop another approximation.

The LCFA for the Breit-Wheeler process is expected to be more accurate for use with LUXE parameters, because the minimum harmonic of the laser frequency, $n_{\ast}$, already fulfills $n_{\ast}\gg1$ (e.g. for $\xi=2$, $n_{\ast}>50$). Also, higher field intensities are required for the Breit-Wheeler process to measurable. It is known that in the high-harmonic, high-intensity limit, SFQED processes are generally well-approximated by the LCFA result. The LCFA for the Breit-Wheeler process is given by:
\bea
\rate^{\tsf{LCFA}}_{\gamma} = \frac{\alpha}{\eta}\int_{0}^{1} d\tea \left\{\Ai_{1} (z_{e}) + \left[\frac{2}{z_{e}}-\chi_{\gamma}\sqrt{z_{e}}\right]\Ai'(z_{e})\right\} \qquad z_{e} = \left(\frac{1}{\chi_{\gamma}}\frac{1}{\tea(1-\tea)}\right)^{2/3} \label{eqn:RlcfaNBW}
\eea
where $\tea=P'^{-}/K^{-}$ and $P'$ is the momentum of the produced \emph{electron}.

\noindent\tbf{LMA: }$\quad$ The Locally Monochromatic Approximation \cite{Heinzl:2020ynb} perturbs around the solution for SFQED processes in an infinite monochromatic plane wave by making a local expansion in the pulse envelope. Consider the example of a circularly-polarised plane wave pulse with vector potential:
\[
a^{\mu} = m\xi f\left(\frac{\vphi}{\Phi}\right) \left[\varepsilon^{\mu}\cos\vphi + \beta^{\mu}\sin\vphi\right],
\]
where $\varepsilon$, $\beta$ are spacelike vectors normalised to $-1$ perpendicular to the plane wave propagation direction and $\Phi$ is the pulse phase duration. One ``cycle'' refers to $\vphi$ increasing by $2\pi$.  The LMA treats the fast oscillation scale given by the carrier frequency in the square brackets exactly, but keeps only the leading-order term of a derivative expansion of the pulse envelope \cite{Narozhnyi:1996qf,Seipt:2010ya,Seipt:2014yga,Seipt:2016rtk}. Since $\Phi\gg 1$ for the laser pulses used at LUXE, this should be a good approximation. The advantage over the LCFA is that it captures harmonic structure, is exact in the low-energy limit (whereas the LCFA is inaccurate), and is more accurate for arbitrary $\xi$. The disadvantage is that by specifying to a plane wave it is not as versatile as the LCFA. However, with the weak focussing, long pulse durations and negligible plasma effects planned at LUXE, the LMA will be the central approximation used in numerical codes. Different to the LCFA, the form of the LMA depends on the polarisation of the background laser pulse. For the Compton process in a circularly-polarised background, the rate is $\rate^{\tsf{LMA}}_{\gamma} = \sum_{n=1}^{\infty} \rate^{\tsf{LMA}}_{\gamma,n}$, where the harmonic rate is:
\bea
\tsf{r}^{\tsf{LMA}}_{\gamma,n} =  \frac{\alpha}{\eta}\int_{0}^{u_{n}(\phi)} du \left\{J_{n}^{2}(z)+g\xi^{2}\left[2J_{n}^{2}(z)-J_{n+1}^{2}(z)-J_{n-1}^{2}(z)\right]\right\} \label{eqn:Rnlma}
\eea
\[
z=\frac{2n\xi(\phi)}{\sqrt{1+\xi^2(\phi)}}\left[\frac{\bar{\ess}}{\bar{\ess}_{n}}\left(1-\frac{\bar{\ess}}{\bar{\ess}_{n}}\right)\right]^{1/2};\quad \bar{\ess}=\frac{\ess}{1-\ess}; \quad \bar{\ess}_n=\frac{2n\eta}{1+\xi^2}=\frac{u_{n}}{1-u_{n}};
\]
$\xi(\phi)=\xi f\left(\frac{\phi}{\Phi}\right)$ contains the functional dependency of the pulse envelope and $J_n$ is the $n$th order Bessel function of the first kind. The integrand of Eq.~(\ref{eqn:Rnlma}) for fixed $\xi$ is the integrand to calculate the probability of the Compton process in an infinite monochromatic EM background.

The LMA rate for the Breit-Wheeler process, $\rate^{\tsf{LMA}}_{e}$, is then given by $\rate^{\tsf{LMA}}_{e} = \sum_{n=\lceil n_{\ast}(\phi)\rceil}^{\infty} \tsf{r}^{\tsf{LMA}}_{e,n}$, where the harmonic rate is:
\bea
\tsf{r}^{\tsf{LMA}}_{e,n} =  \frac{\alpha}{\eta_\gamma}\int_{\tea_{-,n}}^{\tea_{+,n}} dt  \left\{J_{n}^{2}(z_{e})-\frac{\xi^{2}(\phi)}{2}\left[\frac{1}{2\tea(1-\tea)}-1\right]\left[2J_{n}^{2}(z_{e})-J_{n+1}^{2}(z_{e})-J_{n-1}^{2}(z_{e})\right]\right\} \label{eqn:NBWlma}
\eea
\[
z_e = \frac{2 n \xi}{ \sqrt{1+\xi^2}} \sqrt{
\frac{n_{\star}(\phi)}{n} \frac{1}{4 \tea (1-\tea)} \left( 1 - \frac{n_{\star}}{n} \frac{1}{4\tea(1-\tea)}\right)},
\]
the threshold harmonic is $\lceil n_{\ast}(\phi) \rceil$, where
\[
n_{\ast}(\phi) = \frac{2(1+\xi^{2}(\phi))}{\eta_\gamma},
\]
and the integration limits are given by $\tea_{\pm,n}=(1\pm(1-n_{\ast}(\phi)/n)^{1/2})/2$.

\subsubsection{Trident}
In the main text, we commented on a separation of the trident process into parts that depend on the pulse duration $\Phi$ in different ways. Here, we mention a second useful separation one can make, in terms of the exchange symmetry present in the (electron-probed) trident process due to identical outgoing electrons. This means that the amplitude has two terms with a relative minus sign, i.e. $A_{12}-A_{21}$, where $A_{21}$ is obtained from $A_{12}$ by changing place of the two electrons. On the probability level, the ``direct'' part, $\prob_{\tri}^{\tsf{dir}}\sim|A_{12}|^2+|A_{21}|^2$ gives $\prob_{\tri}^{(2)}$ and one part of $\prob_{\tri}^{(1)}$, and the ``exchange'' part $\prob^{\tsf{ex}}_{\tri}\sim2\text{Re}A_{12}^*A_{21}$ gives the rest of $\prob^{(1)}_{\tri}$. The separation is useful because $\prob^{\tsf{ex}}_{\tri}$ is very challenging to evaluate. 

By direct numerical calculation, it can be shown, that neither the exchange part, $\prob^{\tsf{ex}}_{\tri}$, nor the total one-step part, $\prob_{\tri}^{(1)}$, contribute significantly to the total probability for trident in the parameter regime studied at LUXE.  Therefore, we assume $\prob_{\tri}=\prob_{\tri}^{(2)}$ and benchmark the LCFA and the LMA against the exact QED calculation for this term.

Since LUXE will consider a range of $\xi$ values, it is useful to consider the scaling of the trident probability with $\xi$ and $\eta_e$. When $\chi_{e}=\xi\eta_e\ll1$, one can reason using the saddle-point method. In this case, $\prob_{\tri}\sim\exp[-F(\xi)/\eta_e]$~\cite{Dinu:2017uoj},
where the function $F(\xi)$ depends on the field shape. This exponential suppression means that the probability falls rapidly as $\eta_e$ is reduced. For example, with $\xi=\sqrt{2}$ (in this regime the field maximum sets the relevant scale for $\xi$) and $\eta=0.192$ (for the $16.5\,\trm{GeV}$ case) the maximum of the spectrum is $\sim10^{-13}$ (at $s_1=s_2=s_3=1/3$), which is several orders of magnitude below the peaks in Fig.~\ref{tridentSpectrum}. The exponential suppression makes the field effectively shorter, because only the field maxima that are very close to the the global maximum contribute significantly. The effectively shorter pulse length means that the one-step terms become more important for small $\eta_e$. In this regime, one can compensate for the relatively small $\eta$ (e.g. compared to~\cite{Bamber:1999zt}) by increasing $\xi$, and a larger $\xi$ favours the two-step part.

Although the factorisation in Eq.~(\ref{eqn:triLCFA1}) is straightforward for a linearly-polarised background, for other polarisation choices the factorisation is more subtle and requires the development of a ``gluing approach''. A formalism has been developed based upon the Stokes vectors for the polarisation of the photons and fermions. For the unpolarised trident process, one only needs the photon polarisation dependence of the first-order sub-processes, which can be expressed as ${\sf P}_\gamma={\bf M}_\gamma\cdot {\bf S}_\gamma$ and ${\sf P}_e={\bf M}_e\cdot {\bf S}_\gamma$, where ${\bf M}_\gamma$ and ${\bf M}_e$ are ``strong-field-QED Mueller matrices'' (or vectors in this case) and ${\bf S}_\gamma$ is the Stokes vector for the intermediate photon. The two-step part of trident is now, for arbitrary field polarisation, given by ${\sf P}_{\tri}\sim{\bf M}_\gamma\cdot {\bf M}_e$. More details can be found in~\cite{Dinu:2019pau}.

Therefore, for the parameters that LUXE will use to generate pairs via the non-linear trident process, it is a good approximation to consider just the ``two-step'' process of non-linear  Compton scattering followed by non-linear Breit-Wheeler pair-creation.

\section{Appendix on Simulation}
\label{app:sim}
\subsection{Method of Simulating SFQED at the Interaction Point} \label{app:simtheo}

It can be shown that the motion of ultrarelativistic particles in plane wave backgrounds is semiclassical. Therefore, the objective is to take into account realistic pulse shapes and multiple processes. To achieve this objective, numerical simulations are essential when applying theoretical calculations to experimentally relevant situations.
As a numerical approach to the complete problem---obtaining the scattering amplitude for all possible final states, given an initial state containing a single electron and a prescribed electromagnetic field~\cite{dipiazza.rmp.2012}---is beyond present capability, simulations must use a simplified framework, in which the scattering problem is approached indirectly.
Motivated by the observation that the QED scattering probability in a plane-wave background depends on the \emph{classical} kinetic momentum of an electron in a plane EM wave (see e.g. Eq.~(\ref{eqn:Volkov1})), simulations take as their foundation a classical treatment of charged particle dynamics.
In general, the electromagnetic field that drives those dynamics must be determined self-consistently, accounting for both externally injected fields (e.g. lasers) and those generated by the charged particles themselves.
(Both are essential in the modelling of laser-plasma interactions~\cite{ridgers.jcp.2014,gonoskov.pre.2015}.)
For LUXE parameters, however, the dominant contribution comes from the externally injected field and simulations may take the electromagnetic field to be prescribed.
The existence of a definite trajectory in this field allows for the modelling of scattering events by the use of local probability `rates' outlined in Section \ref{sec:theo1}.
The object of simulations of the interaction point at LUXE, is then to predict final-state particle spectra, given as realistic a set of initial conditions as possible.

The simulation code of the interaction point must propagate particles through the intense laser background field, generate scattering events with the correct probability distribution, and apply conservation laws to determine post-scattering momenta. For the beam of $N_e$ electrons, which is characterised by its six-dimensional (invariant) distribution function $f_e(t, \mbf{x}, \mbf{p})$, i.e. $d N_e = f_e \, d^3 \mbf{x}\, d^3 \mbf{p}$, 
the evolution of this distribution function is controlled by a set of kinetic equations:
    \begin{equation}
    \frac{\partial f_e}{\partial t}
        + \frac{\mbf{p}}{\gamma m} \cdot \frac{\partial f_e}{\partial \mbf{x}}
        - e \left( \mbf{E} + \frac{\mbf{p} \times \mbf{B}}{\gamma m} \right) \cdot \frac{\partial f_e}{\partial \mbf{p}} =
    \left. \frac{\mathrm{d} f_e}{\mathrm{d} t} \right|_\text{scattering}
    \label{eq:KineticEquations}
    \end{equation}
The left-hand side of Eq.~(\ref{eq:KineticEquations}) determines the classical evolution of the distribution under the action of the electric and magnetic fields $\mbf{E}$ and $\mbf{B}$; the right-hand side is a collision operator that accounts for the dynamical effect of scattering events.
For example, for the Compton process,
    \begin{equation}
    \left. \frac{\mathrm{d} f_e}{\mathrm{d} t} \right|_{e \to e \gamma} =
    -f_e \int\! w_{e \to e\gamma}(\mbf{p},\mbf{k}) \,d^3\mbf{k}
    + \int\! f_e' w_{e \to e\gamma}(\mbf{p}', \mbf{p}' - \mbf{p}) \,d^3\mbf{p}',
    \end{equation}
where $w_{e \to e\gamma}(\mbf{p},\mbf{k}) \,d^3\mbf{k}$ is the probability rate that an electron with momentum $\mbf{p}$ emits a photon with momentum $\mbf{k}$.
(where $f_e = f_e(t, \mbf{x}, \mbf{p})$ and $f'_e = f_e(t, \mbf{x}, \mbf{p}')$.)
This framework can be extended to include spin and polarisation effects, as additional degrees of freedom in the distribution function.

\subsubsection{Particle propagation and event generation}
To simulate the LUXE interaction point using Eq.~(\ref{eq:KineticEquations}), the distribution function is represented statistically and the collision operator is implemented via Monte Carlo methods. The trajectories of individual particles, the initial momenta of which are sampled from the distribution function, are obtained numerically by solution of the appropriate equation of motion.
The probability that a scattering event is generated at a point $x^\mu$ along the trajectory is given by $\prob = \rate(x^\mu, p^\mu, ...) \Delta t$, where the rate $\rate$ depends on ``local'' quantities such as the momentum and field strength (see e.g. Eq.~(\ref{eqn:Rlcfa}) and \ref{eqn:Rnlma}), and $\Delta t$ is the relevant time interval.
The momentum (spin, etc.) of a daughter particle is obtained by pseudo-random sampling of the appropriate spectrum.
Particles created within the strong-field region are propagated and tracked for further scattering events in much the same way as the primary particles.
Like all Monte Carlo approaches, final results are subject to statistical fluctuations, which scale with the number of particles, $N$, with which the distribution function is sampled, as $1/\sqrt{N}$.

The structure of the probability rates determines what dynamical quantities must be extracted from the classical trajectory.
In the LCFA (e.g. Eq.~(\ref{eqn:Rlcfa})), the rate is controlled by the quantum nonlinearity parameter $\chi_{e}$, i.e. the instantaneous values of the field strength $F_{\mu\nu}$ and kinetic momentum $\pi_\mu$.
Furthermore, momentum conservation in the plane wave background can be written in a way that the kinetic, not the asymptotic, momentum enters the conservation of momentum.
Thus the classical trajectory must be defined in terms of kinetic momentum, which evolves according to the Lorentz force equation $d \pi_\mu/ d \tau = -e F_{\mu\nu} \pi^\nu / m$ ($\tau$ is the lab time and $F$ the Faraday tensor).
This framework is applied in particle-in-cell codes~\cite{ridgers.jcp.2014,gonoskov.pre.2015}, as well as in particle-tracking codes used to simulate `all-optical' laser-beam interactions~\cite{cole.prx.2018,poder.prx.2018} and beam-beam collisions~\cite{Chen:1994jt,guinea.pig}.
While it may be applied to arbitrary field configurations, provided that the coherence length is smaller than the timescale of the external field, the fact that the rate is local means that interference phenomena are missed.
This causes global failure of the rate at $\xi \lesssim 1$ and the loss of harmonic structure at small momentum transfer even as $\xi \gg 1$ if Eq.~(\ref{eqn:lcfacond}) is not fulfilled  \cite{harvey.pra.2015,dipiazza.pra.2018,blackburn.pop.2018}.
(``Post-LCFA'' methods improve on these properties~\cite{dipiazza.pra.2019,ilderton.pra.2019,king.pra.2020}, but do not successfully reproduce that harmonic structure.)

To bypass this problem, in Eq.~(\ref{eq:KineticEquations}), we use the LMA (plane-wave) rate (e.g. Eq.~(\ref{eqn:Rnlma})) instead.
These retain interference effects at the scale of the laser wavelength, thereby capturing harmonic structure, and ensuring accuracy even if $\xi \sim 1$~\cite{Heinzl:2020ynb}.
The LMA depends on the energy and intensity parameters $\eta$ and $\xi$ \emph{independently} whereas the LCFA only depends on their product. Also, the polarisation of the background modifies the form of the rate, in contrast to the LCFA.
The key dynamical quantity that enters the rate, and also the conservation of momentum, is the, cycle-averaged, \emph{quasi}-momentum $q$, which satisfies $q^2 = m^2 (1 + \langle a^2 \rangle)$ ($\langle f \rangle$ denotes a cycle average of $f$).
Thus the classical trajectory must be defined in terms of $q$ and a cycle-averaged position $\mbf{r}$, which evolve according to the `ponderomotive' force equations: $\dot{\mbf{q}} = -m^2 \mbf{\nabla} \langle a^2 \rangle /(2 q^0)$ and $\dot{\mbf{r}} = \mbf{q} / q^0$, where dots denote differentiation with respect to lab time~\cite{quesnel.pre.1998}.
Similar concepts were used to simulate SLAC experiment 144~\cite{Bamber:1999zt} (codes \textsc{NUMINT}~\cite{numint} and \textsc{MCSCAT}~\cite{mcscat}), and are also used in CAIN \cite{Chen:1994jt} and GUINEA-PIG \cite{guinea.pig}.
In Appendix~\ref{sim:montecarlo} we discuss \ipstrong, the Monte Carlo code that is used to simulate photon and pair creation at the LUXE interaction point.

\subsubsection{Extensions and limitations}

Simulations can, under certain conditions, extend theoretical calculations of a single-vertex process in a plane-wave background.
For LUXE it is necessary to use rates calculated for this background, rather than for a constant, crossed background, as the coherence length is comparable to the laser wavelength.
This requires that the fields are purely transverse and null, which is only approximately true for focused laser pulses.
These exhibit longitudinal fields near focus of size $E_\parallel / E_\perp \sim O(\varepsilon)$, where $\varepsilon = \lambda / (\pi w_0)$ is the diffraction angle of a Gaussian beam with waist $w_0$ and wavelength $\lambda$~\cite{salamin.apb.2007}.
The tighter the focusing, the greater the departure from a plane wave.
On one hand, this maximises the field strength and so the strength of nonlinear effects (at fixed pulse energy); on the other, the overlap between electron beam and laser field is reduced and so the strength of the desired signal.
The most significant effect of focusing is likely to be the variation of the peak laser amplitude $\xi$ across the spatial and temporal profile of the electron beam.
This is straightforwardly accounted for within a simulation.

The LCFA and LMA rates in Eqs.~(\ref{eqn:Rlcfa}) and (\ref{eqn:Rnlma}) are defined via $\prob = \int \rate(\vphi)\,d\vphi$, where $\prob$ is the probability of a  (dressed) single-vertex process (Compton or Breit-Wheeler). However, for the many-cycle pulse with $\xi > 1$ employed at LUXE, the probability of Compton scattering can easily exceed unity. In this case, $\prob$ may be interpreted as the mean number of emissions, $\prob \to \langle N_\gamma \rangle$~\cite{dipiazza.prl.2010}.
This holds exactly in the limit of zero recoil (the classical case), where individual emission events are independent of one another and the probability of $n$-photon emission is given by $\prob^{\trm{cl.}}_n = \langle N_\gamma \rangle^n e^{-\langle N_\gamma \rangle} / n!$.
In general, non-negligible recoil modifies the probability of subsequent scattering events, by modifying the momentum and polarisation of the intermediate state.
Simulations model these higher-order contributions to scattering (e.g. double or triple Compton scattering) as the incoherent product of first-order processes. This is based upon results of calculations of higher-order process such as trident (see below), for which part of the total probability can be factorised in terms of polarised, lower-order processes. In simulation, only these parts of the probability are therefore included - those where the  intermediate state propagates, which is expected to be a good approximation if $\xi^2\Phi \gg 1$, where $\Phi$ is the pulse phase duration. In this way, simulations can model higher order  QED processes, such as cascades~\cite{Blackburn:RevModPlasma2020}.

\subsection{QED Monte Carlo with \ipstrong}
\label{sim:montecarlo}

The strong field particle physics processes depend chiefly on two dimensionless strong field parameters $\xi,\chi$. The intensity parameter, $\xi$ 
depends only on the properties of the laser pulse and the recoil parameter $\chi$ which depends essentially on the field intensity in the rest frame 
of the oncoming particle.

\begin{figure}[htbp]
\centering\begin{subfigure}[t]{0.5\textwidth}
\centerline{\includegraphics[width=\textwidth]{ComptonRate.pdf}}
\vspace{0.1cm}
\end{subfigure}\begin{subfigure}[t]{.5\textwidth}
\centerline{\includegraphics[width=\textwidth]{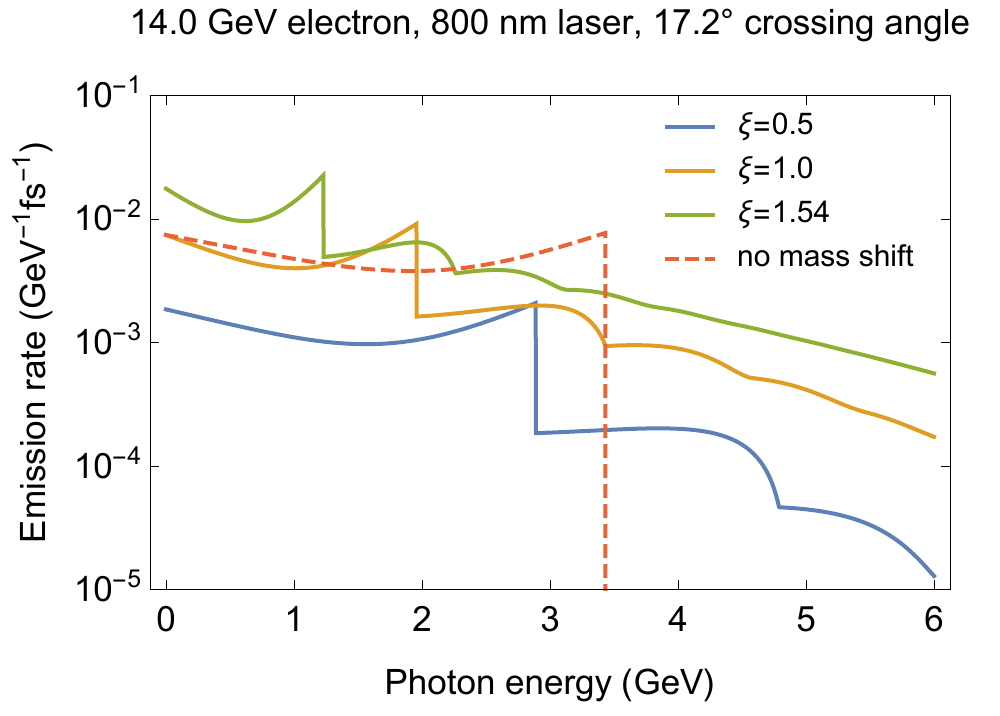}}
\end{subfigure}
\caption{Theoretical calculations of the Compton photon spectrum.}\label{fig:hics}\end{figure}

The radiated photon spectrum of the Compton process shows a series of Compton edges related to multiphoton contributions from the external field. Each 
Compton edge is shifted by an amount proportional to the laser intensity, which can be interpreted as an effective rest mass shift (Fig.~\ref{fig:hics}). 

\begin{figure}[htbp]
\centering\begin{subfigure}[t]{0.5\textwidth}
\centerline{\includegraphics[width=\textwidth]{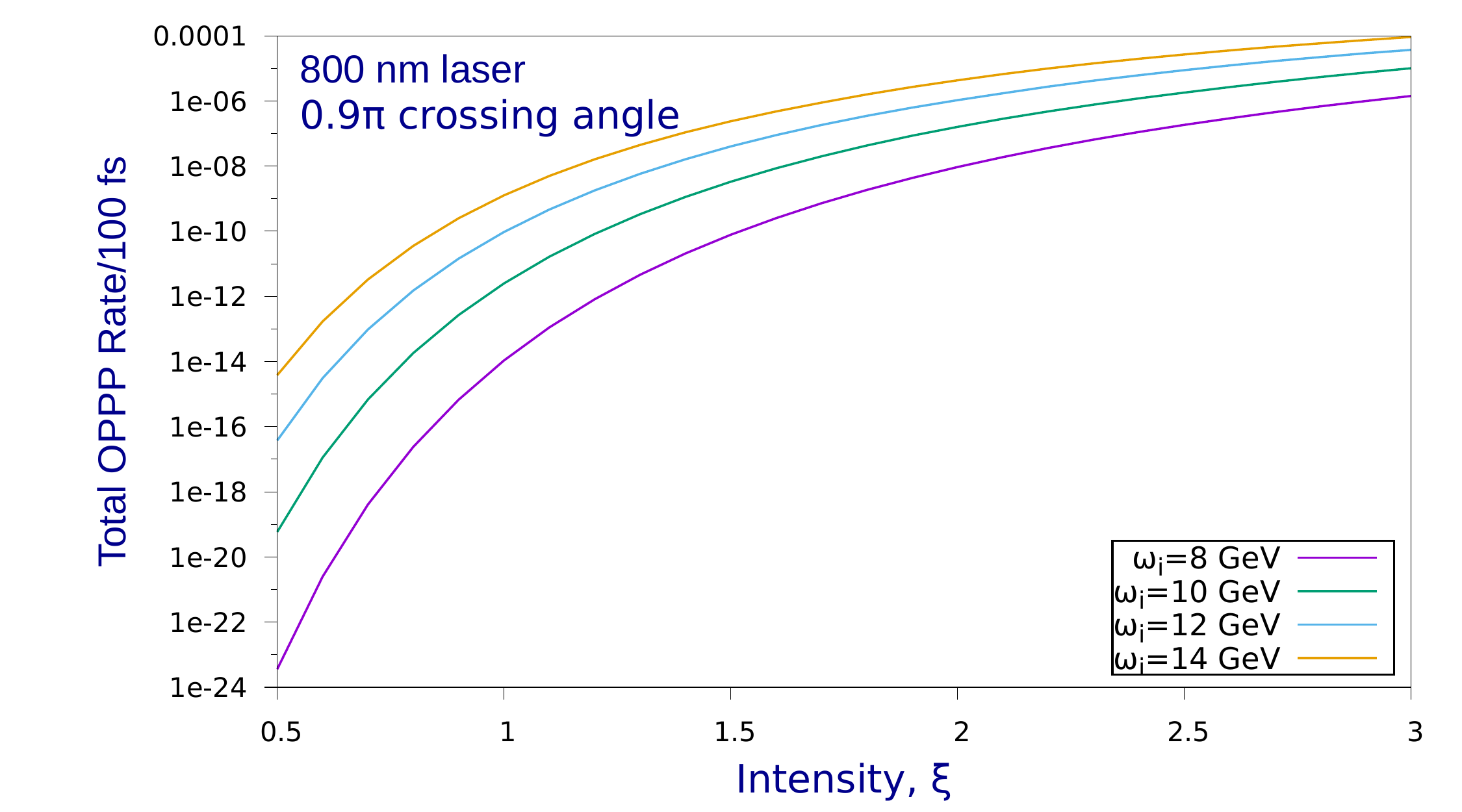}}
\vspace{0.1cm}
\end{subfigure}\begin{subfigure}[t]{.5\textwidth}
\centerline{\includegraphics[width=\textwidth]{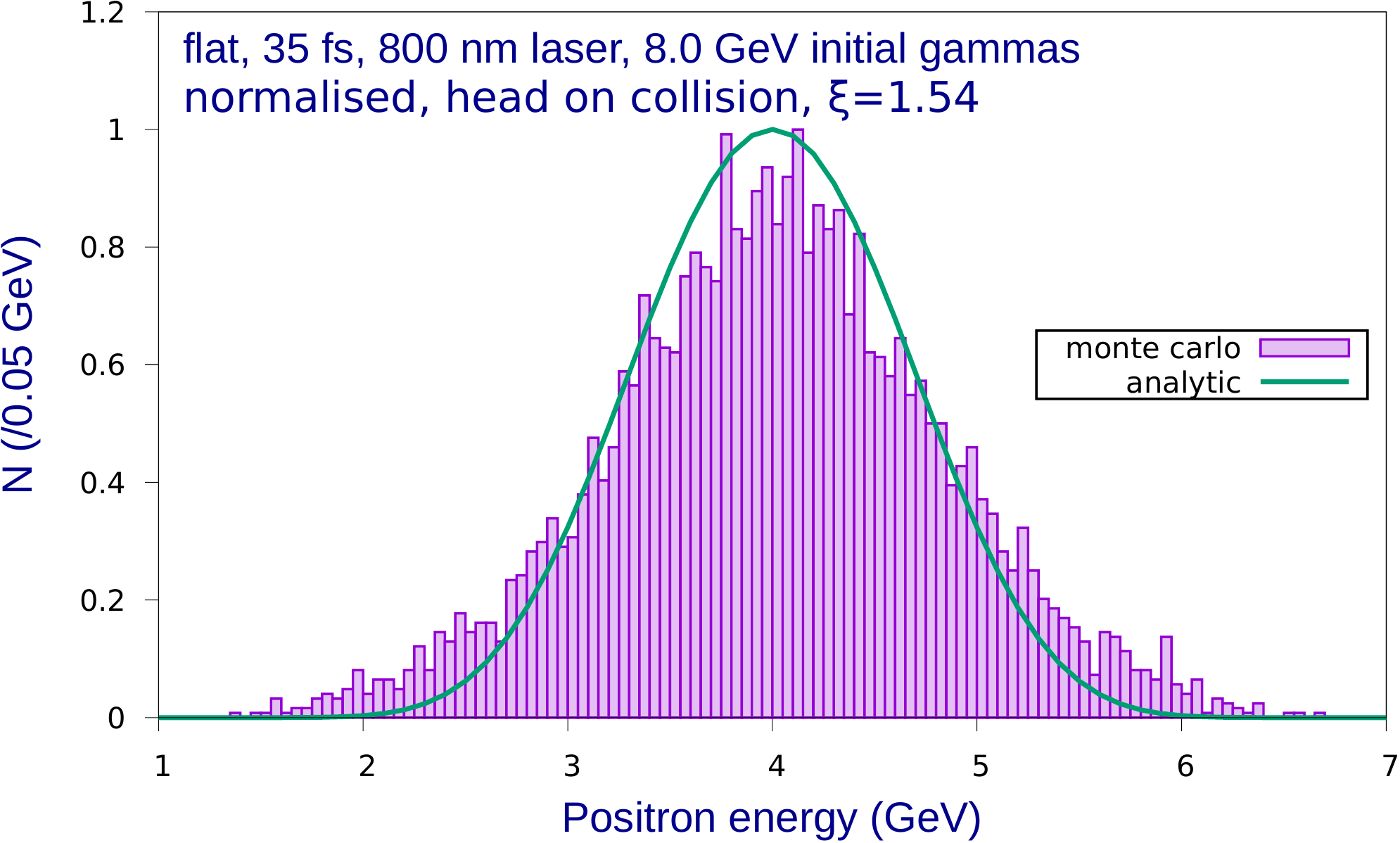}}
\end{subfigure}
\caption{Breit-Wheeler total rate and positron spectrum from \ipstrong.}\label{fig:oppp}\end{figure}

The Breit-Wheeler process will produce positrons which will be detected as precisely as possible. The total transition rate for this process increases with laser 
intensity, though declining below a linear increase as the effective rest mass of the produced positron in the laser pulse increases (see Fig.~\ref{fig:oppp}). 
By fitting an exponential to the total Breit-Wheeler rate it will be possible to make an experimental measurement of the Schwinger critical field~\cite{Hartin:2018sha}. 
The positron spectrum is expected to be peaked at about half the original bremsstrahlung photon energy (see Fig.~\ref{fig:oppp}) and the simulation agrees reasonably well with the analytic expectation.

The final state of all particles is a result of successive interactions of both primary processes combined with the local particle momenta and laser pulse 
intensity where they occur. A review of the theory, relevant processes, references and simulation strategy can be found in~\cite{Hartin:2018egj}.

The LUXE experiment is based on two experimental configurations, with electrons as the primary particles interacting with the laser pulse, and a 
configuration which has bremsstrahlung photons as the primary particles. Different configurations have been considered for the beam parameters.

A custom built strong field QED Monte Carlo computer code, named {\bf \ipstrong}, was used to simulate the strong field interactions for LUXE. 
{\bf \ipstrong} implements the transition rates described in the previous section, namely Eqs.~(\ref{eqn:Rnlma}) and (\ref{eqn:NBWlma}). {\bf \ipstrong} is a FORTRAN2008 code which implements a strong field Monte Carlo event generator, within a particle-in-cell approach that includes the full form of the laser pulse.
Historically, this MC code was used for the development of the LUXE design; however, in the final CDR the results were based on \textsc{PTARMIGAN} as described in the \ref{sec:simulation}.

The task of {\bf \ipstrong} is to obtain an accurate representation of the strong field transition rates within the realistic collision of a charge bunch 
and a laser pulse. First, a time step is chosen on the basis that the process rate probability be not too large (Prob$<0.1$ per random trial). Then, 
the overlap of the interacting beams is divided longitudinally into time slices and then transversely into 3D voxels.

\begin{figure}[htbp]
\centering\begin{subfigure}[t]{0.3\textwidth}
\centerline{\includegraphics[width=\textwidth]{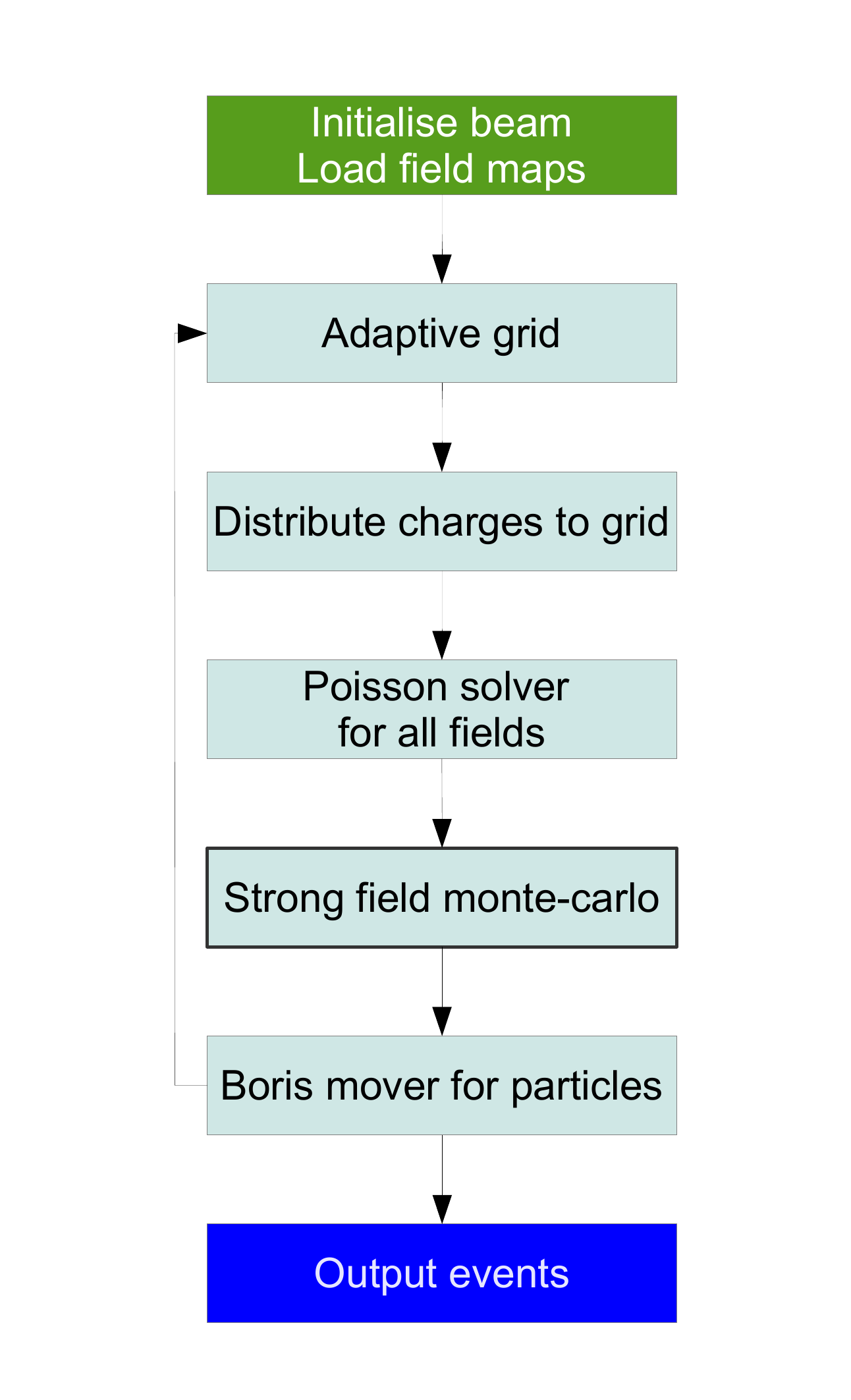}}
\vspace{0.1cm}
\end{subfigure}\begin{subfigure}[t]{0.3\textwidth}
\centerline{\includegraphics[width=\textwidth]{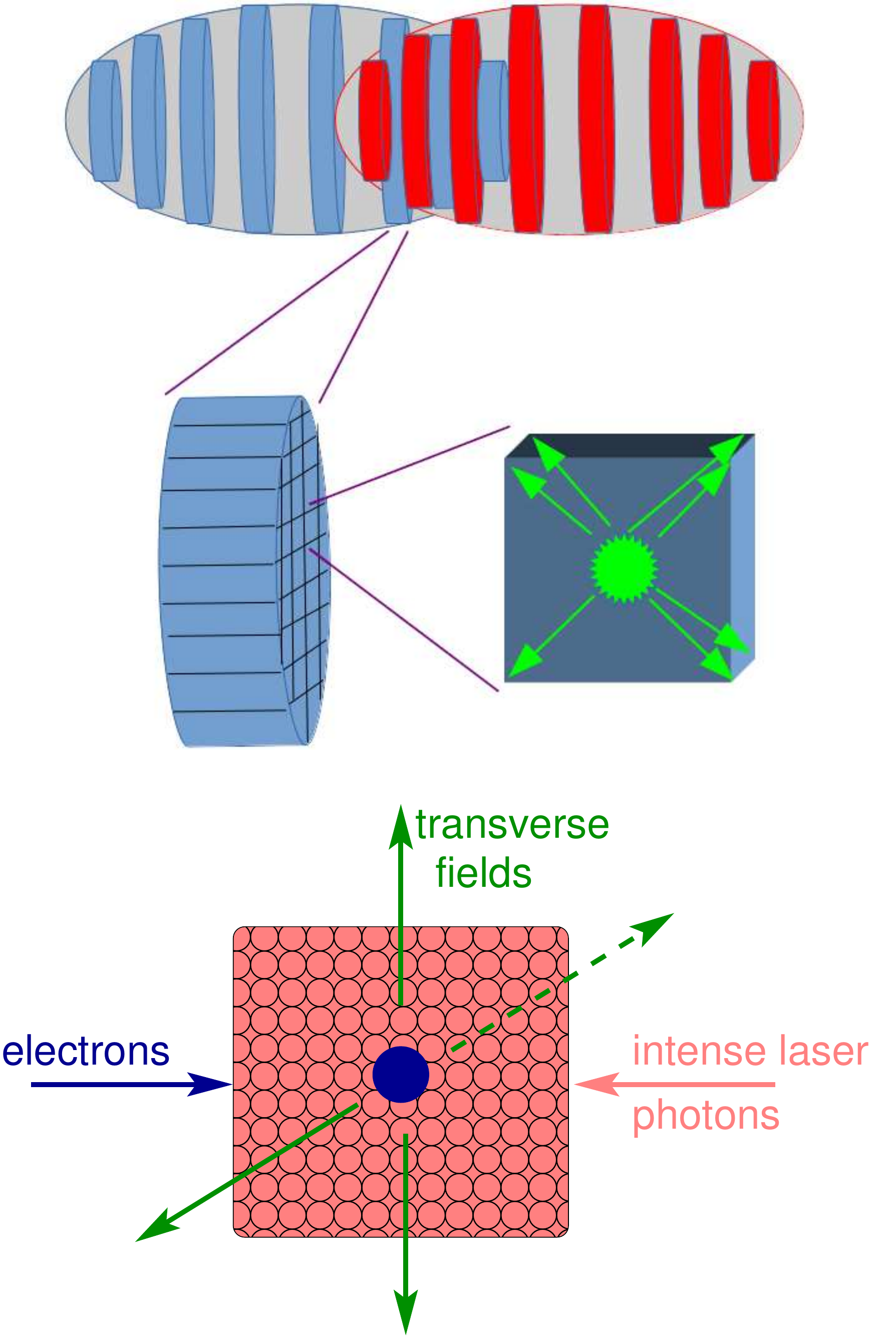}}
\end{subfigure}
\caption{\ipstrong~    program flow and interaction voxels.}\label{fig:ipstrong}\end{figure}

The field strengths are calculated accurately at each macroparticle, which overall represent the distribution of initial states in the interacting beams. 
The field strength at the macroparticle position, combined with the macroparticle 4-momentum, provides the value of the strong field parameters 
$\xi,\chi$ required to calculate transition rates (see Fig.~\ref{fig:ipstrong}). 

For accurate electromagnetic evolution of the interacting beams, {\bf \ipstrong} implements a custom built, Lorentz invariant particle pusher. The 
contribution to the electromagnetic field by charges in any particular voxel is determined by distributing the charge to the voxel vertices and then 
solving the 3D Poisson equation. \ipstrong simulates primary beam bunches, with realistic emittances and energy spread. The starting bunches 
can be generated internally, or loaded externally.


\end{document}